\pdfoutput=1
\documentclass[11pt,twoside,a4paper]{book}

\usepackage{juantxo}

\definecolor{webred}{rgb}{0.75,0,0}

\definecolor{shadecolor}{rgb}{0.4,0.7,0.1}

\makeindex

\makeglossary
\renewcommand{\chaptermark}[1]{\markboth{\textbf{#1}}{}}

\begin{document}

\begin{titlepage}

\begin{textblock*}{297mm}(53mm,200mm)
   \includegraphics[width=11cm]{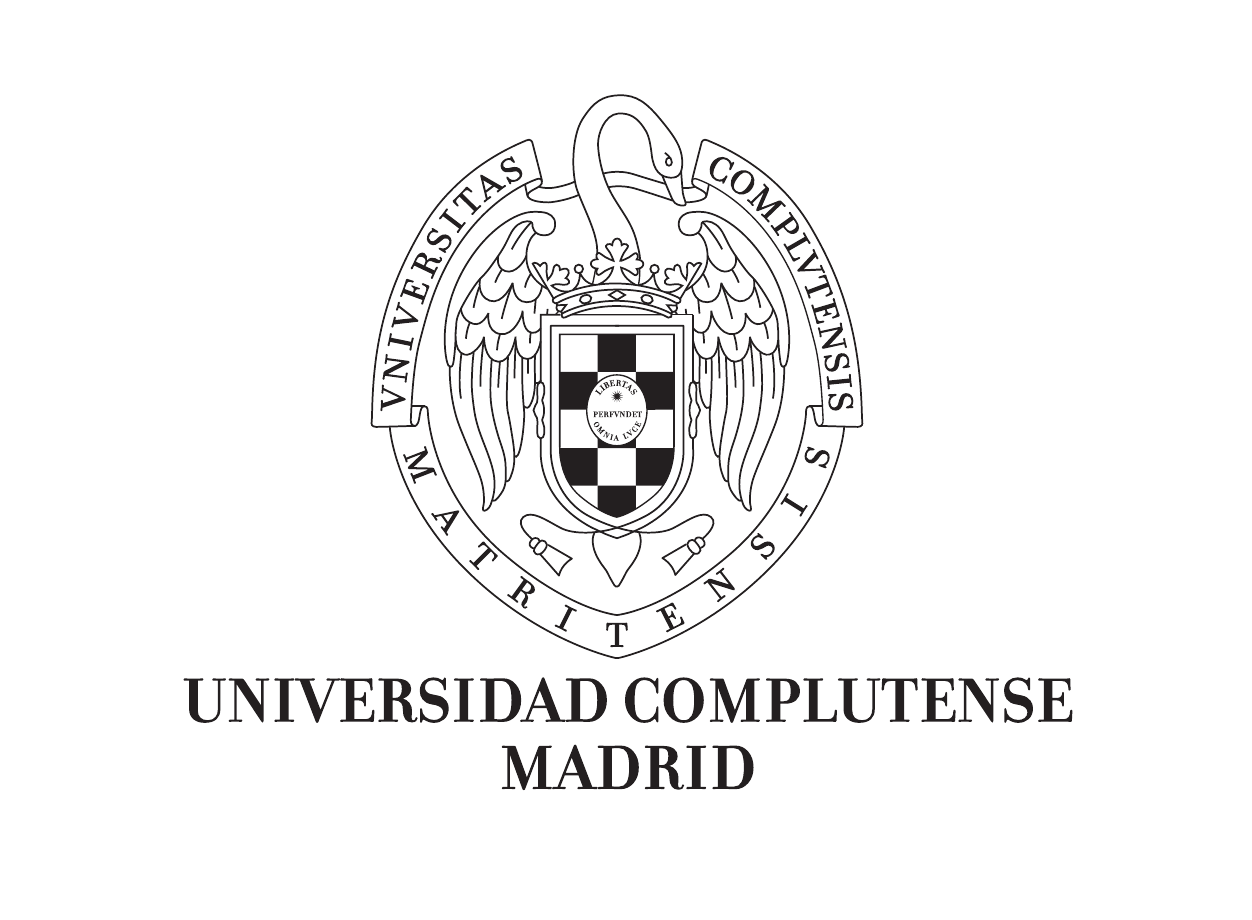}
\end{textblock*}

\vspace*{1cm}

\begin{center}
\Huge \bfseries \textcolor{black}{Hadronic Transport Coefficients \\ from Effective Field Theories }
\end{center}

\vspace{2.3cm}

\begin{center}
 \LARGE \bfseries \textcolor{black}{Juan M. Torres-Rincon}
\end{center}

\vspace{12mm}

\begin{center}
 \Large \bfseries \textcolor{black}{under the supervision of \\ Dr. Antonio Dobado Gonz\'alez and \\ Dr. Felipe J. Llanes-Estrada}
\end{center}

\vspace{10mm}

\begin{center}
 \large \bfseries \textcolor{black}{A dissertation submitted in partial fulfillment of the requirements for the Degree of Doctor of Philosophy in Physics.}
\end{center}

\vspace{7mm}

\begin{center}
\LARGE \textcolor{black}{Facultad de Ciencias F\'isicas} \\ \large
\textcolor{black}{Madrid, April 2012}
\end{center}

\end{titlepage}

\titleformat{\chapter}[frame]
{\color{black}\normalfont\huge\bfseries}{\chaptertitlename \ \thechapter}{20pt}{\huge\filcenter}  \

\titleformat{\section}
{\color{black}\normalfont\Large\bfseries}{\thesection}{1em}{}

\titleformat{\subsection}
{\color{black}\normalfont\large\bfseries}{\thesubsection}{1em}{}

\thispagestyle{empty}
\frontmatter

\chapter{Acknowledgements}

  I would like to dedicate a few lines in order to express my gratitude to all those people
who have contributed in some way to complete this dissertation during the last years.

  To begin with, I would like to thank my supervisors Dr. Antonio Dobado and Dr. Felipe J. Llanes-Estrada,
who not only gave me the guidance and support throughout this thesis, but also their critical thinking, 
physical intuition and experience which will accompany me in my future.

  My feelings of gratitude extend to all the staff of the department ``F\'isica Te\'orica I'',
who provided me a nice workplace and the needed tools to properly conclude this work. 
I am also grateful for the economical support of the FPU program from the Spanish Ministry of
Education. I would like to thank Nastassja Herbst for a merciless correction of errata in this manuscript.
Hopefully, she will have understood the concept of a ``relativistic heavy ion collision''.

  I am indebted to many people with whom I have discussed about physics, with the help of
whom I have increased my personal background in the field and shaped the content of this thesis.
Among all of them, I want to thank Luciano Abreu, Anton Andronic, Daniel Cabrera, Dany Davesne,
Daniel Fern\'andez Fraile, \'Angel G\'omez Nicola, Pedro Ladr\'on de Guevara, Eiji Nakano, Jos\'e Ram\'on Pel\'aez
and the researchers working at the GSI Helmholtzzentrum f\"ur Schwerionenforschung (Darmstadt) during my short research stay there in 2010:
Vincent Pangon, Vladimir Skokov and especially to Profs. Bengt Friman and Jochen Wambach.

  Not only the exchange of ideas is needed to complete such a research project but also a pleasant working atmosphere.
For this reason, I also want to thank all my colleagues, comrades and friends in the 
Physics Department with whom I shared a lot of working days but also many hours of amusement and
entertainment.

  Finally, I wish to end these lines by heartily acknowledge my sister Judit and my parents Juan Miguel
and Soledad for their support and for giving me the opportunity to reach this stage of my professional life.

\chapter{List of publications}

The following articles have been published in the context of this dissertation:

\begin{itemize}
 \item \textbf{$\eta/s$ and phase transitions.} A. Dobado, F.J. Llanes-Estrada and J.M. Torres-Rincon. {\it Phys.Rev.D} 79,014002 (2009)
 \item \textbf{Minimum of $\eta/s$ and the phase transition of the linear sigma model in the large-$N$ limit.} A. Dobado, F.J. Llanes-Estrada and J.M. Torres-Rincon. {\it Phys.Rev.D} 80,114015 (2010)
 \item \textbf{Heavy quark fluorescence.} F.J. Llanes-Estrada and J.M. Torres-Rincon, {\it Phys.Rev.Lett.} 105,022003 (2010)
 \item \textbf{Bulk viscosity and energy-momentum correlations in high energy hadron collisions.} A. Dobado, F.J. Llanes-Estrada and J.M. Torres-Rincon. Accepted in {\it Eur.Phys.J.C}
 \item \textbf{Bulk viscosity of low temperature strongly interacting matter.} A. Dobado, F.J. Llanes-Estrada and J.M. Torres-Rincon. {\it Phys.Lett.B} 702,43 (2011)
 \item \textbf{Charm diffusion in a pion gas implementing unitarity, chiral and heavy quark symmetries.} L. Abreu, D. Cabrera, F.J. Llanes-Estrada and J.M. Torres-Rincon. {\it Ann. Phys. 326}, 2737 (2011)
\end{itemize}

The author has also contributed to the following conference proceedings:

\begin{itemize}
 \item \textbf{Heat conductivity of a pion gas.} A. Dobado, F.J. Llanes-Estrada and J.M. Torres Rincon. Proceedings of QNP2006. Berlin. Springer. (2007)
 \item \textbf{The Status of the KSS bound and its possible violations: How perfect can  a fluid be?} A. Dobado, F.J. Llanes-Estrada and J.M. Torres Rincon. {\it AIP Conf. Proc.} 1031,221-231 (2008)
 \item \textbf{$\eta/s$ is critical (at phase transitions).} A. Dobado, F.J. Llanes-Estrada and J.M. Torres-Rincon. {\it AIP Conf. Proc.} 1116,421-423 (2009)
 \item \textbf{Brief introduction to viscosity in hadron physics.} A. Dobado, F.J. Llanes-Estrada and J.M. Torres-Rincon. {\it AIP Conf. Proc.} 1322,11-18 (2010)
 \item \textbf{Viscosity near phase transitions.} A. Dobado, F.J. Llanes-Estrada and J.M. Torres-Rincon. Gribov-80 Memorial Volume. World Scientific (2011)
 \item \textbf{Bulk viscosity of a pion gas.} A. Dobado and J.M. Torres-Rincon. {\it AIP Conf.Proc.} 1343, 620 (2011)
 \item \textbf{Franck-Condon principle applied to heavy quarkonium.} J.M. Torres-Rincon. {\it AIP Conf.Proc.} 1343, 633 (2011)
 \item \textbf{Coulomb gauge and the excited hadron spectrum.} F.Llanes-Estrada, S.R. Cotanch, T. van Cauteren, J.M. Torres-Rincon, P. Bicudo and M. Cardoso. {\it Fizika B} 20, 63-74 (2011)
 \item \textbf{Transport coefficients of a unitarized pion gas.} J.M Torres-Rincon. To appear in {\it Prog. Part. Nucl. Phys.} (2012)
\end{itemize}

\tableofcontents

\newpage
\mbox{}
\thispagestyle{empty} 

\newpage
\begin{flushright}
 \vspace*{4in}
{\it Es ist nicht genug zu wissen, man mu\ss auch anwenden. \\
 Es ist nicht genug zu wollen, man mu\ss auch tun.} \\
J.W. von Goethe (Wilhelm Meisters Wanderjahre)
 \end{flushright}
\thispagestyle{empty} 

\newpage
\mbox{}
\thispagestyle{empty} 

\mainmatter

\chapter{Relativistic Heavy Ion Collisions\label{ch:1.intro}}

   In this chapter we will motivate the study of transport coefficients in hadronic matter, their relevance in relativistic heavy-ion 
collisions and its importance within theoretical and experimental high energy physics. Afterwards, we will define the fundamental properties and variables that characterize a relativistic heavy-ion collision.
We will discuss a typical heavy-ion collision taking place at the Relativistic Heavy Ion Collider (RHIC) \glossary{name=RHIC,description={Relativistic Heavy Ion Collider}}
or at the Large Hadron Collider (LHC)\glossary{name=LHC,description={Large Hadron Collider}}. We start by describing the collision variables and the observables 
relevant to extract the transport coefficients.

\section{Introduction}

    A large number of studies in high energy physics focus on the analysis of the ultimate components of matter and how they interact. Quantum chromodynamics describes rather well the interactions
between quarks and gluons. Also the hadronic degrees of freedom are expected to be well described by QCD, although a rigorous analytical way for obtaining the hadronic formulation from the QCD Lagrangian is
still lacking. In terms of quarks and gluons or in terms of hadrons, collective phenomena of these components have attracted a great interest in the field. One of the reasons is
that the details of the collective behaviour eventually describe the phase diagram of QCD. Quarks and gluons lie on the high temperature and density part of the phase diagram. In the opposite limit,
hadrons are the relevant degrees of freedom. Some other possible phases may exist but we are not concerned about them in this dissertation. 
   
     The phase diagram has been widely studied and its aspect is well established within the physics community. However, from the experimental point of view, very little is known about it.
The experimental way of accessing the properties of the QCD phase diagram is through relativistic heavy-ion collisions, where a couple of nuclei are boosted at relativistic velocities,
and collided in order to break their internal structure finally producing a large number of products. 
The experimental effort to produce such a collision is impressive. One should take into account that the size of the incoming nuclei is of the order of Fermis, an so is the typical reaction time.
Moreover, the final state of the collision is composed by hundreds (or even thousands) of particles from which we can only measure their energy and momenta. The work to provide a single conclusion
about the hadron dynamics from this information is immense. 

   The entire phase diagram is of physical interest: the phase boundary, the critical end-point or the new exotic phases. In particular, the zone at zero baryonic chemical potential is of great relevance.
It is accepted that the early structure of the Universe has cooled down through this zone, after the Big Bang explosion. The recreation of this stage at the experimental facilities is of huge interest in order
to get some information about the primordial structure of the Universe and its composition. 

   It is not difficult to accept that non-equilibrium phenomena play an important role in such heavy-ion collisions. During the fireball expansion there exist both chemical and thermal non-equilibrium.
Pressure, temperature and momentum gradients are also present from the very beginning of the collision. That makes the non-equilibrium physics a decisive tool in order to understand
the processes appearing in the fireball expansion. The presence of these gradients, together with the existence of conserved quantities in the medium imply the manifestation of the transport coefficients, that control the
relaxation process towards equilibrium.

The interest of this dissertation, is to theoretically access to these transport coefficients in order to gain more insight of the non-equilibrium properties of the hadronic medium created in these relativistic heavy-ion collisions.

\section{Variables of a heavy-ion collision}

  We will review the fundamental variables used to characterize a relativistic heavy-ion collision.
These variables include those belonging to the initial state, namely the two colliding protons or nuclei
and the kinematic variables related to the final particle yield that is detected in an experimental facility.

\subsection{Initial state variables\label{sec:kin_var}}

Consider one nucleus colliding with another in the laboratory frame. Each projectile can be as simple as a single proton or as complex as a gold or lead nucleus.
The relevant variables that define the initial stage of the collision are:
\begin{itemize}
 \item The mass number $A$ \index{mass number}or the number of nucleons inside the nucleus. For proton-proton (p+p) collision
$A=1$ but it can be as large as $A=208$ for a lead-lead (Pb+Pb) collision at the LHC.
  \item  The CM\glossary{name=CM,description={center of mass}} energy of the collision $\sqrt{s}$. This CM energy
can be given per nucleon or as the total energy of the collision. For p+p collisions at the LHC this energy 
has been risen up in the early stages of the facility from the value $\sqrt{s}=0.9$ TeV up to the higher energy of $\sqrt{s}=7$ TeV. For Pb+Pb collisions the typical CM energy per nucleon at LHC is $\sqrt{s_{NN}}=2.76$ TeV.
  \item $z$-axis. Defined by the collision axis. At the moment of the impact, the momenta of the two nuclei are
oriented with along this collision axis.
   \item Impact parameter $\mathbf{b}$. Defined \index{impact parameter} as the vector pointing from the geometrical center of one nucleus to the center of the other at the moment of collision.
The direction of $\mathbf{b}$ defines the $x-$axis of the collision. The modulus of the impact parameter vector can go from zero (central collision) to twice the nuclear radius (ultraperipheral collision).
The nuclear radius is given by the simple formula $R=1.2 A^{1/3}$ fm.
   \item Reaction plane\index{reaction plane}: The plane generated by the collision axis and $\mathbf{b}$, i.e. the $OXZ$ plane. 
The impact parameter and the reaction plane are not known {\it a priori} and they vary from event to event. 
\end{itemize}

\begin{figure}[t]
\begin{center}
\includegraphics[scale=0.35]{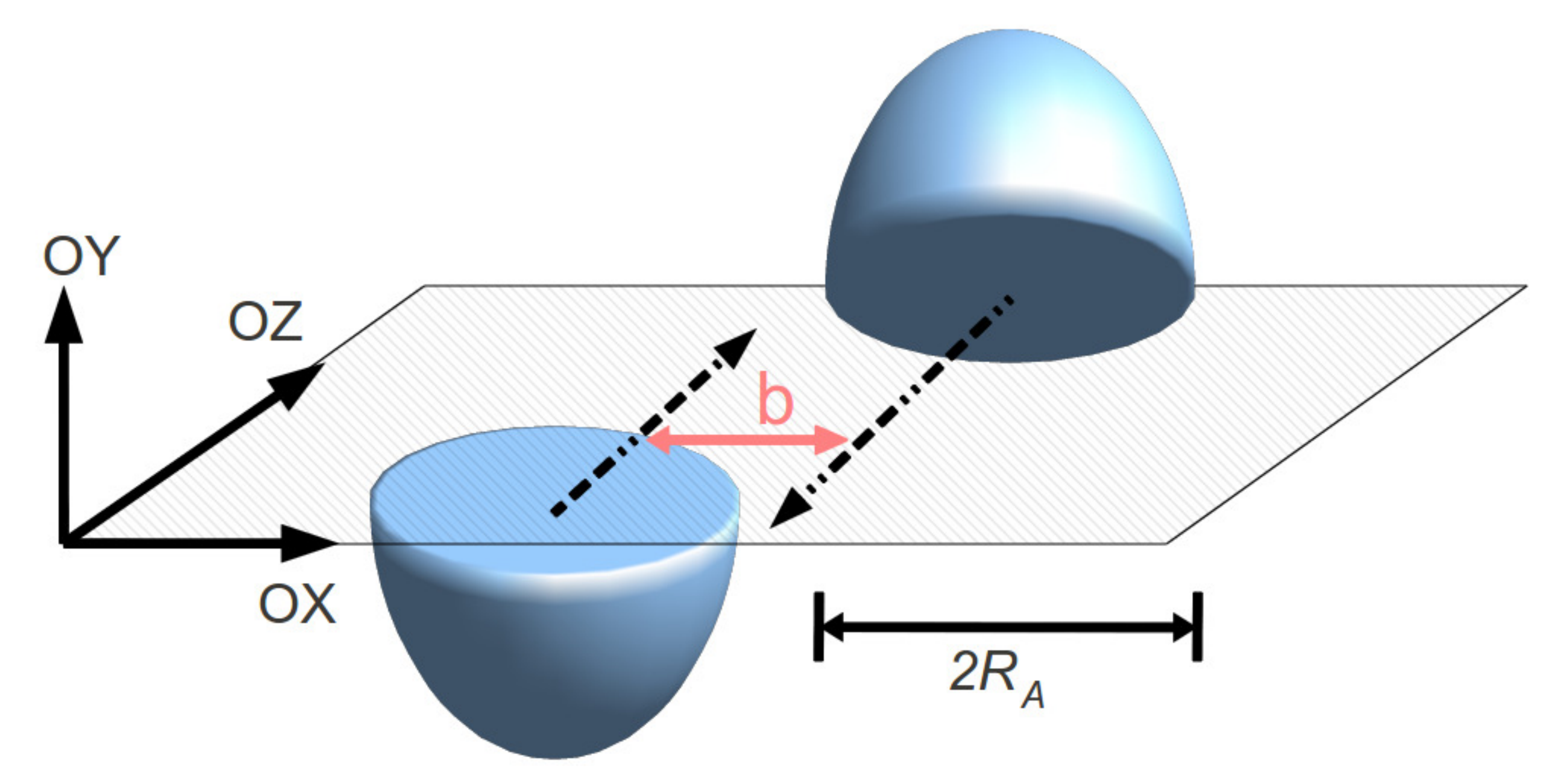}
\caption{\label{fig:in_var} Position of the collision axis\index{collision axis} (OZ), the impact parameter\index{impact parameter} ($b$) and the reaction plane (generated by the OX and OZ axes) of a nucleus-nucleus collision.}
\end{center}
\end{figure}

A schematical view of the geometrical variables that we have defined is shown in Fig.~\ref{fig:in_var}.

\subsection{Variables in the final state}

Due to the huge amount of degrees of freedom (thousands of detected particles), the number of variables in the final state are larger than those for the initial state.
We start by distinguishing between the kinematic variables of a fluid element (or fluid cell)
and variables corresponding to a given particle in the fluid element \cite{Heinz:2004qz}. We will use velocity variables \index{velocity!fluid element}
$u^0,u^i$ (see Appendix \ref{app:hydro} for the definitions) for those characterizing the fluid element, and momentum variables $\mb{p}$, and energy $p_0$ for those that belong to a single particle.
In case the momentum or the energy of the fluid element is needed we will denote them with capital letters, $\mb{P}$ and $P^0$, respectively.

The fluid we are analyzing is a nuclear fireball expanding from the collision point. Taking a space-time point $x$, we separate its components into the direction perpendicular to the
collision beam $\mb{x}_{\perp}=(x,y)$ and the $z$ axis along the collision beam:
\be x \equiv x^{\mu}=(t,\mb{x}_{\perp},z) \ . \ee

We consider the infinitesimal volume of the fluid element centered at $x$, composed of a swarm of particles. The total energy of the fluid element $P^0(x)$ is the sum of all
the individual energies of the particles contained in this volume. The same prescription is also applied to the total three-momentum $\mb{P}(x)$. Using these two variables one can construct the velocity of the fluid
element as $\mb{v}(x)=\mb{P}(x)/P^0(x)$. The velocity field can be decomposed into the transverse plane of the collision $v_{\perp}(x)$ (transverse flow) and in a component along the collision axis $v_z(x)$ (longitudinal flow).

The four-velocity field of the fluid element is constructed from $\mb{v}(x)$ as:
\be u^{\mu}(x) = \gamma(x) \ (1, \mb{v}(x)) \ , \ee
with $\gamma(x)=1/\sqrt{1-\mb{v}(x)^2}$.

A relativistic particle of mass $m$ escapes from the heavy-ion collision with momentum $\mathbf{p}$ and on-shell energy $E_p=\sqrt{m^2+p^2}$.
The momentum of the particle has the same decomposition in the transverse plane and along the collision axis.
The former is called transverse momentum \index{tranverse momentum} $\mathbf{p}_{\perp}$ (sometimes denoted in the literature as $\mathbf{p}_{T}$) and the later is the
longitudinal momentum $p_z$. The angle between these two components is the polar angle $\theta$ and it can be directly measured in the experiment.
The angle between $\mb{p}_{\perp}$ and the $x-$axis is called the azimuthal \index{azimuthal angle} angle $\phi$.

In practice, in order to describe the longitudinal boost of the particle, the $p_z$ variable is inconvenient because it transforms non linearly
under a Lorentz transformation. For this reason, a new variable called rapidity\footnote{This variable should be called ``longitudinal rapidity'' but we follow common usage as there will be no ambiguity.} \index{rapidity!particle} $y_p$ is introduced:
\be y_p = \frac{1}{2} \log \left( \frac{E_p+p_z}{E_p-p_z} \right) \ . \ee
The inverse transformation reads:
\be \label{eq:rapidityinverse} E_p = m_{\perp} \cosh y_p, \quad p_z= m_{\perp} \sinh y_p \ , \ee
where $m_{\perp}$ is the tranverse ``mass''
\be \label{eq:trans_mass} m_{\perp} = \sqrt{m^2 + p^2_{\perp}} \ . \ee

Analogous formulae can be defined for the fluid element's rapidity $Y$\index{rapidity!fluid cell}:
\be Y = \frac{1}{2} \log \left( \frac{1+v_z}{1-v_z} \right) \ , \ee
and the inverse relation to obtain the longitudinal velocity of the fluid element:
\be v_z = \tanh Y \ . \ee

In the nonrelativistic limit velocities and rapidities coincide, $v_z \rightarrow Y$.
However in the ultrarelativistic limit $v_z \rightarrow 1$, while the rapidities go to infinity.

The transformation law of the rapidity under a Lorentz boost is simply additive:
\be y_{LAB} = y_{CM} + Y \ , \ee
where $y_{LAB}$ is the rapidity seen in the laboratory frame and $y_{CM}$ is the rapidity seen in the CM of the fluid cell.

Note that the four-velocity of the fluid element admits the following parametrization:
\be u^{\mu} = \gamma_{\perp} (\cosh Y, \mathbf{v}_{\perp}, \sinh Y) \ , \ee
where 
\be \gamma_{\perp}\equiv\frac{1}{\sqrt{1-v^2_{\perp}}} \ .\ee
It is evident that the relativistic normalization holds $u_{\mu} u^{\mu}=1$.

The four-momentum of a particle can be analogously written as
\be p^{\mu}=(m_{\perp} \cosh y_p, \mb{p}_{\perp}, m_{\perp} \sinh y_p) \ , \ee
with the standard relativistic normalization $p_{\mu} p^{\mu}=m^2$.

The use of the rapidity variable requires to have the particle well identified because it explicitly depends on its mass. However, the identification of a
 particle is always done {\it a posteriori}, after all the kinematic variables have been extracted from the collision. Therefore it is desirable to use a
new variable that only contains geometrical information, without knowing the nature of the particle. For this reason, one often introduces
the concept of pseudorapidity \index{pseudorapidity!particle}$\eta_p$. It is defined as
\be \label{eq:pseudorapidity} \eta_p = \frac{1}{2} \log \left( \frac{p+p_z}{p-p_z} \right) \ , \ee
with $p=\sqrt{\mathbf{p}^2_{\perp} + p_z^2}$. The inverse transformation reads
\be p = p_{\perp} \cosh \eta_p, \quad p_z = p_{\perp} \sinh \eta_p \ . \ee

This variable can be extracted from the particle track geometry without any information of its mass because the particle pseudorapidity can be related
with the polar emission angle as
\be \eta_p =\frac{1}{2} \log \left( \frac{1+\cos \theta}{1-\cos \theta} \right) = \log \left( \cot \frac{\theta}{2} \right) \ . \ee

It is easy to check that in the ultrarelativistic limit the rapidity and pseudorapidity coincide. The general relation between these two variables is \cite{letessier2002hadrons}:
\be \eta_p = \frac{1}{2} \log \left( \frac{\sqrt{m^2_{\perp} \cosh^2 y_p -m^2} + m_{\perp} \sinh y_p}{\sqrt{m^2_{\perp} \cosh^2 y_p - m^2} - m_{\perp} \sinh y_p}\right) \ . \ee

The same set of equations can be derived for the pseudorapidity of the fluid cell\index{pseudorapidity!fluid cell}, that we will denote by $\eta$. Table~\ref{tab:kin_var} summarizes the notation for all the
defined kinematic variables.

\begin{table}
 \begin{center}
\begin{tabular}[ht]{|c||c|c|}
\hline
Variable & Particle & Fluid element \\
\hline
Momentum & $\mb{p}$ & $\mb{P}(x)$ \\
Energy & $E_p$ & $P^0(x)$ \\
Velocity & $\mb{p}/E_p$ & $\mb{v}$ \\
Rapidity & $y_p$ & $Y$ \\
Pseudorapidity & $\eta_p$ & $\eta$ \\
\hline
\end{tabular}
\end{center}
\caption{Notation used in this dissertation for the relevant kinematical variables of a relativistic particle in a heavy-ion collision and for a fluid cell belonging to the expanding fireball.\label{tab:kin_var}}
\end{table}


\section{Multiplicity distribution \label{sec:multiplicity}}

We will introduce the concept of centrality in a heavy-ion collision and its relation with the final multiplicity.

\subsection{Centrality classes}

The centrality \index{centrality} is a key concept in a heavy-ion collision. At the moment of collision, the velocity vectors of the two incoming nuclei
are antiparallel but slightly displaced. The centers of the nuclei are separated by a finite quantity given by the modulus
of the impact parameter, $b$\index{impact parameter}. An ideal head-on collision would have $b=0$, but there exist a whole distribution of events from the central collisions
to the most peripheral ones, where the impact parameter is $b \simeq 2 R_A$.

The most central collisions have a larger number of participating nucleons than the most peripheral ones. In the latter there exists a number of nucleons which do not contribute
to the collision (they are spectator nucleons). In general, the larger the number of spectators in the initial state, the lesser number of individual binary collisions and the lesser
number of produced particles in the final state. Therefore, it is assumed that those events with higher multiplicity are more central than those collisions with a small final multiplicity.

The way for characterizing the centrality of the event is to collect all the measured charged particles that arrive to the detector in a single event. Then, this process is repeated for
all the recorded events for which one obtains a different charged multiplicity depending on the centrality. Afterwards, a histogram is elaborated with the number of events as a function
of the total charged multiplicity (that is mainly composed of charged pions). After that, one sets different bins in the histogram that contains similar
multiplicity (usually measured as a percentage). For instance, one divides the histograms in ten bins, each one containing 10 \% of the charged particles around a fixed multiplicity.

\begin{figure}[t]
\begin{center}
\includegraphics[scale=0.35]{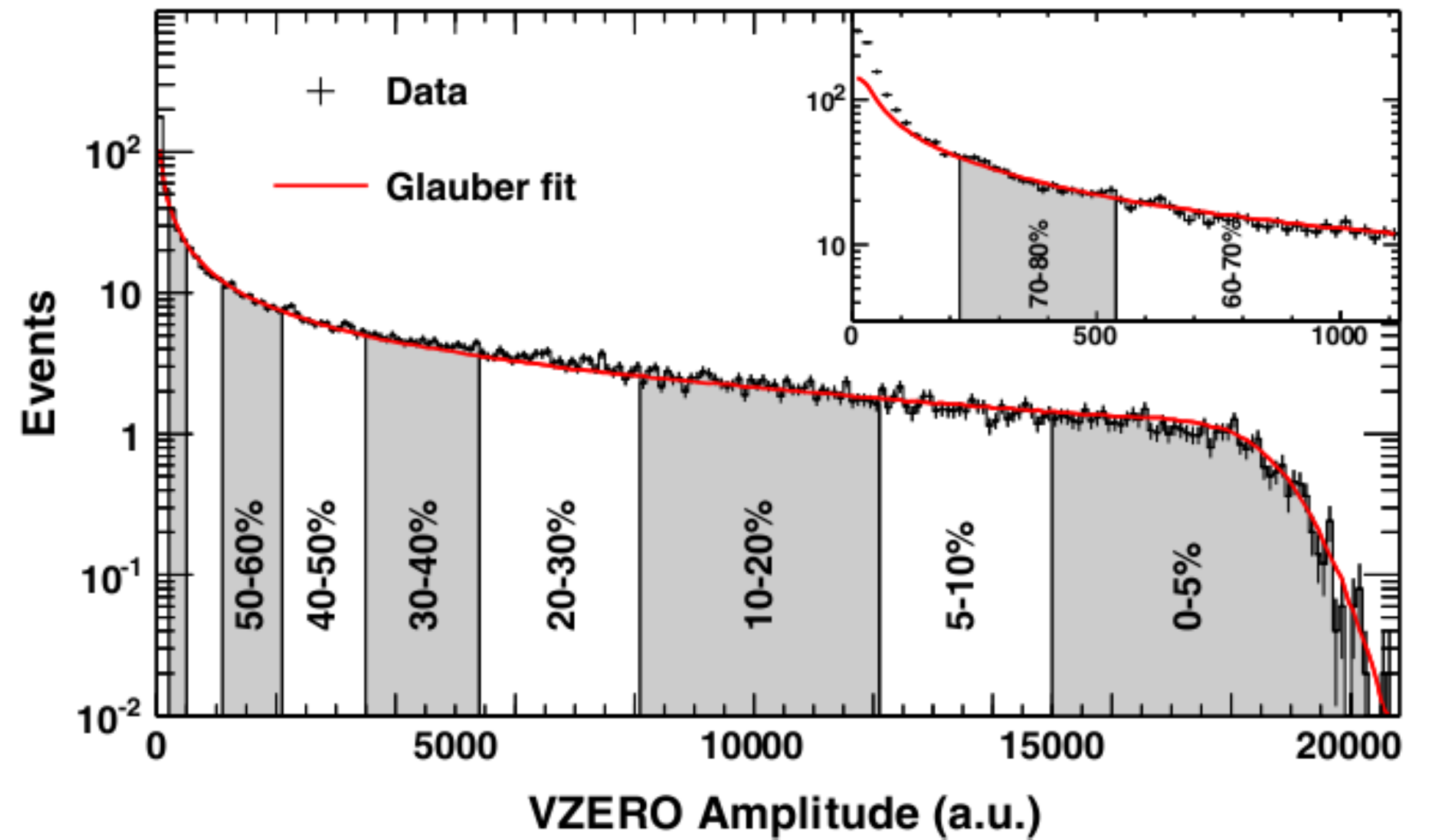}
\caption{\label{fig:centrality} Histogram of total charged multiplicity for 65000 events divided in centrality bins at $\sqrt{s_{NN}}=2.76$ TeV in ALICE. The $x-$axis is proportional to the total
charged multiplicity measured by the detector. Plot taken from \cite{Aamodt:2010cz}. Copyright 2011 by the American Physical Society. }
\end{center}
\end{figure}

A typical multiplicity distribution is reproduced in Fig.~\ref{fig:centrality}. The data \cite{Aamodt:2010cz} is taken from the ALICE collaboration with a total number of 65000 events of lead-lead collisions 
at $\sqrt{s_{NN}}=2.76$ TeV. Because the number of total events decreases with multiplicity, i.e. central collisions are scarce, the centrality bins become wider as the multiplicity increases.

Assuming that the total cross section is $\sigma_{tot} \sim \pi (2R_A)^2$ one can deduce a handy formula relating the centrality bin and the impact parameter \cite{teaney2010viscous}:
\be 100 \times \left( \frac{b}{2R_A} \right)^2 \simeq \% \textrm{ centrality} \ . \ee

\subsection{Initial conditions: Glauber theory}

The Glauber model \index{Glauber model} tries to describe the initial nucleon density profile by means of simple geometrical arguments.
The nucleon distribution inside the nuclei is characterized by the nuclear density distribution\index{nuclear density distribution} $\rho_A (\mb{x})$.
The Woods-Saxon potential\index{Woods-Saxon potential} shape can serve as a good choice to parametrize the nuclear density distribution.
\be \label{eq:nuc_den} \rho_A (\mb{x}) = \frac{\rho_0}{1+\exp \left( \frac{|\mb{x}|-1.2 A^{1/3} \textrm{ fm}}{c} \right)} \ , \ee
where $\rho_0$ is an overall normalization constant, and $c$ is the skin thickness.
Integrating $\rho_A (\mb{x})$ along the $z-$axis (this direction is not relevant due to the Lorentz contraction in the collision axis) one obtains the nuclear thickness function\index{nuclear thickness function}:
\be T_A ( \mb{x_{\perp}}) = \int_{-\infty}^{\infty} dz \ \rho_A (\mb{x}) \ , \ee
where the parameter $\rho_0$ in Eq.~(\ref{eq:nuc_den}) needs to be normalized by the mass number
\be A=\int d^2 \mb{x}_{\perp} \ T_A (\mb{x_{\perp}}) \ . \ee

Consider two incoming nuclei with $A$ nucleons each and an impact parameter\index{impact parameter} $\mb{b}$. The number of participating nucleons per unit area in the collision $n_{part}$ is given by the Glauber model and reads\cite{Luzum:2008cw},\cite{teaney2010viscous}:
\begin{eqnarray}
\nonumber \label{eq:glauber1} n_{part} (\mb{x}_{\perp},\mb{b}) & = & T_A (\mb{x}_{\perp} + \mb{b}/2) \  \left[ 1 - \left( 1- \frac{\sigma_{in} T_A (\mb{x}_{\perp} - \mb{b}/2)}{A}\right)^A \right] \\
 & & + \ T_A (\mb{x}_{\perp} - \mb{b}/2) \ \left[ 1 - \left( 1- \frac{\sigma_{in} T_A (\mb{x}_{\perp} + \mb{b}/2)}{A}\right)^A \right] \ , \end{eqnarray}
that can be well approximated (when $A \gg 1$) to 
\begin{eqnarray}
   \nonumber n_{part} (\mb{x}_{\perp},\mb{b}) & \simeq & T_A (\mb{x}_{\perp} + \mb{b}/2) \left[ 1 - e^{-\sigma_{in} T_A (\mb{x}_{\perp} - \mb{b}/2)} \right] \\
 & & + T_A (\mb{x}_{\perp} - \mb{b}/2) \left[ 1 - e^{ - \sigma_{in} T_A (\mb{x}_{\perp} + \mb{b}/2)}\right] \ , 
\end{eqnarray}
where $\sigma_{in}$ is the inelastic nucleon-nucleon cross section ($\sigma_{in}=42$ mb for Au+Au at $\sqrt{s_{NN}}=200$ GeV \cite{Luzum:2008cw} and $\sigma_{in}=64$ mb for Pb+Pb at $\sqrt{s_{NN}}=2.76$ TeV \cite{Aamodt:2010jd}).

The total number of participating \index{number of participants} nucleons as a function of the impact parameter is
\be N_{part} (b) = \int d\mb{x}_{\perp} \ n_{part} (\mb{x}_{\perp},b) \ , \ee
and this number is used to quantitatively characterize the centrality class of the collision. A Glauber fit to the experimental data corresponds to the red line of Fig.~\ref{fig:centrality}.

The number of binary collisions per unit area \index{number of binary collisions} $n_{coll}$ is

\be \label{eq:glauber2} n_{coll} (\mb{x}_{\perp},b) = \sigma_{in} \ T_A (\mb{x}_{\perp} + \mb{b}/2) \ T_A(\mb{x}_{\perp} - \mb{b}/2) \ .\ee

\begin{figure}[t]
\begin{center}
\includegraphics[scale=0.35]{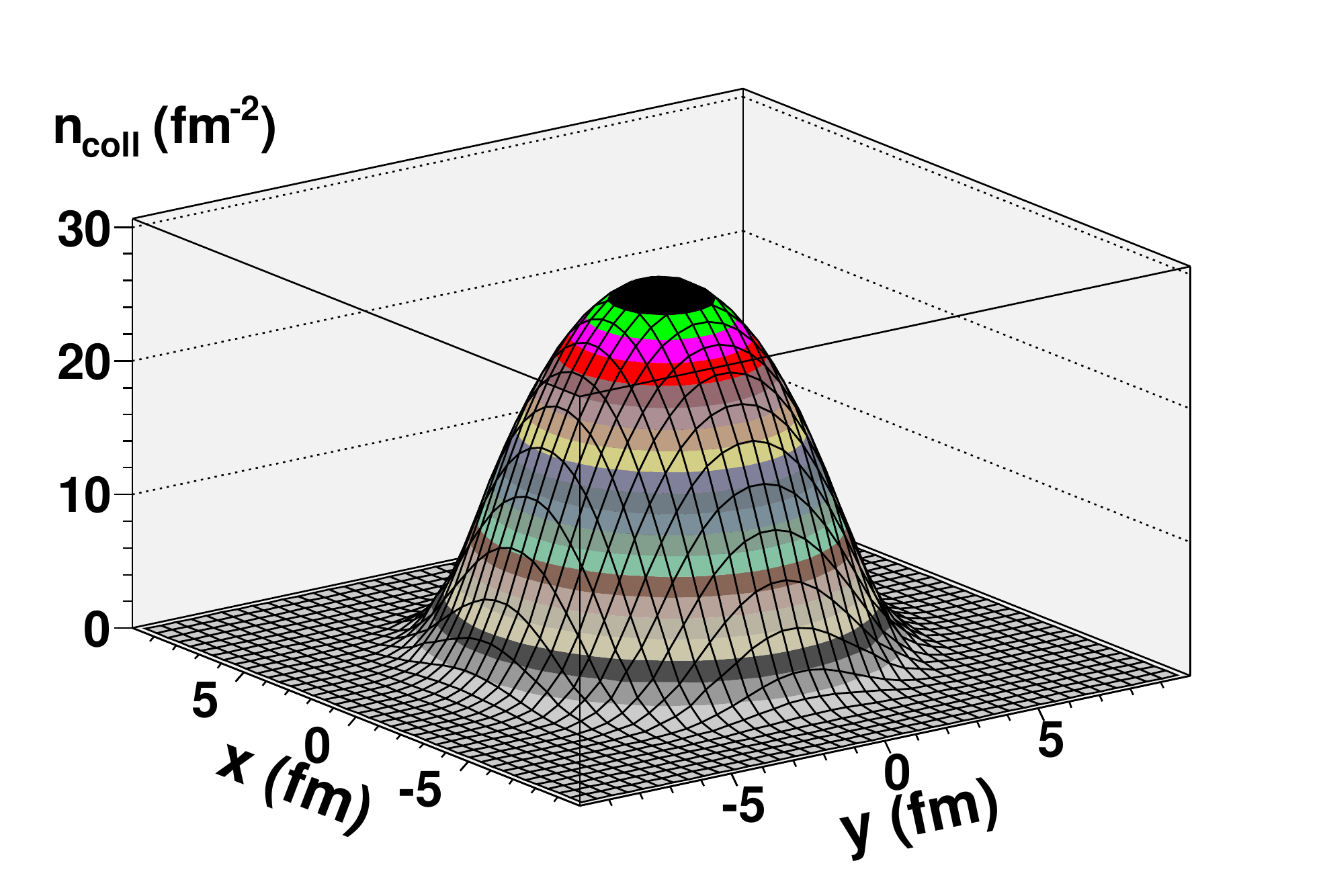}
\caption{\label{fig:ncoll} Number density of binary collisions in a typical heavy-ion collision at LHC. The energy density profile used in a typical hydrodynamic simulation with Glauber
 initial conditions\index{Glauber model!initial conditions} is this function times a multiplicative factor.}
\end{center}
\end{figure}

This function is plotted in Fig~\ref{fig:ncoll} for a relativistic heavy-ion collision of Pb+Pb at $\sqrt{s_{NN}}=2.76$ TeV and $b=2.4$ fm.

\begin{figure}[t]
\begin{center}
\includegraphics[scale=0.35]{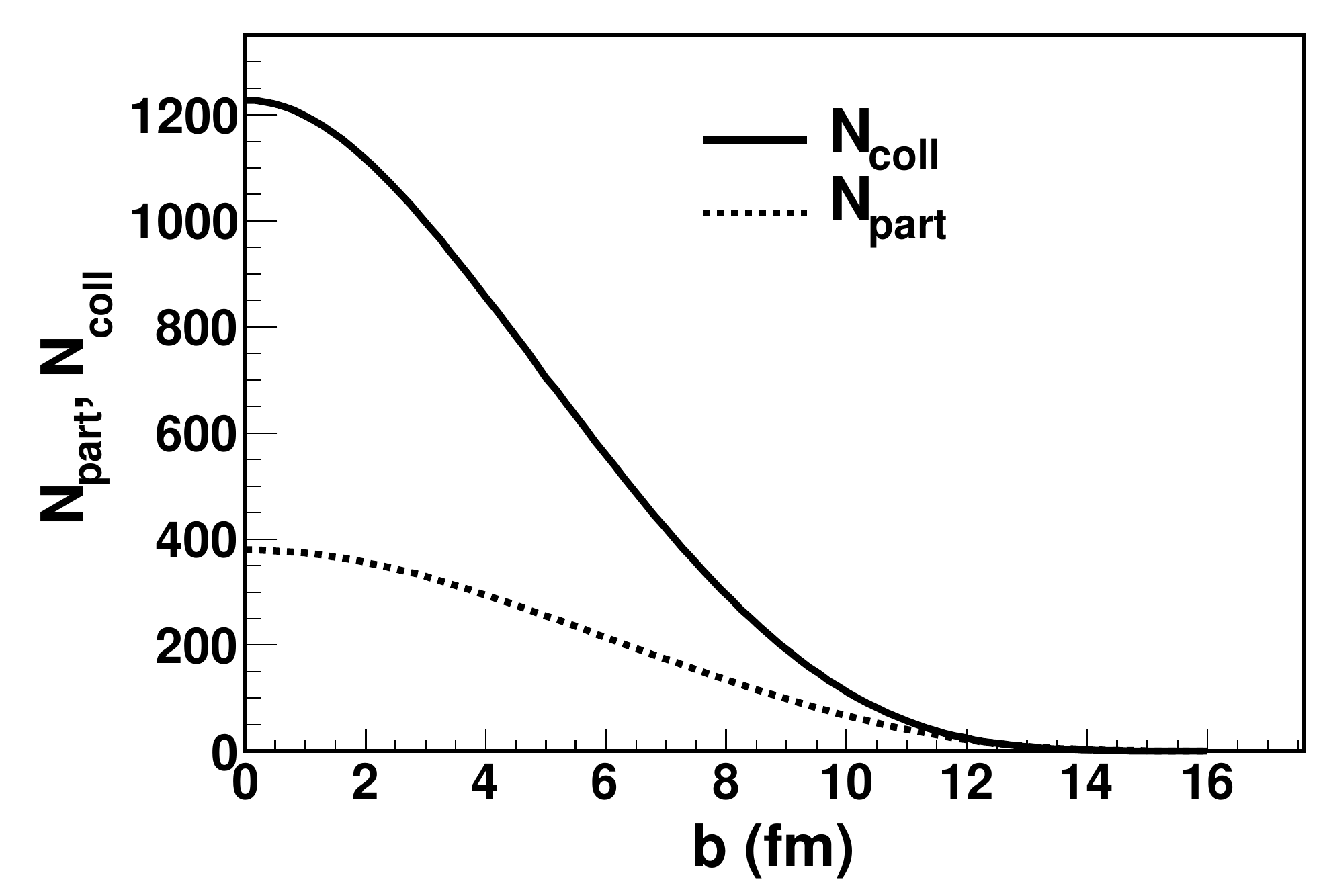}
\includegraphics[scale=0.35]{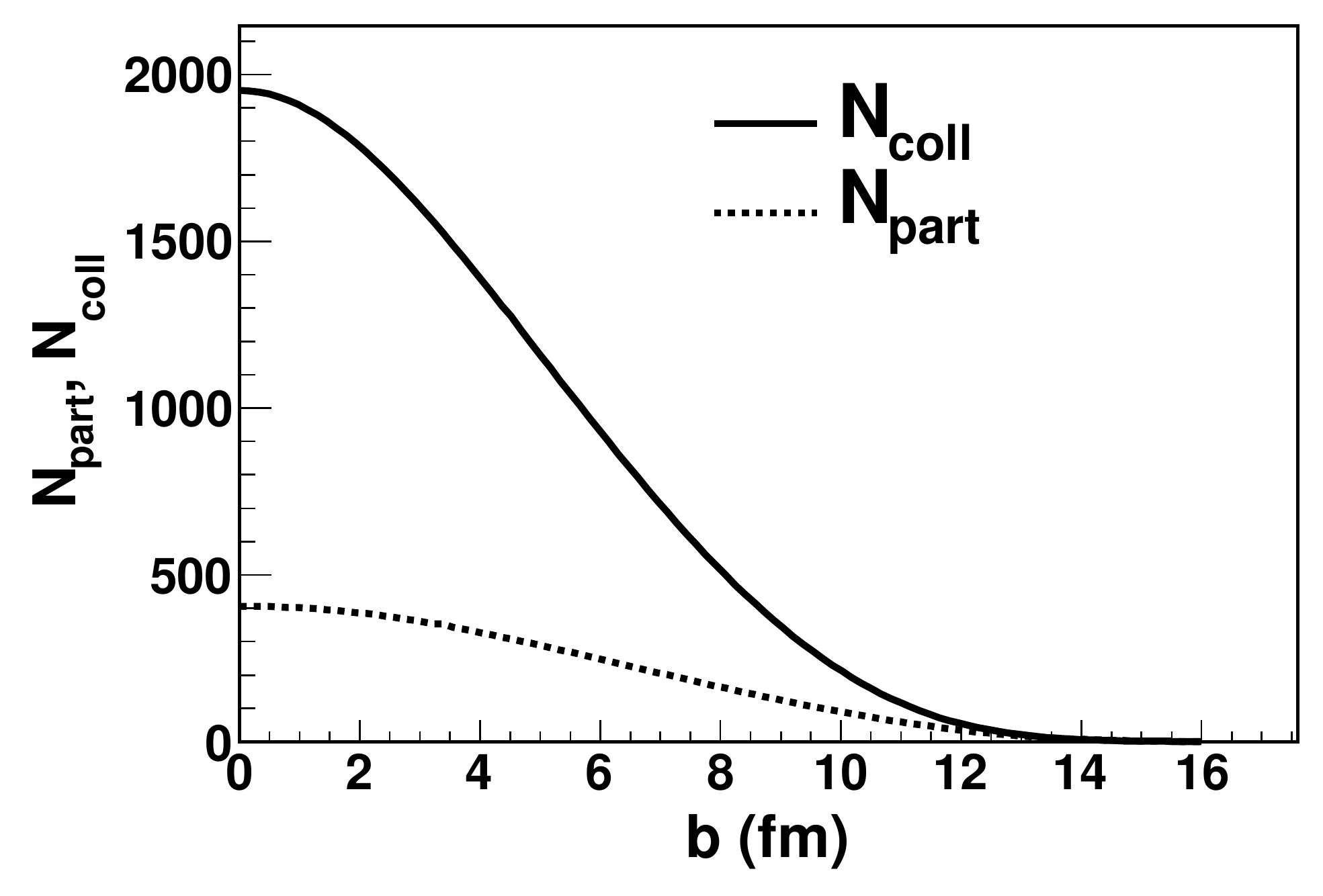}
\caption{\label{fig:glauber} Number of participating nucleons and number of binary collisions in a typical relativistic heavy-ion collision at RHIC (left) and at LHC (right).}
\end{center}
\end{figure}

In Fig.~\ref{fig:glauber} we show the number of participating nucleons and the number of binary collisions that take place in the collision as obtained by direct evaluation of Eqs.~(\ref{eq:glauber1}) and (\ref{eq:glauber2}).
The parameters of the Woods-Saxon potential for gold nuclei at RHIC are taken from \cite{Luzum:2008cw} and for lead nuclei at the LHC are taken from \cite{Aamodt:2010jd}.

The equations from (\ref{eq:glauber1}) to (\ref{eq:glauber2}) correspond to the so-called optical Glauber model, where the nucleon density profiles --given from the Woods-Saxon potential-- are smooth functions of $\mb{x}$.

\section{Energy\index{energy density!of the fireball} and entropy densities\index{entropy density!of the fireball}}

We are going to derive the total transverse energy per unit of rapidity produced in one event, $dE_{\perp}/dy_p$. This quantity can be estimated as
\be \frac{dE_{\perp}}{dy_p} \simeq  \frac{dN_{ev}}{dy_p} \langle e_{\perp} \rangle , \ee
where $\langle e_{\perp} \rangle$ is the average tranverse energy per particle in the final state and $N_{ev}$ is the total number of particles detected in experiment.
For a head-on collision the volume of the fireball is expressed as \cite{Wiedemann:2008zz}
\be V = \pi R^2 \tau_0 \approx \pi \ 1.2^2 A^{2/3} \ \tau_f \textrm{ fm}^2 \ ,\ee
where $\tau_0$ is the time duration of the expansion traded by the freeze-out time $\tau_f$\footnote{In this chapter we consider thermal or kinetic freeze-out where
not only the particle abundances are fixed, but also the momentum distribution.}. 
The Bjorken \cite{Bjorken:1982qr} estimation of the energy density at that time is
\be \epsilon(\tau_f) = \frac{1}{\pi R^2} \frac{1}{\tau_f} \frac{dE_{\perp}}{dy_p} \ . \ee

For the most central Pb+Pb collisions at ALICE at $\sqrt{s_{NN}}=2.76$ TeV, a recent value is \cite{Collaboration:2011rta}
\be \frac{1}{\pi R^2} \frac{dE_{\perp}}{dy_p} = 15 \textrm{ GeV/fm}^2 \ . \ee

\begin{figure}[t]
\begin{center}
\includegraphics[scale=0.35]{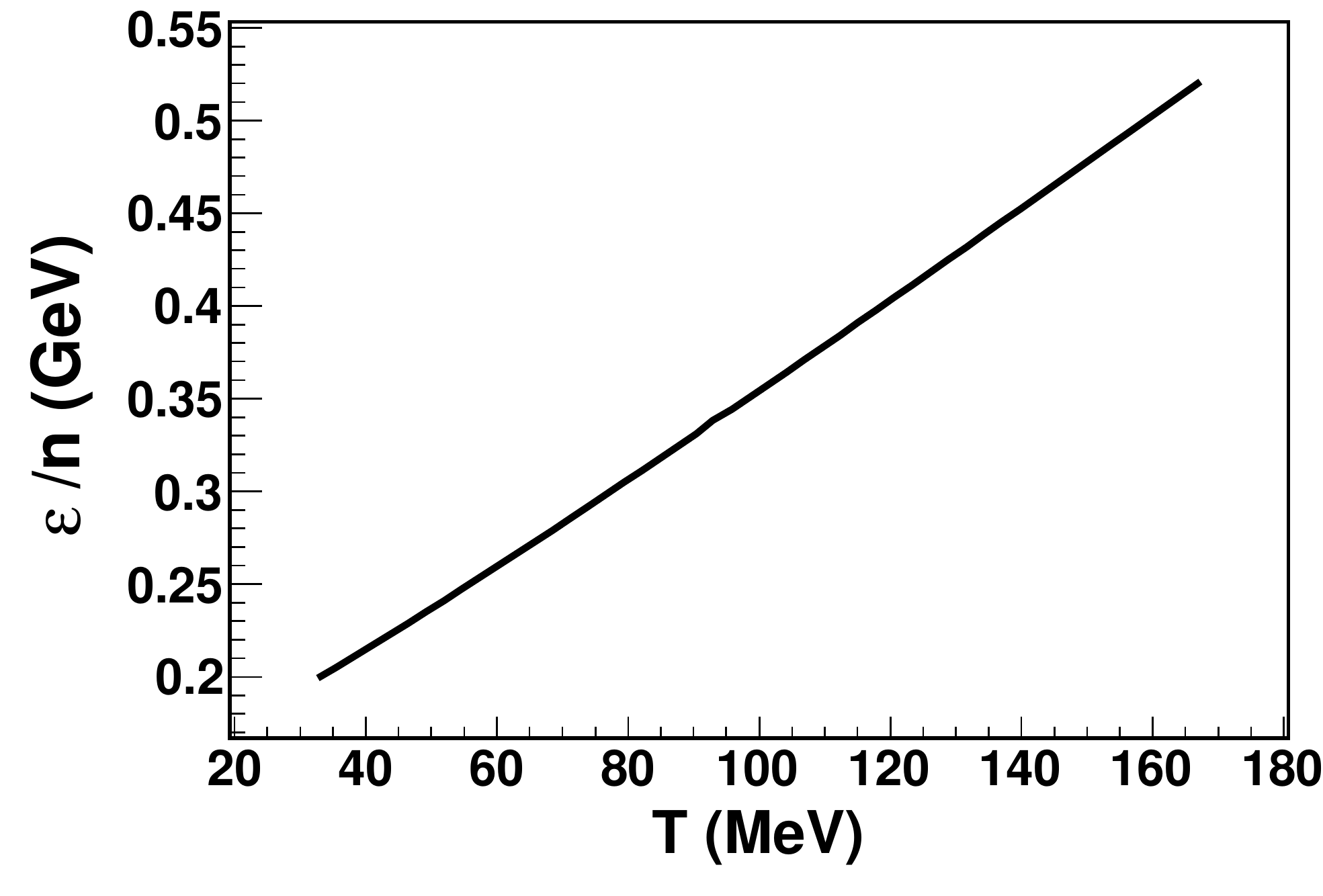}
\includegraphics[scale=0.35]{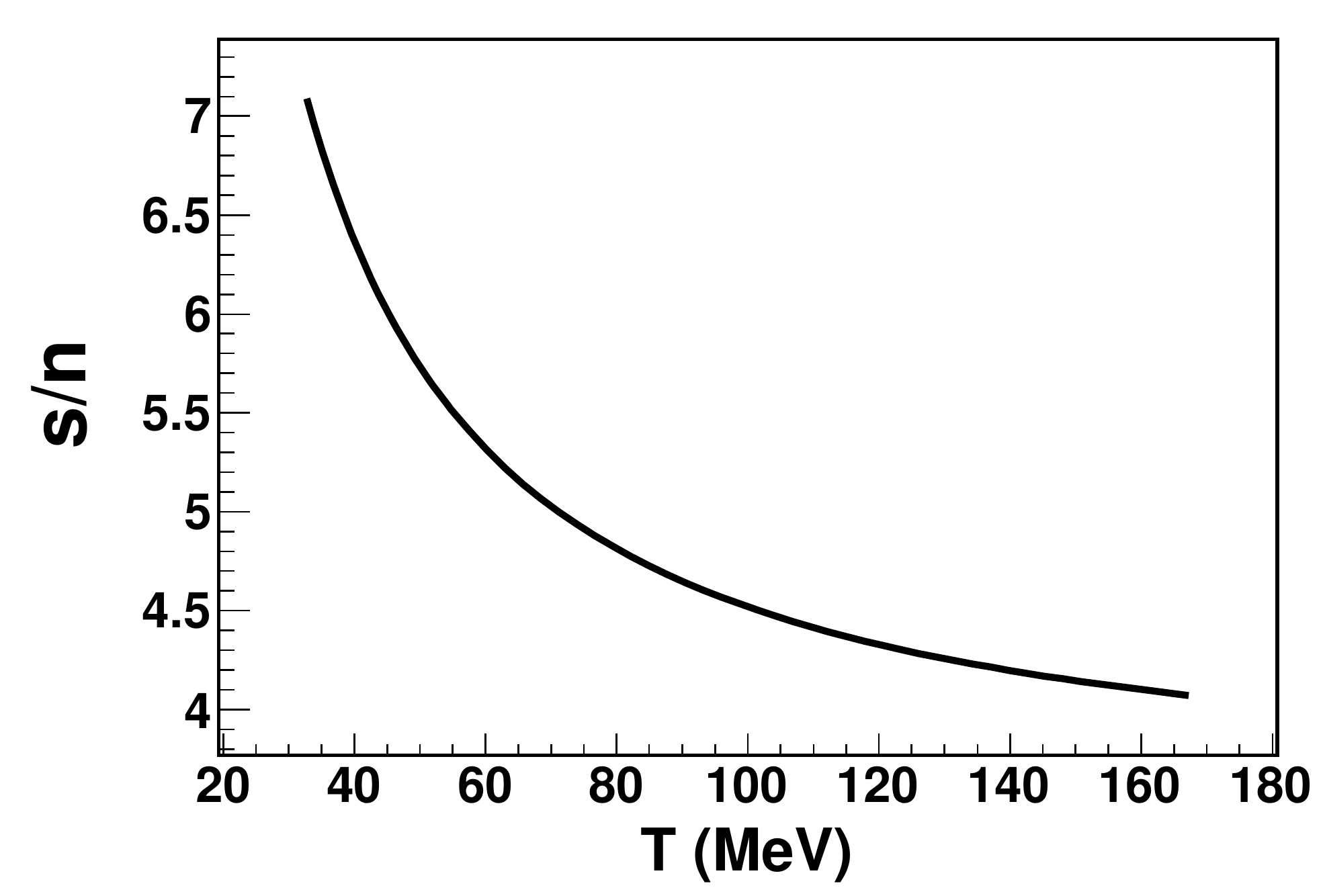}
\caption{\label{fig:sovernpion} Energy\index{energy!per particle} and entropy per particle\index{entropy!per particle} for an ideal pion gas in equilibrium as a function of temperature. The formulae given in Appendix~\ref{app:moments} have been used.}
\end{center}
\end{figure}

Taking $\langle e_{\perp} \rangle \simeq 0.4$ GeV at the freeze-out time (see Fig.~\ref{fig:sovernpion}) and trading the rapidity distribution by the pseudorapidity distribution (as they are similar
in the ultrarelativistic limit):
\be \label{eq:energytf} \epsilon(\tau_f)= \frac{1}{\pi 1.2^2 A^{2/3}} \frac{1}{\tau_f}  0.4 \frac{3}{2} \frac{dN_{ch}}{d\eta_p} \textrm{ GeV/fm}^3 , \ee
where the factor $3/2$ takes into account that only the charged pions ($\pi^+,\pi^-$) have been efficiently detected in the $\frac{dN_{ch}}{d\eta_p}$ distribution.

A similar equation can be obtained for the entropy density \cite{letessier2002hadrons}. Under the same assumptions, the entropy density reads
\be \frac{dS}{dy_p} \simeq 4 \frac{3}{2} \frac{dN_{ch}}{dy_p} \ , \ee
where each particle is taken to have four units of entropy density at the freeze-out time. In Fig.~\ref{fig:sovernpion} we show the temperature dependence of this coefficient, where this value for $s/n$
is a reasonable one for freeze-out temperatures around $120-140$ MeV.

The entropy density finally becomes
\be \label{eq:entropytf} s(\tau_f) = \frac{1}{\pi 1.2^2 A^{2/3}} \frac{1}{\tau_f} 4 \frac{3}{2} \frac{dN_{ch}}{d\eta_p} \ 1/\textrm{fm}^{3} \ , \ee
where we have traded $dN_{ch}/dy_p$ by $dN_{ch}/d\eta_p$ for relativistic particles. The average charged multiplicity from ALICE at $\sqrt{s_{NN}}=2.76$ TeV for the 5 \% most central events is $\langle dN_{ch}/d\eta_p \rangle = 1601 \pm 60$ \cite{Aamodt:2010cz}.

\section{Hanbury-Brown-Twiss interferometry}

   After the collision between the two incoming nuclei has taken place, the fireball expands in space cooling down to the freeze-out time\index{freeze-out time} $\tau_f$, where it reaches the freeze-out 
temperature\index{freeze-out!temperature} $T_f$. It is important to have an idea of the spatial extension of this fireball when hadronization has occurred as well as an estimate of the $\tau_f$, needed for example
in order to constraint the equation of state of the system, for the energy and entropy densities in Eqs.~(\ref{eq:energytf})-(\ref{eq:entropytf}) and for the experimental extraction
of the bulk viscosity described in Chapter~\ref{ch:10.corre}.

The size of the fireball at $\tau_f$ can be accessed by performing Hanbury-Brown-Twiss (HBT)\glossary{name=HBT,description={Hanbury-Brown-Twiss}} interferometry \index{HBT interferometry} over the pions
after the kinetic freeze-out. This method is based on the Bose-Einstein enhancement of identical bosons coming from close points in the phase-space.

The symmetrized wave function of a pair of pions produced at $\mb{x}_1$ and $\mb{x}_2$ with momenta $\mb{p}_1$ and $\mb{p}_2$ can be written as
\be \Psi(\mb{x}_1,\mb{x}_2)_{\mb{p}_1,\mb{p}_2} = \frac{1}{\sqrt{2}} \left[ e^{i (\mb{x}_1 \mb{p}_1 + \mb{x}_2 \mb{p}_2) } + e^{ i (\mb{x}_1 \mb{p}_2 + \mb{x}_2 \mb{p}_1)} \right] \ , \ee
where all the strong and electromagnetic interactions have been neglected.

The probability amplitude is the square of the wave function:
\be |\Psi(\mb{x}_1,\mb{x}_2)_{\mb{p}_1,\mb{p}_2} |^2 = 1 + \cos (\mb{q} \cdot \mb{r}) \ , \ee
where $\mb{r} \equiv \mb{x}_1 - \mb{x}_2$ and $\mb{q} \equiv \mb{p}_1 - \mb{p}_2$. This probability amplitude is thus enhanced if the two pions are produced with similar momenta.

In general, one can define a source function that describes the distribution of the pions produced at different space-time points $S(x)$. In that case, the probability amplitude $| \Psi(\mb{x}_1,\mb{x}_2)_{\mb{p}_1,\mb{p}_2} |^2$ will
contain the Fourier transform of $S(x)$. The two-particle correlation function is defined as \cite{Yagi:2005yb,Gramling:2011hc}
\be C(\mb{p}_1, \mb{p}_2) = \frac{\int dx_1 dx_2 \ S(x_1) S(x_2) \  | \Psi(\mb{x}_1,\mb{x}_2)_{\mb{p}_1,\mb{p}_2} |^2}{\int dx_1 S(x_1) \ \int dx_2 S(x_2)} \ , \ee
that will be of the form
\be  C(\mb{p}_1, \mb{p}_2) = 1 + |\tilde{S}(\mb{q})|^2 \ , \ee
where $\tilde{S} (\mb{q})$ is the Fourier transform of the source function.

This two-particle correlation function is experimentally obtained by measuring the distribution of the difference between the momenta of two detected particles coming form the same event, $\mb{p}_1-\mb{p}_2$
(and conveniently normalized to the same distribution of particles coming from different events).
Taking a Gaussian shape for the source function:
\be S(x,y,z,t) \propto \frac{1}{4 \pi^2 R_x R_y R_z \sigma_t} \exp \left[ - \frac{1}{2} \left( \frac{x^2}{R_x^2} + \frac{y^2}{R_y^2} + \frac{z^2}{R_z^2} +\frac{t^2}{\sigma_t^2} \right) \right]\ee
the correlation function turns out to be
\be  C(\mb{p}_1, \mb{p}_2) = 1 + N \exp \left[ -\frac{1}{2} \left( R_x^2 q_x^2 + R_y^2 q_y^2 + R_z^2 q_z^2 + \sigma_t^2 q_t^2 \right)\right] \ , \ee
where $R_i$ are the Gaussian HBT radii, that encode the dimensions of the source.

A more convenient parametrization of the shape of the fireball is the Pratt-Bertsch parametrization \cite{Pratt:1986cc,Bertsch:1989vn}, in which $R_{long}$ is the direction along the beam axis, $R_{out}$ is the
direction of the pair transverse momentum and $R_{side}$ is perpendicular to both. The correlation function is slightly modified (Sinyukov formula \cite{Sinyukov:1998fc}):
\be C (\mb{q}) =  N (1- \lambda) +N \lambda K(q) \left\{ 1 + \exp \left[ - R^2_{out} q^2_{out} - R^2_{side} q^2_{side} - R^2_{long} q^2_{long} \right] \right\} \ , \ee
where $\lambda$ is the correlation strength and $K(q)$ is the squared Coulomb wave function because of the presence of electromagnetic effects in the correlations.

\cite{Aamodt:2011mr} reports the measured HBT radii as a function of the mean perpendicular momentum $\langle k_{\perp} \rangle$. For the $5 \%$ most central collisions in ALICE
at $\sqrt{s_{NN}}=2.76$ TeV, the values of the radii at $\langle k_{\perp} \rangle \simeq 0.75$ GeV are quite similar (with a small hierarchy $R_{out} < R_{side} < R_{long}$) and of the order of $4.5$ fm. The product of the 
three radii gives a source's volume of $94$ fm$^{3}$. 

\begin{figure}[t]
\begin{center}
\includegraphics[scale=0.35]{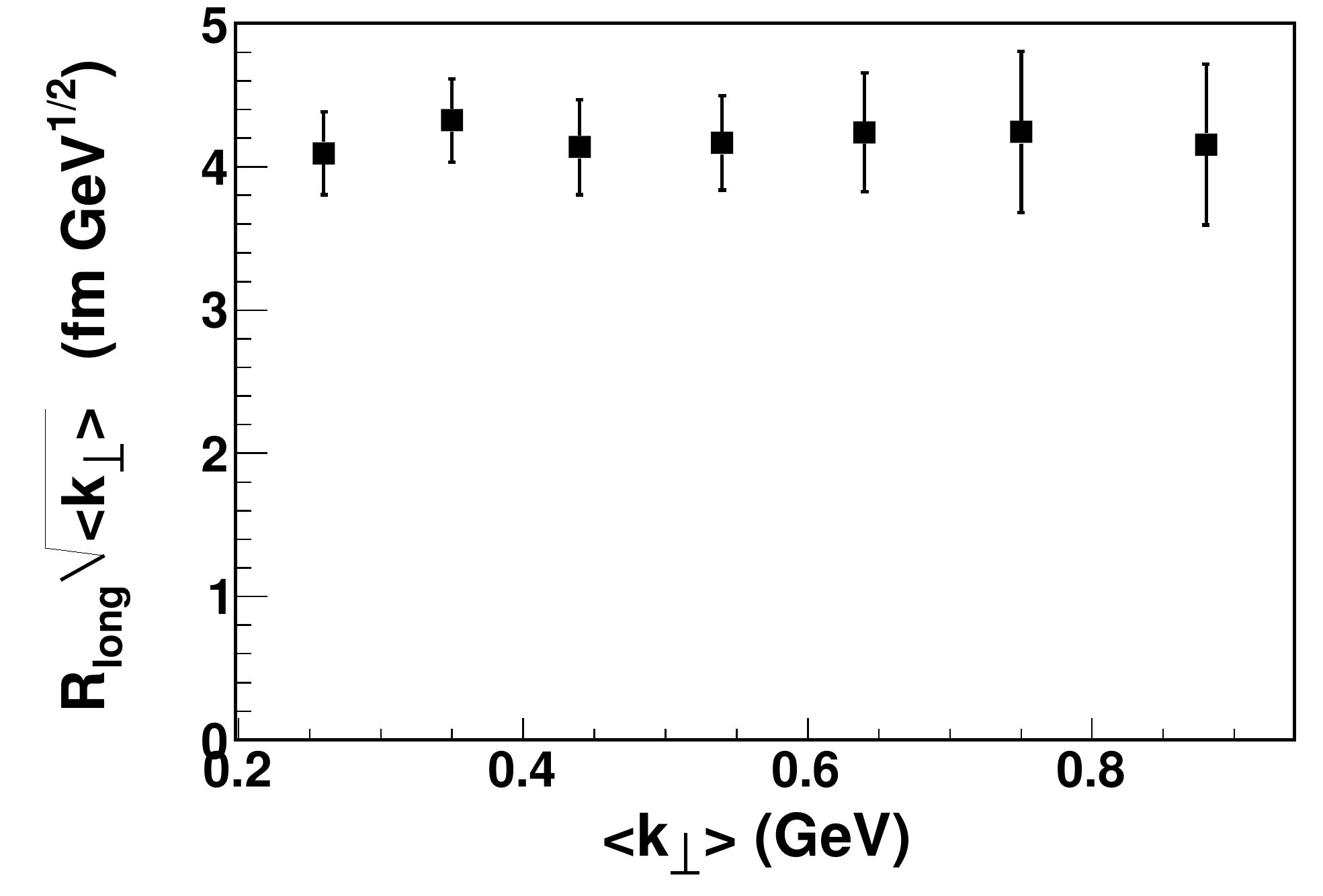}
\caption{\label{fig:rlong} We show the scaling $R_{long} \propto \langle k_{\perp} \rangle^{-1/2}$ of formula (\ref{eq:taufrlong}) with the data of the HBT radii from \cite{Aamodt:2011mr}.}
\end{center}
\end{figure}

Using hydrodynamics we will show in Sec.~\ref{sec:chapman-enskog} that the size of the homogeneity region $h$ is inversely proportional to the velocity gradient 
of the system, that decreases with $1/\tau$. Therefore, $R_{long}$ is proportional to the duration of the longitudinal expansion along the axis, i.e. the decoupling time $\tau_f$. The exact relation 
between $R_{long}$ and $\tau_f$ is given in \cite{Aamodt:2011mr}:
\be R_{long}^2 (k_{\perp})= \frac{\tau_f^2 T_f}{m_{\perp}} \frac{K_2 (m_{\perp}/T_f)}{K_1(m_{\perp}/T_f)}  \ , \ee
with the transverse mass $m_{\perp}$ defined in Eq.~(\ref{eq:trans_mass}) and $K_1,K_2$ the modified Bessel functions of the second kind. Assuming that $m_{\perp} \gg T_f$ a handy formula can be obtained ($k_{\perp} \simeq m_{\perp}$):
\be \label{eq:taufrlong} \tau_f \simeq \sqrt{\frac{\langle k_{\perp} \rangle}{T_f}} R_{long} \ . \ee

For a temperature of $T_f=0.12$ GeV a value of $\tau_f \simeq 10$ fm is found. It is not difficult to see from the experimental data that the product $\sqrt{T_f} \tau_f$ or equivalently $R_{long} \sqrt{\langle k_{\perp} \rangle}$ is essentially a constant, independent of $\langle k_{\perp} \rangle$.
We exemplify this scaling in Fig.~\ref{fig:rlong} where we have used the data given in \cite{Aamodt:2011mr}. Finally, within the assumptions we have made, we obtain the following simple relation between $\tau_f$ and $T_f$:
\be \label{eq:tauf} \tau_f \simeq 4 \sqrt{\frac{\textrm{GeV}}{T_f}} \textrm{ fm} \ , \ee
where the freeze-out temperature $T_f$ is expressed in GeV.

\section{Particle thermal spectra \label{sec:thermalspec}}

In order to obtain the particle spectra one must count the number of particles that reach the detectors during all the expansion time. The three-dimensional hypersurface where the particles reach the detector is defined as $\Sigma (x)$.
In the simplest case it can be a two-dimensional spherical surface containing the detector walls plus the temporal dimension. In the general case is a complicated hypersurface containing the future light cone emerging from the collision.

The infinitesimal element of this hypersurface at the point $x$ is $d\sigma_{\mu}$ and defines a four-vector pointing outwards the hypersurface $\Sigma(x)$. The number of particles of species $i$ that cross the hypersurface $\Sigma$
is just the scalar product of $d\sigma_{\mu}$ with the particle four-current $j_i^{\mu}(x)$ \cite{Heinz:2004qz}:
\be N_i = \int_{\Sigma} d^3 \sigma_{\mu} (x) j^{\mu}_i (x) \ , \ee
where (see Appendix~\ref{app:hydro})
\be j_i^{\mu} (x)= \int \frac{d^3p}{(2 \pi)^3 E_p} p^{\mu} f_i (x,p) \ . \ee
with $f_i (x,p)$ being the one-particle distribution function of the species $i$.

Assuming that the momentum distribution of particles at the kinetic freeze-out remains the same as the distribution of the particles that are detected, one obtains the ``Cooper-Frye formula'' \cite{Cooper:1974mv}
\index{Cooper-Frye prescription}for the final multiplicity of a given species detected at the hypersurface $\Sigma$.

\be \label{eq:partdensity} E_p \frac{dN_i}{ d^3p} = \frac{dN_i}{dy_p p_{\perp} dp_{\perp} d \phi} = \frac{1}{(2\pi)^3} \int_{\Sigma} p^{\mu} d^3 \sigma_{\mu} (x) \ f_i (x,p) \ , \ee
where we hace used the relation $dp_z=E_p dy_p$ that follows from Eq.~(\ref{eq:rapidityinverse}) taking $p_{\perp}$ constant.
It is possible to prove \cite{Cooper:1974mv} that integrating the previous formula over two different hypersurfaces $\Sigma$ and $\Sigma'$, they give the same particle number $N_i$ if between the two hypersurfaces the distribution
function evolves via the Boltzmann equation with a number-conserving integral. In addition, one obtains the same form of the distribution function if and only if it evolves between
the two hypersurfaces through a colissionless Boltzmann equation, i.e. by free streaming.

The Cooper-Frye prescription tells us that in order to get the particle momentum spectrum one can continously deform the hypersurface describing the detector shape, towards the approximate surface in which the particles
suffered last scattering. This surface is called ``kinetic freeze-out surface'' $\Sigma_f$ and it is characterized by the \index{freeze-out!time}freeze-out time $\tau_f(\mathbf{x})$.

Introducing the freeze-out time, the radial variable $r_{\perp}$ and substituting $f_i (x,p)$ by the equilibrium distribution function, Eq.~(\ref{eq:partdensity}) can be reduced to \cite{Heinz:2004qz}:
\begin{eqnarray}
 \frac{dN_i}{dy_p m_{\perp} dm_{\perp}} & = & \frac{g_i}{\pi^2}  \sum_{n=1}^{\infty} (\mp)^{n+1} \int_0^{\infty} r_{\perp} dr_{\perp} \ \tau_f \ e^{n \mu_i/T} \ \left[ m_{\perp} K_1 (n \beta_{\perp}) I_0 (n \alpha_{\perp}) \right. \nonumber \\
& &\left. -p_{\perp} \frac{\pa \tau_{f}}{\pa r_{\perp}} K_0 (n\beta_{\perp}) I_1 (n \alpha_{\perp})\right] \ , \end{eqnarray}
where $\mu_i$ and $g_i$ are the chemical potential and degeneracy of the species $i$ and the $I_0,I_1$ and $K_0,K_1$ are the modified Bessel functions of the first and second kind, respectively.
The summation is nothing but a virial expansion where the $\mp$ sign should be taken in case of fermions or bosons, respectively.
The new variables $\alpha_{\perp}$ and $\beta_{\perp}$ read:
\be \alpha_{\perp} = \frac{\gamma_{\perp} v_{\perp} p_{\perp}}{T} \ ,	 \qquad \beta_{\perp} = \frac{\gamma_{\perp} m_{\perp}}{T} \ . \ee

When considering only the first term in the series (valid for all mesons except for pions, for which Bose-Einstein statistics should apply) the final formula for the particle spectrum is:
\begin{eqnarray}
  \nonumber \frac{dN_i}{dy_p m_{\perp} dm_{\perp}} & = & \frac{g_i}{\pi^2} \int_0^{\infty} r_{\perp} dr_{\perp} \tau_f (\mb{x}_{\perp})  e^{\frac{\mu_i (\mb{x}_{\perp})}{T(\mb{x}_{\perp})}} 
\left[ m_{\perp} K_1 \left( \frac{m_{\perp} \cosh \rho(\mb{x}_{\perp})}{T(\mb{x}_{\perp})}\right)
 I_0 \left( \frac{p_{\perp} \sinh \rho(\mb{x}_{\perp})}{T(\mb{x}_{\perp})} \right) \right.\\
& &\left.  - p_{\perp} \frac{\partial \tau_f}{\partial |\mb{x}_{\perp}|}
K_0 \left( \frac{m_{\perp} \cosh \rho(\mb{x}_{\perp})}{T(\mb{x}_{\perp})}\right)
I_1 \left( \frac{p_{\perp} \sinh \rho(\mb{x}_{\perp})}{T(\mb{x}_{\perp})} \right) 
  \right] \  , 
\end{eqnarray}
where the radial flow rapidity $\rho=\arctan u_{\perp}$ has been introduced \cite{Heinz:2004qz}.

Finally, assuming that the temperature, the freeze-out time and the $\rho$ do not depend on $\mb{x}_{\perp}$ it is possible to extract the result \cite{Schnedermann:1993ws,Heinz:2004qz}
\be \label{eq:partmulti} \frac{dN_i}{dy_p m_{\perp} dm_{\perp}} \propto m_{\perp} K_1 \left( \frac{m_{\perp} \cosh \rho}{T}\right)
I_0 \left( \frac{p_{\perp} \sinh \rho}{T} \right) \ . \ee

It gives important information of the thermal particle spectra in terms of the temperature and under the presence of transverse flow $u_{\perp} = \tan \rho$.

\subsection{Radial flow and freeze-out temperature}

Consider a central collision in which we will assume that there is no tranverse flow $\mb{v}_{\perp}=0$ or $\rho=0$. From Eq.~(\ref{eq:partmulti}) one has 
\be \frac{dN_i}{dy_p m_{\perp} dm_{\perp}} \propto m_{\perp} K_1 \left( \frac{m_{\perp}}{T} \right) \ . \ee
Thus, written in terms of the variable $m_{\perp}$, the particle spectrum is universal for all hadrons. This is called ``$m_{\perp}$ scaling''. Using the fact that $m_{\perp} > T$ for all the hadrons (except maybe for the pions), the 
spectrum can be simplified by using the asymptotic properties of the modified Bessel functions.
\be \frac{dN_i}{dy_p m_{\perp} dm_{\perp}} \sim \sqrt{T m_{\perp}} \ e^{ - m_{\perp}/T } \ . \ee
The only dependence on the hadron species is the range in which $m_{\perp}$ is defined (its minimum value is the hadron mass) and the corresponding degeneracy factor.
Besides these differences, the spectrum is an exponential whose slope (in a semilogarithmic plot) gives directly the freeze-out temperature \index{freeze-out temperature}. 

The approximation $\rho=0$ is only acceptable for a p+p collision where there are no flow effects. However, for Pb+Pb collisions the assumption $\rho=0$ is hardly sustainable.
Calling $T_{is}$ the inverse log slope of Eq.~(\ref{eq:partmulti}), one can obtain \cite{Schnedermann:1993ws}:

\be \label{eq:tinvslope} T^{-1}_{is}= \frac{d}{dm_{\perp}} \log \left( \frac{dN^i}{dy_p m_{\perp} dm_{\perp}}\right) = \frac{I_1 \left( \frac{p_{\perp} \sinh \rho}{T_f}\right) }{I_0 \left( \frac{p_{\perp} \sinh \rho}{T_f} \right)}
 \frac{m_{\perp}}{p_{\perp}} \frac{\sinh \rho}{T_f}
- \frac{K_0 \left(\frac{m_{\perp} \cosh \rho}{T_f} \right)}{K_1 \left( \frac{m_{\perp} \cosh \rho}{T_f} \right)} \frac{\cosh \rho}{T_f} , \ee
where $T_f$ is the actual freeze-out temperature. In the limit of low $p_{\perp} \ll T_f$ and $\rho \ll 1$ one gets:
\be T_{is}^{-1} = \frac{-m_i \mathbf{u}_{\perp}^2}{2T_f^2} + T_f^{-1}\ee
or
\be \label{eq:tis} T_{is} \simeq T_f + \frac{m_i}{2} \mathbf{u}_{\perp}^2 \ . \ee

The collective flow breaks the $m_{\perp}$ scaling. The kinetic energy due to the velocity of the flow affects the particle spectrum, especially
at low $m_{\perp}$.

\begin{figure}[t]
\begin{center}
\includegraphics[scale=0.5]{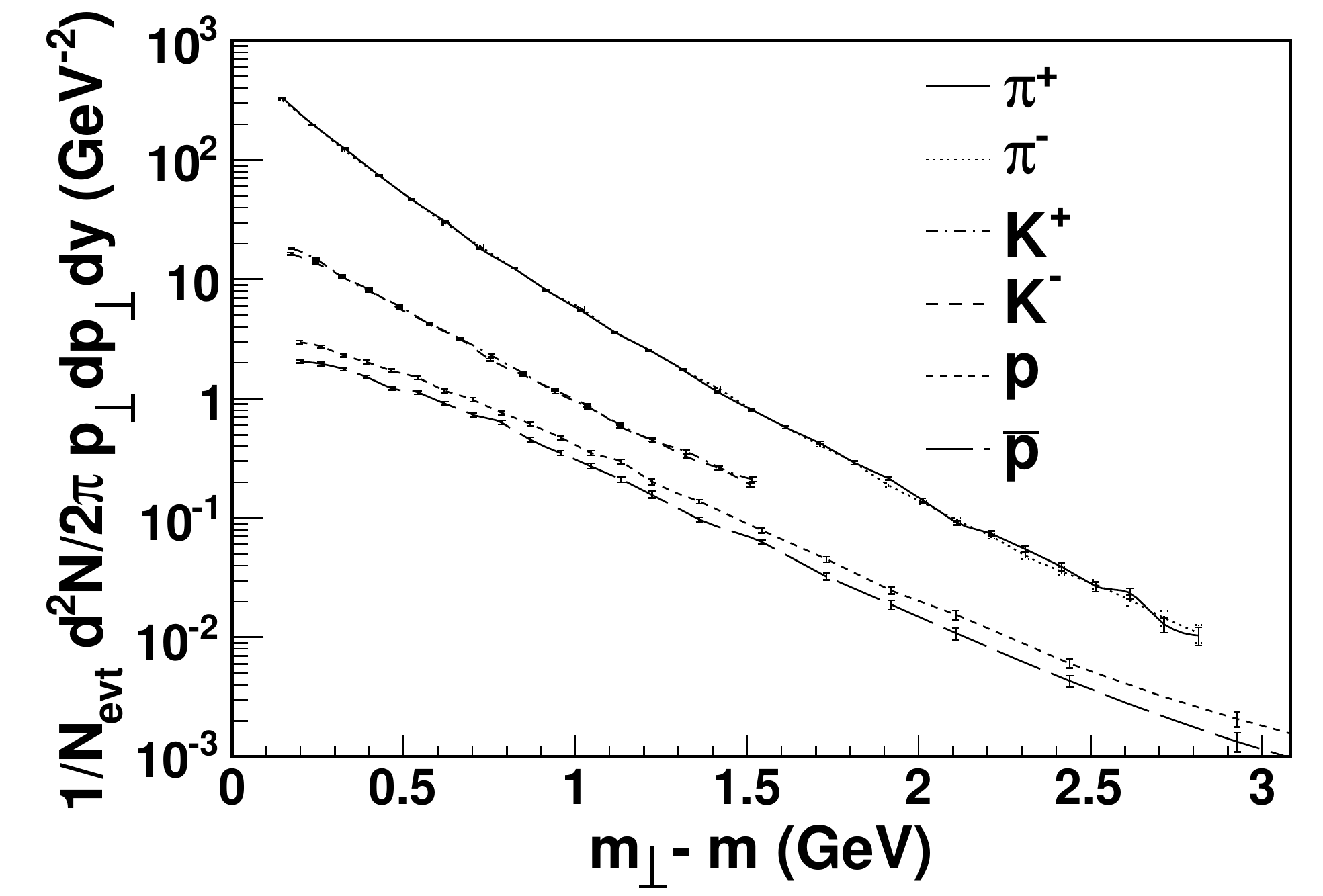}
\caption{\label{fig:pikpspectra} Multiplicity of positive pions, kaons and protons (and their antiparticles) as a function of $m_{\perp}-m_i$ from Au+Au collisions at $\sqrt{s_{NN}}=200$ GeV as
measured by the PHENIX collaboration. Data provided in \cite{Adler:2003cb}.}
\end{center}
\end{figure}

In Fig.~\ref{fig:pikpspectra} we show the charge spectra of pions, kaons and protons as measured by the PHENIX collaboration at RHIC~\cite{Adler:2003cb}. The data is taken from Au+Au collisions at $\sqrt{s_{NN}}=200$ GeV/nucleon,
where the fluid flow is not negligible. The effect of the flow ($u_{\perp} \neq 0$) causes the multiplicities not to be parallel with respect to each other, showing a particle mass dependence following Eq.~(\ref{eq:tis}). 
The effect of the mass-dependent term in (\ref{eq:tis}) is larger in the low $m_{\perp}$ part of the spectrum. This produces a positive contribution to the inverse slope, and therefore
a flattening of the spectra. For the most massive particles (protons and antiprotons) this effect is naturally larger. For pions, this effect is not seen due to the accumulation of slow pions coming from resonance decays, showing
an increase of the pion multiplicity at low $p_{\perp}$. 

From the results in Fig.~\ref{fig:pikpspectra} an important conclusion can be extracted. The number of positive pions is practically the same as the number of negative pions. The same fact occurs for the kaons.
Therefore, the assumption of isospin symmetry is fairly well established.
Note that this is not the case for the proton-antiproton spectra, where the number of antiprotons is slightly smaller. This is nothing but a signature
that the net baryon number is not exactly zero (due to the initial colliding nuclei, this asymmetry should be absent at the proton-antiproton collisions at the Tevatron).


\section{Collective flow and viscosities}

We now lift the restriction of central collisions and consider an arbitrary event with a finite impact parameter. 
In this case, azimuthal symmetry is lost and the particle multiplicity distribution admits a $\phi$ dependence.

At the moment of the collision, the overlap region (that contains the participant nucleons) presents
an almond shape characterized by the spatial eccentricity\index{spatial eccentricity} parameter $\varepsilon_x$:
\be \varepsilon_x(b) =\frac{\langle y^2-x^2\rangle_{\epsilon}}{\langle y^2+x^2 \rangle_{\epsilon}} \ , \ee
where the average is weighted by the energy density $\epsilon$ defined in Eq.~(\ref{eq:energydensity}).

\begin{figure}[t]
\begin{center}
\includegraphics[scale=0.35]{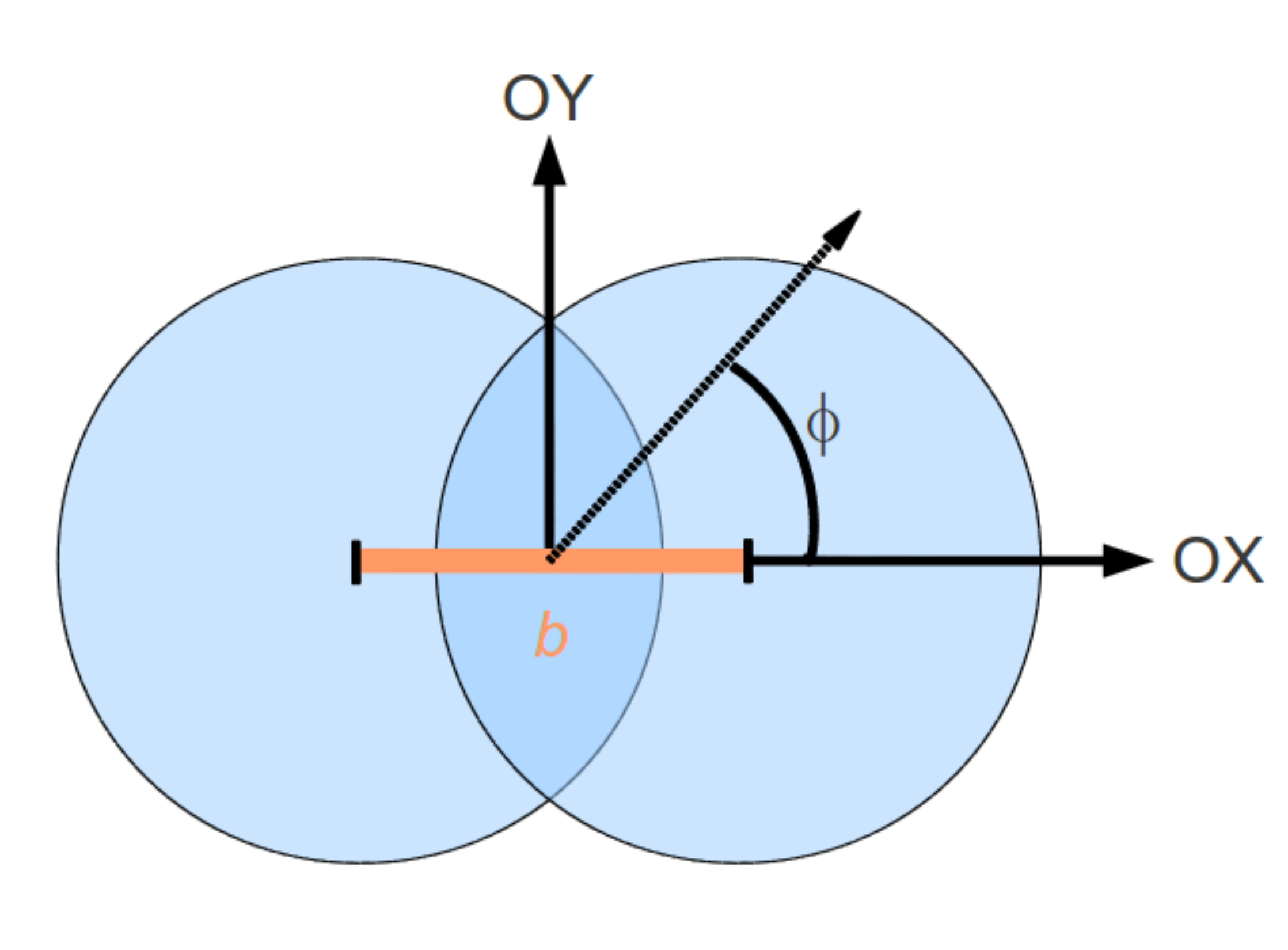}
\caption{\label{fig:eccen} Typical non-central heavy-ion collision projected onto the plane perpendicular to the beam axis.}
\end{center}
\end{figure}

In Fig.~\ref{fig:eccen} we show a typical non-central collision defined by its impact parameter\index{impact parameter}. The inner region is 
composed by the participant nucleons and due to its geometrical anisotropy it has a non-zero value of $\varepsilon_x(b)$. Once the spectator nucleons have
gone away the pressure in the inner region is much higher than the outside of the reaction zone. Due to the spatial anisotropy, the pressure gradient along
the $x$-direction is much larger than the gradient along the perpendicular direction. The response of the system is to create a hydrodynamical boost which is greater
in the $x$-direction than in the $y$-direction and producing a momentum anisotropy in the fluid. The collective motion of the system
converts the initial non-zero spatial asymmetry into momentum anisotropy in the transverse plane and the former tends to decrease at the expense of the latter. 
The experimental evidence of this momentum anisotropy is an azimuthal
anisotropy in the final particle spectrum \cite{Ollitrault:1992bk}.\footnote{In a non-interacting gas, the anisotropy of the almond-shaped source could be detected by Bose-Einstein
HBT correlations and the difference with real data is thus ascribed to interactions.}

\subsection{Flow coefficients}

The particles emitted in a given event follow an azimuthal distribution that can be expressed as a 
sum over Fourier components. The most general expansion for this distribution is

\be \label{eq:partdistriphi} E_p \frac{dN}{d^3p} = \frac{1}{2\pi} \frac{dN}{p_{\perp} dp_{\perp} dy_p} \left[ 1 + \sum_{n=1}^{\infty} v_n(p_{\perp},y_p) \cos (n \phi - n \Psi_R) \right] \ ,\ee

where $v_n$ is the $n^{th}$ flow or harmonic coefficient \index{flow coefficients}and the $\Psi_R$ is the reaction plane\index{reaction plane} (the $OXZ$ plane defined in Sec.~\ref{sec:kin_var}).
The flow coefficients depend on the transverse momentum $p_{\perp}$, the rapidity $y_p$, the centrality and the particle species.
The first flow coefficients are called the ``direct flow'' ($n=1$), the ``elliptic flow'' ($n=2$)\index{elliptic flow} and the ``triangular flow'' ($n=3$)\index{triangular flow}. 
As a Fourier coefficients in the expansion (\ref{eq:partdistriphi}) they can be extracted as
\be \label{eq:Fou_comp} v_n = \langle \cos [n (\phi- \Psi_R)] \rangle \ , \ee
where the average is taken over all the considered particles in a particular event.

The effect of momentum anisotropy is mainly seen in the elliptic flow, that is usually the dominant flow coefficient. 
Moreover, the odd harmonics are in principle forbidden by reflection symmetry with respect to the reaction plane. This is true in the optical Glauber model, where 
the combination of two Woods-Saxon distributions gives an smooth nucleon distribution (see Fig.~\ref{fig:ncoll}). These considerations, made the elliptic flow the only relevant flow coefficient over years.

However, event-by-event fluctuations appear at the positions of the participating nucleons \cite{Alver:2010gr}. These fluctuations in the initial state give non-zero odd harmonics.
They can be computationally generated by the use of a Monte Carlo Glauber model. This model generates random initial positions of the nucleons following the Woods-Saxon distribution.
Since the publication of \cite{Alver:2010gr}, much attention has been paid to the higher order flow coefficients, especially to the next dominant one, the triangular flow $v_3$\index{triangular flow}.

Some unusual structures appeared in the two particle azimuthal correlations at RHIC \cite{Adams:2005ph,Adare:2007vu}. They are typically referred to as the ``ridge'' (an anomalous peak at
 $\Delta \phi \simeq 0$) and the ``shoulder'' (a dip in the away-side peak at $\Delta \phi \simeq \pi$) and they did not show up in p+p collisions. These phenomena appear even at
large pseudorapidity intervals, ruling out the possibility of an origin from the jet quenching. In \cite{Alver:2010gr} they suggest that the presence of the higher order flow coefficients could naturally
explain these two effects. Nowadays, this is the most accepted explanation \cite{Li:2011mp,Aamodt:2011by} and it has been checked for instance by the reconstruction of the two particle
correlation from the measured $v_n$ in ATLAS collaboration up to $n=6$, with a very good agreement \cite{collaboration:2011hfa}.

\subsection{Experimental measurement}

The flow coefficients $v_n$ can be experimentally extracted by different methods. For completeness, we will describe the most common:

\begin{itemize}
 \item Event plane method\index{event plane method}, $v_n \{EP\}$ \glossary{name=EP,description={event plane}}
   
  The event plane method makes direct application of Eq.~(\ref{eq:Fou_comp}). It estimates the $n$-th flow coefficient as (taking the continuum limit)
\be v_n = \frac{\int f_1 (\mathbf{p}) \cos [n (\phi-\Psi_R)] d^3p }{\int f_1(\mathbf{p}) d^3p} \ , \ee
where $f_1(\mathbf{p})= dN/d^3p$ is the one-particle distribution function.

However, one needs to know the orientation of the reaction plane, which is not known {\it a priori} and it varies from event to event. This method replaces the unobservable reaction plane $\Psi_R$ by the
reconstructed event plane $\Psi_n$. The event plane is determined by histogramming the angular distribution of final particles and choosing the angular direction
in which the recorded particle number is maximum. More specifically, taking all the particles in an event one forms the two-component vector:
\be \mathbf{Q}=\left( \sum_i \cos 2\phi_i , \sum_i \sin 2 \phi_i \right) \ . \ee
The event plane angle is defined as
\be (\cos 2 \Psi_n, \sin 2 \Psi_n ) \equiv \frac{\mathbf{Q}}{|\mathbf{Q}|} \ . \ee
One expects that $\Psi_n \simeq \Psi_R$, the difference between these to planes being due to statistical fluctuations, which systematically underestimate the flow coefficients.

 \item Two particle correlations\index{cumulant method}, $v_n\{2\}$. 

   It is possible to access the flow coefficient without resolving the reaction plane. This can be done by computing multiparticle
correlations, which is the basic ingredient of the so-called ``cumulant methods''. In the simplest case one makes use of the two particle correlations. In spite of measuring
angular distributions with respect to the reaction plane, one can combine the relative azimuthal distribution of two particles to cancel the dependence of the reaction plane. One measures

\be \label{eq:twopion} \langle \cos [n (\phi_1 - \phi_2)] \rangle = \frac{\int f_2 (\mathbf{p}_1,\mathbf{p}_2) \cos [n (\phi_1-\phi_2)] d^3p_1 d^3p_2 }{\int f_2(\mathbf{p}_1,\mathbf{p}_2) d^3p_1 d^3p_2} \ , \ee

where the two-particle distribution function $f_2 (\mathbf{p}_1, \mathbf{p}_2)$ describes the probability of finding a pair of particles in the same event, one with $\mb{p}_1$ and the other with $\mb{p}_2$.
The two-particle distribution function contains an uncorrelated part which is a product of two independent one-particle distribution functions and also
a correlated part that accounts for processes in which the two particles are correlated but not through the reaction plane,
\be f_2 (\mathbf{p}_1, \mathbf{p}_2) = f(\mathbf{p}_1) f(\mathbf{p}_2) + f_c (\mathbf{p}_1,\mathbf{p}_2) \ . \ee
The last term takes into account correlations not described by collective motion, but by statistical processes that would be present even in the absence of the reaction plane. 
These correlations can come from resonance decays, jets... and they are irrelevant for the collective motion.

The main idea of the method is that the correlated part of the two-particle distribution function is suppressed by $1/N_{ev}$, where $N_{ev}$ is the event
multiplicity. The argument can be stated as follows\cite{Wiedemann:2008zz}: suppose that in the final state there are $N_{ev}$ pions coming from $N_{ev}/2$ 2-2 processes
like $\rho$ decays, for instance. Each one of these pions would have one decay partner with which it is evidently correlated, and $N_{ev}-2$ pions with which it is not 
correlated through this decay process. However, one pion would be correlated with all the other pions through the reaction plane, due to collective motion.

Thus, the average in Eq.~(\ref{eq:twopion}) contains a correlated term that goes suppressed by $1/N_{ev}$:

\be \langle \cos [n (\phi_1 - \phi_2)] \rangle = v_n^2 + \mathcal{O} \left( \frac{1}{N_{ev}} \right) \ . \ee

The second term is referred to as ``non-flow'' contribution and it contains the effects of jets, resonance and weak decays, etc.

The two particle correlation is therefore a good method if the condition $v_n \gg 1/\sqrt{N_{ev}}$ is fulfilled. At RHIC, the elliptic flow $v_2$ \index{elliptic flow}
reaches a maximum value of around $0.2$. The number of particles in the selected final phase-space is around $N_{ev} \sim 100$, 
so this condition is hardly satisfied at RHIC \cite{Wiedemann:2008zz}, concluding that in the elliptic flow there is a non-negligible contamination of
non-flow effects.

 \item Many particle correlations, $v_2 \{4\},v_2 \{6\},...$. 

     The way to disentagle the non-flow effects in the harmonic coefficients consists on doing appropriate correlations on a larger number of particles. For example,
performing four particle correlations one can measure the following average \cite{Wiedemann:2008zz}:
\begin{eqnarray} \langle \langle \cos [n (\phi_1 + \phi_2 - \phi_3 - \phi_4)] \rangle \rangle & \equiv & \langle \cos [n (\phi_1 + \phi_2 - \phi_3 - \phi_4)] \rangle \\
\nonumber & & -\langle \cos [n (\phi_1 - \phi_3)] \rangle \langle \cos [n (\phi_2 - \phi_4)] \rangle \\
\nonumber & &-\langle \cos [n (\phi_1 - \phi_4)] \rangle\langle \cos [n (\phi_2 - \phi_3)] \rangle \ .
\end{eqnarray}

This average gives the fourth power of the flow plus some suppressed terms
\be  \langle \langle \cos [n (\phi_1 + \phi_2 - \phi_3 - \phi_4)] \rangle \rangle = -v_n^4 + \mathcal{O} \left( \frac{1}{N_{ev}^3}\right) + \mathcal{O} \left( \frac{v_{2n}^2}{N^2_{ev}} \right) \ . \ee
Taking into account that the higher order flow coefficients $v_{2n}$ are much smaller than $v_n$, the condition to suppress the non-flow effects is
\be v_n \gg \frac{1}{N_{ev}^{3/4}} \ , \ee
that is now fulfilled by the RHIC data.

In this direction, one expects that the fourth order cumulant method gives a more accurate description of the flow coefficients with the non-flow effects minimized. It is possible to extend this method
in order to include correlations between six, eight,... particles that suppress even more the contribution of these effects. At the LHC, because the beam energy is larger than RHIC,
the expected number of particles in an event is increased and non-flow effects are more suppressed by the use of the cumulant methods.

\end{itemize}

\begin{figure}[t]
\centering
\includegraphics[scale=0.55]{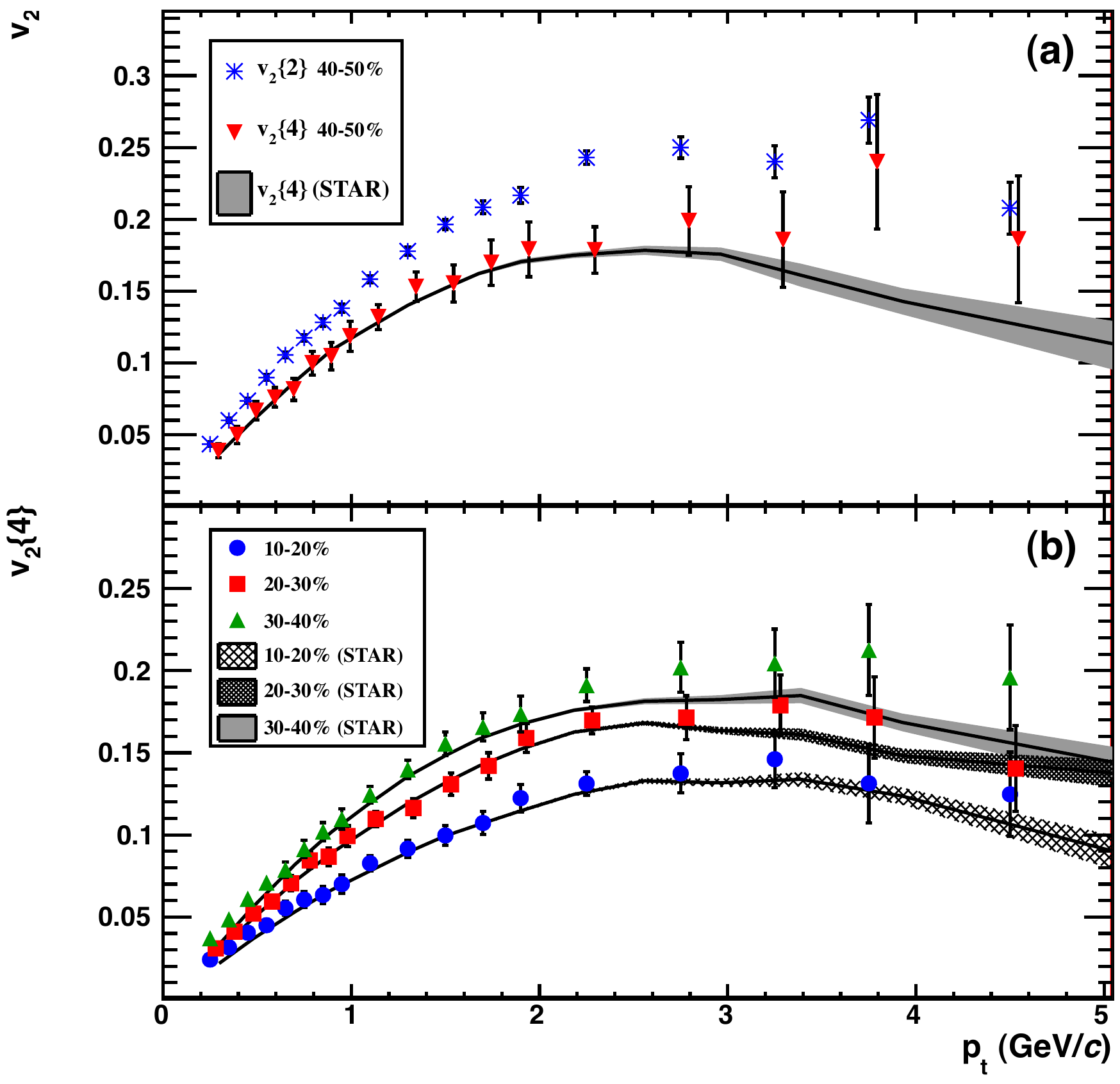}
\caption{\label{fig:ALICE_diff_v2} Differential elliptic flow $v_2$ as a function of $p_{\perp}$ as measured by the ALICE collaboration. Top panel: Results for midperipheral events
when using two-particle correlations (blue asterisks) and the four-particle correlations (red triangles) where the 'non-flow' effects are suppressed. The grey band is the result of STAR collaboration. Bottom panel:
Results for the elliptic flow \index{elliptic flow} for different centralities calculated with four-particle correlations. The elliptic flow increases with the centrality, having larger values for peripheral events. Figures taken from \cite{Aamodt:2010pa}. 
Copyright 2010 by The American Physical Society.}
\end{figure}

In the top panel of Fig.~\ref{fig:ALICE_diff_v2} we show the ALICE results \cite{Aamodt:2010pa} for the differential elliptic flow as a function of $p_{\perp}$ for those events
with centrality $40-50 \%$. The CM energy is $\sqrt{s_{NN}} = 2.76$ TeV and the charged multiplicity can be as large as 500 for this centrality bin.
The blue asterisks are the extracted elliptic flow by using the two-particle cumulant method, that contains non-flow effects. The red triangles correspond to the $v_2$ calculated through
four-particle correlations where the non-flow is negligible \cite{Aamodt:2010pa}. The result for the same centrality bin at STAR experiment ($\sqrt{s_{NN}}=200$ GeV) is also included.
The non-flow corrections always tend to decrease the numerical value of the elliptic flow\index{elliptic flow}.
In the bottom panel of the figure, the elliptic flow using four-particle correlations is shown for different centrality bins. It is evident that when increasing the centrality bin (more peripheral events)
the spatial anisotropy of the initial state is greater and the elliptic flow becomes larger.

\begin{figure}[t]
\centering
\includegraphics[scale=0.42]{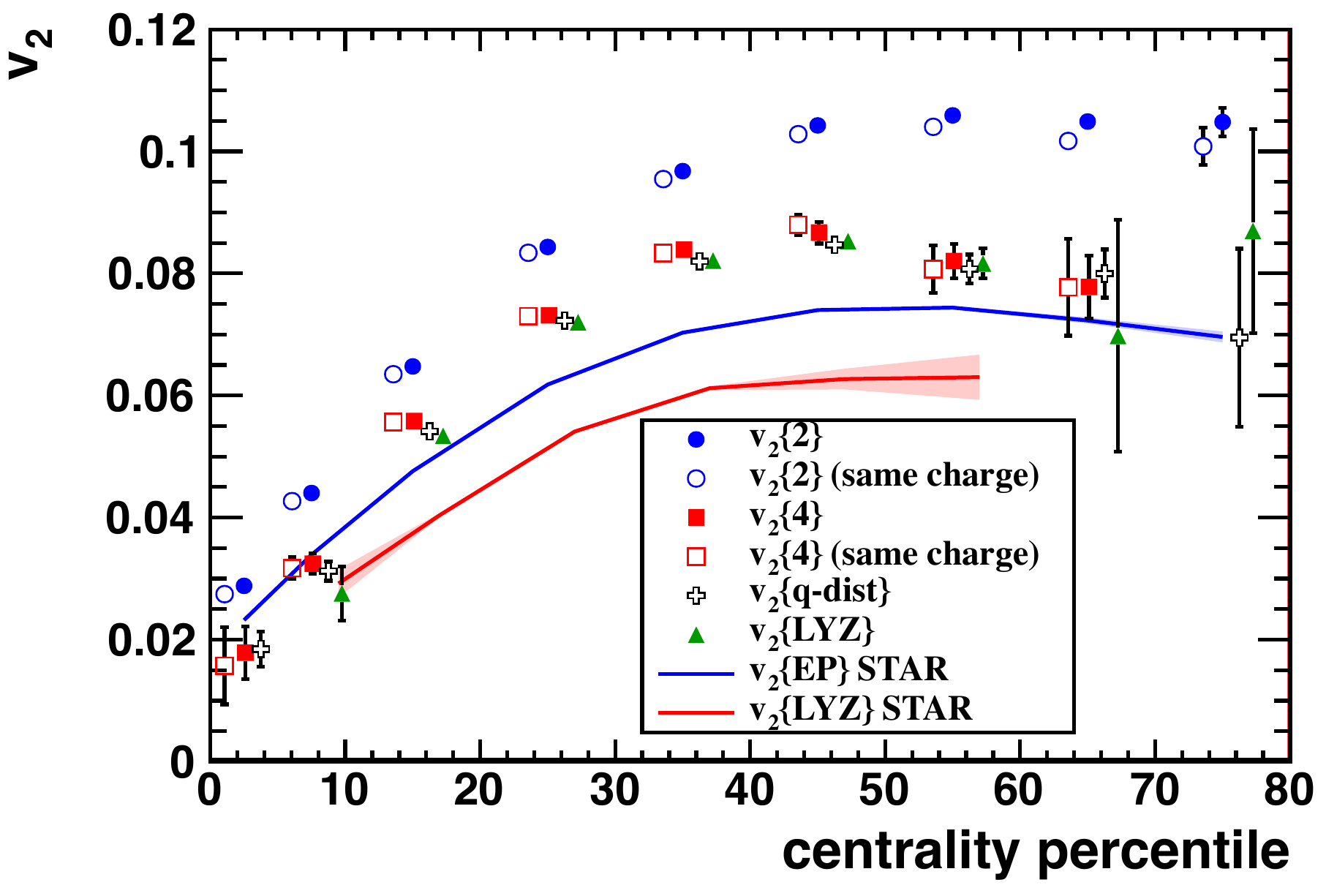}
\includegraphics[scale=0.42]{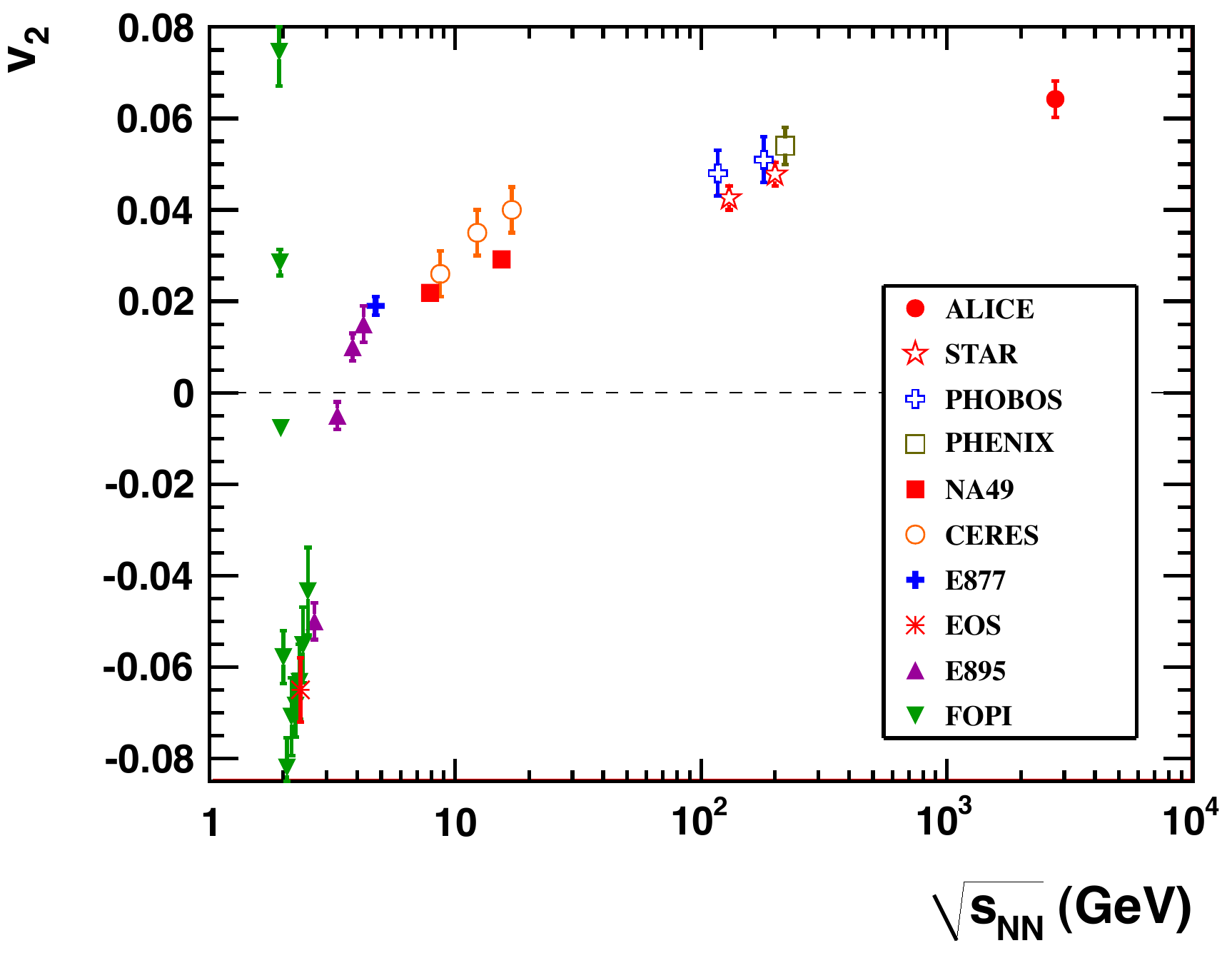}
\caption{\label{fig:ALICE_int_v2} Top panel: Integrated elliptic flow $v_2$ as a function of the centrality. The blue dots are extracted using two-particle correlations and the red dots using four-particle correlations. Some results obtained by 
using other methods are also shown. The blue and red lines are the STAR results. The integrated elliptic flow is larger for collisions with higher collision energy. Bottom panel: Integrated elliptic flow for the centrality class $20\%-30\%$ as a
function of the beam CM energy. The integrated elliptic flow increases with $\sqrt{s_{NN}}$. The last point is the result of the ALICE collaboration with an $\sqrt{s_{NN}}=2.76$ TeV. The cluster around $\sqrt{s_{NN}}=150$ GeV correspond to the
results of three of the experiments at RHIC (STAR, PHOBOS and PHENIX). Figures taken from \cite{Aamodt:2010pa}. Copyright 2010 by The American Physical Society.}
\end{figure}

In the top panel of Fig.~\ref{fig:ALICE_int_v2} we show the integrated $v_2$ between $p_{\perp} \in (0.2,5.9)$ GeV as a function of the centrality. The elliptic flow is estimated by using some different methods, trying
to minimize the non-flow effects. They agree quite well with the results from the four-particle cumulant. The full and open markers show repectively the differences when doing the multiparticle correlations among all particles and among particles with the same
charge. In the bottom panel we reproduce the elliptic flow for a centrality bin of $20-30 \%$ measured by several collaborations at different CM energies. 

\begin{figure}[t]
\centering
\includegraphics[scale=0.45]{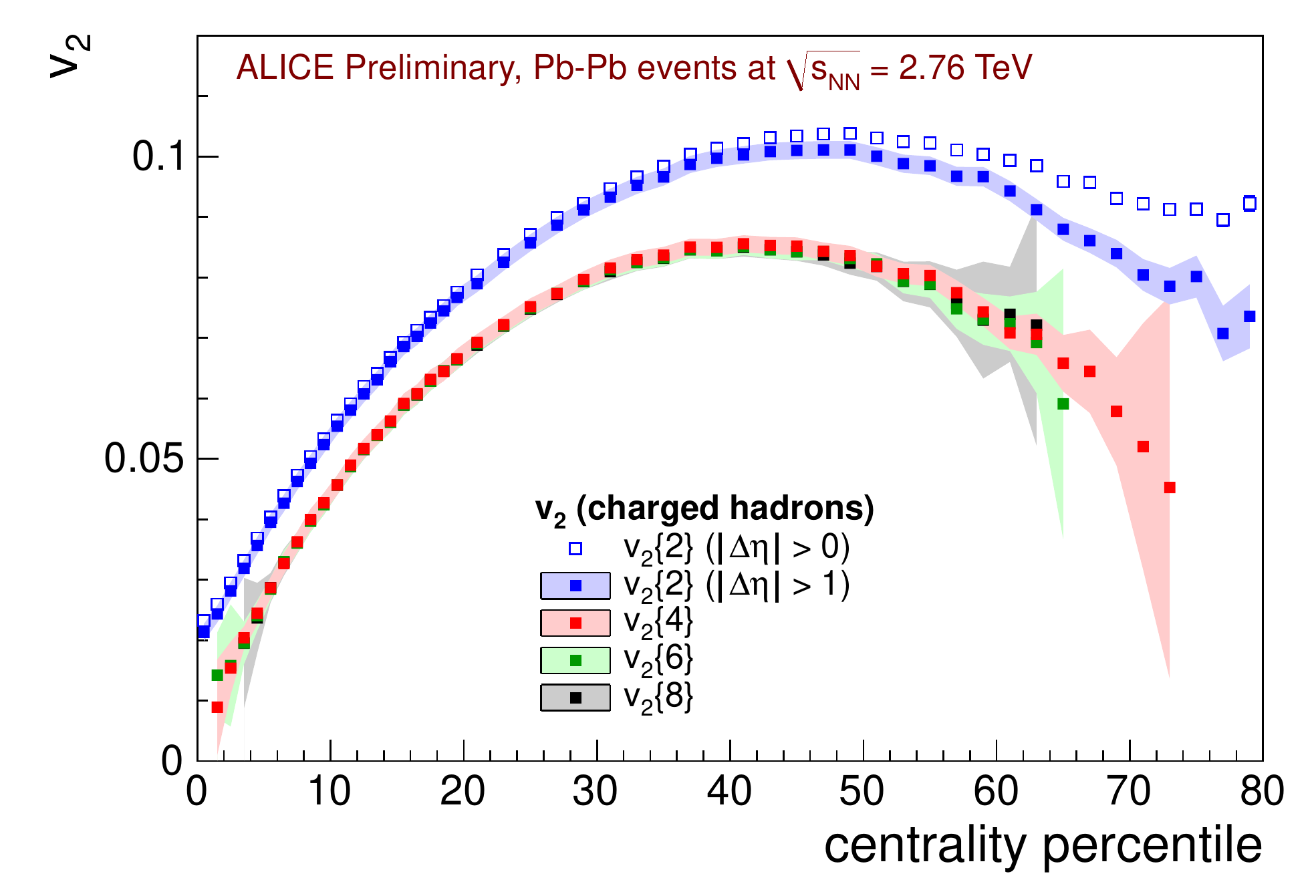}
\caption{\label{fig:bilandzic} Integrated elliptic flow as a function of centrality for different cumulant methods, two-, four-, six- and eight-particle correlations. One can immediately see that for these
collisions at ALICE the four-particle correlations suppress all the 'non-flow' effects with respect to the two-particle correlations. Figure courtesy of A. Bilandzic from \cite{Bilandzic:2011ww}. }
\end{figure}

To prove how the multiparticle correlations converge to the same value of the elliptic flow\index{elliptic flow} (free of ``non-flow'' effects) we show in Fig.~\ref{fig:bilandzic} the preliminary results from the ALICE collaboration \cite{Bilandzic:2011ww}.

\begin{figure}[t]
\begin{center}
\includegraphics[scale=0.6]{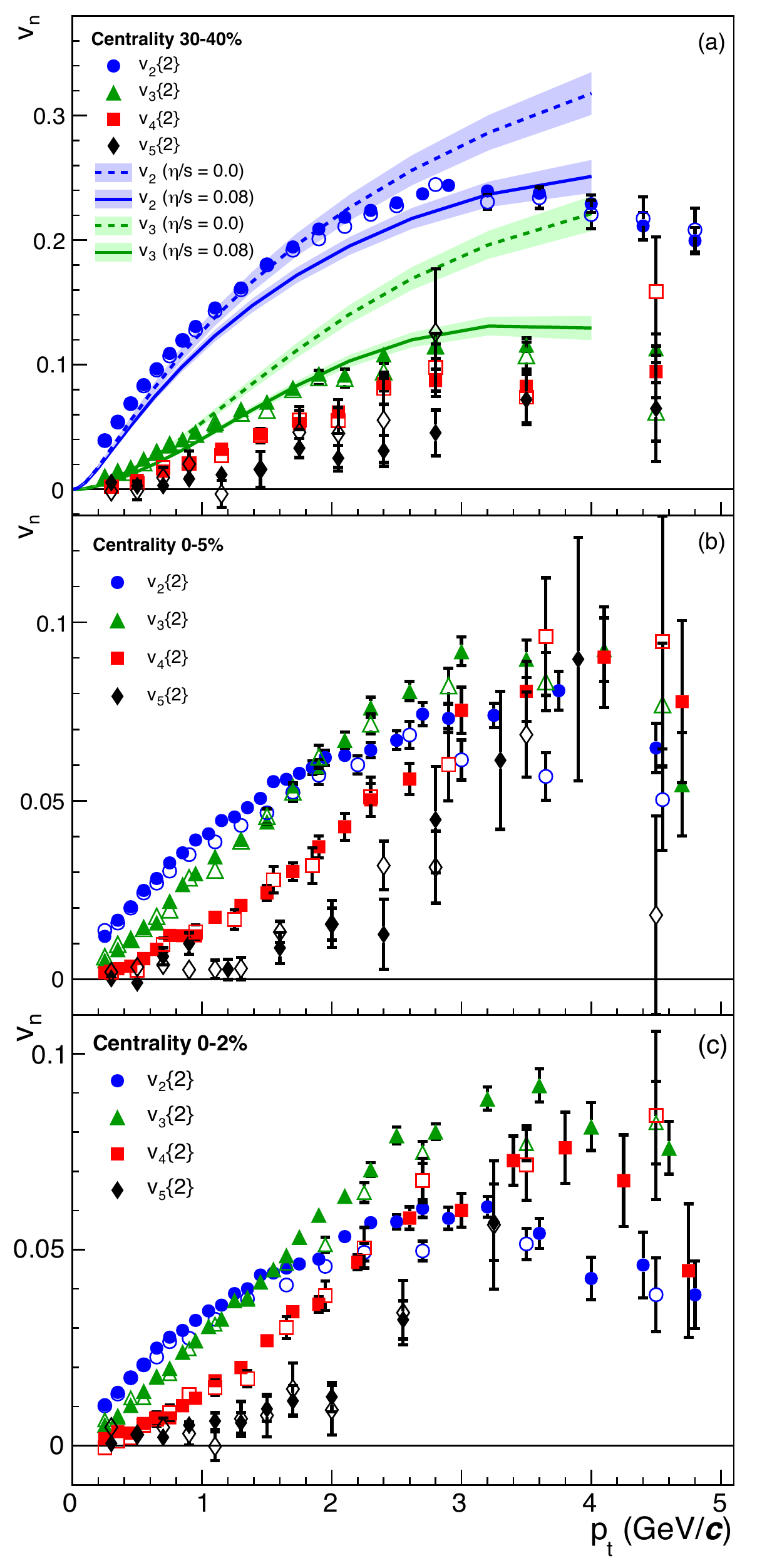}
\caption{\label{fig:ALICE_vn} Differential flow coefficients up to $v_5$ as measured by ALICE for three different centrality bins. A good resolution is achieved for all of them. The last two panels correspond to very central
collisions where the triangular flow is the dominant one at moderate $p_{\perp}$. Figures taken from \cite{Alice:2011vk}. Copyright 2011 by The American Physical Society.}
\end{center}
\end{figure}

Higher order harmonics can be measured as well. In Fig.\ref{fig:ALICE_vn} we show the ALICE results \cite{Alice:2011vk} for the extraction of different higher order harmonics as a function of $p_{\perp}$ and centrality.
One can appreciate the important role of $v_3$ in central collisions, that can be larger than the elliptic flow for higher values of $p_{\perp}$. The fourth and fifth harmonics are also shown in the same plot. The
integrated triangular flow is also the dominant one for central collisions showing an important effect of the fluctuations in the initial state.

\subsection{Viscous hydrodynamic simulations\label{sec:hydrocodes}}

   The dynamics of the expanding system at relativistic heavy-ion collisions can be reproduced by using hydrodynamical \index{hydrodynamical simulations} simulations on a computer.
These simulations try to numerically solve the equations of fluid's hydrodynamics and reproduce the final state momentum distribution as observed in the detector.
If the ideal hydrodynamics (without energy dissipation) is used then the input parameters for the code are fixed in order to properly describe the experimental data for the radial flow.
The initial energy density of the system is fixed such that the final particle multiplicity coincides with the experimental value. The two most used models to describe the initial energy
density are the Glauber model\index{Glauber model! initial conditions} and the Color Glass Condensate (CGC)\glossary{name=CGC,description={Color Glass Condensate}} model\index{CGC initial conditions}.

In a nutshell, the Glauber initial condition \index{Glauber model!initial conditions} takes in the initial time $\tau_0$ the energy density profile to be proportional to the number of binary collisions
\be \rho (\tau_0, \mb{x}_{\perp},b)  \propto n_{coll} (\mb{x}_{\perp},b) \ , \ee
that means that the initial energy density in a heavy-ion collision follows the nucleon distribution. Using the Glauber model, we have plotted the number density of binary collisions
(using the LHC data for Pb+Pb collision) in Fig.~\ref{fig:ncoll}. The Glauber initial condition assumes that the energy density profile is just proportional to the distribution shown in that figure.

This model has been widely used for describing the initial state of the fireball, both in the optical Glauber model (with smooth distribution coming from the Woods-Saxon potential) and in the Monte 
Carlo Glauber model (where the positions of the nucleons are randomly distributed). However, the CGC \index{CGC initial conditions}initial condition has attracted much attention
because it includes physical information about QCD at high energies~\cite{Gelis:2010nm,Armesto:2006bv}. When the compression of the nuclei is as huge as in a heavy-ion collision, the gluonic density is expected to
saturate due to the strong color fields.

This model uses the number density of gluons in a binary collision $\frac{dN_g}{d^2 \mb{x}_{\perp} dy_g}$, where $\mb{x}_{\perp}$ are the perpendicular directions and $y_g$ is the rapidity of the produced
gluons in the collision. The initial energy density profile is then defined as \cite{Romatschke:2009im}:
\be \rho(\tau_0, \mb{x}_{\perp},b) \propto \left[ \frac{dN_g}{d^2 \mb{x}_{\perp} dy_g} \right]^{4/3} \ . \ee

To describe the collective phenomena, dissipative (or viscous) hydrodynamics should be taken into account. At first order in hydrodynamical gradients, the shear viscosity\index{shear viscosity}, the bulk viscosity\index{bulk viscosity}
and the heat conductivity\index{heat conductivity} enter in the hydrodynamic equations of motion. In practice, the shear viscosity (usually normalized by the entropy density\index{entropy density})
is the most important coefficient (at least, out of the critical region) and it is responsible for some collective properties of the fluid. As we have discussed, collective effects generate
 non-vanishing flow coefficients, which can be extracted from the results of the hydrodynamic simulations. 

The hydrodynamic codes use the so-called second-order hydrodynamics \index{hydrodynamics!second order} where gradients up to second order must be included in the expression for the entropy density 
(see Appendix~\ref{app:hydro}). This must be done in order to avoid numerical problems when the short wavelength modes are included. This problem is associated with the loss of causality that the
 Navier-Stokes equation\index{Navier-Stokes equation} presents when considering these high frequency modes. We briefly describe this issue in Appendix~\ref{app:second-order}.

\subsection{Extraction of $\eta/s$}

\begin{figure}[t]
\begin{center}
\includegraphics[scale=0.25]{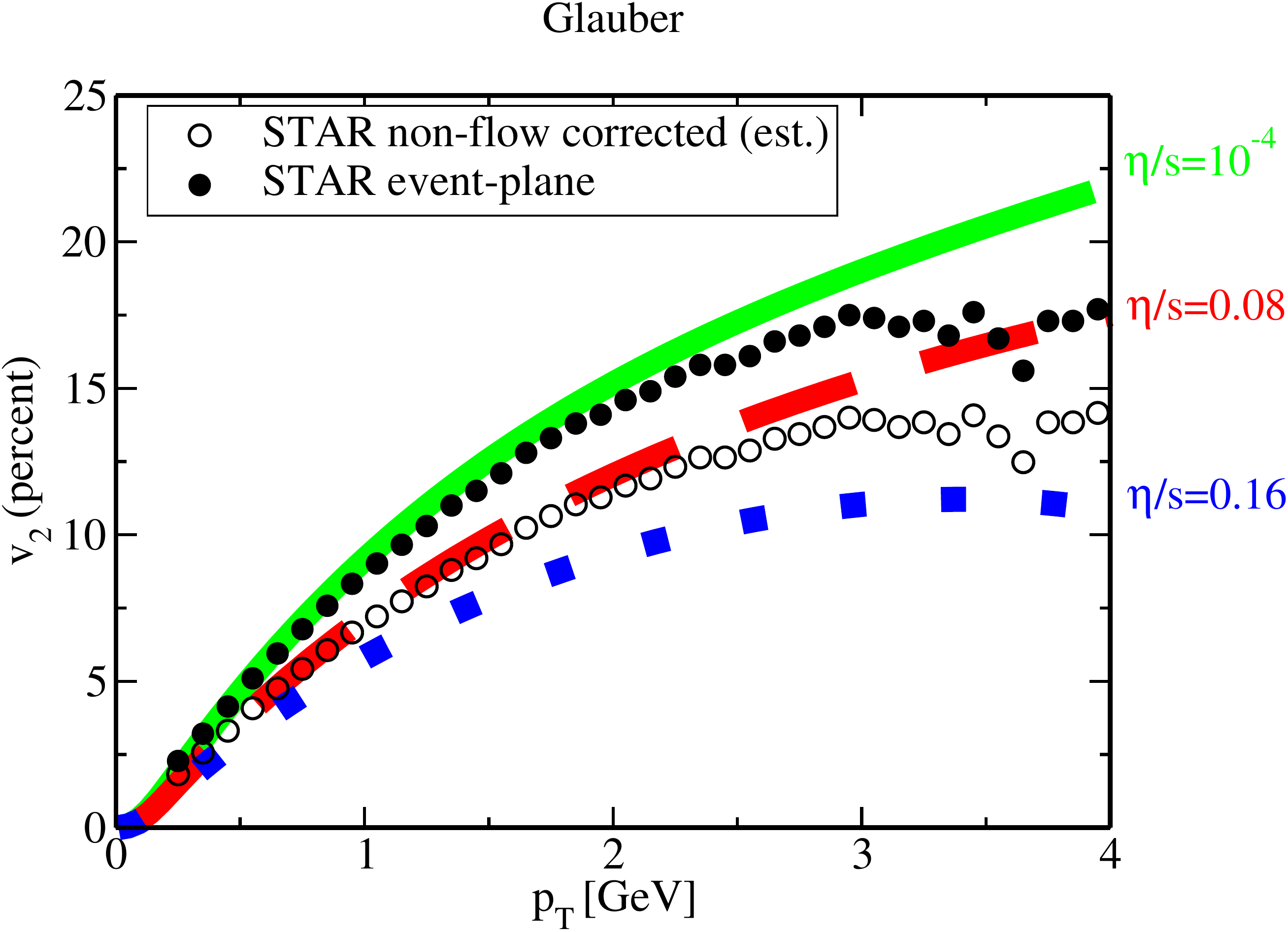}
\includegraphics[scale=0.25]{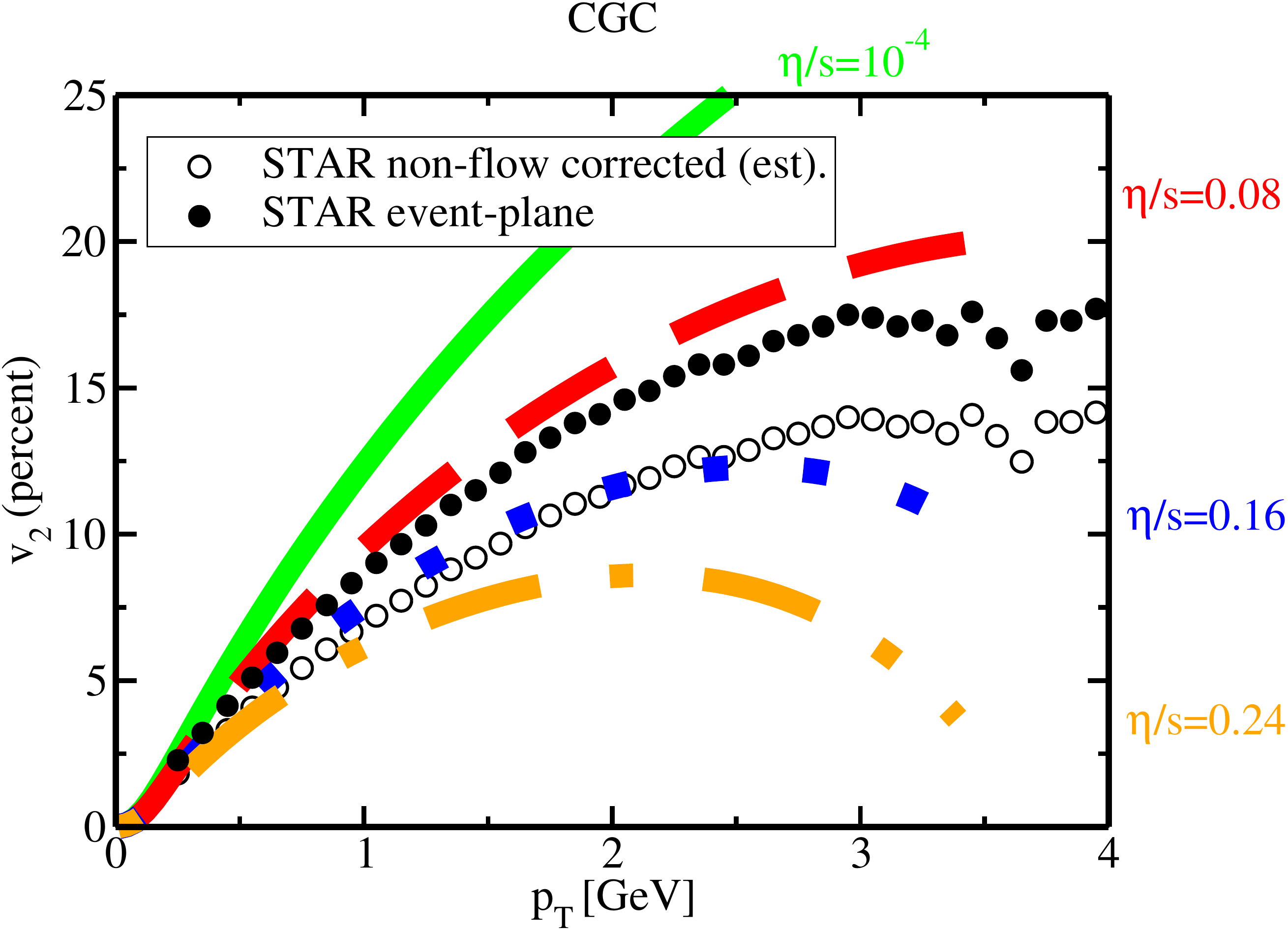}
\caption{\label{fig:v2Luzum} Differential elliptic flow as a function of $p_{\perp}$ for gold-gold collisions as measured by the STAR collaboration. The solid dots correspond to the result obtained by the event plane method \index{event plane method}.
This result contains non-flow effects that are substracted in the empty dots. The result from the hydrodynamic calculation for several values of $\eta/s$ is also shown. Figures courtesy of M. Luzum and P. Romatschke from \cite{Luzum:2008cw}. Copyright 2008 by
The American Physical Society.}
\end{center}
\end{figure}

The determination of the shear viscosity over entropy density combines experimental techniques with hydrodynamic simulations. The estimation of this coefficient is made by matching the
experimental dependence of the flow coefficients (especially the elliptic flow as the dominant one in non-central collisions) to the numerical results from the hydrodynamic codes, that use the
$\eta/s$ coefficient as an input. As an example we show in Fig.~\ref{fig:v2Luzum} the results from the simulation in \cite{Luzum:2008cw} of the elliptic flow at RHIC energies as a function
of $p_{\perp}$. The $v_2$ coefficient is plotted for several values of the shear viscosity over entropy density. Both Glauber and CGC initial conditions have been used in the simulations. Finally, a comparison
with the experimental value of the elliptic flow from the STAR collaboration is made. The full dots correspond to the measurement of elliptic flow by the event plane method and the open dots to the estimate
of the elliptic flow where the non-flow effects have been removed.

In principle, ideal hydrodynamics ($\eta/s \sim 10^{-4}$) shows a good description of the data but the effect of the shear viscosity \index{shear viscosity!from hydro codes} is needed to better explain
the experimental curve. The value of $\eta/s \sim 0.08$ seems to be the optimal one for the Glauber initial conditions\index{Glauber model!initial conditions} whereas the best value for the CGC initial conditions turns out to be $\eta/s \sim 0.16$.
An important conclusion can be extracted. The matter created at heavy-ion collisions behaves like an ideal fluid with a very low shear viscosity/entropy density. The strongly coupled quark-gluon plasma
 (sQPG) \index{quark-gluon plasma}\glossary{name=QGP,description={quark-gluon plasma}} is therefore a collective state with a very low $\eta/s$ near the KSS bound $1/(4\pi)\simeq 0.08$\index{KSS bound}\glossary{name=KSS,description={Kovtun-Son-Starinets}}
\cite{Kovtun:2004de}.

The viscosity dependence is stronger in higher harmonics like $v_3$ as can be seen in Fig.~\ref{fig:ALICE_vn}. The triangular flow can give a better estimate of
$\eta/s$ and it can serve as a probe for initial state assumptions. Moreover, the determination of $\eta/s$ can help distinguish between initial state models. For example, once the value of
the $\eta/s$ coefficient has been determined by matching the elliptic flow for both Glauber and CGC models, one can predict the value of the higher flow coefficients (with $\eta/s$ fixed) and compare the results with the data provided
 by the experiment. The triangular flow, very sensitive to initial fluctuations, can help distinguish between one model and another as proposed by the PHENIX in \cite{Adare:2011tg} favouring the Monte Carlo Glauber model
rather that the CGC-inspired model.

\subsection{Bulk viscosity effects}

The bulk viscosity $\zeta$ \index{bulk viscosity!from hydro codes}has usually been neglected in hydrodynamic calculations due to the general belief that it should numerically be much smaller
than the shear viscosity. This idea came from experience with ordinary fluid such as water where $\zeta \ll \eta$. Also from the perturbative calculations of $\zeta$ in the quark-gluon plasma as in \cite{Arnold:2006fz}, where the ratio between
the bulk and shear viscosities for $N_f=3$ massless quarks and in the range of strong coupling constant $\alpha_s \in (0.02,0.3)$ is
\be \frac{\zeta}{\eta} \simeq 0.3 \ \alpha^4_s \sim \ 10^{-3}-10^{-8} \ . \ee
\begin{figure}[t]
\begin{center}
\includegraphics[scale=0.44]{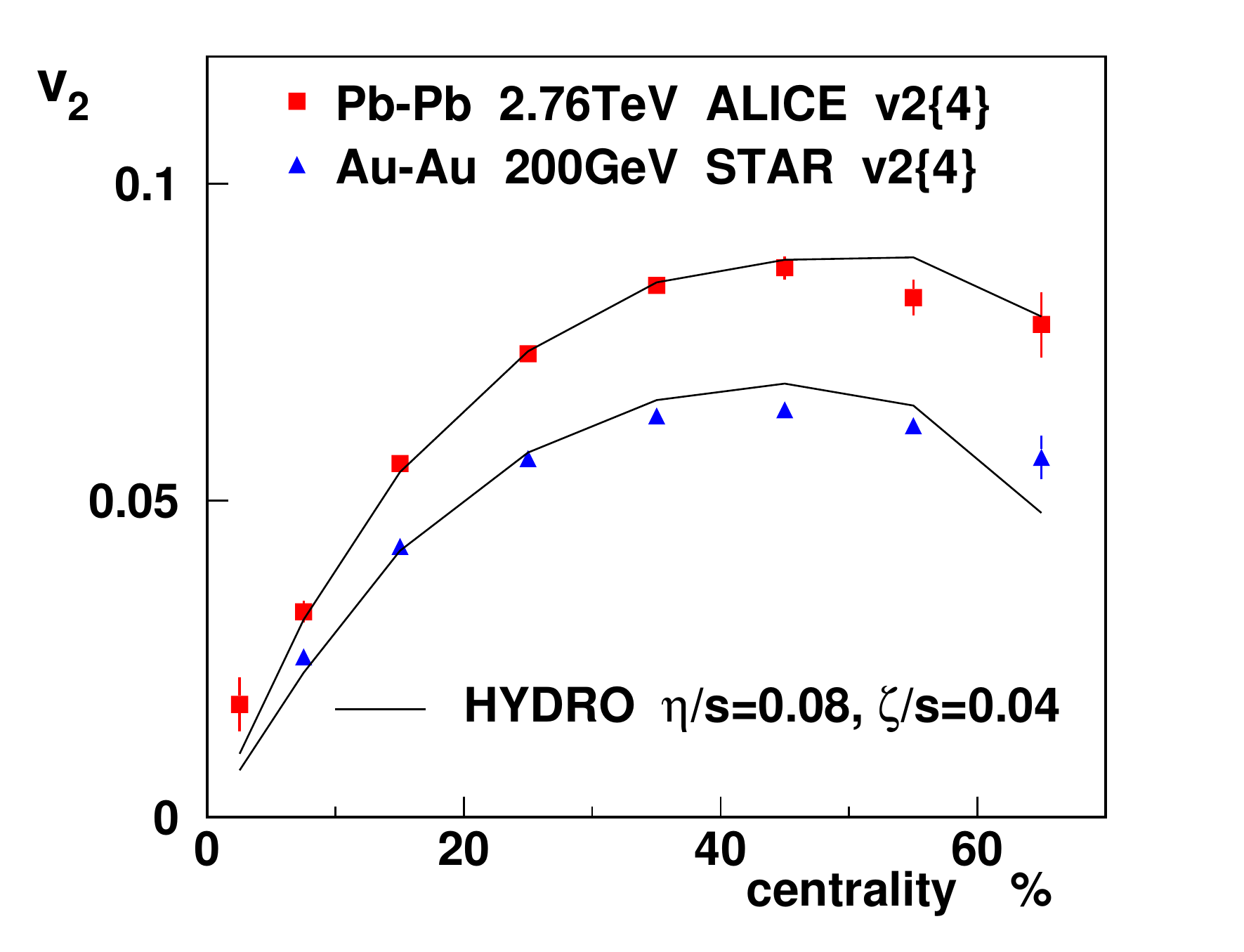}
\caption{\label{fig:v2Bozec} Integrated elliptic flow as a function of centrality for lead-lead collisions at ALICE (red squares) and Au+Au collisions at STAR (blue triangles). Solid line: The results
coming from the hydrodynamic calculation with $\eta/s=0.08$ and $\zeta/s=0.04$. Figure from \cite{Bozek:2011fm} courtesy of P. Bozek.}
\end{center}
\end{figure}

However, this picture can change in the strong coupling regime, where the hadronic states are the relevant degrees of freedom. Actually, the bulk viscosity
can be much larger than the shear viscosity around the critical temperature where the conformal anomaly peaks. Lattice QCD calculations \cite{Karsch:2007jc} have shown that 
the bulk viscosity over entropy density diverges at the critical temperature as we reproduce in Fig.~\ref{fig:lattice_bulk}.

The latest hydrodynamical simulations are incorporating the $\zeta/s$ coefficient in order to include its effects in the collective flow observables.
As an example we show in Fig.~\ref{fig:v2Bozec} a recent calculation~\cite{Bozek:2011fm} showing the integrated elliptic flow as a function of centrality. The red points are data 
from the ALICE collaboration and the blue dots from the STAR collaboration at lower energy. In both cases, the elliptic flow is calculated by four-particle correlations in order to
minimize the non-flow effects. The two curves on the plot are the output from the hydrodynamic calculations using Glauber initial conditions, with an optimal value for $\eta/s$ and $\zeta/s$ of
0.08 and 0.04, respectively, showing that $\zeta/\eta$ could be of the order of $0.5$.

A limitation of these calculations is that they include a fixed value of $\eta/s$ and $\zeta/s$ for the whole evolution of the fireball. Some hydrodynamical
computations have also included the entire temperature dependence of the viscosities providing a more detailed description of the hydrodynamical expansion. 
The temperature dependence of the bulk viscosity has been qualitatively added in \cite{Song:2008hj} and more recently in \cite{Roy:2011pk}. The temperature dependence of $\eta/s$ has been added
in \cite{Niemi:2011ix} showing that the viscosity in the hadronic side gives the most relevant contribution at RHIC energies, whereas the QGP viscosity is more important at LHC energies.

In Table~\ref{tab:viscoesti} we summarize some values of the shear and bulk viscosities in the hadronic phase that have appeared in recent literature.

\begin{table}[t]
\begin{center}
\begin{tabular}{|c|c|c|c|} \hline 
Reference & Temperature & $\eta/s$ & $\zeta/s$  \\
\hline \hline
\cite{Song:2008hj} & $T<T_c$ & $1/(4\pi)$ & 0 \\
\cite{Song:2008hj} & $T_c$ & $1/(4\pi)$ & 0.04-0.08 \\
\cite{Niemi:2011ix} & 100 MeV & 0.6 &  - \\
\cite{Niemi:2011ix} & 180 MeV & 0.08& - \\
\cite{Dusling:2011fd} & $T\sim T_f$ & - & $\lesssim 0.05$  \\
\cite{Bozek:2011fm} & $T<T_c$ & 0.08 & 0.04  \\
\cite{Roy:2011pk} & $T<T_{cross}=175$ MeV & 0.08-0.16 & 0.023  \\
\cite{Roy:2011pk} & $T_{cross}=175$ MeV & 0.08-0.16 & 0.038  \\
\cite{Plumari:2011re} & $T<T_c$ & 0.64 & - \\
\hline
\end{tabular}
\caption{\label{tab:viscoesti}Some data-based estimates of $\eta/s$ and $\zeta/s$ in the hadronic side at different temperatures.}
\end{center}
\end{table}

\chapter[BUU Equation]{Boltzmann-Uehling-Uhlenbeck \\ Equation \label{ch:2.buuq}}

   When an external perturbation is applied to a system, this leaves the equilibrium state. The particles of the system, described
by the one-particle distribution function, suffer from many collisions producing a transport of the conserved quantities (energy, momentum and others).
The one-particle distribution function is then modified and its time evolution is described by a kinetic equation.
By knowing the one-particle distribution function out of equilibrium, and connecting with the hydrodynamic formalism
one is able to extract the transport coefficients that govern the relaxation of the fluid to the equilibrium state. We will start by introducing the 
Boltzmann-Uehling-Uhlenbeck kinetic equation 
and the method derived by Chapman and Enskog to solve it. For the sake of simplicity, in this chapter we will set the formalism for a one-species
gas, ignoring the conserved internal (or flavor-like) charges. Later, in Chapters~\ref{ch:7.strangeness} and ~\ref{ch:8.charm} we will attend to these.

\section[Distribution functions and BBGKY hierarchy] {$n$-particle distribution functions and \\
Bogoliubov-Born-Green-Kyrkwood-Yvon hierarchy}

\subsection{Classical description}

Consider a gas with a large number of particles $N$ inside a volume $V$ \cite{landau1984cours},\cite{liboff2003kinetic}.
Each of these particles is specified at time $t$ by its position $\mb{r}_i(t)$ (with $i=1,...,N$) and its momentum $\mb{q}_i(t)$.
Therefore the whole system is entirely described by a set of $6N$ coordinates in the so-called $\Gamma$-phase space\index{$\Gamma$-phase space}. 
A microstate of the system (ensemble of particles with definite positions and momenta at a given time) is specified by a
representative point in the $\Gamma$-phase space. This microstate must be compatible with the macrostate, defined by some thermodynamic
functions as $N$, $V$, $E$, $T$...
The evolution of the system is determined by the Hamilton's equation (the Hamiltonian of the system is a function of $t$, $\mb{r}_i$ and $\mb{q}_i$)
\be \label{eq:hamilton} \dot{\mb{q}}_i=- \frac{\pa H(t,r_i(t),q_i(t))}{\pa \mb{r}_i} \ , \qquad \dot{\mb{r}}_i = \frac{\pa H (t,r_i(t),q_i(t))}{\pa \mb{q}_i}  \ee
and the evolution of the system follows a curve in the $\Gamma$-phase space.

For systems with constant total energy, their microestates are restricted to those representative points in the phase space 
$(\mb{r}_i,\mb{q}_i)$ which $H(t,\mb{r}_i,\mb{q}_i)=E$.
For open systems there is no such a restriction (apart from the condition of being compatible with the macrostate) and some microstates are easier
to reach in the $\Gamma$-phase space than others. 
One can define the phase-space density\index{phase space density} $\rho(t,\mb{r}_i,\mb{q}_i)$ 
which is the probability for a system to be in a given microstate in the $\Gamma$-phase space.

The quantity \be \prod_{i=1}^N d^{3} \mb{r}_i d^{3} \mb{q}_i \ \rho(t,\mb{r}_i,\mb{q}_i ) \ee
 is the number of accesible microstates that at time $t$
are contained in the phase-space volume element $\prod_{i=1}^N d^{3}\mb{r}_i d^{3} \mb{q}_i$ centered at $(\mb{r}_i,\mb{q}_i)$.

The evolution of the phase-space density along a phase-space trajectory is given by the Liouville's \index{Liouville equation}equation:
\be \frac{d}{dt} \rho(t,\mb{r}_i,\mb{q}_i) = \frac{\pa}{\pa t} \rho(t,\mb{r}_i,\mb{q}_i) + \{ \rho(t,\mb{r}_i,\mb{q}_i),H\} =0\ , \ee
where $H$ is the Hamiltonian of the system. This equation can be obtained from Hamilton's equations (\ref{eq:hamilton}).\\

Any observable $O$, a function of the $\mb{r}_i$ and $\mb{q}_i$ can be averaged over the ensemble as follows:

\be \label{eq:average} \langle O(t) \rangle = \frac{\int \prod_{i=1}^N d^{3} \mb{r}_i \ d^{3} \mb{q}_i \ O(\mb{r}_i, \mb{q}_i) \ \rho (t,\mb{r}_i,\mb{q}_i )}
{\int \prod_{i=1}^N d^{3} \mb{r}_i \ d^{3} \mb{q}_i \ \rho (t,\mb{r}_i,\mb{q}_i )} \ . \ee \\
 

The building blocks of kinetic theory are the $n$-particle distribution functions \index{$n$-particle distribution functions}
or joint-probability distributions, $f (t,\mb{x}_1, \mb{p}_1, ..., \mb{x}_n,\mb{p}_n)$.
A $n$-particle distribution function represents the probability of finding at given time $t$ the particle 1 at $(\mb{x}_1,\mb{p}_1)$,
the particle 2 at $(\mb{x}_2,\mb{p}_2)$, and so on up to the particle $n$.

These distribution functions can be obtained from the phase-space density function $\rho(t,\mb{r}_i,\mb{q}_i)$ by integrating the appropriate remaining coordinates
of the phase-space. For example, the one-particle distribution function is defined as:
\be f (t,\mb{x},\mb{p}) = \langle \sum_i^N \delta(\mb{x}-\mb{r}_i ) \delta(\mb{p}-\mb{q}_i) \rangle \ . \ee

To have access to all the distribution functions would provide all the physical information of the system, completely equivalent
to knowing the phase-space density function. They can be obtained by solving their equations of evolution (as hard as solving the Liouville equation).
 The equation for $f$ is generated by integrating the Liouville equation \index{Lioville equation}
over the phase-space coordinates $\mb{r}_n, \mb{q}_n, ..., \mb{r}_N, \mb{q}_N$ by assuming a particular form of the Hamiltonian (with one- and two- particle
interaction potential).

After integrating the Liouville equation, one realizes that the equation of evolution for the $n$-particle distribution function
is non-linearly coupled with the $n+1$-distribution function (and so on up to $n=N$). This set of coupled integro-differential equations is called the BBGKY\glossary{name=BBGKY,description={Bogoliubov-Born-Green-Kyrkwood-Yvon}} (Bogoliubov-Born-Green-Kyrkwood-Yvon) hierarchy
of equations\index{BBGKY hierarchy}.

\section{Kinetic equation \label{sec:kin_eq}}

Taking only the first equation of the BBGKY hierarchy for the one-particle distribution function and performing
the approximation of substituting the two-particle distribution function by a product of two one-particle
distribution functions one obtains a closed equation for $f(t,\mb{x},\mb{p})$. The resulting equation is called a kinetic equation.

The assumptions for obtaining the kinetic equation are:
\begin{itemize}
 \item The fluid is a dilute medium in which only binary collisions occur. Therefore multiple collisions are neglected. However, the binary collisions
       may be, in principle, inelastic ($\pi \pi \rightarrow KK$) as well as elastic ($\pi \pi \rightarrow \pi \pi$).
 \item The collision time is much smaller than the mean free time between consecutive collisions. This assumption is valid when the gas is dilute enough. In terms of typical lengths this is expressed as
     \be L \gg \lambda_{mfp} \gg R \ , \ee
      where $L$ is the size of the system, $\lambda_{mfp}$ is the mean free path or length between succesive collisions and $R$ is the range of interaction
(typically a scattering length or the radius of the particles in the hard-sphere approximation).
 \item {\it Stosszahlansatz} \index{\emph{Stosszahlansatz}} or molecular chaos hypothesis. This implies the absence of particle correlations before the collision
       process takes place. This assumption entails decoupling the first equation in the BBGKY hierarchy by the replacement:
     \be f (t, \mb{x}_1, \mb{p}_1,\mb{x}_2, \mb{p}_2) \simeq f (t, \mb{x}_1, \mb{p}_1) f (t, \mb{x}_2, \mb{p}_2) \ . \ee
\end{itemize}

Because from now on we will only work with the one-particle distribution function we will change the notation:
\be f_p(t,\mathbf{x}) \equiv f(t,\mb{x},\mb{p}) \ , \ee
where the subindex $p$ denote both the momentum dependence of the distribution function and also a label for the particle entering in the elastic scattering process.
Considering a classical scattering between two particles, we will denote
by $k_1$ and $k_2$ the momenta of the two incoming particles; and $k_3$ and $p$ the momenta of the outgoing particles.
 
The time evolution of the one-particle distribution function $f_1(t,\mb{x})$ is given by the kinetic equation, which is of the following type:
\be \label{eq:kin_eq} \frac{df_1}{dt} = C [f_1,f_2] \ . \ee
The classical kinetic equation is called the Boltzmann equation and is known since L. Boltzmann derived it in 1872 for a gas of classical particles.
The collision operator of the Boltzmann equation in the right-hand side of (\ref{eq:kin_eq}) reads explicitly \cite{landau1981physical,de1980relativistic,liboff2003kinetic}
\be C[f_1,f_2]= \frac{1}{(2\pi)^3} \int d\Omega d\mathbf{k}_2 \ v_{rel} \ \frac{d\sigma_{12}}{d\Omega}  \ [f_3 f_p  -f_1 f_2 ] \ , \ee
where $v_{rel}$ is the relative velocity between the incoming particles, and $\frac{d\sigma_{12}}{d\Omega}$ the differential cross section of the process.
The existence of an equilibrium solution to this equation was proven by Boltzmann in the form of the $H-$theorem\index{Boltzmann's $H$-theorem}. 

The local $H$-theorem follows from that the entropy production at any time-space point is never negative. The entropy production is only vanishing when the solution of the kinetic equation is the local equilibrium
distribution function or equilibrium Maxwellian:
\be \label{eq:localequilibrium} f_p (t,\mb{x})= n_p (t,\mb{x}) \equiv \frac{1}{e^{\frac{ p^{\alpha} u_{\alpha} (t,\mb{x})-\mu(t,\mb{x})}{T(t,\mb{x})}}} \ , \ee
that satisfies the \index{detailed balance} detailed balance equation
\be n_1 (t,\mb{x}) n_2(t,\mb{x}) = n_3(t,\mb{x}) n_p(t,\mb{x}) \ . \ee

\subsection{Wigner function}

So far, the discussion has been purely classical. In quantum theory an analogous derivation can be made, using quantum-mechanical averages instead of (\ref{eq:average}).
The analogue to the one-particle distribution function is called Wigner function \cite{Wigner:1932eb} and it formally coincides with the classical distribution function.
Moreover, a global factor due to the quantum mechanical formulation appears in this function:
\be f^{C}_p(t,x)  \rightarrow f^{Q}_p(t,x) \ \frac{g}{(2\pi \hbar)^3} \ , \ee
where $g$ accounts for the quantum degeneracy of the particle ($g$=2 for electrons due to spin, $g$=3 for pions due to isospin or $g=2$ for photons due to the polarization states) and
the factor $1/(2\pi \hbar)^3$ comes from the fact that $d\mb{x} d\mb{p} /h^3$ is the number of quantum states in the infinitesimal phase-space volume.
Additionally, the collision operator is not written in terms of the cross section but in terms of the scattering matrix elements\index{S-matrix}.

When the Bose-Einstein must be applied (as is the case for pion at moderate temperatures) the kinetic equation is called the Boltzmann-Uehling-Uhlenbeck (BUU)
 \index{BUU equation} equation.\glossary{name=BUU,description={Boltzmann-Uehling-Uhlenbeck}} It contains some extra factors that accounts for the Bose-Einstein
nature of the particles and that produce an enhancement of the phase space in the available states. To be consistent with our later references we will focus on the BUU
equation for $f_p$:
\be \frac{df_p}{dt} = C[f_3,f_p] \ , \ee
where the collision operator of the BUU equation reads explicitly
\be C[f_3,f_p]= \frac{g_3}{1+\delta_{3,p}} \int d \Gamma_{12,3p} \ [f_1 f_2 (1 +f_3)(1+f_p) -f_3 f_p (1+f_1)(1+f_2)] \ , \ee
where $g_3$ is the degeneracy of the particle 3 and $1+\delta_{3,p}$ factor accounts for the possible 
undistinguishable particles in the final state. The scattering measure is
\be d \Gamma_{12,3p} \ \equiv \frac{1}{2E_p} \overline{|T|^2} \prod_{i=1}^3 \frac{d\mathbf{k}_i}{(2\pi)^3 2 E_i} \ (2\pi)^4 \delta^{(4)} (k_1 + k_2 -k_3 -p) \ .\ee
The local equilibrium distribution function is the Bose-Einstein function:
\be \label{localbe} f_p (t,\mb{x})= n_p (t,\mb{x}) \equiv \frac{1}{e^{\frac{ p^{\alpha} u_{\alpha} (t,\mb{x})-\mu(t,\mb{x})}{T(t,\mb{x})}}-1} \ . \ee
This function satisfies the detailed balance condition as well:
\be \label{eq:bosedetailedbalance} n_1 (t,\mb{x}) n_2(t,\mb{x}) [1+n_3(t,\mb{x})] [1+n_p(t,\mb{x})] = n_3(t,\mb{x}) n_p(t,\mb{x}) [1+n_1(t,\mb{x})] [1+n_2(t,\mb{x})] \ . \ee

In the following, we will denote as $x$ the space-time four-vector $(t,\mb{x})$ on which the hydrodynamic fields and distribution functions depend.

\section{\label{sec:chapman-enskog}Chapman-Enskog expansion}

The so-called Chapman-Enskog expansion \index{Chapman-Enskog expansion} is one of the several classical methods to obtain an approximate solution of the BUU equation.

In addition to the three length scales defined in Sec.~\ref{sec:kin_eq} one can introduce a characteristic hydrodynamic length $h$ which is
the typical size of the inhomogeneities of the system \cite{de1980relativistic},\cite{garzo2003kinetic}. The separation of scales are the following:
A particle suffers from a collision with another in a charateristic length being the range of interaction, $R$. After that, the particle moves freely
a distance of the order of $\lambda_{mfp}$ until it encounters another particle and collides again. Inside $h$, the particle suffers from many collisions.
Due to these scatterings the distribution function becomes close to the local equilibrium one. This local equilibrium state is characterized
by $\mu$, $\mb{u}$ and $T$ that vary from one region to another. In a larger time, the particle has travelled distances greater than $h$ and 
the differences in the three hydrodynamical fields smooth across the whole system. The gas reaches a state of global equilibrium defined by $\mu$, $\mb{u}$
and $T$ which do not depend on $x$.

We can summarize the hierarchy of scales as
\be \label{eq:lengths} L \gg h \gg \lambda_{mfp} \gg R \ ,\ee
where $L$ is the typical size of the system.
In terms of characteristic times, one can divide the previous inequalities by the thermal velocity $v \sim \sqrt{T/m}$.
\be L/v \gg \tau_h \gg \tau_{mft} \gg \tau_R \ ,\ee
where $\tau_h=h/v$ is the characteristic time of travel through the inhomogeneities of the system, $\tau_{mft}$ is the mean free time, and $\tau_R=R/v$ is the duration of a collision. 

In this scenario there are two main time scales governed by a fast and a subsequent slow processes:
1) The fast relaxation from the non-equilibrium initial state to a local equilibrium state, due to many collisions inside $h$. The time of local equilibration
is of the order of $\tau_{mft}$. This stage is called the kinetic regime, sensitive to initial state.
2) The slow relaxation from local to global equilibrium, at distances of several $h$. The time needed for this process is of the order of $\tau_h$.
This process is called the hydrodynamic regime. It does not depend on the initial state but only on the hydrodynamic fields $T(x),\mu(x), \mb{u}(x)$ that depend on
the time-space variables.

According to this, one expects that the one-particle distribution function in the second stage depends on space-time through a 
functional in the hydrodynamical variables:
\be f_p(x)= f_p [T(x),\mu(x),\mb{u}(x) ] \ . \ee
A solution of the BUU equation\index{BUU equation} of this type is called a {\it normal solution}\index{normal solution}.

The Chapman-Enskog \index{Chapman-Enskog expansion}procedure is a systematic way of constructing a normal solution to the BUU equation in powers of
the Knudsen number (Kn $= \lambda_{mfp}/h$).

To proceed, take the BUU equation (we drop the argument $(t,\mb{x})$ of the distribution function to ease the notation)
\be \frac{d f_p}{dt} = C[f_3,f_p] \ , \ee
separate the convective time derivative and the gradient operator to get
\be \pa_t f_p = - v^i \nabla_i f_p + C[f_3,f_p] \ee
 and divide the right-hand side of the equation by $f_p$. The first term of the right-hand side is the inverse characteristic length for the inhomogeneities
\be \left| \frac{p^i}{E_p} \nabla_i \ln f_p \right| \simeq h^{-1} v \ee
and the second is the inverse mean free path 
\be \left| \frac{C[f_3,f_p]}{f_p} \right| \simeq \lambda_{mfp}^{-1} v \ . \ee
Taking into account the inequalities in Eq.~(\ref{eq:lengths}) we deduce that $C[f_p,f_p]/f_p$ is much smaller than $v^i \nabla_i \ln f_p$. 
Since $h^{-1} \simeq |\nabla \ln f_p |$, the expansion in powers of the Knudsen number\index{Knudsen number} is actually equivalent to an expansion in powers of hydrodynamical gradients.

The way of translating the separation of scales into the BUU equation is the following.
The normal solution to the kinetic equation is expanded:
\be f_p=f_p^{(0)} + \epsilon f_p^{(1)} + \epsilon^2 f_p^{(2)} + \cdots \  , \ee
where $\epsilon$ is the so-called non-uniformity parameter and it measures the relative strength of the gradient. For consistency, it is set to one at
the end of the calculation, so it is nothing but a book-keeping parameter that can be interpreted as the Knudsen number, that controls the order of 
the approximation.

The spatial gradient is formally substituted by, $\nabla_i \rightarrow \epsilon \nabla_i$, and the time derivative of the normal solution is expanded
\be \pa_t f_p = \epsilon (\pa_t )^{(1)} f_p + \epsilon^2 (\pa_t)^{(2)} f_p+ \cdots \ .  \ee
The action of $\pa_t^{(i)}$ occurs through the dependence on the hydrodynamic fields
\be \label{eq:temp_deri} \pa_t^{(i)} f_p = 
\pa_t^{(i)} T \frac{\pa f_p}{\pa T}+\pa_t^{(i)} \mu \frac{\pa f_p}{\pa \mu} +\pa_t^{(i)} \mb{u} \cdot \frac{\pa f_p}{\pa \mb{u}} \ . \ee
The operators $\pa_t^{(i)} \mu, \pa_t^{(i)} T$ and $\pa_t^{(i)} \mb{u}$ are obtained from the macroscopic conservation laws
performing the same expansion and equating the terms with equal powers of $\epsilon$. Up to order $\epsilon^1$ they explicitly read 

\be \left\{ 
\begin{array}{lll} \label{eq:euler}
\pa_t^{(0)} T  &=&  \pa_t^{(0)} \mu = \pa_t^{(0)} \mb{u}= 0 \ , \\
\pa_t^{(1)} T & = & -T \left(\frac{\pa P}{\pa \epsilon} \right)_n \mb{\nabla} \cdot \mb{V} \ , \\
\pa_t^{(1)} \mu  & = & - \left[\mu \left( \frac{\pa P}{\pa \epsilon} \right)_n + \left(\frac{\pa P}{\pa n} \right)_{\epsilon} \right] \mb{\nabla} \cdot \mb{V} \ , \\
\pa_t^{(1)} \mb{u} & =& -\frac{\mb{\nabla} P}{w} \ . \\
\end{array} \right.
\ee

One sees that is enough to consider the ideal gas approximation. For higher orders the situation is more complicated and
one should take into account the $\epsilon$-expansion inside the energy-momentum tensor and four-particle flux.

Then, one substitutes all the previous equations into the BUU equation and identifies terms with equal powers in $\epsilon$.
One obtains the following hierarchy of equations
\be \left\{ 
\begin{array}{lll}
0 & = & C [ f_3^{(0)} ,f_p^{(0)} ] \ , \\
\pa_t^{(1)} f_p^{(0)} + \frac{p^i}{E_p} \nabla_i f_p^{(0)} & = &  C [f_3^{(0)},f_p^{(1)}]+C [f_3^{(1)},f_p^{(0)}] \ . \\
\end{array} \right. \ee

The solution of the zeroth-order equation is the local Bose-Einstein function of Eq.~(\ref{localbe}) with arguments depending on time and
space (sometimes called Juetner distribution function). This zeroth order approximation reads
\be f_p^{(0)} (x)= n_p (x) \ . \ee
The next order gives the first nontrivial contribution to the distribution function.
In this dissertation we will stop at first order:
\be  \label{eq:chapman-enskog} f_p (x)= n_p (x)+ f_p^{(1)} (x) \ . \ee
The first-order kinetic equation reads
\begin{eqnarray}
\label{eq:firstce} \pa_t^{(1)} f_p^{(0)} + \frac{p^i}{E_p} \nabla_i f_p^{(0)} & = & -\frac{g}{2} \int d \Gamma_{12,3p} \ (1+n_1)(1+n_2)n_3 n_p   \\
 & & \times  \left( \frac{f^{(1)}_p}{n_p(1+n_p)} + \frac{f^{(1)}_3}{n_3 (1+n_3)}-\frac{f^{(1)}_1}{n_1(1+n_1)}-\frac{f^{(1)}_2}{n_2 (1+n_2)}\right) \nonumber \ , 
\end{eqnarray}
where $g=3$ for the pion isospin degeneracy.

Observing the form of the collision operator (with Bose statistics) the {\it ansatz} for $f_p^{(1)}$ is conveniently parametrized as
\be \label{eq:param} f_p^{(1)} = - n_p (1+n_p) \Phi (\mb{p}) \ . \ee
$\Phi(\mb{p})$ is an adimensional function of $\mb{p}$ that will contain an appropriate hydrodynamic gradient depending on the transport coefficient.

One realizes here the main feature in the Chapman-Enskog expansion at first order. The left-hand side of Eq.~(\ref{eq:firstce})
only depends on derivatives of the local equilibrium distribution function (through the hydrodynamic fields) and therefore it does not depend on $\Phi(\mb{p})$.

\subsection{Left-hand side of the BUU equation}

At order $\epsilon^0$ we have seen that the equation for $f_p^{(0)}$reads
\be 0 = C[f_3^{(0)},f_p^{(0)}] \ee
and the solution is the local Bose-Einstein distribution function. In an arbitrary frame it reads
\be n_p [ T(x),\mu(x),u^{\alpha}(x) ] = \frac{1}{e^{\frac{ p^{\alpha} u_{\alpha}(x) -\mu(x)}{T(x)}}-1} \ , \ee
that reduces to the usual Bose-Einstein distribution function in the comoving frame ($\mathbf{V} \neq 0$, $\gamma=1$).

One important remark is that the five independent hydrodynamical fields $T(x),\mu(x)$ and  $u^{i}(x)$ are not necessary the same as in equilibrium
and one should fix them appropriately. The prescription to do this is to impose the conditions of fit, that we will describe in the next section.

With the parametrization used in Eq.~(\ref{eq:param}), $f_p = n_p  - n_p (1+n_p) \Phi (\mb{p})$ the order $\epsilon$ reads

\be \nonumber E_p \pa_t^{(1)} n_p(x) + p^i \ \nabla_i n_p (x) =\ee
\be \label{eq:linear_equation} \frac{gE_p}{2} \int d \Gamma_{12,3p} \ (1+n_1)(1+n_2)n_3 n_p \left( \Phi_p + \Phi_3  - \Phi_1 - \Phi_2 \right).\ee
Note that we have multiplied the equation by $E_p$ for convenience.

The collision operator has became a linearized operator in $\Phi_p$.
We focus on the left-hand side of Eq.~(\ref{eq:linear_equation}) to obtain the final form of this term.

One uses first Eq.~(\ref{eq:temp_deri}) for the temporal derivative, together with the Euler equations\index{Euler's equation} Eq.~(\ref{eq:euler}),
for the derivatives of the hydrodynamical field and the Leibniz rule for the derivative of $n_p$ with respect to these fields
\be \pa_A n_p(x)= -n_p(x) [1+ n_p(x)] \pa_A \left[ \beta (p^{\alpha} u_{\alpha} -\mu)\right] \ , \ee
with $A=T,\mu,u^i$.
The gradient is obtained by using the Leibniz rule with
\be \nabla_i \beta = - \beta^2 \nabla_i T \ee
and the Gibbs-Duhem relation 
\be \label{eq:gibbs-duhem} \nabla_i \mu = -\frac{s}{n} \nabla_i T + \frac{1}{n} \nabla_i P \ . \ee

As an intermediate step, we perform the separation into a traceless and ``traceful'' parts.
\be p^i p^j \pa_i V_j = p^i p^j \left( \frac{1}{2} \left( \pa_i V_j + \pa_j V_i \right) - \frac{1}{3} \delta_{ij} \mathbf{\nabla} \cdot \mathbf{V} +\frac{1}{3} \delta_{ij} \mathbf{\nabla} \cdot \mathbf{V} \right) = p^i p^j ( \tilde{V}_{ij} + \frac{1}{3} \delta_{ij} \diver{V} ). \ee
The first term goes to the shear viscosity and is the ``traceless'' part because it satisfies $\sum_i \tilde{V}_{ii} =0$. The part that goes with the bulk viscosity is the ``traceful'' part $\frac{1}{3} \delta_{ij} \diver{V}$.

The final solution reads:
\begin{eqnarray}
 \nonumber E_p \frac{df_p}{dt} & =&  \beta n_p(x) (1+n_p (x))  \left\{ p^i p^j \tilde{V}_{ij}  + \left( \frac{1}{3} \mathbf{p}^2 -  E_p^2 v^2_n -E_p  \kappa^{-1}_{\epsilon} \right) \mathbf{\nabla} \cdot \mathbf{V} \right. \\
& & \left.  + \beta \left( E_p - \frac{w}{n}  \right) \mathbf{p} \cdot  \left( \nabla T - \frac{T}{w} \nabla P \right) \right\} \ , \label{eq:finallhs}
\end{eqnarray}
where we have defined the isochorus sound speed and the compressibility\index{isochorus speed of sound}\index{compressibility} at constant energy density

\be v^2_n= \left( \frac{\pa P}{\pa \epsilon} \right)_{n}, \quad \kappa^{-1}_{\epsilon} = \left( \frac{\pa P}{\pa n} \right)_{\epsilon} \ . \ee

These two quantities can be calculated from derivatives of the pressure by the following formulae

\begin{eqnarray}
 v^2_n & = &  \frac{s\chi_{\mu \mu}-n \chi_{\mu T}}{C_V \chi_{\mu \mu}} \ , \\
\kappa^{-1}_{\epsilon} & = & \frac{nT \chi_{TT} + (n\mu -sT) \chi_{\mu T} - s\mu \chi_{\mu \mu}}{C_V \chi_{\mu \mu}} \ , 
\end{eqnarray}

where $n=\left(\frac{\pa P}{\pa \mu} \right)_T$, $s=\left(\frac{\pa P}{\pa T}\right)_{\mu}$, the susceptibilities \index{thermal susceptibilities} 
\be \label{eq:susceptibilities} \chi_{xy}= \frac{\pa^2 P}{\pa x \pa y} \ee
and the specific heat \index{specific heat}
\be C_V= T \left( \frac{\pa s}{\pa T} \right)_{V} = T \left( \chi_{TT} - \frac{\chi^2_{\mu T}}{\chi_{\mu \mu}}\right) \ .\ee

One important remark to take into account is that from the previous lines one can see that at first order in the Chapman-Enskog expansion
all the thermodynamic functions that appear in the linearized Boltzmann equation should be defined as in equilibrium. That means, that the
functions appearing in the left-hand side of the BUU equation~(\ref{eq:finallhs}) should be taken as those for an ideal Bose-Einstein gas.

\subsubsection{Shear Viscosity}
The shear viscosity \index{shear viscosity} appears when the perturbation of the fluid is exclusively by a shear perturbation of the velocity field.
For this reason the term associated with the shear viscosity in Eq.~(\ref{eq:finallhs}) is

\be \label{eq:lhsshear}\left. p_{\mu} \pa^{\mu} n_p (x) \right|_{\eta}= \beta n_p (x) (1+n_p(x)) \ p^i p^j \tilde{V}_{ij} \ . \ee

\subsubsection{Bulk Viscosity}
The part of Eq.~(\ref{eq:finallhs}) describing uniform compression or expansion of the fluid is related with the bulk viscosity: \index{bulk viscosity}
\be \label{eq:lhsbulk} p_{\mu} \pa^{\mu} n_p(x) |_{\zeta}=  \beta n_p(x) (1+n_p(x)) \left( \frac{1}{3} \mathbf{p}^2 -  E_p^2 v_n^2- E_p \kappa^{-1}_{\epsilon} \right) \mathbf{\nabla} \cdot \mathbf{V} \ . \ee

\subsubsection{Heat Conductivity}
Finally, the terms of Eq.~(\ref{eq:finallhs}) coming with the heat conductivity \index{heat conductivity} are those related with gradients in temperature and pressure:
\be \label{eq:lhsheat} p_{\mu} \pa^{\mu} n_p(x) |_{\kappa}= \beta^2 n_p(x) (1+n_p(x)) \left( E_p - \frac{w}{n} \right) \mathbf{p} \cdot  \left(\nabla T - \frac{T}{w} \nabla P \right) \ . \ee

\section{Conditions of fit \label{sec:condsfit}} 

As pointed out before, the hydrodynamic functions are not necessarily the same as in equilibrium but it is convenient to 
make this identification. The formal way of doing that is to claim that the energy density, particle density and the particle
 or energy flux are entirely defined in equilibrium.  
The three conditions needed for fully specify the hydrodynamical variables are called ``conditions of fit'' \index{conditions of fit} and
we detail them here:
\begin{itemize}
 \item { \it Condition of fit no.1:} The energy density is defined as being the same as in equilibrium. In the local reference frame this quantity corresponds to the 00 component
of the energy-momentum tensor. Therefore, this amounts in
\be \label{eq:cond1} T^{00} = T^{00}_{eq} \rightarrow \tau^{00}=0 \ . \ee
Implicitly, this condition defines the temperature field in non-equilibrium to be the same as in equilibrium.
 \item {\it Condition of fit no.2:} The particle density is defined as being the same as in equilibrium. This condition makes only sense if
the system has a conserved current (particle, charge...). In the local reference frame this corresponds to the 0 component of the four-particle flow
\be \label{eq:cond2} n^{0} = n^{0}_{eq} \rightarrow \nu^0=0 \ . \ee 

This condition fixes the chemical potential associated with the conserved current in the non-equilibrium as being the same as in equilibrium.
\end{itemize}

The hydrodynamical velocity posseses a similar condition to fix. However, one can define it to be parallel to the particle flow (Eckart's choice) or to the 
energy flow (Landau's choice) because in a relativistic theory both quantities are not necessarily parallel. Even an intermediate choice can be used.
\begin{itemize}
   \item {\it Condition of fit no. 3a (Landau or Landau-Lifshitz condition\index{Landau-Lifshitz condition of fit|see{conditions of fit}}):}  The energy flux is defined as being the same as in equilibrium. That makes the velocity field to be parallel as
the energy flux. This condition can be applied to a system with or without a conserved current.
\be \label{eq:cond3a} T^{0i}=T^{0i}_{eq} \rightarrow \tau^{0i}=0 \ . \ee
 Moreover, in the local rest frame this implies that there is no energy flux at all
  \be T^{0i}=0 \ .\ee

  \item {\it Condition of fit no. 3b (Eckart condition):}  The particle flux is defined as in equilibrium. That defines the velocity field to be parallel to
the particle flux. This condition can only be applied to a system with a conserved current.
\be \label{eq:cond3b} n^{i}=n^{i}_{eq} \rightarrow \nu^{i}=0 \ . \ee
 Moreover, in the local frame this implies that there is no particle flux at all
  \be n^{i}=0 \ .\ee

\begin{center}
\begin{tabular}{|cc|cc|} 
\hline
Variable to & Invariant & Expression & Expression \\
be defined& quantity & (arb. frame) & (local rest frame) \\
\hline
$T$ & Energy density & $u_{\nu} u_{\nu} \tau^{\mu \nu}=0$ & $\tau^{00}=0$ \\
$\mu$ & Particle density & $u_{\mu} \nu^{\mu}=0$ & $\nu^0=0$ \\
$u^i$ & Energy flux & $\Delta^{\mu \nu} T_{\nu \sigma} u^{\sigma} =0$ & $T^{0i}=0 \rightarrow \tau^{0i}=0^*$ \\ 
$u^i$ & Particle flux & $\Delta^{\mu \nu} n_{\nu} =0$ & $n^i=0 \rightarrow \nu^i=0^*$ \\ 

\hline
\end{tabular}
\end{center}
*) In equilibrium $T_{eq}^{0i}=0$ and $n_{eq}^{i}=0$ because of the symmetry of the integrand in (\ref{eq:ndensimicro}) and (\ref{eq:energymommicro}).

\end{itemize}

\chapter[$\eta$ and KSS Coefficient]{Shear Viscosity and \\ 
KSS Coefficient\label{ch:3.shear}}

The experimental results in relativistic heavy-ion collisions at RHIC and LHC indicate
that the description of the expanding system as a nearly ideal fluid is well-suited. The use
of viscous hydrodynamics should provide a more accurate and detailed picture of the system
by taking into account the leading corrections to the ideal fluid description.

As seen in Chapter \ref{ch:1.intro} the most relevant transport coefficient for understanding
some of the properties of the collective flow in relativistic heavy-ion collisions is the shear viscosity
over entropy density (the KSS number). For example, because of its dependence over the behaviour of the flow coefficients as a function
of $p_{\perp}$ and centrality.

The shear viscosity has also been calculated in the color-flavor locked phase of dense quark matter at low temperature in~\cite{Manuel:2004iv}.
This result is of interest for describing the properties of rotating compact stars.

In this chapter we will calculate the shear viscosity of a pion gas and the KSS coefficient\index{KSS coefficient} showing that it is plausible
 to have a minimum in the crossover temperature of deconfinement.

\section{Shear viscosity \label{sec:shearvisco}}

We read the left-hand side of the linearized equation for the shear viscosity from Eq.~(\ref{eq:lhsshear}):

\be \left. p_{\mu} \pa^{\mu} n_p (x) \right|_{\eta}= \beta \ n_p (x) [1+n_p(x)] \ p^i p^j \ \tilde{V}_{ij} \ , \ee
with $\tilde{V}_{ij}$ being
\be \tilde{V}_{ij} = \frac{1}{2} \left( \pa_i V_j + \pa_j V_i \right) - \frac{1}{3} \delta_{ij} \nabla \cdot \mathbf{V} \ .\ee

The linearized right-hand side reads:

\be \frac{g_{\pi}E_p}{2} \int d\Gamma_{12,3p} \ (1+n_1)(1+n_2)n_3 n_p \left( \Phi_p+ \Phi_3 -\Phi_1 - \Phi_2 \right) \ .\ee

We will use the following parametrization for the function $\Phi_a$:
\be \label{eq:phishear} \Phi_a = \beta^3 \  B_a^{ij} \ \tilde{V}_{ij} \ , \ee
where $B_a^{ij}$ is a function of $\mb{k}_a$. In \cite{dobado2004viscosity} the following parametrization was used
\be B_a^{ij} = \left( k_a^i k_a^j - \delta^{ij} k^2_a \right) B(k_a) \ ,\ee
where $B(k_a)$ is an adimensional function of $k_a$. In \cite{teaney2010viscous}, they use (with a different normalization)
an alternative factorization
\be \label{Teaney} B_a^{ij} = k_a^i k_a^j \ B(k_a) \ , \ee
but one can easily check that both are equivalent due to the fact that $\tilde{V}_{ij} \delta^{ij} = 0$. 

The kinetic equation must hold for any components of the tensor $\tilde{V}_{ij}$ and therefore the equation that we must solve for 
the function $B(p)$ is
\begin{eqnarray} \nonumber n_p (1+n_p) p^i p^j &= & \frac{g_{\pi} E_p}{2T^2} \int d \Gamma_{12,3p}  (1+n_1)(1+n_2)n_3 n_p \\
  \label{eq:shear_buu} && \times\left[ p^i p^j B(p) + k_3^i k_3^j B(k_3) -  k_1^i k_1^j B(k_1) - k_2^i k_2^j B(k_2) \right] \ .  
\end{eqnarray}


We still need to connect the shear viscosity with the unknown function $B(p)$. To do so, one employs hydrodynamics and kinetic theory.
The spatial components of the shear-stress tensor for an isotropic gas read (see Eq.~(\ref{eq:tau})):
\be \tau_{ij} = -2 \ \eta\ \tilde{V}_{ij}, \ee
where $\eta$ is the shear viscosity.

From kinetic theory, the shear-stress tensor is expressed as an average over the non-ideal distribution function. In the first order Chapman-Enskog expansion, 
it reads
\be \tau_{ij} =g_{\pi} \int \frac{d^3 p}{(2\pi)^3} \ f_p^{(1)} \ \frac{p_i p_j}{E_p} .\ee

The last step is to connect the two expressions for the shear-stress tensor with the help of the following identity \cite{chapman1991mathematical} to eliminate $\tilde{V}_{ij}$:
\be \int d^3 p \ f(p) p_i p_j p^k p^l \ W_{kl} = \frac{2}{15} W_{ij} \int d^3p \ f(p) p^4, \ee
with the arbitrary tensor $W_{kl}$ independent of $p^i$.

We finally obtain the desired expression of the shear viscosity:
\be \label{eq:viscofinal} \eta = \frac{g_{\pi}}{30 \pi^2 T^3} \int \frac{dp}{E_p} n_p (1+n_p) p^6  B(p). \ee

\subsection{Integration measure}

We will use adimensional variables to express all the integrals. We have found that the most convenient change of variables is
\be \label{eq:new_var} x=\frac{E_p}{m}; \quad p=m\sqrt{x^2-1}; \quad y=\frac{m}{T} \ .\ee

This choice has to do with the existence of zero modes\index{zero modes} in the bulk viscosity and conductivities, as will be seen in Sections~\ref{sec:zeromodes} and \ref{sec:conducinteg}.
The zero modes appear because of
the presence of conserved quantities. Due to the tensorial structure inside the collision operator, the shear viscosity does not suffer of zero modes. 

When expanding the perturbation function inside the collision integral in powers of this variable $x$, it is possible to identify and extract
these zero modes. The choice of a different variable can hide the zero modes provoking an inconsistency in the solution of the linearized equation.
For the shear viscosity, this choice of variables or another (for example, in the work of Llanes-Estrada and Dobado \cite{dobado2004viscosity}) are equally
 valid, but this is not the case for transport coefficients that present zero modes in the linearized
collision integral. The discussion will be clear in Chapters~\ref{ch:4.bulk} and \ref{ch:5.conductivities} when looking at the bulk viscosity and the conductivities.

The expression for the shear viscosity becomes

\be \eta = \frac{g_{\pi} m^6}{30 \pi^2 T^3}\int dx \ (x^2-1)^{5/2} \frac{z^{-1} e^{y (x-1)}}{\left[z^{-1}
e^{y(x-1)}-1 \right]^2} \ B(x) \ . \ee

This expression naturally defines an integration measure that will be characteristic of $\eta$, and a similar one will appear for each one of the transport coefficients.
The integration measure reads

\be d\mu_{\eta} (x;y,z) = dx \ (x^2-1)^{5/2} \frac{z^{-1} e^{y (x-1)}}{\left[z^{-1}
e^{y(x-1)}-1 \right]^2}  \ , \ee
and it has all the properties to be a consistent integration measure.
Written in physical variables it reads:
\be d\mu_{\eta}= d^3 p \frac{1}{4\pi m^6} \frac{p^4}{E_p} n_p (1+n_p) \ .\ee

One defines an inner product in the following way
\be \label{eq:product_shear}\langle A(x) | B(x) \rangle (y,z) \equiv \int_1^{\infty} d\mu_{\eta} (x; y,z) \ A(x) B(x) \ . \ee

With the notion of perpendicularity given by the inner product, one can construct a polynomial basis in powers of $x$.
We choose it to be a monic orthogonal basis. Its first elements read:
\be 
\begin{array}{lll}
P_0(x) & = & 1 \, \\
P_1(x) & = & x+P_{10} \, \\
P_2(x) & = & x^2+P_{21}x+P_{20} \ , \\
\end{array}
\ee

with the coefficients $P_{10} = \frac{-K_1}{K_0}$, $P_{21}=\frac{K_0 K_3 - K_1 K_2}{K_1^2-K_0 K_2}$ and $P_{20}=\frac{K_2^2-K_1K_3}{K_1^2-K_0K_2}$, ... conveniently 
obtained on a computer by a generalization of the Gram-Schmidt's method. Defined in this way, the elements of the basis satisfy:

\be \langle P_i | P_j \rangle = ||P_i||^2 \delta_{ij} \ . \ee

For convenience we also define the functions $K_i (y,z)$ with $i=n+m$:
\be \label{eq:Kintegrals} K_i \equiv \langle x^n | x^m \rangle = \int d\mu_{\eta} \ x^i = \int dx \ (x^2-1)^{5/2} \frac{z^{-1} e^{y (x-1)}}{\left[z^{-1}
e^{y(x-1)}-1 \right]^2} x^i.\ee

The shear viscosity is simply expressed as
\be \label{eq:shear_prod} \eta = \frac{g_{\pi} m^6}{30 \pi^2 T^3}\langle B(x) | P_0 \rangle \ .\ee

We project the kinetic equation Eq.~(\ref{eq:shear_buu}) by multiplying it by 
\be \frac{1}{4\pi m^6} \frac{p_i p_j}{E_p}\ee
and contracting the indices $i$ and $j$ to obtain

\be \nonumber \frac{1}{4\pi m^6} n_p (1+n_p) \frac{p^4}{E_p} P_0(x)  = \frac{g_{\pi}}{8 \pi m^6 T^2} \int d \Gamma_{12,3p}  (1+n_1)(1+n_2)n_3 n_p \ee
\be \times p_i p_j \left[ p^i p^j B(p) + k_3^i k_3^j B(k_3) -
 k_1^i k_1^j B(k_1) - k_2^i k_2^j B(k_2) \right] \ .\ee

Now, multiply by $P_l(x)$ and integrate on $d^3p$ on both sides

\be \delta_{l0} ||P_0||^2 \nonumber \ee
\be = \frac{g_{\pi} \pi^2}{m^6 T^2} \int \prod_{m=1}^4 \frac{d^3 k_m}{(2\pi)^32E_m} \overline{|T|^2} (2\pi)^4 \delta^{(4)} (k_1+k_2-k_3-p)  (1+n_1)(1+n_2)n_3 n_p \nonumber \ee
\be \times p_i p_j P_l(x) \left[ p^i p^j B(p) + k_3^i k_3^j B(k_3) -
 k_1^i k_1^j B(k_1) - k_2^i k_2^j B(k_2) \right] \ ,\ee
with $||P_0||^2=K_0$.

Then, one can expand the $B(x)$ function in the polynomial basis we defined before:
\be \label{eq:shear_exp} B(x) = \sum_{n=0}^{\infty} b_n P_n (x) \ . \ee

Symmetrizing the kinetic equation with the help of the detailed balance condition (\ref{eq:bosedetailedbalance}) one gets the final matricial equation:
\be \delta_{l0} K_0 \nonumber \ee
\be = \sum_{n=0}^{\infty} b_n \frac{g_{\pi} \pi^2}{4m^6 T^2} \int \prod_{m=1}^4 \frac{d^3 k_m}{(2\pi)^32E_m} \overline{|T|^2} (2\pi)^4 \delta^{(4)} (k_1+k_2-k_3-p)  (1+n_1)(1+n_2)n_3 n_p \nonumber \ee
\be \times \left[ p_i p_j P_l(p) + p_{3i} p_{3j} P_l(p_3) - p_{1i} p_{1j} P_l (p_1) - p_{2i} p_{2j} P_l (p_2) \right] \nonumber \ee
\be \times \left[ p^i p^j P_n(p) + p_3^i p_3^j P_n(p_3) -
 p_1^i p_1^j P_n(p_1) - p_2^i p_2^j P_n(p_2) \right] \ ,\ee
that is solved order by order in the indices $l$ and $n$.

Defining the following collision matrix,

\begin{eqnarray}
  \mathcal{C}_{nl} & = & \frac{g_{\pi} \pi^2}{4m^6 T^2} \int \prod_{m=1}^4 \frac{d^3 k_m}{(2\pi)^32E_m} \overline{|T|^2} (2\pi)^4 \delta^{(4)} (k_1+k_2-k_3-p)  (1+n_1)(1+n_2)n_3 n_p \nonumber \\
 & & \times \left[ p_i p_j P_l(p) + k_{3i} k_{3j} P_l(k_3) - k_{1i} k_{1j} P_l (k_1) - k_{2i} k_{2j} P_l (k_2) \right] \nonumber \\
 & &  \times \left[ p^i p^j P_n(p) + k_3^i k_3^j P_n(k_3) - k_1^i k_1^j P_n(k_1) - k_2^i k_2^j P_n(k_2) \right] \ , 
\end{eqnarray}

we can write the linear system as
\be \label{eq:shear_system} \sum_{n=0}^N \mathcal{C}_{nl} b_n=K_0 \delta_{l0} \ . \ee

Truncating at $N=0$ ($1\times1$ problem):
\be b_0= K_0 \mathcal{C}_{00}^{-1} \ , \ee
with
\begin{eqnarray} 
\mathcal{C}_{00} & = & \frac{g_{\pi}\pi^2}{4m^6 T^2} \int \prod_{m=1}^4 \frac{d^3 k_m}{(2\pi)^32E_m} \overline{|T|^2} (2\pi)^4 \delta^{(4)} (k_1+k_2-k_3-p)  (1+n_1)(1+n_2)n_3 n_p \nonumber \\
& & \label{eq:collision00} \times \Delta [k_i k_j]  \Delta [k^i k^j] \ , 
\end{eqnarray}
where 
\be \Delta [k_i k_j] \equiv \left[ p_i p_j + k_{3i} k_{3j} - k_{1i} k_{1j} - k_{2i} k_{2j} \right].\ee
The shear viscosity reads then
\be \eta^{(0)} =\frac{g_{\pi}m^6}{30 \pi^2 T^3} b_0 \langle P_0 | P_0 \rangle = \frac{g_{\pi}m^6}{30 \pi^2 T^3} \frac{K_0^2}{\mathcal{C}_{00}} \ .\ee

Truncating at $N=1$ ($2\times2$ system):
\be
\begin{array}{lll}
 \mathcal{C}_{00} b_0 + \mathcal{C}_{01} b_1 & = & K_0 \ , \\
 \mathcal{C}_{10} b_0 + \mathcal{C}_{11} b_1 & = & 0 \ , \\
\end{array} \ee
where $\mathcal{C}_{10} = \mathcal{C}_{01}$. To obtain the shear viscosity coefficient we only need the solution for $b_0$, because of Eq.~(\ref{eq:shear_prod}):
\be b_0=\frac{\mathcal{C}_{11} K_0}{\mathcal{C}_{00} \mathcal{C}_{11}-\mathcal{C}^2_{01}} =
 \frac{K_0}{\mathcal{C}_{00}} \left( 1+ \frac{\mathcal{C}^2_{01}}{\mathcal{C}_{00} \mathcal{C}_{11}-\mathcal{C}^2_{01}} \right) \ ,\ee
to finally obtain
\be \eta^{(1)} =\frac{g_{\pi}m^6}{30 \pi^2 T^3} b_0 \langle P_0 | P_0 \rangle = \frac{g_{\pi}m^6}{30 \pi^2 T^3} \frac{K_0^2}{\mathcal{C}_{00}}
\left( 1+ \frac{\mathcal{C}^2_{01}}{\mathcal{C}_{00} \mathcal{C}_{11}-\mathcal{C}^2_{01}} \right) \ .\ee

The difference between the two first orders can be obtained explicitly:
\be \frac{\eta^{(1)}}{\eta^{(0)}}= 1 + \frac{\mathcal{C}^2_{01}}{\mathcal{C}_{00} \mathcal{C}_{11}-\mathcal{C}^2_{01}}  \ . \ee

\begin{figure}[t]
\begin{center}
\includegraphics[scale=0.5]{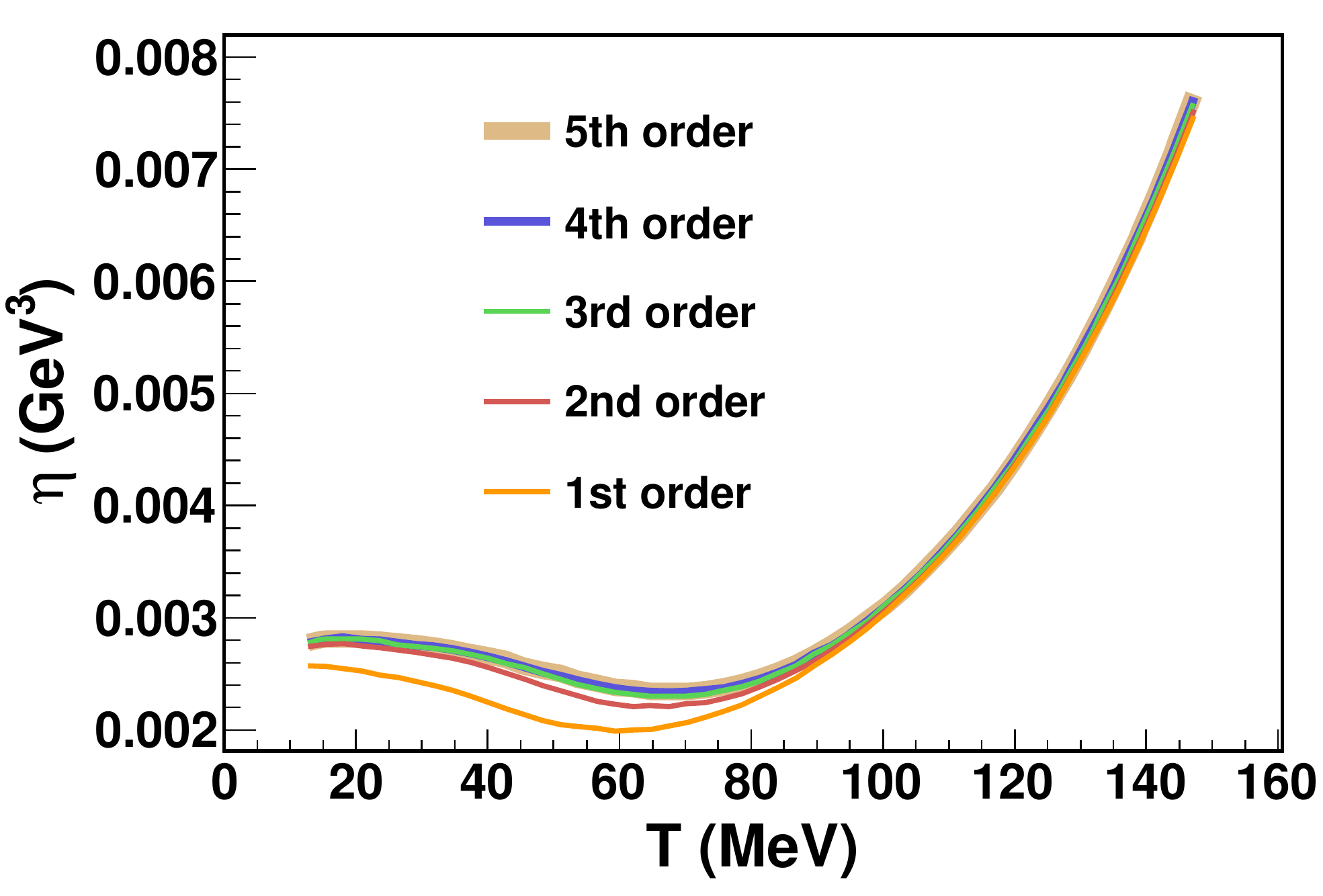} 
\caption{\label{fig:shearorders} Shear viscosity of a pion gas with $\mu=0$ for different orders in the power expansion (\ref{eq:shear_exp}) showing fast convergence. }
\end{center}
\end{figure}

In the computational program, we automatically solve the matricial system~(\ref{eq:shear_system}) for a given order. Very good convergence is achieved 
in the polynomial expansion, as seen in Fig.~\ref{fig:shearorders} up to fifth order.

\begin{figure}[t]
\begin{center}
\includegraphics[scale=0.35]{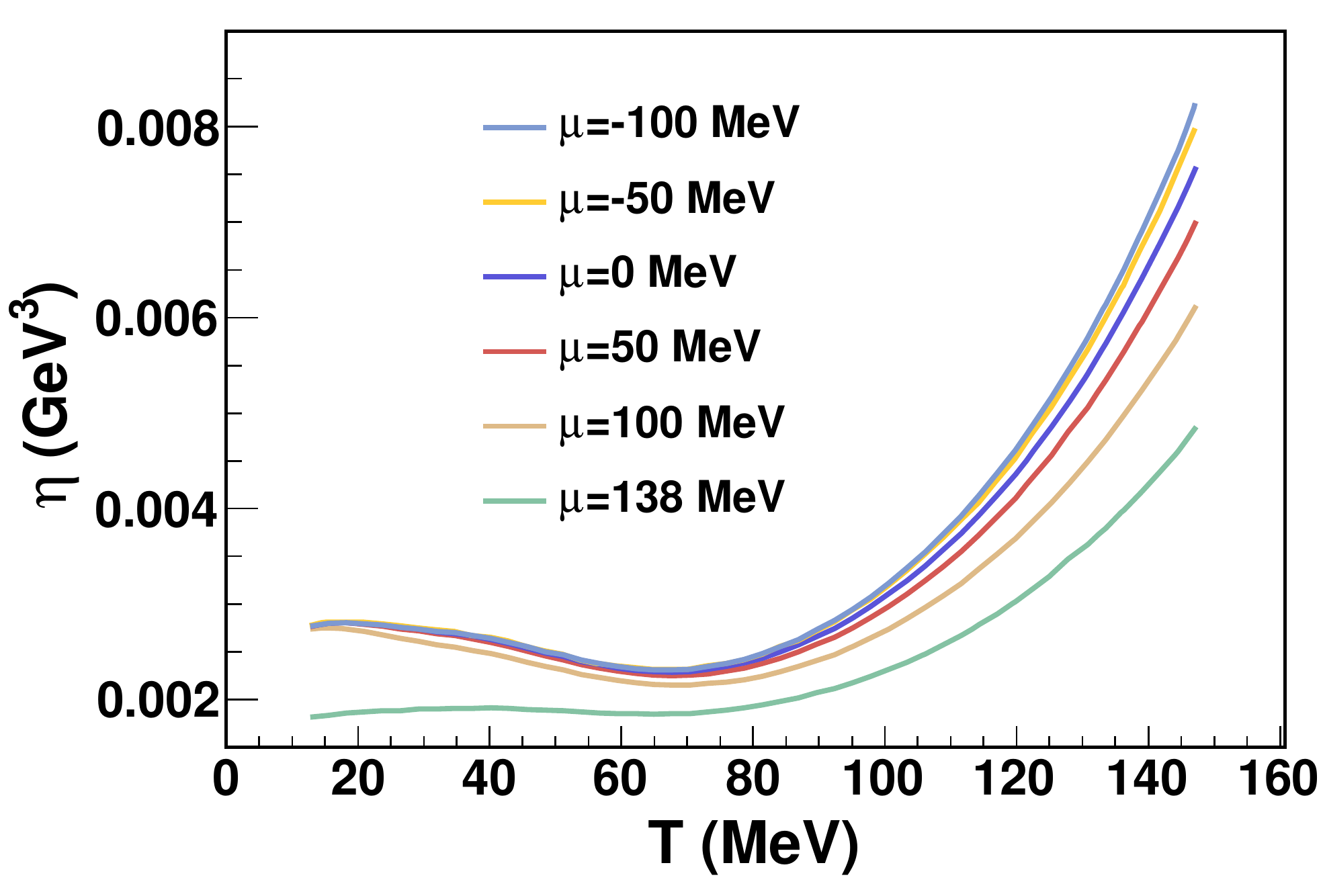}
\includegraphics[scale=0.35]{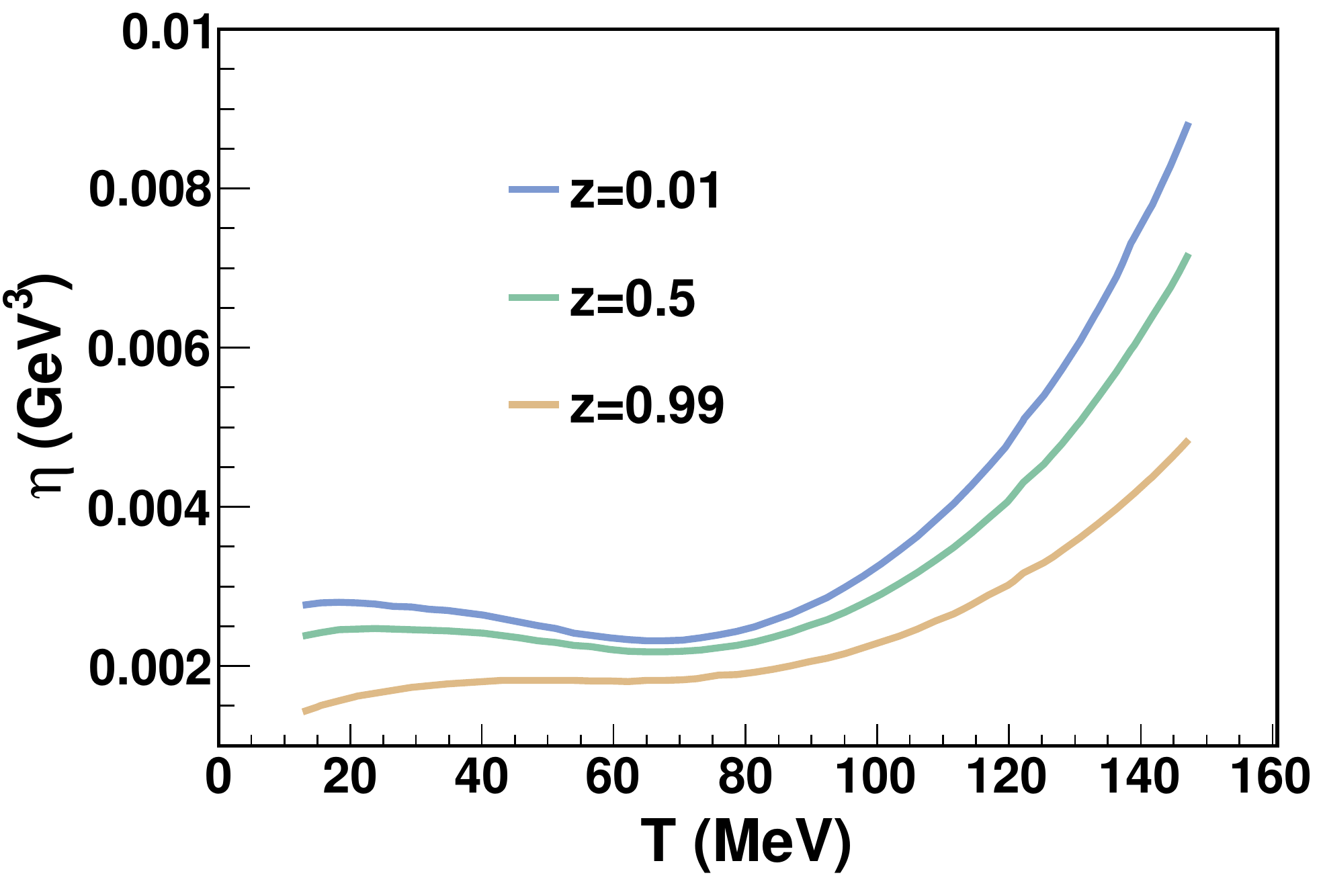}  
\caption{\label{fig:shear_chem} Shear viscosity of a pion gas at several chemical potentials and fugacities. We use the inverse amplitude method in order to unitarize the pion scattering amplitudes.}
\end{center}
\end{figure}

We plot the shear viscosity at third order for low temperatures up to $T \simeq 150 \sim m$ MeV. In Fig.~\ref{fig:shear_chem} we show the shear
viscosity for different pion chemical potentials and fugacities.

Beyond this temperature the calculation is not reliable any more.
Several reasons that explain why this is so, are the following:

\begin{itemize}
 \item Relevance of inelastic scattering channels: We have neglected inelastic processes like $\pi \pi \rightarrow \pi \pi \pi \pi$ due to the Boltzmann suppression $e^{-2m/T}$ in the final state. 
       This channel becomes more important when the temperature becomes moderately high (around $T=150$ MeV). This point will be detailed in the next chapter due to its relevance in the bulk viscosity.
 \item Interplay of new degrees of freedom: At moderate temperature the effects of kaons and $\eta$ meson are important. This implies extending the interaction to $SU(3)$ chiral perturbation \index{chiral perturbation theory}
       theory (ChPT)\glossary{name=ChPT,description={chiral perturbation theory}}. We have accomplished this in reference \cite{Dobado:2008vt}.
 \item Failure of unitarized ChPT: The unitarized interaction used for describing the pion is valid up to pion momentum of $p \simeq 1.2-1.4$ GeV, \cite{gomezpelaez01}. Choosing this UV cutoff to be ten times the 
       most probable pion momentum, which is of the order of $p \sim \sqrt{Tm}$, this amounts to having a limiting temperature of the order of the pion mass.
 \item Loss of dilute gas assumption: Increasing the temperature makes larger the particle density of the gas larger and eventually one deals with a dense gas in which the
       condition $\lambda_{mfp} \ll  1/n\sigma$ does not hold anymore. 
\end{itemize}

\begin{figure}[t]
\begin{center}
\includegraphics[scale=0.35]{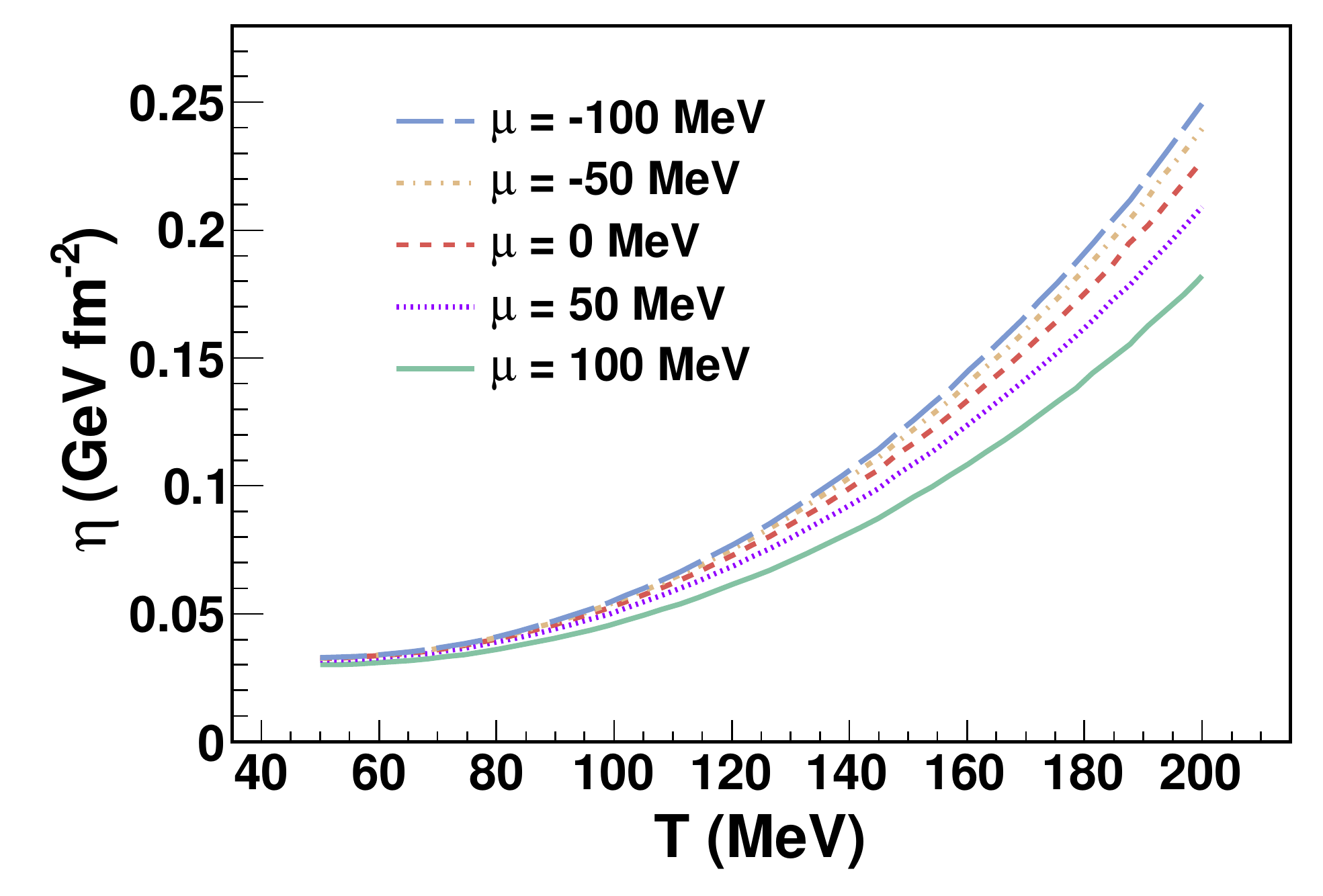}
\includegraphics[scale=0.35]{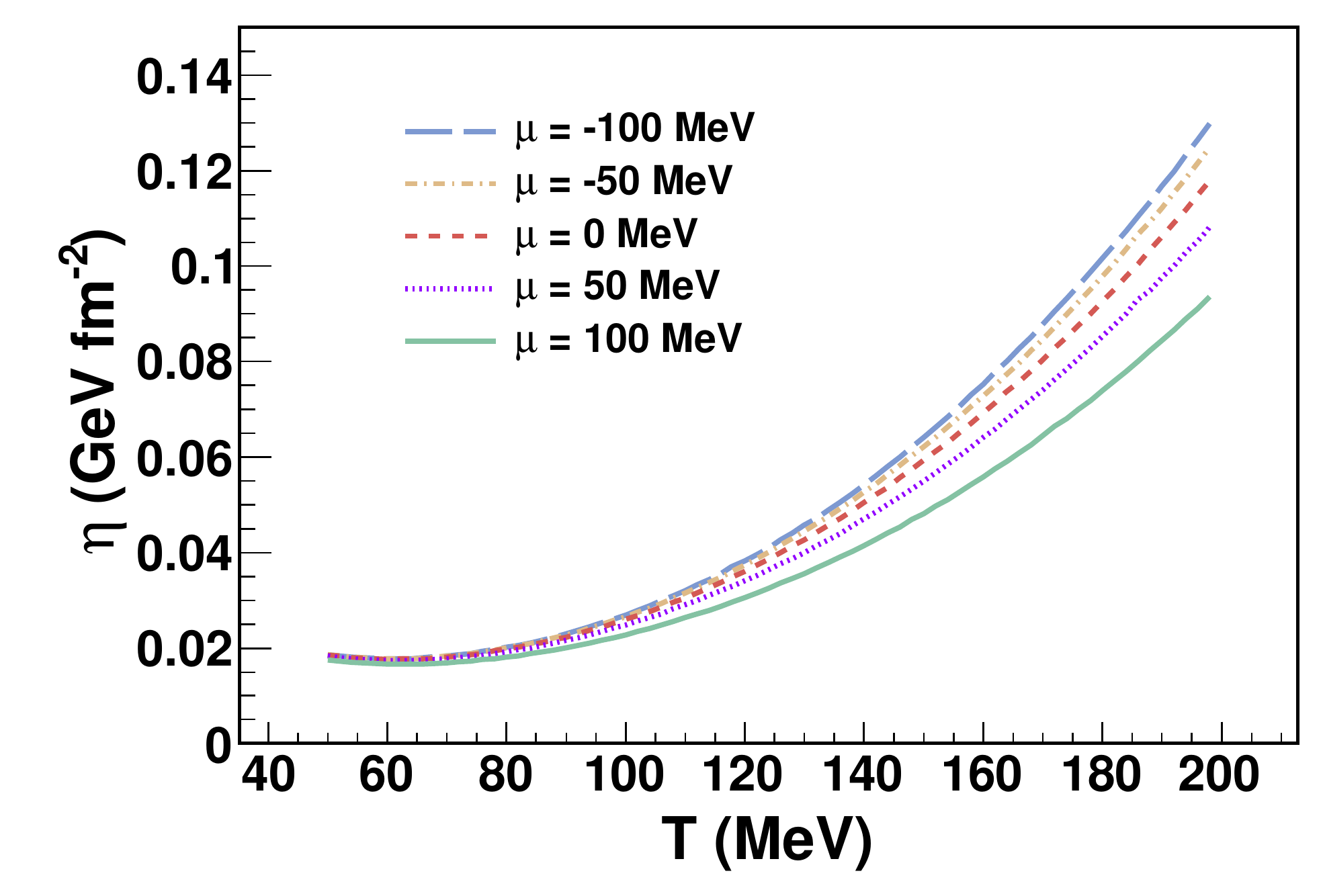}  
\caption{\label{fig:davcomparison} Top panel: Shear viscosity of a pion gas at several chemical potentials and fugacities with the phenomenological phase-shifts of \cite{Prakash:1993bt}. Bottom panel: Results 
for the shear viscosity of \cite{Davesne:1995ms}. Data kindly provided by D. Davesne.}
\end{center}
\end{figure}

Finally, in Fig.~\ref{fig:davcomparison} we compare the results obtained using the phenomenological phase-shifts in \cite{Prakash:1993bt} with the same transport coefficients computed by Davesne~\cite{Davesne:1995ms}.

\section{KSS coefficient}
   
   In fluid mechanics it is common to construct dimensionless ratios that allow similarity analysis between different systems. In the previous section
we introduced one of them, the Knudsen number, defined as the ratio of the mean free path $\lambda_{mfp}$ over a characteristic fluid length $L$. In our case, this characteristic length was 
the size of inhomogeneities in the system $L=h$.
\be \nonumber \textrm{Kn} = \frac{\lambda_{mfp}}{h} \ . \ee
Perhaps the best known of these is the Reynolds number,
\be \textrm{Re} = \frac{m n L V}{\eta} \ , \ee
that quotients the mass density ($mn$), characteristic fluid size and velocity, by the shear viscosity. High values of this ratio (low viscosities) imply turbulent, unstable
flows, whereas small values (large viscosities) allow laminar, stationary flows. 

The definition of the Reynolds number\index{Reynolds number} includes the inverse of the kinematic viscosity $\nu=\eta/mn$. The kinematic viscosity is used for measuring dissipation in a nonrelativistic system
\cite{landau1987fluid}. However, in relativistic theory the presence of the particle number is problematic, because this number is usually not conserved. The generalization
of the Reynolds number for relativistic theories is to replace the mass density by the relativistic enthalpy density $w=\epsilon+P$:
\be \textrm{Re} = \frac{(\epsilon+P) L V}{\eta} \, \ee
and the kinematic viscosity is substituted by $\eta/(\epsilon+P)$ where the denominator can be substituted by $Ts$ when the chemical potentials are set to zero.
 
One can obtain the linearized equations of motion by expanding the Navier-Stokes equations (\ref{eq:navier-stokes1}), (\ref{eq:navier-stokes2}) and the continuity equation (\ref{eq:eqcontinuity}) (up to first order 
when a small perturbation) around the equilibrium values of the particle density $n_0 + \delta n$, velocity $\delta u_i$ and temperature $T_0 + \delta T$. Taking Fourier transform one can obtain 
a linear system that couples all the perturbations. This linear system is charaterized by the ``hydrodynamical matrix''. Diagonalizing this matrix one can obtain the form of the dispersion relations $\omega=\omega(k)$.
In the case of no conserved charge ($\mu_i=0$), four dispersion relations are obtained \cite{Schafer:2009dj}, \cite{Dobado:2008ri}, a pair of tranverse diffusive modes
\be \omega (k) = - i \frac{\eta}{Ts} \mb{k}^2 \ee
and a pair of sound modes
\be \omega (k) \simeq c_s k - i \frac{\Gamma_s \mb{k}^2}{2} \ , \ee
with
\be \Gamma_s = \frac{\frac{4}{3} \eta +\zeta}{Ts} \ . \ee

The imaginary part of the dispersion relations entails a damping of the corresponding modes, for which the kinematic viscosity plays a very important role.

The adimensional combination $\eta/s$ was calculated for a very large class of four dimensional conformal quantum field theories at finite
temperature in \cite{Kovtun:2004de} by using the Anti de Sitter/Conformal Field Theory\index{conformal field theory} (AdS/CFT) \glossary{name=AdS/CFT,description={anti-de Sitter/conformal field theory correspondence}}correspondence \index{AdS/CFT correspondence}
introduced by Maldacena in \cite{Maldacena:1997re}.
These CFT's are typically strongly coupled $SU(N)$ supersymmetric Yang-Mills theories, whose gravity duals are string theories define on $AdS \times S^5$ space.
This amounts to calculating the absorption cross section
of a graviton polarized in the $x-y$ direction propagating perpendicularly to a black brane.

Because of the optical theorem this cross section measures in the dual CFT the imaginary part of the retarded Green's function
of the operator coupled to the metric, i.e. the energy-momentum tensor (Kubo's formula).

Therefore the shear viscosity can be expressed as
\be \eta = \frac{\sigma(0)_{\textrm{abs}}}{16 \pi G_5} \ , \ee
where $\sigma(0)_{\textrm{abs}}$ is the graviton absorption cross section at zero energy and $G_5$ is the five-dimensional Newton constant.

Taking the area of the black brane horizon $a=A/V_3$ --which is equal to the graviton cross section in the low energy limit-- to calculate the Bekenstein entropy of the black brane, 
\be s = \frac{A}{4G_5 V_3} = \frac{\sigma(0)_{\textrm{abs}}}{4G_5} \, \ee 
where $V_3$ is the spatial volume along the three infinite dimensions of the horizon. The authors of \cite{Kovtun:2004de} obtained the temperature-independent result:
\be \frac{\eta}{s} = \frac{1}{4\pi} \ . \ee
After the works \cite{Kovtun:2004de,Son:2006em} this coefficient is also named the KSS coefficient\index{KSS coefficient}.
  
From this result, the authors of \cite{Kovtun:2004de} proposed a conjecture that for a very wide class of systems (those described by a sensible and UV finite quantum field theories), the above ratio has the lower bound:
\be \frac{\eta}{s} \ge \frac{\hbar}{4\pi k} \ , \ee
where we have reintroduced the Planck and Boltzmann's constants.
   It is remarkable that the existence of a lower bound for $\eta/s$ can be obtained by using the Heisenberg uncertainty principle\index{Heisenberg uncertainty principle}~\cite{Danielewicz:1984ww}:
The viscosity of a plasma is proportional to the energy density $\epsilon$ times the mean free time $\tau_{mft}$. On the other hand, the entropy density is proportional
to the number density times the Boltzmann's constant $n k$. Therefore, $\eta/s \sim E \tau_{mft}/k$, where $E$ is the average particle density. Thus, from the time-energy Heisenberg uncertainty principle
one can obtain the mentioned bound modulo the numerical factor.
   All the known fluids in nature do not undercome this bound and for the most common gases and liquids this ratio is much larger. One of the systems for which the KSS coefficient seems to be
very close to $1/4 \pi$ is the quark-gluon plasma created in relativistic heavy-ion collissions (see Chapter~\ref{ch:1.intro}). For this reason, the QGP is expected to be a strongly coupled collective system instead of a weakly 
coupled plasma, as perturbative QCD suggests\footnote{However, a different interpretation is suggested in \cite{Mrowczynski:2005ki} where a weakly coupled plasma presents instabilities from magnetic plasma modes,
producing a momentum isotropization speeding up the equilibration.}.

   Finally, it has been pointed out that the KSS coefficient has a minimum at a phase transition. This is the case for the common fluids in the liquid-gas phase
transition~\cite{Csernai:2006zz}. Empirically, $\eta/s$ seems to have a discontinuity at a first order phase transition, but it is continuous and has an extremum
at a second order phase transition or at a crossover. Other types of phase transitions (like in superfluid helium-4 \index{helium-4}or the BCS-BEC\glossary{name=BCS,description={Bardeen-Cooper-Schrieffer}}
\glossary{name=BEC,description={Bose-Einstein condensate}} transition in Fermi gases\index{Fermi gas}) also present a minimum in $\eta/s$~\cite{Dobado:2009ek}. 

To explicitly show this behaviour we have calculated the $\eta/s$ coefficient in both the gas and liquid phases of atomic Argon\index{Argon}. We have chosen Argon due to its sphericity and  closed-shell atomic structure.

For the gas phase we have described the interaction by a hard-sphere model. Neglecting correlations between successive scatterings the formula for the shear viscosity in the hard-sphere\index{hard-sphere approximation} approximation reads:
\be \eta_{\textrm{gas}} = \frac{5}{16 d^2} \sqrt{\frac{mT}{\pi}} \ , \ee
where $d=3.42 \cdot 10^5$ fm is the diameter of the atomic Argon and $m=37.3$ GeV its mass. The entropy density \index{entropy density}is given by Eq.~(\ref{eq:entropyden}) (with $g_{\pi}=1$). In the experimental data, the pressure\index{pressure} is fixed,
so that we keep the pressure of the gas constant. When increasing the temperature, a variation of the chemical potential is extracted and then inserted again in the equation for the entropy density.
Bose-Einstein corrections are not taken into account because the gas liquefies before these effects are relevant.

The liquid phase is more complicated because the momentum transfer mechanism is quite complex. There is no rigorous theory for it. To calculate the shear viscosity we have used the Eyring vacancy theory\index{Eyring's theory}
in which each molecule in the liquid can have gas-like or solid-like degrees of freedom. It has gas-like degrees of freedom when jumping into a vacant hole in the liquid, and a solid-like
degree of freedom when fully surrounded by other molecules. For the Argon \index{Argon} liquid, the partition function\index{partition function} reads:
\be \nonumber Z = \left\{ \frac{e^{E_s/N_A T}}{(1-e^{-\theta/T})^3} 
\left( 1 + n \frac{V-V_s}{V_s} e^{-\frac{aE_s V_s}{(V-V_s)N_AT}} \right) \right\}^{\frac{N_AV_s}{V}}  \left\{ \frac{e(2 \pi mT)^{3/2} V}{(2\pi)^3 N_A} 
\right\}^{\frac{N_A(V-V_s)}{V}} \ , \ee
where $e$ is the Napier's number, $E_s$ is the sublimation energy of Argon (expressed in electronvolts per particle), $\theta$ is the Einstein characteristic temperature of the solid, $a=a'$ is a model parameter
that controls the molecular ``jump''	between sites, or activation energy and $nV/V_s$ is the number of nearest vacancies to which an atom can jump. The used values are shown in Table~\ref{tab:eyring}.

\begin{table}[t]
\begin{center}
\begin{tabular}{|cc|}
\hline 
 Parameter & Value \\
\hline \hline
$\theta$ & $5.17$ meV \\
$n$ & $10.80$ \\
$a=a'$ & $0.00534$ \\
$\kappa$ & $0.667$ \\
$E_s$ & $0.082$ eV/particle \\
$V_s$ & $4.16 \times 10^{16}$ fm$^3$/particle \\
\hline
\end{tabular}
\end{center}
\caption{Parameters used in the liquid Argon Eyring theory. \label{tab:eyring} }
\end{table}

The shear viscosity of the liquid is also a weighted average between the viscosity of solid-like and gas-like degrees of freedom of the liquid's particles:
\be \eta_{\textrm{liq}} = \frac{N_A 2\pi}{V} \frac{1}{1-e^{-\theta/T}} \frac{6}{n\kappa} \frac{V}{V-V_s} e^{\frac{a' E_s V_s}{(V-V_s)N_A T}} + \frac{V-V_s}{V} \frac{5}{16d^2}\sqrt{\frac{mT}{\pi}} \ , \ee
where $N_A$ is the Avogadro's number\index{Avogadro's number}.

The entropy can be calculated taking the temperature derivative of the Helmholtz free energy\index{Helmholtz free energy}
\be S = -\frac{\pa A}{\pa T} = \frac{\pa (T \log Z)}{\pa T} \ . \ee	

\begin{figure}[t]
\begin{center}
\includegraphics[scale=0.4]{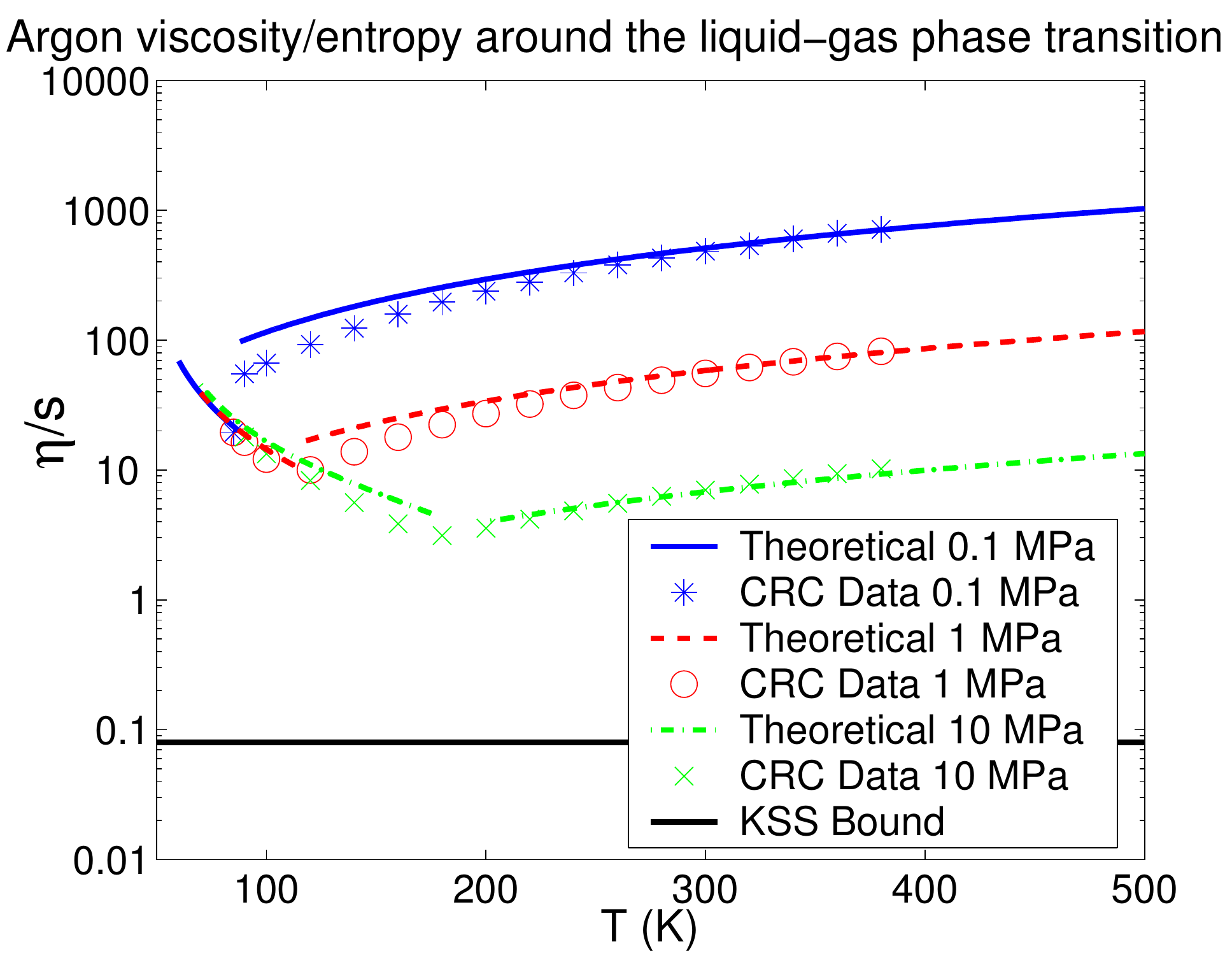} 
\caption{\label{fig:kss_argon} $\eta/s$ around the liquid-gas phase transition of atomic Argon.}
\end{center}
\end{figure}

As the partition function does not depend on a chemical potential we cannot compute the particle density as a derivative over it. One way out is to use the liquid density obtained by the 
van der Waals equation of state. This equation takes into account the volume excluded by the particles and also the attractive force between them. In its simplest form, the van der Waals equation\index{van der Waals equation} reads:
\be \label{eq:vdWaals} \left( n_{\textrm{gas}} + n_{\textrm{liq}}^2 \frac{a}{T} \right) (1- n_{\textrm{liq}} b) = n_{\textrm{liq}} \ ,   \ee
where $n_{\textrm{gas}}$ and $n_{\textrm{liq}}$ are the particle densities of gaseous and liquid Argon, respectively. $T$ is the temperature, $2b=4\pi d^3/3$ is the covolume, i.e. the excluded volume
by the particle and $a=27T_c/64P_c$ is a measure of the particle attraction related to the properties at the critical point ($T_c=150.87$ K and $P_c=4.898$ MPa). In spite of the simplicity of
Eq.~(\ref{eq:vdWaals}) it gives very good results. The final graph of this calculation is shown in Fig.~\ref{fig:kss_argon}.

\begin{figure}[t]
\begin{center}
\includegraphics[scale=0.4]{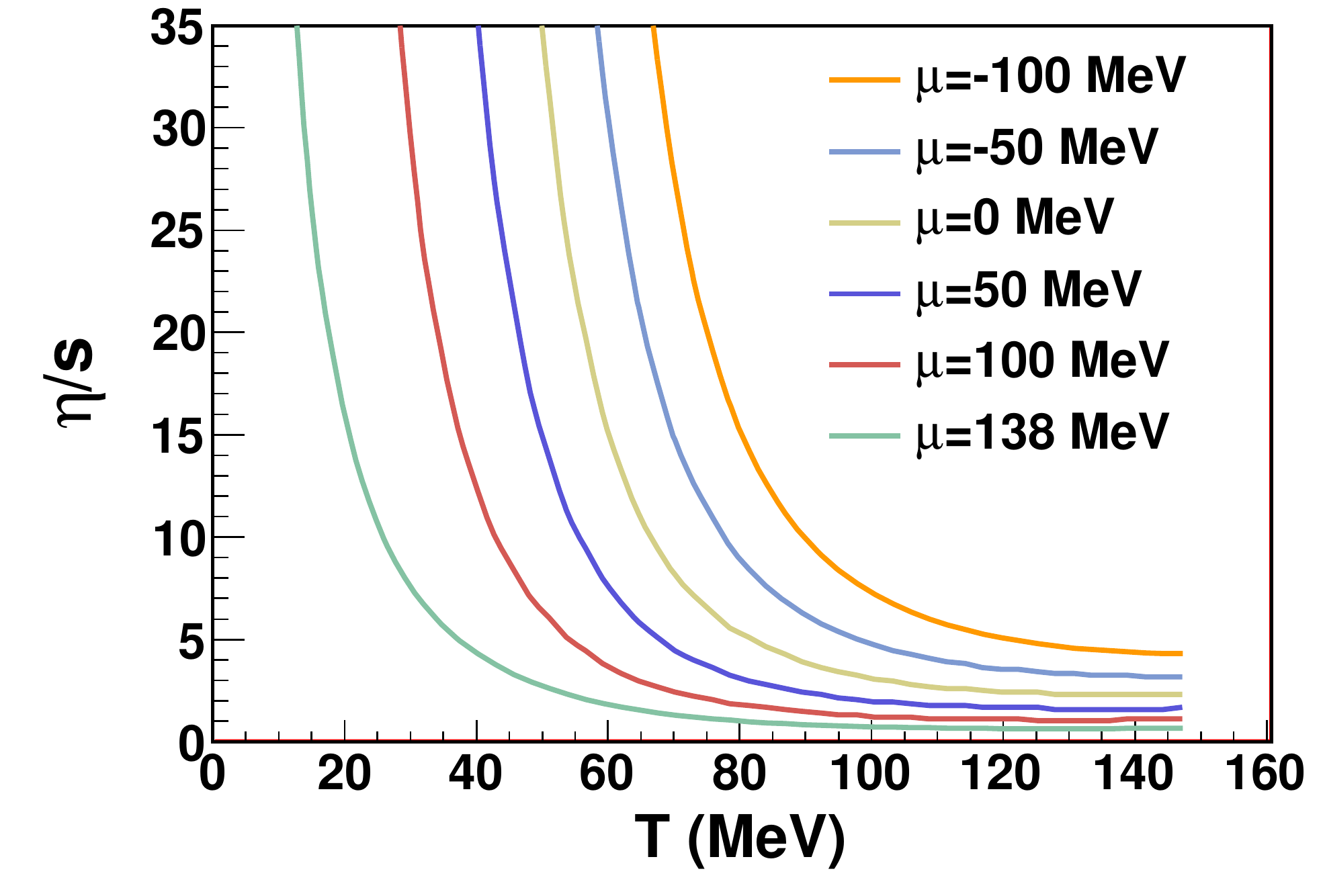} 
\caption{\label{fig:kss_mu} $\eta/s$ for a pion gas at several pion chemical potentials.}
\end{center}
\end{figure}

Turning back to the pion gas, we can plot the KSS coefficient just dividing the shear viscosity over the ideal gas entropy density \index{entropy density}:
\be \label{eq:entropyden} s = \frac{g_{\pi}}{6 \pi^2 T^2} \int_0^{\infty} dp \ p^4 \frac{E-\mu}{E} \frac{e^{\beta(E-\mu)}}{\left[ e^{\beta(E-\mu)}-1\right] ^2} \ , \ee
or, in terms of an integral over adimensional variables,
\be s = \frac{g_{\pi}m^5}{6 \pi^2 T^2} \int dx \ (x^2-1)^{3/2} \frac{z^{-1} e^{y(x-1)}}{\left[z^{-1} e^{y(x-1)} -1\right]^2} \ \left[ x-\left( 1 + y^{-1} \log z \right) \right]  \ . \ee 
The resulting curve is plotted in Fig.~\ref{fig:kss_mu} for several pion chemical potentials.

\begin{figure}[t]
\begin{center}
\includegraphics[scale=0.35]{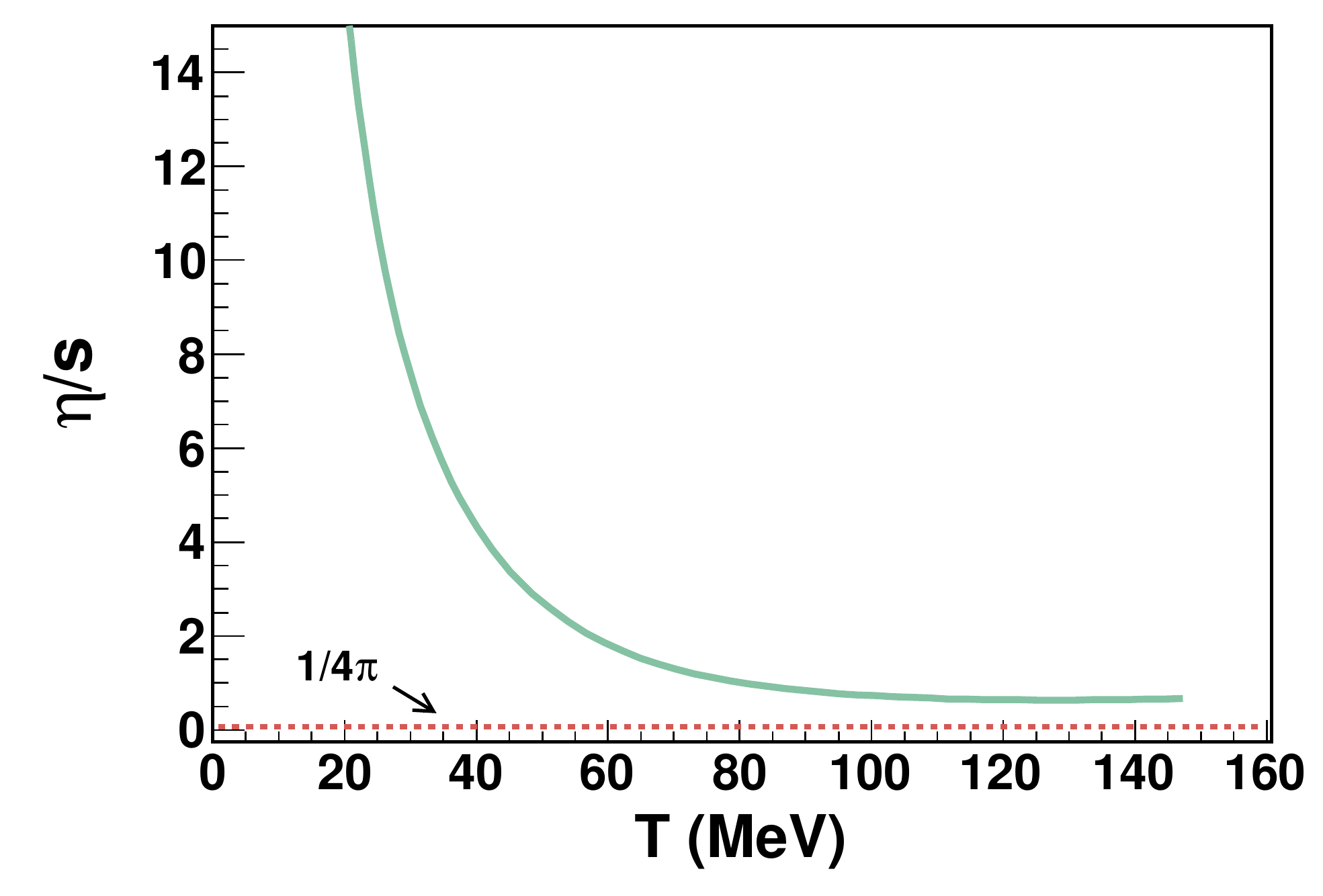}
\caption{\label{fig:kss_mu_mpi} $\eta/s$ for a pion gas at the limiting chemical potential $\mu=138$ MeV. The KSS bound is also shown in dashed line.}
\end{center}
\end{figure}

The KSS bound $1/4\pi$ \index{KSS bound}is not violated as claimed in other works \cite{Chen:2006iga}. The reason is that in ChPT \index{chiral perturbation theory}
the cross section grows unchecked, eventually violating the unitarity bound, which induces a very small viscosity. However, with unitarized phase-shifts, the scattering
amplitude satisfies elastic unitarity and the viscosity presents a softer decrease due to the saturation of the cross section.

\subsection{$\eta/s$ in QGP and deconfined phase transition}

As indicated before, one cannot trust the calculation of the transport coefficients beyond the temperature $150$ MeV. Moreover, at temperatures not too far away from this, we
expect the liberation of quark and gluon degrees of freedom and the formation of the quark-gluon plasma phase. To provide a view of the high-temperature behaviour of $\eta/s$ 
we include the perturbative calculation \index{pQCD} of the KSS coefficient in the QGP\index{shear viscosity!in the perturbative plasma} phase taken from \cite{Arnold:2000dr},\cite{Arnold:2003zc} and the thermal strong coupling
constant from \cite{letessier2002hadrons}.

For $SU(3)$ and $N_f=2$ the KSS number reads 
\be \frac{\eta}{s} = \frac{5.328}{g^4 \ln(2.558 g^{-1})} \ , \ee
with the thermal strong coupling constant \index{$\alpha_s$ thermal} up to two-loops
\be \label{eq:qcd_coupling2} g^{-2} (T)= \frac{29}{24\pi^2} \log \left( \frac{T}{T_0} \right) + \frac{115}{232 \pi^2} \log \left( 2 \log \left( \frac{T}{T_0} \right) \right) \ , \ee
where $2 \pi T_0 = \Lambda_{QCD} \approx 200$ MeV, with $\Lambda_{QCD}$ the scale of QCD at which the strong coupling constant becomes very large and the perturbative physics is not valid anymore.

\begin{figure}[t]
\begin{center}
\includegraphics[scale=0.35]{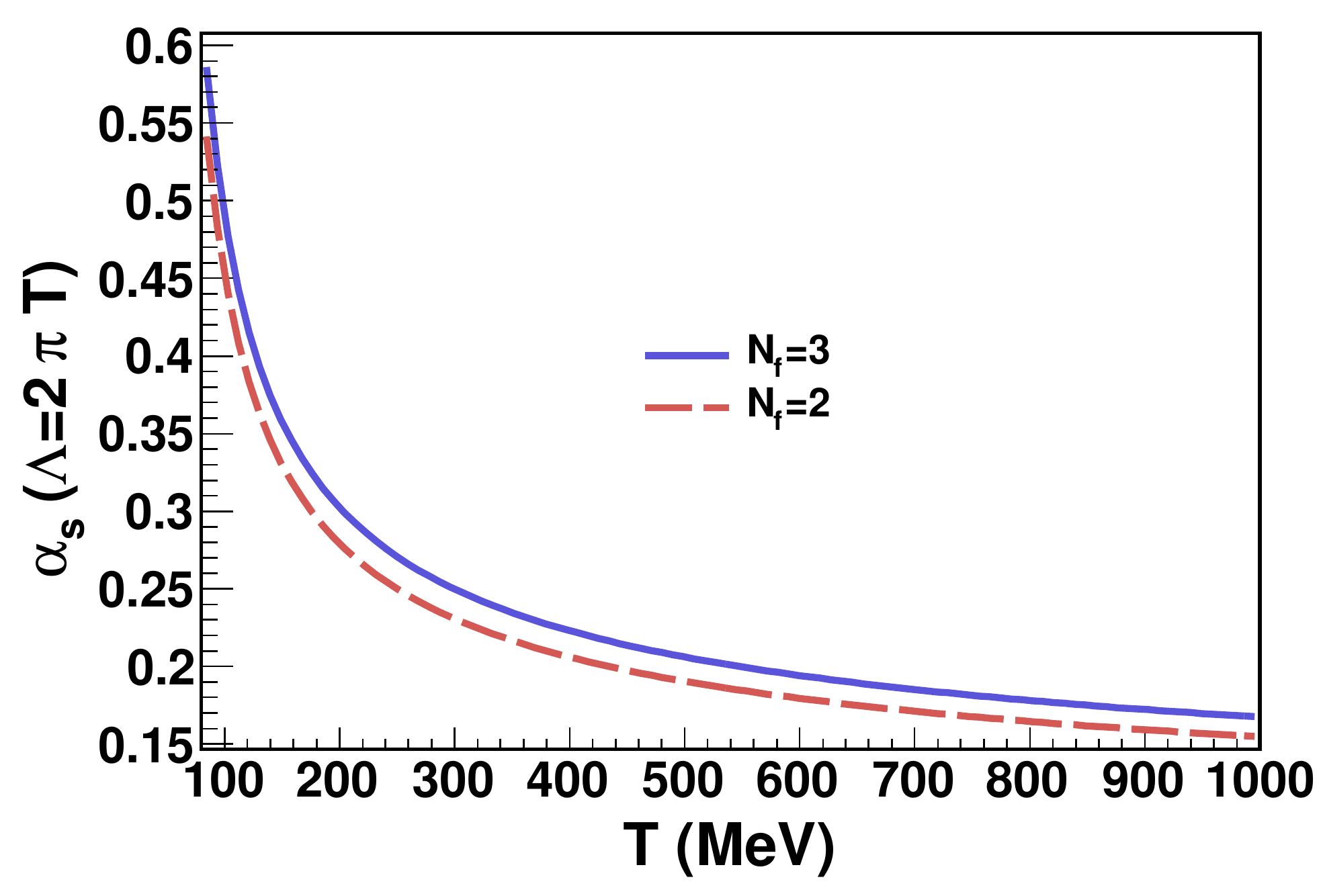} 
\caption{\label{fig:alphas} Two-loop strong thermal coupling constant $\alpha_s(T)$ as a function of temperature for $N_f=2$ and $N_f=3$.}
\end{center}
\end{figure}

For $N_f=3$ the coefficient reads
\be \frac{\eta}{s} = \frac{5.119}{g^4 \ln(2.414 g^{-1})} \ , \ee
with
\be \label{eq:qcd_coupling3} g^{-2} (T)= \frac{9}{8\pi^2} \log \left( \frac{T}{T_0} \right) + \frac{4}{9 \pi^2} \log \left( 2 \log \left( \frac{T}{T_0} \right) \right) \ . \ee
Note (see Fig.~\ref{fig:alphas}) that for example $\alpha_s (T=150 \textrm{ MeV}) \simeq 0.4$ so that good convergence of the perturbative calculations is relegated to significantly higher temperatures.

\begin{figure}[t]
\begin{center}
\includegraphics[scale=0.34]{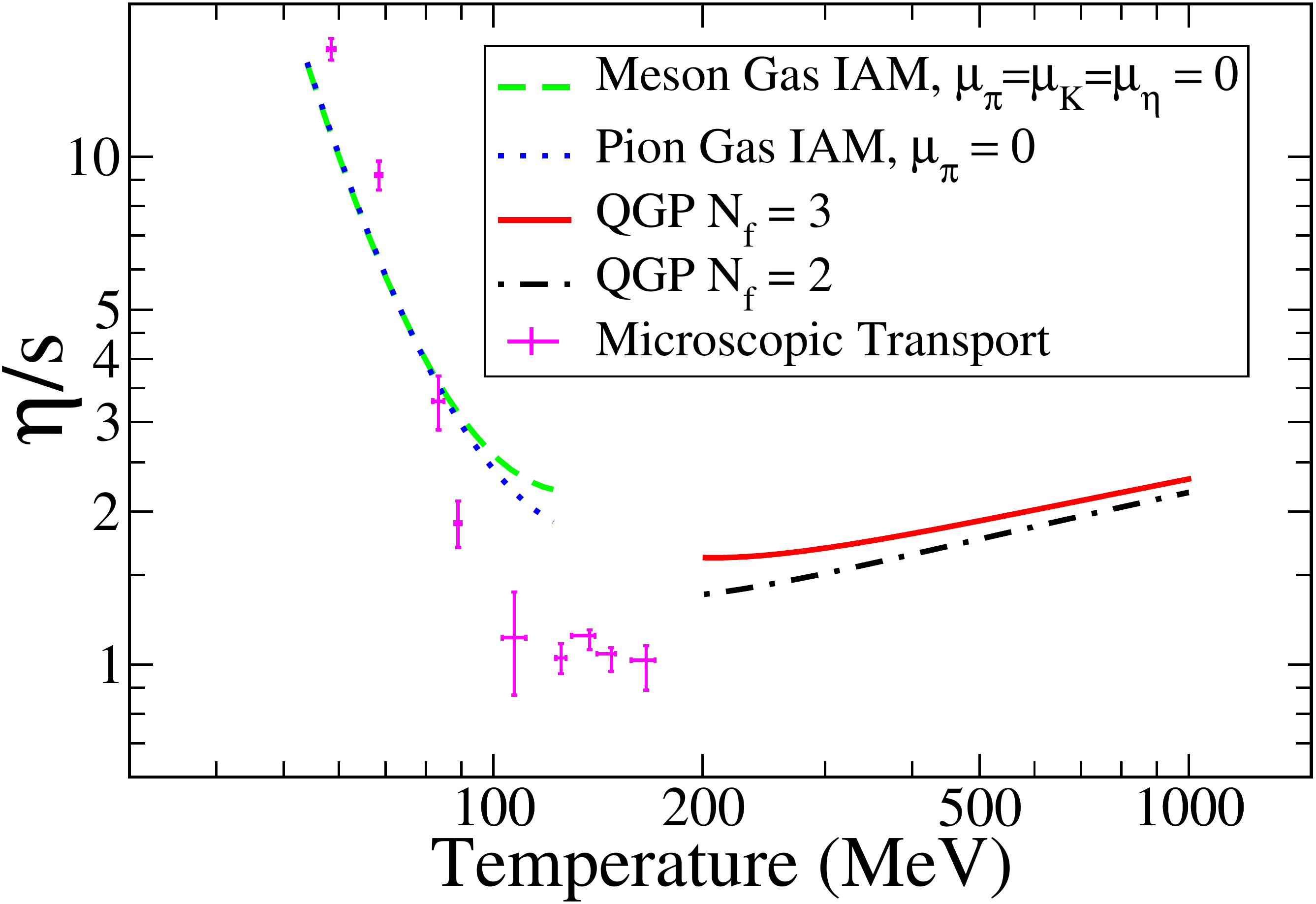} 
\caption{\label{fig:shear_qgp} Shear viscosity of a pion gas with $\mu=0$ and for a $SU(3)$ quark-gluon plasma with two and three flavors in the high temperature limit.}
\end{center}
\end{figure}

We summarize the situation in Fig.~\ref{fig:shear_qgp} where in the low temperature phase we plot the KSS coefficient for a gas of pions at vanishing chemical potential taken over from Fig.~\ref{fig:kss_mu}.
To have a more detailed insight into the hadronic phase, we also add a computation of the KSS coefficient for a mixture of mesons. As discussed in \cite{Dobado:2009ek} we employ
unitarized $SU(3)$ ChPT to solve the BUU equation for the eight pseudo-Goldstone bosons. The result from the microscopic transport model calculation (UrQMD) of \cite{demir2009shear}
 is included as well.

In the high temperature phase we show the $N_f=2$ case, consistent with a gas of pions where only $u$,$d$-quarks enter as valence quarks
and the $N_f=3$ calculation where the strange quark is included in the calculation. This perturbative calculation with massless quarks at next-to-leading log is described in detail in \cite{Arnold:2003zc}.
The numerical values of $\eta/s$ in this phase are not sensitive to the effect of including the quark masses. We have studied this modification in \cite{Dobado:2008vt}.

One clearly sees that the possibility of having a minimum value for $\eta/s$ near the crossover temperature $T_c \sim 150$ MeV is well-founded. However, around
this temperature the unitarized ChPT calculation breaks down, and also the perturbative QCD calculation has lost its validity at much higher temperatures. The presence of a minimum in
the KSS coefficient at the critical temperature will be treated in Chapter~\ref{ch:9.LSM}, where we study the linear sigma model in the large-$N$ limit. In this model one can
study both low- and high-temperature phases from the same partition function.

\chapter{Bulk Viscosity \label{ch:4.bulk}}

\begin{figure}[t]
\begin{center}
\includegraphics[scale=0.45]{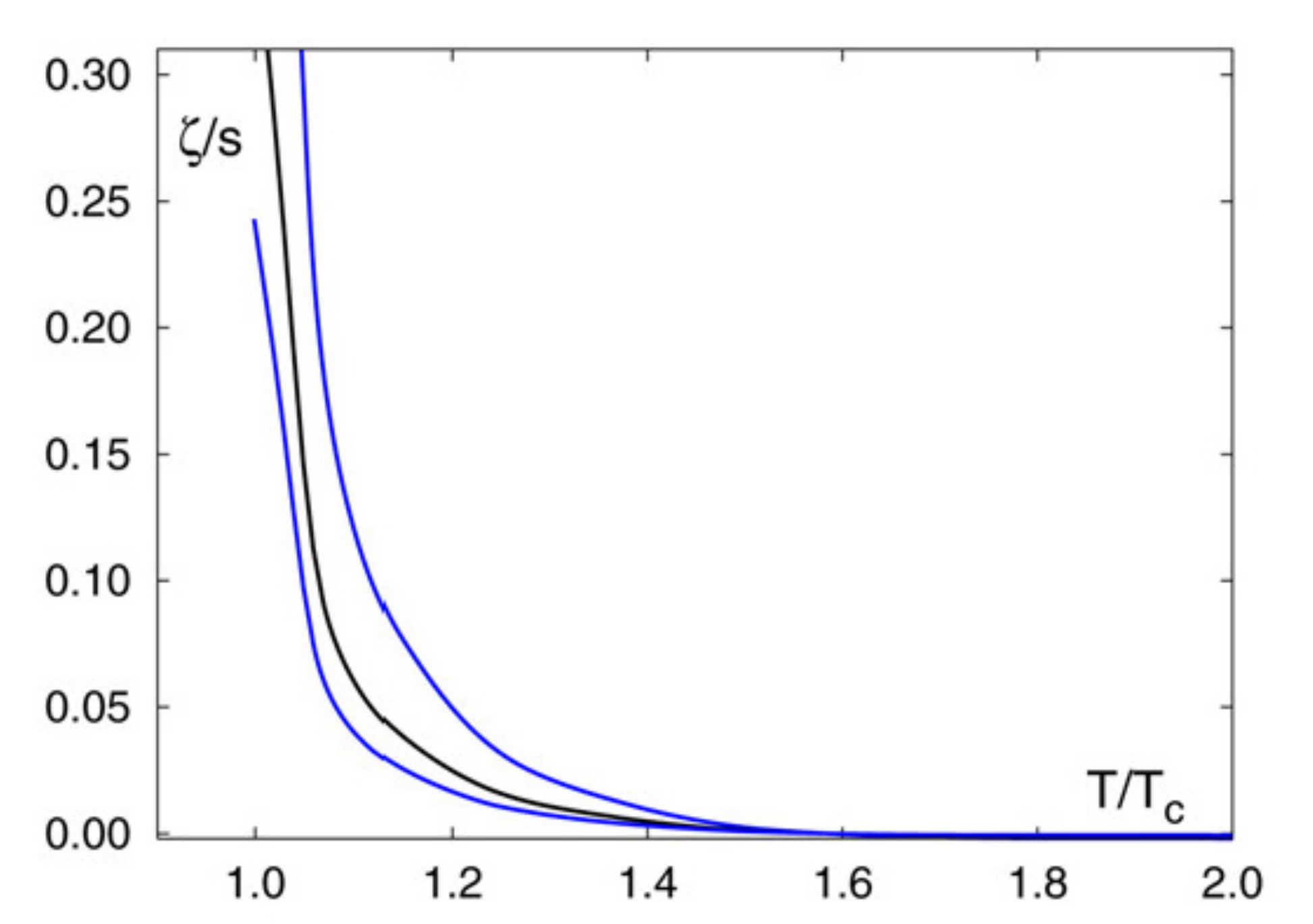} 
\caption{\label{fig:lattice_bulk} Bulk viscosity over entropy density above the deconfinement critical temperature as computed in lattice QCD\index{lattice QCD}~\cite{Karsch:2007jc}.}
\end{center}
\end{figure}

   The bulk viscosity \index{bulk viscosity} (also called second\index{second viscosity|see {bulk viscosity}} or volume viscosity\index{volume viscosity|see {bulk viscosity}} $\zeta$ is the transport coefficient responsible for the equilibration of
a fluid subject to a small dilatation or compression. This coefficient is usually smaller than the shear
viscosity and as we have shown in Chapter \ref{ch:1.intro} it is generally neglected in hydrodynamical
simulations. As a matter of fact, as derived in lattice QCD calculations \cite{Karsch:2007jc} the bulk viscosity over entropy density
could be much larger than the KSS number near the deconfinement phase transition.
As shown in Fig.~\ref{fig:lattice_bulk} it can even diverge at the critical temperature with a critical exponent near one.

  In the context of relativistic heavy-ion collisions, we have proposed that the bulk viscosity can be experimentally accessed
by means of fluctuations of the two-point correlation of the energy-momentum tensor \cite{Dobado:2011wc}. 
This method will be described in detail in Chapter~\ref{ch:10.corre}.

  From the point of view of symmetries it is a very interesting coefficient because it reflects the loss of dilatation invariance,
through the trace anomaly\index{trace anomaly} in quantum field theory:
\be \zeta \propto T_{\mu}^{\mu} \ . \ee

 Because of this, $\zeta$ vanishes for those fluids described by conformal quantum field theories\index{conformal field theory}. For instance, a gas of ultrarelativistic particles with
equation of state $\epsilon=3P$\index{equation of state}, or massless theories with vanishing $\beta$ function.

The bulk viscosity is also vanishing for nonrelativistic monoatomic molecules interacting through two-body collisions like rigid spheres~\cite{landau1981physical} (claim 
originally atributted to J.C. Maxwell). For intermediate temperatures, the bulk viscosity of such a simple gas does not vanish.
Moreover, the gases composed of molecules with internal degrees of freedom have a finite bulk viscosity. One example is a gas of diatomic molecules for which,
due to the exchange of energy between translational and rotational degrees of freedom, the bulk viscosity may be sizeable, and plays an important role in sound absorption.
The bulk viscosity has also been calculated in the color-flavor locked phase of quark matter in \cite{Manuel:2007pz}.

\section{Irrelevance of inelastic pion scattering}

The term in the left-hand side of the kinetic equation\index{BUU equation} (\ref{eq:linear_equation}) that takes into account expansion or compression perturbations is (\ref{eq:lhsbulk}):

\be \left. p_{\mu} \pa^{\mu} n_p (x) \right|_{\zeta}=
 \beta \ n_p (x) [1+n_p(x)] \left( \frac{p^2}{3} - E_p^2 v_n^2 - E_p \kappa^{-1}_{\epsilon} \right) \ \nabla \cdot \mb{V}\ . \ee

Turning to the collision operator in (\ref{eq:linear_equation}), we will use the following parametrization for the perturbation function $\Phi_a$:
\be \label{eq:phiforzeta} \Phi_a= \beta \  \nabla \cdot \mathbf{V} \ A (k_a) \ , \ee
where $A(k_a)$ is an adimensional function of $k_a$.

Cancelling the factor $\nabla \cdot \mb{V}$, the kinetic equation reads
\be \nonumber n_p (1+n_p) \left( \frac{1}{3} p^2 -E_p^2 v_n^2  -E_p \kappa^{-1}_{\epsilon} \right)  \ee
\be \label{eq:bulk_buu} = \frac{g_{\pi} E_p}{2} \int d \Gamma_{12,3p}  (1+n_1)(1+n_2)n_3 n_p \ [ A(p) + A(k_3) -A(k_1) - A(k_2)] \ . \ee

In contrast to the case of shear viscosity this collision operator presents zero modes\index{zero modes}. The zero modes 
can be identified from the structure $A(p) + A(k_3) -A(k_1) - A(k_2)$. They correspond to $A_0 (p)=1$
and $A_1 (p)=x=E_p/m$ and they make the right-hand side of (\ref{eq:bulk_buu}) vanish. The former is associated to particle number conservation and
the latter to the energy conservation law. Introducing particle number-changing processes in the collision one can remove the first zero mode.
However, the zero mode associated with energy conservation is present in the collision even when inelastic collisions are included.

In chiral perturbation theory the number-changing processes among pions are allowed, but only with an even number of participants, because of $G$-parity conservation.
At leading order, the relevant processes are
\be \pi \pi \pi \pi \rightarrow \pi \pi; \quad \pi \pi \rightarrow \pi \pi \pi \pi \ .\ee

The first process, in which four particles interact between themselves, is unlikely to occur in a dilute gas. Such a process is not in the
spirit of the Boltzmann's asumptions for deriving kinetic theory, where only binary collisions can occur.

The second process is suppressed at low temperatures with respect to the elastic one $\pi \pi \rightarrow \pi \pi$. The thermal supression
factor is $e^{-2 m/T}$ \cite{Dobado:2011qu}. Note that this process requires an extra amount of energy of twice the pion mass from the two incoming
particles in order to create two more pions in the system. The available phase-space for the final state is therefore reduced.
For these reasons the inelastic pion scattering can be neglected in our calculation at low temperatures and for
the physical value of the pion mass (in the chiral limit the situation is drastically different).

In this scenario the number of pions is effectively conserved and a pseudo-chemical potential for them must be introduced. The pion chemical potential must satisfy $\mu \le m$.

\section{Kinetic theory calculation of $\zeta$}

In the local rest reference frame the trace of the stress-energy tensor\index{stress-energy tensor} reads from Eq.~(\ref{eq:pretau})
\be \tau^i_i = 3 \zeta \ \mb{\nabla} \cdot \mb{V} \ , \ee
that relates the trace with the bulk viscosity. From Eq.~(\ref{eq:stressmicro}) we also have:
\be \tau^i_i = -\int d^3p \frac{p^2}{E_p} f_p^{(1)} (t,\mb{x}) \ , \ee
with $f_p^{(1)} (t,\mb{x})= - n_p (1+n_p) \Phi_{p}$.
Equating the two traces and inserting Eq.~(\ref{eq:phiforzeta}) we can provide the following microscopic formula for the bulk viscosity\index{bulk viscosity}:

\be \zeta =  \frac{g_{\pi}}{T} \int \frac{d^3 p}{(2\pi)^3 E_p} n_p (1+n_p) A(p) \ \frac{p^2}{3} \ .\ee

Inspired by the Eq.~(\ref{eq:lhsbulk}) it is convenient to use the two conditions of fit for adding two (vanishing) terms to the previous equation. Taking the
Eqs. (\ref{eq:cond1}) and (\ref{eq:cond2}) in the local rest reference frame we get:
\be \label{eq:twoconditionsoffit} \tau^{00}=\int \frac{d^3p}{(2 \pi)^3} n_p (1+n_p) \ A(p) E_p = 0 \ ; \quad \nu^0=\int \frac{d^3p}{(2\pi)^3} n_p (1+n_p) \ A(p) =0 \ . \ee

Inserting them in the formula for the bulk viscosity we obtain:
\be \zeta =  \frac{g_{\pi}}{T} \int \frac{d^3 p}{(2\pi)^3 E_p} n_p (1+n_p) A(p) \ \left( \frac{p^2}{3} - E_p^2 v_n^2 - E_p \kappa^{-1}_{\epsilon} \right) \ .\ee

Transforming the previous equation with the help of the adimensional variables defined in (\ref{eq:new_var}) we obtain
\be \zeta = \frac{g_{\pi} m^4}{2\pi^2 T} \int dx \ (x^2-1)^{1/2} \frac{z^{-1} e^{y (x-1)}}{\left[z^{-1}
e^{y(x-1)}-1 \right]^2} \left[ \left( \frac{1}{3} - v_n^2 \right) x^2 - \frac{\kappa^{-1}_{\epsilon}}{m} x- \frac{1}{3} \right] A(x) \ . \ee

The natural integration measure for this transport coefficient is

\be d\mu_{\zeta} = dx \ (x^2-1)^{1/2} \frac{z^{-1} e^{y (x-1)}}{\left[z^{-1} e^{y(x-1)}-1 \right]^2} \ee

or equivalently
\be d\mu_{\zeta}= d^3 p \frac{1}{4\pi m^2} \frac{1}{E_p} n_p (1+n_p) \ee
in terms of physical quantities. The inner product is defined in analogy with Eq.~(\ref{eq:product_shear}), the new measure being $d\mu_{\zeta} (x; y,z)$. The bulk viscosity 
is expressed as the following inner product:

\be \label{eq:zeta_prod}\zeta = \frac{g_{\pi} m^4}{2 \pi^2 T}\langle A(x) | \left[ \left( \frac{1}{3} - v_n^2 \right) x^2 - \frac{\kappa^{-1}_{\epsilon}}{m} x- \frac{1}{3} \right] \rangle \ .\ee

In analogy to the $K_i$, we define for convenience the following integrals $I_i$ with $i=n+m$:
\be \label{eq:Iintegrals} I_i=\langle x^n | x^m \rangle = \int d\mu_{\zeta} \ x^i = \int_1^{\infty} dx \ (x^2-1)^{1/2} \frac{z^{-1} e^{y (x-1)}}{\left[z^{-1}
e^{y(x-1)}-1 \right]^2} \ x^i \ .\ee

All thermodynamic functions of the ideal gas are expressible as various integrations over the Bose-Einstein\index{Bose-Einstein distribution function}
distribution functions (some details are commented in Appendix \ref{app:moments}). For example, the two functions $v_n^2$ and $\kappa^{-1}_{\epsilon}/m$ read:
\begin{eqnarray}
\label{eq:vn}  v_n^2 & = &  -\frac{1}{3} \frac{(I_0-I_2)I_2-I_1(I_1-I_3)}{I_2^2-I_1I_3} \ , \\
\label{eq:kappa} \frac{\kappa^{-1}_{\epsilon}}{m} & = & -\frac{1}{3}\frac{I_1I_2-I_0I_3}{I_2^2-I_1I_3} \ .
\end{eqnarray}

The next step is to construct the polynomial basis. The first two elements $P_0(x)$ and $P_1(x)$ must span the
 zero modes\index{zero modes} of the collision operator, i.e. they
should be linear combinations of $1$ and $x$. We will use the convention of monic polynomials, so that:
\be
\begin{array}{lll}
 P_0(x) & = & 1 \ , \\
 P_1(x) & = & x+P_{10} \ , \\
\end{array}
\ee
with $P_{10}$ such that they are orthogonal, i.e. $P_{10}=-I_1/I_0$. $P_2(x)$ is conveniently chosen to be the
inhomogeneous term of the collision operator, that is the source function. This brings about a certain simplification.
\be \label{eq:P2} P_2(x)  =  \left( \frac{1}{3} - v_n^2 \right) x^2 - \frac{\kappa_{\epsilon}^{-1}}{m} x- \frac{1}{3} \ . \ee

Although $P_2(x)$ has been fixed without employing the Gram-Schmidt method, one can check that $P_0$ and $P_1(x)$ are indeed perpendicular to $P_2(x)$, without further orthogonalization:
\be \langle P_0 | P_2 \rangle = \left( \frac{1}{3} - v_n^2 \right) I_2 - \frac{\kappa_{\epsilon}^{-1}}{m} I_1- \frac{1}{3} I_0 =0 \ , \ee
\be \langle P_1 | P_2 \rangle = \left( \frac{1}{3} - v_n^2 \right) I_3 - \frac{\kappa_{\epsilon}^{-1}}{m} I_2- \frac{1}{3} I_1 =0 \ , \ee
by using the expressions (\ref{eq:vn}) and (\ref{eq:kappa}).

The rest of the polynomial basis elements are chosen monic and orthogonal to these three first elements.

With the help of Eq.~(\ref{eq:P2}), the bulk viscosity in (\ref{eq:zeta_prod}) is expressed as the inner product between the function $A(p)$ and the second element of the basis $P_2(x)$:
\be \label{eq:bulk_final} \zeta = \frac{g_{\pi} m^4}{2 \pi^2 T} \langle A(x) | P_2(x) \rangle \ .\ee

On the other hand, taking the kinetic equation, multiplying it by $1/(4 \pi m^4 E_p)$ and projecting it on to $P_l(x)$ one gets

\be \nonumber \label{eq:bulk_lin_buu} \langle P_l(x) | P_2(x) \rangle = \frac{g_{\pi}}{8 \pi m^4} \int d^3 p \int d \Gamma_{12,3p}  (1+n_1)(1+n_2)n_3 n_p \ P_l(x) \ee
\be \times [ A(p) + A(p_3) - A(p_1) - A(p_2) ],\ee

Expanding the solution $A(p)$ in the polynomial basis we constructed before we have:
\be A(x) = \sum_{n=0}^{\infty} a_n P_n (x) \ . \ee
The first two terms in the expansion are proportional to the zero modes of the collision operator $1$ and $x$. If we now symmetrize
the right-hand side of (\ref{eq:bulk_lin_buu}) we end up with the following system

\be \nonumber  \langle P_l(x) | P_2(x) \rangle = \sum_{n=0}^{\infty} a_n \frac{g\pi^2}{4 m^4} \int \prod_{i=1}^4 \frac{d^3 k_i}{(2\pi)^32E_i} \overline{|T|^2} (2\pi)^4 \delta^{(4)} (k_1+k_2-k_3-p) \ee
\be \label{eq:bulk_lin_buu2} \times  (1+n_1)(1+n_2)n_3 n_p \ \Delta [P_l(p) ] \ \Delta [P_n(p)] \ , \ee
with the notation
\be \Delta [P_l(p)] \equiv P_l(p) + P_l (k_3) - P_l (k_1) - P_l (k_2) \ . \ee

\subsection{Zero modes and Fredholm's alternative \label{sec:zeromodes}}

   Note that in the right-hand side of Eq.~(\ref{eq:bulk_lin_buu2}) the terms with $l=0,1$ make the collision
integral vanish because the corresponding polynomials are precisely the zero modes\index{zero modes}. This is a case in which the Fredholm's
alternative for integral equation applies. It states that the integral equation has a solution if and only if the source
function is perpendicular to the zero modes of the integral operator.

   First, note that we have two zero modes that are spanned by the two first elements of the basis:
\be \{1,x\} = \textrm{ span } \{P_0,P_1(x)\} \ . \ee
Secondly, note that the inhomogeneous term or source function in Eq.~(\ref{eq:bulk_lin_buu2}) is the third element of the basis $P_2(x)$ by definition.
And finally, note that the basis is orthogonal, i.e. 
\be \langle P_i(x)|P_j(x) \rangle = \delta_{ij} ||P_i||^2 \ .  \ee

From these assertions one immediately deduces that the source function is actually orthogonal to the two zero modes.
That means that the BUU equation\index{BUU equation} is compatible in the whole linear space spanned by $\{ P_i \}$.

The Fredholm's alternative\index{Fredholm's alternative} is evident when returning to Eq.~(\ref{eq:bulk_lin_buu2}). The right-hand side vanishes when $l=0$ or $l=1$.
To have a consistent equation that is solvable in the whole space, one must have a vanishing left-hand side for $l=0$ and $l=1$.
This condition is fulfilled because our basis is orthogonal. Otherwise, the equation would be incompatible and one would need to restrict the
space of solutions to the perpendicular subspace to the zero modes.

However, one sees that the first two elements of the solution's expansion are not fixed by the Eq.~(\ref{eq:bulk_lin_buu2}) because
the first two equation are of the type $0=0$. The system is indeed compatible but indeterminate and these two remaining components 
of the solution function, $a_0$ and $a_1$ must be determined by the use of the two
conditions of fit (\ref{eq:twoconditionsoffit}). However, for the sole purpose of calculating the bulk viscosity these two components are not needed.

\subsection{Bulk viscosity and conditions of fit}

Denoting by $\mathcal{C}_{ln}$ the collision integral:
\begin{eqnarray}
  \nonumber \mathcal{C}_{ln} &=& \frac{g_{\pi} \pi^2}{4m^4} \int \prod_{i=1}^4 \frac{d^3 k_i}{(2\pi)^32E_i} \overline{|T|^2} (2\pi)^4 \delta^{(4)} (k_1+k_2-k_3-p)  (1+n_1)(1+n_2)n_3 n_p \\
 & & \times \Delta [P_l(p) ] \Delta [P_n(p)] \ , \end{eqnarray}
we can write down the matricial system as
\be \sum_{n=2}^N \mathcal{C}_{ln} a_n=\langle P_l(x) | P_2(x) \rangle \ , \ee
where only starting at $N=2$ is the system compatible and determinate. The solution truncated at $N=2$ ($1\times1$ problem) is
\be a_2= \frac{||P_2(x)||^2}{\mathcal{C}_{22}} \ ,\ee
with
\begin{eqnarray}
  \nonumber \mathcal{C}_{22} & = & \frac{g_{\pi} \pi^2}{4m^4} \int \prod_{i=1}^4 \frac{d^3 k_i}{(2\pi)^32E_i} \overline{|T|^2} (2\pi)^4 \delta^{(4)} (k_1+k_2-k_3-p)  (1+n_1)(1+n_2)n_3 n_p \\
 & & \label{eq:c22bulk} \times \Delta [\left( \frac{E_p}{m}\right)^2]  \Delta [ \left( \frac{E_p}{m} \right)^2] \left( \frac{1}{3} - c_s^2\right)^2 \ . 
\end{eqnarray}

The bulk viscosity is\index{bulk viscosity}
\be \label{eq:finalbulk} \zeta =\frac{g_{\pi} m^4}{2 \pi^2 T} \  \frac{B_2}{ \mathcal{C}_{22}} \langle P_2 | P_2 \rangle  \ .\ee

For completeness --although not needed for the bulk viscosity-- we clarify that the two first components of the solution, $a_0$ and $a_1$,
are fixed by the conditions of fit. Up to $N=2$ the two conditions of fit read:
\be
\begin{array}{lll}
 \int \frac{d^3p}{E_p} n_p (1+n_p) E_p^2 \left[ a_0 P_0 + a_1 P_1 + a_2 P_2 \right]  & =  &  0 \ , \\
 \int \frac{d^3p}{E_p} n_p (1+n_p) E_p \left[ a_0 P_0 + a_1 P_1 + a_2 P_2 \right] & = & 0 \ , \\
\end{array}
\ee
that is converted into a non-homogeneous linear system for the coefficients $a_0$ and $a_1$:
\be \left\{ 
\begin{array}{lll}
  a_0 \langle x^2 | P_0 \rangle + a_1 \langle x^2 | P_1 \rangle  & = & -a_2 \langle x^2 | P_2 \rangle \ , \\
 a_0 \langle x | P_0 \rangle + a_1 \langle x | P_1 \rangle & = & -a_2 \langle x | P_2 \rangle \ . \\
\end{array} \right. 
\ee

\begin{figure}[t]
\begin{center}
\includegraphics[angle=90,width=200pt,height=150pt]{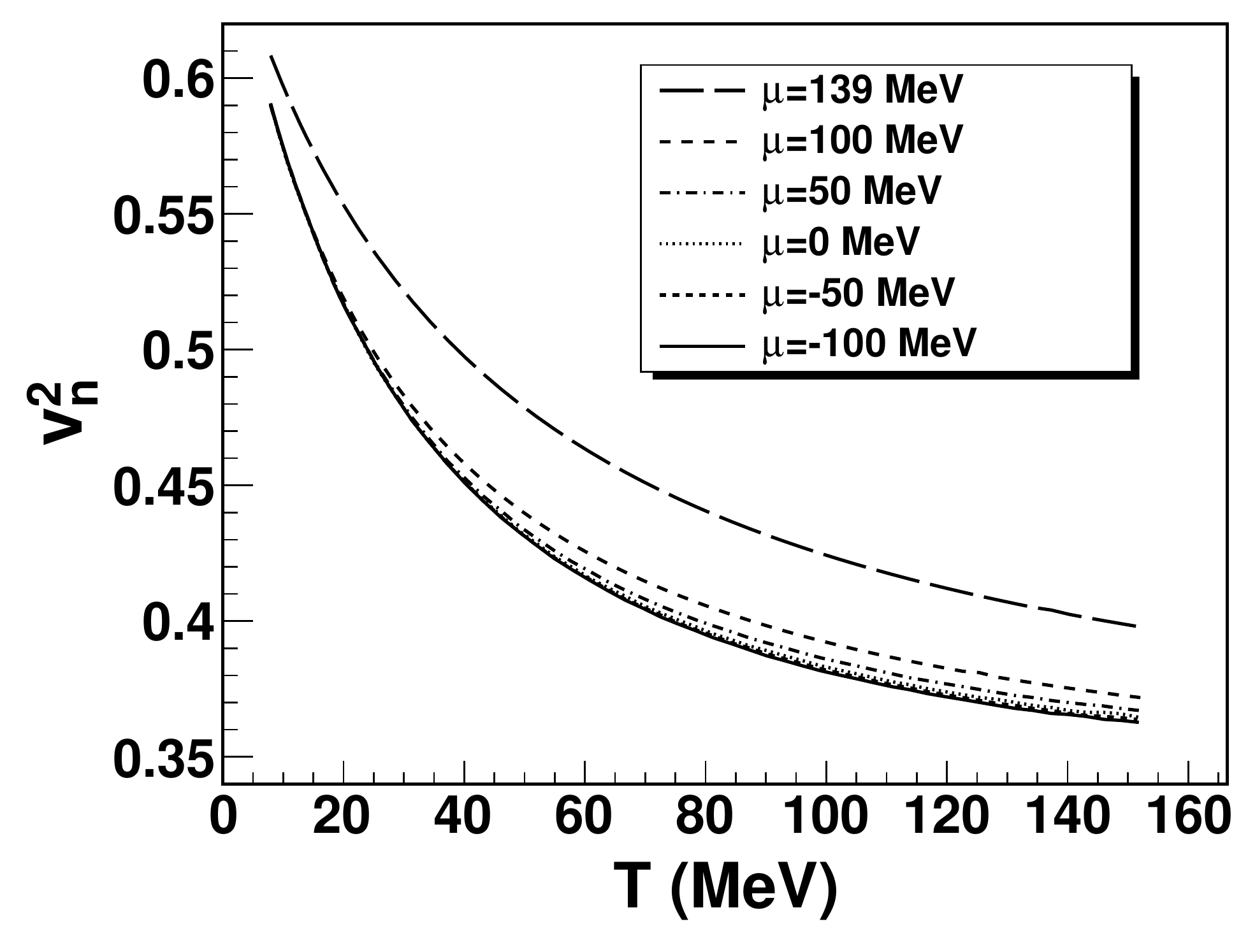} 
\includegraphics[angle=90,width=200pt,height=150pt]{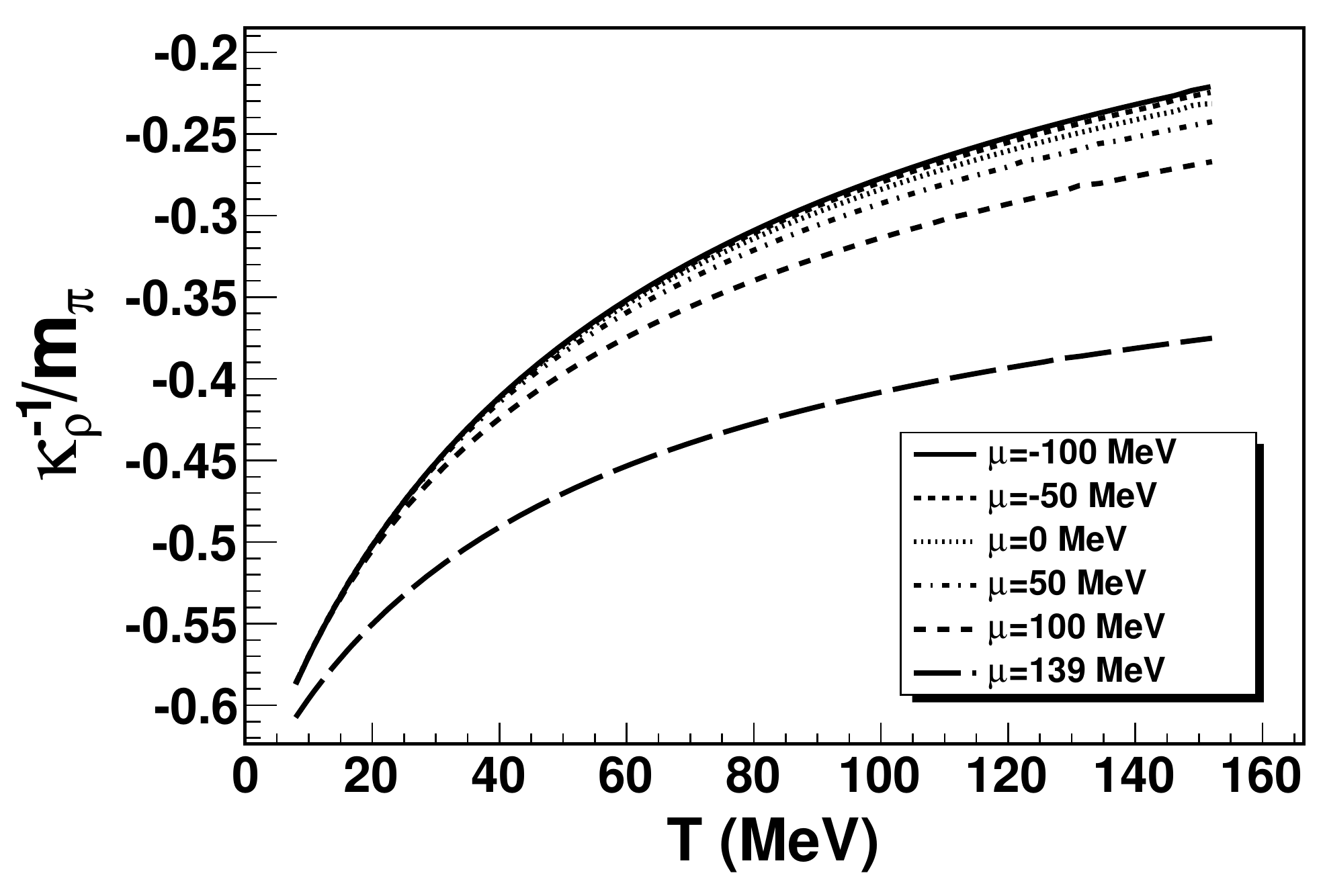} 
\caption{\label{fig:vnandkappa} Speed of sound at constant particle density and inverse compressibility at fixed energy density for a pion gas.}
\end{center}
\end{figure}

Finally, we plot the two thermodynamic functions $v_n^2$ \index{speed of sound!isochorus}and $\kappa_{\epsilon}^{-1}$\index{compressibility} as a function 
of temperature and pion chemical potential in Fig.~\ref{fig:vnandkappa}.

\begin{figure}[t]
\begin{center}
\includegraphics[scale=0.35]{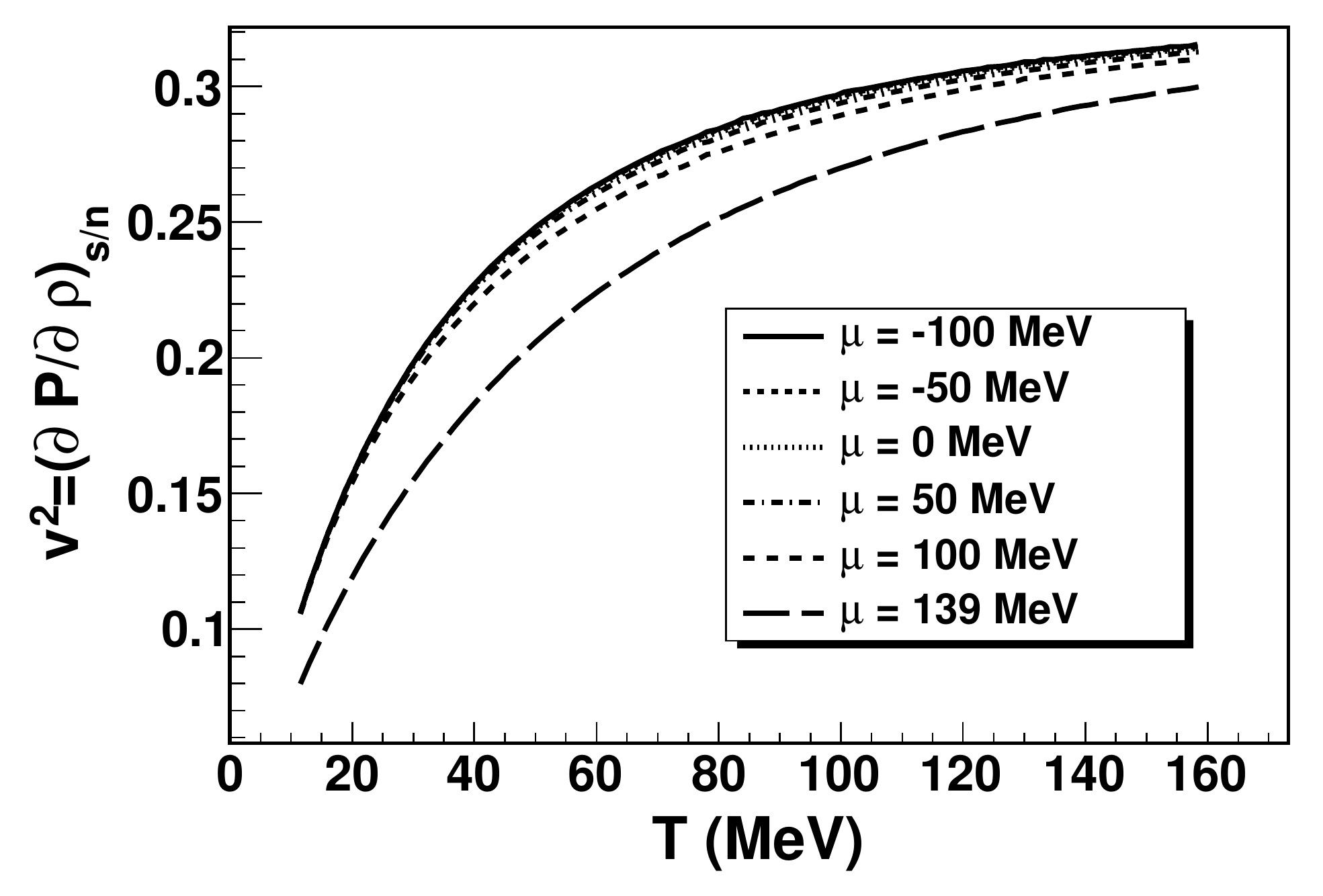} 
\caption{\label{fig:cs} Adiabatic speed of sound of an ideal pion gas as a function of temperature for several pion chemical potentials.}
\end{center}
\end{figure}

The adiabatic speed of sound, that controls the propagation of sound modes in the fluid is related with the previous functions by
\be v_s^2 = \left( \frac{\pa P}{\pa \epsilon} \right)_{s/n} = v_n^2 + \frac{n}{w} \kappa^{-1}_{\epsilon}\ ,\ee
and it is shown in Fig.~\ref{fig:cs} as a function of $T$ and $\mu$.

\begin{figure}[t]
\begin{center}
\includegraphics[scale=0.35]{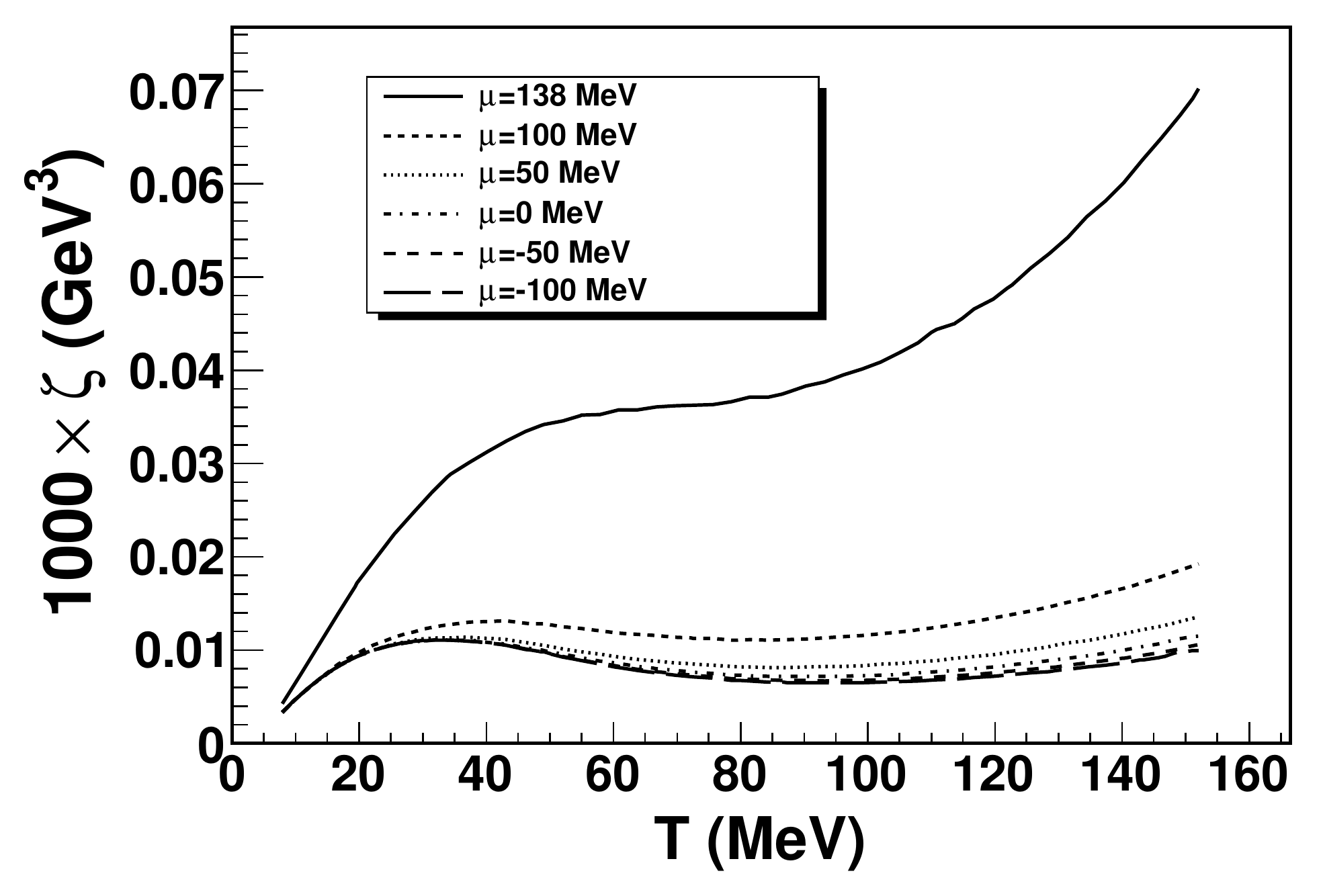} 
\includegraphics[scale=0.35]{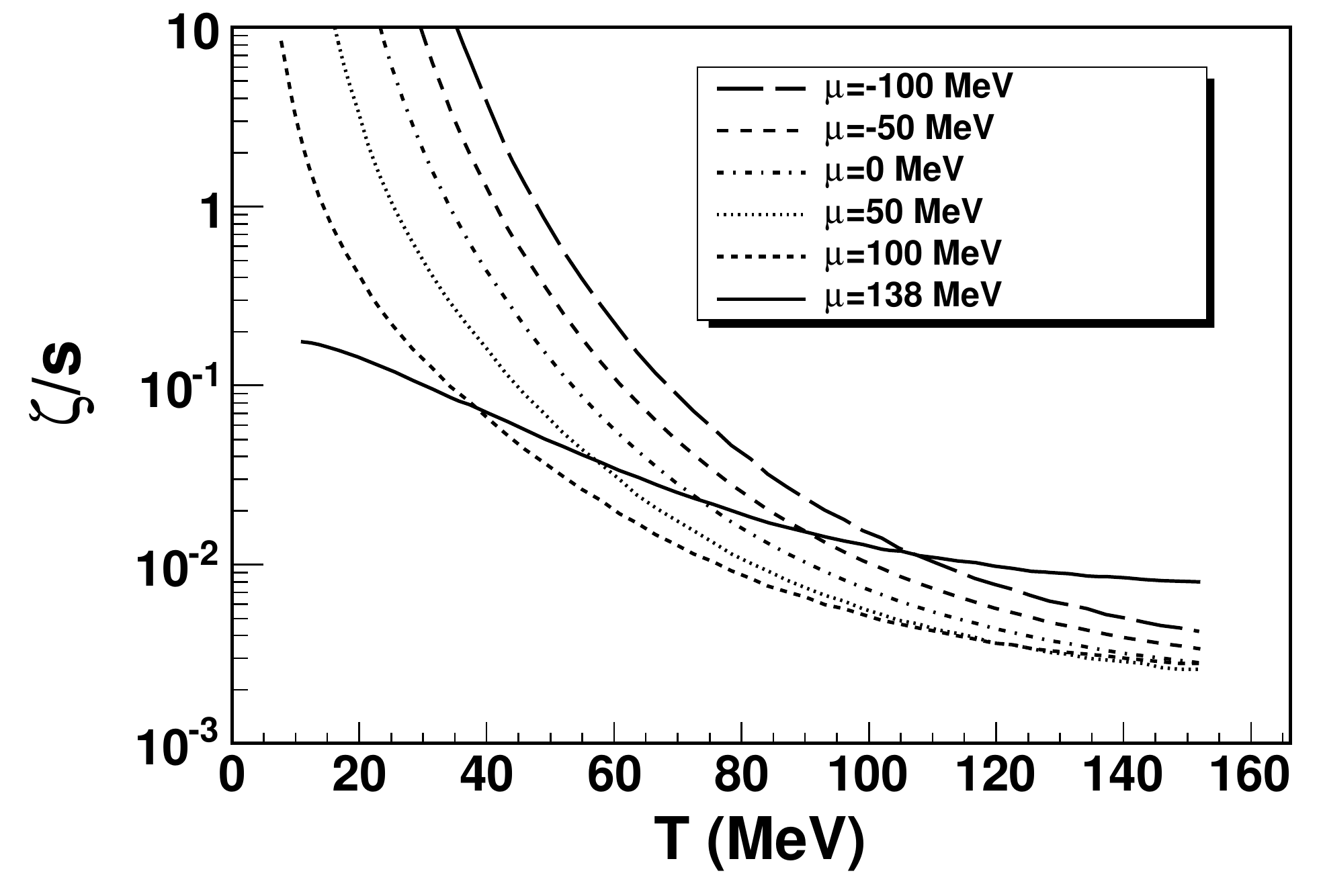} 
\caption{\label{fig:bulk_mu} Bulk viscosity of a pion gas in the Inverse Amplitude Method as a function of temperature for several pion chemical potentials. In the lower panel
we normalize the bulk viscosity to the entropy density.}
\end{center}
\end{figure}

In Fig.~\ref{fig:bulk_mu} we plot the bulk viscosity as a function of the pion
chemical potential up to $\mu=m$. In the lower panel of the same figure we normalize the result to the entropy density to create an adimensional
coefficient analogous to the KSS number.

\begin{figure}[t]
\begin{center}
\includegraphics[scale=0.35]{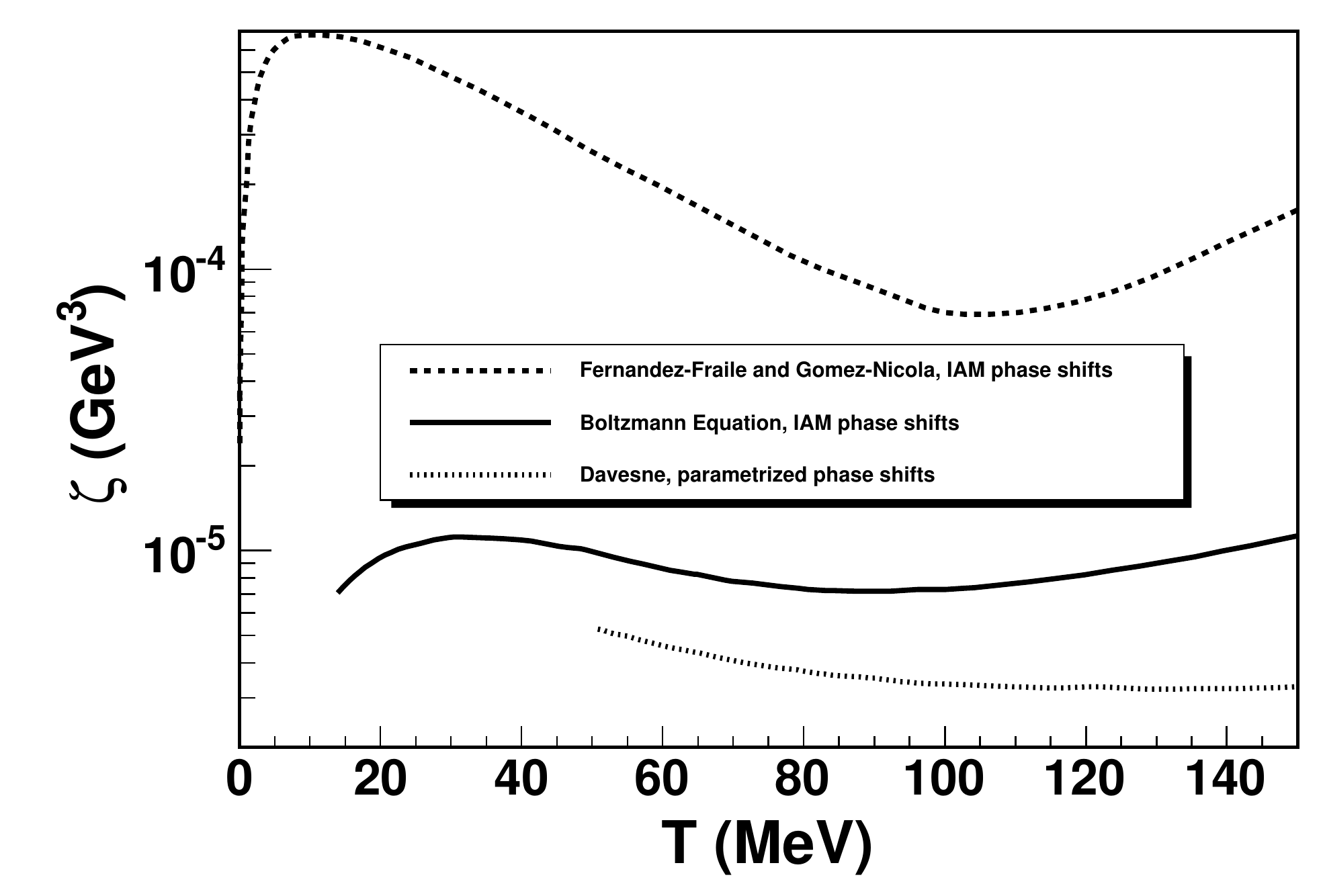} 
\caption{\label{fig:bulk_comp} Comparison of our computation with prior evaluations at $\mu=0$. }
\end{center}
\end{figure}

In Fig.~\ref{fig:bulk_comp} we compare our numerical results based on the $SU(2)$ inverse amplitude method \index{inverse amplitude method} with prior approaches based on the elastic pion-pion interactions. 
The first work \cite{Davesne:1995ms} employs a pion scattering amplitude that fits the experimental phase-shifts but has no connection to chiral
perturbation theory. Our calculation is numerically similar but somewhat higher. The second computation \cite{FernandezFraile:2008vu} is a field theory evaluation based on a certain ladder resummation, and is numerically off our result based on the
physical phase shifts. However, the qualitative features and saliently the low-temperature limit coincide with our findings.

\begin{figure}[t]
\begin{center}
\includegraphics[scale=0.35]{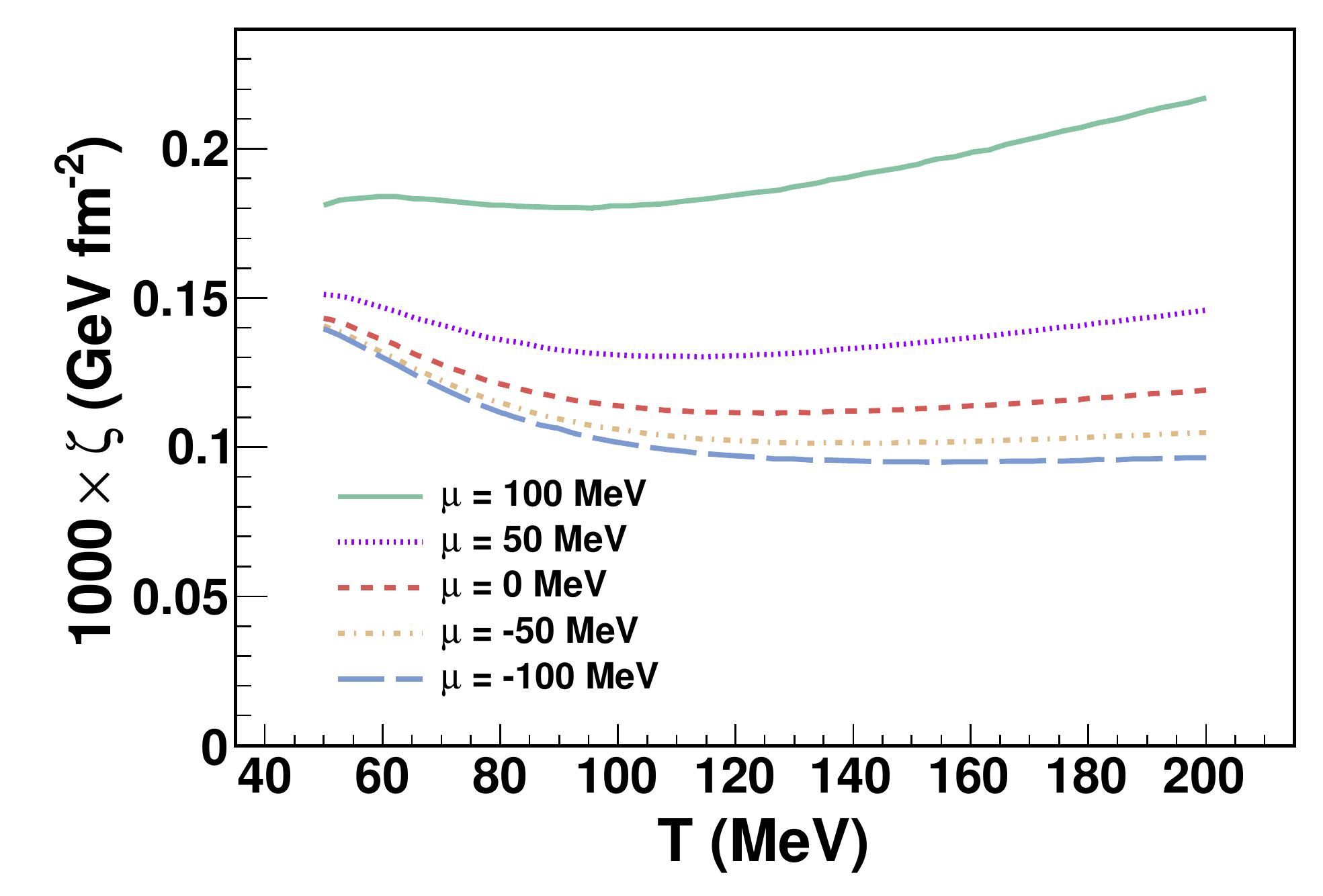}
\includegraphics[scale=0.35]{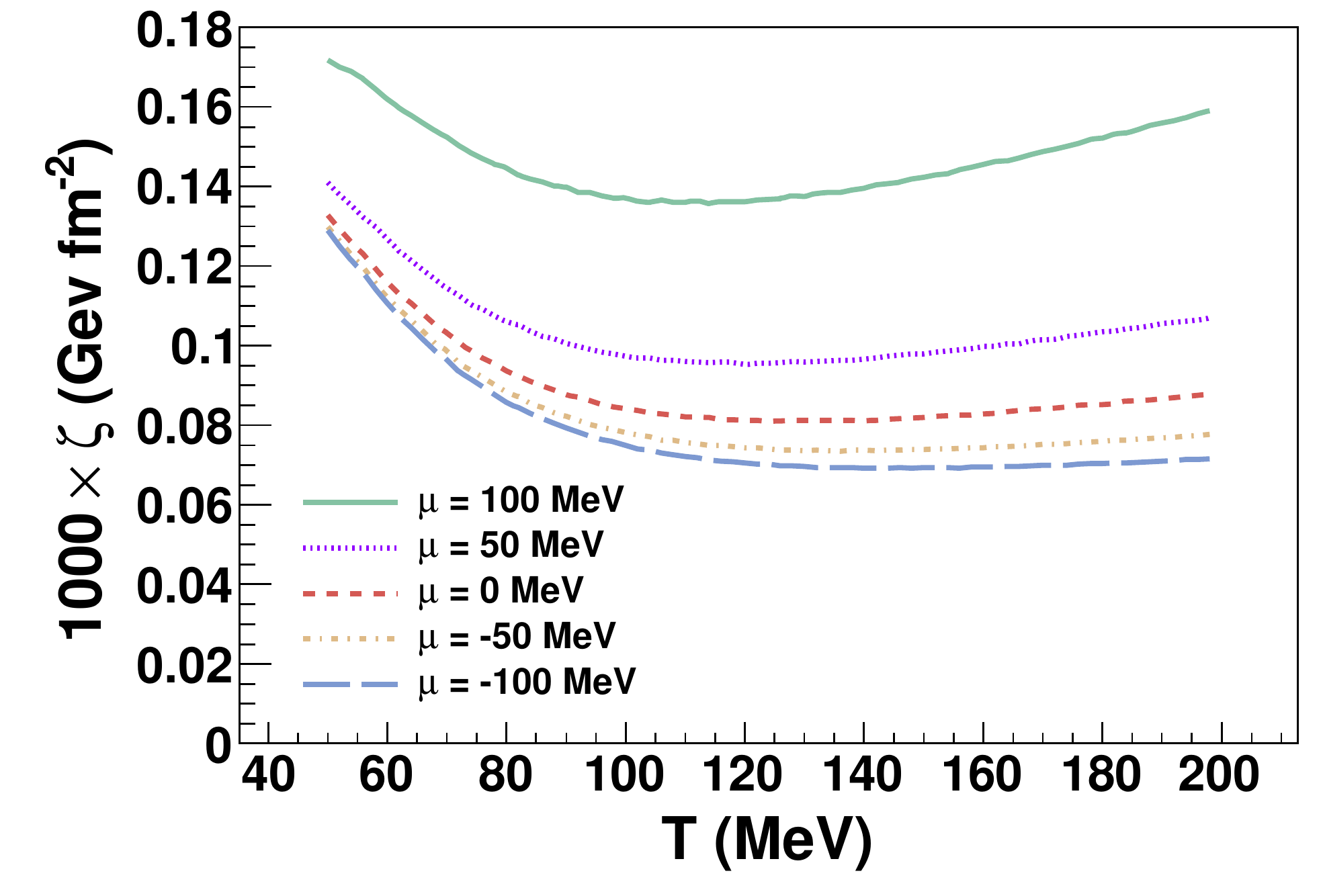}  
\caption{\label{fig:bulk_comp2} Comparison of our computation with the phase-shifts from \cite{Prakash:1993bt} (top panel) and the result of \cite{Davesne:1995ms} (bottom panel) for several pion chemical potentials. Data kindly provided by D. Davesne.}
\end{center}
\end{figure}

Finally, in Fig.~\ref{fig:bulk_comp2} we compare our results with the phenomenological phase-shifts in \cite{Prakash:1993bt} and those obtained by \cite{Davesne:1995ms} with the same phase-shifts.

\newpage

\section{$\zeta/s$ in perturbative QGP}

\begin{figure}[t]
\begin{center}
\includegraphics[scale=0.35]{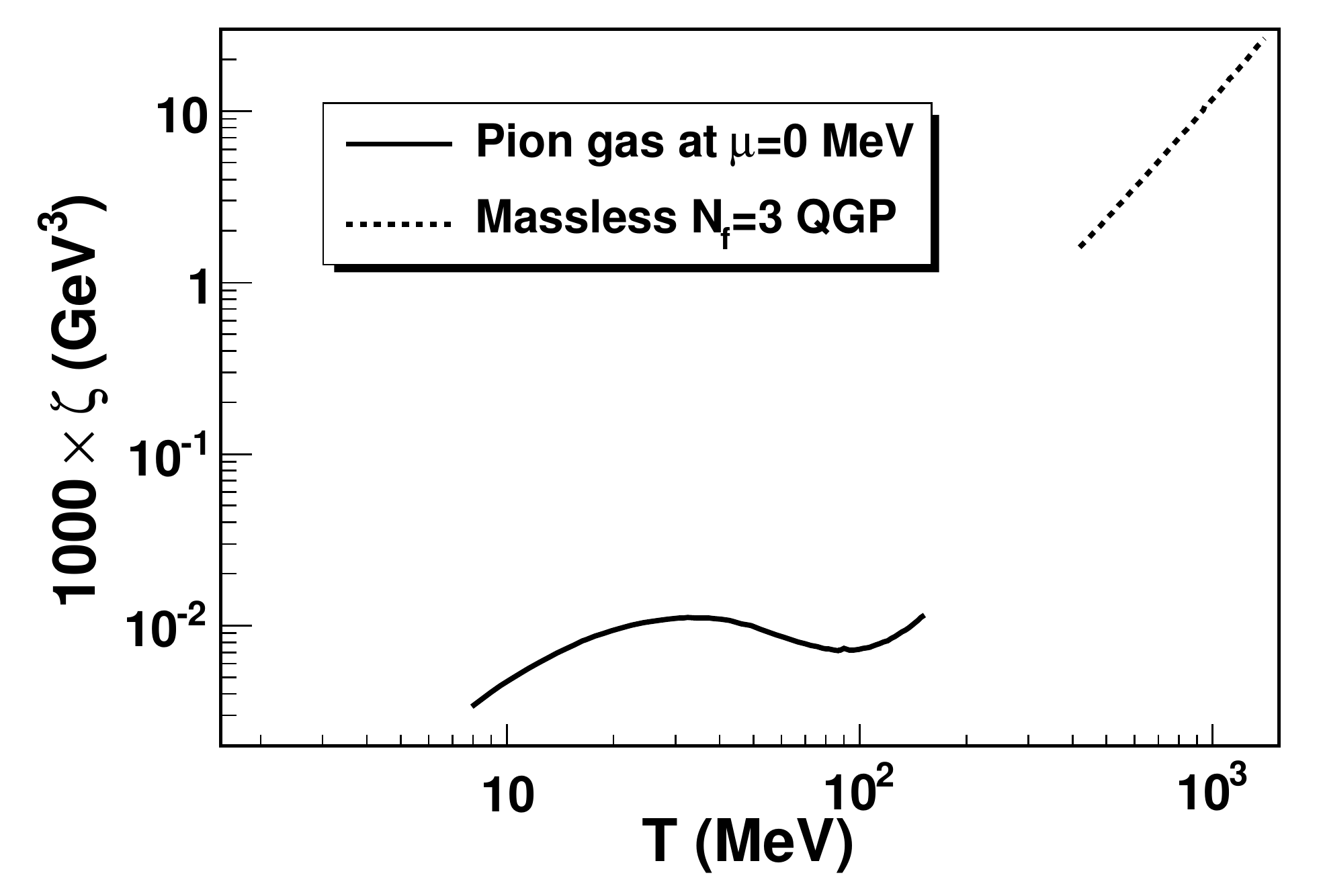} 
\includegraphics[scale=0.35]{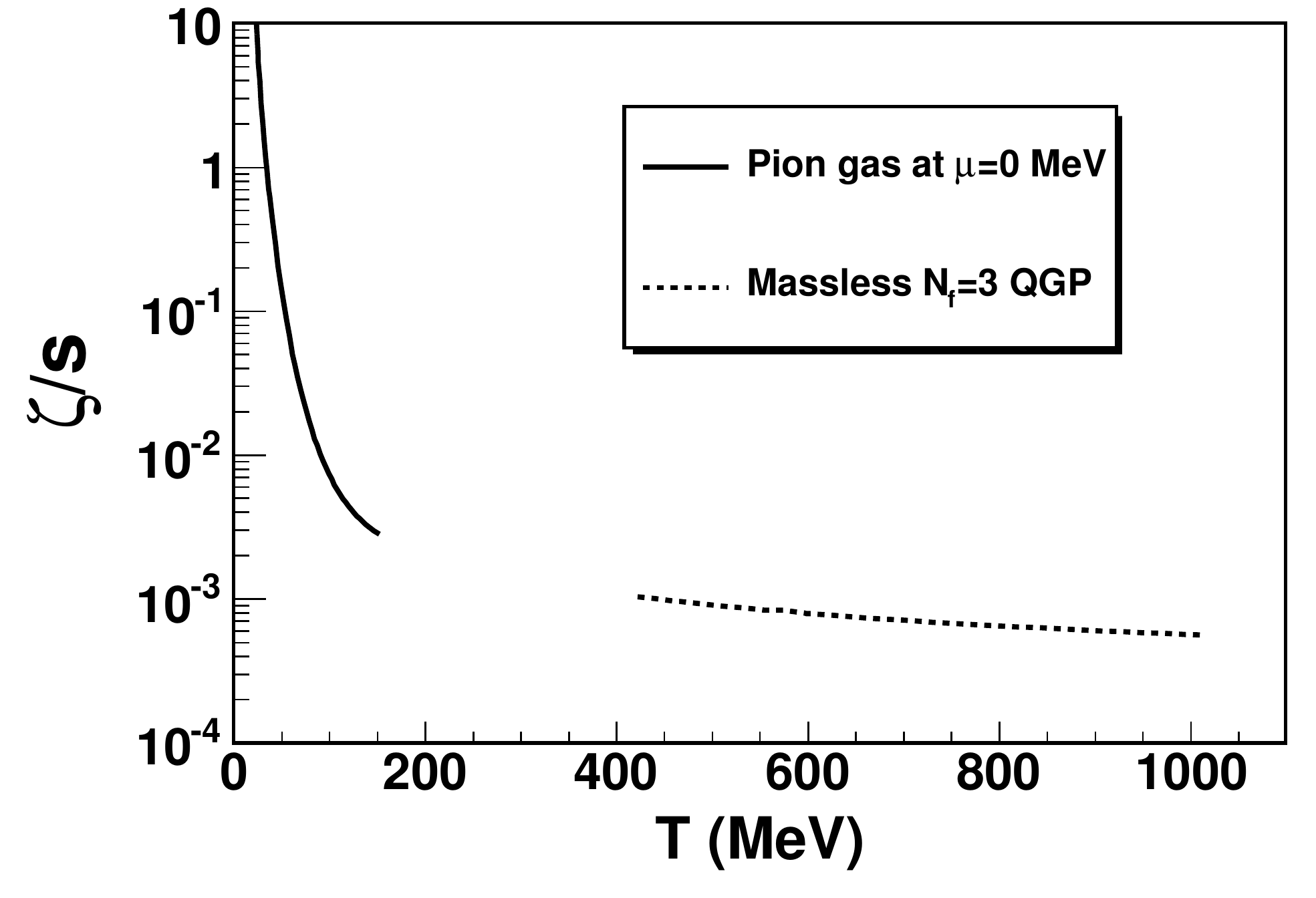} 
\caption{\label{fig:bulk_qgp} Bulk viscosity of a pion gas in the inverse amplitude method and the high temperature result of perturbative QCD for a quark-gluon plasma phase. In the lower panel
we normalize the bulk viscosity over the entropy density.}
\end{center}
\end{figure}

One can try to make a connection with the quark-gluon plasma at higher temperatures.
The bulk viscosity\index{bulk viscosity!in perturbative plasma} can be calculated for a perturbative gas of massless quarks and gluons to leading order in $\alpha_s(T)$\index{$\alpha_s$ thermal}. This was done in
\cite{Arnold:2006fz} for $N_c=3$ taking into account number changing processes (vanishing chemical potential). 
For $N_f=2$ the expression for $\eta/s$ is
\be \frac{\zeta}{s} = 2.52 \cdot 10^{-4} \ \frac{g^4}{\log \left( \frac{6.556}{g} \right)} \ ,\ee
and for $N_f=3$: 
\be \frac{\zeta}{s} = 2.00 \cdot 10^{-4} \ \frac{g^4}{\log \left( \frac{6.344}{g} \right)} \ ,\ee
with the thermal coupling constants given in (\ref{eq:qcd_coupling2}) and (\ref{eq:qcd_coupling3}) respectively.
We plot this result together with the pion gas bulk viscosity at vanishing chemical potential in Fig.~\ref{fig:bulk_qgp}.

As in the case of the shear viscosity it is not clear what happens to $\zeta/s$ near the crossover temperature. None of the two limiting theories (ChPT and perturbative QCD) for this gas
can provide a description of a deconfined phase-transition. A possible maximum at $T_c$ could be expected when looking at
lattice QCD\index{lattice QCD} results in \cite{Karsch:2007jc}.

\chapter{Thermal and Electrical Conductivities \label{ch:5.conductivities}}

The thermal and electrical conductivities are the last transport coefficients which we will calculate in the pure pion gas. These coefficients have not
received much attention in the pion gas because they only appear in a system with a conserved number of particles. As we have discussed before,
the pion interaction does not conserve the pion number as the effective Lagrangian allows for particle number changing processes. However, in the
low temperature limit, these processes are strongly suppressed and the pion number is effectively conserved. This conservation permits us
to have a well defined thermal conductivity \index{thermal conductivity} in the system, that appears when a gradient of temperature and pressure
is applied to the system. As the pion gas is composed of electrically charged pions one can consider
the electrical conductivity \index{electrical conductivity} when a small external electric field \index{electric field}is applied.

\section{Thermal conductivity}

We start with the thermal or heat conductivity that we have investigated in~\cite{Dobado:2007cv}. However, we implement several changes and improvements with respect to that reference.

The left-hand side of the BUU equation that enters in the description of the thermal conductivity contains those terms carrying a gradient of temperature and pressure:

\be \label{eq:lhsthermal} \left. p_{\mu} \pa^{\mu} n_p (x) \right|_{\kappa} =\beta^2 n_p(x) [1+n_p(x)] \left( E_p -\frac{w}{n} \right) \mathbf{p} \cdot \left( \mathbf{\nabla} T - \frac{T}{w} \mb{\nabla} P\right) \ .\ee

The parametrization for the unknown perturbation function $\Phi_a$ analogous to Eq.~(\ref{eq:phishear}) is chosen to be

\be \label{eq:pert_heat} \Phi_a = \beta^3 \ \mb{k}_a \cdot \left( \mb{\nabla} T - \frac{T}{w} \mathbf{\nabla} P \right) C(k_a) \ , \ee
where the scalar function $C(k_a)$ is an adimensional function of $|\mb{k}_a|$. We do not include the factor $E_p-w/n$ in the parametrization in order
to be able to catch the zero mode inside the collision operator when expanding the function $C(p)$ in powers of $x$. This will be clearer in the following step.

Cancelling the term $\left( \mb{\nabla} T - \frac{T}{w} \mb{\nabla} P \right)$ between left-hand side and right-hand side, the linearized BUU equation reads

\begin{eqnarray} \nonumber  n_p (1+n_p) \left( E_p -\frac{w}{n} \right) \ p^i & = & \frac{g_{\pi} E_p}{2T} \int d \Gamma_{12,3p}  (1+n_1)(1+n_2)n_3 n_p \\
 \label{eq:buuforkappa} & & \times  \left[ p^i C(p) + k_3^i C(k_3) - k_1^i C(k_1) 
- k_2^i C(k_2) \right] \ . \end{eqnarray}

One appreciates the presence of one zero mode \index{zero modes} in the collisional operator. It appears when $C(k_a)$ is proportional to $1$
and is associated with the momentum conservation law. This zero mode is responsible for the indeterminacy of one component of the final solution $C(p)$.
The BUU equation does not determine the component proportional to this zero mode. However, this component can be fixed by using the Landau-Lifshitz condition of fit.
When the zero mode is not properly identified --for example,
by using a different parametrization for $\Phi_p$ in which the zero mode appears hidden-- one has problems of convergence as in~\cite{Dobado:2007cv}
at low temperatures.

\subsection{Heat flow and Fourier's law \label{sec:heatflow}}

The heat flow \index{heat flow} is defined in an arbitrary reference frame as the difference between the energy flow and the enthalpy flow \cite{de1980relativistic}:
\be \label{eq:heatflux} I_q^{\mu}= \left(u_{\nu} T^{\nu \sigma} - \frac{w}{n} n^{\sigma} \right) \Delta_{\ \ \sigma}^{\mu} \ , \ee
where $w$ is the enthalpy density, $n$ is the particle density and $n^{\sigma}$ is the particle number four-flux. The projector orthogonal to the velocity 
$\Delta_{\ \ \sigma}^{\mu}$ is defined in Appendix~\ref{app:hydro}. From its definition it follows that
\be I_q^{\mu} u_{\mu}=0 \ . \ee

Different choices of reference frame are used in the literature.
We feel that the easiest to compute is the local rest reference frame as in \cite{Gavin:1985ph,Prakash:1993bt,Davesne:1995ms}.
In this frame the two conditions of fit that define the temperature and chemical potential out of equilibrium read:
\begin{eqnarray}
\nu^0 & = & 0 \ , \\
\tau^{00} & = & 0 \ . 
\end{eqnarray}
These conditions of fit\index{conditions of fit} were relevant for the calculation of the bulk viscosity in Chapter~\ref{ch:4.bulk}. The relevant condition of fit for the thermal conductivity is the
condition that fixes the velocity field. We will use that of Landau-Lifshitz associating the velocity of the fluid with the direction of the energy flow.
In the local rest frame it reads:
\be \label{eq:llcf} \tau^{0i}=0  \ , \ee
whence the same components of the energy-momentum tensor \index{energy-momentum tensor} vanish in such frame
\be T^{0i} = \gamma^2 w u^i + \tau^{0i} =0 \ . \ee
In equilibrium, the particle flux is proportional to the fluid element's velocity as well:
\be n_{eq}^i=nu^i \ . \ee
Therefore as the energy flow \index{energy flow}and the particle flow \index{particle flow} are proportional to $u^i$, they both vanish in the local rest frame.

However, this is not true out of equilibrium. Whereas the energy flow vanishes, the particle flux is not necessarily zero due to the nonequilibrium effects.
\be n^i = nu^i + \nu^i = \nu^i \ . \ee

In the local reference frame the zero component of $I_q^{\mu}$ vanishes, $I_q^0=0$ as is evident from the formula
\be I_q^{\mu}=u_{\nu} T^{\nu \mu} - u_{\nu} u_{\sigma} u^{\mu} T^{\nu \sigma} - \frac{w}{n} n^{\mu} + w u^{\mu} \ , \ee
coming directly from Eq.~(\ref{eq:heatflux}). And the spatial components are

\be \label{eq:heatflow} I_q^i= T^{0i} -\frac{w}{n} n^i = - \frac{w}{n} \nu^i \ .\ee
As the $T^{0i}$ vanish, the heat flux \index{heat flow}is proportional to the particle flux,
with the enthalpy per particle as proportionality constant\index{enthalpy density}.

The explicit form of $\nu^i$ is given in Appendix \ref{app:hydro} and it reads \cite{landau1987fluid}:
\be \nu^{\mu}= -\kappa \left( \frac{nT}{w} \right)^2 \left[ \pa^{\mu} \left( \frac{\mu}{T} \right) - u^{\mu} u^{\nu} \pa_{\nu} \left( \frac{\mu}{T} \right) \right] \ , \ee
where $\kappa$ is the heat conductivity coefficient.
In the local rest frame, the zero component is trivially zero (as expected) and the spatial components are
\be  \nu^i= -\kappa \left( \frac{nT}{w} \right)^2 \pa^i \left( \frac{\mu}{T} \right) = \kappa \frac{n}{w} \left(\pa^i T - \frac{T}{w} \pa^i P \right) \ ,  \ee
where we have used the Gibbs-Duhem relation (\ref{eq:gibbs-duhem}). Finally, the expression for the heat flow is
\be \label{eq:fourierlaw} I_q^i = - \frac{w}{n} \nu^i = -\kappa \left(\pa^i T - \frac{T}{w} \pa^i P \right) \ .\ee
This is the well-known Fourier's law \index{Fourier's law} where the last factor, proportional to the pressure\index{pressure} gradient, is a relativistic generalization.


In Appendix~\ref{app:hydro} we make the connection to kinetic theory in order to express the non-equilibrium particle flux
through an integration over the distribution function:
\be \nu^i =  g_{\pi} \int \frac{d^3p}{(2\pi)^3} \frac{p^i}{E_p} \ f_p^{(1)} \ . \ee
The heat flux is therefore obtained from Eq.~(\ref{eq:heatflow})
\be I_q^i =  - g_{\pi} \frac{w}{n} \int \frac{d^3p}{(2\pi)^3} \frac{p^i}{E_p} \ f_p^{(1)}  \ . \ee

Using the Landau-Lifshitz condition of fit (\ref{eq:llcf}) we can add a convenient zero contribution \index{conditions of fit}:
\be \label{eq:landau_cond} \tau^{i0}= g_{\pi} \int \frac{d^3p}{(2 \pi)^3} \frac{p^i}{E_p} E_p \ f_p^{(1)} =0 \ , \ee
to get from (\ref{eq:fourierlaw}):

\be \label{eq:heat_con} g_{\pi} \int \frac{d^3p}{(2\pi)^3}  f_p^{(1)} \frac{p^i}{E_p} \left( E_p - \frac{w}{n} \right) = - \kappa \left( \pa^i T - \frac{T}{w} \pa^i P \right) \ . \ee

Substituting the {\it ansatz} for $f_p^{(1)}=-n_p (1+n_p) \Phi_p$ with $\Phi_p$ given in Eq.~(\ref{eq:pert_heat}) we get

\be g_{\pi} \int \frac{d^3p}{(2\pi)^3} n_p(1+n_p) \beta^3 C(p) \ p_j \left( \pa^j T - \frac{T}{w} \pa^j P\right)
\frac{p^i}{E_p} \left( E_p - \frac{w}{n} \right) = \kappa \left(\pa^i T - \frac{T}{w} \pa^i P \right) \ .\ee

Invariance under rotations of the integrand allows to substitute

\be \label{eq:integvector} \int d^3 p \ f(p) \mathbf{p} \cdot \mathbf{a} \  p_j = \frac{1}{3} a_j \int d^3p \ f(p) p^2, \ee
for any fixed vector $a_{j}$ independent of momentum.

The expression for the thermal conductivity \index{thermal conductivity} is therefore

\be \kappa  = \frac{g_{\pi}}{3T^3} \int \frac{d^3p}{(2\pi)^3} n_p (1+n_p) \frac{p^2}{E_p} C(p) \left( E_p - \frac{w}{n} \right). \ee

\subsection{Integration measure \label{sec:conducinteg}}

Using the variables defined in Eq.~(\ref{eq:new_var}) we can express the heat conductivity as

\be \label{eq:conducescprod} \kappa  = \frac{g_{\pi}m^5}{6\pi^2 T^3} \int_1^{\infty} dx \
\frac{z^{-1} e^{y(x-1)}}{[z^{-1} e^{y(x-1)}-1]^2} (x^2-1)^{3/2} \ C(x)
 \left( x - \frac{w}{mn} \right) \ . \ee
This integral expression defines a natural integration measure
\be d\mu_{\kappa} \equiv dx \ \frac{z^{-1} e^{y(x-1)}}{[z^{-1} e^{y(x-1)}-1]^2} (x^2-1)^{3/2} \ , \ee
that in terms of physical variables reads
\be d\mu_{\kappa}= d^3p \frac{p^2}{E_p} n_p (1+n_p)  \frac{1}{4\pi m^4} \ . \ee
This measure induces an inner product that we can use to express the heat conductivity as
\be \kappa  = \frac{g_{\pi}m^5}{6\pi^2 T^3} \langle C(x)| \left(x-\frac{w}{mn} \right) \rangle \ . \ee

We can also define the integrals $J_i$ with $i=n+m$:
\be \label{eq:Jintegrals} J_i = \langle x^n | x^m \rangle = \int d\mu_{\kappa} x^i = \int dx \ \frac{z^{-1} e^{y(x-1)}}{[z^{-1} e^{y(x-1)}-1]^2} (x^2-1)^{3/2} x^i \ . \ee

For the polynomial basis we choose $P_0$ to be the generator of the zero mode\index{zero modes}, i.e. proportional to one.
\be \label{eq:condp0} P_0 = 1 \ . \ee
The next element $P_1(x)$ is defined to be the source function in the BUU equation.
\be \label{eq:condp1} P_1 (x)=x - \frac{w}{mn} \ , \ee
where the entalphy density $w$ and particle density $n$ are defined in the ideal gas. It is possible to express this ratio
in terms of the $J_i$ integrals (see Eqs.~(\ref{eq:somequan}) and (\ref{eq:JrekJ}) of Appendix \ref{app:hydro} for more details):
\be \frac{w}{mn} = \frac{J_1}{J_0} \ . \ee
The rest of the basis is constructed in such a way that each element is monic and orthogonal to the others. For instance, the next element reads
\be \label{eq:condp2} P_2 (x)=x^2+P_{21} x + P_{20} \ ,  \ee
with
\be P_{21} = \frac{J_0J_3-J_1J_2}{J_1^2-J_0J_2};\quad P_{20}=\frac{J_2^2-J_1J_3}{J_1^2-J_0 J_2} \ . \ee

Indeed, it is possible to check that the whole basis is orthogonal, due to the fact that $P_0$ and $P_1(x)$ are perpendicular
\be \label{eq:p0ortop1} \langle P_0 | P_1 \rangle =  J_1 - \frac{w}{mn} J_0 = 0 \ . \ee

The relation (\ref{eq:p0ortop1}) is crucial to have a solution of the linearized BUU equation.
The left-hand side of the BUU equation is nothing but $P_1$ written in adimensional variables. The left-hand side plays the role of the source function
from the point of view of the integral equation. As the vector that generated the zero mode is perpendicular to the source function, 
the Fredholm's alternative states that the linearized BUU equation is therefore solvable in the whole space generated by the basis
(even in the subspace containing the zero mode\index{zero modes}). We multiply the BUU equation \index{BUU equation} 

\be \nonumber  n_p (1+n_p) \left( E_p -\frac{w}{n} \right) \ p^i \ee
\be \label{eq:buu_thermal}= \frac{g_{\pi}E_p}{2T} \int d \Gamma_{12,3p}  (1+n_1)(1+n_2)n_3 n_p \left[ p^i C(p) + k_3^i C(k_3) - k_1^i C(k_1) 
- k_2^i C(k_2) \right] \,  \ee

by $\frac{1}{4 \pi m^5} p_i \frac{1}{E_p} P_l(x)$ and integrate over $d^3p$ to get:

\begin{eqnarray} \nonumber  \langle P_l(x) | P_1(x) \rangle& =& \frac{g_{\pi}}{8 \pi m^5 T} \int d^3 p \ P_l(x)\int d \Gamma_{12,3p} (1+n_1)(1+n_2)n_3 n_p  \\
 & & \times p_i \left[ p^i C(p) + k_3^i C(k_3) - k_1^i C(k_1) - k_2^i C(k_2) \right] \ . 
\end{eqnarray}

Now we substitute the expansion of the solution $C(p)$ as a linear combination of the element of the basis:

\be \label{eq:heatexpansion} C(x)=\sum_{n=0}^{\infty} c_n P_n(x) \ee

and symmetrize the right-hand side to obtain 

\ba
 \nonumber  \delta_{l1} ||P_1(x)||^2   & = & \sum_{n=0}^{\infty} c_n \frac{g_{\pi} \pi^2}{4 m^5 T}  \prod_{j=1}^4 \frac{d^3k_j}{(2\pi)^32E_j} \ \overline{|T|^2} (2\pi)^4 \delta^{(4)} (k_1+k_2-k_3-p)  \\
& & \nonumber  \times (1+n_1)(1+n_2)n_3 n_p \ \left[ p_i P_l (p) + k_{3i} P_l(k_3) - k_{1i} P_l (k_1) - k_{2i} P_l (k_2) \right] \\
\label{eq:libuuheat} &  & \times \left[ p^i P_n(p) + k_3^i P_n(k_3) - k_1^i P_n(k_1) - k_2^i P_n(k_2) \right], 
\ea

with
\be ||P_1(x)||^2 = \frac{J_2 J_0 - J_1^2}{J_0} \ .\ee

The term with $n=l=0$ gives a trivial $0=0$ equation that does not
fix the coefficient $c_0$. This coefficient can be fixed by the Landau-Lifshitz condition of fit:
\be g_{\pi} \int \frac{d^3p}{(2\pi)^3 E_p} \ f_p^{(1)} E_p \ p^i =0 \ , \ee
that under some algebra turns out to be
\be c_0 = - c_1 \frac{\langle x | P_1(x) \rangle}{\langle x | P_0(x) \rangle} \ , \ee
at first order in the expansion (\ref{eq:heatexpansion}) \footnote{However this coefficient is not needed for the heat conductivity in Eq.~(\ref{eq:conducescprod}) because of the orthogonality between $P_0$ and $P_1$.}.

The first coefficient to be set by the kinetic equation (\ref{eq:libuuheat}) is therefore $c_1$.

The set of equations above can be expressed as a matricial system,
\be \sum_{n=1}^N \mathcal{C}_{ln}  c_n = \delta_{l1} ||P_1(x)||^2 \ , \ee
where
\ba
\mathcal{C}_{nl} & \equiv  & \frac{g_{\pi} \pi^2}{4 m^5 T}  \prod_{j=1}^4 \frac{d^3k_j}{(2\pi)^3 2E_j} \ \overline{|T|^2} (2\pi)^4 \delta^{(4)} (k_1+k_2-k_3-p)  \nonumber \\
& & \times (1+n_1)(1+n_2)n_3 n_p \ \Delta [ p_i P_l (p)] \ \Delta[ p^i P_n(p) ] \ , \ea
with 
\be \Delta [p_i P_l(p)] \equiv \left[ p_i P_l(p) + k_{3i} P_l(k_3) - k_{1i} P_l(k_1) - k_{2i} P_l(k_2) \right] \ .\ee

The lowest order solution is:
\be c_1= \frac{||P_1(x)||^2}{\mathcal{C}_{11}} \ . \ee
with
\ba 
\nonumber \mathcal{C}_{11} & = & \frac{g_{\pi} \pi^2}{4 m^5 T} \int \prod_{j=1}^4 \frac{d^3 k_j}{(2\pi)^32E_j} \overline{|T|^2} (2\pi)^4 \delta^{(4)} (k_1+k_2-k_3-p)  (1+n_1)(1+n_2)n_3 n_p \\
& & \label{eq:c11thermal} \times \Delta [p_i \frac{E_p}{m}] \ \Delta [p^i \frac{E_p}{m}] \ . \ea

The heat conductivity finally reads:
\be \label{eq:heatfirst} \kappa =\frac{g_{\pi} m^5}{6 \pi^2 T^3} \ c_1 \langle P_1 | P_1 \rangle
=\frac{g_{\pi} m^5}{6 \pi^2 T^3} \frac{||P_{1}(x)||^4}{\mathcal{C}_{11}} \ .\ee

\begin{figure}[t]
\begin{center}
\includegraphics[scale=0.35]{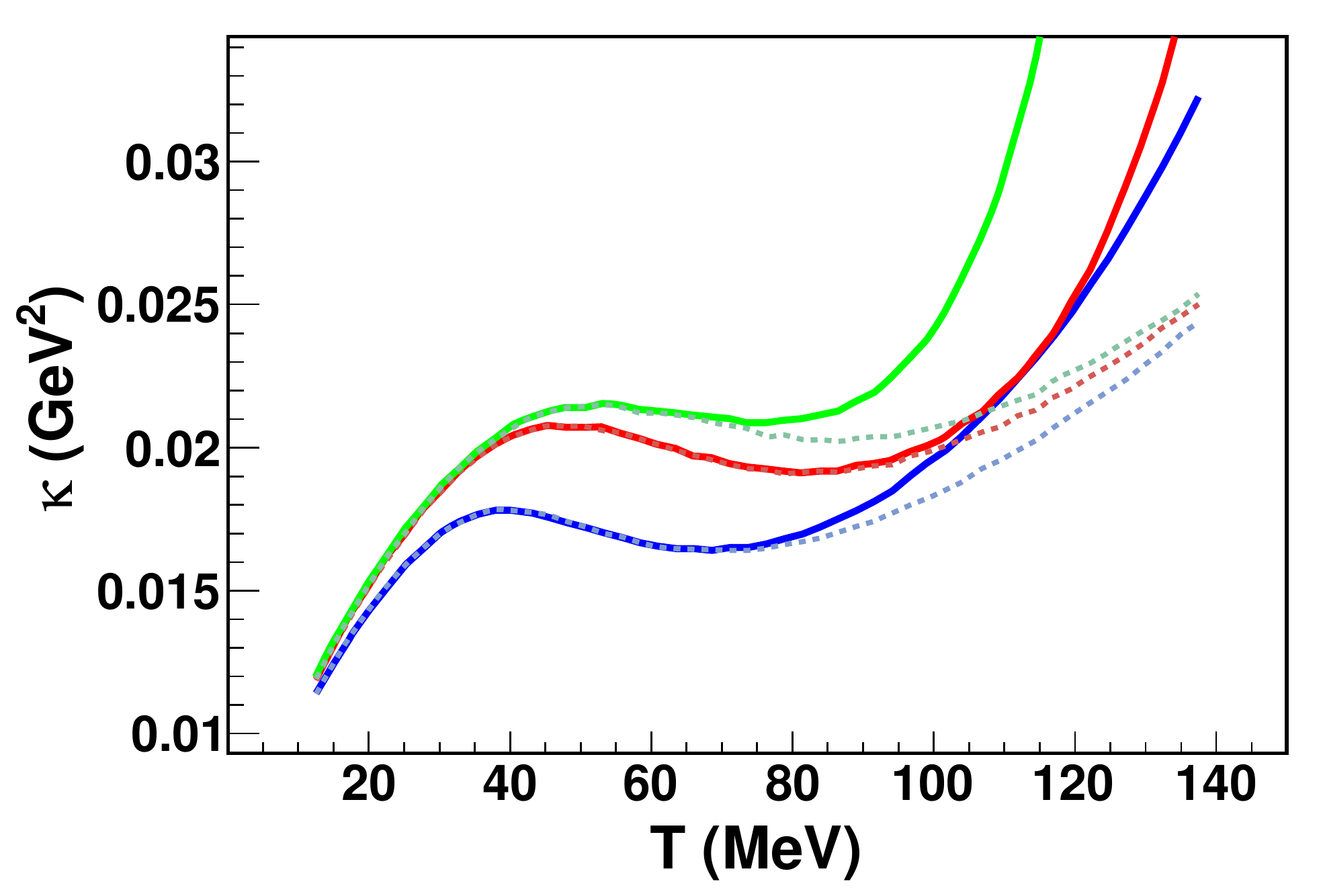}
\includegraphics[scale=0.35]{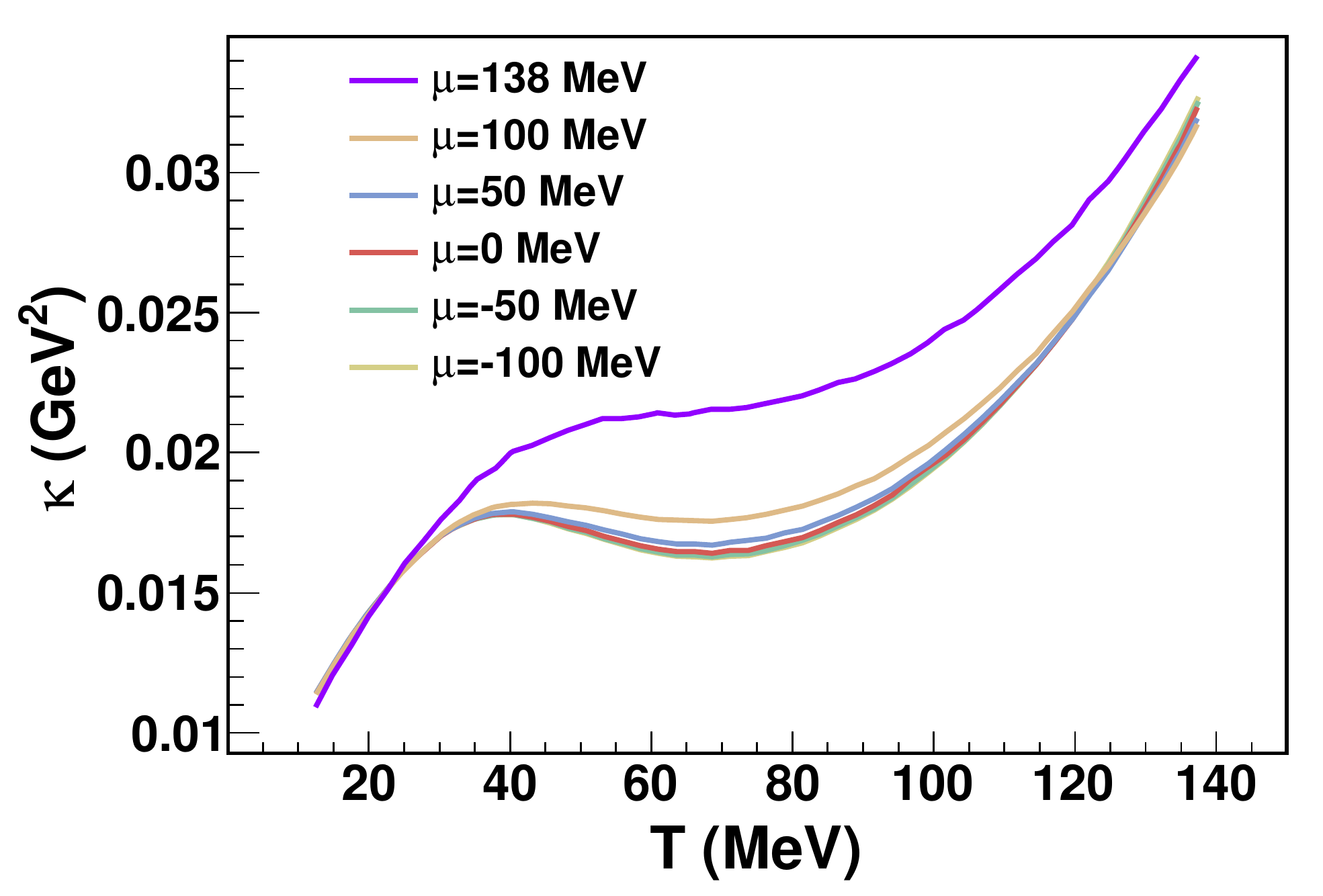}
\caption{\label{fig:heat1pra} Upper panel: Heat conductivity of a pion gas at vanishing chemical potential. The blue curves are a first order calculation, the red curves second order
and the green ones third order. The solid curve were computed with a momentum cutoff of $1.2$ GeV and the dashed ones without cutoff, all with the phenomenological phase-shifts in \cite{Prakash:1993bt}. Lower panel: Heat conductivity at first order with
several pion chemical potentials.}
\end{center}
\end{figure}

In the upper panel of Fig.~\ref{fig:heat1pra} we show the heat conductivity of a pion gas at $\mu=0$. For all curves we use the phenomenological phase-shifts given in \cite{Prakash:1993bt} (results from the inverse amplitude method will be shown shortly).
The two blue curves correspond to the first order calculation given 
by Eq.~(\ref{eq:heatfirst}), the second and third orders are given by the red and green curves, respectively. 

The solid curves are the calculations with a momentum cutoff of $1.2$ GeV, physically mandatory because of the lack of knowledge of the interaction beyond this scale.

By ignoring the issue and blindly extending the $p-$integrals with a constant cross section above $p=1.2$ GeV one can see that the convergence is rather poor at higher temperatures. 
This problem appears always when the momentum cutoff is not high enough
for the thermal distribution functions to force good integral convergence. Of course, this effect is affecting at high temperatures of the order of $T > 100$ MeV.
At low temperatures, the distribution functions force convergence before $p=1.2$ GeV.
\begin{figure}[t]
\begin{center}
\includegraphics[scale=0.35]{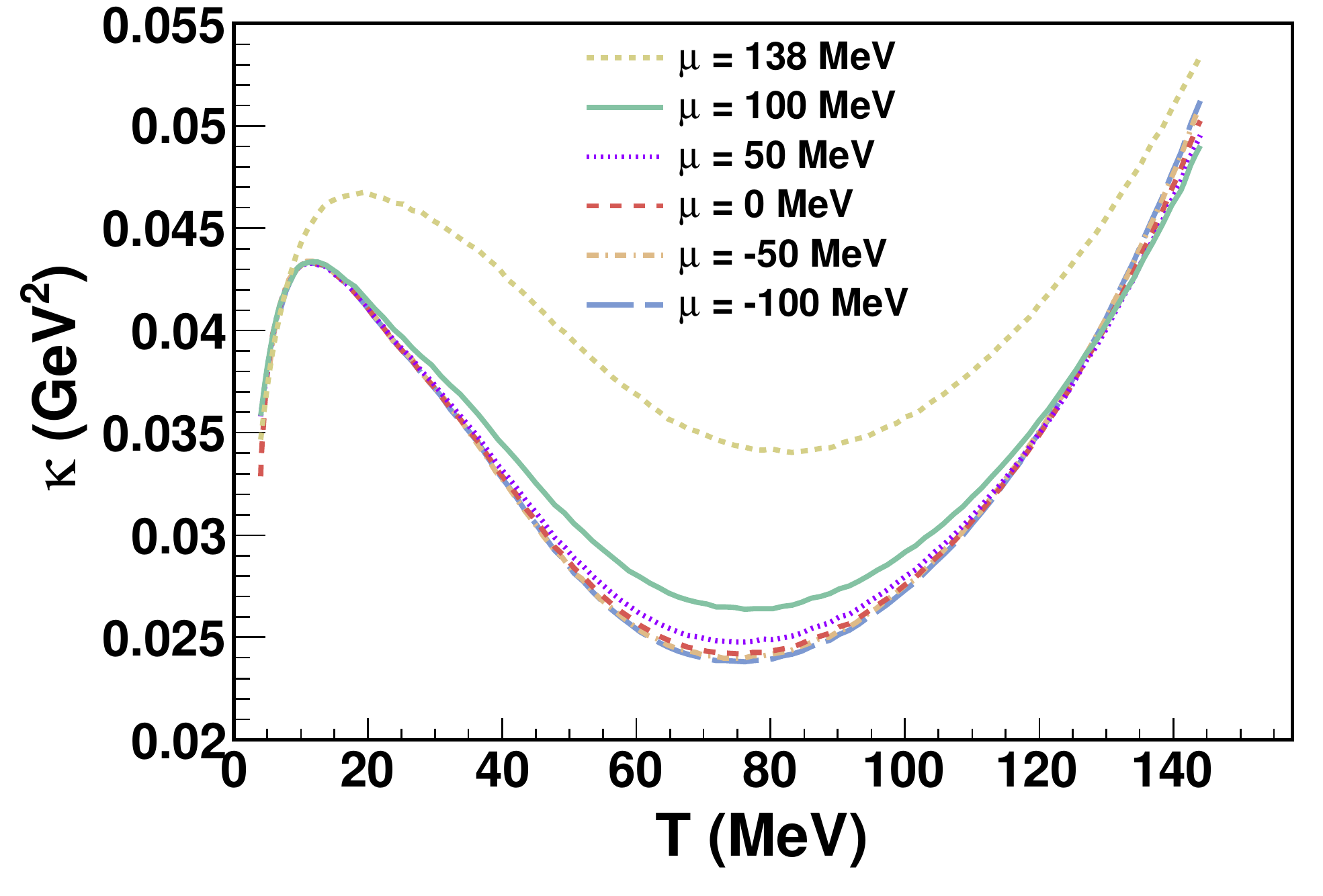}
\caption{\label{fig:heat1iam} Heat conductivity of a pion gas at different chemical potentials. We impose a momentum cutoff of $1.2$ GeV with the $SU(2)$ inverse amplitude method phase-shifts.}
\end{center}
\end{figure}

Working in first order approximation, we vary the pion chemical potential as represented in the lower panel of Fig.~\ref{fig:heat1pra}.
All the curves are very close to each other except the limiting value of $\mu=m$. We repeat the calculation with the phase-shifts from the $SU(2)$ inverse amplitude method with a momentum
cutoff and plot the results in Fig.~\ref{fig:heat1iam}. 

For low temperatures we recover the nonrelativistic limit $\kappa \sim T^{1/2}$. This is obtained by using the nonrelativistic result \cite{landau1981physical}
\be \kappa \propto \frac{\lambda_{mfp} \ v n}{T} \left(\frac{w}{n}-\frac{\epsilon}{n} \right) \ , \ee
with $\lambda_{mft} \simeq 1/(\sigma n)$, $v \simeq \sqrt{T/M}$ and the cross section is a constant by Weinberg's theorem. In the nonrelativistic limit one should extract the mass
contribution to the energy and enthalpy densities. The nonrelativistic limit for them is
\be \frac{\epsilon}{n} = \frac{3}{2} T + \cdots \ee
\be \frac{w}{n} = \frac{5}{2} T + \cdots \ee
Combining these results we therefore obtain that $\kappa \propto \sqrt{T}$ at low temperatures.

At high temperatures (without momentum cutoff) we obtain the expecting ultrarelativistic scaling $\kappa \sim T^2$
(as in the UV limit the only relevant scale is the temperature).

\begin{figure}[t]
\begin{center}
\includegraphics[scale=0.32]{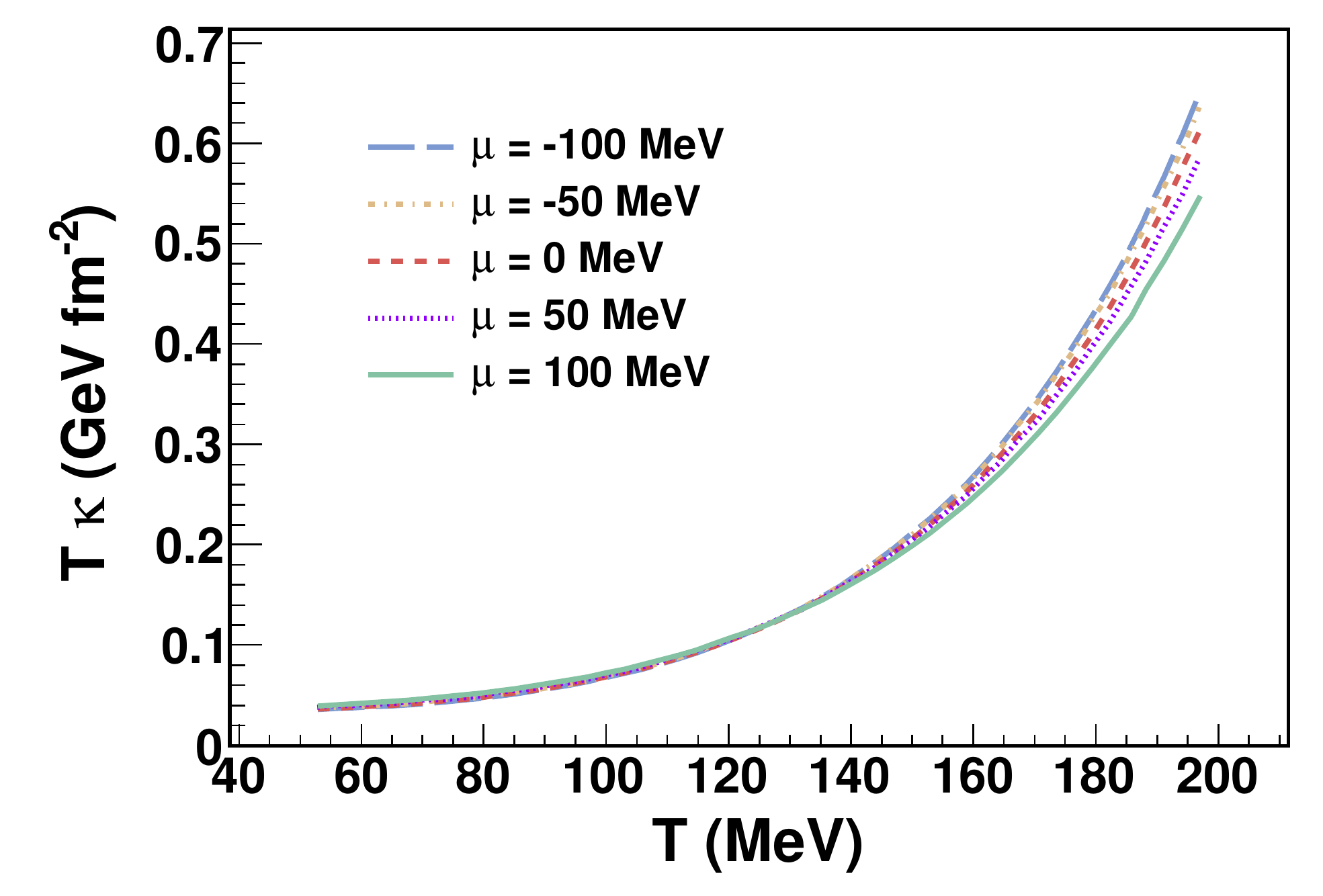}
\includegraphics[scale=0.32]{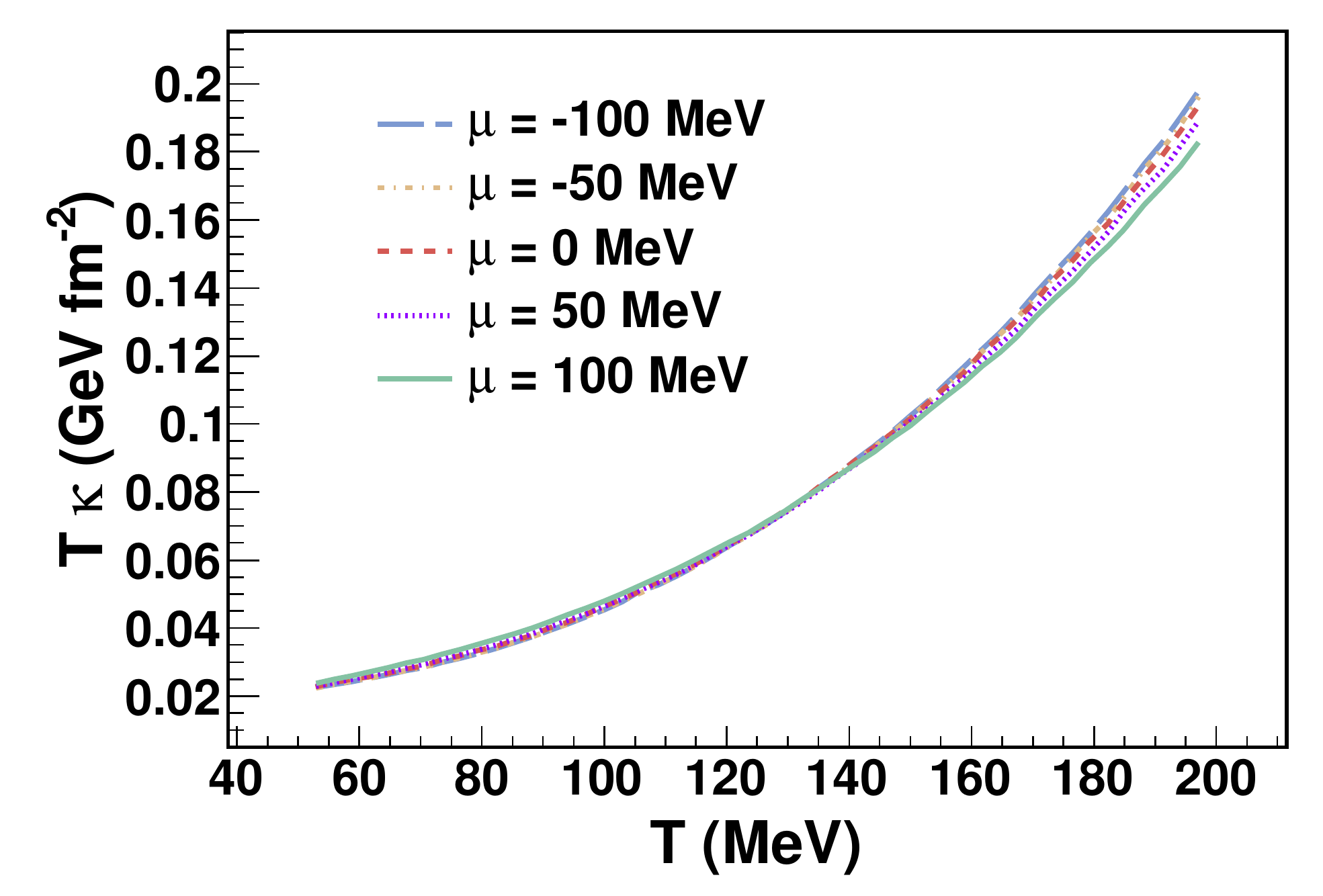}
\includegraphics[scale=0.32]{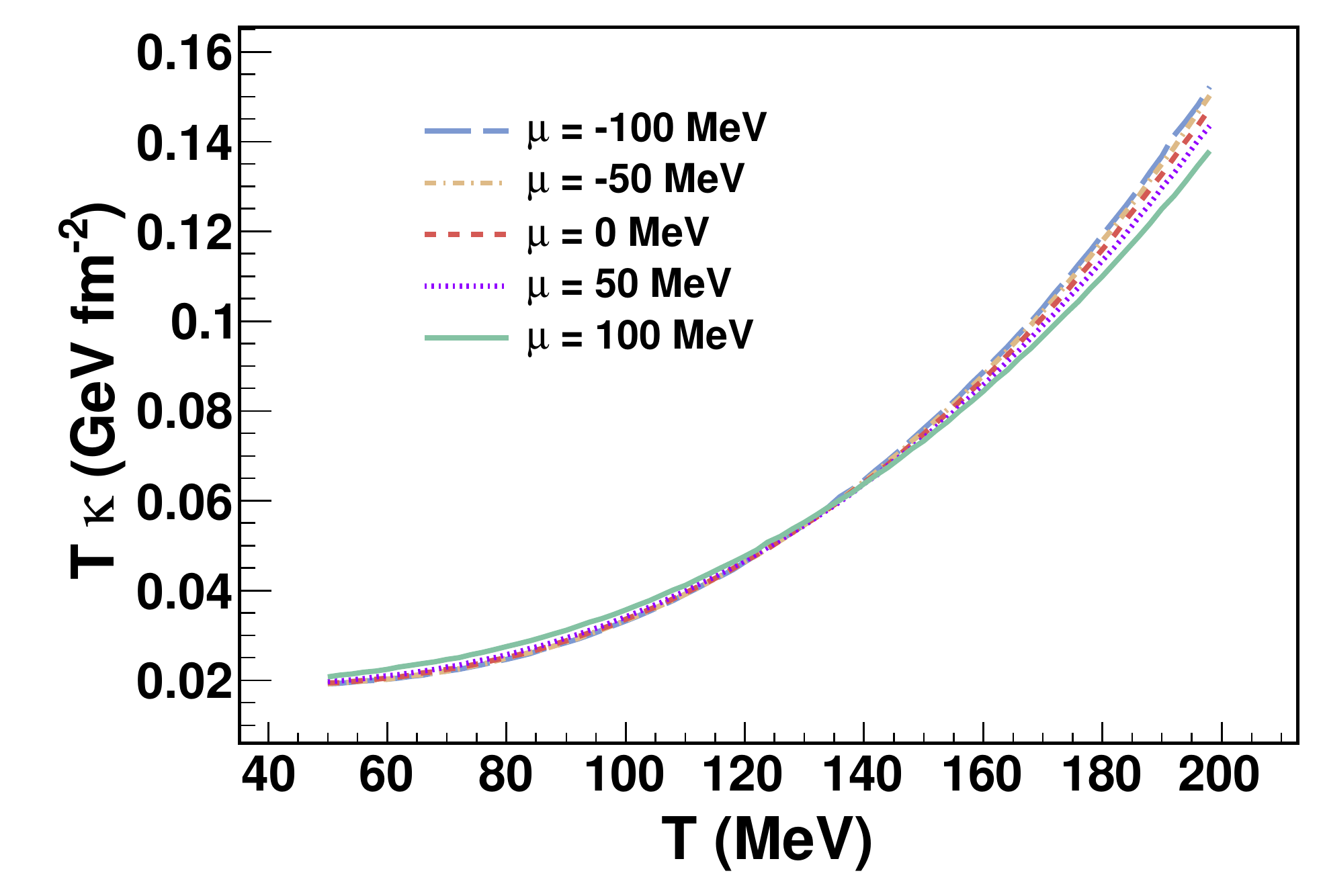}
\caption{\label{fig:heatnocut} Heat conductivity of a pion gas at high temperatures. Top-left panel: $SU(2)$ inverse amplitude method phase-shifts. Top-right panel: Phenomenological phase-shifts. Bottom panel: Result given in \cite{Davesne:1995ms}.
Data kindly provided by D. Davesne}
\end{center}
\end{figure}

Finally, in order to compare with previous approaches we extend the temperature range. In the top left panel of Fig.~\ref{fig:heatnocut} we show the inverse amplitude method calculation up to $T=200$ MeV with
a momentum cutoff of $1.2$ GeV. The next plot is the result with the fitted phase-shifts of \cite{Prakash:1993bt} and no limiting momentum cutoff. This calculation has to be compared with the
curves in~\cite{Davesne:1995ms} that we show in the bottom panel. Good agreement is achieved with this reference. However, a factor of 3 still remains with the results where the cutoff is kept.
To obtain a good convergence at temperature $T=200$ MeV, one must numerically count pions with a momentum of $p\sim 3$ GeV. Obviously, to use the low energy pion interaction up to momenta as high as
that value makes the calculation completely unphysical. Therefore, the cutoff is necessary and the temperature cannot be larger than 150 MeV to avoid probing the cutoff phase-space.

\newpage 
\section{Electrical conductivity}

To obtain the electrical conductivity\index{electrical conductivity} of a gas of pions one must introduce an external electric force $F^i$ to the BUU equation:
 
\be \label{eq:transportelec} \frac{\pa f_p(x)}{\pa t} + \mathbf{v} \cdot \mathbf{\nabla} f_p(x) + \mathbf{F} \cdot \mathbf{\nabla}_p f_p (x)=  C[f_3,f_p] , \ee

where the electrical force is related to the electric field\index{electric field} as $F^i = q E^i$.

The presence of this electric field (that is nothing but a gradient of the electrical potential) creates an electric current in the system whose
magnitude depends on the coefficient of electrical conductivity. This effect is the content of Ohm's law\index{Ohm's law}:
\be \label{eq:ohmslaw} j_Q^i = \sigma E^i \ . \ee

The left-hand side of the BUU equation is calculated according to the Chapman-Enskog expansion. A first term comes from the momentum-derivative
of the local equilibrium distribution function
 
\be \frac{\pa n_p(x)}{\pa p_i}= -n_p(x) [1+ n_p(x)] \frac{\pa}{\pa p_i} \left[ \beta (u_{\alpha} p^{\alpha} -\mu)\right] = -n_p(x) [1+ n_p(x)] \beta \frac{p_i}{E_p} \ ,\ee

so that the explicit term with the electrical force reads
\be \mathbf{F} \cdot \mathbf{\nabla}_p f_p (x)= -q n_p (1+n_p) \beta \frac{E^i p_i}{E_p} \ . \ee
The second contribution comes from the term with temporal derivative:
  \be \frac{\pa f_p}{\pa t} = \beta n_p (1+n_p) \mathbf{p} \cdot \pa_t \mathbf{V} \ ,\ee
with the Lorentz force\index{Lorentz force} (see for instance \cite{de1980relativistic}):
\be \pa_t \mathbf{V} = \frac{n}{w} q \mathbf{E} \ . \ee
 
Combining the two terms we arrive at the following expression for the left-hand side of the transport equation (\ref{eq:transportelec}):
 \be n_p (1+n_p) q \frac{p^i}{E_p} \frac{E_i}{T} \frac{n}{w} \left( E_p - \frac{w}{n} \right) \ . \ee

Recall that the electric current can be written in terms of the distribution function in a similar manner to the particle flux:
\be \label{eq:electricflux} j_Q^i = g_{\pi_c} q \int \frac{d^3p}{(2\pi)^3 E_p} p^i f_p^{(1)}(x) \ , \ee
where $g_{\pi_c}$ is the degeneration factor of charged pions. Note that the flux of positive and negative pions is opposite in direction, when the electric field
is switched on. However, as the electrical current is multiplied by the pion charge, the effect of the two fluxes is added. For this reason we prefer to use a global
degeneracy of charged pions $g_{\pi_c}=2$ in the expression (\ref{eq:electricflux}).

We can add a zero contribution to Eq.~(\ref{eq:electricflux}) by using the Landau-Lifshitz condition of fit (\ref{eq:landau_cond}):
 
\be \label{eq:el_current} j_Q^i = g_{\pi_c} \int \frac{d^3p}{(2\pi)^3 E_p}  p^i q \  f_p^{(1)} \left( 1 - \frac{n}{w} E_p \right) \ . \ee

The right-hand side of Eq.~(\ref{eq:transportelec}) in the presence of a perturbation like $f_p=n_p + f_p^{(1)}$ becomes
 
\begin{eqnarray}
\nonumber -\frac{g_{\pi_c}}{2} \int d \Gamma_{12,3p} \ (1+n_1)(1+n_2)n_3 n_p & & \left( \frac{f_p^{(1)}}{n_p(1+n_p)} + \frac{f_3^{(1)}}{n_3 (1+n_3)} \right.  \\
  & & \left. -\frac{f_1^{(1)}}{n_1(1+n_1)} -\frac{f_2^{(1)}}{n_2 (1+n_2)}\right)  \ . 
\end{eqnarray}
 
The {\it ansazt} for $f_a^{(1)}$ is chosen to be very similar to that for the heat conductivity:
 \be f_a^{(1)} = -n_p (1+n_p) \beta^3 \ \mathbf{k_a} \cdot \frac{\mb{E}}{q} \frac{w}{mn} Z(k_a) \ , \ee
 with $Z(k_a)$ a dimensionless function of $k_a$. To follow the mass dimension of the various electric quantities, we summarize them in
Table~\ref{tab:cond_unit}.

\begin{table}[t]
\begin{centering}
\begin{tabular}{|c||c|}
\hline
Physical Quantity & Energy dimension \\
\hline
\hline
Electrical charge & 0 \\
Force & 2 \\
Electric field & 2 \\
Electrical conductivity & 1 \\	
Electric current density& 3 \\
\hline 
\end{tabular}
\caption{\label{tab:cond_unit} Energy dimensions in natural units.}
\end{centering}
\end{table}

 Equating Eqs.~(\ref{eq:el_current}) and (\ref{eq:ohmslaw}):
 
 \be \label{eq:elecconduct} \sigma E^i = -\frac{g_{\pi_c}}{mT^3} \int \frac{d^3p}{(2\pi)^3 E_p} p^i n_p (1+n_p) \ Z(p) \mb{p} \cdot \mb{E} \left(\frac{w}{n}-
  E_p\right)\ . \ee
 After using the relation (\ref{eq:integvector}) to eliminate the electric field, the equation finally transforms to
 \be \sigma = \frac{g_{\pi_c}}{3T^3} \int \frac{d^3p}{(2\pi)^3 E_p} p^2 n_p (1+n_p) \ Z(p) \left( \frac{E_p}{m} - \frac{w}{mn}\right)\ .\ee
 
 
 This expression for the electric conductivity suggests the same measure we used for the heat conductivity in Eq.~(\ref{eq:elecconduct}):
 \be d\mu_{\sigma} = d\mu_{\kappa}=dx \frac{z^{-1} e^{y(x-1)}}{[z^{-1}e^{y(x-1)}-1]^2} (x^2-1)^{3/2} \ . \ee
 The polynomial basis is also inherited from the heat conductivity in Eqs.~(\ref{eq:condp0}), (\ref{eq:condp1}) and (\ref{eq:condp2}).
 
 
 One can simply express the electric conductivity as a scalar product
 \be \sigma = \frac{g_{\pi_c} m^4}{6 \pi^2 T^3} \langle Z(p) | P_1(p) \rangle \ . \ee
 
The BUU equation\index{BUU equation}
\begin{eqnarray}
\nonumber n_p (1+n_p) \left( E_p- \frac{w}{n}\right) p^i & = & \frac{g_{\pi_c}E_p}{2q^2T^2}\frac{w^2}{mn^2} \int d \Gamma_{12,3p} \left\{ (1+n_1)(1+n_2)n_3 n_p \right. \\
  & & \left. \times \left[ Z(p) p^i + Z(k_3) k_3^i- Z(k_1) k_1^i - Z(k_2)  k_2^i \right] \right\}
\end{eqnarray}
is very similar to the transport equation for the heat conductivity\index{heat conductivity} in Eq.~(\ref{eq:buu_thermal}). Comparing the two expressions
one deduces that the solution $Z(p)$ is related to the solution for $C(p)$ as
 
\be Z(p)= \frac{g_{\pi}}{g_{\pi_c}} \frac{n^2mTq^2}{w^2} C(p) \ , \ee
 
and the electrical conductivity\index{electrical conductivity} is therefore proportional to the heat conductivity:
 
\be \label{eq:wiedemannfranz} \frac{\kappa}{\sigma} = \frac{g_{\pi}}{g_{\pi_c}}\frac{w^2}{q^2 T n^2} \ . \ee
Eq.~(\ref{eq:wiedemannfranz}) is nothing but the Wiedemann-Franz law\index{Wiedemann-Franz law} for the pion gas.

\begin{figure}[t]
\begin{center}
\includegraphics[scale=0.34]{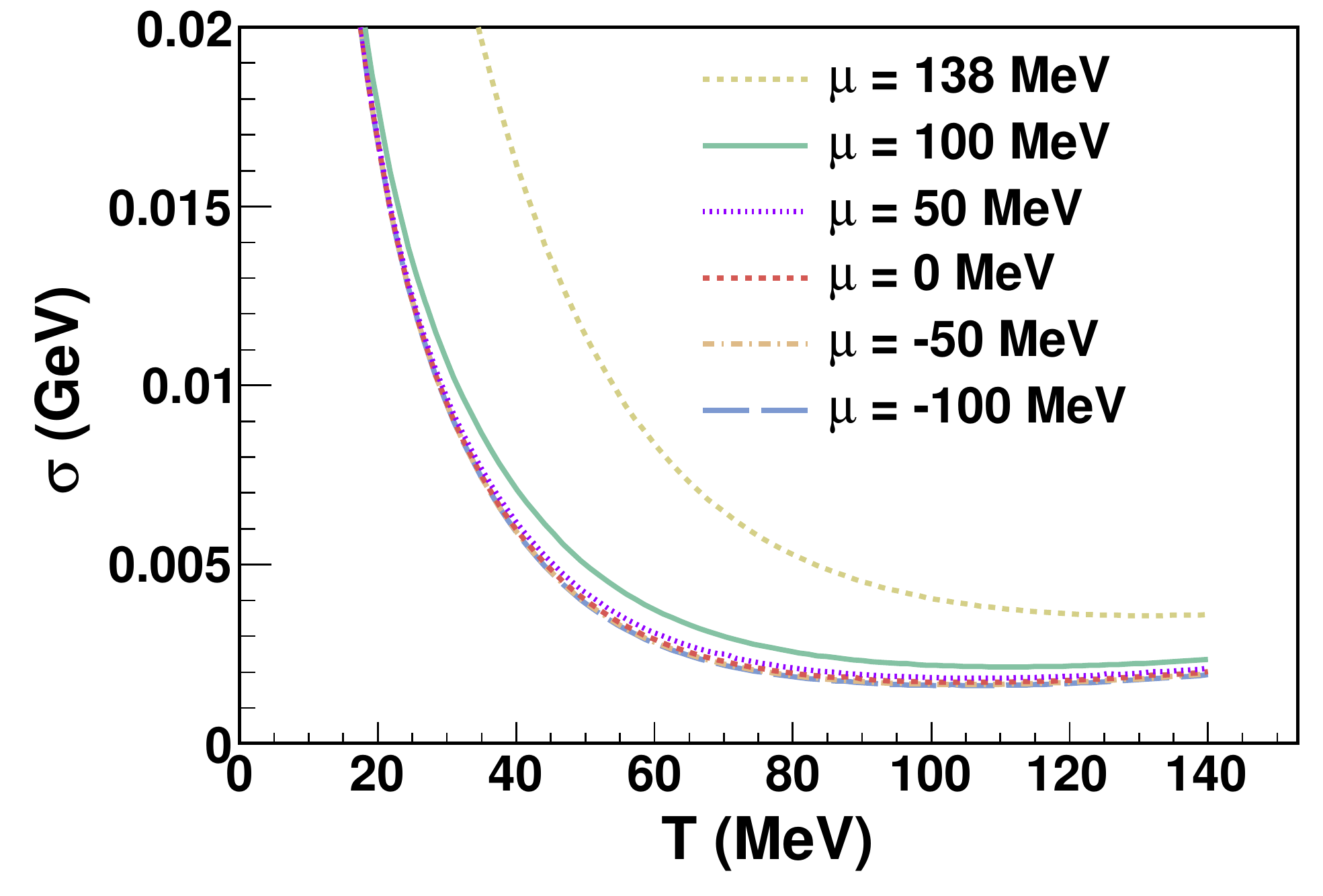}
\includegraphics[scale=0.34]{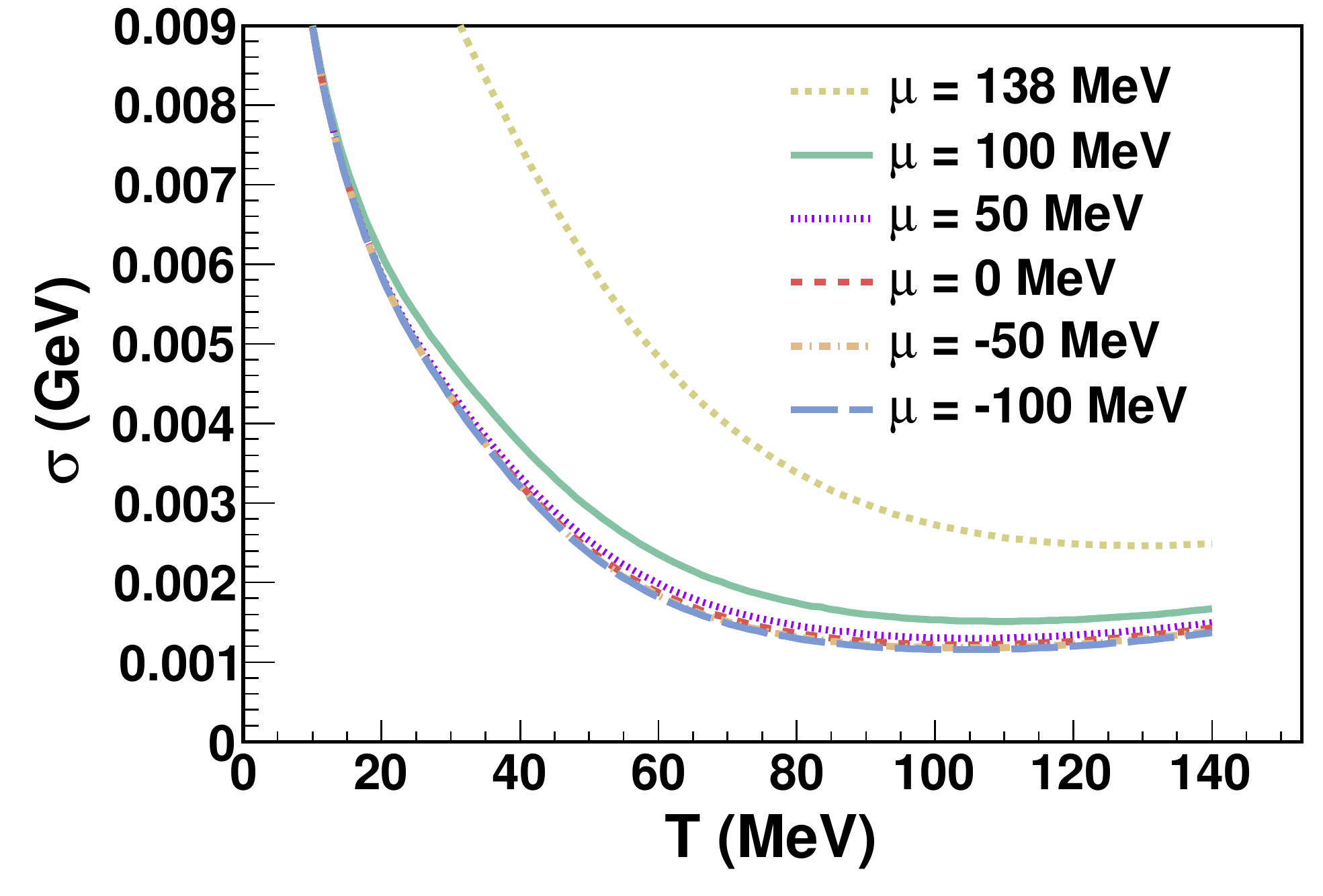}
\caption{\label{fig:electrical} Electrical conductivity of a pion gas with $SU(2)$ inverse amplitude method phase-shifts (upper panel) and phenomenological phase-shifts in \cite{Prakash:1993bt} (lower panel).}
\end{center}
\end{figure}

In Figure \ref{fig:electrical} we plot the electric conductivity for a pion gas at different chemical potentials. The upper panel shows the result from the inverse amplitude method and the lower panel from the phase-shifts in \cite{Prakash:1993bt}.

\begin{figure}[t]
\begin{center}
\includegraphics[scale=0.44]{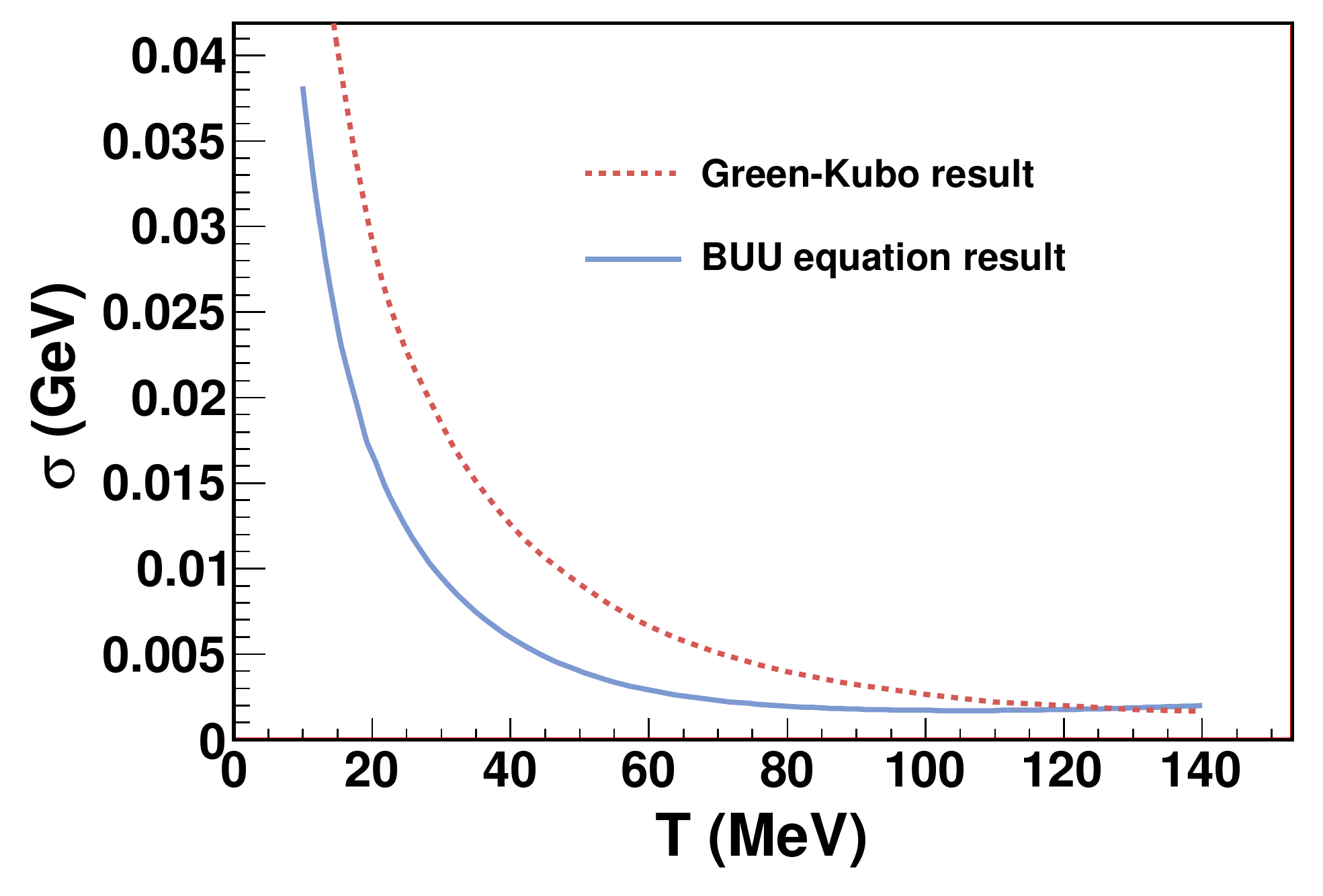}
\caption{\label{fig:eleccomp} Comparison between the electrical conductivity for the pion gas obtained by the BUU equation and the result obtained by the Green-Kubo formula
as represented in \cite{FernandezFraile:2005ka}. Both calculations use unitarized phase-shifts for the pion elastic interaction. The data for the calculation in \cite{FernandezFraile:2005ka} have been kindly provided by D. Fernandez-Fraile.}
\end{center}
\end{figure}

In Figure \ref{fig:eleccomp} we compare the result from the BUU equation (obtained by using the Wiedemann-Franz law (\ref{eq:wiedemannfranz}) at zero chemical potential and the result
obtained in \cite{FernandezFraile:2005ka} from the Green-Kubo equation. Both calculations use the unitarized pion interaction by the inverse amplitude method.

In the high temperature limit one has
 \be \epsilon \rightarrow \frac{g_{\pi} T^4 \pi^2}{30}; \quad P \rightarrow \frac{g_{\pi}T^4 \pi^2}{90}; \quad n=\frac{g_{\pi} T^3 \zeta(3)}{\pi^2} \ ,\ee
where $\zeta(3) \simeq 1.202...$ is Apery's constant.
 That means that the Wiedemann-Franz law reads
 \be \frac{\kappa}{\sigma} \rightarrow \frac{g_{\pi}}{g_{\pi_c}} \frac{4 \pi^8}{2025 \zeta^2(3)} \frac{T}{q^2} \simeq 19.5  \frac{T}{q^2} \ . \ee
To obtain the nonrelativistic limit one has to be aware of the relativistic convention of counting the pion mass in the energy density. The nonrelativistic
limit of the $J_i$ integrals is
\be J_i \rightarrow \frac{T^{5/2}}{m^{5/2}} e^{-\frac{\mu -m}{T}} + \cdots \ . \ee
That makes the energy per particle behave as (we have checked numerically all these limits)
\be \frac{\epsilon}{mn} \rightarrow 1 + \frac{3}{2} \frac{T}{m} + \cdots \ee
Where the first term is nothing but the rest particle mass and the second term is the well known value of the nonrelativistic energy per particle. 
In the same limit the enthalpy per particle is:
\be \frac{w}{mn} = \frac{J_1}{J_0} \rightarrow 1 + \frac{5}{2} \frac{T}{m} + \cdots \ee
To connect with the usual nonrelativistic result we need to eliminate by hand the rest mass contribution in the previous expression.
Therefore we will use the nonrelativistic value of 
\be \frac{w}{n} \simeq \frac{5}{2} T \ee

The quotient of the heat conductivity over the electrical conductivity reads in the nonrelativistic limit: 
\be \frac{\kappa}{\sigma} \rightarrow \frac{25}{4} \frac{g_{\pi}}{g_{\pi_c}} \frac{T}{q^2} = 9.4 \frac{T}{q^2} \ , \ee
that, when combining with the result $\kappa \simeq \sqrt{T}$ makes the electrical conductivity diverge as $1/\sqrt{T}$ as reproduced by the numerical computation in Fig.~\ref{fig:eleccomp}.

\chapter{Bhatnagar-Gross-Krook or 
Relaxation Time Approximation \label{ch:6.RTA}}

When the departure from equilibrium is small, the collision side of the kinetic equation can be simplified 
by the introduction of a relaxation time\index{relaxation time}. The relaxation time accounts for the characteristic time of change of the distribution function.
A way to extract the relaxation times once the transport coefficients are known, is to use the Bhatnagar-Gross-Krook\glossary{name=BGK,description={Bhatnagar-Gross-Krook}}
approximation \index{BGK approximation|see {relaxation-time approximation}} or relaxation time approximation (RTA)\index{relaxation-time approximation}\glossary{name=RTA,description={relaxation time approximation}} \cite{Bhatnagar:1954zz}.

We have separated the solution of the kinetic equation into an equilibrium Bose-Einstein term and a small perturbation $\delta f_p \ll n_p$:
\be f_p = n_p + \delta f_p \ . \ee
In the RTA approximation, all the non-zero eigenvalues of the collision term $C[f_3,f_p]$ are further taken to have a common value $-1/\tau_R (E_p)$, 
where $\tau_R$ is the relaxation time. Note that the non-positivity of the eigenvalues of the collision operator follows from Boltzmann's H-theorem\index{Boltzmann's $H$-theorem} \cite{liboff2003kinetic}.

In effect, the collision operator is substituted by the simpler expression
\be C[f_3,f_p] = - \frac{1}{\tau_R (E_p)} \delta f_p \ .\ee

Combining this approximation with the Chapman-Enskog expansion we identify $\delta f_p$ with $f_p^{(1)}$ and the kinetic equation reads:
\be \label{eq:RTA} \frac{p^{\mu}}{E_p} \pa_{\mu} n_p = - \frac{1}{\tau_R (E_p)}  \delta f_p \ , \ee
where the left-hand side only depends on the equilibrium distribution function with hydrodynamical fields depending on spacetime. We still identify these fields $T(x)$, $\mu(x)$ and 
$u^i(x)$ as the same as in equilibrium so that the conditions of fit still hold. In the local rest reference frame they are
\begin{eqnarray}
\int d^3p \ \delta f_p E_p &=& 0 \ , \\
\int d^3p \ \delta f_p & = & 0 \ , \\
\int d^3p \ \delta f_p p^i & = & 0 \ .
\end{eqnarray}

From Chapters \ref{ch:3.shear}, \ref{ch:4.bulk} and \ref{ch:5.conductivities} we know the explicit expression of the left-hand side for the different transport coefficients. Moreover we know the microscopical expressions
to calculate the transport coefficients from an integration of the functions $ \delta f_p$. We will calculate $\tau_R$ in two different approximations, the 
energy-independent RTA \index{relaxation-time approximation!energy-independent}and using the ``quadratic {\it ansatz}'' \index{relaxation-time approximation!quadratic ansatz} for the relaxation time.

\section{Energy-independent RTA \label{sec:eirta}}

The simplest approximation is to consider that the relaxation time has no energy dependence at all\index{relaxation-time approximation!energy-independent}:
\be \tau_R(E_p)=\tau_R \ . \ee
This approximation is rather crude but it will provide the comparison with previous works \cite{Hosoya:1983xm,Gavin:1985ph,Prakash:1993bt,Davesne:1995ms}. It also provides a simple interpretation of this coefficient.
If $\tau_R$ is independent of energy, then the solution to the transport equation
\be \frac{df_p}{dt} = - \frac{f_p- n_p}{\tau_R} \ , \ee
for $f_p$ is an exponential approach to equilibrium
\be f_p(t)-n_p = (f_p(t=0) - n_p ) \exp \left( - \frac{t}{\tau_R} \right) \ , \ee
and $\tau_R$ corresponds to the time in which the distribution function decreases a factor $1/e$ towards the equilibrium distribution function $n_p$.

In the dilute limit the relaxation time is nearly equal to the collision time \cite{Prakash:1993bt}, that
gives the inverse of the mean-free path.

Let us start with the shear viscosity. The function $\delta f_p$ is given by the substitution of the left-hand side of (\ref{eq:lhsshear}) into Eq.~(\ref{eq:RTA}):
\be \delta f_p^{\eta} = - \tau_R^{\eta}  \frac{\beta}{E_p} n_p (1+n_p) p^i p^j \tilde{V}_{ij} \ .\ee

Now identify $\delta f_p^{\eta}$ with the shape of $f_p^{(1)}=-n_p (1+n_p) \Phi_p$ for the shear viscosity given in Eq.~(\ref{eq:phishear}) to obtain
\be B(p) = \tau_R^{\eta}  \frac{T^2}{E_p}  \ee
and finally insert this function into Eq.~(\ref{eq:viscofinal}) to get:
\be \eta= \frac{g_{\pi} \tau_R^{\eta}}{15T} \int \frac{d^3p}{(2\pi)^3} \frac{p^4}{E_p^2} n_p (1+n_p) \ . \ee
Thus one can extract the relaxation time\index{relaxation time} associated with the shear viscosity if we know the value of $\eta$ as
a function of the temperature and pion chemical potential.

We perform the same calculation for the bulk viscosity with the left-hand side given in Eq.~(\ref{eq:lhsbulk}):
\be \delta f_p^{\zeta} = - \tau_R^{\zeta} \frac{\beta}{E_p} n_p (1+n_p) \left( \frac{p^2}{3} - E_p^2 v_n^2 - E_p \kappa^{-1}_{\epsilon}\right) \mb{\nabla} \cdot \mb{V} \ .\ee
Inserting this function into the microscopic definition of the bulk viscosity:
\be \zeta = \tau_R^{\zeta} \ \frac{g_{\pi} }{3T} \int \frac{d^3p}{(2\pi)^3} n_p (1+n_p) \frac{p^2}{E_p^2} \left( \frac{p^2}{3} - E_p^2 v_n^2 - E_p \kappa^{-1}_{\epsilon}\right) \ . \ee
Using the conditions of fit one can manipulate this expression and convert it into the final form:
\be \zeta= \tau_R^{\zeta} \ \frac{g_{\pi} }{T} \int \frac{d^3p}{(2\pi)^3} n_p (1+n_p) \frac{1}{E^2_p} \left( \frac{p^2}{3} - E_p^2 v_n^2 - E_p \kappa^{-1}_{\epsilon}\right)^2 \ . \ee

Analogously, we use the left-hand side for the thermal conductivity in Eq.~(\ref{eq:lhsthermal}):
\be \delta f_p^{\kappa} = - \tau_R^{\kappa} \frac{\beta^2}{E_p} n_p (1+n_p) \left( E_p - \frac{w}{n} \right) \mb{p} \cdot \left( \mb{\nabla} T - \frac{T}{w} \mb{\nabla}P \right) \ . \ee

And the thermal conductivity coefficient reads
\be \kappa = \tau_R^{\kappa} \ \frac{g_{\pi} }{3T^2} \int \frac{d^3p}{(2\pi)^3} n_p (1+n_p) \frac{p^2}{E_p^2}  \left( E_p - \frac{w}{n} \right)^2 \ . \ee

In the energy-independent RTA one can compare directly with the equations obtained in \cite{Hosoya:1983xm} or \cite{Gavin:1985ph}.
The formulae in this references are the same as ours. However, there is a discrepancy with respect to the formula given in \cite{Hosoya:1983xm}
for the heat conductivity. The term $w/n$ does not appear in their equation. This is so because they do not introduce a chemical potential
in the calculation and therefore they do not take into account the corresponding terms coming from the spacetime derivatives of this variable. 

The relaxation times finally read: \index{relaxation time}

\begin{eqnarray}
(\tau_R^{\eta})^{-1} & = & \eta^{-1} \ \frac{g_{\pi} m^5}{30 \pi^2T} \int_1^{\infty} dx \ \frac{(x^2-1)^{5/2}}{x} \ \frac{ z^{-1} e^{y (x-1 )} }{ \left( z^{-1} e^{y(x-1 )} -1\right)^2} \ , \\
(\tau_R^{\zeta})^{-1} & = & \zeta^{-1} \ \frac{g_{\pi} m^5}{2 \pi^2 T} \int_1^{\infty} dx \ \frac{(x^2-1)^{1/2}}{x} 
\ \left[ \left( \frac{1}{3} - v_n^2 \right) x^2 - \frac{\kappa^{-1}_{\epsilon}}{m} x - \frac{1}{3} \right] ^2 \nonumber \\
& & \times   \frac{ z^{-1} e^{y (x-1 )} }{ \left( z^{-1} e^{y (x-1 )} -1\right)^2} \ , \\
 (\tau_R^{\kappa})^{-1} & = & \kappa^{-1} \ \frac{g_{\pi} m^5}{6 \pi^2 T^2} \int_1^{\infty} dx \  \frac{(x^2-1)^{3/2}}{x} 
\ \left( x - \frac{w}{nm} \right)^2 \nonumber \\
& & \times \frac{ z^{-1} e^{y (x-1 )} }{ \left( z^{-1} e^{y (x-1 )} -1\right)^2} \ . 
\end{eqnarray}

\begin{figure}[t]
\begin{center}
\includegraphics[scale=0.34]{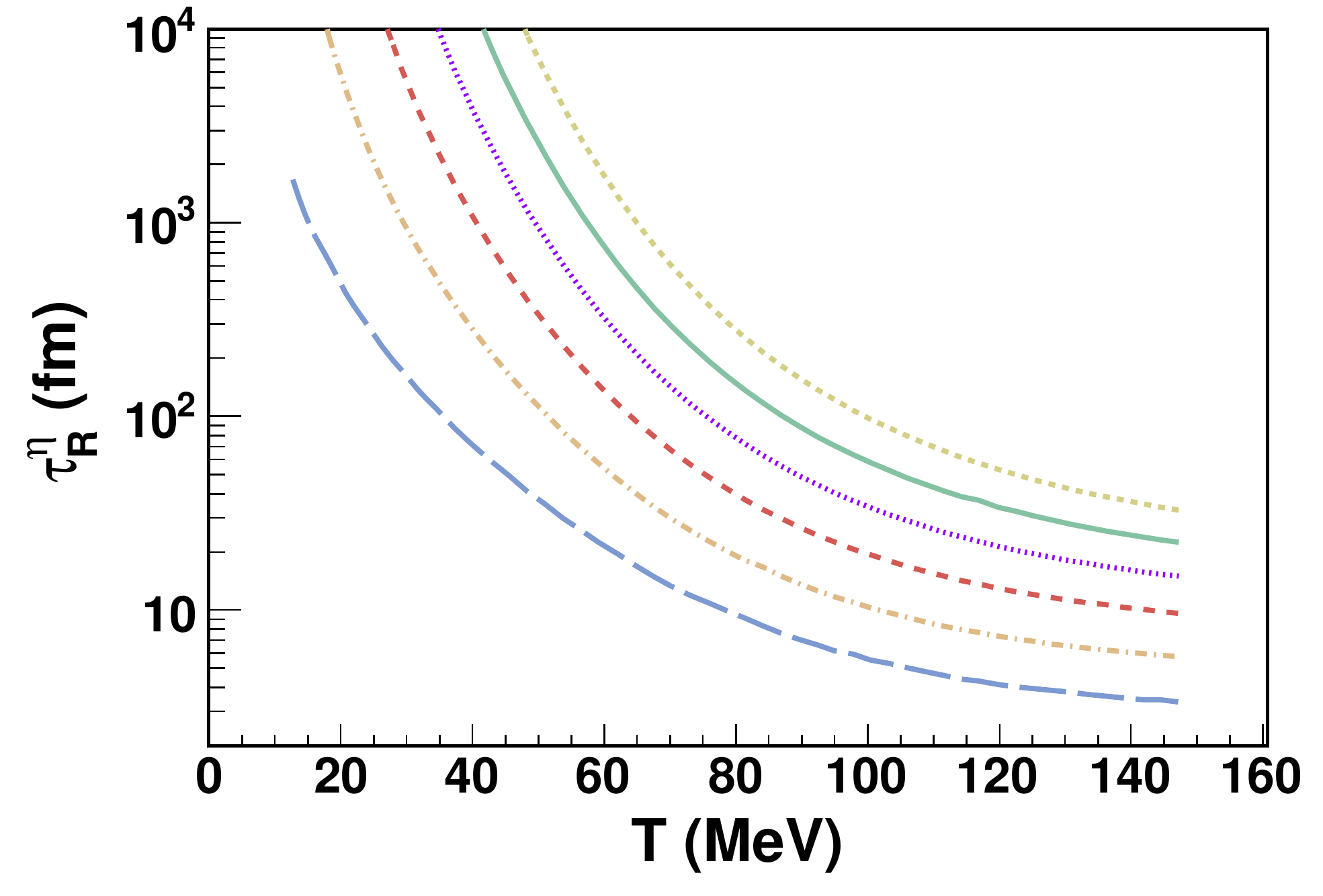}
\includegraphics[scale=0.34]{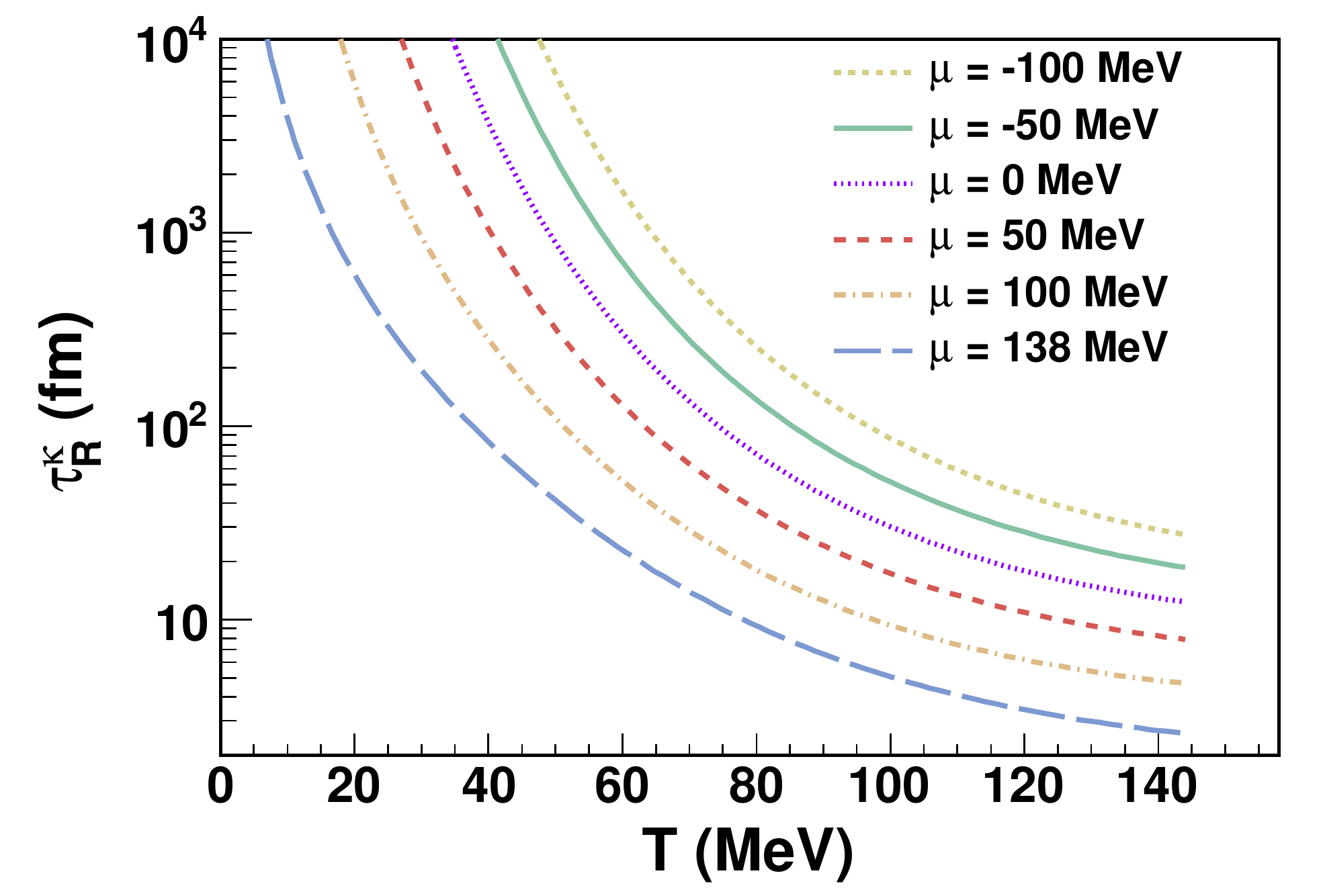}
\includegraphics[scale=0.34]{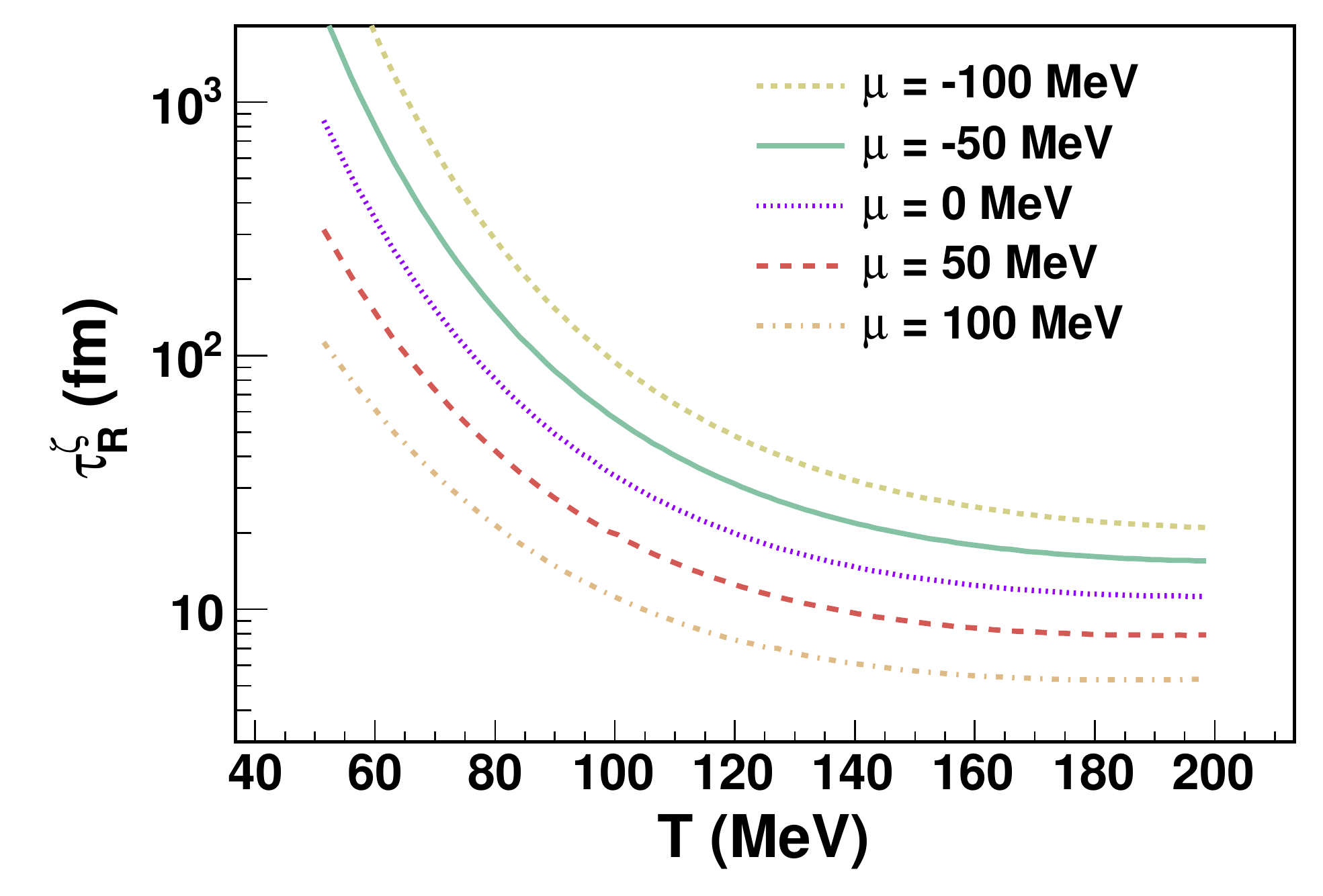}
\caption{\label{fig:reltimes} Relaxation times for the shear viscosity, thermal conductivity and bulk viscosity in the energy-independent RTA approximation for the pion gas with $SU(2)$ IAM phase-shifts.}
\end{center}
\end{figure}

We plot the results for the three different relaxation times in Fig.~\ref{fig:reltimes}. We use the results coming from the $SU(2)$ IAM interaction at several chemical potentials.

\section{Quadratic {\it ansatz}}

The energy-independent RTA --although widely used by some authors especially in the past-- is not the best simple approximation we can make for the relaxation time.
Comparing the expression obtained for $\delta f_p$ in Sec.~\ref{sec:eirta} and the left-hand side of the BUU equation $p_{\mu} \pa^{\mu} n_p(x)|_i$, one can immediately deduce that
the natural choice (for the three coefficients) is to consider a linear energy dependence to the relaxation time as $\tau_R (E_p) \propto E_p$.
This approximation is called the ``quadratic {\it ansatz}''\index{relaxation-time approximation!quadratic ansatz}.

We will repeat the calculation assuming that
\be \tau_R (E_p)= \tau_0 \frac{E_p}{T}  \ , \ee
where $\tau_0$ is a constant. 

In terms of our adimensional variables it simply reads $\tau_R (E_p) = \tau_0 xy$. We the obtain the following equations for the transport coefficients:

\begin{eqnarray}
\eta & = & \tau_0^{\eta} \ \frac{g_{\pi} }{15T^2} \int \frac{d^3p}{(2\pi)^3} \frac{p^4}{E_p} n_p (1+n_p) \ , \\
\zeta & = &  \tau_0^{\zeta} \ \frac{g_{\pi}}{T^2} \int \frac{d^3p}{(2\pi)^3} n_p (1+n_p) \frac{1}{E_p} \left( \frac{p^2}{3} - E_p^2 v_n^2 - E_p \kappa^{-1}_{\epsilon}\right)^2 \ , \\
\kappa & = & \tau_0^{\kappa} \ \frac{g_{\pi} }{3T^3} \int \frac{d^3p}{(2\pi)^3} n_p (1+n_p) \ p^2 \left( E_p - \frac{w}{n} \right)^2 \ . 
\end{eqnarray}

Inverting these relations we obtain the values of $\tau_0^i$, that in terms of adimensinal variables read,
\begin{eqnarray}
(\tau_0^{\eta})^{-1} & = & \eta^{-1} \ \frac{g_{\pi} m^6}{30 \pi^2T^2} \int_1^{\infty} dx \ (x^2-1)^{5/2} \ \frac{ z^{-1} e^{y (x-1 )} }{ \left[ z^{-1} e^{y(x-1 )} -1\right]^2} \ , \\
(\tau_0^{\zeta})^{-1} & = & \zeta^{-1} \ \frac{g_{\pi} m^6}{2 \pi^2 T^2} \int_1^{\infty} dx \ (x^2-1)^{1/2}
\ \left[ \left( \frac{1}{3} - v_n^2 \right) x^2 - \frac{\kappa^{-1}_{\epsilon}}{m} x - \frac{1}{3} \right] ^2 \nonumber \\
& & \times \ \frac{ z^{-1} e^{y (x-1 )} }{ \left( z^{-1} e^{y (x-1 )} -1\right)^2} \ , \\
(\tau_0^{\kappa})^{-1} & = & \kappa^{-1} \ \frac{g_{\pi} m^6}{6 \pi^2 T^3} \int_1^{\infty} dx \  (x^2-1)^{3/2}
\ \left( x - \frac{w}{nm} \right)^2 \\
& & \times \frac{ z^{-1} e^{y (x-1 )} }{ \left( z^{-1} e^{y (x-1 )} -1\right)^2} \ .
\end{eqnarray}

These integrals can be expressed as combinations of the $I_i,J_i$ and $K_i$ functions defined in Eqs.~(\ref{eq:Iintegrals}), (\ref{eq:Jintegrals}) and (\ref{eq:Kintegrals}) respectively.

\begin{eqnarray}
(\tau_0^{\eta})^{-1} & =& \eta^{-1} \frac{g_{\pi} m^6}{30 \pi^2T^2} \langle P_0 | P_0 \rangle_{\mu_{\eta}} = \eta^{-1} \frac{g_{\pi} m^6}{30 \pi^2T^2} \ K_0 \ , \\
(\tau_0^{\zeta})^{-1} &=&  \zeta^{-1} \frac{g_{\pi} m^6}{2 \pi^2 T^2} \langle P_2 |P_2 \rangle_{\mu_{\zeta}} = \zeta^{-1} \frac{g_{\pi} m^6}{2 \pi^2 T^2} \ \frac{I_4 + I_3 \Delta_2 + I_2 \Delta_1}{9 \Delta_1^2}  \ , \\
(\tau_0^{\kappa})^{-1} &=&  \kappa^{-1} \frac{g_{\pi} m^6}{6 \pi^2 T^3} \langle P_1 | P_1 \rangle_{\mu_{\kappa}} =\kappa^{-1} \frac{g_{\pi} m^6}{6 \pi^2 T^3} \ \frac{J_2 J_0 - J_1^2}{J_0}  \ ,
\end{eqnarray}

with
\be \Delta_1 \equiv \frac{I_2^2-I_1I_3}{I_1^2-I_0I_2} \ee
and
\be \Delta_2 \equiv \frac{I_0 I_3 -I_1 I_2}{I_1^2-I_0I_2} \ . \ee

If we work only at first order in the polynomial expansion of the solutions we can obtain a very simple expression for the $\tau_0$ in terms of the collision operators:

\begin{eqnarray}
(\tau_0^{\eta})^{-1} &=& T \frac{\mathcal{C}_{00}}{K_0} \ , \\
(\tau_0^{\zeta})^{-1} &=&  \frac{m^2}{T} \frac{\mathcal{C}_{22}}{\frac{I_4 + I_3 \Delta_2 + I_2 \Delta_1}{9 \Delta_1^2}  } \ , \\
(\tau_0^{\kappa})^{-1} & =&  m \frac{\mathcal{C}_{11}}{ \frac{J_2 J_0 - J_1^2}{J_0}} \ , \\
\end{eqnarray}
where the collision operators $\mathcal{C}_{00},\mathcal{C}_{22}$ and $\mathcal{C}_{11}$ are defined differently according to the transport coefficient they belong.

\begin{figure}[t]
\begin{center}
\includegraphics[scale=0.34]{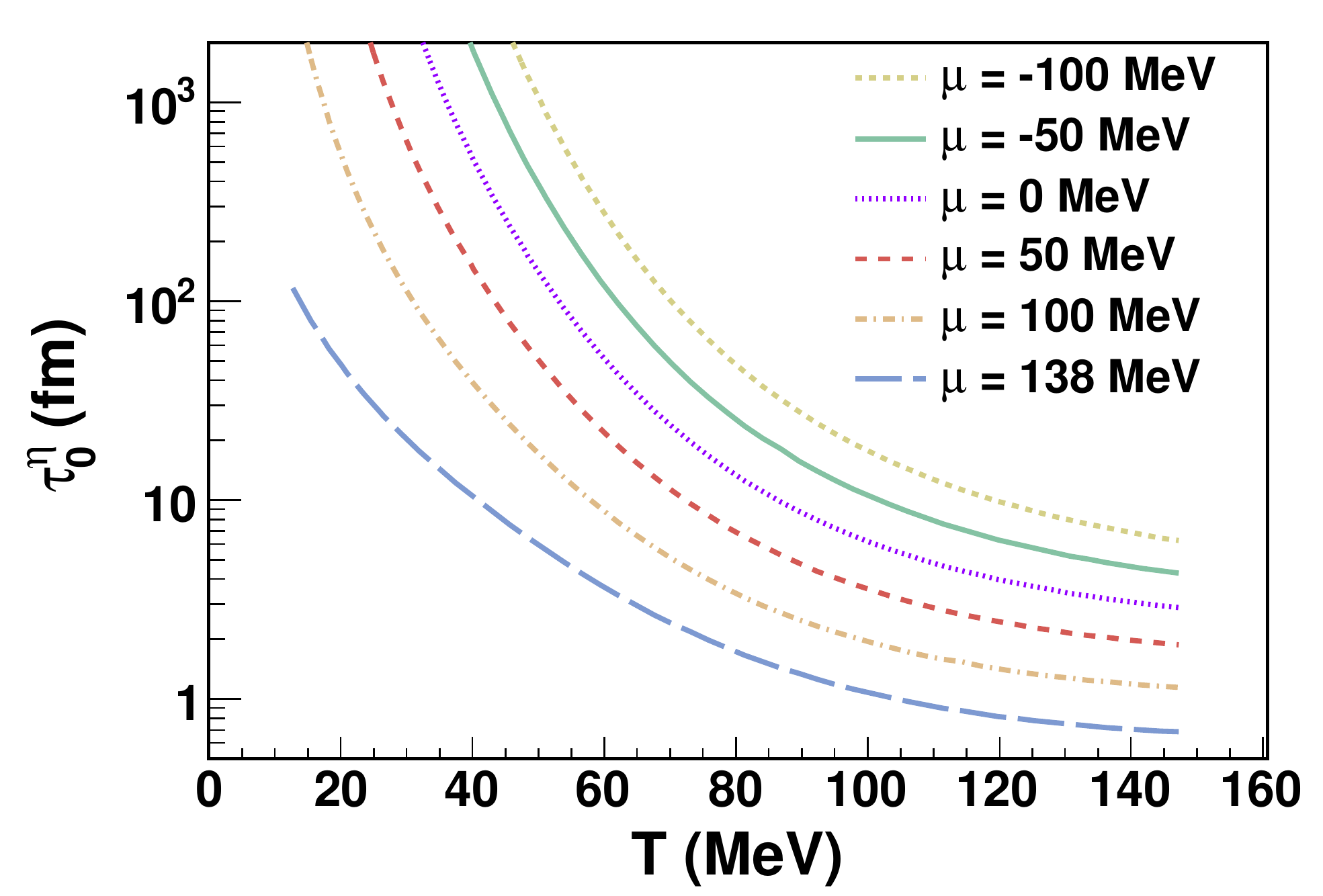}
\includegraphics[scale=0.34]{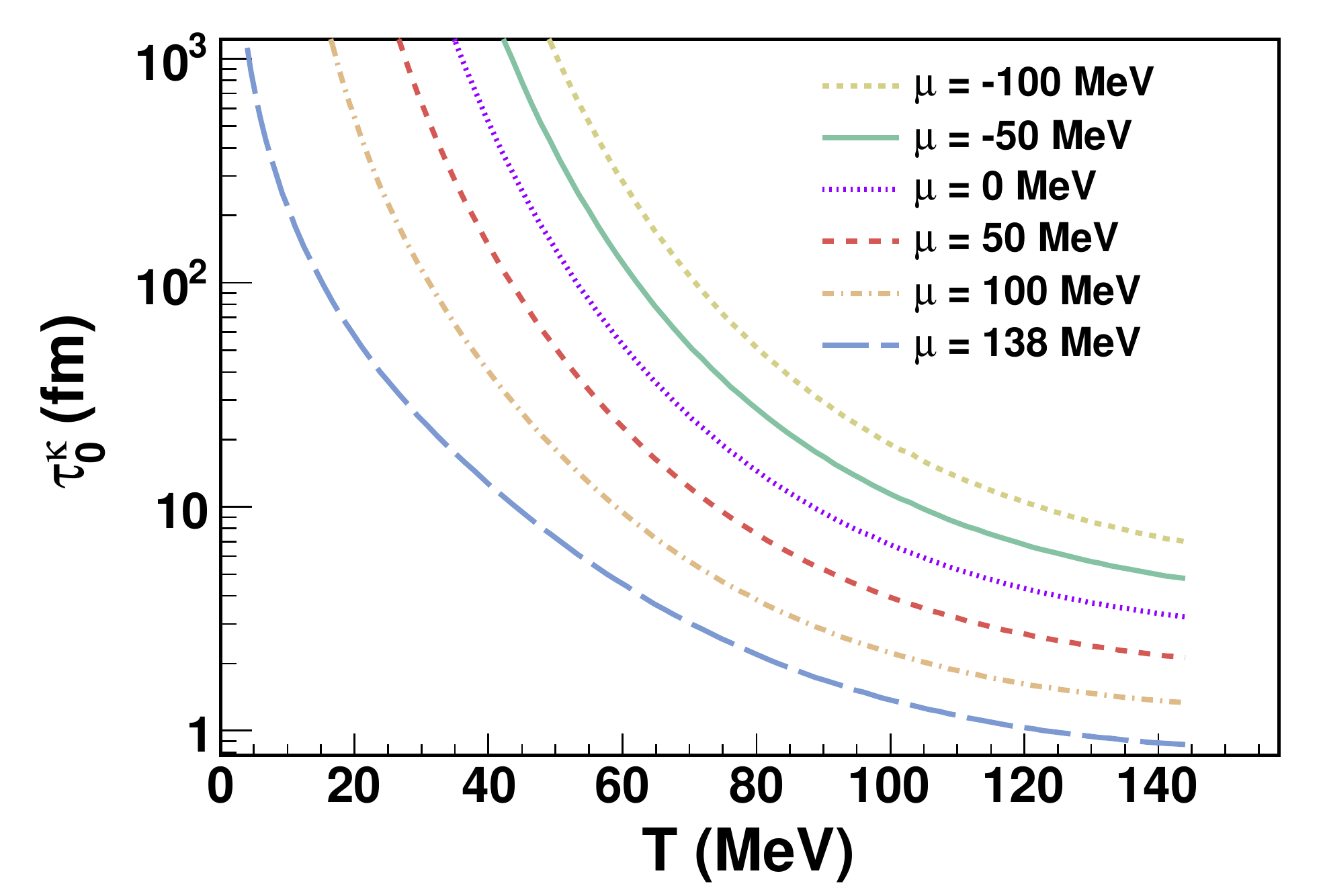}
\includegraphics[scale=0.34]{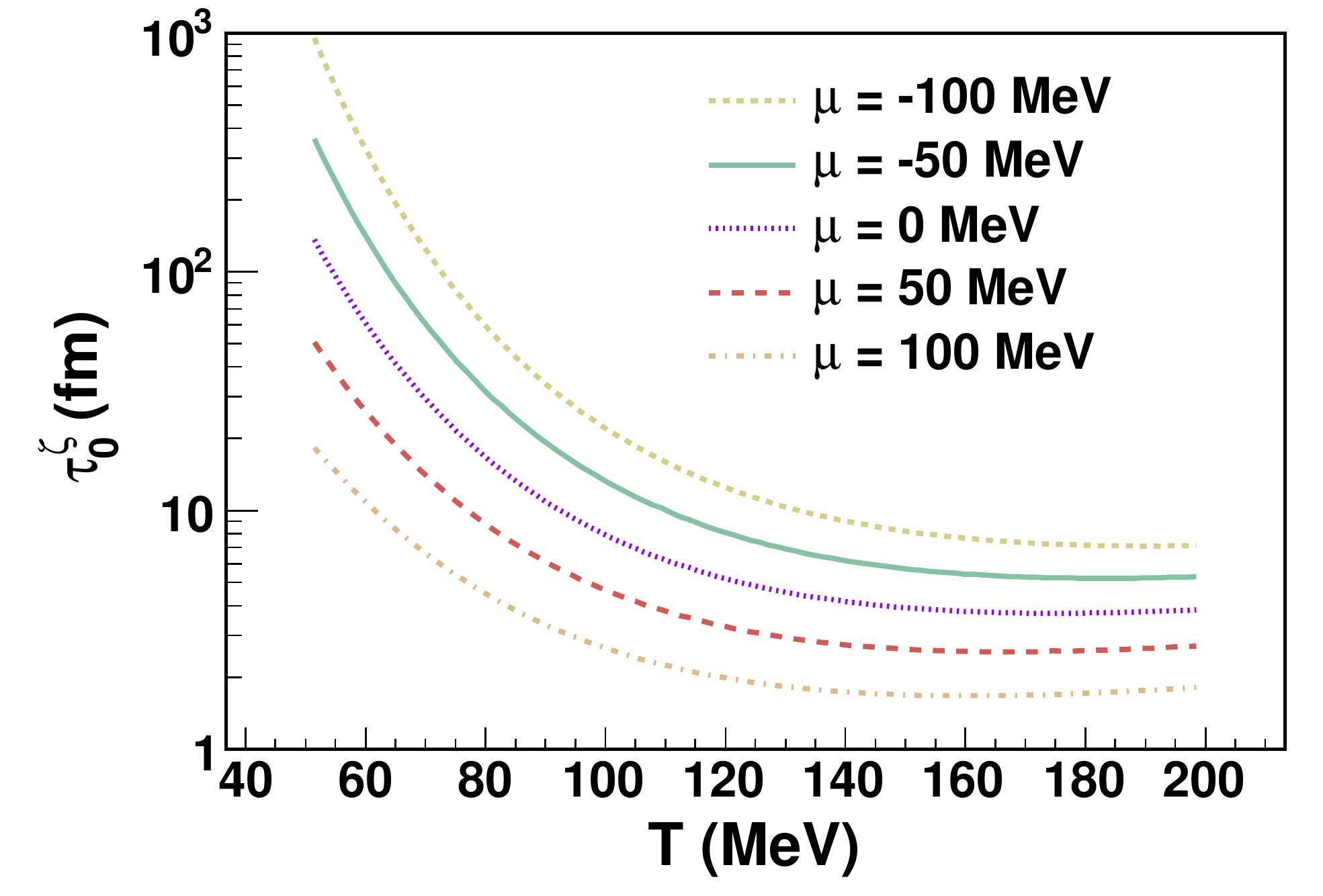}
\caption{\label{fig:reltimes2} Relaxation times\index{relaxation time} $\tau_0$ for the shear viscosity, thermal conductivity and bulk viscosity in the ``quadratic {\it ansatz}'' approximation and for the pion gas within $SU(2)$ IAM phase-shifts.}
\end{center}
\end{figure}

We show the results for the coefficients $\tau^i_0$ in Fig.~\ref{fig:reltimes2}.

The relaxation times follow the same behaviour as the transport coefficients (as they are proportional). The conclusion from both computations is that the relaxation times at
typical $\mu \sim 50-100$ MeV and $T \sim 120-140$ MeV are of the order 4-5 fm. Under these conditions the authors of \cite{Prakash:1993bt} give a collision time $\tau_R \sim 1$ fm (see Sec.~\ref{sec:chapman-enskog} for its definition).
We conclude that in order to produce the equilibration of the gas the pions suffer 4-5 collisions, and a significant diffusive increase of entropy does occur in the pion gas.

\chapter{Strangeness Diffusion \label{ch:7.strangeness}}

In this chapter and the next one, we include flavor-like charges and consider the strangeness\index{strangeness} and charm diffusion coefficients. 
For the former we will solve the BUU equation \index{BUU equation} for the kaon meson distribution function. An extension to $SU(3)$ ChPT is used for the
pion-kaon interaction and the inverse amplitude method is used to unitarize the scattering amplitudes. For the charm diffusion coefficient,
we will use an alternative effective field theory that incorporates both chiral and heavy quark symmetries, again with unitarized scattering amplitudes.
In that case, the BUU equation is transformed into a Fokker-Planck equation\index{Fokker-Planck equation}.

\section{Mixed hadron gas with pions and kaons}

At low temperatures the meson gas is well described by a system containing pions as the unique degree of freedom. Due to its larger mass, the next mesons (the kaon\index{kaon} and the $\eta$ meson\index{$\eta$ meson})
are mostly suppresed and only appear as important components at moderate temperatures. These two new states carry strangeness\index{strangeness} (as the $s$-quark enters in its quark composition). 
The effective field theory for describing such a system is no longer the $SU(2)$ ChPT. However, a natural extension to $N_f=3$ is possible (see App.~\ref{sec:su3chpt} for more details).
Strangeness is also a conserved quantum number of the strong interactions so it is possible to ask how the $s$-quark \index{$s$-quark} diffuses in the medium due to collisions with the pions.

From the quark model point of view, both the kaon\index{kaon} and the $\eta$ meson\index{$\eta$ meson} carry $s$-quarks (or antiquarks). However, the quark composition of the latter is
\be \eta: \frac{u\overline{u}+ d\overline{d} +s \overline{s}}{\sqrt{3}} \ , \ee
and it contains both a strange quark and a strange antiquark, so the total strangeness flux carried by this meson is zero. The $\eta$ meson does not contribute to the strangeness diffusion.
In contrast, the kaon does carry a non-vanishing strangeness and it will be the only contribution to the strangeness diffusion. We will work in the isospin limit where the number of $K^+$ ($u\overline{s}$) is equal
to the number of negative kaons $K^-$ ($s\overline{u}$) and equal to the number of $K^0$ ($d\overline{s}$) and $\overline{K}_{0}$ ($s \overline{d}$). In this limit, the flux
of $s$-quarks is, up to a sign, the same as the flux of $s$-antiquarks (they are opposite in charge). 

Although we are going to speak about ``strangeness diffusion'' we are going to trade the strange degree of freedom by the kaonic one. This is so, because the $s$-quark in a hadron gas is always confined 
into a kaon, and the strangeness flow is driving by the kaons inside the medium. Therefore we will identify the distribution function of an $s$-quark by the distribution function of a kaon (note that 
one can identify their chemical potentials because the energy needed for creating an $s$-quark in the system is the same as the energy for creating a $K^-$, just because the former is always confined into the latter).

\section{Diffusion equation}

The BUU equation for the dilute one-particle distribution function of a kaon evolving under elastic scattering with the pion gas $f_p^K$ reads:
\be \frac{df_p^K}{dt} = g_\pi \int d\Gamma_{12,3p} [f_1^\pi f_2^K (1+f_3^\pi) (1+f_p^K) - f_3^\pi f_p^K (1+f_1^\pi) (1+f_2^K)]\ee
 
We will consider the pion gas at equilibrium whereas the kaons will be near equilibrium. This is so, because the relaxation time for the kaons is larger than
for the pions and so the former equilibrates earlier \cite{Prakash:1993bt}. 
Moreover, we will consider low temperatures where the kaon density is much lower than the pion density $n_p^K \ll n_p^{\pi}$. As a consequence, only the scattering between kaons and pions matters,
the scattering between two kaons being highly improbable.
When inserting the Chapman-Enskog expansion (\ref{eq:chapman-enskog}) into the BUU equation to linearize it, we will use the previous statements to supress quadratic terms in $f^{(1)}$ and neglect
$n_p^{K} f^{(1) \pi} $ with respect to $n_p^{\pi} f^{(1) K}$ in the collision integral. The linearized BUU equation reads: 
\be \frac{d f_p^K}{d t} =- g_{\pi} \int d\Gamma_{12,3p} n_3^{\pi} (1+n_1^{\pi}) n_p^K (1+n_2^K) \left( \frac{f_p^{(1)K}}{n_p^K(1+n_p^K)} - \frac{f_2^{(1)K}}{n_2^K(1+n_2^K)}\right) \ , \ee
where $n_p^K$ and $n_p^{\pi}$ are the equilibrium distribution function for the kaons and for the pions respectively.
The equilibrium distribution function reads:
\be n_p^K = \frac{1}{\exp \left( \beta(x) [u_{\alpha} (x) p^{\alpha} -\mu_K(x)]\right)-1} \ . \ee

The particle number density of kaons is expressed as an integral of this equilibrium distribution function:
\be \label{eq:partnumbden} n_{K} = g_{K} \int \frac{d^3p}{(2\pi)^3} n_p^K \ , \ee
with $g_K=4$.

In the left-hand side of the BUU the only space-time dependence inside $n_p^K$ that is relevant for the strangeness diffusion is that of the kaon chemical potential.
\be \frac{df_p^K}{dt} = \frac{p^i}{E^K_p} \nabla^i n_p^K (x) = \frac{p_i}{E_p^K} n_p^K (1+n_p^K) \beta \nabla^i \mu_K \ . \ee
The BUU equation in the first order Chapman-Enskog expansion reads

\begin{eqnarray}
 \nonumber n_p^K (1+n_p^K) p_i \nabla^i \mu_K & = & -g_\pi E_p^K T \int d\Gamma_{12,3p} n_3^{\pi} (1+n_1^{\pi}) n_p^K (1+n_2^K) \\
& &  \times \label{eq:buuforstrange} \left[\frac{ f_p^{(1)K}}{n_p^K (1+n_p^K)} - \frac{ f_2^{(1)K}}{n_2^K (1+n_2^K)} \right] \ .  
\end{eqnarray}

To extract the strangeness diffusion we must make the connection between the kinetic equation and Fick's diffusion law, that reads
\be \label{eq:numflow1} j^i_s=-D_s \nabla^i n_K = - D_s \nabla^i \mu_K \int g_K \frac{d^3p}{(2\pi)^3} n_p^{K} (1+n_p^K) \beta  \ , \ee
where we have computed the gradient of the particle number density in Eq.~(\ref{eq:partnumbden}).

Microscopically, the particle flux can be expressed as an integration over the one-particle distribution function
\be \label{eq:numflow2} j^i_s=g_K \int  \frac{d^3p}{(2\pi)^3} \frac{p^i}{E_p^K} f_p^{(1)K}  \ , \ee
where $E_p^K = \sqrt{p^2+m^2_K}$.
We will conveniently choose the {\it ansatz} for $f_p^{(1)K}$ as
\be f_p^{(1)K} = - n_p^K (1+n_p^K) p^i \ \nabla_i \mu_K \ \beta^3 \ H(p) \ , \ee
with $H(p)$ an adimensional function of the kaon momentum $p$. This function will be obtained by inversion of the BUU equation.

Substituing the {\it ansatz} for $f_p^{(1)}$ into the BUU (\ref{eq:buuforstrange}):
\be n_p^K (1+n_p^K) p^i =g_\pi E^K_p \beta^2 \int d\Gamma_{12,3p} n_3^{\pi} (1+n_1^{\pi}) n_p^K (1+n_2^K) \left[ p^i  H(p)  - k_2^i  H(k_2) \right] \ .  \ee

On the other hand, equating Eqs.~(\ref{eq:numflow1}) and (\ref{eq:numflow2}):
\be  D_s \nabla^i \mu_K \int \frac{d^3p}{(2\pi)^3} n_p^{K} (1+n_p^K) =  \int  \frac{d^3p}{(2\pi)^3} \frac{p^i}{E_p}  n_p^K (1+n_p^K) p^j \nabla_j \mu_K \beta^2 H(p) \ , \ee
and using the relation (\ref{eq:integvector}) we get
\be  D_s = \frac{1}{3T^2} \frac{ \int  \frac{d^3p}{(2\pi)^3} \frac{p^2}{E_p^K}  n_p^K (1+n_p^K)  H(p) }
{ \int \frac{d^3p}{(2\pi)^3} n_p^{K} (1+n_p^K) } \ . \ee

The expression for the diffusion coefficient is naturally expressed with an integration measure which is analogous to that for the heat conductivity:
\be d\mu_D=dx \frac{z^{-1} e^{y(x-1)}}{\left[z^{-1} e^{y(x-1)} \right]^2} \ (x^2-1)^{3/2} = d^3p \frac{p^2}{E^K_p} n_p (1+n_p) \frac{1}{4\pi m_K^4} \ . \ee
With the help of this integration measure we define the integrals ($n=i+j$)
\be L^n = \int d\mu_D x^n = \langle x^i | x^j \rangle \ . \ee

The strangeness diffusion can be written as
\be D_s = \frac{1}{3T^2} \frac{\int d\mu_D H(p) 1}{\int d\mu_D \frac{E^{2K}_p}{p^2}} \ . \ee

In terms of the adimensional variables $x=E_p^K/m_K$ and $y=T/m_K$ it reads

\be D_s = \frac{m_K}{3T^2} \frac{\int d\mu_D H(x) 1}{\int d\mu_D \frac{x}{x^2-1}} \ . \ee

The denominator is related to the susceptibility
\be \chi_{\mu \mu}=\left( \frac{\pa n_K}{\pa \mu_K}\right)_T \ . \ee
From this definition it is not difficult to get
\be \chi_{\mu \mu} = \frac{4 m_K^3 \pi g_K}{T} \int d\mu_D \frac{x}{x^2-1} \  \ee
and the diffusion coefficient reads
\be D_s = \frac{4\pi}{3} \frac{m^4_K}{T^3} \frac{g_K}{\chi_{\mu \mu}} \ \langle H(x) | 1 \rangle \ . \ee

We introduce the standard family of monic orthogonal polynomials:
\begin{eqnarray}
P_0 & = & 1 \ , \\
P_1 & = & x - \frac{L^1}{L^0} \ , \\
P_2 & = & \cdots
\end{eqnarray}

and assume that we can expand the function $H(x)$ in a linear combination of that polynomial basis
\be \label{eq:soldiff} H(p)=H(x)=\sum_{n=0}^{\infty} h_n P_n(x) \ . \ee

Due to the orthogonalization properties, the diffusion coefficient contains only one term of the series (\ref{eq:soldiff}):

\be D_s = \frac{4\pi}{3} \frac{m^4_K}{T^3} \frac{g_K}{\chi_{\mu \mu}} h_0 ||P_0||^2 \ , \ee
with
\be ||P_0||^2=L^0 \ . \ee

The coefficient $h_0$ is obtained by inverting the BUU equation.
Projecting this equation by multiplying it by $P_n(p) p^i/(E^K_p 4\pi m_K^4) d^3p$ and integrating over the momentum we obtain:

\be \nonumber  \int d\mu_D P_n(p) 1 = \frac{g_{\pi} 2\pi^2}{m_K^4T^2} \sum_{m=1}^N h_m \ \int \prod_{j=1}^4 \frac{d^3k_j}{2E_j (2\pi)^3} (2\pi)^4 \delta^{(4)} (k_1+k_2-k_3-p) \ee
\be \times \overline{|T|^2} \ n_3^{\pi} (1+n_1^{\pi}) n_p^K (1+n_2^K) P_n(p) p_i\left[ p^i  P_m(p)  - k_2^i  P_m(k_2) \right] \ .  \ee



Symmetrizing the right-hand side we finally obtain:
\be \nonumber \int d\mu_D P_n(p) P_0(p)= \frac{g_{\pi} \pi^2}{m_K^4T^2} \sum_{m=1}^N h_m \ \int \prod_{j=1}^4 \frac{d^3k_j}{2E_j (2\pi)^3} (2\pi)^4 \delta^{(4)} (k_1+k_2-k_3-p) \ee
\be \label{eq:linbuustrange}\times \overline{|T|^2}\ n_3^{\pi} (1+n_1^{\pi}) n_p^K (1+n_2^K) \left[ p_i  P_n(p)  - k_{2i}  P_n(k_2) \right] \left[ p^i  P_m(p)  - k_2^i  P_m(k_2) \right] \ .  \ee

Note that there is no zero mode (for $m=n=0$ the right-hand side does not vanish) but neither does the left-hand side vanish for $n=0$, so the linearized BUU equation is compatible and determined.

For simplicity we can write the Eq.~(\ref{eq:linbuustrange}) as a matricial equation
\be \mathcal{B}_n =\sum_{m=1}^N h_m \mathcal{C}_{nm} \ , \ee
where
\be \mathcal{B}_n \equiv \langle P_n(p) | P_0(p) \rangle  = L_0 \delta_{n0} \ , \ee
and
\be \nonumber \mathcal{C}_{nm} \equiv \frac{g_{\pi} \pi^2}{m_K^4T^2} \ \int \prod_{j=1}^4 \frac{d^3k_j}{2E_j (2\pi)^3} (2\pi)^4 \delta^{(4)} (k_1+k_2-k_3-p) \ee
\be \times \overline{|T|^2} n_3^{\pi} (1+n_1^{\pi}) n_p^K (1+n_2^K) \left[ p_i  P_n(p)  - k_{2i}  P_n(k_2) \right] \left[ p^i  P_m(p)  - k_2^i  P_m(k_2) \right] \ . \ee

Now, one can solve this matricial system up to some finite order in the expansion (\ref{eq:soldiff}), at first order:
\be h_0 = \frac{\mathcal{B}_0}{\mathcal{C}_{00}} \ , \ee
where $\mathcal{B}_0 = L_0$ and
\begin{eqnarray}
\nonumber \mathcal{C}_{00} & \equiv & \frac{g_{\pi} \pi^2}{m_K^4T^2} \ \int \prod_{j=1}^p \frac{d^3k_j}{2E_j (2\pi)^3} (2\pi)^4 \delta^{(4)} (k_1+k_2-k_3-p) \overline{|T|^2} \\
& & \label{eq:collstrange} \times n_3^{\pi} (1+n_1^{\pi}) n_p^K (1+n_2^K) \left[ p_i  - k_{2i}  \right]  \left[ p^i  - k_2^i  \right] \ ,
\end{eqnarray}
Finally, the diffusion coefficient reads
\be D_s = \frac{4\pi}{3} \frac{m^4_K}{T^3} \frac{g_K}{\chi_{\mu \mu}} \frac{L^2_0}{\mathcal{C}_{00}} \ . \ee

\section{Scattering amplitude from $SU(3)$ IAM}

We need to describe the interaction between pions and kaons that enters in Eq.~(\ref{eq:collstrange}) in the form of the averaged scattering amplitude.
As in the case of the pure pion gas we will use an effective field theory to calculate the partial scattering amplitudes.
The $SU(3)$ ChPT (see Appendix~\ref{a.uchpt}) provides the needed scattering amplitudes in the different isospin-spin channels that are $IJ=\frac{1}{2}0,\frac{1}{2}1,\frac{3}{2}0$.
For these three channels only the scattering amplitude in the isospin channel $I=3/2$ is needed $T^{3/2} (s,t,u)$. The $I=1/2$ amplitude is obtained by crossing symmetry:
\be T^{1/2} (s,t,u) = \frac{3}{2} T^{3/2} (u,t,s) - \frac{1}{2} T^{3/2} (s,t,u) \ .\ee
The form of $T^{3/2} (s,t,u)$ from the $SU(3)$ ChPT at one loop is obtained from \cite{GomezNicola:2001as}. This amplitude consists of one term coming from the tree-level 
LO Lagrangian (\ref{eq:losu3chpt}) and reads
\be T^{3/2}_{(0)} (s,t,u)=\frac{m_K^2 + m_{\pi}^2 -s}{2 f^2_{\pi}} \ , \ee
where $f_{\pi}=93$ MeV. 
The NLO part of the amplitude (that we do not reproduce here for shortness) contains tree level terms of the NLO Lagrangian (\ref{eq:nlosu3chpt}) and the one-loop corrections at $\mathcal{O} (p^4)$. 
\begin{figure}[t]
\begin{center}
\includegraphics[scale=0.26]{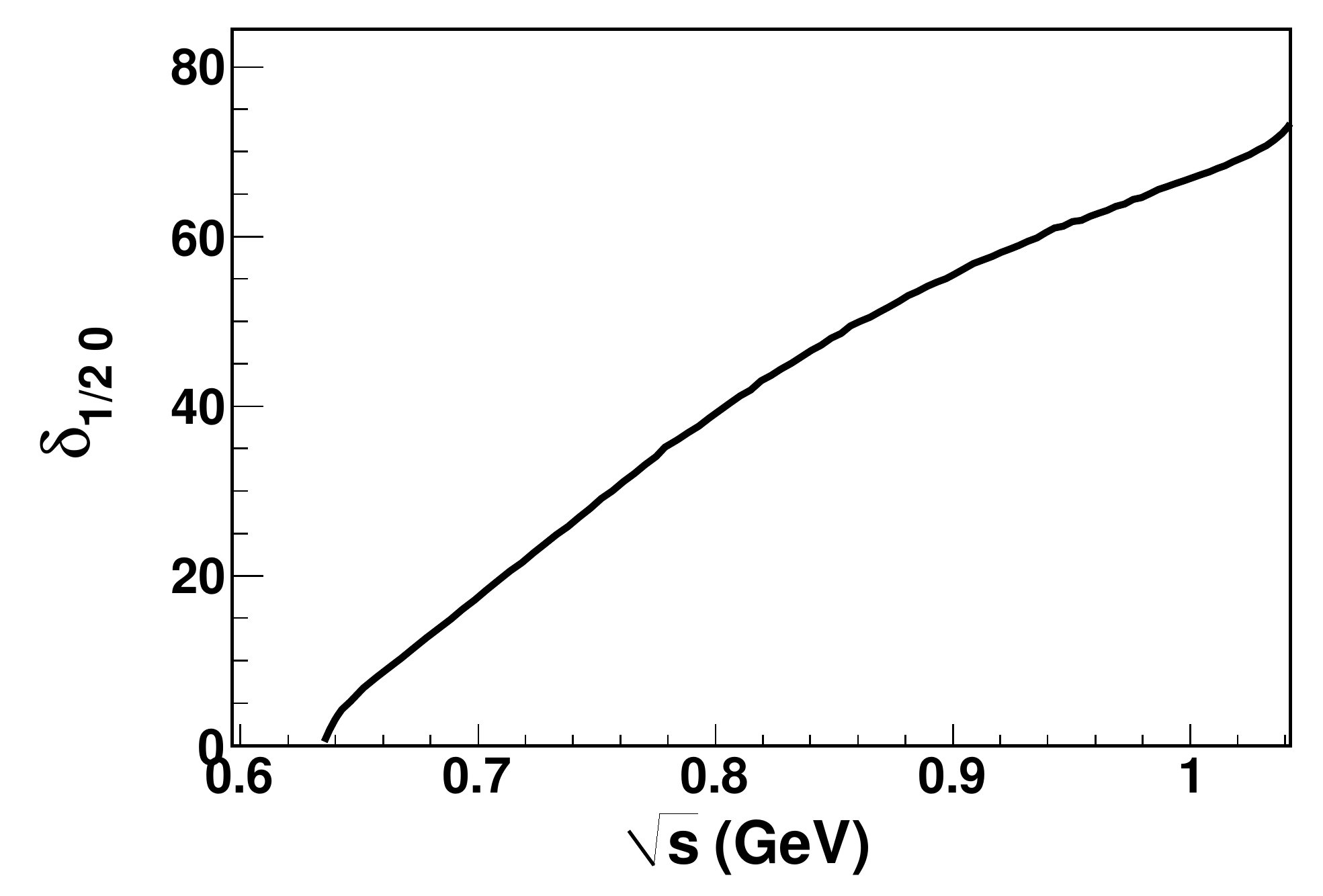}
\includegraphics[scale=0.26]{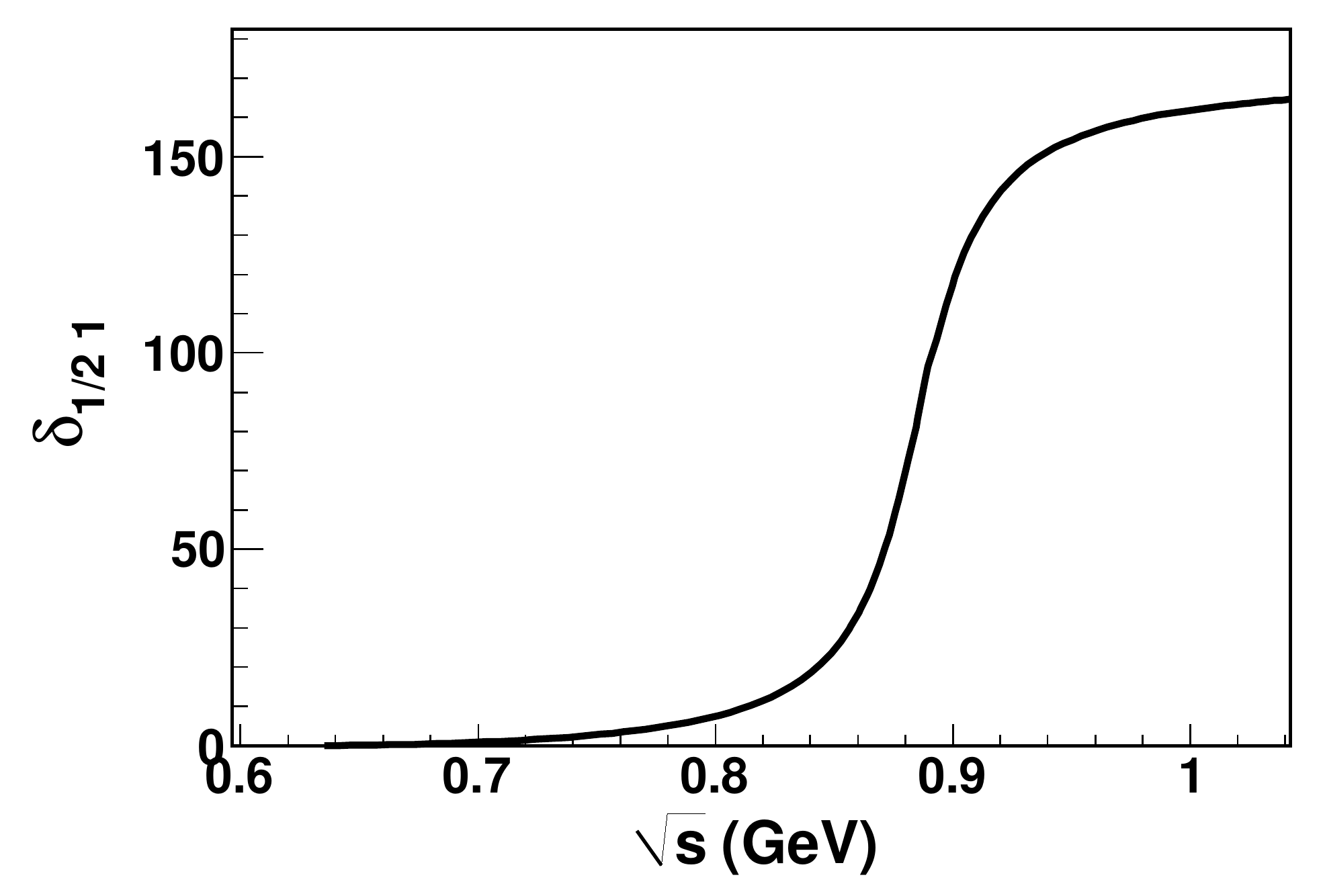}
\includegraphics[scale=0.26]{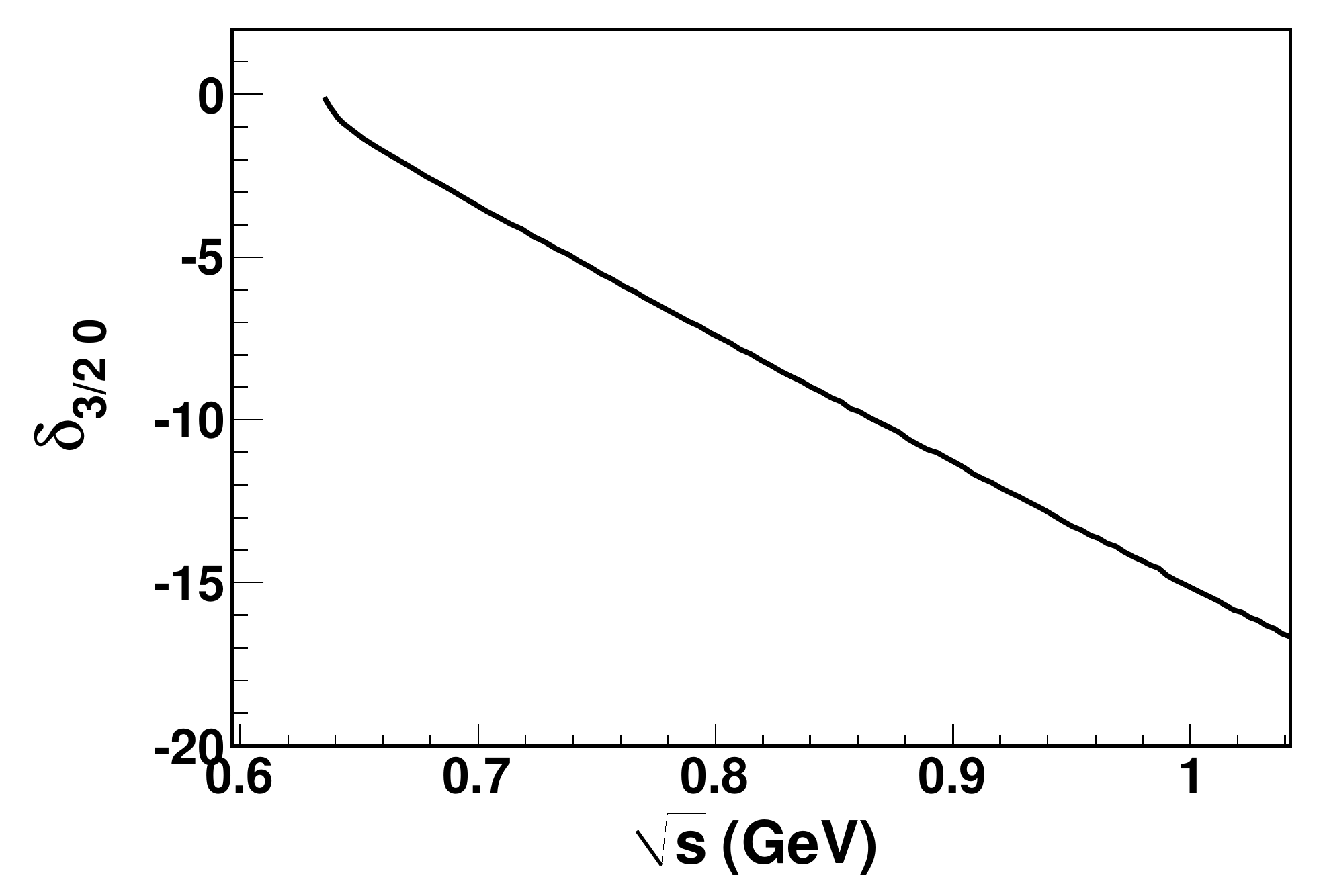}    
\caption{\label{fig:iamsu3} Kaon-pion phase-shifts obtained from the $SU(3)$ IAM in the three relevant channels at low energy. We use these phase-shifts as an input for the calculation of the strangeness diffusion. In the
top right panel the narrow $K^{*}(892)$ is visible. The exotic channel $IJ=\frac{3}{2}0$ is repulsive.}
\end{center}
\end{figure}
We project the two amplitudes $T^{1/2} (s,t,u)$ and $T^{3/2} (s,t,u)$ into definite spin channels with the help of an analogous formula to Eq.~(\ref{eq:partampli}) that for pion-kaon scattering reads
\be t_{IJ} (s) = \frac{1}{32 \pi} \int_{-1}^1 dx \ P_J (x) T^I (s,t(s,x),u(s,x)) \ .  \ee

Once the three partial amplitudes $t_{\frac{3}{2} 0} (s)$,$t_{\frac{1}{2} 1} (s)$,$t_{\frac{3}{2} 0} (s)$ are obtained
we proceed to unitarize them with the inverse amplitude method described in Sec.~\ref{sec:iam}.
The key equation is (\ref{eq:iamamplit}) in which the index $(0)$ refers to the LO amplitude and the index $(1)$ to the NLO amplitude.
In the Figure \ref{fig:iamsu3} we show the result of the phase-shifts (obtained from the partial wave amplitudes by using Eq.~(\ref{eq:amplphase})) 
in the channels as a function of the CM energy.

We use the values of the low energy constants\index{low energy constants} that appear in \cite{GomezNicola:2001as}, that we show in Table \ref{tab:lecforsu3}.

\begin{table}[t]
\begin{center}
\begin{tabular}{|c|c|c|c|c|c|c|c|} 
\hline
$L_1$ & $L_2$ & $L_3$ & $L_4$ & $L_5$ & $L_6$ & $L_7$ & $L_8$ \\
\hline \hline
$0.56$ & $1.21$  & $-2.79$ & $-0.36$ & $1.4$ & $0.07$ & $-0.44$ & $0.78$ \\
\hline
\end{tabular}
\end{center}
\caption{Set of low energy constants used in the $\pi-K$ scattering needed for the calculation of the strangeness diffusion. All of them have been multiplied by $10^3$.\label{tab:lecforsu3}}
\end{table}

\section{Diffusion coefficient}

\begin{figure}[t]
\begin{center}
\includegraphics[scale=0.4]{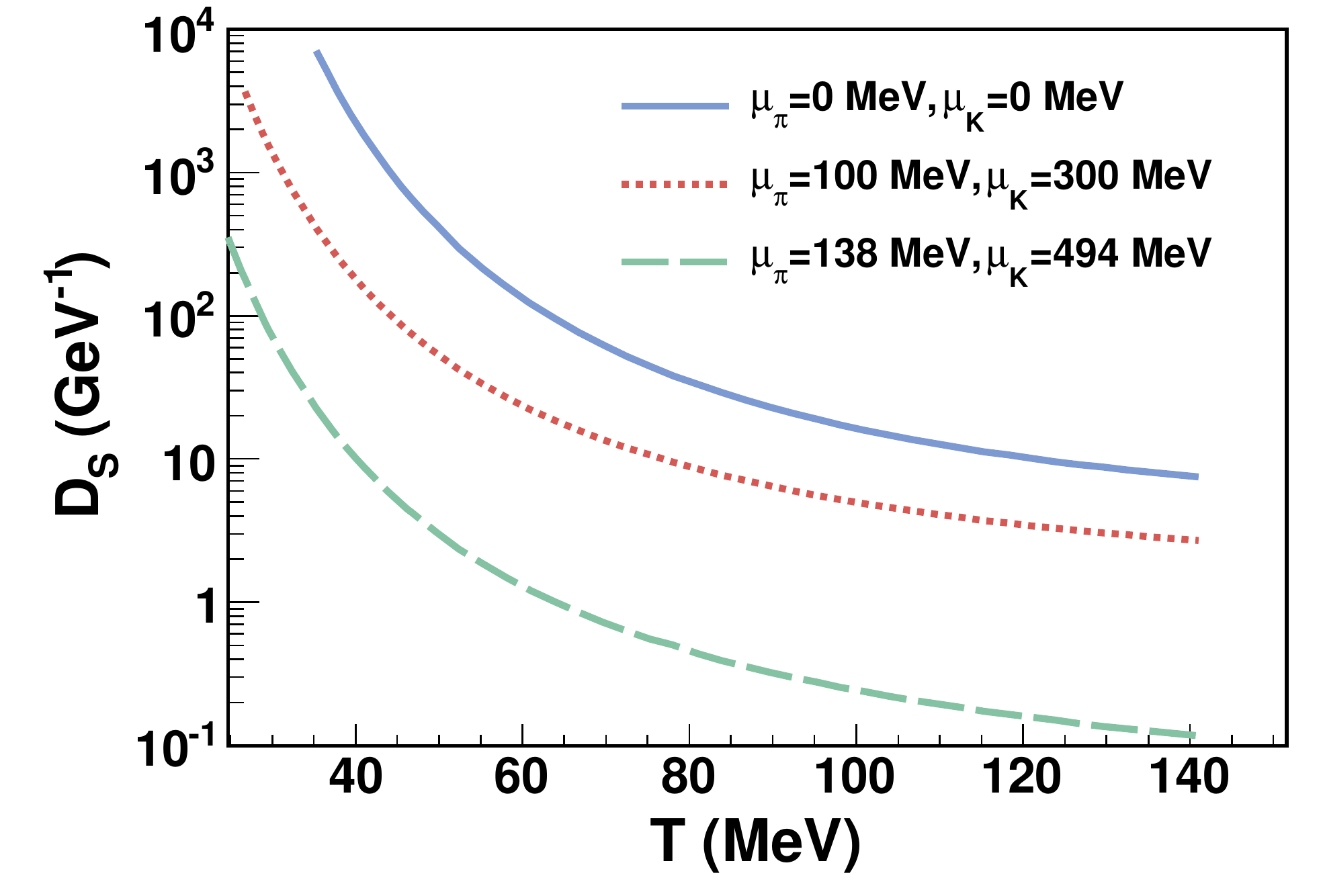}
\caption{\label{fig:s_diffusion} Strangeness diffusion coefficient for a gas of pions and kaons as a function of temperature. Three combinations of meson chemical potentials are used.}
\end{center}
\end{figure}

In Figure~\ref{fig:s_diffusion} we show the result for the strangeness diffusion coefficient as a function of temperature. We have used different combinations of the pion and kaon
chemical potentials up to the limiting value of $\mu_\pi \simeq m_{\pi}$ and $\mu_K \simeq m_K$.

\chapter{Charm Diffusion \label{ch:8.charm} }

Heavy flavored hadrons are interesting because the hadron medium is not hot enough to excite charm pairs.
They are produced by hard gluons in the initial stages of the collision and their spectra carry a memory of it,
unlike pions and kaons that can be produced in the hadronic thermal medium at later stages, and thus show a spectrum close to
black-body without much information from the initial configuration of fields.

Charmed mesons interact with the hadron gas after the crossover from the quark-gluon plasma phase. The corrections to their
properties due to this cooler medium require their scattering cross section with the medium pions.
This cross section can be theoretically accessed combining chiral perturbation theory, heavy quark effective theory and unitarity \cite{Abreu20112737}.
Given the scattering amplitudes one can proceed to kinetic simulations following individual particles, or employ kinetic theory
to compute transport coefficients that can be input into bulk hydrodynamics simulations.

As the charm transport coefficients are concerned, we will consider the drag or friction force $F$\index{charm drag coefficient},
and the two $\Gamma_0$ and $\Gamma_1$ momentum-space diffusion coefficients\index{charm diffusion coefficients}. Other works have considered only isotropic drag 
and diffusion, in which case there is only one diffusion coefficient also denoted as $\kappa$. We do not make this hypothesis of isotropy 
(because of the interesting elliptic flow observable) and provide both coefficients corresponding to parallel and shear momentum transfers.
Finally, in the $p\to 0$ limit, we make contact with the traditional kinetic theory and compute the space diffusion coefficient $D_x$ \index{spatial diffusion coeficient}. 

We employ the Fokker-Planck formalism for a heavy Brownian particle \index{Brownian particle} subject to the bombardement of the light pions in the medium.
This heavy particle will be either a $D$ meson or a $D^*$ meson. The later, although unstable through the decay $D^*\rightarrow D \pi$,
has a small width (given its closeness to threshold) and for the duration of the hadron gas it propagates as a stable mode.
We will justify this statement showing the properties of the $D-$meson spectrum in Sec.~\ref{sec:Dspectrum}.
Our approximations will be sensible as long as the momentum of the heavy particle remains smaller than its mass in natural units, 
so that $p\ge 2$ GeV is not accessible by our computation (although we show plots at higher momentum for ease of comparison with
future investigations addressing hard heavy flavors).

\section{D-meson spectrum \label{sec:Dspectrum}}

A charm quark \index{D mesons} propagating in the medium below the deconfinement crossover must do so confined in a hadron. We consider heavy-ion collisions
at energies of RHIC or LHC in which the baryon number is very small and can be neglected. Therefore, the charm quark is expected to
form a $D$ meson or an excitation thereof.

The ground state of the $D$-spectrum is the pseudoscalar $D$ meson ($J^P=0^-$) with four charge states $+,-,0,\overline{0}$. 
Since we neglect isospin-breaking terms, we can average the masses over this quartet to obtain $M_D \simeq 1867$ MeV. 
This meson cannot decay by any strong process and we will take it to be absolutely stable.

The first excitation is the vector $1^-$ $D^*$ meson whose mass average is
$M_{D^*}=2008.5$ MeV. In the heavy quark limit this meson should degenerate with the
$D$, (and in fact this is seen by glancing higher to the $B$-meson whose splitting to the $B^*$ is much smaller).
This mass is barely above $D\pi$ threshold, so there is only this one strong decay
channel, and it is very suppressed.

The width of the charged $D^*$ is estimated at $1$ MeV, and that for the neutral 
partners has not been measured but is consistent with $\Gamma \le 2$ MeV.
This means that a $D^*$ has a mean lifetime in vacuum of order $100-200$
fm. Since the typical freeze-out time of a heavy-ion collision is about 20 fm
it is not a bad first approximation to take the $D^*$ meson as also stable during
the fireball's lifetime: the decay time is an order of magnitude larger than the freeze-out time.

\begin{table}[t]
\centering
\begin{tabular}{|c c||c c|}
\hline 
Meson & $J^P$ & $M$ (MeV) & $\Gamma$ (MeV) \\
\hline
$D $  & $0^-$ & 1867     & -       \\
$D^*$ & $1^-$ & 2008     & 1       \\
$D_0$ & $0^+$ & 2360(40) & 270(50) \\
$D_1$ & $1^+$ & 2422     & 22(5)   \\
$D_1$ & $1^+$ & 2427(40) & 380(150)\\
$D_2$ & $2^+$ & 2460     & 30      \\
\hline
\end{tabular}
\caption{Charged-average masses and experimental estimates~\cite{Nakamura:2010zzi} for the strong 
widths of the $D$-meson resonances. Errors not quoted are about $1$ MeV or less. \label{tab:Dspectrum}}
\end{table}

\begin{figure}[t]
\centering
\includegraphics[scale=0.32]{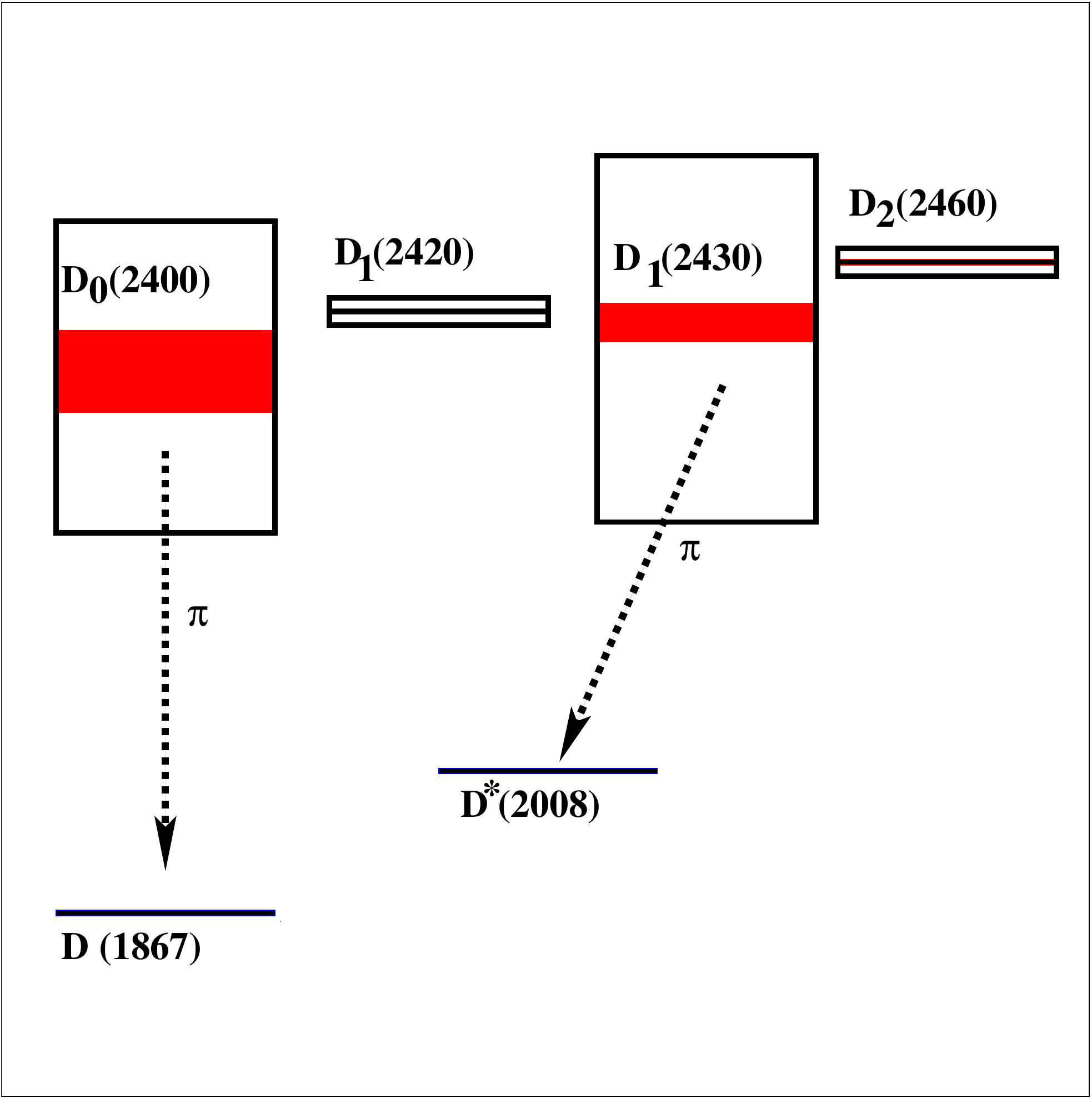}
\caption{\label{fig:Dmesonspectrum} 
Low-lying $D$-meson spectrum. The blue lines represent the states $D$ and $D^*$. The red areas represent the position and uncertainty of the four excited states $D_0 (2400)$, $D_1(2420)$, $D_1(2430)$, $D_2(2460)$.
The black boxes represent their widths.
}
\end{figure}

In agreement with quark model expectations, the next-higher excitations of the
$D$ system seem to be a triplet and a singlet of positive parity, with spins
$0^+,1^+,2^+$ and $1^+$ respectively, corresponding to $ ^{2S+1}L_J= 
^3P_J$ and $ ^1P_1$. 
The two mesons with spin 1 and positive parity must mix, and they do so in an 
interesting manner: the one with lowest mass, $D_1(2420)$ becomes narrow and hence
decoupled from the natural $s$-wave decay channel $D^*\pi$, whereas the 
higher member $D_1(2430)$ is very broad and seen in that configuration.
The situation can be seen in Table~\ref{tab:Dspectrum} and depicted in Fig.~\ref{fig:Dmesonspectrum}.

The remaining low-lying resonance, the $D_2$, is again narrow. Since its mass at 
2460 MeV is 600 MeV above the ground-state $D$ meson, and it is quite decoupled
due to its moderate width of about $40$ MeV, we do not expect this (nor the $D_1(2420)$
to play an important role at small temperatures.

Thus a sensible approach to charm propagation in a heavy-ion collision after the
phase transition to a hadron gas has occurred, is to take the $D$ and $D^*$ 
mesons as absolutely stable degrees of freedom for the $c$-quark, that in 
collision with the in-medium pions rescatter into the resonances $D_0$ and
$D_1(2430)$.

\section{Fokker-Planck equation \index{Fokker-Planck equation}}

The one-particle distribution function of a charmed meson with momentum $p$, $f_c (t,\mb{x},\mb{p})$, is not in equilibrium when the hadron
 phase of a heavy-ion collision forms, and must relax via a Boltzmann-Uehling-Uhlenbeck equation (\ref{eq:kin_eq})\index{BUU equation}:
\be \nonumber  \frac{df_c (t,\mathbf{x}, \mathbf{p})}{dt} = C [f_{\pi} (t,\mathbf{x}, \mathbf{k}_3),f_c (t,\mathbf{x}, \mathbf{p})] \ , \ee
We have slightly changed notations with respect to the previous sections. Because we are dealing with two different
species we will use the subindex $c$ to denote the charmed meson, $\pi$ for the pion and keep all the phase-space variables as explicit arguments.
Notice that, because of the scarcity of strange quarks in the heavy-ion collision debris (kaon multiplicity is 10\% of typical pion multiplicity, see Fig.~\ref{fig:pikpspectra}) we
are interested only in channels involving the scattering between charmed mesons and pions with total strangeness equal to zero. 

As in the case for pions, the left-hand side, in the absence of external forces, is the advective derivative
\be \frac{\partial f_c (t,\mathbf{x}, \mathbf{p})}{\pa t} + \frac{\mathbf{p}^i}{E_p} \cdot {\nabla}_i f_c (t,\mathbf{x}, \mathbf{p}) = 
\left[ \frac{\pa f_c (t,\mathbf{x}, \mathbf{p})}{\pa t} \right]_{coll} \ .\ee

Because the density of $D$ and $D^*$ mesons is very small, one can neglect collisions between $D$ mesons themselves and concentrate
only on the interaction of these charmed mesons with the pion bath, assumed in thermal equilibrium.

The bath's distribution function $f_{\pi} (\mathbf{q})$  is hence the Bose-Einstein function. Moreover, the gas is assumed homogeneous and the
 distribution does not depend on $\mathbf{x}$. For this reason one can average the Boltzmann equation over the collision volume and understand 
the one-particle distribution function for the charmed mesons as the average
\be f_c (t,\mb{p})\equiv \frac{1}{V} \int d \mathbf{x} \ f_c (t,\mathbf{x}, \mathbf{p}) \ .\ee
The averaged BUU equation then becomes
\be \frac{\pa f_c (t,\mb{p})}{\pa t} = \left[ \frac{\pa f_c (t,\mb{p})}{\pa t} \right]_{coll} \ .\ee

Charmed mesons may enter and exit the momentum element $d{\bf p}$ around ${\bf p}$ by collisions with the pion bath, so the collision term has two parts associated with gains and losses.

Gains in the momentum distribution around ${\bf p}$ are proportional to the probability density around $(\mathbf{p}+\mathbf{k})$ times the probability of transferring momentum $\mathbf{k}$ from the charmed meson to the bath.
It is therefore convenient to define a collision rate $w(\mathbf{p},\mathbf{k})$ for a charmed meson with initial and final momenta $\mathbf{p}$, $\mathbf{p}-\mathbf{k}$.

Conversely, losses are proportional to the distribution function around 
$\mathbf{p}$ times the probability of transferring momentum $\mathbf{k}$ to the pion bath.

The BUU equation should take into account Bose enhancement effects in the final state, with factors $(1+f_c)$ that encode 
the increased probability of a charmed meson scattering into an already occupied state, 
\begin{eqnarray}
\nonumber \frac{\pa f_c (t,\mb{p})}{\pa t} & = &
\int d\mathbf{k} \left\{ f_c (t,\mathbf{p}+\mathbf{k}) w (\mathbf{p}+ \mathbf{k}, \mathbf{k}) 
\left[1+f_c (t,\mathbf{p}) \right] \right. \\
  & & \left. -  f_c(t,\mathbf{p}) w (\mathbf{p}, \mathbf{k} ) \left[ 1+f_c (t, \mathbf{p}-\mathbf{k}) \right] \right\} \ . 
\end{eqnarray}

However, as the number of charmed mesons is very small, we can approximate 
$1+f_c(t, \mathbf{p}) \approx 1$ inside the collision operator in practice. This is equivalent to use a classical Boltzmann measure inside the collision operator.
\be 
\frac{\pa f_c (t , \mathbf{p})}{\pa t} = \int d\mathbf{k} \left[ f_c(t,\mathbf{p}+\mathbf{k}) w (\mathbf{p}+ \mathbf{k}, \mathbf{k}) -  f_c(t,\mathbf{p}) w (\mathbf{p}, \mathbf{k} ) \right]
\ .\ee

This approximation however is probably not valid for the pion distribution function and we keep the $(1+f_\pi)$ factor in Eq.~(\ref{eq:probdist}) below. In turn the collision rate 
can be spelled out in terms of the Lorentz invariant charm quark-pion scattering amplitude,

\begin{eqnarray}
 \nonumber  w (\mathbf{p}, \mathbf{k} ) & =& g_{\pi} \int \frac{d \mathbf{q}}{(2\pi)^9} f_{\pi} (\mathbf{q}) \left[ 1+ f_{\pi} (\mathbf{q}+\mathbf{k}) \right]
 \frac{1}{2E_q^{\pi}} \frac{1}{2E_p^{c}} \frac{1}{2 E_{q+k}^{\pi}} \frac{1}{2 E_{p-k}^c} \\
& & \label{eq:probdist}  \times (2\pi)^4 \delta (E_p^c + E_q^{\pi} - E^c_{p-k} -E_{q+k}^{\pi} )  \sum |\mathcal{M}_{\pi c}(s,t,\chi)|^2 \ ,
\end{eqnarray}

\noindent
where $\chi$ denotes the possible spin degrees of freedom, active if the $c$ quark finds itself inside a $D^*$ meson. 
 The scattering amplitude $\mathcal{M}_{\pi c}$ is normalized according to standard covariant convention~\cite{Nakamura:2010zzi}. 
Note that
 Eq.~(7) of~\cite{He:2011yi} differs by the Bose-enhancement factor $(1+f_\pi)$ for the pion exiting the collision. We believe that in the 
temperature range of $m_\pi\simeq T \simeq 150$ MeV that we (and those authors) treat, this enhancement should not be neglected.

The BUU equation in this case reduces to a much simpler Fokker-Planck equation \index{Fokker-Planck equation} because the mass of the $D$ and $D^*$ mesons carrying the $c$-quark is much 
greater than the mass of the pions and the temperature of the heat bath.
Then, the scale of momentum for which there is a significant change of $f_c(t,\mb{p})$ with the momentum of the $D$ meson $|\mathbf{p}|$ is greater than the typical
 transfered momentum $|\mathbf{k}|$, that is of the order of $T$:
\be |\mathbf{p}|_{\rm{f_c}} \gg |\mathbf{k}| \sim T \sim 150 \textrm{ MeV} \ . \ee

Because of this separation of scales, it is natural to expand the collision rate inside the collision operator respect to its first argument ${\bf p}+{\bf k}$,
\ba \nonumber wf & \equiv &  w(\mathbf{p}+\mathbf{k},\mathbf{k}) \ f_c(t, \mathbf{p} + \mathbf{k}) \\
 \label{eq:colintegral} &  = & w (\mathbf{p},\mathbf{k}) f_c (t,\mathbf{p}) + k_i \frac{\pa}{\pa p_i} (wf) + \frac{1}{2} k_i k_j \frac{\pa ^2}{\pa p_i \pa p_j} (wf) \dots
\ea
with $i,j=1,2,3$.
The collision integral reads, with this substitution,
\be \left[ \frac{\pa f_c (t , \mathbf{p})}{\pa t} \right]_{coll} = \int d\mathbf{k} \left[ k_i \frac{\pa}{\pa p_i} + \frac{1}{2} k_i k_j \frac{\pa ^2}{\pa p_i \pa p_j} \right] \left( wf \right)\ .
\ee
This suggests defining two auxiliary functions, 
\begin{eqnarray}
 F_i (\mathbf{p}) & = & \int d\mathbf{k} \ w(\mathbf{p},\mathbf{k}) \ k_i \ ,\\
\Gamma_{ij} (\mathbf{p}) & = & \frac{1}{2} \int d\mathbf{k} \ w(\mathbf{p},\mathbf{k}) \ k_i k_j\ .
\end{eqnarray}

Eq.~(\ref{eq:colintegral}) reduces to the Fokker-Planck equation \index{Fokker-Planck equation}
\be \label{eq:FKPL}
\frac{\pa f_c (t, \mathbf{p})}{\pa t} = \frac{\pa}{\pa p_i} \left\{ F_i (\mathbf{p}) f_c (t,\mathbf{p}) + \frac{\pa}{\pa p_j} \left[ \Gamma_{ij} (\mathbf{p})  f_c (t,\mathbf{p}) \right] \right\} \ ,
\ee
where we can see that $F_i$ behaves as a friction term \index{charm drag coefficient}representing the average momentum change 
of the $D$ meson and $\Gamma_{ij}$ acts as a diffusion coefficient \index{charm diffusion coefficients}in momentum space, as it forces a broadening of
 the average momentum distribution of the $D$ meson.

The goal of this chapter is to calculate the coefficients $F_i$ and $\Gamma_{ij}$ that encode the physics of charm drag and 
diffusion.

In the ideal case where the pion gas is homogeneous and isotropic, and because the coefficients $F_i$ and $\Gamma_{ij}$ only depend on $p^i$, they can be expressed in terms of three scalar functions by means of
\begin{eqnarray} \label{eq:defFandG}
F_i (\mathbf{p}) & = & F(p^2) p_i \ , \\ \nonumber
\Gamma_{ij} (\mathbf{p}) & = & \Gamma_0 (p^2) \Delta_{ij} + \Gamma_1 (p^2) \frac{p_i p_j}{p^2} \ ,
\end{eqnarray}
where
\be \Delta_{ij} \equiv \delta_{ij} - \frac{p_i p_j}{p^2}
\ee
satisfies the handy identity $\Delta_{ij} \Delta^{ij} =2$.

We choose the momenta of the elastic collision between a charmed meson $D$ or $D^*$ and a pion as
\be D (\mathbf{p}) + \pi (\mathbf{q}) \rightarrow D(\mathbf{p}-\mathbf{k}) + \pi (\mathbf{q} +\mathbf{k}). \ee

The three scalar coefficients in Eq.~(\ref{eq:defFandG}) are then simple integrals over the interaction rate
\begin{eqnarray} \label{Transportintegrals}
 \nonumber F(p^2) & = & \frac{p^i F_i}{p^2} = \int d\mathbf{k}\  w(\mathbf{p},\mathbf{k})  \ \frac{k_ip^i}{p^2} \ , \\ 
 \nonumber \Gamma_0 (p^2) & = & \frac{1}{2} \Delta_{ij} \Gamma^{ij} = \frac{1}{4} \int d\mathbf{k}\ w(\mathbf{p},\mathbf{k}) \left[ \mathbf{k}^2 - \frac{(k_i p^i)^2}{p^2} \right] \ , \\ 
 \Gamma_1(p^2) & = & \frac{p_i p_j}{p^2} \Gamma^{ij} = \frac{1}{2} \int d\mathbf{k}\  w(\mathbf{p},\mathbf{k}) \ \frac{(k_i p^i)^2}{p^2} \ , 
\end{eqnarray}
where the dynamics is fed-in by the scattering matrix elements $|\mathcal{M}_{\pi c}|$.

In Appendix~\ref{app:langevin} we show how the interpretation of the friction coefficient times the quark momentum $F(p^2) p$ is that of an energy loss \index{heavy quark!energy loss}
per unit length upon propagation of the charm quark in the plasma, and the loss of momentum \index{heavy quark!momentum loss} per unit length is simply $F(p^2) E_p$ in terms
of energy and momentum of the charmed particle.

\section{Effective Lagrangian with ChPT and HQET\glossary{name=HQEF,description={heavy quark effective theory}}}

Here we sketch we construction of the chiral Lagrangian density that describes \index{effective Lagrangian} the interactions between the spin-0 and spin-1 $D$-mesons and
pseudoscalar Goldstone bosons\index{Goldstone bosons}.
The Lagrangian is elaborated by writing down all the possible terms compatible with
Lorentz and $C$, $P$ and $T$ invariances. It must also respect the chiral and heavy quark symmetries at lowest order, and break them in a controlled power-series expansion. 
The non-linear realization of the chiral symmetry is based on the exponential parametrization of the Goldstone bosons ($N_f=3$) \cite{Yan:1992gz}:
\be
 U = \exp{\left( \frac{2i \Phi}{\sqrt{2} F_0} \right)} \ ,
\label{eq:u_field}
\ee
with $F_0$ being the Goldstone boson decay constant in the chiral limit and
\be
\Phi = \left( \begin{array}{ccc}
  \frac{1}{\sqrt{2}}\pi ^{0}+\frac{1}{\sqrt{6}} \eta  & \pi ^{+} & K^{+} \\
    \pi^{-} & -\frac{1}{\sqrt{2}}\pi ^{0}+\frac{1}{\sqrt{6}} \eta & K^0 \\
    K^{-}   & \bar{K}^{0}  &  -\frac{2}{\sqrt{6}} \eta
\end{array}
    \right).
\ee

This field $U$ transforms under the $SU(3)_L \times SU(3)_R$ symmetry as
\be U \rightarrow U' = LUR^{\dag} \ , \ee
where $L$ and $R$ are global transformations under $SU(3)_L$ and $SU(3)_R$ respectively.
The kinetic term for the Goldstone bosons that is invariant under this chiral transformation is the canonical one
\be \mathcal{L} = \frac{F^2_0}{4} \textrm{Tr } \pa_{\mu} U^{\dag} \pa^{\mu} U \ .\ee

For conveniency one introduces the matrix
\be u = \sqrt{U} \ , \ee
wich under an $SU(3)_L \times SU(3)_R$ transforms as
\be u \rightarrow u' = L u W^{\dag} = W u R^{\dag} \ , \ee
where $W$ is a unitary matrix expressible as a complicated combination of the matrices $R$,$L$ and $\Phi$.

With the matrix $u$ one can construct the a vector and an axial vector fields as:
\be \Gamma _{\mu}  =  \frac{1}{2} \left( u^{\da} \partial _{\mu} u +  u\partial _{\mu} u^{\da} \right) \ , \ee
\be u _{\mu}  =  i \left( u^{\da} \partial _{\mu} u -  u\partial _{\mu} u^{\da} \right) \ .\ee
The transformation rule for these vectors is:
\begin{eqnarray}
\Gamma_{\mu} & \rightarrow & \Gamma_{\mu}' = W \Gamma_{\mu} W^{\dag} + W \pa_{\mu} W^{\dag} \ , \\
 u_{\mu} & \rightarrow & u_{\mu}' = W u_{\mu} W^{\dag} \ . 
\end{eqnarray}
With $\Gamma_{\mu}$ one defines the covariant derivative as
\be \nabla_{\mu}   =  \partial _{\mu} - \Gamma _{\mu} \ . \ee

With all these pieces one constructs the leading order (LO) chiral Lagrangian $\mathcal{L}^{(1)}$. It is given by \cite{Lutz:2007sk,Guo:2009ct,Geng:2010vw},   
 \ba
\nonumber \mathcal{L}^{(1)} & = &  \textrm{Tr } [\nabla ^{\mu} D \, \nabla _{\mu}D^{\da}] - M_D ^2 \textrm{Tr } [D D^{\da}] -  \textrm{Tr } [\nabla ^{\mu} D^{\ast \nu} \,\nabla _{\mu} D^{\ast \da}_{\nu}]
 + m_{D}^2 \textrm{Tr } [D^{\ast \mu}  D^{\ast \da}_{\mu}]   \\
\nonumber & &  + i \ g \ \textrm{Tr } [\left(  D^{\ast \mu} u_{\mu}  D^{ \da} - D u^{\mu} D^{\ast
\da}_{\mu} \right) \\
& & +  \frac{g}{2 M_D} \textrm{Tr } [\left(  D^{\ast}_{\mu} u_{\alpha}   \nabla_{\beta} D^{\ast \da}_{\nu} -  \nabla_{\beta} D^{\ast}_{\mu} u_{\alpha}  D^{\ast \da}_{\nu} \right) \varepsilon ^{\mu \nu \alpha \beta} ]  \ , 
\label{eq:lag1}
\ea
where $D=(D^0,D^{+}, D^{+}_{s})$ and  $D^{\ast} _{\mu}=(D^{\ast 0}, D^{\ast + },D^{\ast +}_{s})_{\mu}$ are the $SU(3)$ anti-triplets of spin-zero and spin-one $D$-mesons with the chiral limit mass $M_D$, respectively.
Under the chiral transformation the covariant derivative $\nabla_{\mu}$ simply transforms as
\be \nabla_{\mu} D \rightarrow \nabla_{\mu} D' = \nabla_{\mu} D W^{\dag}  \ . \ee
It is not difficult to check that the Lagrangian $\mathcal{L}^{(1)}$ is indeed invariant under the chiral group $SU(3)_L \times SU(3)_R$.

The NLO chiral Lagrangian $\mathcal{L}^{(2)}$ reads
 \ba
\mathcal{L}^{(2)} & = & - h_0  \textrm{Tr } [D D^{\da}] \textrm{Tr } [\chi_+] + h_1  \textrm{Tr } [D \chi_+ D^{\da}] + h_2  \textrm{Tr } [D D^{\da}] \textrm{Tr } [u^{\mu}u_{\mu}]   \nonumber \\
& & + h_3 \textrm{Tr } [ D u^{\mu}u_{\mu} D^{\da}] + h_4  \textrm{Tr } [\nabla _{\mu} D  \,  \nabla _{\nu} D^{\da}]  \textrm{Tr } [u^{\mu}u^{\nu}] + h_5 \textrm{Tr } [ \nabla _{\mu} D   \{ u^{\mu}, u^{\nu} \} \nabla _{\nu}D^{\da}] \nonumber \\
& & + \tilde{h}_0 \textrm{Tr } [ D^{\ast \mu} D^{\ast \da}_{\mu}] \textrm{Tr } [ \chi_+]  - \tilde{h}_1 \textrm{Tr } [ D^{\ast \mu} \chi_+ D^{\ast \da}_{\mu}]  - \tilde{h}_2 \textrm{Tr } [ D^{\ast \mu}  D^{\ast \da}_{\mu}]  \textrm{Tr } [u^{\nu} u_{\nu}]
   \nonumber \\
& & - \tilde{h}_3 \textrm{Tr } [ D^{\ast \mu}  u^{\nu}u_{\nu} D^{\ast \da}_{\mu}] - \tilde{h}_4  \textrm{Tr } [\nabla _{\mu}D ^{\ast \alpha}  \,  \nabla _{\nu}D^{\ast \da}_{\alpha} ] \textrm{Tr } [u^{\mu}u^{\nu}] \nonumber \\
& & - \tilde{h}_5 \textrm{Tr } [\nabla _{\mu}D ^{\ast \alpha}  \{ u^{\mu}, u^{\nu} \}\nabla _{\nu}D^{\ast \da} _{\alpha}] \ . 
\label{eq:lag2}
\ea
where
\be
\chi _{+} =  u^{\da} \chi u^{\da} +u \chi u \ ,
\ee
with $\chi = \mathrm{diag}(m^2_{\pi}, m^2_{\pi}, 2 m^2_{K} -m^2_{\pi})$ being the mass matrix. The twelve parameters $h_i, \tilde{h}_i (i=0,...,5)$ are the
 low-energy constants (LECs), to be determined. However, we can make use of some constraints to reduce the set of free LECs. First, it should be noticed that in the \glossary{name=QCD,description={quantum chromodynamics}}
limit of large number of colors ($N_c$) of QCD~\cite{'tHooft:1973jz}, single-flavor trace interactions are dominant.
So, we fix $h_0 = h_2= h_4 = \tilde{h}_0 = \tilde{h}_2 = \tilde{h}_4 = 0$ henceforth. Besides, by imposing the heavy-quark symmetry
(as will become clear in subsection~\ref{sec:hqs} below), it follows that $\tilde{h}_i \simeq  h_i $. 

In the following, the lowest order of the perturbative expansion of the quantities $\Gamma _{\mu}$, $u _{\mu}$ and $\chi _{+} $ in Eqs. (\ref{eq:lag1}) 
and (\ref{eq:lag2}) is considered to construct the scattering matrix of the interactions between the charmed mesons and the pseudoscalar Goldstone bosons.

\subsection{$D-\pi$ tree-level scattering amplitudes}

\begin{figure}[t]
\centering
\includegraphics[width=8.5cm]{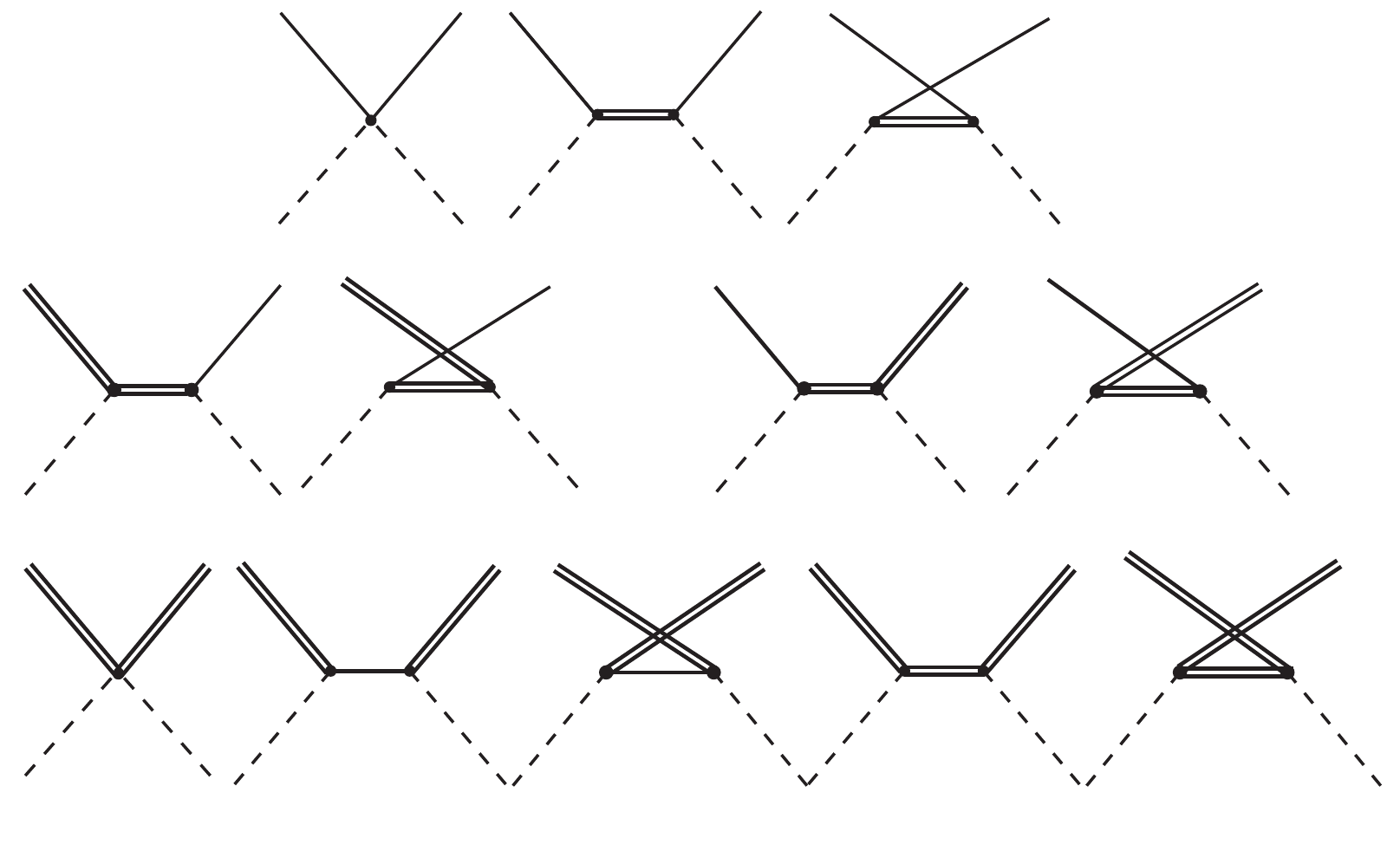}
\caption{Tree-level diagrams relevant to the scattering amplitudes in Eq. (\ref{eq:ampl1}). The solid, double and dashed lines represent the $D$-mesons,
 $D^{\ast}$-mesons and pions, respectively.}
\label{fig:tree}
\end{figure}

From the Lagrangian in Eq.~(\ref{eq:lag1}) we are able to obtain the scattering amplitudes\index{scattering amplitudes ($D\pi$ system)} $V$ for $D (D^{\ast}) \pi \rightarrow D (D^{\ast}) \pi$
processes. In Fig.~\ref{fig:tree} we show the tree-level diagrams constructed from the LO \glossary{name=LO,description={leading order}} and NLO interactions\glossary{name=NLO,description={next-to leading order}}. 
These include both contact interactions and Born exchanges. The different scattering channels are labeled as $V_{a}$ through $V_{d}$, where the subscripts refer to the scattering channels as follows 

\ba
&(a)&:  D \pi \rightarrow D \pi \ , \nonumber \\
&(b)&:  D ^{\ast}\pi \rightarrow D \pi \ ,  \nonumber \\
&(c)&:  D \pi \rightarrow D ^{\ast}\pi \ , \nonumber \\
&(d)&:  D ^{\ast} \pi  \rightarrow D^{\ast} \pi \ .  
\label{eq:proc}
\ea

For clarity in this section we will denote the incoming and outgoing four-momenta as
\be D( p_1^{\mu} ) + \pi (p_2^{\mu}) \rightarrow D(p_3^{\mu}) + \pi (p_4^{\mu}) \ . \ee
Nevertheless, when introducing the scattering amplitudes into Eq.~(\ref{Transportintegrals}) we will make the identification:
\begin{eqnarray}
\nonumber (p_1^{0},\mb{p_1}) &\rightarrow& (\sqrt{M_D^2+p^2},\mb{p}) \ ,\\
\nonumber (p_2^{0},\mb{p_2}) &\rightarrow& (\sqrt{m_{\pi}^2+p^2},\mb{q}) \ , \\
\nonumber (p_3^{0},\mb{p_3}) &\rightarrow& (\sqrt{M_D^2+(\mb{p-k})^2},\mb{p-k}) \ , \\
\nonumber (p_4^{0},\mb{p_4}) &\rightarrow& (\sqrt{m_{\pi}^2+(\mb{q+k})^2},\mb{q+k}) \ .
\end{eqnarray}

The four tree-level scattering amplitudes read:
\ba
V_{a} & = &  \frac{C_{0}}{4 F^2} (s - u)  + \frac{2C_{1}\,m_{\pi }^2}{ F^2} h_1   +  \frac{ 2 C_{2}}{ F^2} h_3 (p_2 \cdot p_4 ) +  \frac{2 C_{3}}{ F^2} h_5 \left[ (p_1 \cdot p_2 ) (p_3 \cdot p_4 )  \right.
  \nonumber \\
& & \left. + (p_1 \cdot p_4 )(p_2 \cdot p_3 ) \right] + \frac{2 i\,g^2}{ F^2}  p_2^{\mu} \left[   C_{4}\,D_{\mu \nu } (p_1 + p_2)+  C_{5}\, D_{\mu \nu } (p_2 - p_3) \right] p_4 ^{\nu} \ ,\nonumber \\
V_{b} & = & \frac{ i \,g^2}{ M_D F^2} \left[  C_{4} \, p_2^{\alpha } \left( 2 p_1^{\beta } + p_2  ^{\beta }\right) p_{4\rho} 
D^{\nu \rho}(p_1 + p_2)   +  C_{5}\, p_4^{\alpha } \left( p_2^{\beta } - p_3  ^{\beta } -p_1^{\beta } \right) p_{2\rho} 
  D^{\nu \rho}(p_2 - p_3)  \right] \nonumber \\
& & \times \ \varepsilon _{\alpha \beta \mu \nu} \ \mathbf{\epsilon }^{\mu}(p_1) \ , \nonumber \\
 V_{c} & = & \frac{ i \,g^2}{ M_D F^2} \left[  C_{4}\, p_4^{\alpha } \left( p_1^{\beta } + p_2^{\beta } + p_3  ^{\beta }  \right) 
p_{2\rho}    D^{ \rho \nu }(p_1 + p_2)  +  C_{5} \, p_2^{\alpha } \left( p_2^{\beta } -2  p_3  ^{\beta } \right) p_{4\rho}  
D^{\nu \rho}(p_2 - p_3)  \right] \nonumber \\ 
& & \times \ \varepsilon _{\alpha \beta \mu \nu} \  \mathbf{\epsilon } ^{\ast \mu}(p_3) \ , \nonumber \\
V_{d} & = &  -\left\{ \frac{C_{0}}{4 F^2} (s - u)  + \frac{2C_{1}\,m_{\pi }^2}{ F^2} \tilde{h}_1   +  
\frac{ 2 C_{2}}{ F^2}\tilde{h}_3 (p_2 \cdot p_4) +  \frac{2 C_{3}}{ F^2} \tilde{h}_5 \left[ (p_1 \cdot p_2 ) (p_3 \cdot p_4 ) \right. \right. \nonumber \\
& & \left. \left.  + (p_1 \cdot p_4 )(p_2 \cdot p_3 ) \right] \right\} \epsilon ^{\mu}(p_1)\epsilon^{\ast} _{ \mu}(p_3)\nonumber \\ 
& & + \frac{2 i \,g^2}{ F^2}  \left[ C_{4}\, D (p_1 + p_2) +  C_{5} \, D(p_2 - p_3) \right] \ p_2^{\mu}\epsilon _{\mu}(p_1)  p_4 ^{\nu}
\epsilon^{\ast} _{ \nu}(p_3) \nonumber \\
& & +  \frac{ i g^2}{3 M_D^2 F^2} \left[  C_{6} \, p_2^{\alpha } \left( 2 p_1^{\beta } 
+ p_2  ^{\beta }\right) p_{4} ^{\rho} \left( p_1^{\sigma } + p_2  ^{\sigma } +  p_3  ^{\sigma }\right)  
D^{\nu \gamma}(p_1 + p_2) \right. \nonumber \\
& & \left. +   C_{7} \,  p_2^{\alpha } \left(  p_2^{\beta } - 2 p_3  ^{\beta }\right) p_{4} ^{\rho} \left( p_2^{\sigma }
 - p_3  ^{\sigma } -  p_1  ^{\sigma }\right) D^{\nu \gamma}(p_2 - p_3)   \right]\nonumber \\
& & \times \ \varepsilon _{\alpha \beta \mu \nu} \ \varepsilon _{\rho \sigma \gamma \delta} \ \mathbf{\epsilon } ^{\mu}(p_1) \ \mathbf{\epsilon }^{\ast \delta}(p_3) \ ,
\label{eq:ampl1}
\ea  

\noindent
where $ C_{i}\;(i=0,...,7)$ are the coefficients of the scattering amplitudes for $D\pi ,D^{\ast} \pi$  channels with total isospin $I$, done in Table \ref{tab:table2},
 and $D (p)$, $D_{\mu \nu } (p)$ are the propagators of $D$ and $D^{\ast}$-mesons, respectively, 
\ba D (p) & = & \frac{i}{p^{2} - M_D^2} \ , \nonumber \\
D^{\mu \nu }(p) & = & \frac{-i}{p^{2} - M^2_{D^{*}}} \left( \eta^{\mu \nu} - \frac{p^{\mu}p^{\nu}}{M^2_{D^{*}}}\right)\ .
\ea
As the two particles in all amplitudes are distinguishable, there is no $t$-channel type contribution (as e.g. in Compton scattering) with our relevant fields
 (open charm mesons and pions), and only $s$ and $u$-channel interactions appear. Between a $D$ and a $\pi$ one could exchange additional, closed flavor resonances
 in the $t$-channel, but a quick examination makes clear that these contributions are totally negligible. For example, $f_0$ exchange, while having strong coupling
 to two pions, has negligible coupling to two $D$ mesons, so one of the vertices makes the amplitude very small. Similarly, $J/\psi$ $t$-channel exchange is suppresed 
because of the small two-pion coupling of the very narrow state (and similar for other, closed flavor resonances). It doesn't make sense to include these resonances while
 neglecting higher order chiral and heavy quark corrections to the $D\pi$ Lagrangian with the basic fields.

Finally $\mathbf{\epsilon } ^{\mu}(p)$ is the polarization vector of the vector $D^{\ast}$-meson. If we were to write the polarization indices explicitly, 
 $\mathbf{\epsilon } ^{\mu}(p)\equiv  \mathbf{\epsilon}^{\mu}_\lambda(p)$,  
$V_b\equiv V_{b\lambda}$, $V_c\equiv V_{c\lambda}$, $V_d\equiv V_{d\lambda \lambda'}$, while $V_a$ remains a scalar as no vector mesons appear.

The amplitudes $V_b$ and $V_c$ must be related by time reversal, since they encode $D^*\pi \to D\pi$ and $D\pi\to D^*\pi$ respectively. Indeed, if one exchanges
 $p_1$ by $p_3$ and $p_2$ by $p_4$, and employs energy-momentum conservation $p_1+p_2=p_3+p_4$, they map onto each other as $V_b \to V_c$, $V_c\to V_b$.

\begin{table}[t]
\centering
\begin{tabular}{|c ||c c|}
\hline 
Constants &  $I= \frac{1}{2}$ & $I=\frac{3}{2} $   \\
\hline
$C_0 $  & $-2$ & 1       \\
$C_1 $  & $-1$ & $-1$    \\
$C_2 $  & $1$ & $1$       \\
$C_3 $  & $1$ & $1$       \\
$C_4 $  & $3$ & $0$       \\
$C_5 $  & $\frac{1}{3}$ & $\frac{2}{3}$       \\
$C_6 $  & $3$ & $0$       \\
$C_7 $  & $\frac{1}{3}$ & $\frac{2}{3}$ \\
\hline
\end{tabular}
\caption{Coefficients of the scattering amplitudes for the $D\pi ,D^{\ast} \pi$ channels with total isospin $I$ in Eq. (\ref{eq:ampl1}).  \label{tab:table2}}
\end{table}

\subsection{On-shell unitarization}

Chiral perturbation theory amplitudes are by construction a series expansion (albeit with logarithmic corrections and, in our case, Born terms with an
 intermediate propagator due to the $DD^*\pi$ coupling) and by their very nature are unable to describe excited elastic resonances (in our case, $D_0$ and $D_1$). \\
The key to understanding this limitation is to note that, at fixed order, ChPT violates unitarity \index{unitarity} as momentum is increased. Therefore several strategies have been
 adopted to bypass the shortcoming, such as the $N/D$ method, the inverse amplitude method, or the K-matrix method.

We pursue the simplest partial-wave unitarization by employing on-shell factorization~\cite{Oller:1997ti} \index{on-shell unitarization} which is a nice 
feature of polynomial expansions and leads to algebraic formulae for the unitarized partial wave amplitudes, capable of reproducing resonances. 
Our conventions for the expansion of the perturbative $V_a$ and unitarized $T_a$ amplitudes in terms of Legendre polinomials are
\be 
V^l_a =\frac{1}{2} \int_{-1}^1 dx \  P_l (x) V_a(s,x) \ , 
\ee
\be T^l_a =\frac{1}{2} \int_{-1}^1 dx \  P_l (x)  T_a (s,x)  \ ,
\ee
where $x \equiv \cos \theta$ and $P_0 (x)=1$ and $a$ is a channel index.

We proceed by projecting the perturbative amplitude into the $s$-wave, 
that dominates at low energies because of the $k^{2l+1}$ suppression of higher waves, and is resonant at the $D_0$ (for $D\pi$ scattering) and $D_1$ (for $D^*\pi$ scattering), thus
 dominating the entire amplitude at moderate heavy-quark velocities (at higher velocities, boosting to the moving center of mass frame kinematically induces higher waves).
Thus the perturbative amplitude is substituted for
\be 
V^{l=0}_a (s)= \frac{1}{2} \int_{-1}^1 dx V_a (s,t(x),u(s,t(x))) \ P_{0} (x) \ .
\ee
The unitarized scalar amplitudes $T_a$ decouple in leading order HQET and read \cite{Roca:2005nm}:
\be \label{Unitarizedampl}
T^{l=0}_a (s) = \frac{-V^{l=0}_a (s)}{1- V_a^{l=0} (s) \ G_{l=0} (s)}\ .
\ee
This equation manifestly is a relativistic generalization of the Lippmann-Schwinger equation.

The factorized resolvent function is the standard one-loop integral
\be 
G_{l=0} (s) = i \int \frac{d^4 q}{(2\pi)^4} \frac{1}{(P-q)^2- M_D^{2}+i\epsilon}\frac{1}{q^2 - m_{\pi}^2 + i\epsilon} \ .
\ee
We employ dimensional regularization of the divergent integral to read from Ref.~\cite{Roca:2005nm}
\begin{eqnarray}
G_{l=0} (s) & = & \frac{1}{16 \pi^2} \left\{ a(\mu) + \ln \frac{M_D^2}{\mu^2} + \frac{m_\pi^2-M_D^2 + s}{2s} \ln \frac{m_\pi^2}{M_D^2} \right. \label{propdr}  \\
 \nonumber & & \left. + \frac{q}{\sqrt{s}} \left[ \ln(s-(M_D^2-m_\pi^2)+2 q\sqrt{s}) + \ln(s+(M_D^2-m_\pi^2)+2 q\sqrt{s}) \right. \right.  \\
 & & \left.\left. +\hspace*{-0.3cm}- \ln(s-(M_D^2-m_\pi^2)-2 q\sqrt{s}) -\ln(s+(M_D^2-m_\pi^2)-2 q\sqrt{s}) -2\pi i \right] \phantom{\frac{1}{1}} 
\right\} \nonumber \ ,
\end{eqnarray}
where the imaginary part of the logarithms above $D\pi$ threshold reads
\be 
\Im \ G_{l=0} (s) = - \frac{q}{8 \pi \sqrt{s}} \ ,
\ee
with $q$ the modulus of the pion's three-momentum in the CM frame.

Introducing the conventional two-body phase space \index{two-body phase space}
\be \label{eq:twobody}
\rho_{ \pi D} (s)= \sqrt{\left( 1+ \frac{ (m_{\pi} + M_D )^2}{s}\right) \left(1- \frac{(m_{\pi} -M_D)^2}{s} \right)} 
\ee
or, in terms of $q$,
\be 
\rho_{ \pi D} (s)= \frac{2q}{\sqrt{s}}\ ,
\ee
this imaginary part is
\be 
\Im \ G_{l=0} (s) = - \frac{\rho_{\pi D} (s)}{16 \pi} \ .
\ee

With these ingredients it is straightforward to show that, by construction, the complex $T_a$'s satisfy single--channel unitarity relations
\be 
\Im \ T^{l=0}_a (s) = - |T^{l=0}_a(s)|^2 \frac{\rho_{\pi D} (s)}{16 \pi^2} 
\ee
(providing a convenient numerical check of our computer programmes).
The amplitude can be parametrized in terms of the phase-shift
\be 
T^{I0}_a (s)= \frac{ \sin \delta_{I0}(s) e^{i \delta_{I0}(s)}}{\rho_{\pi D}(s)} \ , 
\ee
that is then extracted via
\be 
\tan \delta_{I0} (s) = \frac{\Im \ T^{I0} (s)}{\Re \ T^{I0} (s)} 
\ee
with $I=1/2, 3/2$. 

Finally, the isospin averaged amplitude for the LO-HQET decoupled single-channel problem becomes
\be 
|\ov{T}_a|^2 = \frac{1}{6} \left( 2 |T_a^{1/2,0}|^2 + 4 |T_a^{3/2,0}|^2 \right) \ . 
\ee
Heavy-quark spin symmetry dictates that, whether $D$ or $D^*$ in any spin state, the scattering cross section will be the same, and since an s-wave cannot flip the spin upon interaction, no further spin averaging is needed in leading order HQET. 
One can then use
\be 
\sum | \mathcal{M}_{\pi c} (s,t,\chi)|^2 = |\ov{T}_a|^2 
\ee
in Eq.~(\ref{eq:probdist}).

Going beyond LO in HQET we need to distinguish between $D\pi\to D\pi$ and $D^*\pi\to D^*\pi$ scattering. 
To implement it, we assume that a charm quark propagates as a linear combination of both states
\be
| c \ra = \alpha | D\ra + {\vec{\beta}}\cd | {\bf D}^*\ra \ .
\ee
The moduli of the complex numbers $\alpha$ and $\beta_i$ are determined by thermal Bose-Einstein distribution factors, since the mass difference between $D$ and $D^*$ slightly suppresses the latter. We then average over the relative (quasi-random) phases of $\alpha$ and ${\vec{\beta}}$ upon squaring to construct $\sum | \mathcal{M}_{\pi c} (s,t,\chi)|^2$.

The cross section is given by
\be
\sigma(s)_{\pi D} = \frac{1}{16\pi s} | {\mathcal{M}_{\pi D}} |^2
\ee
and
\be
\sigma(s)_{\pi D^*} = \frac{1}{16\pi s} | {\mathcal{M}_{\pi D^*}} |^2 \ .
\ee

\section{Results}

\subsection{Low-energy constants and cross sections \label{sec:hqs}}

In the philosophy of low-energy effective theories\index{low energy constants}, after all the symmetries have been used to constrain the Lagrangian density,
the remaining free constants have to be fit to experimental data. Our choices are widely discussed in \cite{Abreu20112737}. We quote here the values we have used.

1) The pion decay constant in the chiral limit $F_0$ can be approximated by its physical value,
$f_\pi=92$ MeV, the difference being of one higher order in the chiral expansion. 

2) The renormalization scale for the NLO ChPT constants is to be understood as 
$\mu=770$ MeV, and the scheme is such that the subtraction constant $a(\mu)=1.85$ is fixed as in \cite{Roca:2005nm}.

3) We adopt the value $g=1177 \pm 137$ MeV for the heavy-light pseudoscalar-vector coupling constant by reproducing the decay of $D^{*+}$. 
From our Lagrangian we obtain 
\be
\Gamma_{D^{*+}} = \frac{g^2 |p_{\pi}|^3}{12 \pi F^2 M^{2}_{D^*}} \ .
\ee

The value of $g$ deduced by matching the decay rate is the same as in \cite{Geng:2010vw} and consistent with the value of $g_{\pi}$ given in \cite{Laine:2011is}. 

4) Turning now to the NLO constants, we have stated that $\tilde{h}_i=h_i$ is a requirement of heavy quark symmetry tying the $D$ and $D^*$ amplitudes
 at LO in HQET.

5) Likewise we have set $h_0=h_2=h_4=0$ based purely on large-$N_c$ counting. These constants well deserve being revisited in future work, but we
 are content here with accepting a $1/N_c$ systematic error as customary in the current literature.

6) The mass differences between the $D$-mesons \cite{Lutz:2007sk,Geng:2010vw}, fixes $h_1 \approx -0.45$. 

7) Finally, we fix $h_3$ and $h_5$ through the $D\pi$ channel corresponding to the $T_a$ scattering amplitude, two pieces of known data (the $D_0$ mass and width) 
to which we can tie $h_3$ and $h_5$. We find that reasonable values are
$(h_3,h_5)=(7\pm 2,-0.5\pm 0.2 \textrm{ GeV}^{-2})$ with correlated errors, that is, an increased $h_3$ needs to be used with a more negative $h_5$.

We now present numeric computations of the unitarized and squared amplitudes in Eq.~(\ref{Unitarizedampl}), and of the cross section.

\begin{figure}[t]
\centering
\includegraphics[scale=0.27]{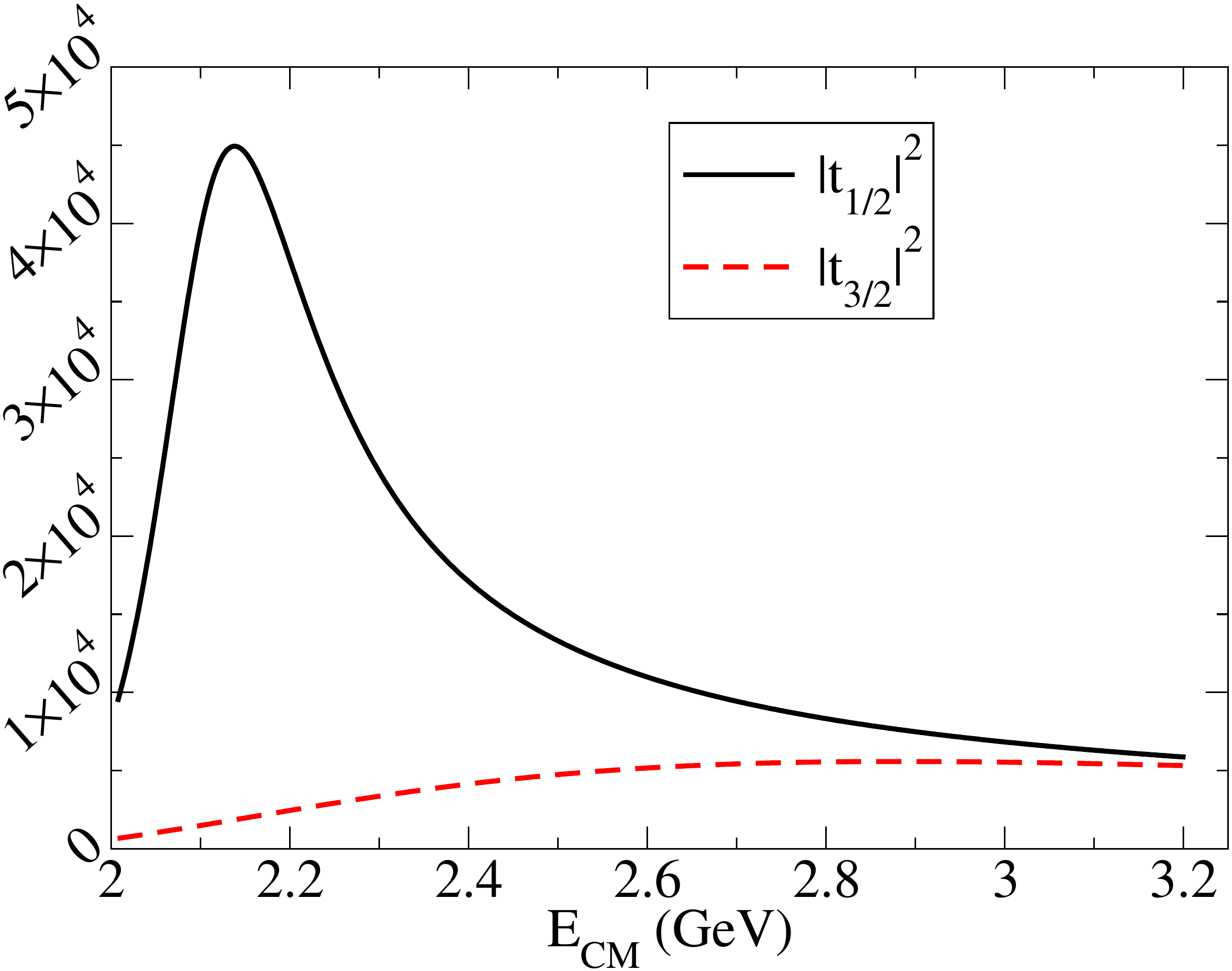}
\includegraphics[scale=0.27]{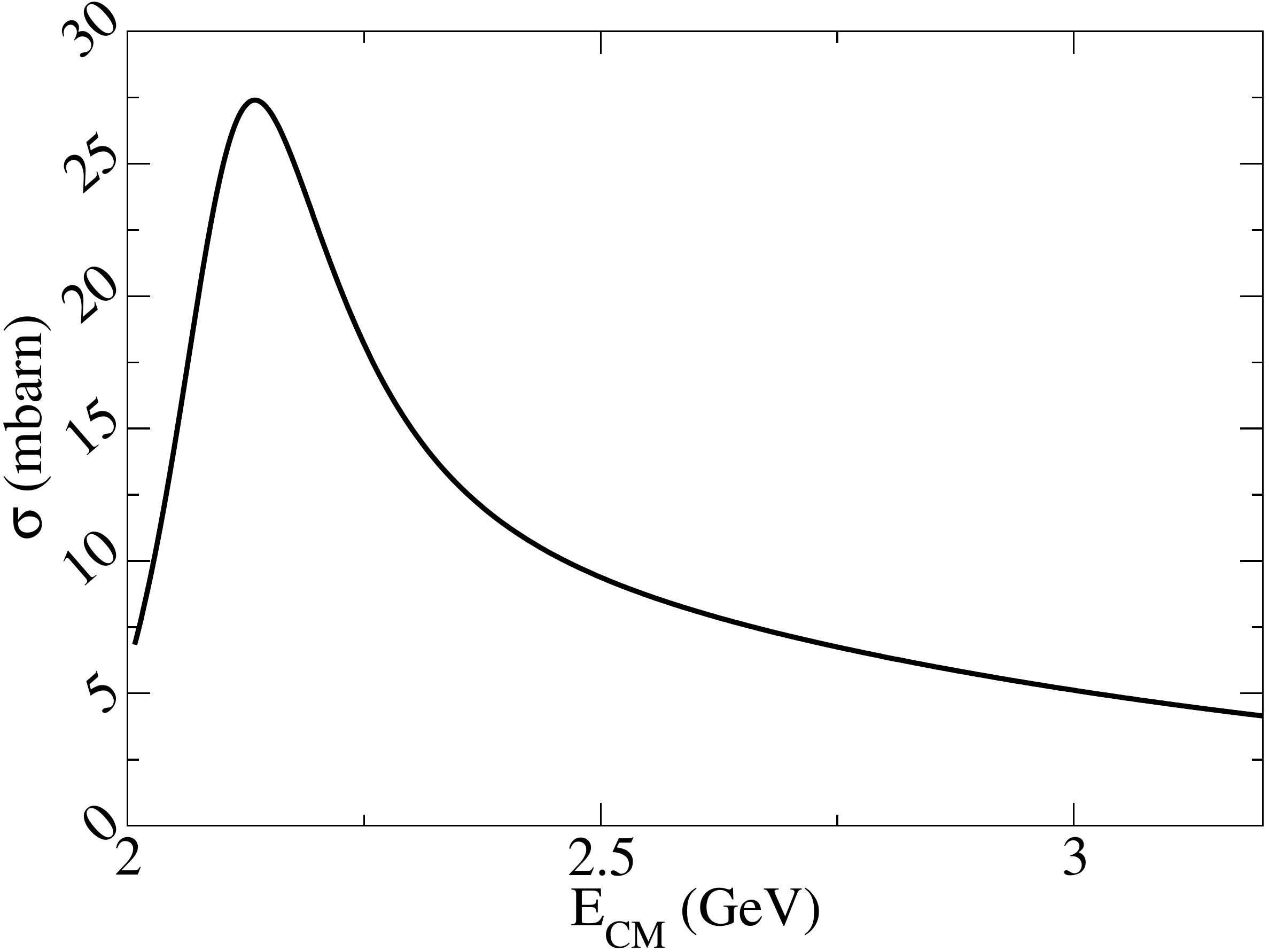}
\caption{\label{fig:sminusuonly} 
Left panel: square amplitudes for $D\pi$ $s$-wave elastic scattering employing only the $(s-u)$ term of the interaction potential $V_a$ (as in \cite{Gamermann:2007fi}).
Right panel: isospin averaged cross section associated to those amplitudes.
}
\end{figure}

In the first place, and to compare with the work of \cite{Gamermann:2007fi}, we keep only the $(s-u)$ term in the $D\pi$ elastic
amplitude $V_a$. The square amplitudes with isospin $I=1/2$ and $I=3/2$ and $l=0$ are depicted in Fig.~\ref{fig:sminusuonly}.
The figure shows how the exotic $I=3/2$ is non-resonant (this will also be the case for all the calculations presented below), which could have been guessed
because no $q\ov{q}$ state exists with such isospin, so there is no intrinsic strength at low energies in exotic waves.
The non-exotic $I=1/2$ channel presents a clear $s$-wave resonance, with approximate mass and width $M\simeq 2140$ MeV and $\Gamma\simeq 170$ MeV. These values
are somewhat too low if compared with the experimental $M_{D_0}=2360(40)$ MeV and $\Gamma_{D_0}=270(50)$ MeV taken from the Review of Particle Physics \cite{Nakamura:2010zzi}. 
We do not deem this a problem since there is room for the NLO terms containing the $h_i$ constants to modify the computation.

Next we add one by one the NLO constants $h_1$, $h_3$, $h_5$.
Because the $h_1$ term does not increase with momentum, but is multiplied by a small $m_\pi^2$ constant, it does not change the amplitudes appreciably.
We include it with a value $h_1 \approx -0.45$ as commented before but do not discuss it any further.

\begin{figure}[t]
\centering
\includegraphics[scale=0.32]{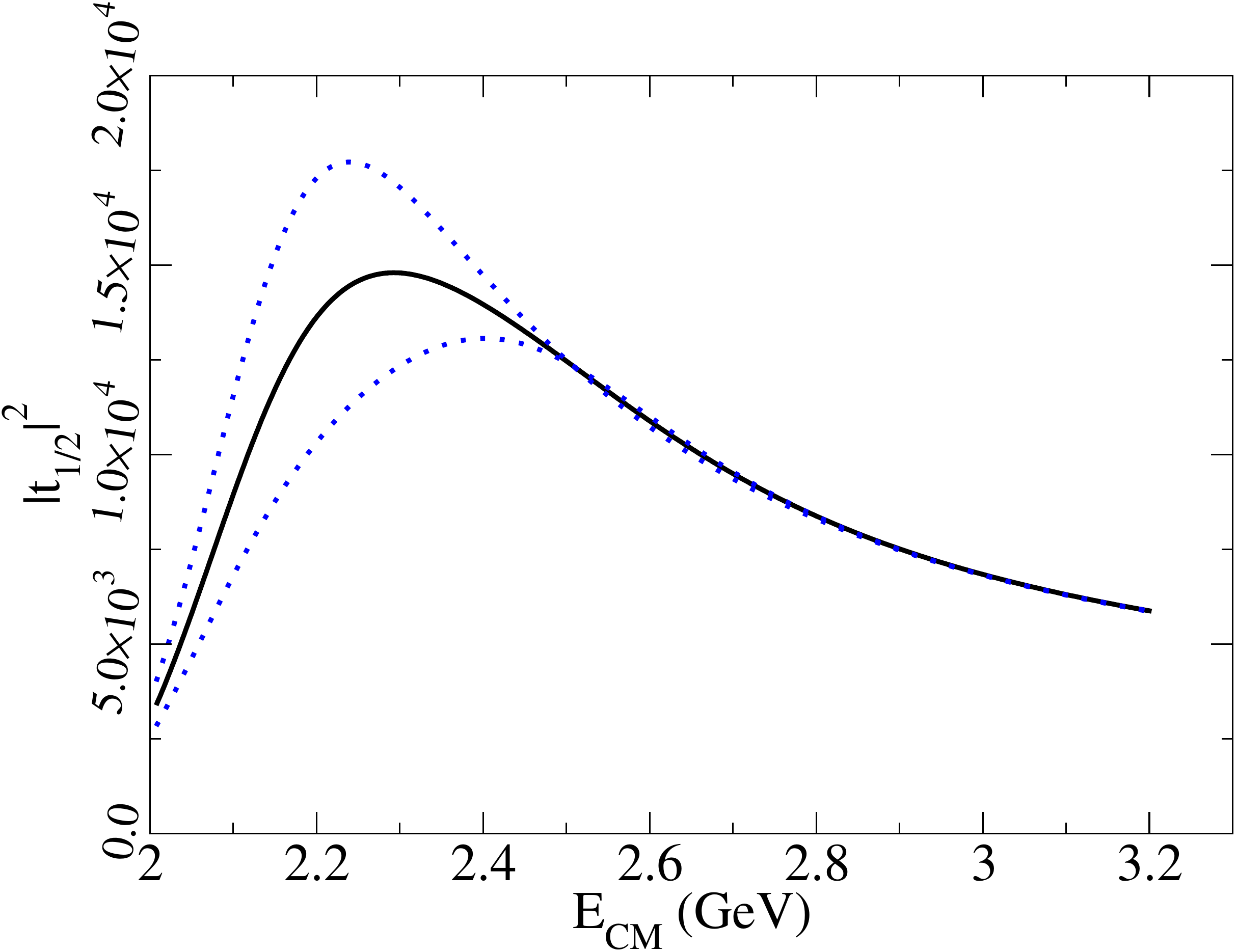}
\caption{\label{fig:withh5} 
Squared isospin $1/2$ amplitude for $D\pi$ scattering with $(h_3,h_5)=(7,-0.5$ GeV$^{-2})$ (solid line). The dotted lines give the limits of the error band.
}
\end{figure}

In \cite{Abreu20112737} we study the sensitivity to $h_3$ and $h_5$. For small, positive values of $h_3$ the $D_0$ peak moves to larger masses
and it becomes broader. Adding the $h_5$ term, we observe that its presence (if the sign is chosen negative as in \cite{Guo:2009ct}, 
for example $h_5=-0.25$ GeV$^{-2}$) narrows the resonance shifting it to slightly lower masses.Therefore, a strategy to improve agreement with 
the experimental $D_0$ data is to combine a positive $h_3$ with a negative $h_5$ to increase the resonance mass without distorting the line-shape unacceptably.
Our best computation is then shown in Fig.~\ref{fig:withh5}.
 
The maximum of the squared amplitude, employing $(h_3,h_5)=(7,-0.5$ GeV$^{-2})$ as central value, gives a reasonable $M_{D_0}=2300$ MeV, just slightly below the experimental
value, and a width just slightly above $\Gamma=350$ MeV. The two parameters are very correlated, so that varying one significantly requires varying the other
 simultaneously to maintain reasonable agreement with the experimental resonance. Shown in the figure are two more lines with the error band $\Delta h_3=\pm 2$ and $\Delta h_5 = \pm 0.2$ GeV$^{-2}$.
It is this squared amplitude, leading order in Heavy Quark Effective Theory, that we adopt in our Fokker-Planck equation for the transport coefficients.

\begin{figure}[t]
\centering
\includegraphics[scale=0.32]{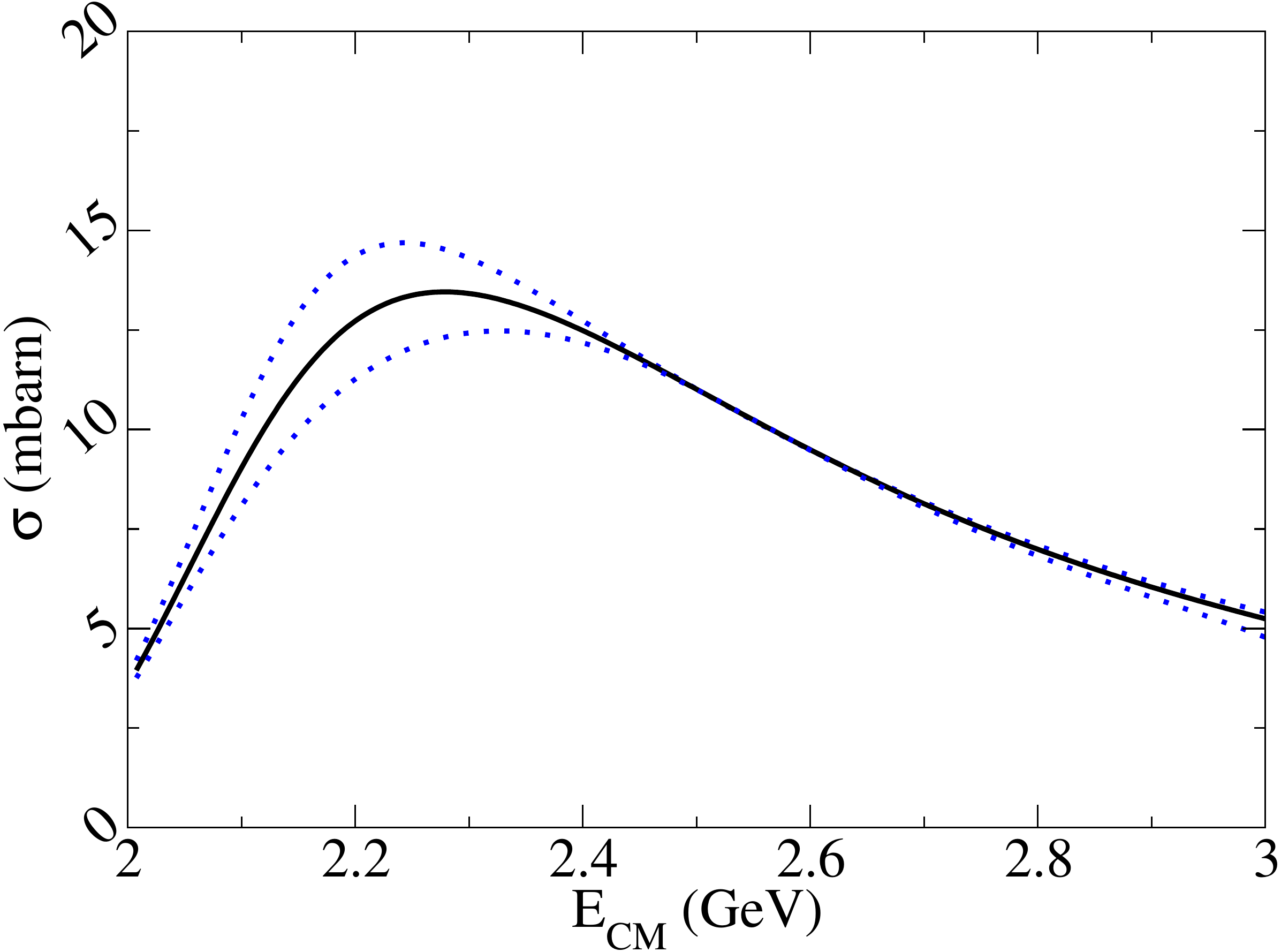}
\caption{\label{fig:withh5cross} 
Cross section for $D\pi$ elastic scattering with $(h_3,h_5)=(7,-0.5$ GeV$^{-2})$ (solid line). The dotted lines give the limits of the error band.
}
\end{figure}

Although the diffusion and drag coefficients require the $|\mathcal M |^2$ square amplitude, it is convenient for the discussion to also plot the resulting cross section,
which we do in Fig.~\ref{fig:withh5cross}.

The maximum of the cross section is about $13.5\pm 1$ mbarn, and for the entire range of center of mass energies $\sqrt{s}\in(2-3)$ GeV we find 
$\sigma\ge 5$ mbarn. In fact, for a large window between 2.1 and 2.5 GeV we have $\sigma \ge 10$ mbarn, which is slightly larger but in reasonable
agreement with the guess by the authors of~\cite{He:2011yi}, that assume $7-10$ mbarn, or by Svetitsky and Uziel~\cite{Svetitsky:1996nj} of 9 mbarn based on quark constituent counting.

\begin{figure}[t]
\centering
\includegraphics[scale=0.32]{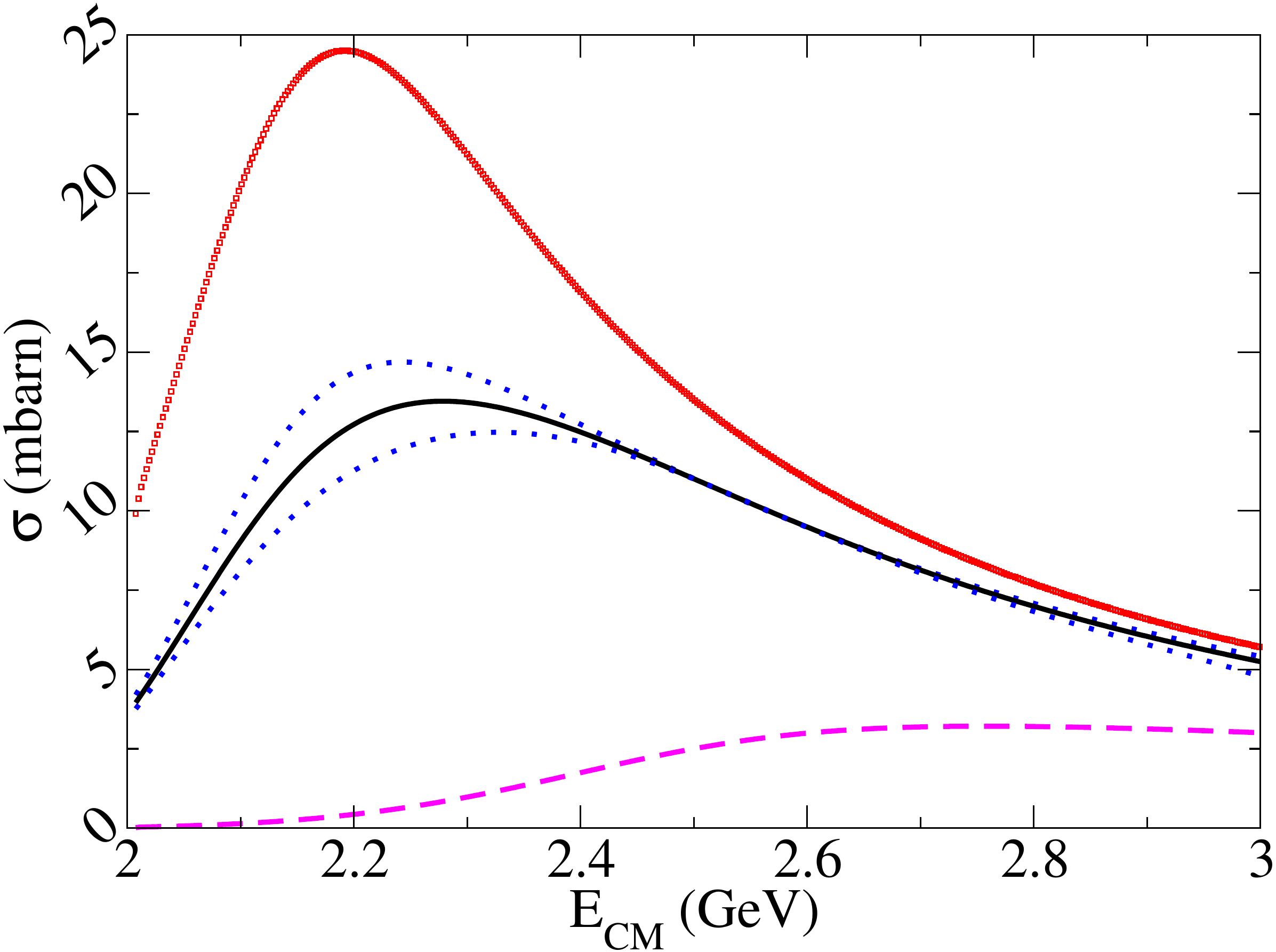}
\caption{\label{fig:Born1} 
Effect of including the Born terms associated with the $D^*$. 
  The bottom line (purple) is the cross section associated to the Born term alone, as in the model of~\cite{Ghosh:2011bw}. The top line (red squares) is
the resulting cross section combining the Born term with the contact terms, without modifying the $h_i$ constants from Fig.~\ref{fig:withh5cross}, and then unitarizing. 
}
\end{figure}

Next, we proceed to the next-to-leading order in Heavy Quark Effective Theory. We only consider for now the Born $s$ and $t$-channel exchange terms
 due to $D^*$ exchange between the $D\pi$ pair. The effect of adding these terms is akin to making $h_5$ more negative, that is, a narrowing of the $D_0$ resonance,
as shown in Fig.~\ref{fig:Born1}. 

\begin{figure}[t]
\centering
\includegraphics[scale=0.32]{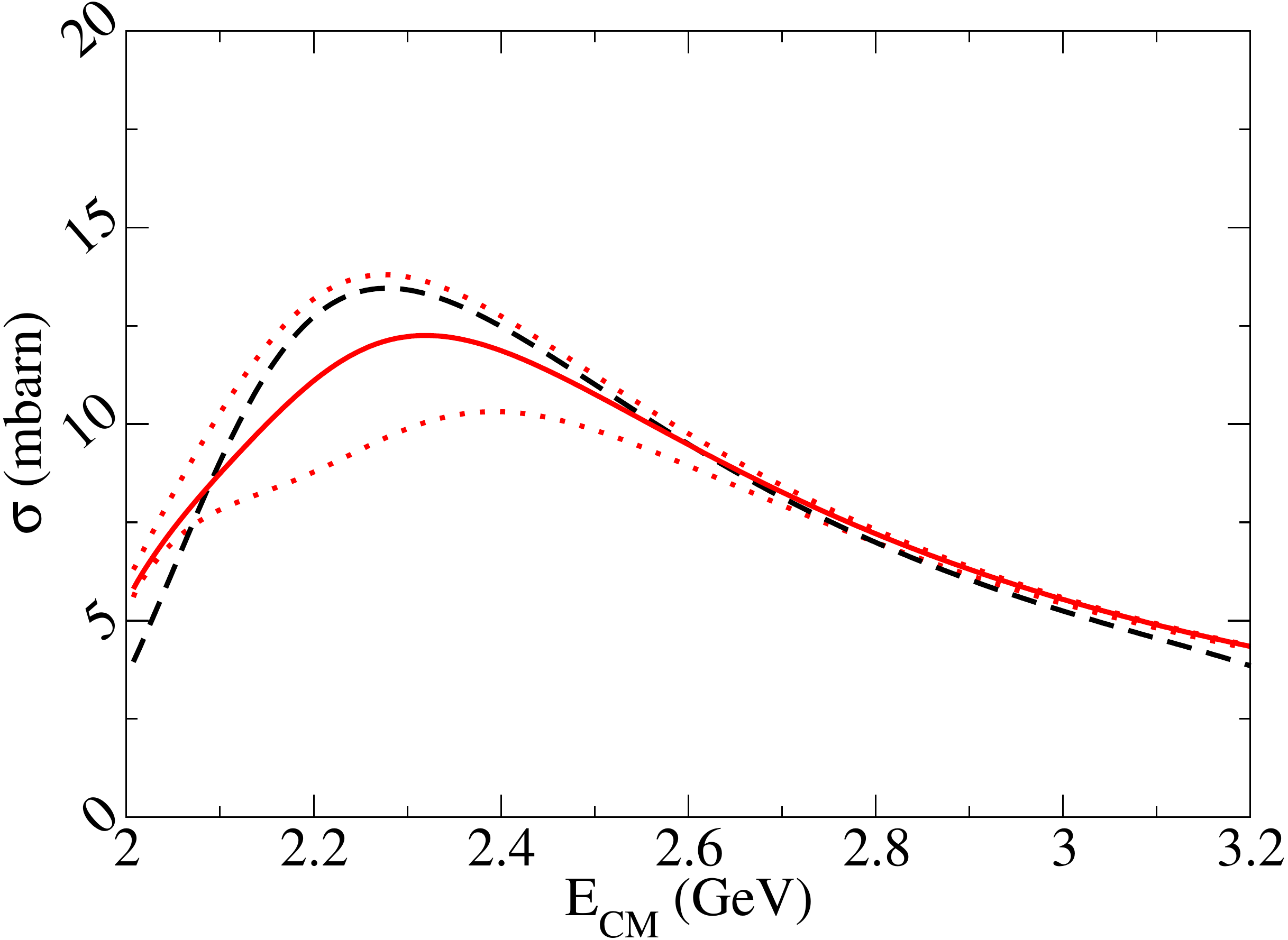}
\caption{\label{fig:Born2} 
Effect of including the Born terms associated with the $D^*$, but leaving the $h_i$ coefficients free. The red, solid line is the central value with $h_3=8$, $h_5=0.35$
 GeV$^{-2}$. The black, dashed line coincides with the cross section in Fig.~\ref{fig:withh5cross} without the Born terms.
}
\end{figure}

However, a renormalization of the $h_i$ constants effectively brings back the pole position in better agreement with experimental data, as seen in Fig.~\ref{fig:Born2}. 

\begin{figure}[t]
\centering
\includegraphics[scale=0.32]{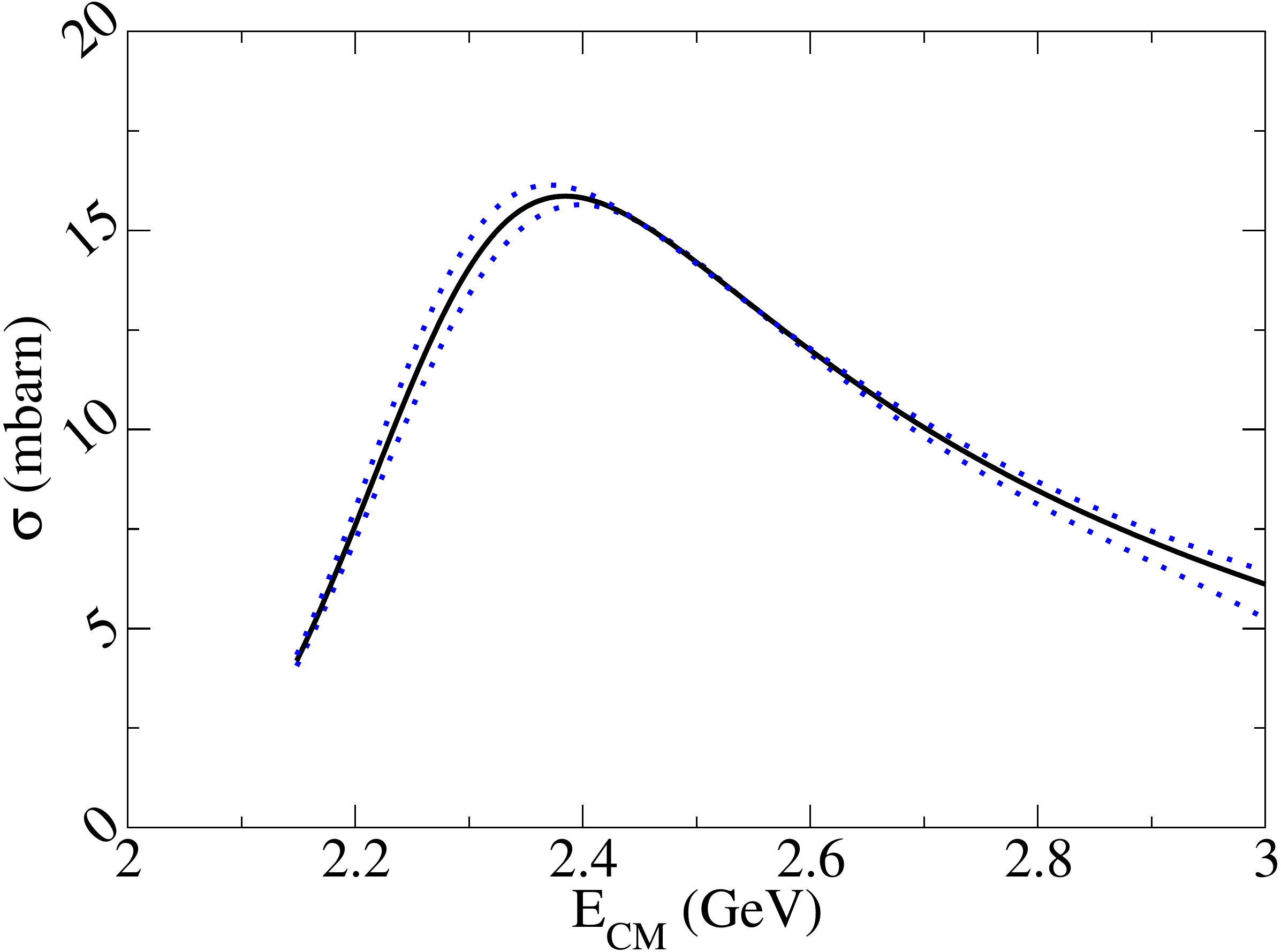}
\caption{\label{fig:Dstarscattering} 
Elastic cross section for $D^*\pi$ scattering computed replacing $M_D$ by
$M_D^*$ in Fig.~\ref{fig:withh5cross}. The resonance should now be interpreted as the broad $D_1(2427)$.
}
\end{figure}

Shown in the figure are lines with $(h_3,h_5)=(7.5\pm 2.5, 0.4\pm 0.3$ GeV$^{-2})$, together with the result of Fig.~\ref{fig:withh5cross} without including the Born terms. As can
 be seen, the effect of the $D^*$ exchanges can be largely absorbed in the $h_i$ counterterms (for fixed $m_c$ mass of course, since they scale differently) so we will ignore the 
Born terms in this computation. However a certain uncertainty should be understood, of order 30\% in the cross section, that could be larger than our estimate in the region of the $D_0$ resonance.

Finally, we return to the computation in Fig.~\ref{fig:withh5cross}, but substitute $M_D$ by $M_D^*$ (an NLO effect in HQET) as only modification to obtain $V_d$ instead of $V_a$. We
 interpret the resulting cross section as that corresponding to $D^*\pi$ scattering, and plot the result in Fig.~\ref{fig:Dstarscattering}.

The cross section including both $1/2$ and $3/2$ isospin channels is clearly resonant, with the $D_1$ well visible. As was the case for the $D_0$,
the mass is slightly below the data.
The cross section peak is about 15 mbarn.

Thus we have performed an exhaustive study of the LO-HQET interaction and now proceed to compute transport coefficients equipped with the interaction leading to
 Figs.~\ref{fig:withh5cross} and~\ref{fig:Dstarscattering}.
The conclusion of this section is that reproducing the correct parameters of the $D_0,D_1$ resonances (masses and widths) essentially fixes the cross section in the region of interest, because
of unitarity not leaving much room for model dependence.

\subsection{Diffusion and drag coefficients \index{charm diffusion coefficients} \index{charm drag coefficient}}

\begin{figure}[t]
\centering
\includegraphics[width=7.7cm]{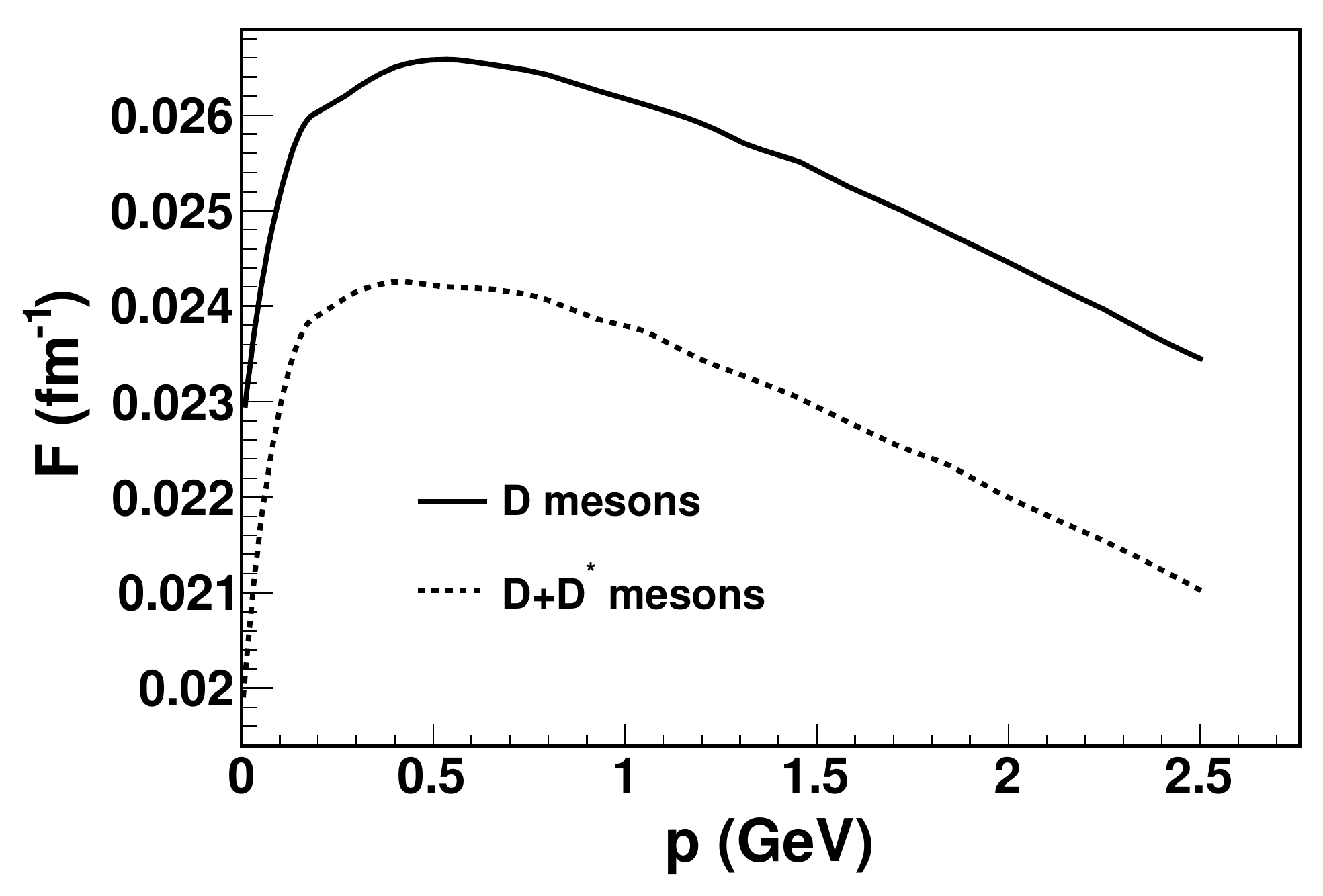}
\includegraphics[width=7.7cm]{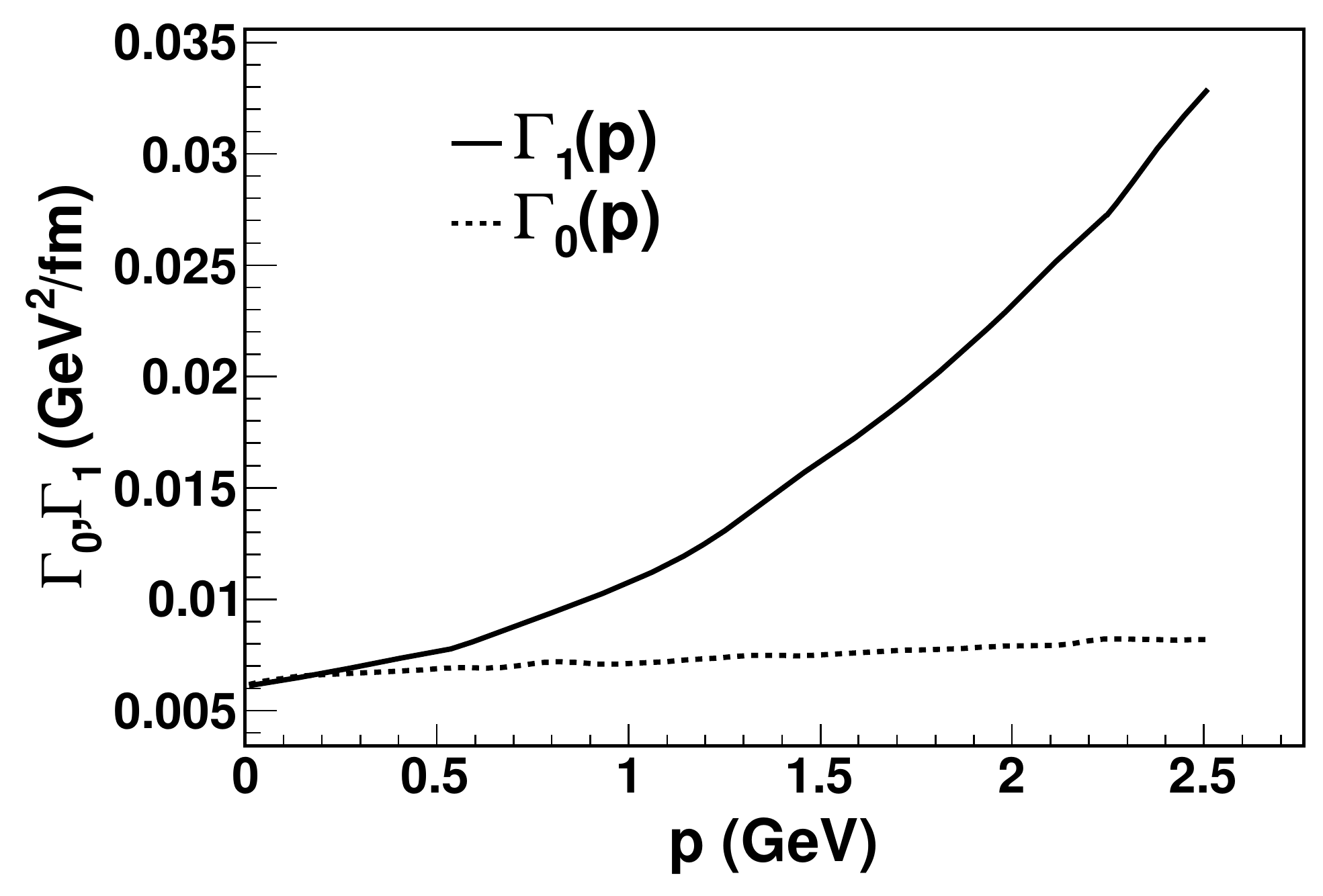}
\includegraphics[width=7.7cm]{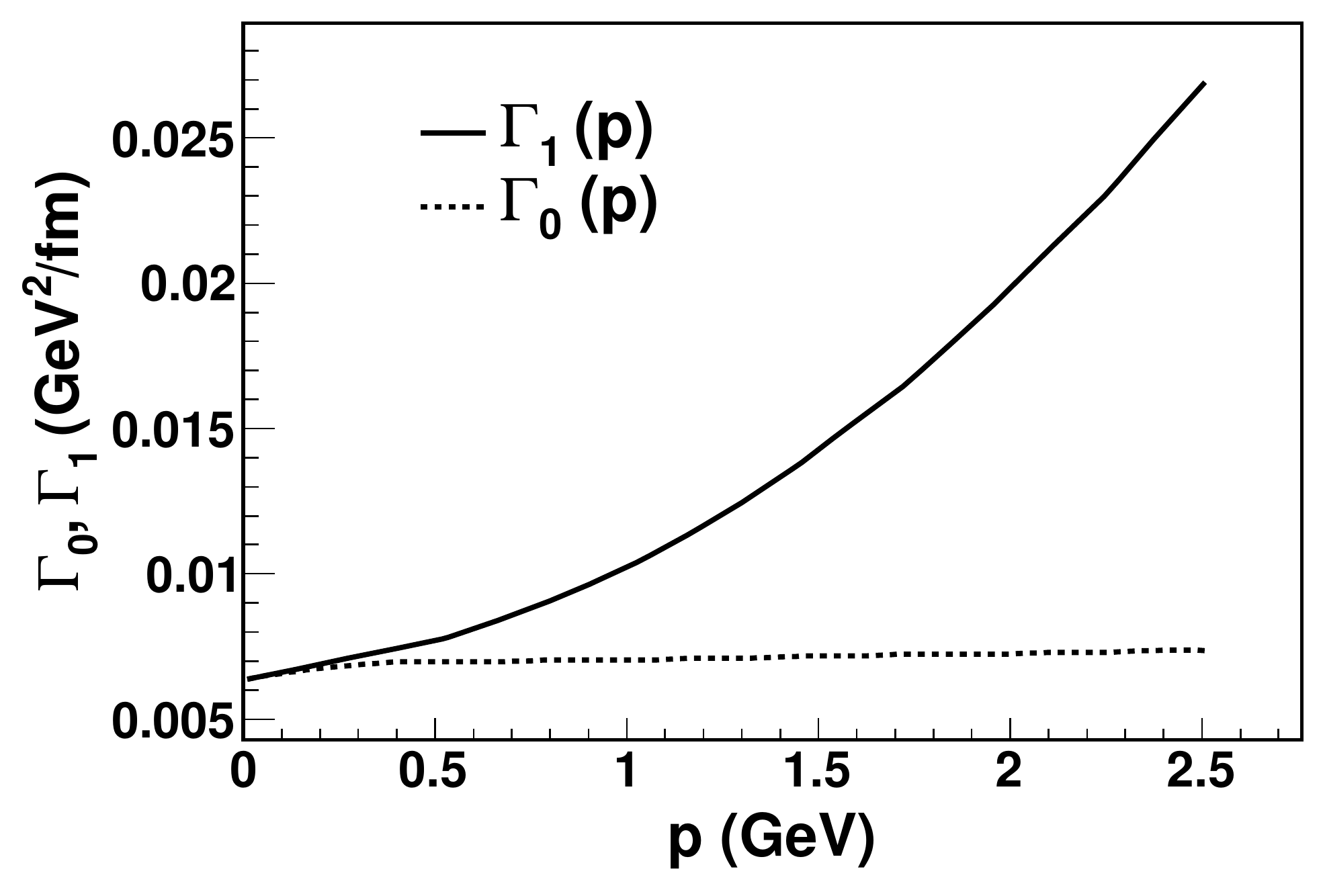}
\caption{ \label{fig:threcoeffs} 
We show the three coefficients in the Fokker-Planck equation as a function of charm-quark momentum, at a reference temperature of 150 MeV in the pion gas.
The low-energy constants in the $D \pi \rightarrow D \pi$ amplitude are fixed to $h_1 = - 0.45$, $g=1177$ MeV, and $h_3$ and $h_5$ fit to describe the
mass and width of the $D_0$ resonance. Top: $F$ including and not including the possible propagation of the $c$ quark as a $D^*$ meson. Middle: $\Gamma_0$ and $\Gamma_1$
including $D$-like propagation alone. Bottom: $\Gamma_0$ and $\Gamma_1$ including also propagation as a $D^*$ meson. }
\end{figure}

We now proceed to the calculation, with the square amplitude so numerically computed, the $F$, $\Gamma_0$ and $\Gamma_1$ transport coefficients in Eq.~\ref{Transportintegrals}.
The three pannels of Fig.~\ref{fig:threcoeffs} show them as function of squared momentum $p^2$ for fixed temperature $T=150$ MeV. One should
not trust these results above charm momenta of order $p=1.5$ GeV, but we spell them out for completeness.

In the top panel of this figure we show the drag coefficient $F(p)$ in fm$^{-1}$, which exhibits a modest momentum variation of about $10-20 \%$ within the range of $p\in(0,2.5)$ GeV.
From this coefficient one can extract the relaxation length for a charm quark propagating in the pion medium that turns out to be around $40$ fm at $p=$1 GeV.

Quite strikingly, one can see in the figure that $\Gamma_0$ has a very mild momentum dependence, its value can very well be approximated by a constant for the entire momentum range. $\Gamma_1$ is
 seen to grow with momentum, increasing the difference $\Gamma_1-\Gamma_0$, and thus favoring diffusion at higher typical momenta.

\begin{figure}[t]
\centering
\includegraphics[width=8.5cm]{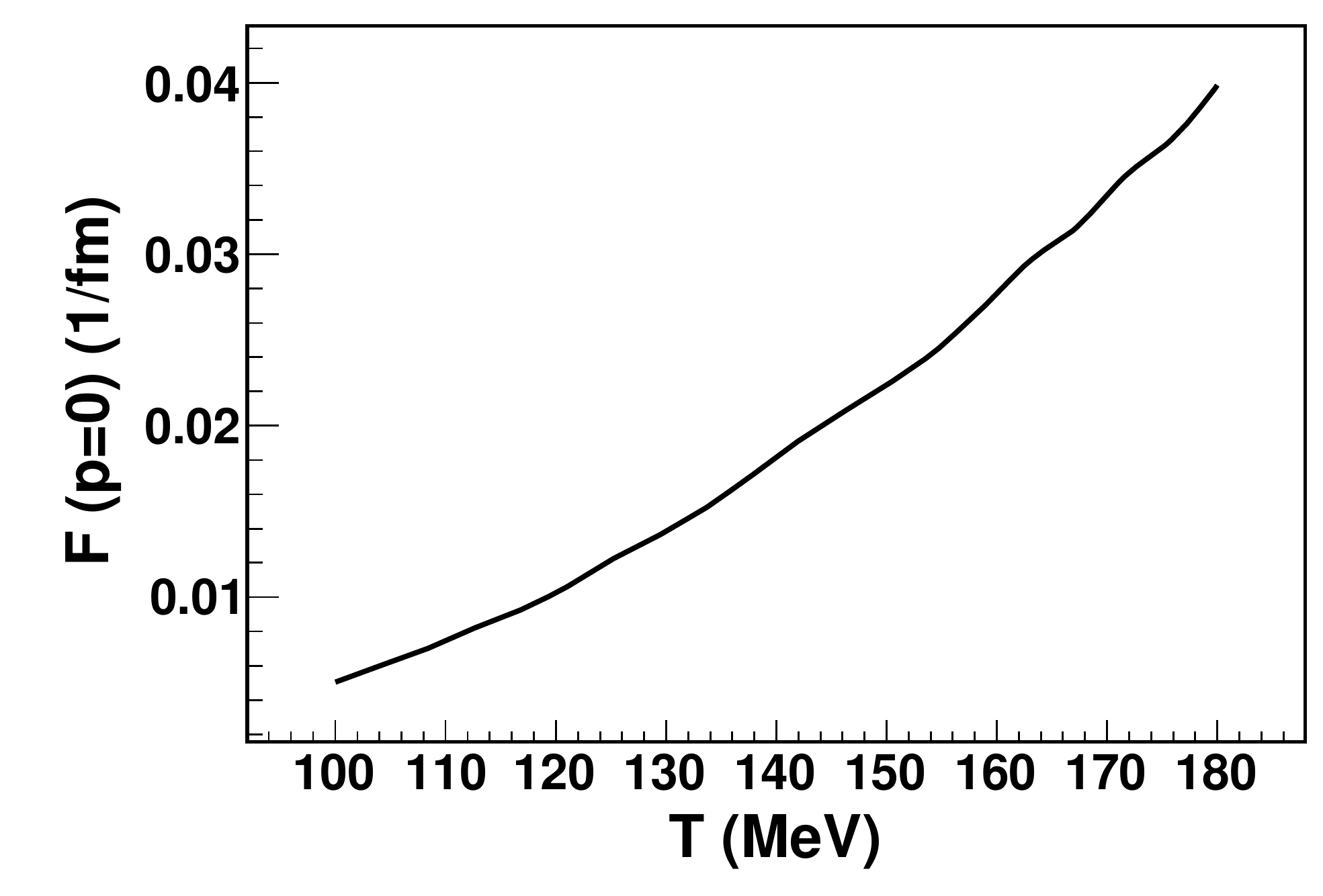}
\caption{\label{fig:dragfixedP0} 
Momentum-space drag coefficient as function of temperature for a stopped charm quark in the hadron gas. We obtain the coefficient by employing the Einstein relation taking the limit of $p\to 0$.}
\end{figure}

\begin{figure}[t]
\centering
\includegraphics[width=8.5cm]{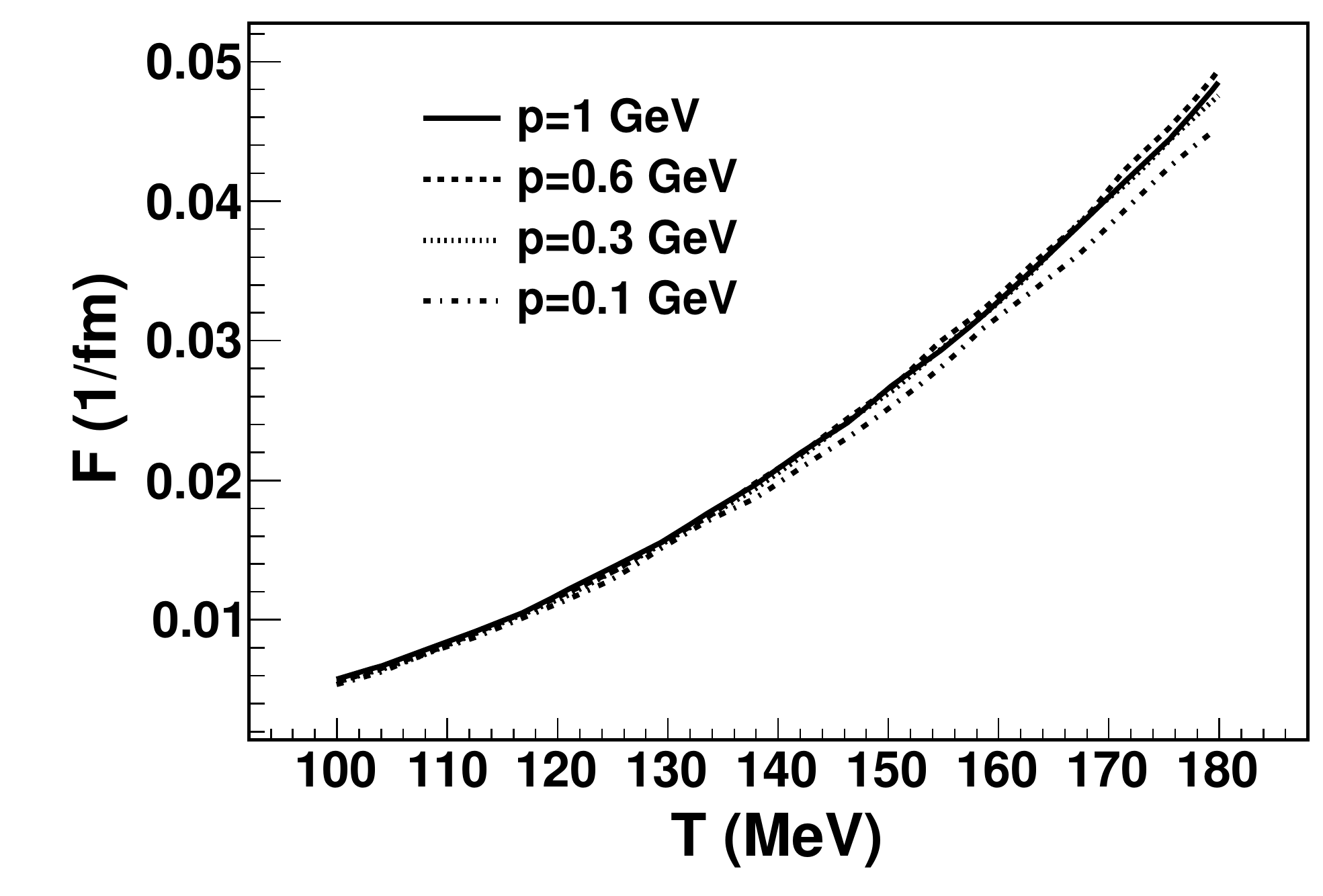}
\includegraphics[width=8.5cm]{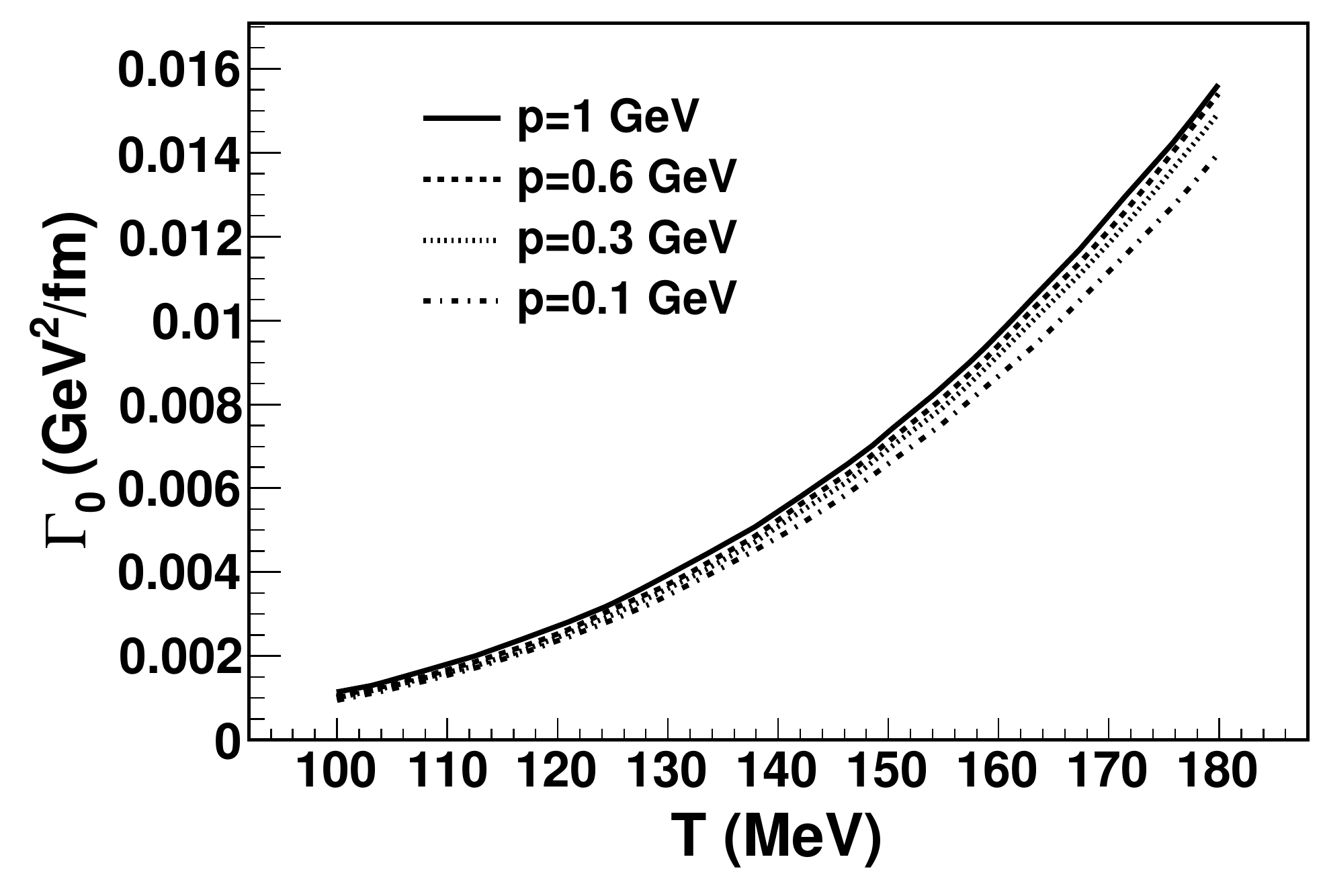}
\includegraphics[width=8.5cm]{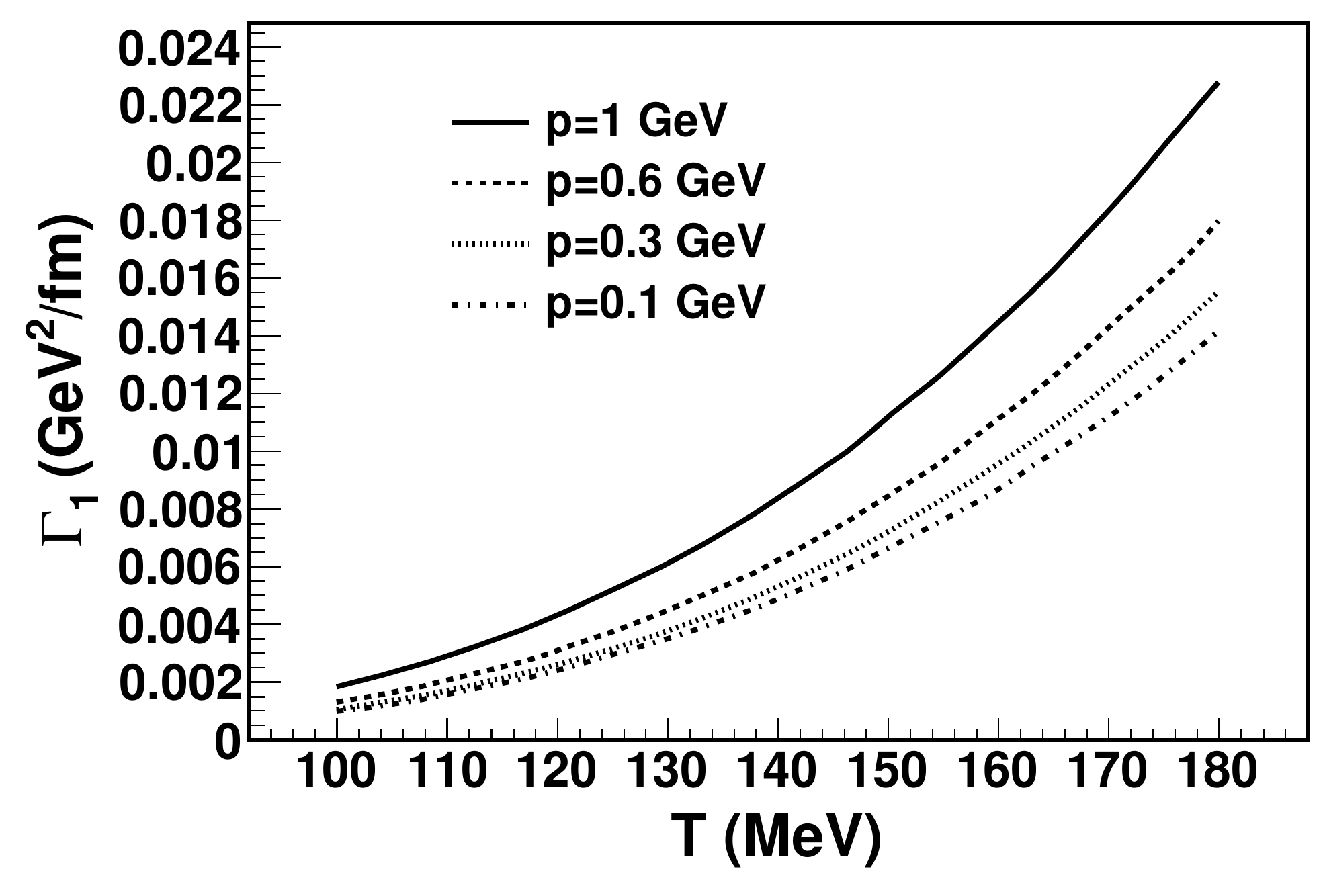}
\caption{\label{fig:dragfixedP} 
Momentum-space drag and diffusion coefficients as function of temperature for a slow charm quark with momentum $p=1$ GeV, 0.6 GeV, 0.3 GeV and 0.1 GeV.
Note that the intensity of the drag force is roughly proportional to the temperature. 
}
\end{figure}

In Figs.~\ref{fig:dragfixedP0} and \ref{fig:dragfixedP} we show the dependence with the temperature of the drag coefficient at fixed momentum. Since the direct computation of $F(p^2\to 0)$ is 
rather unstable, for the plot in Fig.~\ref{fig:dragfixedP0}, $F$ is computed from $\Gamma$ by employing the Einstein relation\index{Einstein relation}:
\be F(p^2 \rightarrow 0)= \frac{\Gamma(p^2 \rightarrow 0)}{MT} \ . \ee

The drag coefficient is seen to increase by a factor of about 4 in the range from 100 to 150 MeV, so that most of the drag in a heavy-ion collision is expected in the hotter stages, with the charm 
quarks freezing out progressively until they freely stream outwards till they decay.

We compare with other authors, choosing a reference temperature of 100 MeV where all existing works make a statement, and show the drag coefficient for each recent work in Table~\ref{tab:drag}.

\begin{table}[t]
\begin{center}
\begin{tabular}{|cc|} \hline
Authors             & $F(\textrm{fm}^{-1}$) \\ \hline \hline
Laine, \cite{Laine:2011is}              & $0.05\times 10^{-3}$ \\ 
He, Fries, Rapp, \cite{He:2011yi}    & $5\times 10^{-3}$ \\
Ghosh {\it{et al.}}, \cite{Ghosh:2011bw} & 0.11 \\
Our estimate, \cite{Abreu20112737}          & ${\mathbf{3.5\times 10^{-3}}}$ \\
\hline
\end{tabular}
\end{center}
\caption{Value of the drag coefficient at $p\rightarrow 0$ and $T=100$ MeV. \label{tab:drag}}
\end{table}

It can be seen that the phenomenological model of \cite{He:2011yi} is of the same order of magnitude of our result, with \cite{Ghosh:2011bw} quoting an extremely large value in their Fig.~2, and
Laine a smaller value by one order of magnitude. We believe that we have a larger control of the charm-pion scattering amplitudes at moderate temperatures, but the reader would be cautious to employ
a factor 2 as error band to our result.

\begin{figure}[t]
\centering
\includegraphics[width=8.5cm]{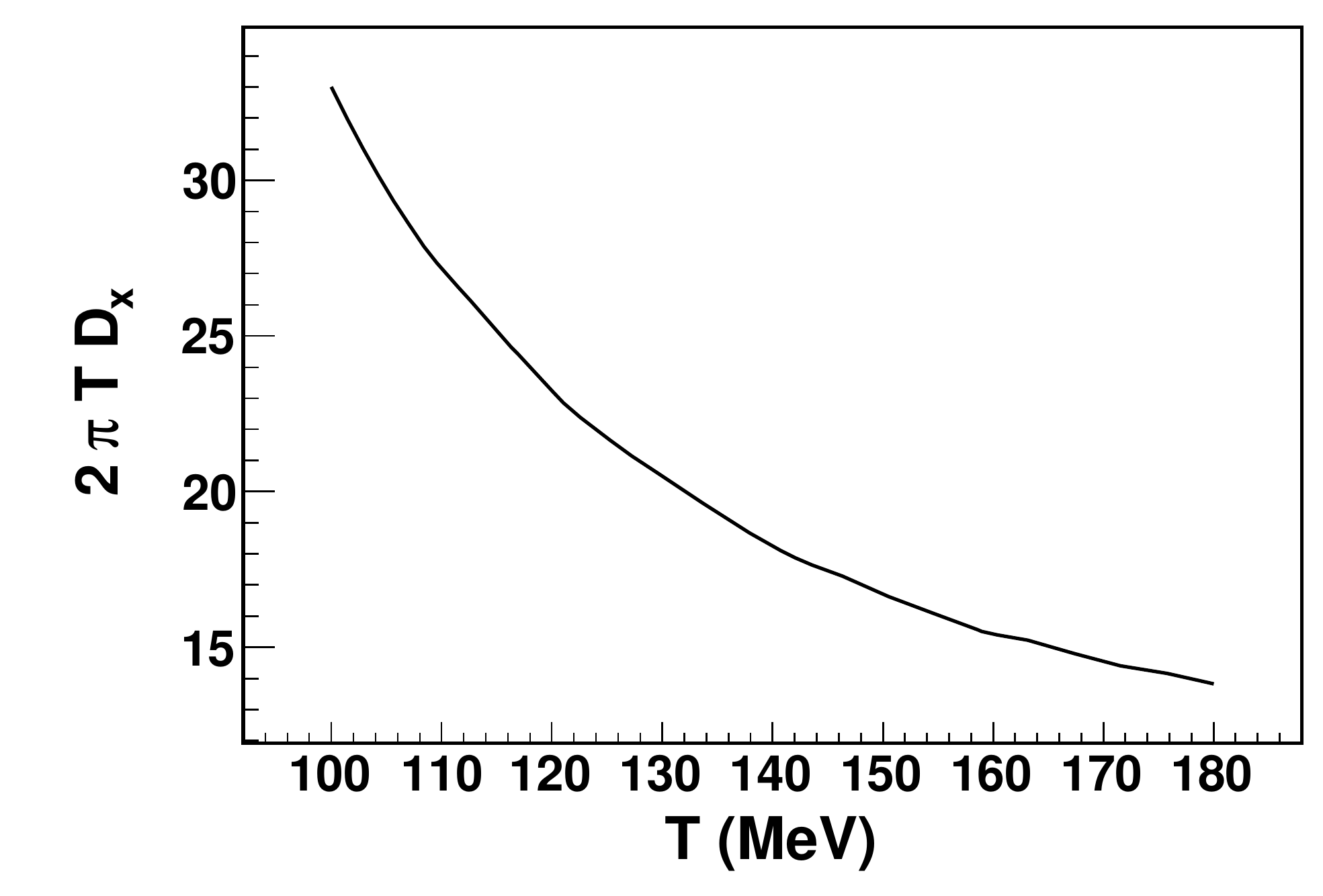}
\caption{\label{fig:spacediff} 
Spatial diffusion coefficient as a function of temperature.
}
\end{figure}

The spatial diffusion coefficient is then plotted in Fig.~\ref{fig:spacediff} as function of temperature.

\begin{figure}[t]
\centering
\includegraphics[width=8.5cm]{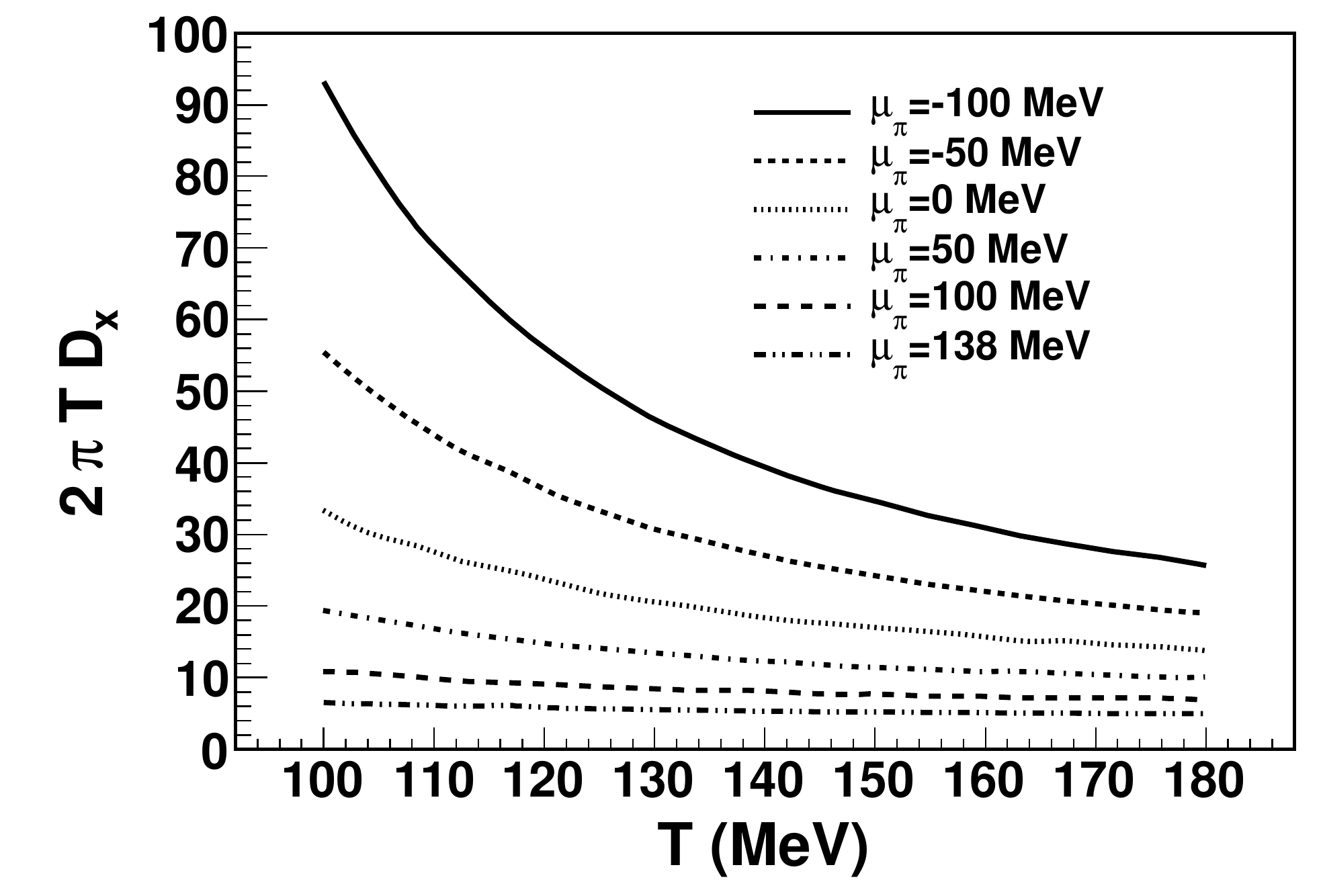}
\caption{\label{fig:spacediff_mu} 
Same as in Fig.~\ref{fig:spacediff} but as a function of the chemical potential 
}
\end{figure}

At low temperatures it correctly takes the nonrelativistic limit 
\be
D_x = \frac{3T^{3/2}}{\sigma P\sqrt{m}}
\ee
with $m$ the particle mass, $\sigma$ the cross section, and $P$ the pion gas pressure, that is temperature dependent.
We also note that, during the lifetime of the pion gas after the crossover from the quark-gluon plasma phase, the interactions between pions are almost entirely elastic, so that pion number is
 effectively conserved and one should introduce a pion chemical potential, not included in the very recent works by other groups. 
Introducing this approximate pion chemical potential $\mu$,
\be
P\propto m_\pi^{3/2} T^{5/2} e^{\frac{\mu-m_\pi}{T}}
\ee 
makes the product $TD_x$ diverge at low temperature and vanishing chemical potential
(which just means that gas particles are too cold and slow to stop the charm quark from diffusively moving inside the pion gas).
However, at chemical equilibrium with $\mu\to m_\pi$ (that is not expected in the hadron phase of a heavy-ion collision, but is relevant to make contact with the nonrelativistic limit), the
 exponential becomes unity and $TD_x$ becomes a constant at low temperature.
We further show the effect of this pion chemical potential in Fig.~\ref{fig:spacediff_mu}.

We find the effect sizeable. At a reference temperature of 120 MeV, the ratio between $D_x$ at $\mu_\pi=0$ and $\mu_\pi=138$ MeV is a factor of about 5.

\begin{figure}[t]
\centering
\includegraphics[width=8.5cm]{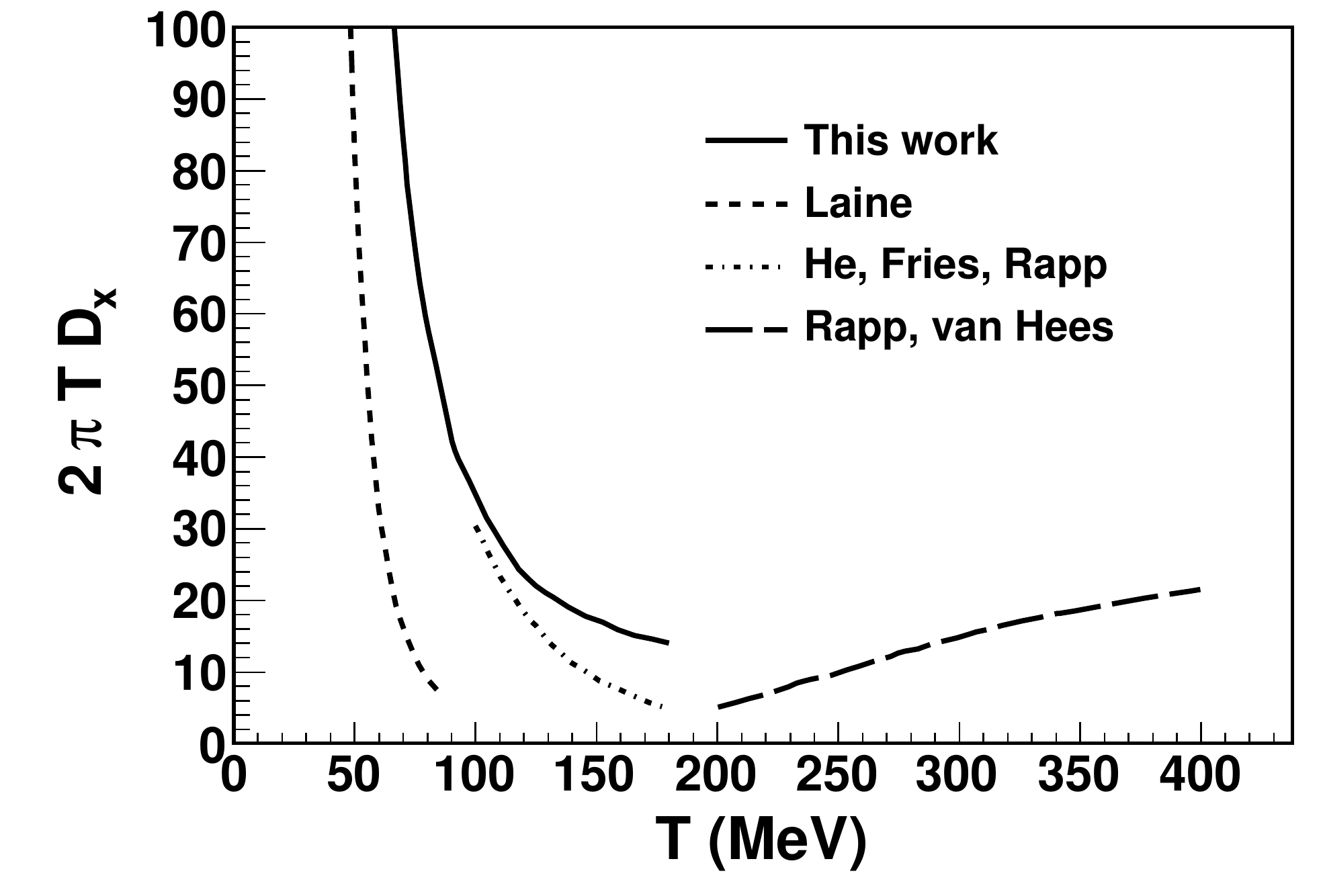}
\caption{\label{fig:charmcomparison}
Comparison of our result for the diffusion coefficient at zero chemical potential with other estimates. The different results are explained in the text.
}
\end{figure}

In Fig.~\ref{fig:charmcomparison} we plot our result (solid line) at zero chemical potential together other results of the same coefficient. The result labeled as ``Laine'' refers to \cite{Laine:2011is}
where Heavy Meson Chiral Perturbation Theory without unitarization is used. The ``He,Fries,Rapp'' curve corresponds to \cite{He:2011yi} where they use empirical elastic scattering amplitudes
and finally the calculation of the charm diffusion in the quark-gluon plasma from \cite{Rapp:2008qc}.

\begin{figure}[t]
\centering
\includegraphics[width=8.5cm]{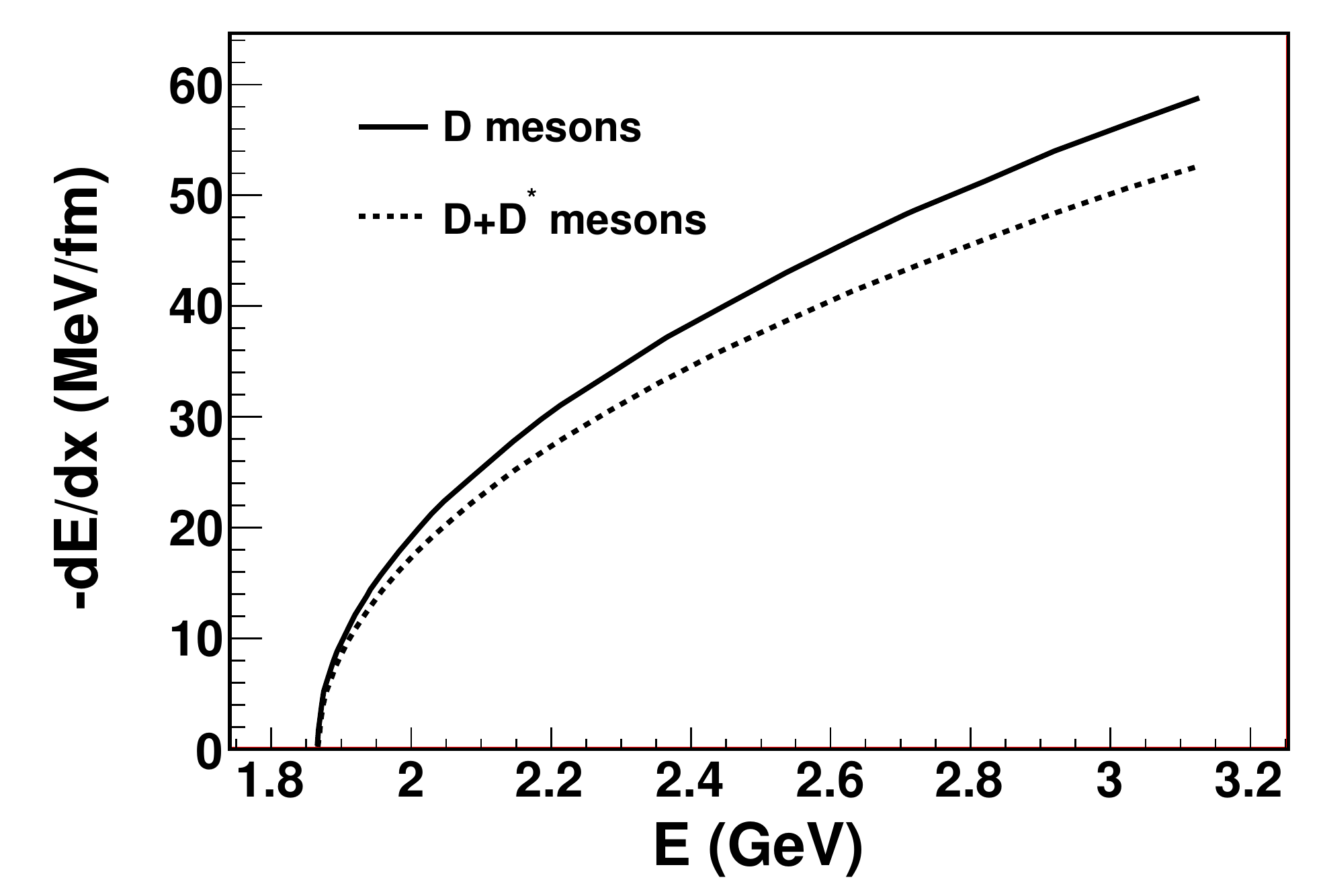}
\caption{\label{fig:energyloss} 
Loss of energy of a charmed meson as function of the energy in a pion gas at a fixed temperature of 150 MeV, assuming it can travel as a $D$ or a $D^*$ meson during the few Fermi of the gas's lifetime.
}
\end{figure}

\begin{figure}[t]
\centering
\includegraphics[width=8.5cm]{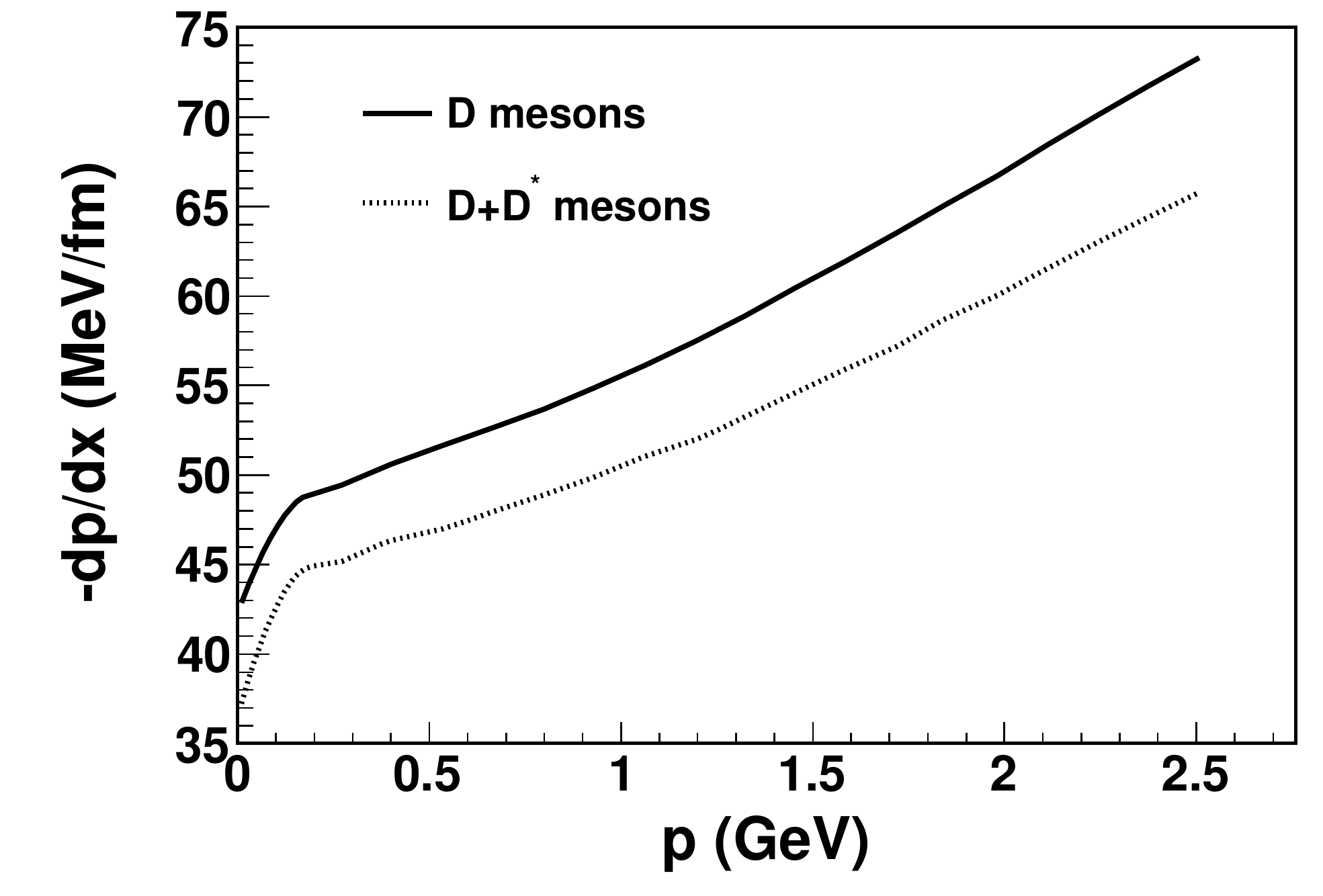}
\caption{\label{fig:momentumloss} 
Loss of momentum per unit length as function of momentum of a charmed meson in a pion gas, same as in Fig.~\ref{fig:energyloss}.
}
\end{figure}

To assist in the physical interpretation of our results, we have plotted in Figs.~\ref{fig:energyloss} and \ref{fig:momentumloss} the loss of energy and momentum per unit length, derived from our
results for the drag coefficient $F$, for various momenta $p^2$. 

From Fig.~\ref{fig:momentumloss} one can estimate that a reference charm quark in a $D$ or $D^*$ meson with momentum 1 GeV measured in the rest frame of the pion fluid surrounding it,
 will deposit about 50 MeV per fm travelled in the fluid. Thus, if the pion gas is in existence for, say, 4 fm, the $D$ meson measured in the final state with a momentum
of 800 MeV will have been emitted from the quark-gluon plasma phase with a GeV. This result is similar to the $20\%$ effect recently quoted by \cite{He:2011yi} and means
 that, while the $D$ and $D^*$ mesons can be used as probes of the quark-gluon plasma, their distributions should be shifted up in momentum (or alternatively both the quark-gluon plasma and
 hadron phases have to be treated in hydrodynamical simulations).

\begin{figure}[t]
\centering
\includegraphics[width=8.5cm]{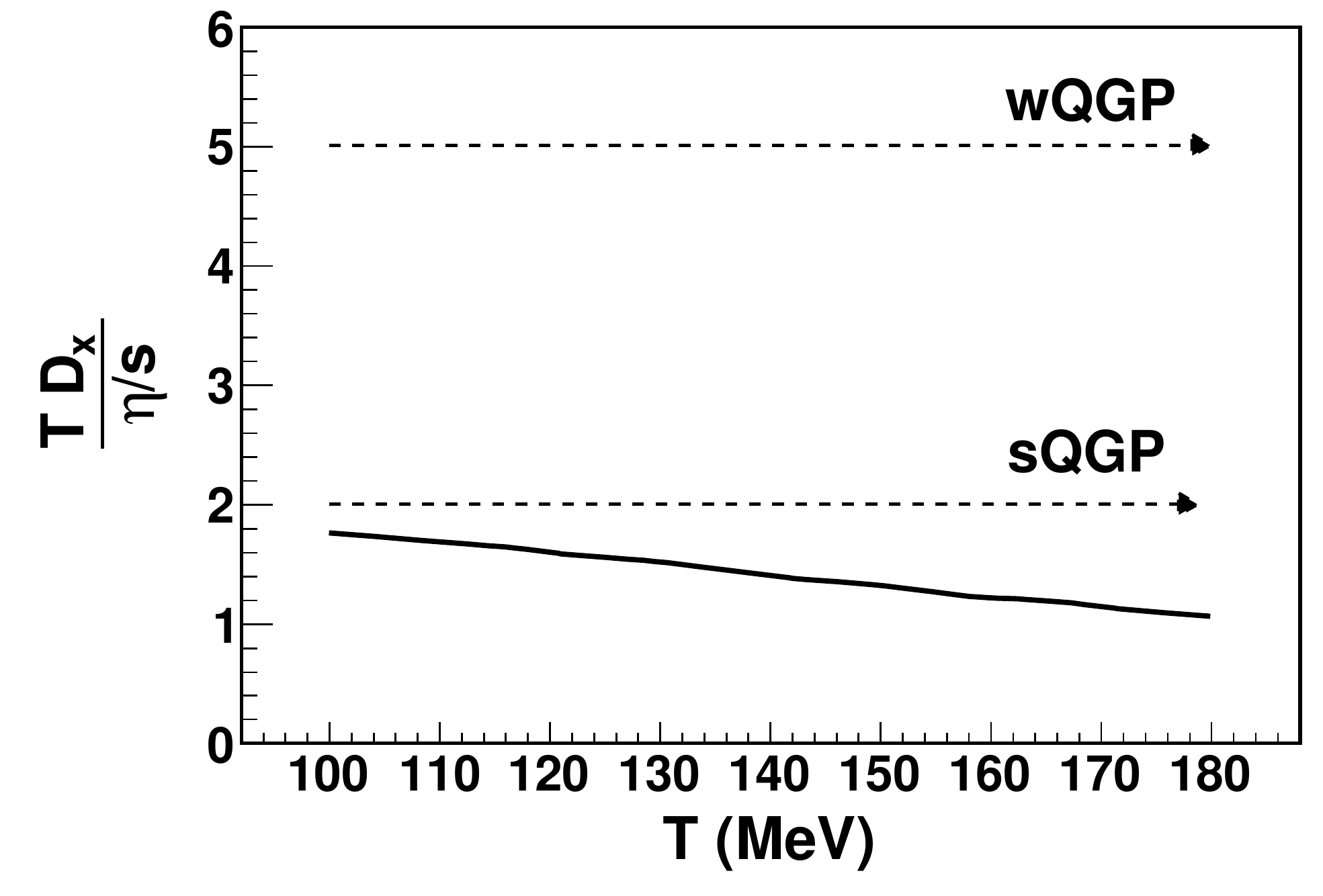}
\caption{\label{fig:stronglycoupled} 
A dimensionless ratio with the viscosity over entropy density, proposed in Ref.~\cite{vanHees:2005wb}. The top dashed line corresponds to a ``weakly coupled'' quark-gluon plasma, 
the bottom line to a ``strongly coupled'' quark-gluon plasma. The solid line at the bottom, for charm propagating in our pion gas, is more suggestive of the second than of the first.
}
\end{figure}

The authors of reference~\cite{vanHees:2005wb} proposed to divide the temperature times the spatial diffusion coefficient by the shear viscosity over entropy density ratio $\eta/s$, producing
 a dimensionless quantity that should give an idea of how strongly coupled is the quark-gluon plasma \index{quark-gluon plasma}, and they quote two estimates based
 on AdS/CFT \index{AdS/CFT correspondence} that we plot in Fig.~\ref{fig:stronglycoupled}.
In the figure we also plot our computation based on charm quarks travelling through the pion gas, together with our computation of viscosity over entropy density
 in the pion gas presented in~\cite{Dobado:2008vt}. It seems that, according to this criterion, the charm quark is somewhat strongly coupled to the pion gas.

\chapter{Linear Sigma Model and Phase Transitions \label{ch:9.LSM}}

In Chapter \ref{ch:3.shear} we have commented about the experimental minimum of $\eta/s$ observed at the phase transition for some fluids. We have supported this fact by explicitly calculating this coefficient for the atomic Argon.
The purpose of this chapter is to gain more insight about this hypothesis by calculating explicitly this coefficient in a system that possesses
a phase transition. However, the deconfinement phase transition of QCD ocurring at a critical temperature \footnote{We will speak about critical temperature both in the case of a critical point, a first order phase transition
or a crossover and call it $T_c$. However one should understand that $T_c$ actually corresponds to the critical temperature or to an approximate crossover temperature depending on the case.} is not accesible
only by chiral perturbation theory nor by perturbative QCD alone. For this reason we have chosen an alternative model for which a
full description of the different thermodynamical phases from the same partition function is possible. The linear sigma model\index{linear sigma model} (L$\sigma$M)\glossary{name=L$\sigma$M,description={linear sigma model}}
 provides a simple model to perform such calculation. Moreover, we will work in the large-$N$ limit in order to simplify the computation of the effective potential and the scattering amplitude.

\section{L$\sigma$M Lagrangian and effective potential at finite temperature}

The bare Euclidean Lagrangian of the L$\sigma$M reads
\be \label{eq:lsmlagrangian} \mathcal{L} [\Phi_i,\pa_{\mu} \Phi_i] = \frac{1}{2} \pa_{\mu} \Phi^T \pa^{\mu} \Phi - \overline{\mu}^2 \Phi^T \Phi + \frac{\lambda}{N} \left(\Phi^T \Phi
 \right)^2 -\epsilon \Phi_{N+1} \ , \ee
where the multiplet $\Phi_i$ contains $N+1$ scalar fields. The parameter $\lambda$ is positive in order to have a potential bounded from below and we consider $\overline{\mu}^2$ to be positive
in order to provide a spontaneous symmetry breaking (SSB)\index{spontaneous symmetry breaking}\glossary{name=SSB,description={spontaneous symmetry breaking}}. The SSB pattern reads (when $\epsilon=0$) $SO(N+1) \rightarrow SO(N)$.
The factor $\epsilon=m^2_{\pi} f_{\pi}$ is responsible for the physical pion mass and when considered, it produces an explicitly breaking of the $SO(N+1)$ symmetry.
We will denote the first $N$ components of the multiplet as the ``pions'' and the last component as the $\sigma$:
\be \Phi_i=(\pi_a , \sigma) \ , \ee
with $i=1, \cdots , N+1$, and $a=1, \cdots, N$.

Let us briefly review the dynamics of the model at $T=0$. In this case as $\overline{\mu}^2 > 0$ the potential possesses a non-zero vacuum expectation value (VEV) \glossary{name=VEV,description={vacuum expectation value}}and
one expects a SSB\index{spontaneous symmetry breaking}. We choose the VEV to be in the $N+1$ direction, i.e in the $\sigma$ direction and call it $f_{\pi}$.

\be \langle \Phi^T \Phi \rangle = \langle \sigma^2 (T=0) \rangle = f^2_{\pi} = NF^2 \ . \ee
This VEV satisfies the equation
\be \label{eq:min} - 2 \overline{\mu}^2 f_{\pi} + \frac{4\lambda}{N} f^3_{\pi} - \epsilon =0 \ . \ee
The solution to this equation is obtained for small explicit breaking term:
\be f_{\pi} = \sqrt{\frac{N \overline{\mu}^2}{2 \lambda}} + \frac{\epsilon}{4 \overline{\mu}^2} = f_{\pi} (\epsilon=0) + \frac{\epsilon}{4 \overline{\mu}^2}
= f_{\pi} (\epsilon=0) + \frac{N \epsilon}{8 \lambda f^2_{\pi} (\epsilon=0)} \ . \ee

In our notation we will call $f_{\pi} (\epsilon=0)$ and $f_{\pi}=\sqrt{N}F$ to the VEV value of the $\sigma$ at $T=0$ for the case without and with explicit symmetry breaking term, respectively.
In what follows, the VEV will be denoted as $v(T)$ for arbitrary temperature, in such a way that $f_{\pi} = v(T=0)$. Moreover, recall
that the $N$-dependence of the parameters of the model reads:
\be \label{eq:Ncounting} \lambda \sim \mathcal{O} (1), \quad \overline{\mu}^2 \sim \mathcal{O} (1), \quad F^2 \sim \mathcal{O} (1), \quad f^2_{\pi} \sim \mathcal{O} (N), \quad \epsilon \sim \mathcal{O} (\sqrt{N}) \ . \ee

\subsection{Spontaneous symmetry breaking at $T=0$}

For simplicity, we start with the $T=0$ case, where the dynamics are governed by the broken phase. The VEV $f_{\pi}$ is chosen to be in the 
$\sigma$ direction. In this case, the VEV for the pions is zero. Recall that when finding the minimum of the potential we have assumed
that the field configuration that minimizes the action is homogeneous. This is a strong assumption but we will maintain it for simplicity.

The degrees of freedom are the quantum fluctuations around the VEV. In the case of the pions, these fluctuations correspond to the Goldstone bosons if $\epsilon=0$
or to the pseudo-Goldstone bosons if $\epsilon \neq 0$.
They are the radial modes and they are massless because the potential is ``flat'' around them (the second derivative of the potential vanishes).
We will maintain the notation $\pi$ for these Goldstone modes. 
\be \langle \pi^a \rangle =0 \ . \ee

Along the $\sigma$ direction, the VEV is $\langle \sigma^2 (T=0) \rangle = f_{\pi}^2= NF^2$, and quantum fluctuations around
this value will be denoted by $\ts$ and correspond to the Higgs, the massive or longitudinal mode, because the potential has curvature around the VEV. 
So that

\be \label{eq:sigma} \sigma = f_{\pi} + \ts \rightarrow \langle  \sigma  \rangle = f_{\pi} \ . \ee

This description is the appropiate one for a saddle point treatment. We will consider that the most important contribution to the quantum partition function
is the classical minimum configuration and the next contribution comes from the quantum fluctuations (Goldstone bosons and Higgs).

Because at $T=0 $ there exists SSB we will rewrite the Lagrangian (\ref{eq:lsmlagrangian}) in terms of the variables $\pi$, $f_{\pi}$ and $\ts$
just substituing (\ref{eq:sigma}) in the original Lagrangian:

\begin{eqnarray}{}
\nonumber  \mathcal{L} & = &  \frac{1}{2} \pa_{\mu} \pi^a \pa^{\mu} \pi^a+\frac{1}{2} \pa_{\mu} \ts \pa^{\mu} \ts - \overline{\mu}^2 \pi^a \pi^a -
\overline{\mu}^2 \ts^2 
 \\
 \nonumber & + &  \frac{\lambda}{N} \left[ 2 f^2_{\pi} \left( \pi^a \pi^a + 3 \ts^2 \right) + (\pi^a \pi^a)^2 + 2 \pi^a  \pi^a \ts^2 + \ts^4 + 4 f_{\pi} \ts^3 +4 \pi^a \pi^a f_{\pi} \ts
 \right]  \\
& + &  \left(-\epsilon + \frac{4 \lambda}{N} f^3_{\pi} - 2 \overline{\mu}^2 f_{\pi} \right) \ts  +
\frac{1}{2} \pa_{\mu} f_{\pi} \pa^{\mu} f_{\pi} - \overline{\mu}^2 f^2_{\pi} + \frac{\lambda}{N} f^4_{\pi} - \epsilon f_{\pi}
\end{eqnarray}

Note that the tadpole term\index{tadpole} vanishes by construction because the term between parenthesis satisfies Eq.~(\ref{eq:min}). 
We are interested in the expression of an effective potential for the variable $f_{\pi}$, where the fluctuations
are integrated out. From such an effective potential one can obtain the vacuum expectation value just imposing $\frac{dV_{eff}}{df_{\pi}} =0$.
Before doing so, we can read the value of the Higgs mass from the obtained Lagrangian
\be M^2_{\ts} = - 2\overline{\mu}^2 + \frac{12 \lambda}{N} f^2_{\pi} \ , \ee
that at $T=0$ and at tree level it is
\be M^2_{\ts} = 4 \overline{\mu}^2+\frac{\epsilon}{f_{\pi}} = \frac{8 \lambda f^2_{\pi}}{N} + \frac{\epsilon}{f_{\pi}} = \frac{8 \lambda f^2_{\pi} (\epsilon=0)}{N} + 3\frac{\epsilon}{f_{\pi}} \ee
The mass of the pions reads
\be m^2_{\pi} = - 2 \overline{\mu}^2 + \frac{4 \lambda}{N} f_{\pi}^2 \ , \ee
that as expected it only depends of the explicit symmetry breaking term
\be m^2_{\pi} = \frac{\epsilon}{f_{\pi}} \ ,\ee
and vanishes at $\epsilon=0$.

In the opposite way, one can choose the value of the bare parameters $\overline{\mu}^2$ and $\lambda$ in such a way that they reproduce
the physical pion and Higgs masses and $f_{\pi}$:

\begin{eqnarray}
\overline{\mu}^2 & = & \frac{M^2_{\ts} - 3 m^2_{\pi}}{4} \ , \\	
\lambda & = & \frac{N}{8 f^2_{\pi}} (M^2_{\ts}-m^2_{\pi}) \ . 
\end{eqnarray}

The last expression can be expressed in terms of the VEV with $\epsilon=0$:
\be \lambda = \frac{N}{8 f^2_{\pi} (\epsilon=0)} \frac{M^2_{\ts}-m^2_{\pi}}{\alpha^2} \ , \ee
where $\alpha$ is a multiplicative constant relating $f_{\pi}$ and $f_{\pi} (\epsilon=0)$
\be f_{\pi} = \alpha f_{\pi} (\epsilon=0) \ , \ee
that explicitly reads
\be \alpha= \frac{M^2_{\ts}-3 m^2_{\pi}}{M^2_{\ts}-4 m^2_{\pi}} \ . \ee

Finally, the tree level propagators for the Goldstone bosons and the Higgs:
\begin{eqnarray}
 D_{\pi}^{-1} & = & k^2 - 2\overline{\mu}^2 +\frac{4 \lambda}{N} f^2_{\pi} \ , \\
 D_{\ts}^{-1} & = & k^2 - 2\overline{\mu}^2 +\frac{12 \lambda}{N} f^2_{\pi} \ .  
\end{eqnarray}

The first method to obtain the effective potential\index{effective potential} for $f_{\pi}$ is the standard calculation described in textbooks. In this case, one neglects the fluctuations 
of the Higgs and performs a mean field approximation\index{mean field approximation} for it $\sigma \equiv f_{\pi}$. Then, expanding the quantum fluctuations
of the Goldstone bosons, one only considers the quadratic terms as being the first correction in the action. The integration of these fluctuations is straightforward as
it is performed by a simple Gaussian integral. The cubic and quartic terms are neglected as they are not so easily integrable in the partition function. 
The fluctuations of the Higgs can be neglected at $T=0$ but around the critical temperature this is a hard assumption, because
the fluctuations of the Higgs are rather relevant. One could even perform the complete Gaussian integration for both the Higgs and the Goldstone boson fluctuations. This would correspond
to a complete 1-loop calculation of the effective potential. 

However, as we want to take the large-$N$ limit, we feel that we can go one step further and perform the quartic integration of the fluctuations. This
is done by using the auxiliary field method. The calculation gets more complicated but
the introduction of the auxiliary field allows for a systematic counting of $N$ factors and gives simplification in the large-$N$ limit. 

All these perturbative approaches perform the 1-loop integration of the fluctuations regardless of its wavelenght. All the frequency modes of the fields are treated at the same footing and this
unorganized integration produces two undesirable features in the effective potential. First, an imaginary part of the effective potential appears. This imaginary part\index{effective potential ! imaginary part of} has been
given the interpretation of a decay rate per unit volume of the unstable vacuum state by Weinberg and Wu in \cite{Weinberg:1987vp}.

The second characteristic is the non-convexitivity of the quantum effective potential, but the effective potential (defined through a Legendre transformation) should be always convex\index{effective potential!convexitivity of}.
This non-convexitivity problem and the imaginary part appear as long as a perturbative method is used to calculate the effective potential \cite{vanKessel:2008ht}. 
  
A solution to these features can be to use a non-perturbative method to generate the effective potential. For example, the Functional Renormalization Group (FRG) \index{functional renormalization group} generates the effective
potential in such a way that an organized integration of the fluctuations is performed. Following the ideas of the renormalization group\index{renormalization group}, only the low wavelength components 
of the quantum fluctuations are integrated-out at each step. The UV components are then integrated infinitesimally step by step and the final effective potential (defined in the infrared scale) does not acquire an
imaginary part and it remains convex at every scale (at the IR point, the Maxwell construction \index{Maxwell construction}can be dynamically generated through renormalization flow) \cite{Alexandre:1998ts}.

In spite of these issues, we believe that it is not necessary to perform a more sofisticated method to obtain the effective potential. The only relevant result for us is the localization
of the minimum of the effective potential, which eventually gives us the position of the critical temperature, and this minimum is always outside of the non-convex region. In any case, the possible presence of an
imaginary part (whose domain in fact coincides with the domain of the non-convex part of the potential) is not even relevant for us.

\subsection{Auxiliary field method}
We start considering the partition function:
\be \mathcal{Z} = \int \mathcal{D} \pi^a \mathcal{D} \sigma \ \exp \left( -\int d^4 x \mathcal{L} \right) \ , \ee
with the Lagrangian in Eq.~(\ref{eq:lsmlagrangian}).
We introduce an auxiliary field $\chi$\index{auxiliary field method} to make the integral Gaussian.
\be \chi \equiv 2 \sqrt{2} \frac{\lambda}{N} \Phi^T \Phi \ . \ee

The quartic coupling is therefore substituted by
\be \exp \left( \int d^4x \frac{\lambda}{N} \Phi^4 \right) = \int \mathcal{D} \chi \exp \left[ - \frac{1}{2} \int d^4x \left( \frac{N}{4 \lambda} \chi^2 - \sqrt{2} \chi \Phi^2 \right) \right] \ee
up to a overall constant. Note that this auxiliary field has introduced a mass term and a coupling with $\Phi^2$ in the Lagrangian. However, there is no kinetic term for it. That means that $\chi$
possesses no true dynamics.

The partition function transforms to
\begin{eqnarray}
  \nonumber \mathcal{Z} & =&  \int \mathcal{D} \pi^a \mathcal{D} \sigma \mathcal{D} \chi \  \exp \left( -\int d^4 x \
\frac{1}{2} \pa_{\mu} \pi^a \pa^{\mu} \pi^a + \frac{1}{2} \pa_{\mu} \sigma \pa^{\mu} \sigma - \overline{\mu}^2 \pi^a \pi^a
- \overline{\mu}^2 \sigma^2 \right. \\
& & \left. - \frac{1}{2} \frac{N}{4\lambda} \chi^2 + \frac{1}{2} \sqrt{2} \chi \pi^a \pi^a + \frac{1}{2} \sqrt{2} \chi \sigma^2 - \epsilon \sigma \ 
\right) \ .
\end{eqnarray}

The action in terms of the $\pi^a$, $\sigma$ and $\chi$ fields reads
\be S = \int d^4 x \ 
\frac{1}{2} \pi^a \left( - \square_E - 2\overline{\mu}^2 + \sqrt{2} \chi \right) \pi^a 
+ \frac{1}{2} \sigma \left( - \square_E - 2\overline{\mu}^2 + \sqrt{2} \chi \right) \sigma
- \frac{1}{2} \frac{N}{4 \lambda} \chi^2 - \epsilon \sigma \ee

Note that before identifying the pion propagator one must get rid of the unphysical $\sigma$ tadpole. We have already seen that this term vanishes at $T=0$.
Now, to see the cancelation of the tadpole we perform a shift of the $\sigma$ field $\sigma = v + \ts$. This separation produces an additional shift of the auxiliary field:
\be \chi =2 \sqrt{2} \frac{\lambda}{N} v^2 + \tilde{\chi} \ . \ee
Inserting the two transformations and using Eq.~(\ref{eq:min}) we see that the tadpole for $\ts$ dissapears:

\begin{eqnarray}
\label{eq:actmassmix} S[\pi^a, v , \ts,\tilde{\chi}] & = &  \int d^4 x \ 
\frac{1}{2} \pi^a \ \left( -\square_E + G^{-1}_{\pi} [0,\chi] \right) \pi^a + \frac{1}{2} \ts \ \left( - \square_E + G^{-1}_{\pi} [0,\chi] \right) \ts \nonumber \\  
& & - \overline{\mu}^2 v^2   +   \frac{1}{2} \sqrt{2} \tilde{\chi} v^2 + 2 \frac{\lambda}{N} v^4- \epsilon v  - \frac{1}{2} \frac{N}{4 \lambda} \tilde{\chi}^2 + \sqrt{2} \tilde{\chi} v \ts \ , 
\end{eqnarray}

where we have introduced for conveniency the following function:
\be G^{-1}_{\pi} [q,\chi] \equiv q^2 - 2\overline{\mu}^2 + \sqrt{2} \chi =  q^2 - 2 \overline{\mu}^2 + \frac{4 \lambda}{N} \Phi^2  \ .\ee

In the action (\ref{eq:actmassmix}) has appeared a mass mixing term between $\tilde{\chi}$ and $\ts$. To avoid such a term we make an extra shift to the field $\tilde{\chi}$,
namely
\be \tilde{\chi} = \chi + 4 \sqrt{2} \frac{\lambda}{N} v \ts \ . \ee

The action in Eq.~(\ref{eq:actmassmix}) transforms to

\begin{eqnarray}  S[\pi^a, v , \ts, G^{-1}_{\pi}[0,\chi]] & = &  \int d^4 x \ 
\frac{1}{2} \pi^a \ (-\square_E + G^{-1}_{\pi}[0,\chi]) \pi^a + \frac{1}{2} \ts \ (-\square_E + G^{-1}_{\pi}[0,\chi] \nonumber \\
& & + 8\frac{\lambda}{N} v^2 ) \ \ts - \overline{\mu}^2 v^2 + \frac{\lambda}{N}v^4 - \epsilon v - \frac{N}{8 \lambda} \chi^2 \ . \end{eqnarray}

The function $G^{-1}_{\pi}$ is nothing but the inverse of the pion propagator in the Fourier space. The inverse propagator of the $\ts$ field is
\be G^{-1}_{\ts} [q,\chi] = G^{-1}_{\pi} [q,\chi] + 8 \frac{\lambda}{N} v^2 \ .  \ee

\section{Effective potential at $T \neq 0$}

Now we procceed to integrate the fluctuations out of the action in order to generate the effective potential for the field $v$.
The integration of the pions is done in the standard way
\begin{eqnarray}
& & \nonumber  \int \mathcal{D} \pi^a \exp \left(- \int d^4x \ \frac{1}{2} \pi^a \left[ - \square_E + G_{\pi}^{-1} [0,\chi] \right] \pi^a \right)
 \rightarrow \\
& &  \int d^4x \exp \left( - \frac{N}{2} \int \frac{d^3 q}{(2\pi)^3} \log G_{\pi}^{-1} [q,\chi] \right)  \ .
\end{eqnarray}

The integration of the Higgs is performed in the same way but taking $N=1$.

The effective potential (density) reads
\be V_{eff} = - \overline{\mu}^2 v^2 + \frac{\lambda}{N}v^4 - \epsilon v - \frac{N}{8 \lambda} \chi^2 + \frac{N}{2} \sumint_{\beta} \log G^{-1}_{\pi} [q,\chi]+\frac{1}{2} \sumint_{\beta} \log G^{-1}_{\ts} [q,\chi] \ , \ee
with
\be  \sumint_{\beta} = T \sum_{n \in \mathcal{Z}} \int \frac{d^3q}{(2\pi)^3} \ .\ee

Note the $N$-counting of the terms of the effective potential. Following (\ref{eq:Ncounting}) one finds that all the terms behave as $\mathcal{O}(N)$ except the last one.
So, in the large-$N$ limit the contribution of the Higgs to the effective potential is suppresed by one power of $N$ with respect to the contribution of the pions.
We neglect that term in what follows. The effective action reads:
\be V_{eff}[v,G^{-1}_{\pi}]= - \overline{\mu}^2 v^2 + \frac{\lambda}{N}v^4 - \epsilon v - \frac{N}{8 \lambda} \chi^2 + \frac{N}{2} \sumint_{\beta} \log G^{-1}_{\pi} [q,\chi]  \ .\ee

Finally, we will write the redundant $\chi$ field in terms of $G^{-1}_{\pi} [0,\chi]$:

\be V_{eff} [v,G^{-1}_{\pi}]= \frac{1}{2} \left( v^2 - N\frac{F^2}{\alpha^2} \right) G^{-1}_{\pi} [0,\chi] - \frac{N}{16 \lambda} (G^{-1}_{\pi} [0,\chi])^2 - \epsilon v + \frac{N}{2} \sumint_{\beta} \log G^{-1}_{\pi} [q,\chi] \ , \ee
where we have dropped an $v$-independent term from the potential.
The last term needs to be regulated because it contains a divergence. Therefore, the Higgs mass and the coupling need to be renormalized.

Doing the integration in the last term one gets
\be \frac{N}{2} \sumint_{\beta} \log G_{\pi}^{-1} [q,\chi] = \frac{N}{2} \left[ \frac{G_{\pi}^{-1} [0,\chi]}{2} I_{G_{\pi}^{-1}} - \frac{(G_{\pi}^{-1} [0,\chi])^2}{4 (4 \pi)^2} -g_0 (T,G_{\pi}^{-1}[0,\chi]) \right] \ee
where
\be I_{G^{-1}_{\pi}} = - \frac{G_{\pi}^{-1}[0,\chi]}{(4 \pi)^2} \left[ N_{\epsilon} +1 - \log \frac{\mu^2}{G^{-1}_{\pi} [0,\chi]} \right] \ee
with
\be N_{\epsilon} = \frac{2}{\epsilon} + \log 4\pi - \gamma_E \ . \ee

The function $g_0(T,M^2)$ is the following finite integral
\be  g_0(T,M^2)=\frac{T^4}{3\pi^2} \int_y^{\infty} dx (x^2-y^2)^{3/2} \frac{1}{e^{x}-1} \ , \ee
with $y=M/T$. The derivative of this function with respect to $M^2$ defines the function $g_1(T,M)$:
\be g_1 = - \frac{dg_0}{dM^2} \ , \ee
that in terms of an integral reads
\be  g_1(T,M^2)=\frac{T^2}{2\pi^2} \int_y^{\infty} dx \frac{\sqrt{x^2-y^2}}{e^{x}-1} \ . \ee
In the limit $y\rightarrow 0$ the two functions take the analytic results:
\begin{eqnarray}
\label{eq:g0limit} g_0 (T,0) & = & \frac{T^4}{3 \pi^2} \Gamma(4) \zeta(4) = \frac{\pi^2 T^4}{45} \ , \\
\label{eq:g1limit} g_1 (T,0) & = & \frac{T^2}{2 \pi^2} \Gamma(2) \zeta(2) = \frac{T^2}{12} \ .
\end{eqnarray}

We have used the dimensional regularization scheme. We are going to define the renormalized coupling as
\be \frac{1}{\lambda_R} \equiv \frac{1}{\lambda} + \frac{1}{4 \pi^2} \left( 1 -\log \frac{\mu^2}{G^{-1}_{\pi} [0,\chi]} - \frac{(4 \pi)^2}{G^{-1}_{\pi} [0,\chi]} I_{G^{-1}_{\pi}} \right)=
\frac{1}{\lambda} +  \frac{1}{2 \pi^2} \left(1 + \frac{N_{\epsilon}}{2} -  \log \frac{\mu^2}{G^{-1}_{\pi} [0,\chi]} \right)   .\ee

The renormalization works by taking the following two terms of the potential:

\begin{eqnarray}
& &   - \frac{N}{16 \lambda} (G^{-1}_{\pi} [0,\chi])^2 + \frac{N}{2} \sumint_{\beta} \log G^{-1}_{\pi} [q,\chi]  = \\
& & \nonumber =  -\frac{N}{16} (G^{-1}_{\pi}[0,\chi])^2 \left[ \frac{1}{\lambda_R} - \frac{1}{4\pi^2} \log \left( \frac{\sqrt{e} G^{-1}_{\pi} [0,\chi] }{\mu^2} \right) \right] - \frac{N}{2} g_0 (T, G^{-1}_{\pi} [0,\chi]) \ ,
\end{eqnarray}

where now all the terms are finite. The renormalized effective potential finally reads:

\begin{eqnarray}
 V_{eff} (v,G^{-1} [0,\chi]) & = & \frac{1}{2} (v^2 - N\frac{F^2}{\alpha^2}) G^{-1}_{\pi} [0,\chi] - \epsilon v - \frac{N}{2} g_0 (T, G^{-1}_{\pi} [0,\chi]) \nonumber \\
& & - \frac{N}{16} (G^{-1}_{\pi} [0,\chi])^2 \left[
\frac{1}{\lambda_R} - \frac{1}{4 \pi^2} \log \left( \frac{\sqrt{e} G^{-1}_{\pi} [0,\chi]}{\mu^2} \right)
\right] \ . 
\end{eqnarray}

One can see that by construction the effective potential does not depend on the renormalization scale:
\be \frac{d V_{eff}}{d \mu} =0 \ . \ee

To obtain the value of $G^{-1}_{\pi} [0,\chi]$ at the minimum of the potential, we simply do:

\be \left. \frac{d V_{eff}}{d G^{-1}_{\pi} [0,\chi]} \right|_{G^{-1}_{\pi} [0,\chi]=G^{-1}_{\pi,0}[0,\chi]} =0 \ee
that explicitly reads
\be \label{eq:extremumG}\frac{1}{2} \left(v^2 - N \frac{F^2}{\alpha^2}\right) - \frac{N}{8} G_{\pi,0}^{-1} [0,\chi] \left(\frac{1}{\lambda_R} - \frac{1}{4 \pi^2} \log \frac{e G_{\pi,0}^{-1} [0,\chi]}{\mu^2} \right) + \frac{N}{2} g_1 (T,G_{\pi,0}^{-1} [0,\chi]) =0 \ . \ee

The solution $G^{-1}_{\pi,0} [0,\chi]$ to this equation is introduced into the expression of the effective potential to obtain a $v$-dependent potential:

\be V_{eff} (v)= V_{eff} (v,G_{\pi,0}^{-1} [0, \chi] (v))\ee

The extremum of the effective potential gives the value of $v_0$, that is the order parameter (dependent on the temperature):

\be \left. \frac{d V_{eff}}{dv} \right|_{v=v_0}=0 \ . \ee

This equations explicitly reads
\be \label{eq:extremumv} v_0 G^{-1}_{\pi, 0} = \epsilon \ . \ee
The mass of the pion can be extracted as
\be \label{eq:pionmass} m^2_{\pi}= G^{-1}_{\pi, 0} [0,\chi] \ee
and the Higgs mass as
\be M^2_{R} = G^{-1}_{\pi,0} [0,\chi] +8 \frac{\lambda}{N} v_0^2 \ . \ee

The following results can be obtained when there is no explicit symmetry breaking $\epsilon=0$.
In the broken phase (low temperatures) the order parameter is known to be different from zero $v_0 \neq 0$, from Eq.~(\ref{eq:extremumv})
one has the solution $G^{-1}_{\pi,0} [0,\chi]=0$ that means that the pion mass is vanishing in the broken phase. This is nothing but the Goldstone Theorem\index{Goldstone theorem}.

Substituing $G^{-1}_{\pi,0}[0,\chi]$ in (\ref{eq:extremumG}) one arrives at ($\alpha=1$)
\be \label{eq:vTsecond} v^2 (T) = NF^2 \left( 1 - \frac{T^2}{12F^2} \right) \ , \ee
where we have used the result (\ref{eq:g1limit}).

So that the critical temperature (at which the order parameter $v(T)$ vanishes) is
\be \label{eq:criticalT} T_c = \sqrt{12} F = \sqrt{\frac{12}{N}} f_{\pi} \ . \ee

Finally, the Higgs mass
\be M^2_R (\mu) = 8 \frac{\lambda_R (\mu)}{N} v^2 (T)\ee
follows the behaviour of $v$ such as at the critical temperature the Higgs becomes massless.

In the symmetric phase (high temperatures) $v$ is expected to be zero and $G^{-1}_{\pi,0}[0,\chi] \neq 0$
is a nontrivial function of the temperature. The thermal masses for the pion and for the Higgs are now degenerate:
\be m_{\pi}^2 = M_R^2 = G^{-1}_{\pi,0} [0,\chi] \ . \ee

\begin{figure}[t]
\begin{center}
\includegraphics[scale=0.21]{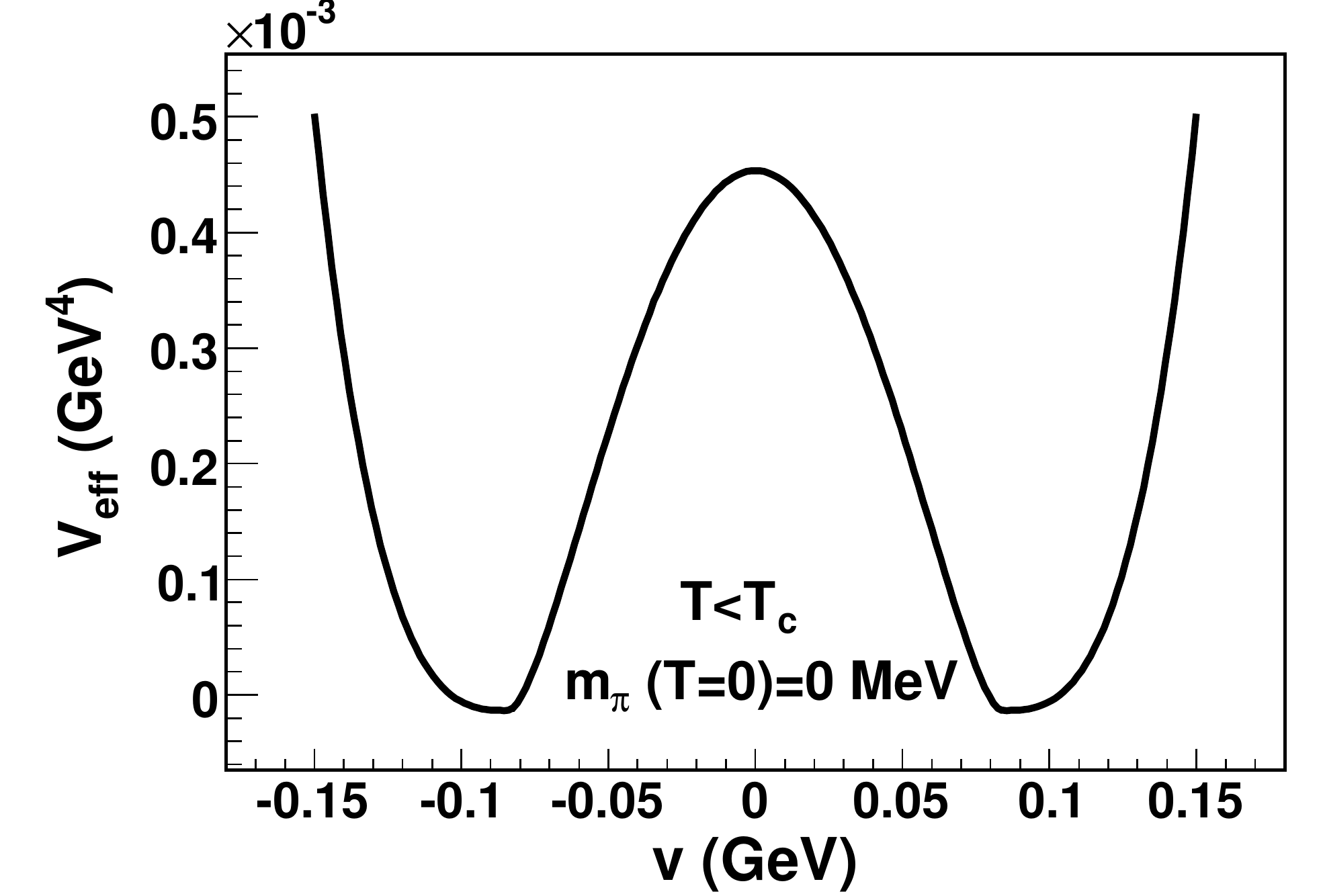}
\includegraphics[scale=0.21]{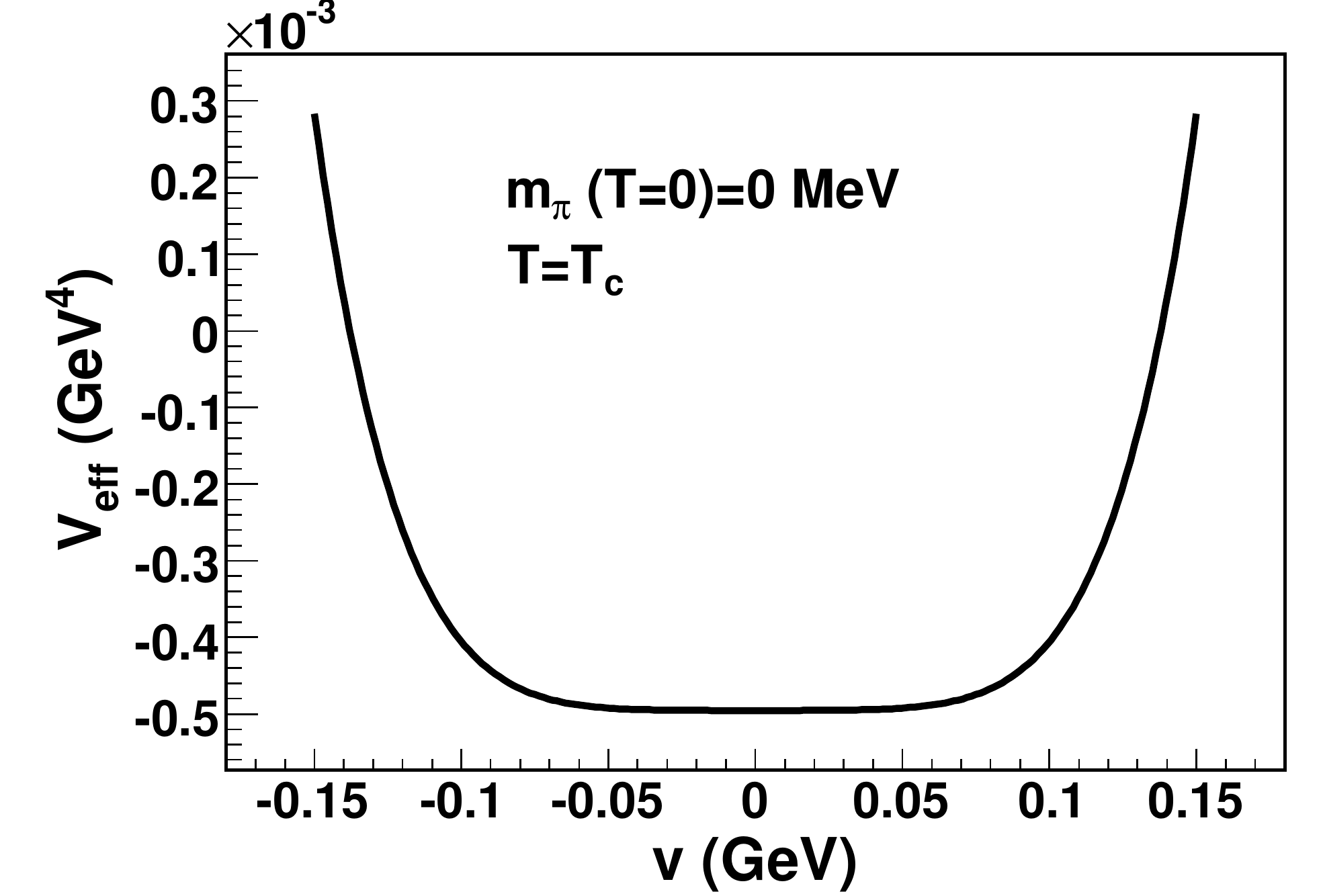}
\includegraphics[scale=0.21]{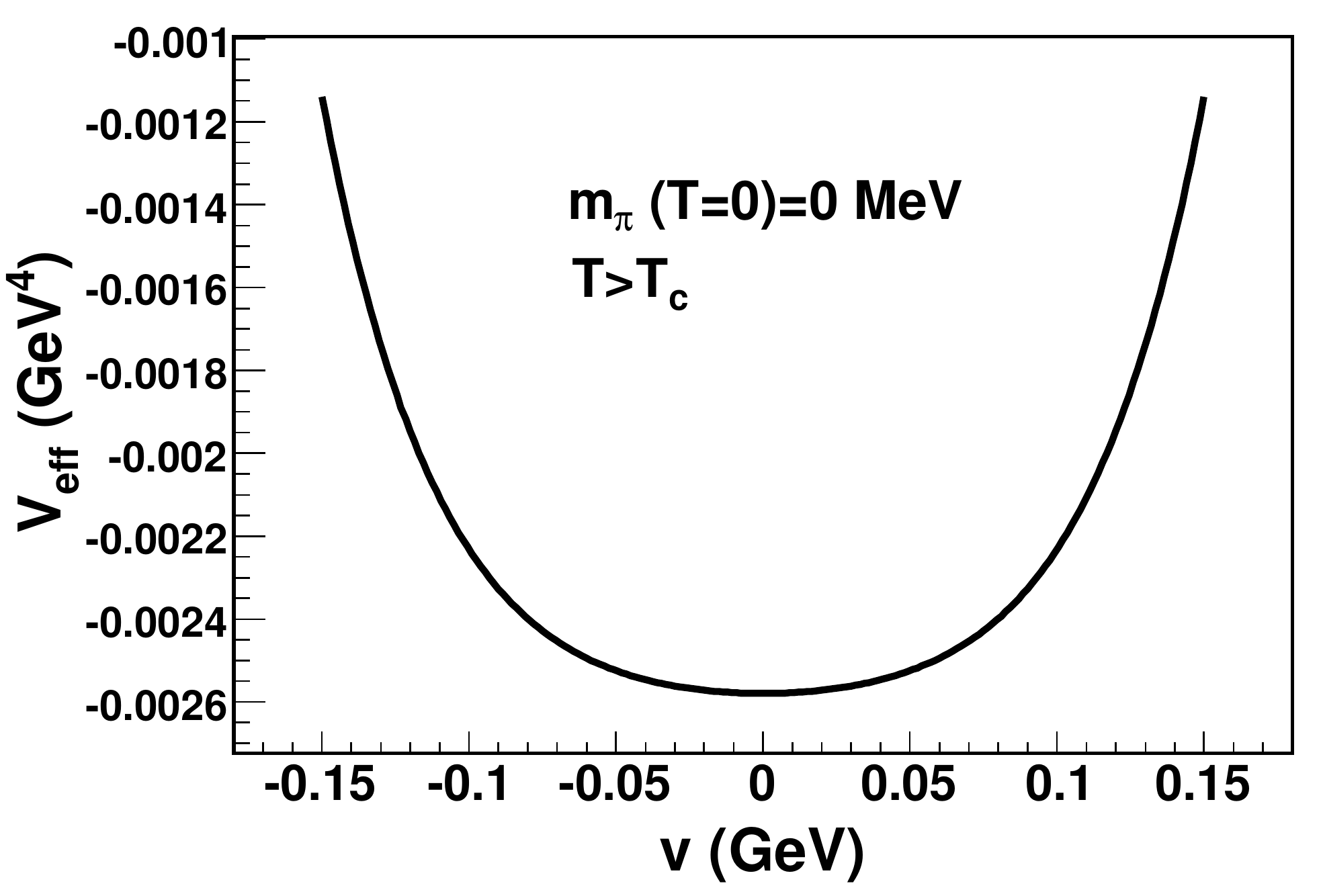}
\includegraphics[scale=0.21]{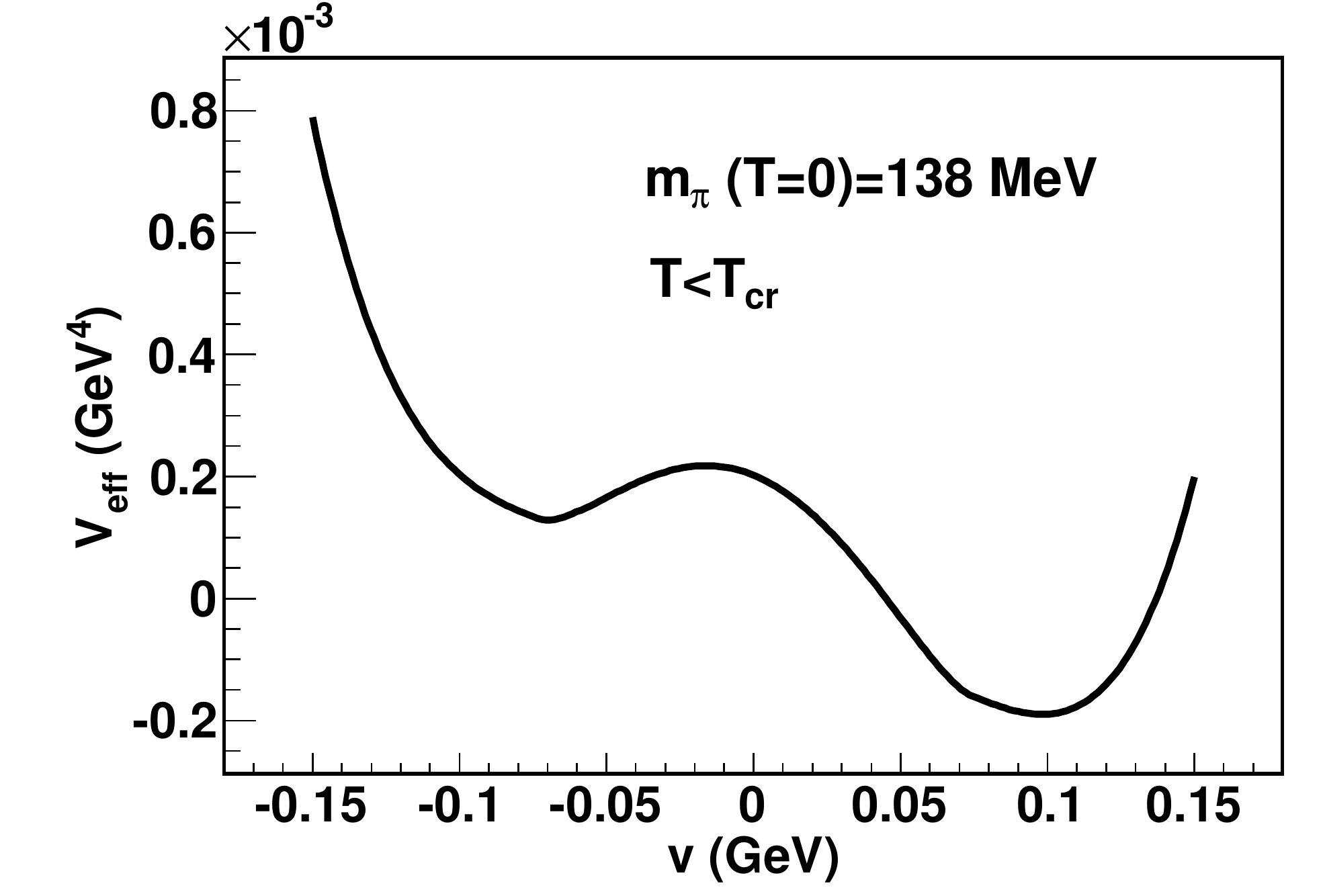}
\includegraphics[scale=0.21]{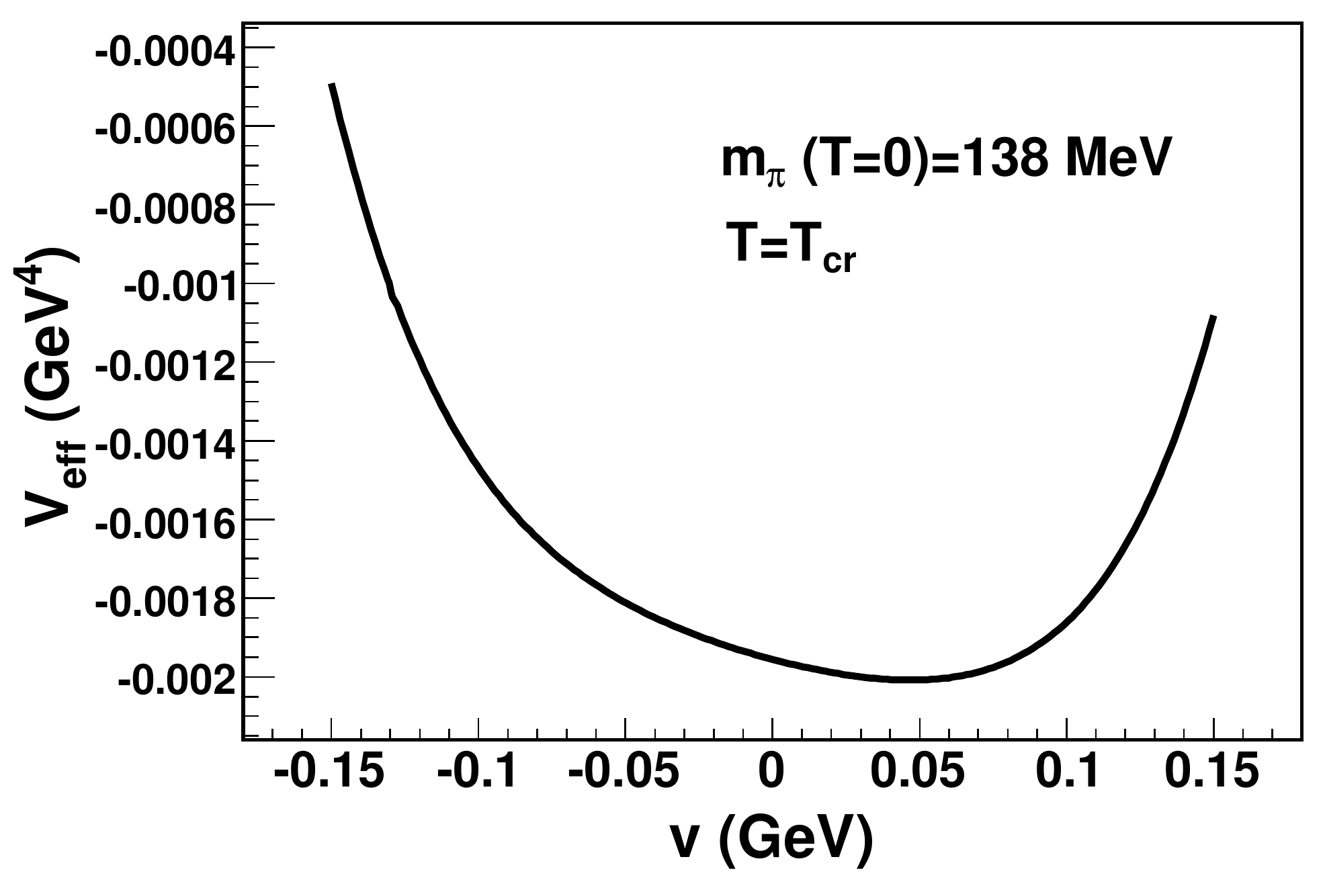}
\includegraphics[scale=0.21]{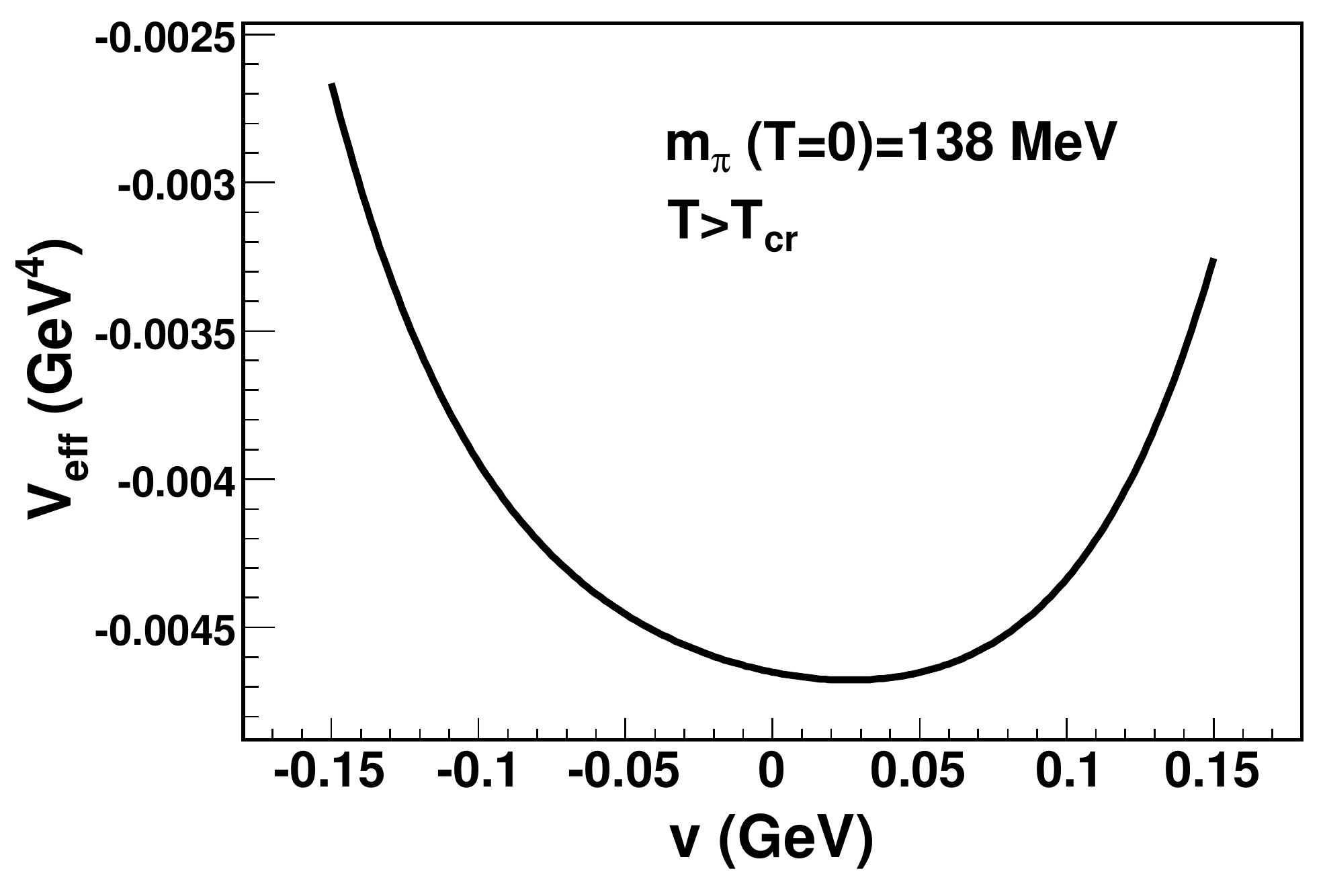}
\caption{\label{fig:eff_pot} Effective potential at different temperatures for the case $\epsilon=0$ (upper panels) and $\epsilon \neq 0$ (lower panels).}
\end{center}
\end{figure}

Turning to the numerical computation, we obtain the following results.
First, we show the shape of the effective potential as a function of $v$ in Fig.~\ref{fig:eff_pot}. The upper panels show the case in which there is no explicit symmetry breaking ($m_{\pi} (T=0)=0$ MeV).
From left to right, the panels show the effective potential at a temperature below the critical one, the potential at $T_c$, and above the critical temperature. In the lower panels we have used an explicitly
symmetry-breaking pion mass at zero temperature of $m_{\pi}=138$ MeV. From left, to right, the panels show $V_{eff}$ for temperatures below $T_c$, at $T_c$ and above $T_c$. Note that the absolute minimum of the effective
potential gives the value of the VEV.

\begin{figure}[t]
\begin{center}
\includegraphics[scale=0.3]{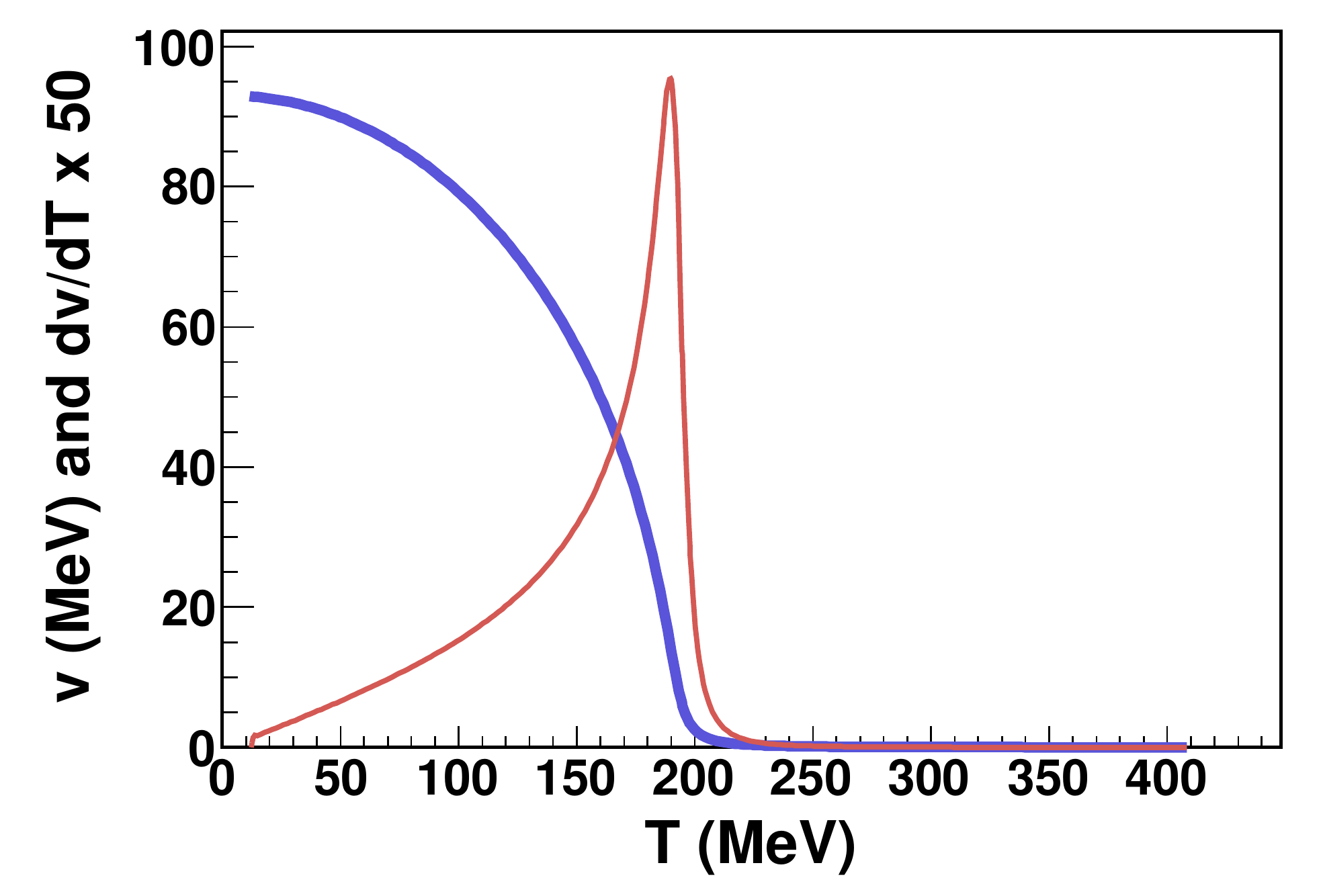}
\includegraphics[scale=0.3]{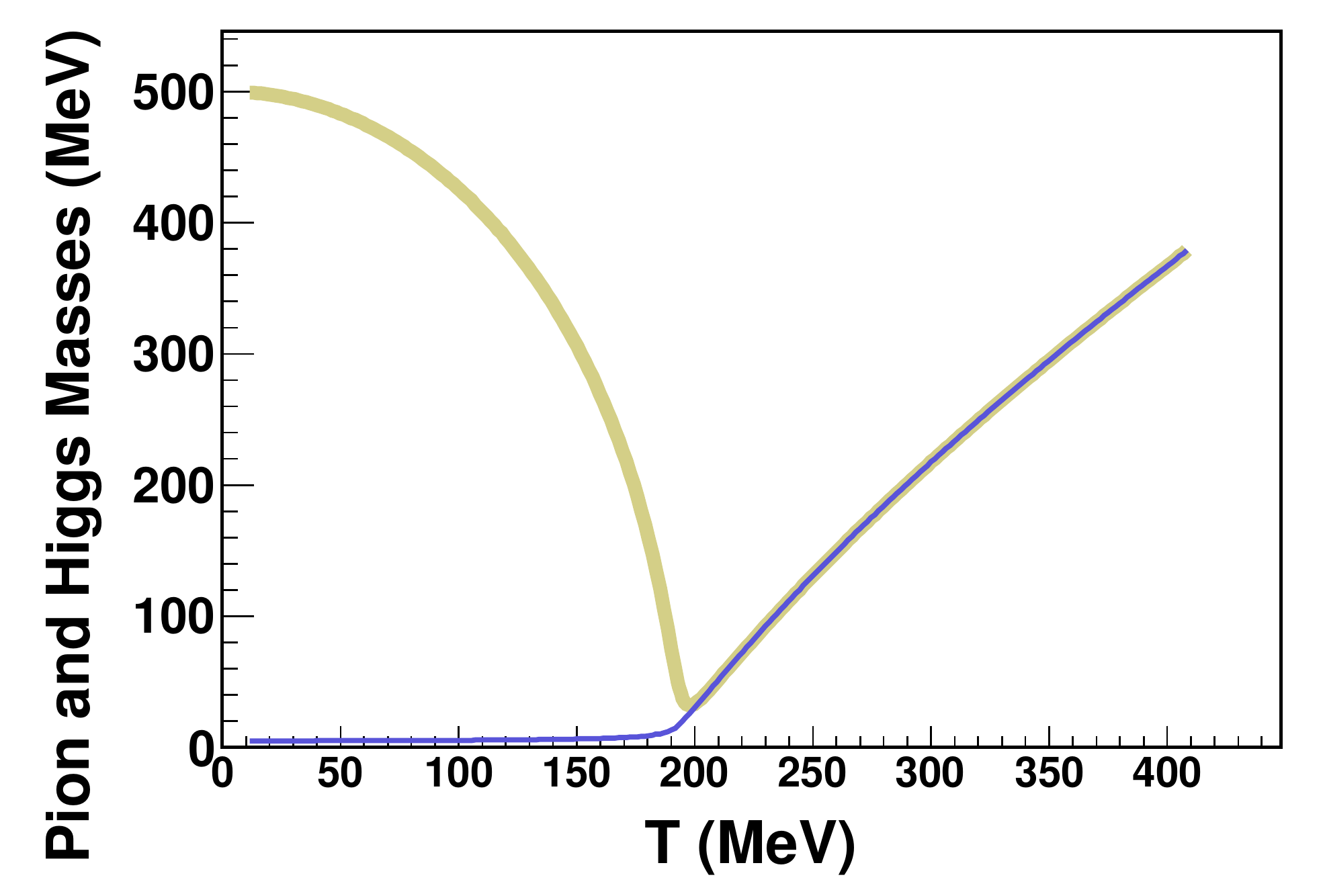}
\caption{\label{fig:sec_ord} Second order phase transition: Left panel: Order parameter or vacuum expectation value of the $\sigma$ (blue broader line) and its derivative (red narrower line). Right panel: Thermal mass
for the pion (blue narrower line) and for the Higgs (yellow broader line).}
\end{center}
\end{figure}

For the case $\epsilon=0$ there is no explicit symmetry breaking term, so we expect to have a 
second order phase transition defined by the critical temperature (\ref{eq:criticalT}). We use a vanishing pion mass at $T=0$ (with $N=3$), a Higgs mass of $M_R=500$ MeV and
a value of $v(T=0)=93$ MeV. We obtain the results appearing in Fig.~\ref{fig:sec_ord}. In the left panel we show in blue line the behaviour of $v(T)$ that follows the analytic 
solution in Eq.~(\ref{eq:vTsecond}). The numerical critical temperature coincides with the theoretical value of $T_c = 2 f_{\pi} \simeq 200$ MeV. In red line we show the susceptibility, defined
as the $T-$derivative of the order parameter. Its peak shows the position of the critical temperature. In the right panel we show the thermal masses as a function of the temperature. The mass of the pions (blue line)
at $T<T_c$ must be always zero according to the Goldstone theorem \index{Goldstone theorem} (numerically it is fixed at $0.5$ MeV in order to avoid computational problems). At $T_c$ it starts growing with temperature 
in the symmetric phase. The mass of the Higgs (yellow line) follows the same pattern as the order parameter becoming zero at $T_c$. For higher temperatures it increases with temperature being degenerate with the thermal pion mass.

\begin{figure}[t]
\begin{center}
\includegraphics[scale=0.3]{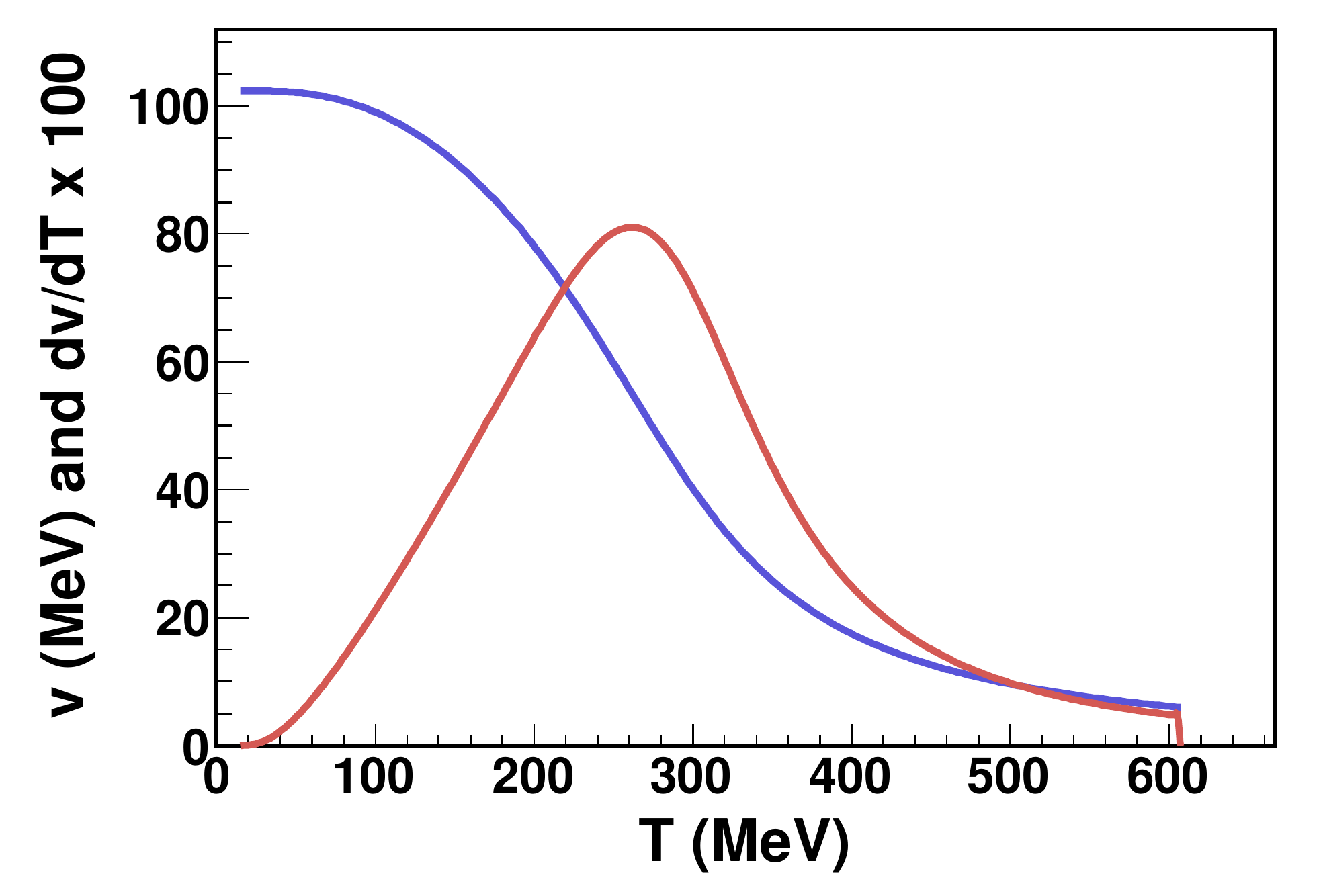}
\includegraphics[scale=0.3]{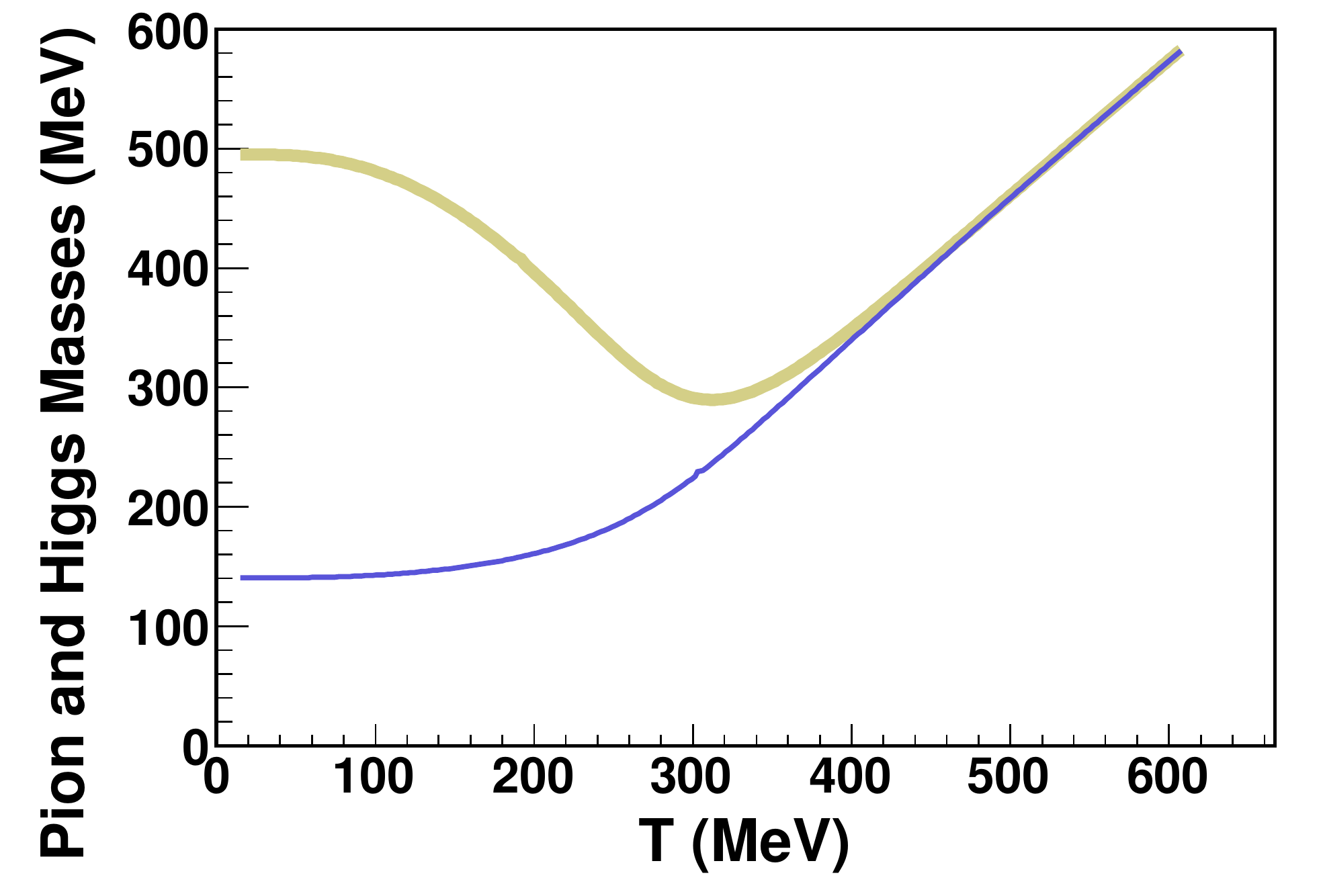}
\caption{\label{fig:crossover} Crossover: Left panel: Order parameter or vacuum expectation value of the $\sigma$ (blue broader line) and its derivative (red narrower line). Right panel: Thermal mass
for the pion (blue narrower line) and for the Higgs (yellow broader line).}
\end{center}
\end{figure}

In the $\epsilon \neq 0$ a crossover effect is expected (this is the same situation as adding an external magnetic field to a ferromagnet). We fix $m_{\pi}=138$ MeV at zero temperature and same values for $M_R$ and $f_{\pi}$.
The results are shown in Fig.~\ref{fig:crossover}. The left panel shows the order parameter $v(T)$ that decreases with temperature and never becomes exactly zero. The crossover temperature can be defined as the position of the peak  
in the susceptibility that we show in red line (although other definitions of the crossover temperature can be made). In the right panel we show the pion thermal mass (blue) and the Higgs mass (yellow). They are degenerate for high temperatures.

\section{Scattering amplitude}

\begin{figure}[t]
\begin{center}
\includegraphics[scale=0.25]{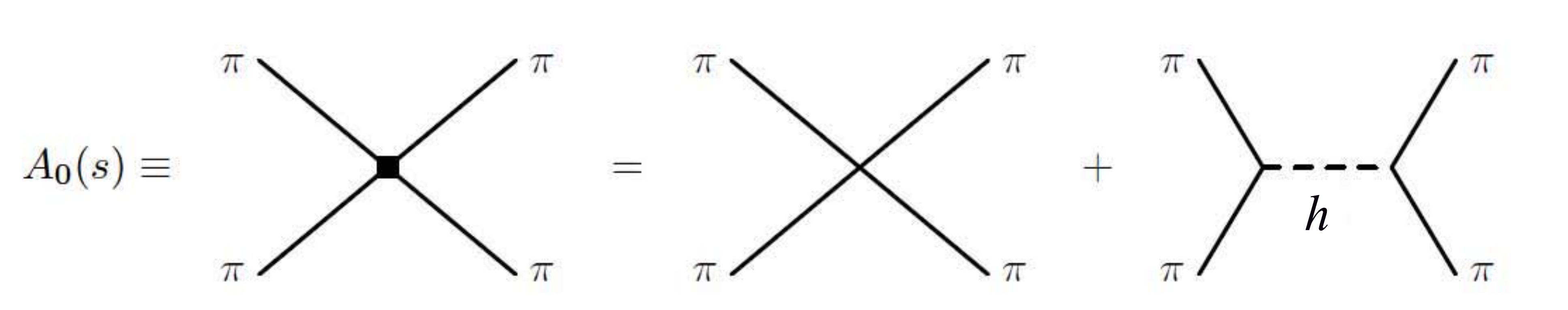}
\includegraphics[scale=0.25]{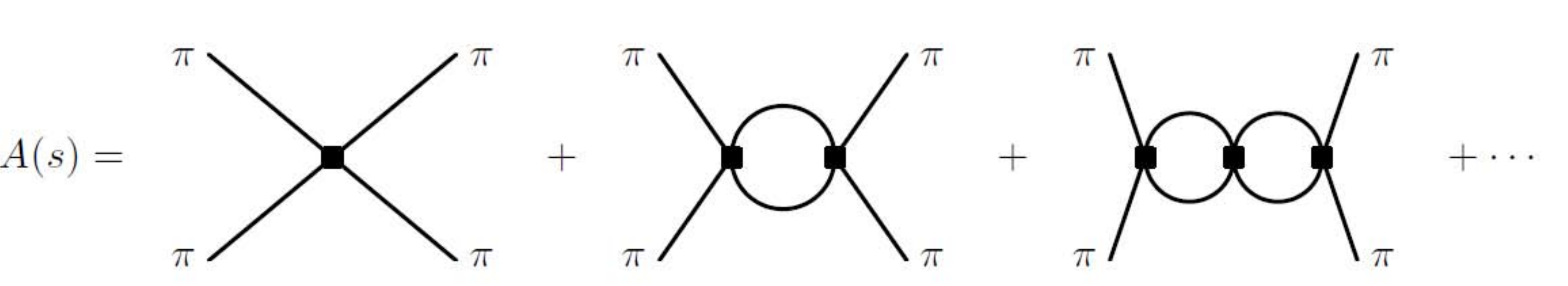}
\caption{\label{fig:amplitudes} Tree-level (left) and resummed (right) amplitude for $\pi-\pi$ scattering at $\mathcal{O}(1/N)$.}
\end{center}
\end{figure}

We now calculate the scattering amplitude for the elastic pion-pion dispersion needed for the computation of the shear viscosity.
The pion-pion scattering amplitude at tree-level is simply
\be \label{eq:amplitchiral} A_0 = \frac{s}{v^2} \frac{1}{1-\frac{s}{M_{\ts}^2}} \ , \ee

where $v(T)$ is the VEV $M_{\ts}$ is the Higgs mass. The amplitude is explicitly $\mathcal{O} (1/N)$.
Although we have traded the coupling constant in terms of other variables this counting is a reminiscence of the $1/N$ behavior of the quartic coupling constant.
In the large-$N$ limit (at fixed $NF^2$) the $s-$channel iteration of the amplitude $A_0$ is also of order $1/N$ so one needs to resum the infinite series containing an increasing number
of pion vertices and pion loops (see Fig.~\ref{fig:amplitudes}). The pion one-loop integral reads
\be I(s)=\frac{1}{16 \pi^2} \left[ N_{\epsilon} +2 + \ \textrm{Log } (-\mu^2/s)\right] \ , \ee
where we have used the dimensional regulation approach. The logarithm is to be understood as complex and the divergence is retained in the $N_{\epsilon}Ç$ factor. This factor is introduced
in the definition of the renormalized mass of the sigma and this provides a finite value for $I(s)$ that depends on the renormalization scale $\mu$.

The resummed amplitude reads
\be \label{eq:amplitresummed} A(s,t,u)=A(s)=\frac{s}{v^2} \frac{1}{1-\frac{s}{M_R^2(s)} + \frac{sN}{32 \pi^2 v^2} \textrm{Log } \left( \frac{-s}{\mu^2} \right)} \ . \ee

As $s>0$, we choose the branch cut of the logarithm along the positive $s$ axis. The complex logarithm reads $\textrm{Log }(-s) = log(s) + i \pi$.

When the pion mass is different from zero, the previous amplitude must be slightly modified:

\be A(s)=\frac{s}{v^2} \frac{1}{1-\frac{s}{M^2_R(\mu)}- \frac{sN}{32 \pi^2 v^2} \left[ \rho_{\pi \pi} \log \left| \frac{\rho_{\pi \pi}-1}{\rho_{\pi \pi}+1}\right| + i \pi \rho_{\pi \pi} - \log \left( \frac{m^2_{\pi}}{\mu^2}\right)\right] } \ , \ee
in order to include the two-body phase space that reads $\rho_{\pi\pi} (s)= \sqrt{1- \frac{4 m_{\pi}^2}{s}}$. In the limit of $m_{\pi} \rightarrow 0$ tends to the Eq.~(\ref{eq:amplitresummed}).

Moreover, the effect of this finite pion mass is to give more vertices that enter in the scattering amplitude, we will call these amplitude $A_m$. We take we expression of this amplitude from the reference \cite{Dobado:1994fd}. It reads:
\be A_m(s)= - \frac{m^2_{\pi}}{v^2} \frac{1+ \frac{2s}{M_R^2} - \frac{sN}{16 \pi^2 v^2} \log \left( \frac{m^2_{\pi}}{\mu^2}\right)}{\left\{ 1- \frac{s}{M_R^2} - \frac{sN}{32 \pi^2 v^2} \left[ \rho_{\pi \pi} \log \left| \frac{\rho_{\pi \pi}-1}{\rho_{\pi \pi}+1}\right| + i \pi \rho_{\pi \pi} - \log \left( \frac{m^2_{\pi}}{\mu^2}\right)\right] \right\}^2} \ . \ee

We now consider the partial amplitudes\index{partial amplitudes} projected on isospin channels, the amplitude $T^0$ is the dominant in the large-$N$ limit:
\be
\left\{
\begin{array}{ccc}
T^0 & = & N A(s) + A(t) + A(u) = N A(s) + \mathcal{O} \left( \frac{1}{N} \right) \ , \\
T^1 & = &  A(t) - A(u) =  \mathcal{O} \left( \frac{1}{N} \right) \ , \\
T^2 & = &  A(t) + A(u) =  \mathcal{O} \left( \frac{1}{N} \right) \ .
\end{array}
\right.
\ee

In terms of the definite isospin-spin partial amplitudes defined in Eq.~(\ref{eq:partampli}) the relevant low-energy amplitudes are the same as in the ChPT pion gas. The $N$-counting for each reads
\be
\left\{
\begin{array}{ccc}
t^{00} (s)& = & \frac{N A(s)}{32 \pi} + \mathcal{O} \left( \frac{1}{N} \right) \ , \\
t^{11} (s)& = &   \mathcal{O} \left( \frac{1}{N} \right) \ , \\
t^{20} (s) & = &   \mathcal{O} \left( \frac{1}{N} \right) \ .
\end{array}
\right.
\ee

The partial amplitude $t^{00}$ reads in terms of the amplitudes
\be t^{00} (s)= \frac{N}{32 \pi} \left[ A (s) + A_m(s) \right] \ . \ee

\begin{figure}[t]
\begin{center}
\includegraphics[scale=0.35]{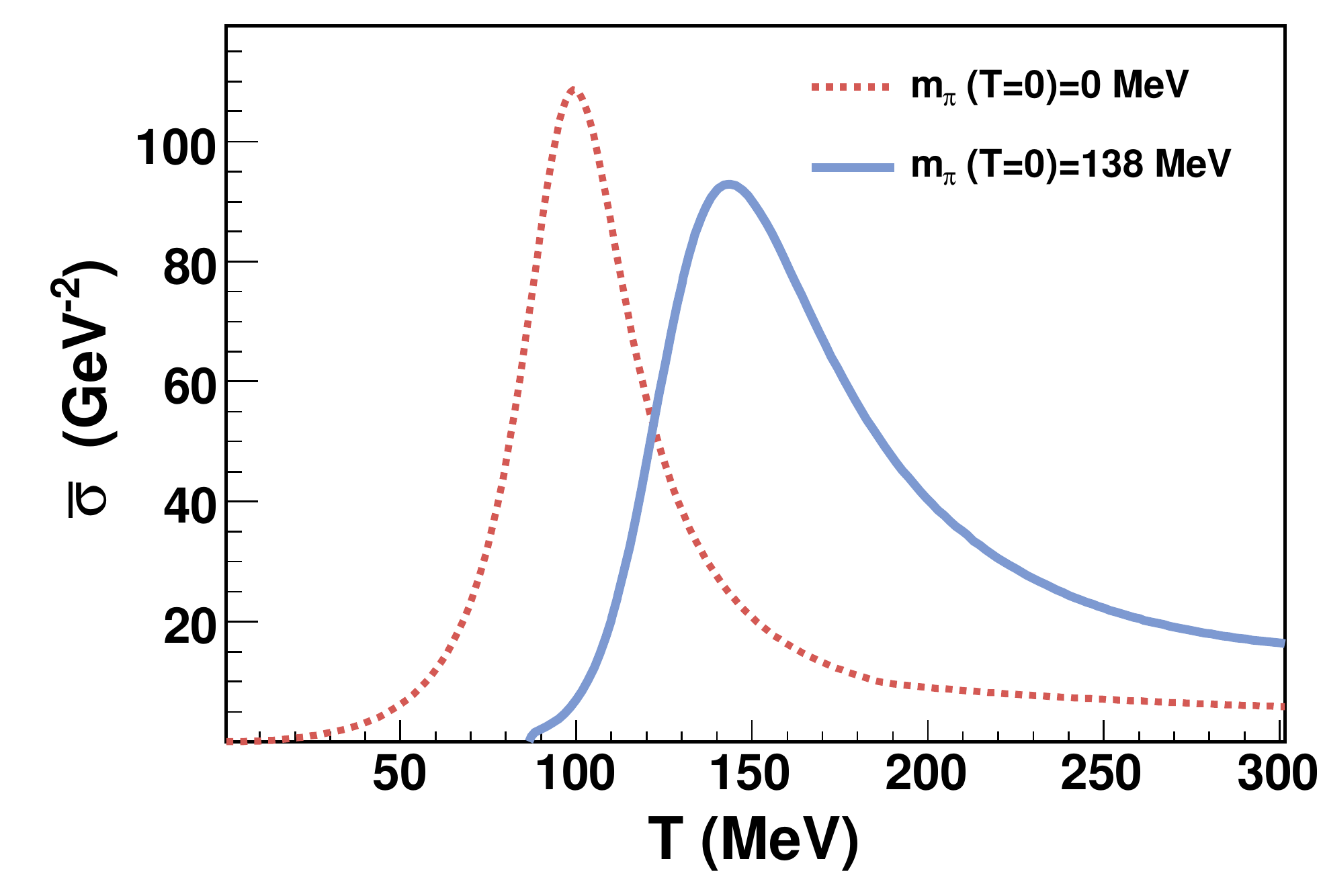}
\caption{\label{fig:lsmcrosss} Cross section for the linear sigma model at large $N$.}
\end{center}
\end{figure}

The total cross section for $m_{\pi}=0$ with averaged initial flavors is \cite{Dobado:2009ek}
\be \overline{\sigma}(s) = \frac{s}{32 \pi v^4} \frac{1}{\left[ 1 - \frac{s}{M_R^2} + \frac{sN}{32 \pi^2 v^2} \log \left( \frac{s}{\mu^2} \right) \right]^2+\left( \frac{sN}{32 \pi v^2}\right)^2 } \ . \ee
This cross section is plotted in Fig.~\ref{fig:lsmcrosss} for the cases with $m_{\pi}(T=0)=0$ (second order phase transition) and $m_{\pi} (T=0)=138$ MeV (crossover).

\section{Shear viscosity over entropy density}

The viscosity for a gas of pions in the L$\sigma$M can be obtained by an analogous derivation of that for the ChPT gas by substituing the pion isospin degeneracy $g_{\pi}$ by $N$.
As we always keep the pion mass finite in the numerical program to avoid complications, we can use the same adimensional variables in the calculation. 
At first order in the expansion of the function $B(x) \simeq b_0$ the viscosity reads
\be \eta = \frac{N m_{\pi}^6}{30 \pi^2 T^3} \frac{K_0}{\mathcal{C}_{00}} \ , \ee
where the pion mass $m_{\pi}$ is now taken from Eq.~(\ref{eq:pionmass}), $K_0$ is defined in Eq.~(\ref{eq:Kintegrals}) and $\mathcal{C}_{00}$ in Eq.~(\ref{eq:collision00}).
Note that as $\mathcal{C}_{00} \sim \mathcal{O} (1/N)$, then the shear viscosity is $\mathcal{O} (N^2)$. This result is expected \cite{Aarts:2004sd}, because the shear viscosity is naively proportional
to the inverse of the coupling constant squared, and the coupling constant is suppressed by one power of $N$.

\begin{figure}[t]
\begin{center}
\includegraphics[scale=0.35]{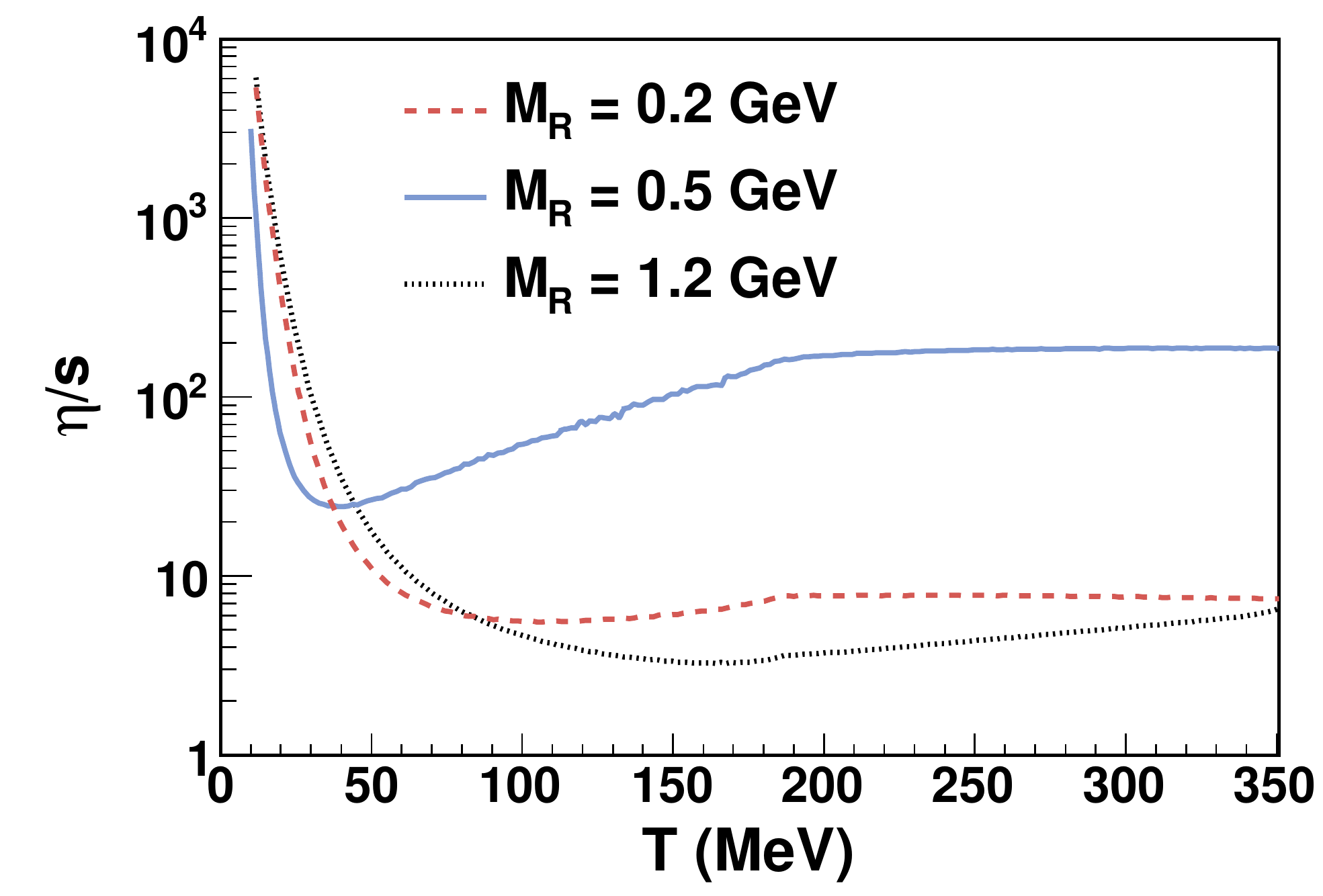}
\caption{\label{fig:kssdiffmr} Viscosity over entropy density in the linear sigma model at large $N$ for different values of $M_R$.}
\end{center}
\end{figure}

The result for $N=3$, $m_{\pi} (T=0)=0$ and different Higgs masses $M_R=0.2,0.5,1.2$ GeV is given in Figure~\ref{fig:kssdiffmr}. The result is similar to that appearing in our reference \cite{Dobado:2009ek} where we have
used a slighty different parametrization for the distribution function. We have found a unique minimum of $\eta/s$ for the three cases, always greater than the KSS value $1/(4\pi)$. However the position of the minimum
depends on the value of $M_R$.

\begin{figure}[t]
\begin{center}
\includegraphics[scale=0.35]{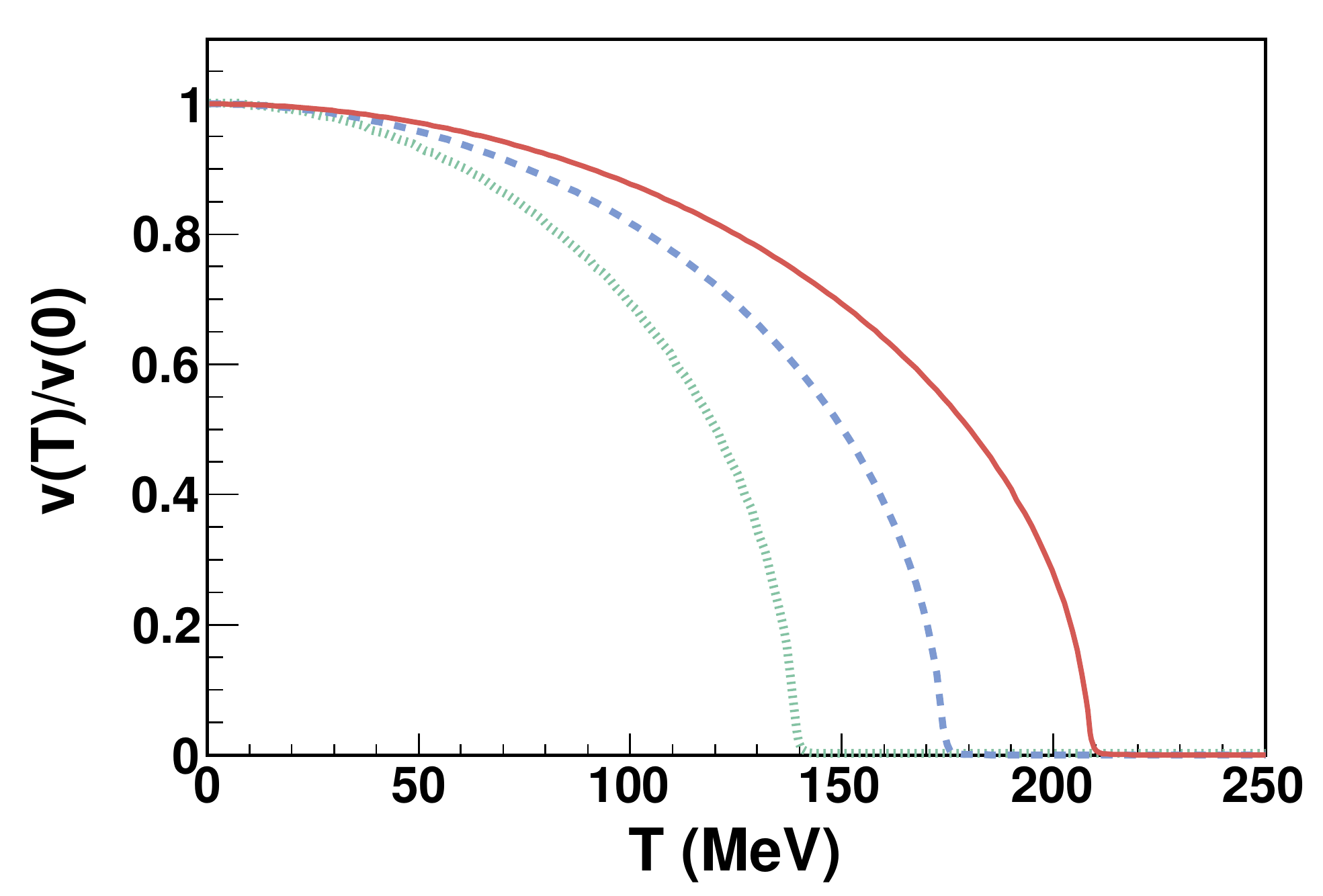}
\includegraphics[scale=0.35]{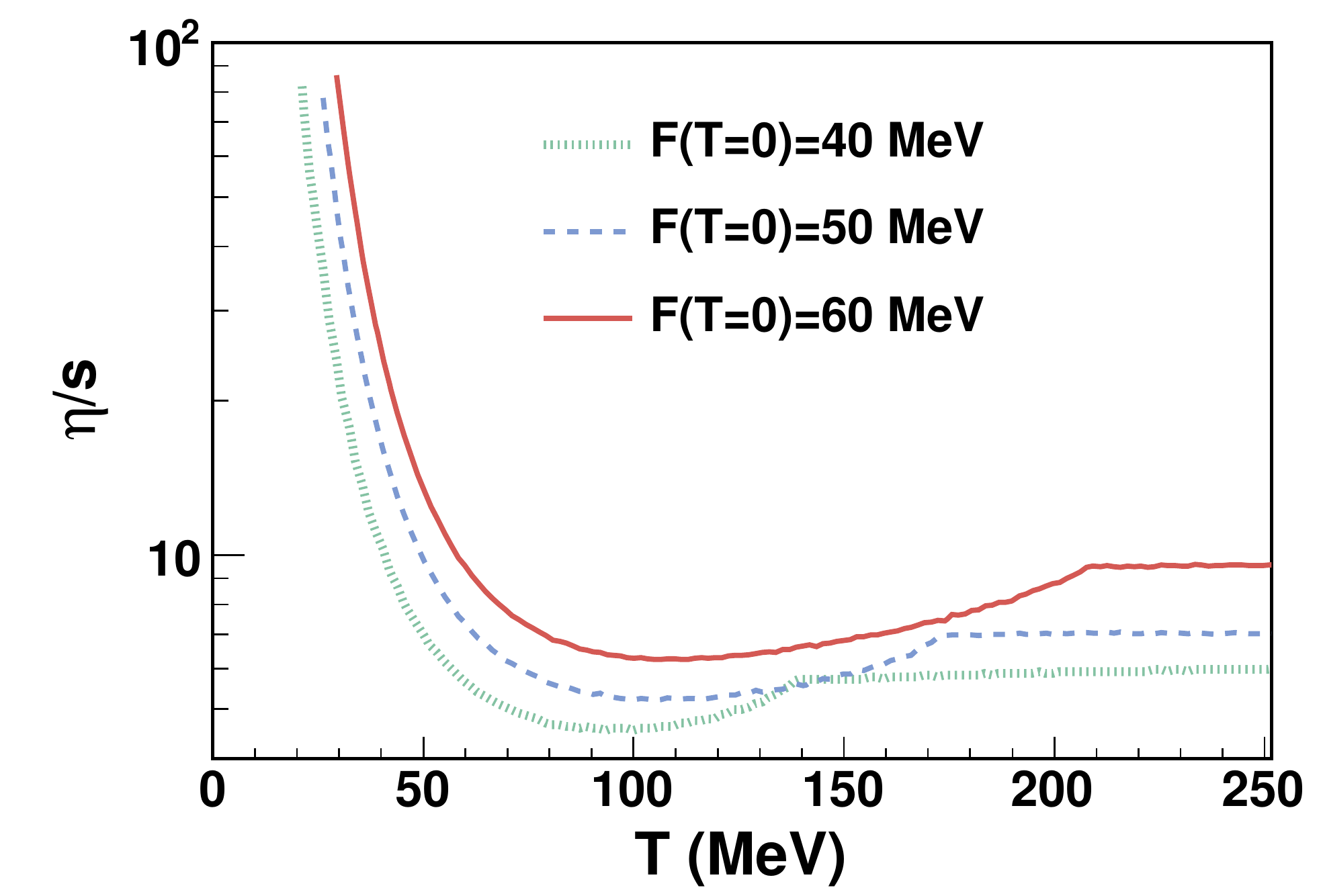}
\caption{\label{fig:ksskey} Comparison between the minimum of the viscosity over entropy density and the position of the critical temperature. The minimum is located close to it but slightly below.}
\end{center}
\end{figure}

To check whether the minimum of $\eta/s$ corresponds to the localization of the critical temperature one must compare the previous figure with the order parameter. We show this comparison in Figure \ref{fig:ksskey} at different
values of $F(T=0)$. In the left panel we show the normalized value of $v(T)$ showing that the position of the critical temperature (where the order parameter takes the value zero) depends linearly on $F$ as shown in Eq.~(\ref{eq:criticalT}).
The same behaviour is followed by the minimum viscosity over entropy density shown in the right panel. Moreover, the exact position of this minimum is near the $T_c$ shown in the left panel. However, the minimum is not exactly at $T_c$ as shown in 
Fig.~\ref{fig:ksskey} but slightly below.

\subsection{Validity of transport equation}

  The second hypothesis for deriving the Boltzmann equation in Section~\ref{sec:kin_eq} was that the mean free path should be much smaller than the interaction range. The mean free path depends inversely on the particle density
and the cross section:
\be \lambda_{mfp} \sim \frac{1}{n \ov{\sigma}} \ . \ee
 The interaction range can be expressed in terms of the scattering lenghts\index{scattering lenght} at low energies. More generically it is of the order of the square root of the cross section. Therefore, the condition
of applicability of the BUU equation reads:
\be \label{eq:lsmappli} \frac{1}{n \ov{\sigma}} \gg \sqrt{\ov{\sigma}} \ \rightarrow n \ov{\sigma}^{3/2} \ll 1 \ . \ee

\begin{figure}[t]
\begin{center}
\includegraphics[scale=0.32]{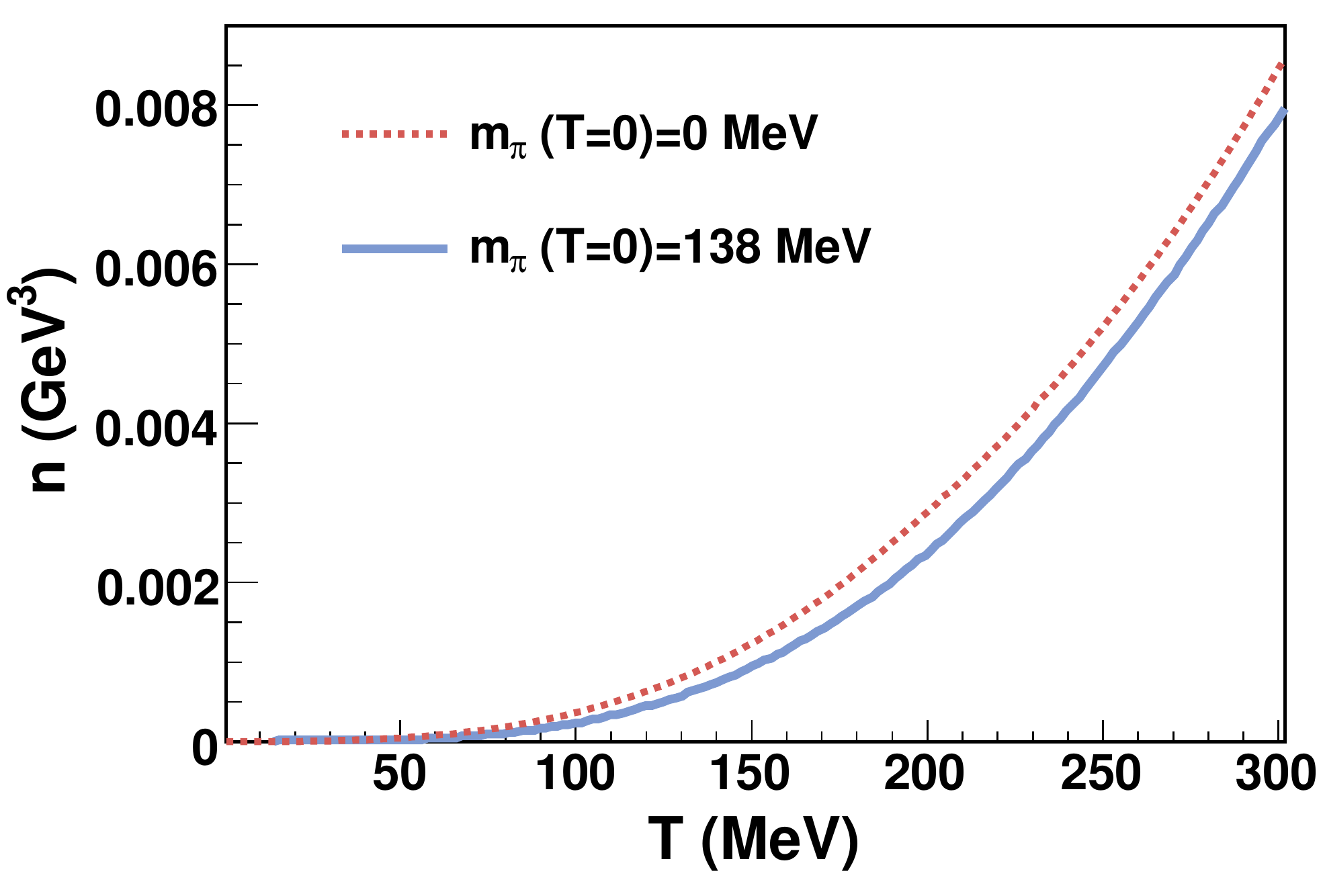}
\includegraphics[scale=0.32]{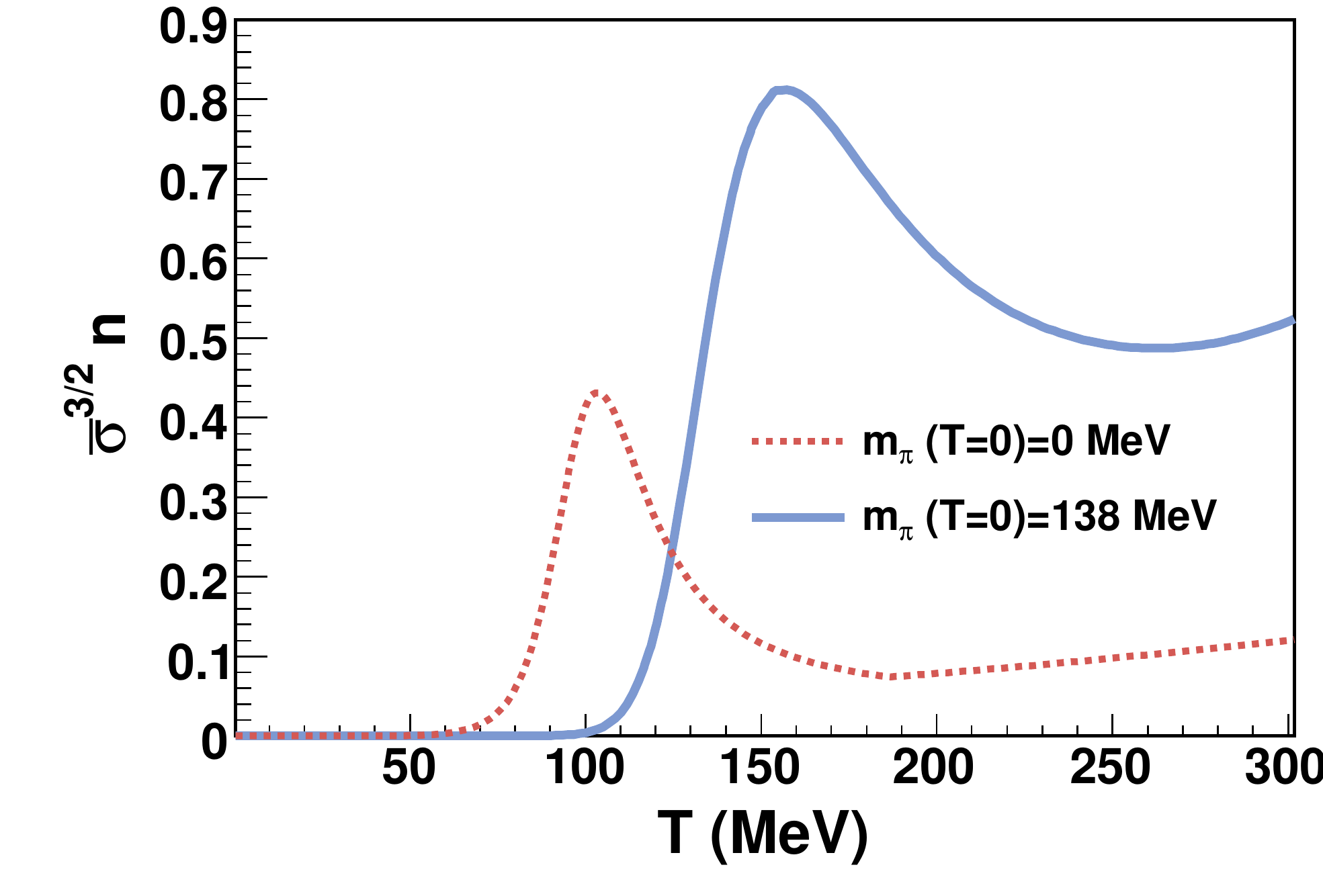}
\caption{\label{fig:lsmnden} Left panel: particle number density at equilibrium for the linear sigma model with $N=3$. Right panel: Product $\overline{\sigma}^{3/2}n$, that should be less than $1$ for the applicability of the BUU equation.}
\end{center}
\end{figure}

In Fig.~\ref{fig:lsmnden} we plot this product and see that the condition is satisfied for all temperatures. The interaction is not weak at the position of the peak of the cross section, at this point the condition (\ref{eq:lsmappli})
is less evident, however the value of the particle density makes the product be smaller than one. This does not happen at higher temperatures where the particle density is large and the dilute gas assumption is not valid anymore.

\chapter{Measurement of the Bulk Viscosity \label{ch:10.corre}}

The estimation of $\zeta/s$ by matching experimental measurements of flow coefficients
and the results from hydrodynamic simulations is hard for the bulk viscosity because of the difficulty of disentangle its effect
with respect to the bigger effect of the shear viscosity. Nevertheless, there have been attemps of adding the
effect of non-vanishing $\zeta/s$ along these research lines \cite{Song:2008hj,Bozek:2011wa,Bozek:2011ph}.
Another novel method \cite{Torrieri:2007fb} consists in extracting the bulk viscosity from the typical size of clusters close to the freeze-out,
 near the hadronization temperature. 

Here we propose a method to extract the bulk viscosity from the particle energy-momentum correlations. The basics of this method
follows a similar treatment developed by Gavin and Abdel-Aziz in \cite{Gavin:2006xd,Gavin:2007zz} to estimate the shear viscosity.

\section{Fluctuations and correlations of the stress-energy tensor}

Consider a system in thermal equilibrium at temperature $T$. An active degree of freedom needs to maintain an
energy of order $kT$ due to the energy equipartition theorem\index{energy equipartition theorem}. But this energy is dissipated according to the transport equations
in the medium. Therefore, the energy is a fluctuating statistical variable according to the macrocanonical description of thermodynamics.
This is the intuitive explanation of the fluctuation-dissipation theorem\index{fluctuation-dissipation theorem}\cite{Callen:1951vq}, as a necessity of energy equipartition.
Extending this fluctuation to all of the elements of the stress-energy tensor, one can expect that the shear and the bulk viscosity 
could be accesed by the fluctuations of some off-diagonal elements and its trace, respectively.
The stress-energy tensor is divided into an ideal and a dissipative part (\ref{eq:viscoustmunu}):
\be T_{\mu \nu} = w u_{\mu} u_{\nu} - P \eta_{\mu \nu} + \tau_{\mu \nu} \ , \ee
where the latter is encoding the shear and bulk viscosities together with first order velocity gradients and a fluctuating term. For a fluid element
at rest (\ref{eq:tau}):
\be \tau_{ij} = - \eta \left( \pa_{i} u_j + \pa_j u_i - \frac{2}{3} \delta_{ij} \mb{\nabla} \cdot \mb{u} \right) - \zeta \delta_{ij} \mb{\nabla} \cdot \mb{u} + t_{ij}\ . \ee

The fluctuating part of the stress-energy tensor vanishes when averaged 
\be \langle t_{ij} \rangle =0 \ . \ee
However the two-point correlation \index{stress-energy tensor!correlations of}function is related to the viscosities \cite{landau1987fluid}:
\be
\langle t_{ij}({\bf r}_1,t_1) t_{kl}({\bf r}_2,t_2)\rangle = 2T \ \delta({\bf r}_1-{\bf r}_2)\delta(t_1-t_2) 
\left[ \eta \left( \delta_{ik} \delta_{jl} + \delta_{il} \delta_{jk} \right) + \left(\zeta - \frac{2}{3} \eta\right) \delta_{ij} \delta_{kl}\right]  \ . 
\ee
To separate the bulk viscosity we make $j=i$ and $l=k$ and sum over $i$ and $k$:

\be \label{eq:bulkcorre} \langle t_{ii} ({\bf r}_1,t_1) t_{kk}({\bf r}_2,t_2)\rangle = 18 T \zeta \ \delta({\bf r}_1-{\bf r}_2)\delta(t_1-t_2) \ . \ee
The bulk viscosity can be extracted by measuring correlations over the fluctuation \index{stress-energy tensor!fluctuations of} of the trace of the stress-energy tensor $t_{ii}$.

\subsection{Local rest frame}

Consider first a fluid element in the rest frame, ignoring relativistic corrections. Taking cylindrical coordinates $(r,\phi,z)$ defined
in Chapter~\ref{ch:1.intro}, the trace of the stress-energy tensor is
\be \tau_{xx}+\tau_{yy}+\tau_{zz} = \tau_{rr} + r^2 \tau_{\phi \phi} + \tau_{zz} \ . \ee
Now, if we assume that there is cylindrical symmetry for perfectly central collisions, one expects that $\pa \mb{v}/ \pa \phi =0$, meaning 
that $\tau_{\phi \phi} =0$.
\be \langle \tau_{ii}({\bf r}_1,t_1) \tau_{kk}({\bf r}_2,t_2)\rangle =
\langle (\tau_{rr}+\tau_{zz})({\bf r}_1,t_1) (\tau_{rr}+\tau_{zz})({\bf r}_2,t_2)\rangle \ . \ee
Finally, we are left with three independent correlations, which will lead to the necessity of full energy reconstruction.

\subsubsection{Average equilibrium hypothesis}

Consider a large number of head-on collisions of Au+Au (RHIC) or Pb+Pb (LHC). As explained in Sec.~\ref{sec:multiplicity} those events with
highest multiplicity are considered to be the most central ones, thus belonging to the lowest centrality class\index{centrality}. If one has a large
database of equally prepared systems, the $\langle \rangle$ average symbol is then undestood as an average over all the recorded central
collisions.

The fundamental hypothesis underlying the analysis is that a state of hydrodynamic equilibrium is reached after the collision.
This is supported by a large body of data from the RHIC experiments and is widely assumed to be a good approximation to reality.
We require, as in \cite{Gavin:2006xd,Gavin:2007zz} that the average taken over all collisions coincides with the equilibrium state, i.e. the dissipative part of the stress-energy
tensor averages to zero $\langle \tau_{ij} \rangle=0$. We refer to this as the ``Average Equilibrium Hypothesis'' \index{average equilibrium hypothesis}and this excludes
systematic deviations from equilibrium that may affect all collisions, and in effect attributes deviations from equilibrium to event-by-event fluctuations.

This hypothesis is in the spirit of the Gibbs ensemble\index{Gibbs ensemble} and the ergodic hypothesis\index{ergodic hypothesis}, where the study of many copies of the same system at fixed time is
equivalent to the study over very large times of the fluctuations of one given system.

\subsubsection{Correlation of particle momenta}

As an example, we will detail the calculation of the correlation $\langle \tau_{rr} \tau_{rr} \rangle$ \index{correlation of stress-energy tensor}in terms of the detected
particle energy and momenta. The calculation of $\langle \tau_{zz} \tau_{zz} \rangle$ and the mixed correlation $\langle \tau_{rr} \tau_{zz} \rangle$
are analogously performed.

The number of particles per unit of phase space is
\be \frac{dN}{dx^3 d^3p} =  \frac{dn}{d^3p}=\frac{f_p}{(2\pi)^3} \ , \ee
where $f_p$ is the distribution function, that is separated in an equilibrium part and an out-of-equilibrium correction $f_p =n_p + \delta f_p$. The number of particles is
\be \label{eq:nopart} dN = (n_p + \delta f_p) \ d^3x \  \frac{d^3p}{(2\pi)^3} \ . \ee

Consider the following correlator
\be \mathcal{C}_{rr} = \langle \sum_{\textrm{all }ij}  \frac{(p_{ri})^2}{E_i} \frac{(p_{rj})^2}{E_j}   \rangle \ . \ee
The sum over $i, j$ runs over all pairs of particles in a given collision event (including the square of the function of each particle).
Note that we include both charged and neutral particles. Additionally, we assume that almost all of the detected particles are pions, in order
to finally obtain the bulk viscosity of a pion gas. After constructing the sum over all particles, one takes the average over all events of the data sample. Using Eq.~(\ref{eq:nopart}) we obtain

\begin{eqnarray}
\mathcal{C}_{rr} & = &  \int dn_1 dn_2  \langle   \frac{(p_{r1})^2}{E_1} \frac{(p_{r2})^2}{E_2} \rangle 
=  \int\frac{d^3p_1}{(2\pi)^3}\frac{d^3p_2}{(2\pi)^3} d^3x_1 d^3 x_2 \langle   \frac{(p_{r1})^2}{E_1} \frac{(p_{r2})^2}{E_2} 
f (1,2) \rangle \nonumber \\
&=& \int\frac{d^3p_1}{(2\pi)^3}\frac{d^3p_2}{(2\pi)^3} d^3x_1 d^3 x_2 
\langle   \frac{(p_{r1})^2}{E_1} \frac{(p_{r2})^2}{E_2} 
(n_{p1}+\delta f_{p1})(n_{p2}+\delta f_{p2}) \rangle \ . 
\end{eqnarray}
In the previous step we have factorized the two-particle distribution function $f (1,2)$ into the product of two one-particle distribution functions.

Additionally, invoking the Average Equilibrium Hypothesis we can ignore the terms linear in $\delta f_p$ since under the average symbol
$\langle \delta f_p \rangle =0$ (no systematic out-of-equilibrium effects). Finally,
\be \label{eq:Crrcorre} \mathcal{C}_{rr} = \langle N \rangle^2 \langle \frac{p^2_r}{E} \rangle^2 + \int d^3 x_1 d^3 x_2 \langle \tau_{rr} ( \mb{x}_1)
\tau_{rr} (\mb{x}_2) \rangle , \ee
with $N$ the total particle multiplicity in an individual event. We have used Eq.~(\ref{eq:stressmicro}) in the last step.

Equation~(\ref{eq:Crrcorre}) relates the fluctuations of the stress-energy tensor to the particle momenta as measured in a detector, but includes both the stochastic force $t_{ij}$
whose correlator reveals the bulk viscosity and the hydrodynamic non-fluctuating part $\tau^{hyd}_{ij}$. One can substitute directly $\tau_{ij}$ by $t_{ij}$ in the terms linear in $t_{ij}$ due to
\be \langle \tau_{rr} (\mb{x}) \rangle = \langle t_{rr} (\mb{x}) \rangle = \langle \int \frac{d^3p}{(2 \pi)^3} \frac{(p_r)^2}{E} \delta f_p \rangle =0\ee
under the Average Equilibrium Hypothesis\index{average equilibrium hypothesis}. This leaves the quadratic terms \\ $\langle \tau^{hyd}_{rr}({\bf x}_1,t_1) \tau^{hyd}_{rr}({\bf x}_2,t_2)\rangle$ and
$\langle t_{rr}({\bf x}_1,t_1) t_{rr}({\bf x}_2,t_2)\rangle $.

For calculating the shear viscosity in \cite{Gavin:2006xd,Gavin:2007zz} they take the correlator of the hydrodynamic part. This correlator satisfies a diffusion equation whose exponential solution
decays in time with a characteristic diffusion time. From this time they read off the shear viscosity. In our case we will take the correlator of the fluctuating force $t_{ij}$ from which one
can have access to the viscosity as well. To separate the two terms we perform a time-integration over a small $\Delta \mathcal{T}$, this allows to separate the stochastic force (that is proportional
to $\delta(t_1-t_2)$ and the hydrodynamic part (that features a mild time dependence $e^{-\lambda t}$). This works as follows.
A general fluctuating force, $x$ that obeys an equation like \cite{landau1984cours}
\be \frac{dx}{dt} = - \lambda x + y \ , \ee
has a time-correlation function with the shape
\be \langle x(t_1) x(t_2) \rangle = \langle x^2 \rangle e^{-\lambda |t_1-t_2|} \ . \ee
Performing a double integration in time,
\be \int_{\Delta \mathcal{T}} dt_1 \int_{\Delta \mathcal{T}} dt_2 \ \langle x(t_1) x(t_2) \rangle = \langle x^2 \rangle (\Delta \mathcal{T})^2 + \mathcal{O} ( (\Delta \mathcal{T})^3) \ , \ee
one sees that the solution is a second order infinitesimal.
However, doing the same integration over the stochastic part of the right-hand side of Eq.~(\ref{eq:bulkcorre}) (equivalent to the correlator $\langle y(t_1) y(t_2) \rangle$ )
\be \int_{\Delta \mathcal{T}} dt_1 \int_{\Delta \mathcal{T}} dt_2 \ \delta(t_1-t_2) = \Delta \mathcal{T} \ee
we obtain that it is a first order infinitesimal. Thus, the hydrodynamic correlation is of one order lesser and it is possible to separate this correlation from the stochastic one. Under the integration 
one can safely exchange $\tau_{rr} (\mb{x}_1) \tau_{rr} (\mb{x}_2)$ by $t_{rr} (\mb{x}_1) t_{rr} (\mb{x}_2)$.

Finally, the expression relating the stress-energy tensor fluctuations and the experimental observable in terms of particle momenta is,

\ba \label{eq:localobser}
\int_{\Delta {\mathcal{T}}} dt_1  \int_{\Delta {\mathcal{T}}} dt_1  
\int_{\Delta V} d^3x_1 \int_{\Delta V} d^3 x_2 \langle t_{rr}({\bf x}_1,t_1) t_{rr}({\bf x}_2,t_2)\rangle 
= \nonumber \\ \Delta {\mathcal{T}} \left(
\langle \sum_{{\rm all\ }ij} \frac{(p_{ri})^2}{E_i} \frac{(p_{rj})^2}{E_j} \rangle 
- \langle N \rangle^2 \langle \frac{p_r^2}{E} \rangle^2 \right) \ ,
\ea
where the average on the last term is a double average over both the particles in an event and the sample of events under certain kinematic cuts. $\langle N \rangle$ is by the average multiplicity, and the particles are supposed
 to have been emitted during the small interval $\Delta \mathcal{T}$.

The experimental observable proposed in Eq.~(\ref{eq:localobser}) can be achieved by measuring particle momenta and energies alone.
The integration over space cannot be extended to the entire collision volume, since different fluid elements have wildly different velocities,
and we have considered the local rest frame of the fluid. We will lift this restriction in the next section.

For the time being, take a fluid element in the small volume $\Delta V$ characterized by small rapidity and tranverse velocity so that the nonrelativistic analysis is a reasonable starting point.
Then, integrating the Eq.~(\ref{eq:bulkcorre}) over this volume and the time duration of the particle emission $\Delta \mathcal{T}$, we have

\ba \nonumber 18T \ \zeta \ \frac{\Delta V}{\Delta {\mathcal T}}  & = & 
\langle \sum_{{\rm all\ }ij} \frac{(p_{ri})^2}{E_i} \frac{(p_{rj})^2}{E_j} \rangle 
- \langle N \rangle^2 \langle \frac{p_r^2}{E} \rangle^2 
+
\langle \sum_{{\rm all\ }ij} \frac{(p_{zi})^2}{E_i} \frac{(p_{zj})^2}{E_j} \rangle \\
& & - \langle N \rangle^2 \langle \frac{p_z^2}{E} \rangle^2 
 +  2 \langle \sum_{{\rm all\ }ij} \frac{(p_{ri})^2}{E_i} \frac{(p_{zj})^2}{E_j} \rangle 
- 2 \langle N \rangle^2 \langle \frac{p_r^2}{E} \rangle \langle \frac{p_z^2}{E} \rangle \ .
\ea

One can express the last equation in terms of each particle's energy and mass by noting that $p_r^2+p_z^2 = E^2-m^2$ as
\be \label{eq:bulkcorre2} \nonumber 18T \ \zeta \ \frac{\Delta V}{\Delta {\mathcal T}}  =
\langle \sum_{{\rm all\ }ij} \frac{(E^2- m^2)_i (E^2- m^2)_j}{E_i E_j}
\rangle -  \langle N \rangle^2 \langle \frac{E^2-m^2}{E} \rangle^2
\equiv \Delta \left( \frac{E^2-m^2}{E} \right) \ .
\ee

Under the assumption of purely radial transverse flow (no vorticity) one can identify $p_r=p_{\perp}$, the perpendicular particle momentum.
In the left-hand side there still remains to extract the emision time and volume (hydrodynamical problem) and the temperature of the system,
that can be obtained by other observations such as photon or particle spectra.

\subsection{Boosted fluid element}

The fluid element in the nuclear explosion is boosted in the laboratory frame. Now we will leave the rest frame assumption to include a 
boost of the fluid element. If the fluid four-velocity is denoted by $u^{\mu}$ the Eq.~(\ref{eq:bulkcorre2}) can be taken to the laboratory frame by introducing the time-dilatation
factor $\gamma=(\sqrt{1-\beta^2})^{-1}$ and noting that $E_i= p^{\mu}_i u_{\mu}$. The result is
\be \label{eq:bulkinv} 18T \ \zeta \ \gamma^2 \frac{\Delta V_{\rm lab}}{\Delta {\mathcal T}_{\rm lab}}  = \Delta \left( \frac{(p\cd u)^2-m^2}{p\cd u} \right) \ , \ee
with
\be \Delta \left( \frac{(p\cd u)^2-m^2}{p\cd u} \right) \equiv
\langle \! \sum_{{\rm all\ }ij}\!\! \frac{((p\cd u)^2\! -\! m^2)_i ((p\cd u)^2\! -\! m^2)_j}{(p\cd u)_i (p\cd u)_j}
\rangle -  \langle \! N \! \rangle^2 \langle\! \frac{(p\cd u)^2\! -\! m^2}{p\cd u} \! \rangle^2 \ . \ee

Let us assume that one has identified a set of kinematic cuts that select a swarm composed of those particles coming from the fluid element $\Delta V_{\rm lab}$ during
the time interval $\Delta \mathcal{T}_{\rm lab}$. The fluid element's rest frame will coincide with the center of mass frame. Therefore its velocity can be obtained from the particle swarm's energy-momentum
in the laboratory frame as
\be \label{eq:defbeta} \mb{\beta} = \frac{\sum_i \mb{p}_i}{\sum_i E_i} \ , \ee
and
\be u^{\mu}= \gamma (1,\mb{\beta})  \ . \ee
Once $u^{\mu}$ corresponding to the fluid element has been so constructed, one can compute all the products $(p \cdot u)_i$ as
\be (p \cdot u)_i=p^{\mu}_i u_{\mu}= \gamma (E_i - \mb{p}_i \cdot \mb{\beta}) \ . \ee 
The four-velocity $u^{\mu}$ satisfies $u^{\mu} u_{\mu} = 1$ and can be parametrized by
\be u^{\mu} = \left( \sqrt{1+u^2_{\perp}} \cosh \eta, u_{\perp} \cos \phi, u_{\perp} \sin \phi, \sqrt{1+u^2_{\perp}} \sinh \eta \right) \ , \ee
where $\eta$ is the pseudorapidity \index{pseudorapidity} variable, analogously defined as the particle pseudorapidity in Eq.~(\ref{eq:pseudorapidity}):
\be \eta = \frac{1}{2} \log \left( \frac{P+P_z}{P-P_z} \right) \ . \ee

Let us address the fluid's element' space and time sizes $\Delta V_{\rm lab}$, $\Delta \mathcal{T}_{\rm lab}$.
This requires understanding of the hydrodynamics of the expanding fireball, and here we will contempt ourselves with the simplest of models,
a spherical expansion characterized by a freeze-out surface\index{freeze-out time} at time $\tau_f$ (this is a valid approximation if the formation radius
 is much smaller, $\tau_0\ll \tau_f$, else the polar caps of the sphere are distorted, and if the elliptic flow is moderately small). The total
 swarm's longitudinal momentum will be $P_z$ in the direction of the heavy-ion beam and is usually traded by pseudorapidity\index{pseudorapidity}.

We will consider pure radial flow, so that the swarm's perpendicular momentum $\mb{P}_{\perp}$ in the transverse plane is parallel to the radial vector
in cylindrical coodinates. The radial direction is automatically determined by the measurement of $P_{\perp}$ for the swarm.
We will express $\Delta V_{\rm lab}$ and $\Delta \mathcal{T}_{\rm lab}$ in terms of the momentum spread of the chosen particle swarm, centered around energy $E$, transverse
momentum $P_{\perp}$, azimuthal angle $\phi$ and pseudorapidity $\eta$.

In the time of kinetic freeze-out $\tau_f$ the particle travelled a distance $\rho=\tau_f \beta_{\perp}$ from the origin ($\beta_{\perp} = P_{\perp} /E$). A particle
arriving at the freeze out distance a time $\Delta \mathcal{T}_{\rm lab}$ later will have lagged by $\rho/ \Delta \beta_{\perp}$. Therefore
\be \Delta {\mathcal T}_{\rm lab} = \tau_f \frac{\Delta \beta_{\perp}}{\beta_{\perp}} = \rho \frac{\Delta \beta_{\perp}}{\beta_{\perp}^2} \ee
and differentiating $E=m\gamma$
\be
\Delta {\mathcal T}_{\rm lab} =\frac{\tau_f}{P_{\perp}^2} \frac{\Delta E}{E} m^2 \ .
\ee
Turning now to spatial cylindrical coordinates,
\be
\Delta V_{\rm lab}\equiv \Delta z  \rho \Delta \rho \Delta \phi\ .
\ee
The longitudinal velocity gives $\Delta z = \tau_f \Delta \beta_z$.
Likewise, $\Delta \rho = \tau_f \Delta \beta_\perp$.
Altogether, employing again the definition of $\beta$ in terms of the total energy and momentum in Eq.~(\ref{eq:defbeta}),
\be \label{eq:finalvolume}
\Delta V_{\rm lab} = \tau_f^3 \Delta \phi \frac{P_{\perp}}{E} \left[ \left( \frac{1}{P_z}-\frac{P_z}{E^2} \right) \Delta E -\frac{1}{E \sinh \eta} \Delta P_{\perp} \right]
\left( \frac{\Delta P_\perp}{E}-\frac{P_\perp}{E^2} \Delta E \right)  \ .
\ee
Finally, eliminating $ \Delta P_z$ in terms of $\Delta P_{\perp}$ and $\Delta E$, we find

\be \label{eq:volandtime}
\frac{\Delta V_{\rm lab}}{\Delta {\mathcal T}_{\rm lab}} =
\tau_f^2 \frac{\Delta \phi}{\Delta E} \frac{P^3_{\perp}}{m^2} \left[ \left( \frac{1}{P_z}- \frac{P_z}{E^2}\right) \Delta E - \frac{1}{E \sinh \eta} \Delta P_{\perp} \right]
\left( \frac{\Delta P_\perp}{E}-\frac{P_\perp}{E^2} \Delta E \right) \ . 
\ee

Note that differentiating the invariant mass of the swarm
$$
M^2=E^2-{\bf P}^2
$$
the three cuts $\Delta E$, $\Delta P_\perp$ and $\Delta \eta$ are not independent, satisfying the constraint

\be E \Delta E = \cosh \eta \sinh \eta P_{\perp}^2\Delta \eta + \cosh ^2 \eta P_{\perp} \Delta P_{\perp} \ . \ee

Putting all together we obtain the final formula for the bulk viscosity. The {\it modus operandi} is the following. Define three appropiate kinematic cuts $\Delta \phi$,
$\Delta P_{\perp}$ and $\Delta E$ defining a swarm of particles centered around $\phi, P_{\perp}$ and $E$ to a set of recorded central collision events.
To choose the appropriate cuts we study the efficiency dependence by means of a Monte Carlo simulation.

Then, the estimate for the bulk viscosity \index{bulk viscosity!measurement} is obtained by substituing Eq.~(\ref{eq:volandtime}) into Eq.~(\ref{eq:bulkinv}) to give
\be \label{eq:zetageneral} \zeta = 
 \frac{E^3 \Delta E m^2}{18T_f\gamma^2 \tau_f^2 \Delta \phi P^3_\perp} 
 \Delta \left( \frac{(p\cd U)^2-m^2}{p\cd U} \right) \frac{1}{\left[ \left( \frac{E}{P_z} - \frac{P_z}{E} \right)
\Delta E- \frac{1}{\sinh \eta} \Delta P_{\perp} \right] \left( E \Delta P_\perp-P_\perp \Delta E \right)} \ ,
\ee
that depends on the freeze-out temperature $T_f$ \index{freeze-out!temperature} and the freeze-out time\index{freeze-out!time} $\tau_f$. These can be obtained from other measurements and then grant access to the bulk viscosity. 
The temperature can be obtained by fitting the low $p_{\perp}$ particle multiplicity to a thermal distribution (Bose-Einstein for pions) as we explained for the
ALICE distribution in Section~\ref{sec:thermalspec}. The freeze-out time can be estimated as indicated in Eq.~(\ref{eq:tauf}) by using the Hanbury-Brown-Twiss interferometry\index{HBT interferometry}. 

For midrapidity\index{midrapidity} ($\eta \simeq 0$) one can use the approximate formula:
\be \label{eq:zetaalice} \zeta \simeq \frac{1}{18T_f\gamma^2 \tau_f^2} \frac{1}{\Delta \phi \Delta \eta} \frac{m^2}{P^3_{\perp}} \frac{E^3}{E^2- P^2_{\perp}} \Delta \left( \frac{(p\cd U)^2-m^2}{p\cd U} \right) \ , \ee
that is independent of $\Delta P_{\perp}$.

In this particular case, one can divide the bulk viscosity over the entropy density in Eq.~(\ref{eq:entropytf}). If all the magnitudes are expressed in terms of GeV the final result reads
\be \label{eq:zetaoversalice} \frac{\zeta}{s}  \simeq 6.4 \cdot 10^{-4} \ \frac{\sqrt{T_f (\textrm{GeV})}}{\gamma^2 \Delta \phi} \frac{m^2}{P^2_{\perp}}  \frac{E^2}{E^2- P^2_{\perp}} \Delta \left( \frac{(p\cd U)^2-m^2}{p\cd U} \right) \ , \ee
where we have particularized some parameters for the ALICE results, viz. $\Delta \eta=1.8$ for the whole pseudorapidity interval of the detector, $dN_{ch}/d\eta=1601$ \cite{Aamodt:2010cz} and we have used the relation (\ref{eq:tauf}) to
simplify the final expression.

In the Sec.~\ref{sec:kinematiccuts}, we study a sample of possible kinematic cuts and their efficiency for the hypothesis of the particles in a swarm of given kinematic cuts come
from the same fluid element created by the thermal distribution in the freeze-out time.

\section{Kinematic cuts \label{sec:kinematiccuts}}

In this section we discuss the choices for the kinematic cuts, particularly 
$\Delta P_\perp$, $\Delta \phi$ that are workable for an experimental collaboration, considering especially the ALICE experiment at the LHC. In devising them, we have to compromise between several constraints.

\begin{itemize}
\item First, since our method calls for the separation of an interval $\Delta \mathcal{T}$
smaller than the lifetime of the entire collision, to isolate the fluctuations, we need to consider a fluid element that is actually in motion and provides us with a clock. Therefore we will need to impose a $p_\perp$ cut that excludes $p_\perp=0$.
\item
Second, not all particles in a swarm move parallel enough to the average velocity $U$ and may end up in a different element of phase space.
To quantify the theory error introduced by this effect we have written a small Monte Carlo program described shortly.
\item
Third, the phase space element chosen for the measurement needs to contain enough particles across the collision data base to make a measurement possible.
\item
Fourth and last, we have to consider that ALICE's pseudorapidity acceptance is limited to the interval $(-0.9,0.9)$ (the barrel spans about 46 degrees in polar angle to each side of the collision point).
\end{itemize}

The crux of the matter is in the second point. The pion emission due to the freeze out of a fluid element at rest can approximately be described by a Bose-Einstein distribution in momentum $p$,
\be \label{eq:MB1}
\frac{dN}{N} = C \frac{p^2dp}{e^{(\sqrt{p^2+m^2}-\mu)/(k_BT)}-1}
\ee
characterized by a temperature $T$ and chemical potential $\mu$.
This emission is isotropic in the rest-frame of the fluid, but if the fluid element is boosted, the boost velocity has to be compounded with the particle velocity (according to the special-relativistic velocity
transformation rule). If the boost velocity is large enough, it dominates the composition. Most particles are emitted aligned with $\beta$.\\
However, if the boost velocity is of order of the Bose-Einstein velocity allowed by this distribution, the emission becomes less beamed and each element of phase space is populated by particles emitted from different fluid elements.

In view of our fourth point above, since the longitudinal boost accepted by the ALICE\index{ALICE collaboration} detector has at most $|\eta| \simeq 0.9$, we will consider the central part of
 the collision, that is, take the entire longitudinal acceptance as one bin with $\eta=0$, $\Delta \eta\simeq 1.8$. Neglect of longitudinal momentum allows to write Eq.~(\ref{eq:MB1}) in terms of the transverse momentum alone as
\be \label{eq:MB2}
\frac{dN}{N} = C \frac{p_\perp^2dp_\perp}{e^{(\sqrt{p_\perp^2+m^2}-\mu)/(k_BT)}-1} \ .
\ee

To assess the kinematic cuts we proceed by writing a Monte Carlo program. Employing Von Neumann's rejection method we generate a sample of several thousands of pions (corresponding to a few simulated collision events) 
distributed at random in $\phi$ and according to the ALICE  experimental $p_\perp$ distribution~\cite{Appelshauser:2011ds} in 900 GeV p+p collisions, that is well fit by an ad-hoc formula
\be 
\label{eq:Alicefit}
 \frac{1}{N_{ev}} \frac{1}{2\pi p_{\perp}} \frac{d^2 N_{ch}}{d \eta dp_{\perp}} = \left\{
\begin{array}{cc} 
 11.47 \ e^{-4.10 p_{\perp}} & p_{\perp} < 1.7 \textrm{ GeV} \\
 0.25 \ p_{\perp}^{-5.95} & p_{\perp} > 1.7 \textrm{ GeV}
\end{array} \ .
\right.
\ee
This we call {\emph{defining sample}} and is only used to construct average boost velocities.\footnote{
Incidently, the same data~\cite{Appelshauser:2011ds} taken at low $p_{\perp}$ can be used to fit the rest-frame Bose-Einstein thermal distribution parameters
(temperature and pion chemical potential) in Eq.~(\ref{eq:MB2}).}

To explore pairs of $(\Delta p_{\perp},\Delta \phi)$ cuts  
we select the pions from the {\emph{defining sample}} whose momenta fall within the so chosen fluid cell. We sum their momenta and energy to construct the cell's velocity according to Eq.~(\ref{eq:defbeta}).

Once the fluid cell has been defined and the average velocity is known, we turn to Eq.~(\ref{eq:MB2}) and generate a second sample of thermally distributed pions in the rest frame, also by Von Neumann's rejection method, the {\emph{thermal sample}}. \\
This sample represents isotropic emission in the fluid's rest frame and we impose no restriction on $p_{\perp}$ or $\phi$ except thermal distribution. 
\\
Finally, we apply the Lorentz boost with the velocity from Eq.~(\ref{eq:defbeta}) corresponding to the fluid cell to each of the pions in the {\emph{thermal sample}}, and examine what fraction of them falls outside of
the initial kinematic cuts that defined the fluid cell. 

We find that a non-negligible but controllable percentage of the sample pions end up into a different fluid cell. The results are listed in Table \ref{tab:MC_results} as percentages of particles appearing with momenta
that would correspond to a fluid cell other than used to generate them.

For completeness we also address ALICE's Pb+Pb data\index{ALICE collaboration} at $\sqrt{s}=2.76$ TeV.
We fit the $p_\perp$ distribution in analogy with Eq.~(\ref{eq:Alicefit}) by
\be \label{eq:Alicefit2}
 \frac{1}{N_{ev}} \frac{1}{2\pi p_{\perp}} \frac{d^2 N_{ch}}{d \eta dp_{\perp}} = \left\{
\begin{array}{cc} 
  8.90 \cdot 10^5 \ e^{-2.93 p_{\perp}} & p_{\perp} < 2.0 \textrm{ GeV} \\
  2.00 \cdot 10^5 \ p_{\perp}^{-6.29} & p_{\perp} > 2.0 \textrm{ GeV}
\end{array}
\right.
\ee
to obtain the corresponding {\emph{defining sample}} and repeat the analysis (obtain each cell's velocity, generate a {\emph{thermal sample}}, boost the pions thereof and examine their final momenta). The corresponding
 result is given in Table~\ref{tab:MC_results2}.

\begin{table}[t]
\begin{centering}
\begin{tabular}{|c|cc|cc|cc|} \hline 
 & \multicolumn{2}{|c|}{All $p_{\perp}$} & \multicolumn{2}{|c|}{$p_{\perp} > 0.3$ GeV} & \multicolumn{2}{|c|}{$p_\perp > 0.5$ GeV}  \\ 
\hline \hline
& $\beta$ & \% & $\beta$ & \%  & $\beta$ & \%  \\ 
\hline \hline
$\Delta \phi=\pm 20 ^{\circ}$ & 0.93& 41.6 & 0.96 & 36.5 & 0.97  & 36.7 \\ \hline
$\Delta \phi=\pm 30 ^{\circ}$ & 0.91& 31.9 & 0.93 & 33.3 & 0.94 & 36.9  \\ \hline
$\Delta \phi=\pm 45 ^{\circ}$ & 0.86& 24.7 & 0.88 & 33.7 & 0.89 & 42.4  \\ \hline
$\Delta \phi=\pm 60 ^{\circ}$ & 0.79& 18.6 & 0.81 & 35.7 & 0.82 & 49.2  \\ \hline \hline
& \multicolumn{2}{|c|}{ $p_{\perp} \in (0.3,2)$ GeV} & \multicolumn{2}{|c|}{$p_{\perp} \in (0.3,3)$ GeV} & &  \\ 
\hline \hline
& $\beta$ & \% & $\beta$ & \%  & &  \\ 
\hline \hline
$\Delta \phi=\pm 20 ^{\circ}$ &  0.96 & 62.0 & 0.96 & 48.8 & & \\ \hline
$\Delta \phi=\pm 30 ^{\circ}$ &  0.93 & 51.0 & 0.93 & 39.9 & & \\ \hline
$\Delta \phi=\pm 45 ^{\circ}$ &  0.88 & 42.2 & 0.88 & 35.6 & & \\ \hline
$\Delta \phi=\pm 60 ^{\circ}$ &  0.81 & 39.6 & 0.81 & 36.3 & & \\ \hline 
\end{tabular}
\caption{We show average velocity $\beta$ of the given swarms of particles within the azimuthal angular $\Delta \phi$ and the transverse momentum $\Delta p_\perp$ kinematic cuts, with the particles distributed according to Eq.~(\ref{eq:Alicefit}) 
corresponding to proton-proton collisions. We also show, for each given binning with velocity $\beta$, the percentage of thermally emitted particles following Eq.~(\ref{eq:MB2}) that are lost from the bin after compounding $\beta$ with the
 particle's thermal velocity. Typical results show that a fourth to a third of particles with well-chosen cuts populate other fluid elements introducing an irreducible theory error. \label{tab:MC_results} }
\end{centering}
\end{table}

Examination of Table~\ref{tab:MC_results} teaches several general 
lessons.
\begin{itemize}
\item
If the boost velocity is generally larger (the average momentum is at higher $p_{\perp}$), pions do not spread out too much and losses from the cell are lowered. 
\item If the azimuthal-angle cut $\Delta \phi$ is larger, losses from the cell are in general smaller because, after boosting the
{\emph{thermal sample}}, most pions remain inside this larger cone.
\item If  on the other hand the azimuthal-angle cut is very small, low momentum particles find it easy to leave the  tiny resulting angular cone.
One can reduce the mixing between fluid cells by proceeding to larger $p_\perp$ so the boost focuses the swarm in the correct direction. 
 \item In the extreme case, if the momentum cut is centered at huge momenta, the cell's $\beta$ is very close to $1$.
Almost independently of the initial thermal configuration most of the particles follow the boost and fall within the {\emph{defining}} momentum cut. By increasing the angular acceptance this proportion is further improved.
However the statistics with real data fall exponentially with $p_\perp$, 
so a balance has to be found between larger momentum and sufficient data. 
(At too large momentum one should not trust thermalization either). 
\end{itemize}

  A reasonable choice would be for instance to take a small angular cut of $60 ^{\circ}$ and identify all pions with
$p_{\perp} > 0.3$ GeV. The number of particles that mix with other fluid cells is then around a third.
This mixing should be considered a systematic theory uncertainty in the measurement of the bulk viscosity.

\begin{table}[t]
 \begin{centering}
\begin{tabular}{|c|cc|cc|cc|} \hline 
 & \multicolumn{2}{|c|}{All $p_{\perp}$} & \multicolumn{2}{|c|}{$p_{\perp} > 0.3$ GeV} & \multicolumn{2}{|c|}{$p_\perp > 0.5$ GeV} \\ 
\hline \hline
 &  $\beta$ & \% & $\beta$ & \% & $\beta$ & \% \\ 
\hline \hline
$\Delta \phi=\pm 20 ^{\circ}$ & 0.95& 38.1 & 0.96 & 36.1 & 0.97  & 36.6  \\ \hline
$\Delta \phi=\pm 30 ^{\circ}$ & 0.92& 30.0 & 0.94 & 33.1 & 0.94 & 36.4  \\ \hline
$\Delta \phi=\pm 45 ^{\circ}$ & 0.87& 23.9 & 0.88 & 33.2 & 0.89 & 40.9  \\ \hline
$\Delta \phi=\pm 60 ^{\circ}$ & 0.80& 18.2 & 0.81 & 34.3 & 0.81 & 45.8  \\ \hline \hline
 & \multicolumn{2}{|c|}{ $p_{\perp} \in (0.3,2)$ GeV} & \multicolumn{2}{|c|}{$p_{\perp} \in (0.3,3)$ GeV}  &  & \\ 
\hline \hline
 & $\beta$ & \% & $\beta$ & \%  & & \\ 
\hline \hline
$\Delta \phi=\pm 20 ^{\circ}$  & 0.96 & 69.9 & 0.96 & 57.2 & & \\ \hline
$\Delta \phi=\pm 30 ^{\circ}$  & 0.94 & 59.7 & 0.94 & 47.3 & & \\ \hline
$\Delta \phi=\pm 45 ^{\circ}$  & 0.88 & 49.9& 0.88 & 39.7 & & \\ \hline
$\Delta \phi=\pm 60 ^{\circ}$  & 0.81 & 44.1 & 0.81 & 36.8 & &\\ \hline 
\end{tabular}
\caption{ Same as in Table~\ref{tab:MC_results} but for lead-lead collisions, with the pion distribution following Eq.~(\ref{eq:Alicefit2}).
\label{tab:MC_results2}}
\end{centering}
\end{table}

We urge the experimental collaboration to perform the measurement\footnote{The small ALICE group in CIEMAT, Madrid, is attempting it.}.

\chapter*{Conclusions}
\markboth{Conclusions}{Conclusions}
\addcontentsline{toc}{chapter}{Conclusions}

   In this dissertation we have presented the calculation of the transport coefficients in hadronic matter
at low temperature with the use of effective field theories. The transport coefficients are enormously 
relevant for the dynamics of the expanding fireball in a relativistic heavy-ion collision. Collective flow, 
particle spectra and the nuclear modification factors are some of the observables that depend on the
transport properties of the fluid created at the RHIC and the LHC. There is one such transport coefficient for each conserved
(or approximately conserved) charge of the fluid.
   
   First, in Chapter~\ref{ch:3.shear} we have computed the coefficient of shear viscosity of a pure pion gas in Eq.~(\ref{eq:shear_prod}).
We have used $SU(2)$ chiral perturbation theory in order to describe the pion-pion interaction at low temperatures, $T \le m_{\pi}$.
We have implemented the inverse amplitude method in order to unitarize the scattering amplitude. This unitarization scheme 
has provided a well-behaved cross section at moderate energies and has supplied a way to describe the $\rho$ resonance employing only pion fields (with contact coupling appropriately chosen to
incorporate the $q\overline{q}$ physics not explicitly included).
Because at low temperatures the inelastic processes are exponentially suppressed we have only considered elastic interactions and included
a pion (pseudo-)chemical potential. The result of the shear viscosity for several chemical potentials is shown in Fig.~\ref{fig:shear_chem}.

We have also calculated the KSS coefficient (shear viscosity over entropy density) in this gas as a function of temperature showing that
for temperatures near the freeze-out ($T_f \simeq 150$ MeV) this coefficient is of order one. This value can even be lowered by the extension to
the $SU(3)$ chiral perturbation theory with the inclusion of kaons and $\eta$ mesons. This coefficient is plotted in Fig.~\ref{fig:shear_qgp} together 
with the perturbative results of the quark-gluon plasma, providing an indication of a minimum around the phase-transition temperature.

    Some other empirical measurements indicate that $\eta/s$ reaches its minimum value at the liquid-gas phase transition. We have confirmed this fact theoretically
in the case of atomic Argon, by using on the one hand the hard-sphere gas approximation plus the ideal thermodynamics and on the other hand the Eyring theory of liquids
together with the van der Waals equation of state. We always obtain a minimum at the phase transition (regardless of whether it being first order, second order or a crossover transition)
 with very good agreement with the experimental values in Fig.~\ref{fig:kss_argon}.

    In Chapter~\ref{ch:4.bulk} we have calculated the bulk viscosity of the pion gas as a function of temperature and chemical potential. We have
properly identified the two zero modes present in the collision operator (corresponding to energy and particle conservation) and solved the kinetic
equation to obtain an estimate of this value in Eq.~(\ref{eq:finalbulk}). We have also calculated the coefficient $\zeta/s$ obtaining a value around
the freeze-out temperature of the order of $\zeta/s=10^{-2}$ as it is shown in Fig.~\ref{fig:bulk_mu}.
We have plotted the connection with the perturbative plasma at higher temperatures in Fig.~\ref{fig:bulk_qgp}.

    We have completed the calculation of the classical transport coefficients of a pure pion gas in Chapter~\ref{ch:5.conductivities} by computing the
thermal and electrical conductivities. Both present a zero mode corresponding to momentum conservation. We have obtained
the generalization of the Wiedemann-Franz law for the pion gas in Eq.~(\ref{eq:wiedemannfranz}).

    With the use of these transport coefficients we have calculated the relaxation times in Chapter~\ref{ch:6.RTA}.
Using the relaxation time approximation we have been able to estimate the numerical value of these relaxation times both in the energy-independent approximation
in Fig.~\ref{fig:reltimes} and in the more realistic ``quadratic {\it ansatz}'' approximation in Fig.~\ref{fig:reltimes2}.

    Additional transport coefficients can be considered allowing the interplay of flavor degrees of freedom. In Chapter~\ref{ch:7.strangeness} we have included
the strange-degree of freedom in the thermalized pion gas and calculated the strangeness diffusion coefficient (that appears in Fick's diffusion law) by solving the
Boltzmann equation corresponding to the kaonic distribution function. The results are shown in Fig.~\ref{fig:s_diffusion}.

    Moreover, in Chapter~\ref{ch:8.charm} we have considered the heavier charm degree of freedom and calculated the charm drag force and the diffusion coefficients. These
transport coefficients appear in the Fokker-Planck equation. This equation (\ref{eq:FKPL}) has been obtained from the Boltzmann-Uehling-Uhlenbeck equation in the heavy mass limit.
Results are shown in Fig.~\ref{fig:threcoeffs}. We have derived the momentum dependent fluctuation-dissipation theorem and the Einstein relation that allow to obtain a single 
diffusion coefficient in the static limit. For the $\pi-D$ meson interaction we have used an effective Lagrangian that contains both chiral and heavy quark
symmetries. We have unitarized the scattering amplitude in order to avoid an unnatural increase of the cross section. The low energy constants of the effective
Lagrangian have been constrained by symmetry arguments and fixed by matching the pole position of the $D_0$ and $D_1$ resonances and the mass difference of the $D$ and $D^*$ mesons, that are generated in our scheme.
The results for the spatial diffusion coefficient, together with other results along these research lines are shown in Fig.~\ref{fig:charmcomparison}.
We have also estimated the energy and momentum losses of one charmed quark in the medium in Figs.~\ref{fig:energyloss} and \ref{fig:momentumloss}.

    To gain more insight on the possible minimum of $\eta/s$ at the phase transition, although with no claim of realism, we have studied this coefficient in the linear sigma model in the large-$N$ limit.
This model presents a second-order phase transition in the chiral limit and a crossover when the physical pion mass is considered. We have computed the 1-loop effective potential
in the large-$N$ limit with the use of the auxiliary field method. This has provided a clear way to pin down the critical and crossover temperatures. The temperature dependence of the 
order parameter is plotted in Figs.~\ref{fig:sec_ord} and \ref{fig:crossover}. We have calculated the shear viscosity by solving the Boltzmann-Uehling-Uhlenbeck equation
for the pions and we have used a scattering amplitude with corrections due to the physical pion mass in the crossover case.
We obtain a minimum of $\eta/s$ slightly below the critical temperature as shown in Fig.~\ref{fig:ksskey}.

    Finally, we have used the fluctuations of the energy-momentum tensor and the correlation between its components to 
provide an experimental method to measure the bulk viscosity in a relativistic heavy-ion collision. By computing correlations among components of momentum and energies of the
detected pions we are able to provide an estimation of the bulk viscosity in the medium with Eq.~(\ref{eq:zetageneral}). Focusing on the ALICE experiment at midrapidity this 
equation can be simplified, in Eq.~(\ref{eq:zetaoversalice}) we show our estimate for $\zeta/s$.

    Tu summarize, we have provided a comprehensive study of transport in the final stage (meson gas) of a relativistic heavy-ion collision. We hope our results will help disentangle
properties of this hadronic medium from those of the early stage quark-gluon phase. While using solidly established theory methods, we are providing practical coefficients and their 
temperature dependence that can be used in hydrodynamic simulations of heavy-ion collisions.

\begin{appendix}

\chapter{Relativistic Hydrodynamics \label{app:hydro}}
\section{Ideal hydrodynamics}
The dynamics of a relativistic fluid is encoded in the description of 
relativistic hydrodynamics. The fluid is described by its energy density $\epsilon(t,\mb{x})$, its 
pressure\index{pressure} field $P(t,\mb{x})$ and its four-velocity\index{velocity} $u^{\mu}$ defined as:
\be u^{\mu} (t,\mb{x}) \equiv \frac{dx^{\mu}}{d\tau} , \ \ee
where $x^{\mu}=(t,x,y,z)$ and $\tau$ is the proper time\index{proper time}. The four-velocity is expressed as
\be u^{\mu}=\gamma (1,\mb{v}) \ , \ee
where $\gamma=\sqrt{1-\mb{v}^2}$ and $\mb{v}$ is the three-velocity of the fluid element.
The four-velocity satisfies the relativistic normalization $u_{\mu} u^{\mu}=1$ and in the local rest frame it takes the particular value:
\be u^{\mu}=(1,\mb{0}) \ . \ee

The two functions $\epsilon$ and $P$ are related through the equation of state $P=P(\epsilon)$ and they enter
in the relativistic description of the energy content of the fluid, i.e. the energy-momentum tensor, $T^{\mu \nu}$.
For an ideal gas, it has the form:
\be \label{eq:energymomentumeq} T^{\mu \nu}= \epsilon u^{\mu} u^{\nu} - P \Delta^{\mu \nu} \ , \ee
where $\Delta^{\mu \nu} \equiv \eta^{\mu \nu} - u^{\mu} u^{\nu}$ is a projector operator orthogonal to $u^{\mu}$ ($\Delta^{\mu \nu} u_{\mu}=0$).
It can also be written in terms of the enthalpy density $w=\epsilon + P$ as:
\be T^{\mu \nu} = w u^{\mu} u^{\nu} - P \eta^{\mu \nu} \ . \ee 
In the nonrelativistic limit ($P \ll \epsilon$) the enthalpy\index{enthalpy density} density reduces to the mass density $w \rightarrow mn$ of the fluid.
In an arbitrary frame, the energy density can be extracted from the energy-momentum\index{energy-momentum tensor} as
\be \label{eq:energydensity} \epsilon = u_{\mu} u_{\nu} T^{\mu \nu} \ee
as can the pressure scalar
\be P= - \frac{1}{3} \Delta^{\mu \nu} T_{\mu \nu} \ .  \ee

In the absence of external currents the energy-momentum tensor is conserved:
\be \label{eq:cons_T} \pa_{\mu} T^{\mu \nu} =0 \ . \ee
Eq.~(\ref{eq:cons_T}) contains four equations. Contracting this set of equations with $u_{\nu}$ and $\Delta_{\nu}^{\alpha}$ we obtain respectively
\begin{eqnarray}
 \label{eq:rel_euler} D \epsilon + w \nabla^{\mu} u_{\mu}  & = & 0 \ , \\
 \label{eq:rel_cont}  w D u^{\alpha} - \nabla^{\alpha} P  & = & 0 \ ,
\end{eqnarray}
where we have defined $D\equiv u^{\mu} \pa_{\mu}$ and $\nabla^{\alpha} \equiv \Delta^{\alpha \mu} \pa_{\mu}$, in such a way that the space-time derivative is separated into a
time-like and space-like components
\be \label{eq:deridecom}\pa^{\mu} = u^{\mu} D + \nabla^{\mu}  \ . \ee
Note that with these definitions $\pa^{\mu} u_{\mu} = \nabla^{\mu} u_{\mu}$ and $u_{\mu} \nabla^{\mu}=0$.

Eq.~(\ref{eq:rel_euler}) corresponds to the relativistic version of the ``equation of energy'' and Eq.~(\ref{eq:rel_cont})
to the relativistic generalization of Euler's equation\index{Euler's equation}.

If the system presents a conserved particle number we can also define a four-particle current
\be \label{eq:particlecurrenteq} n^{\mu}= nu^{\mu} \ ,  \ee
whose first component gives the particle density $n=u^{\mu} n_{\mu}$. This vector is also conserved:
\be \label{eq:cons_n} \pa_{\mu} n^{\mu} = 0 \ , \ee
that can be written as
\be Dn + n \nabla^{\mu} u_{\mu} =0 \ , \ee
that is the continuity equation for an ideal fluid\index{continuity equation}.

\section{Viscous hydrodynamics}

To take into account the dissipative corrections in the hydrodynamics, extra terms should appear in the expressions of the energy-momentum tensor $T^{\mu \nu}$ and the particle four-flow $n^{\mu}$ \cite{landau1987fluid}:
\begin{eqnarray} 
\label{eq:viscoustmunu} T^{\mu \nu} & = & w u^{\mu} u^{\nu} - P \eta^{\mu \nu} + \tau^{\mu \nu} \ , \\
n^{\mu} & = & n u^{\mu} + \nu^{\mu} \ . 
\end{eqnarray}

The form of these dissipative parts depends on the choice we make of the reference frame (see disscusion in Sec.~\ref{sec:condsfit}). We will use the Landau-Lifshitz frame defined by the conditions that
the momentum density should vanish in the local rest reference frame.
\be T^{i0} = w u^i u^0 + \tau^{i0} = 0 \textrm{ if } u_i=0 \rightarrow \tau^{i0}=0 \ . \ee 
As the energy flow ($T^{0i}$) is equal to the momentum density ($T^{i0}$), that also means that the velocity is associated with the energy flow.
For this reason, sometimes the Landau condition is given in the form of a condition over the velocity of the system to be paralell to
the energy flow \cite{de1980relativistic}:
\be \label{eq:condfit0} u^{\mu} = \frac{T^{\mu \nu} u_{\nu}}{u_{\alpha} T^{\alpha \beta} u_{\beta}} \ . \ee
The Landau condition reads in the local rest reference frame
\be \tau^{i0} = 0 \ . \ee

There are still two more conditions in order to define properly the system. The energy and particle densities are defined out of equilibrium in such a way
that they coincide with the equilibrium values i.e. to be out of equilibrium does not change the energy and particle content of the system.
Taking the definitions of these two quantities the conditions read (in the local rest reference frame):

\begin{eqnarray}
\tau^{00} &=&  0 , \\
\nu^0 & = & 0 \ .
\end{eqnarray}

In an arbitrary reference frame they are:
\be \label{eq:condfit1} \tau^{\mu \nu} u_{\mu} = 0 \ , \ee
\be \label{eq:condfit2} \nu^{\mu} u_{\mu} = 0 \ . \ee
We will refer to Eqs.~(\ref{eq:condfit0}), (\ref{eq:condfit1}), (\ref{eq:condfit2}) as the conditions of fit\index{conditions of fit}.

The energy-momentum tensor and the particle flow must obey the conservation laws given in Eqs.~(\ref{eq:cons_T}) and (\ref{eq:cons_n}). 

The equation of continuity \index{equation of continuity} is given from Eq.~(\ref{eq:cons_n})

\be \label{eq:eqcontinuity} Dn + n \nabla^{\mu} u_{\mu} + \nabla^{\mu} \nu_{\mu} - \nu_{\mu} D u^{\mu}  = 0 \ , \ee

and the rest of the equations of fluid motions are obtained by projecting Eq.~(\ref{eq:cons_T}) along $u^{\mu}$ and $\Delta_{\nu}^{\alpha}$ respectively
\be \label{eq:navier-stokes1} D \epsilon + w \pa_{\mu} u^{\mu} - \tau^{\mu \nu} \nabla_{\{ \mu} u_{\nu \}} = 0 \ , \ee
\be \label{eq:navier-stokes2} w D u^{\alpha} - \nabla^{\alpha} P + \Delta_{\nu}^{\alpha} \pa_{\mu} \tau^{\mu \nu} = 0 \ , \ee
where \be A_{\{ \mu} B_{ \nu \}} = \frac{1}{2} \left(A_{\mu} B_{\nu} + A_{\nu} B_{\mu} \right) \ . \ee
Eqs. (\ref{eq:navier-stokes1}) and (\ref{eq:navier-stokes2}) are the relativistic generalization of the Navier-Stokes equations\index{Navier-Stokes equation}.

The form of the tensors $\tau^{\mu \nu}$ and $\nu^{\mu}$ is unique using the law of entropy increase and the equations of motion. The four-entropy flow \index{entropy flow}is defined as

\be \label{eq:defentropy} s^{\mu} = s u^{\mu} -\frac{\mu}{T} \nu^{\mu} \ . \ee
The law of entropy increase reads
\be \label{eq:increntro} \pa_{\mu} s^{\mu} \ge 0 \ . \ee

Introducing (\ref{eq:defentropy}) into (\ref{eq:increntro}) and using that $D=u^{\mu}\pa_{\mu}$ we find
\be  \pa_{\mu} s^{\mu} = Ds +s \pa_{\mu} u^{\mu} - \pa_{\mu} \left( \frac{\mu}{T} \nu^{\mu} \right) \ . \ee
Using now the equation of state ($w=Ts+\mu n$) and the Gibbs-Duhem equation (\ref{eq:gibbs-duhem}) we can transform the previous equation into
\be  \pa_{\mu} s^{\mu} = \frac{D\epsilon}{T} - \frac{\mu}{T} Dn + \frac{w}{T} \pa_{\mu} u^{\mu} - \frac{\mu}{T} n \nabla_{\mu} u^{\mu} - \pa_{\mu} \left( \frac{\mu}{T} \nu^{\mu} \right) \ . \ee
Now we insert the\index{Navier-Stokes equation} Navier-Stokes equation (\ref{eq:navier-stokes1}) in order to simplify the relation together with Eq.~(\ref{eq:condfit2}) that results in the identity
\be \nu^{\mu} \nabla_{\mu} \left( \frac{\mu}{T} \right) =   \nu^{\mu} \pa_{\mu} \left( \frac{\mu}{T} \right) \ . \ee
We finally obtain for the four-divergence of the entropy flow:
\be \label{eq:diventr}  \pa_{\mu} s^{\mu} = \frac{1}{T} \tau^{\mu \nu} \nabla_{\{ \mu} u_{\nu \}} - \nu^{\mu} \nabla_{\mu} \left( \frac{\mu}{T} \right) \ge 0 \ . \ee
Usually, the tensor $\tau^{\mu \nu}$ is separated into a traceless part that we will call $\pi^{\mu \nu}$ and a part with non-vanishing trace \cite{Romatschke:2009im},
\be \tau^{\mu \nu} = \pi^{\mu \nu} + \Delta^{\mu \nu} \frac{1}{3} \tau^{\alpha}_{\alpha} \ . \ee
Analogously the tensor $\nabla_{\{ \mu} u_{\nu \}}$ is separated into a traceless ($\nabla_{\langle \mu} u_{\nu \rangle}$) and a traceful part:
\be \nabla_{\{ \mu} u_{\nu \}} = \frac{1}{2} \nabla_{ \langle \mu} u_{\nu \rangle} + \frac{1}{3} \Delta_{\mu \nu} \nabla_{\alpha} u^{\alpha} \ .\ee
Then, Eq.~(\ref{eq:diventr}) transforms to
\be \label{eq:secondlaw} \pa_{\mu} s^{\mu} = \frac{1}{2T} \pi^{\mu \nu} \nabla_{\langle \mu} u_{\nu \rangle} + \frac{1}{3T} \tau^{\mu}_{\mu} \nabla_{\alpha} u^{\alpha} - \nu^{\mu} \nabla_{\mu} \left( \frac{\mu}{T} \right) \ge 0 \ . \ee

We now choose the form of the dissipative parts in order to satisfy this inequality. We obtain \cite{landau1987fluid}:
\be \label{eq:pretau} \left\{
\begin{array}{rcl}
  \pi^{\mu \nu} & = & \eta  \nabla^{\langle \mu} u^{\nu \rangle} \ , \\ 
  \frac{1}{3} \tau^{\mu}_{\mu} & = & \zeta \nabla_{\alpha} u^{\alpha} \ , \\
\nu_{\mu} & = &  -\kappa \left( \frac{nT}{w} \right)^2 \ \nabla_{\mu} \left(\frac{\mu}{T} \right) \ ,
\end{array} \right. \ee
or
\be \left\{
\begin{array}{rcl}
 \label{eq:tau} \tau^{\mu \nu} & = & 2 \eta  \nabla^{\{ \mu} u^{\nu \}} 
 + \left( \zeta - \frac{2}{3} \eta \right) \pa_{\alpha} u^{\alpha} \Delta^{\mu \nu} \ ,  \\
\nu_{\mu} & = &  -\kappa \left( \frac{nT}{w} \right)^2 \ \nabla_{\mu} \left(\frac{\mu}{T} \right) \ ,
\end{array} \right. \ee
where $\eta, \zeta$ and $\kappa$ should be non-negative coefficients, called shear viscosity, bulk or volume viscosity and thermal or heat conductivity, respectively.
The factor $( nT/w)^2$ in the definition of the thermal conductivity is needed if one wants to match that expression with the relativistic Fourier's law\index{Fourier's law}.
We have detailed this step in Sec.~\ref{sec:heatflow}.
Note, that from the explicit form of the dissipative terms in (\ref{eq:tau}) it is evident that the conditions (\ref{eq:condfit1}) and (\ref{eq:condfit2}) are fullfilled.

\section{Microscopic relations}

  It is essential to derive the equations that relate the microscopical properties of the particles and the macroscopic quantities that characterize the fluid.
These equations are obtained by using the one-particle distribution function $f_p(t,\mb{x})$. For example, the particle four-flow, noting that $u^{\mu}=p^{\mu}/E_p$:
\be n^{\mu} (t,\mb{x}) = g \int d^3 p \frac{p^{\mu}}{(2\pi)^3 E_p} f_p (t,\mb{x}) \ , \ee
where $E_p$ is just the on-shell energy of the particle $E_p=p_0=\sqrt{m^2+p^2}$.
The energy momentum tensor reads:
\be T^{\mu \nu} (t,\mb{x})= g\int d^3p \frac{p^{\mu} p^{\nu}}{(2\pi)^3 E_p} f_p (t,\mb{x}) \ . \ee

  If the system is only slightly out of equilibrium, the one-particle distribution function can be expressed as the equilibrium distribution function plus a deviation
from equilibrium. As in the case of first order Chapman-Enskog expansion \index{Chapman-Enskog expansion}
\be f_p (t,\mb{x}) = n_p (t,\mb{x}) + f_p^{(1)} (t,\mb{x}) \ , \ee
where the equilibrium distribution function for a Bose-Einstein gas reads (in the local rest reference frame)
\be n_p (t,\mb{x}) =  \frac{1}{e^{\beta(E_p-\mu)}-1} \ . \ee

With the help of this factorization the particle four-flow \index{particle flow} can be separated into an equilibrium and a dissipative part:

\be \label{eq:ndensimicro} n u^{\mu} (t,\mb{x})= g \int d^3 p \frac{p^{\mu}}{(2\pi)^3 E_p} n_p (t,\mb{x}) \ee
and
\be \nu^{\mu} (t,\mb{x}) = g \int d^3p \frac{p^{\mu}}{(2\pi)^3 E_p} f_p^{(1)} (t,\mb{x}) \ .  \ee
An analogous factorization can be made for the energy-momentum tensor\index{energy-momentum tensor}. The ideal part reads:

\be \label{eq:energymommicro} T_0^{\mu \nu} (t,\mb{x}) = g\int d^3p \frac{p^{\mu} p^{\nu}}{(2\pi)^3 E_p} n_p (t,\mb{x}) \ee
and the stress-energy tensor reads
\be \label{eq:stressmicro} \tau^{\mu \nu} (t,\mb{x}) = g\int d^3p \frac{p^{\mu} p^{\nu}}{(2\pi)^3 E_p} f_p^{(1)} (t,\mb{x}) \ . \ee

These expressions in terms of the moments of the one-particle distribution function can be generalized to an arbitrary rank.
We will describe some properties of these distribution moments in Appendix~\ref{app:moments}.

An important remark is that the particle density out of equilibrium $n \equiv u_{\mu} n^{\mu}$ is actually the same
as in equilibrium due to the frame choice
\be \label{eq:cof1} u_{\mu} \nu^{\mu} = u_{\mu} \Delta^{\mu}_{\nu} n^{\nu} = 0 \ . \ee

To be consistent one must ensure that in the microscopical relations the particle density and the energy density must be the same as in equilibrium and therefore
it must satisfy the conditions of fit
\begin{eqnarray}
\nu^0 (t,\mb{x}) & = & \frac{g}{(2\pi)^3} \int d^3p \ f_p^{(1)} (t,\mb{x}) = 0 \ , \\
\tau^{00} (t,\mb{x}) & = & \frac{g}{(2\pi)^3} \int d^3p \ E_p \ f_p^{(1)} (t,\mb{x}) =0 \ ,
\end{eqnarray}
and
\be \tau^{0i} (t,\mb{x})  = \frac{g}{(2\pi)^3}\int d^3p \ p_i \ f_p^{(1)} (t,\mb{x}) =0 \ . \ee

\chapter{Unitarized Chiral Perturbation Theory \label{a.uchpt}}

Quantum chromodynamics is asymptotically free, that means that the strong coupling
constant goes to zero in the UV (see for instance the expressions of the thermal strong coupling constant in the $N_f=2$ and $N_f=3$ cases
in (\ref{eq:qcd_coupling2}) and (\ref{eq:qcd_coupling3}), respectively). In this regime the application of a perturbative scheme 
is possible.

On the contrary, at energy scales of the order of $\Lambda_{QCD}\sim 200$ MeV the
running coupling constant becomes much larger than one and perturbation theory ceases to be valid for
lower scales.

The physics of a dilute meson gas are well under this scale ($m = 138$ MeV, $T,\mu \le m$)
and one needs to develop a nonperturbative method to work with it.
One of these methods is the use of an effective field theory, that it is called Chiral Perturbation Theory (ChPT) \index{chiral perturbation theory}and it is based on the spontaneous 
symmetry breaking pattern of chiral symmetry of QCD (if the quarks were massless): \footnote{
For vanishing quark masses, the QCD Lagrangian is invariant under $U(1)_V \times U(1)_A \times SU(N_f)_L \times SU(N_f)_V$. However, at the quantum
level the current associated with the symmetry $U(1)_A$ is not conserved due to the axial anomaly. The symmetry of $U(1)_V$ in the quantum theory
is responsible for the baryon number conservation.}
\be SU(N_f)_L \times SU(N_f)_R \rightarrow SU(N_f)_{V} \ . \ee 
The Goldstone bosons appearing in the symmetry breaking are the coordinates of the coset space $SU(N_f)_L \times SU(N_f)_R/SU(2)_{V}$, whose
dimension is dim($SU(N_f)_L \times SU(N_f)_R$)-dim($SU(N_f)_{V}$)$=N_f^2-1$.
These Goldstone bosons are identified with the three charged pions when $N_f=2$ and with the meson octet composed by pions\index{pion}, kaons\index{kaon} and the $\eta$
meson\index{$\eta$ meson} if $N_f=3$.

The light mesons are not massless in nature and therefore the chiral symmetry is not an exact symmetry of the Lagrangian.
The introduction of non zero quark masses in the QCD Lagrangian gives an explicit symmetry breaking term.
As the quark masses are very small compared to the chiral breaking scale $\Lambda_{\chi}\sim 1$ GeV, the effect of the quark mass is treated as a small perturbation,
and one can still consider the chiral symmetry as an approximate symmetry of the Lagrangian.

The ChPT Lagrangian is constructed by considering all the possible terms compatible with the symmetries of QCD: $C$,$P$,$T$, Lorentz invariance and chiral symmetry.
These terms are organized by the numbers of derivatives (or powers of momentum) acting on the Goldstone bosons.

\be \label{eq:chiralexpansion} \mathcal{L}_{\textrm{ChPT}} = \mathcal{L}_2 + \mathcal{L}_4 + \mathcal{L}_6 + \cdots \ee

The subindex denote the number of derivatives in the terms (note that this number should be even because parity conservation).
Working at low energy one only needs to consider the first terms in the chiral expansion (\ref{eq:chiralexpansion}). We will discuss
the details of these terms for the cases $N_f=2$ and $N_f=3$.

Thus, non-perturbative quark-gluon interactions at small energies control a few parameters that are fit to data, and appear in a perturbative way in terms of hadronic degrees of freedom.

\section{$SU(2)$ ChPT Lagrangian at $\mathcal{O} (p^4)$}

When the strange degree of freedom is not relevant i.e. considering only $u$ and $d$ quarks, one can use
the $SU(2)$ Chiral Perturbation Theory \cite{Gasser:1983yg}. 
The pions ($\pi^+,\pi^-,\pi^0$) correspond to the three (pseudo-)Goldstone bosons of the theory. These fields are represented nonlinearly as
\be U(x)= \exp \left( i \frac{\lambda^a \pi^a}{F_0} \right) \ , \ee
where $\lambda^a$ are the Pauli matrices and $F_0$ will denote the pion decay constant to lowest order. The index $a$ runs from $1$ to $3$.

The field $U$ transforms under the chiral group $SU(2)_L \times SU(2)_R$ as
\be U \rightarrow U'=RUL^{\dag} \ , \ee
where the matrices $R$ and $L$ belong to $SU(2)_R$ and $SU(2)_L$ respectively.

In the absence of external fields --except for the scalar source that includes the quark masses-- the LO Lagrangian reads
\be \mathcal{L}_2 = \frac{F_0^2}{4} \textrm{Tr } [ \pa_{\mu} U \pa^{\mu} U^{\dag} ] +\frac{F_0^2}{4} \textrm{Tr } [ \chi U^{\dag} + U \chi^{\dag} ] \ , \ee
where 
\be \chi = 2B_0 \left(
\begin{array}{cc}
 m & 0 \\
0 & m
\end{array}
\right) = \left(
\begin{array}{cc}
 M_0^2 & 0 \\
0 & M_0^2
\end{array}
\right) \ , \ee
with $m=m_u \simeq m_d$ is the light quark mass (in the isospin limit) and $M_0$ is the lowest order pion mass.

 Under the chiral transformation $SU(2)_L \times SU(2)_R$ only the kinetic term of $\mathcal{L}_2$ is invariant.

Expanding the Lagrangian in powers of the pion field one obtains an infinite number of interaction terms.
\be \mathcal{L}_2 = F_0^2 M^2_0 + \frac{1}{2} \pa_{\mu} \mathbf{\pi} \cdot \pa^{\mu} \mathbf{\pi} - \frac{1}{2} M^2_0 \mathbf{\pi}^2 + \frac{1}{6 F_0^2} \left[ (\pa_{\mu} \mathbf{\pi} \cdot \mathbf{\pi} )^2 -(\pa_{\mu} \mathbf{\pi} \cdot
 \pa^{\mu} \mathbf{\pi}) \mathbf{\pi}^2 \right] + \frac{M_0^2}{24 F_0^2} (\mathbf{\pi}^2)^2  + \cdots \ee
where the next terms correspond to six-,eight-,... particle interaction.

The next-to-leading Lagrangian reads \cite{Gasser:1983yg},\cite{Dobado:1997jx}
\begin{eqnarray}
\nonumber \mathcal{L}_4 & = &  \frac{l_1}{4} \left\{ \textrm{Tr } [\pa^{\mu} U \pa_{\mu} U^{\dag}] \right\}^2 +\frac{l_2}{4} \textrm{Tr } [\pa_{\mu} U \pa_{\nu} U^{\dag}] \textrm{Tr } [\pa^{\mu} U \pa^{\nu} U^{\dag}] \\
\nonumber & & + \frac{l_3}{16} \left\{ \textrm{Tr } [\chi^{\dag} U + \chi U^{\dag}] \right\}^2 + \frac{l_4}{4} \textrm{Tr } [\pa_{\mu} U \pa^{\mu} \chi^{\dag} + \pa_{\mu} \chi \pa^{\mu} U^{\dag}] - \frac{l_7}{16} \left\{ \textrm{Tr } [ \chi U^{\dag}- U \chi^{\dag}] \right\}^2 \\
\nonumber  & & + \frac{h_1+h_3}{4} \textrm{Tr } [\chi \chi^{\dag}] + \frac{h_1-h_3}{16} \left\{ (\textrm{Tr } [\chi U^{\dag}+ U \chi^{\dag}] )^2 +(\textrm{Tr } [\chi U^{\dag} - U \chi^{\dag}])^2 \right. \\
& & \left. - 2 \textrm{Tr } [\chi U^{\dag} \chi U^{\dag} + U \chi^{\dag} U \chi^{\dag}] \right\} \ ,
\end{eqnarray}

where the constants $l_i$ and $h_i$ are called the ``low energy constants'' \index{low energy constants} and they are not known {\it a priori} as symmetry arguments do not fix them. They must be obtained from experiment or from lattice QCD\index{lattice QCD} calculations.

The Lagrangian $\mathcal{L}_4$ also provides terms that correct the pion mass and the 4-point vertex when expanding them in powers of $\mathbf{\pi}^a$:

\be \mathcal{L}_4 =  (l_3 + h_1 ) M_0^4 - \frac{l_3 M_0^4}{F_0^2} \mathbf{\pi}^2 + \frac{l_1}{F_0^4} (\pa_{\mu} \mathbf{\pi} \cdot \pa^{\mu} \mathbf{\pi} )^2 + \frac{l_2}{F_0^4} (\pa_{\mu} \mathbf{\pi} \cdot \pa_{\nu} \mathbf{\pi})^2 +
 \frac{4l_3+h_1-h_3}{12F_0^4} M_0^4 (\mathbf{\pi}^2)^2  + \cdots \ee

\subsection{$\pi-\pi$ scattering}

The pion-pion scattering ($a,b \rightarrow c,d$) amplitude $T_{ab,cd}$ is expressed as a combination of one function $A(s,t,u)$
that depends on the three Mandelstam variables:
\be T_{ab,cd} = A(s,t,u) \delta_{ab} \delta_{cd} + A(t,s,u) \delta_{ac} \delta_{bd} + A(u,t,s) \delta_{ad} \delta_{cb} \ , \ee
where only two Mandelstam variables are independent due to the condition $s+t+u=4 m_{\pi}^2$.
The function $A(s,t,u)$ can be obtained from the ChPT Langrangian \cite{Gasser:1983yg} at $\mathcal{O} (p^4)$ and it consists of a sum of two pieces
\be A(s,t,u)=A^{(2)}(s,t,u) + A^{(4)}(s,t,u) + \mathcal{O} (p^6) \ . \ee
The LO amplitude $A^{(2)}(s,t,u)$ is obtained from $\mathcal{L}_2$ and it coincides with the ``low energy theorem'' derived by Weinberg \cite{Weinberg:1966kf}:
\be A^{(2)}= \frac{s-m^2_{\pi}}{f^2_{\pi}} \ , \ee
where the pion mass and decay constant are the physical ones. The corrections to these variables that are of order $p^4$ and they are included in the NLO amplitude.
The NLO amplitude contains terms coming from tree-level and tadpole contribution in $\mathcal{L}_4$ and one-loop correction of the LO Lagrangian. Written
in terms of the physical mass and pion decay constant as in \cite{Meissner:1993ah}:

\be \nonumber A^{(4)}= \frac{1}{96 \pi^2 f_{\pi}^4} \left\{ 2 \left(\overline{l}_1 - \frac{4}{3}\right) (s-2 m^2_{\pi})^2 + \left(\overline{l}_2 - \frac{5}{6}\right) \left[ s^2 + (t-u)^2\right]
+ 12 m^2_{\pi} s (\overline{l}_4-1) \right. \ee
\be \nonumber  \left. - 3 m^4_{\pi} (\overline{l}_3 + 4 \overline{l}_4 -5)\right\} 
+ \frac{1}{6 f_{\pi}^4} \left\{ 3 (s^2-m^4_{\pi}) \overline{J} (s) + \left[ t (t-u) - 2 m^2_{\pi} t + 4 m_{\pi}^2 u - 2 m^4_{\pi} \right] \overline{J} (t) \right. \ee
\be \left. + \left[ u (u-t) - 2 m^2_{\pi} u + 4 m_{\pi}^2 t - 2 m^4_{\pi} \right] \overline{J} (u) \right\} + \mathcal{O} (p^6) \ . \ee
The terms of order $p^4$ coming from the correction of the pion mass and decay constant are those containing the low energy constants $\overline{l}_3$ and $\overline{l}_4$.
The function $\overline{J} (s)$ comes from the pion loop and reads
\be \overline{J}(s) = \frac{1}{16 \pi^2} \left[ 2 + \rho_{\pi \pi} \log \frac{\rho_{\pi \pi}-1}{\rho_{\pi \pi}+1} \right] \ , \ee
where the two-body phase space factor is\index{two-body phase space}
\be \rho_{\pi \pi} = \sqrt{1-\frac{4 m^2_{\pi}}{s}} \ . \ee

Once the scattering amplitude has been obtained it is natural to project it into definite isospin chanel $T_{I}$. For $\pi-\pi$ scattering three isospin channels ($I=0,1,2$) are
allowed. All of them can be written as combination of a single amplitude $A(s,t,u)$:
\begin{eqnarray}
 T^0 (s,t,u) & = & 3 A(s,t,u) + A(t,s,u) + A(u,t,s) \ , \\
 T^1 (s,t,u) & = &  A(t,s,u) - A(u,t,s) \ , \\
 T^2 (s,t,u) &= & A(t,s,u) + A(u,t,s) \ . \end{eqnarray}

It is customary to project these amplitudes in definite spin channels, thus representing the scattering amplitudes into partial wave amplitudes with definite isospin and spin channels, $t_{IJ}$.
The explicit expression is
\be \label{eq:partampli} t_{IJ} (s) = \frac{1}{64 \pi} \int_{-1}^1 dx \ P_J (x) T^I (s,t(s,x),u(s,x)), \ee
where $x=\cos \theta_{CM}$ and the $P_J(x)$ are the Legendre polynomial of order $J$.

\section{Meson-meson scattering in $SU(3)$ ChPT at $\mathcal{O} (p^4)$\label{sec:su3chpt}}

At moderate temperatures, the presence of the next light mesons (kaon and $\eta$ meson) may be important. The extension
to $N_f=3$ to include the $s$-quark degree of freedom can be done in ChPT. 
The addition of kaons and $\eta$ mesons improves not only the estimation of the transport coefficients in the hadronic
sector (by the increase of the mesonic content of the gas) but also the pion-pion interaction by the introduction 
of more intermediate channels in the $\pi-\pi$ scattering amplitudes.

The construction of the effective Langrangian follows the same rules as described by the $SU(2)$ case, where the field parametrization is chosen to be
exponential
\be U (x)= \exp \left( i \frac{\lambda^a \phi^a}{F_0} \right) \ , \ee
where now the index $a$ runs from $1$ to $8$, $\lambda^a$ are the Gell-Mann matrices and 
\be \frac{1}{\sqrt{2}} \lambda^a \phi_a = \left(
\begin{array}{ccc}
 \frac{\pi^0}{\sqrt{2}}+\frac{\eta}{\sqrt{6}} & \pi^+ & K^+ \\
 \pi^- & -\frac{\pi^0}{\sqrt{2}}+\frac{\eta}{\sqrt{6}} & K^0 \\
 K^- & \overline{K^0} & -\frac{2\eta}{\sqrt{6}} \\
\end{array}
 \right) \ . \ee

The LO Lagrangian is analogous to the $SU(2)$ case:
\be \label{eq:losu3chpt} \mathcal{L}_2 = \frac{F_0^2}{4} \textrm{Tr } [ \pa_{\mu} U \pa^{\mu} U^{\dag} ] +\frac{F_0^2}{4} \textrm{Tr } [ \chi U^{\dag} + U \chi^{\dag} ] \ , \ee
and the LO scattering amplitude coincides with the low energy theorem by Weinberg \cite{Weinberg:1966kf} as in the $SU(2)$ case.

However, the Lagrangian at $\mathcal{O} (p^4)$ contains twelve low energy constants \index{low energy constants} (ten denoted by $L_i$ and two $H_i$) that are not fixed by any symmetry argument and must
be determined by experimental matching of some specific observables. The expression of the Lagrangian at NLO is \cite{Scherer:2002tk}:

\begin{eqnarray}
\mathcal{L}_4 & = & 
L_1 \left\{ \textrm{Tr }[D_{\mu}U (D^{\mu}U)^{\dag}] \right\}^2
+ L_2 \textrm{Tr } \left[ D_{\mu}U (D_{\nu}U)^{\dag} \right]
\textrm{Tr } \left [D^{\mu}U (D^{\nu}U)^{\dag}\right] \nonumber \\
& & +  L_3 \textrm{Tr } \left[ 
D_{\mu} U (D^{\mu}U)^{\dag} D_{\nu} U (D^{\nu}U)^{\dag}
\right] 
+ L_4 \textrm{Tr } \left[ D_{\mu}U (D^{\mu}U)^{\dag} \right]
\textrm{Tr } \left[ \chi U^{\dag}+ U \chi^{\dag} \right]
\nonumber \\
& & + L_5 \textrm{Tr } \left[ D_{\mu}U (D^{\mu}U)^{\dag}
(\chi U^{\dag}+ U \chi^{\dag}) \right]
+ L_6 \left\{ \textrm{Tr } \left[ \chi U^{\dag}+ U \chi^{\dag} \right]
\right\}^2
\nonumber \\
& &+  L_7 \left\{ \textrm{Tr } \left[ \chi U^{\dag} - U \chi^{\dag} \right]
\right\}^2
+ L_8 \textrm{Tr } \left[ U \chi^{\dag} U \chi^{\dag}
+ \chi U^{\dag} \chi U^{\dag} \right]
\nonumber \\
& & - i L_9 \textrm{Tr } \left[ f^R_{\mu\nu} D^{\mu} U (D^{\nu} U)^{\dag}
+ f^L_{\mu\nu} (D^{\mu} U)^{\dag} D^{\nu} U \right]
+ L_{10} \textrm{Tr } \left[ U f^L_{\mu\nu} U^{\dag} f_R^{\mu\nu} \right]
\nonumber \\
\label{eq:nlosu3chpt} & & +H_1 \textrm{Tr } \left[ f^R_{\mu\nu} f^{\mu\nu}_R +
f^L_{\mu\nu} f^{\mu\nu}_L \right]
+ H_2 \textrm{Tr} \left[ \chi \chi^{\dag} \right] \ .
\end{eqnarray}

The NLO meson scattering amplitudes are obtained by using
the corresponding tree-level and tadpole terms of the Lagrangian at $\mathcal{O} (p^4)$ and the 1-loop corrections coming from the LO Lagrangian.
These 1-loop functions contain all the possible intermediate states, for instance $\pi \pi \rightarrow KK \rightarrow \pi \pi$. 
We read the amplitudes at NLO from the formulas given in \cite{GomezNicola:2001as}, where all the $SU(3)$ ChPT scattering amplitudes for any meson-meson dispersion process
are calculated.

\section{Problem of unitarity\index{unitarity}}

The partial scattering amplitudes $t_{IJ}$ at definite isospin $I$ and spin $J$ given from the ChPT Lagrangian
are expressible as even powers of the pion momentum or in powers of the Mandelstam variable $s$ as
\be t_{IJ}(s)=t_{IJ}^{(0)} (s) + t_{IJ}^{(1)} (s) + \cdots \ , \ee
where $t_{IJ}^{(i)} (s)$ is $\mathcal{O} (s^{i})$. The partial amplitudes are basically polynomials in $s$.
The total partial amplitude $t_{IJ} (s)$ must fulfill the unitarity condition for the scattering amplitude. For the 
partial amplitude this condition reads:
\be \label{eq:unitar} \Im t_{IJ} (s)= \rho_{ab} \ |t_{IJ}|^2 \ , \ee
where $\rho_{ab}$ is defined in~(\ref{eq:twobody}).
The perturbative amplitudes satisfy this relation only order by order:
\be \Im t_{IJ}^{(0)} = 0 \ , \ee
\be \Im t_{IJ}^{(0)} + t_{IJ}^{(1)} = \rho_{ab} \ |t_{IJ}^{(0)}|^2 \ , \ee
\be \Im t_{IJ}^{(0)} + t_{IJ}^{(1)} + t_{IJ}^{(2)} = \rho_{ab} \ (|t_{IJ}^{(0)} + 2 t_{IJ}^{(0)} \Re t_{IJ}^{(1)} ) \simeq \rho_{ab} \ |t_{IJ}^{(0)} + t_{IJ}^{(1)}|^2  \ . \ee
These amplitudes do not respect exact unitarity and this violation of Eq.~(\ref{eq:unitar}) produces an unnatural increase of the cross section even at moderate energies.
In addition, the polynomial expansion of the partial amplitudes makes impossible to describe resonances as
they are unable to present poles at any finite order in the expansion.
Some unitarization methods have been developed in order to cure this problem. We will describe the inverse amplitude method that provides a new scattering amplitude that satifies
exact unitarity and it is constructed from the perturbative amplitudes obtained by ChPT.

\section{Inverse amplitude method\label{sec:iam}}

The inverse amplitude method (IAM)\glossary{name=IAM,description={inverse amplitude method}} \cite{Dobado:1989qm,Dobado:1992ha,Dobado:1996ps} is a way of constructing a scattering amplitude that respects exact unitarity and is able to reproduce
the presence of resonances as poles of the partial amplitudes.

Consider the perturbative amplitude at $\mathcal{O}(s^2)$, $t_{IJ}^{(1)}(s)$. When $s \rightarrow \infty$ the amplitude
grows as $s^2$. We can write down an exact dispersion relation for this amplitude if we apply the Cauchy theory to $t_{IJ}^{(1)}(s)/s^3$.
We need three substractions in the dispersion relation to obtain a well behaved function as $s \rightarrow \infty$.
For elastic scattering of two pions:
\be t_{IJ}(s) = C_0 + C_1 s + C_2 s^2 + \frac{s^3}{\pi} \int_{-\infty}^{\infty} \frac{\Im t_{IJ} (s') ds'}{s'^3 (s'-s-i\epsilon)} º . \ee
In the region of integration the amplitude presents a left cut due to crossing and a right cut above the elastic threshold $s=4m^2_{\pi}$
so that we can write
\be t_{IJ}(s) = C_0 + C_1 s + C_2 s^2 + \frac{s^3}{\pi} \int_{4m^2_{\pi}}^{\infty} \frac{\Im t_{IJ} (s') ds'}{s'^3 (s'-s-i\epsilon)}
+\frac{s^3}{\pi} \int_{-\infty}^0 \frac{\Im t_{IJ} (s') ds'}{s'^3 (s'-s-i\epsilon)} \ . \ee

We apply this dispersion relation to the perturbative amplitudes
\begin{eqnarray}
 t_{IJ}^{(0)} & = & a_0 + a_1 s \\
 t_{IJ}^{(1)} & = & b_0 + b_1 s +b_2 s^2 + \frac{s^3}{\pi} \int_{4 m^2_{\pi}}^{\infty} \frac{\rho_{ab} t_{IJ}^{(0)2}}{s'^3 (s'-s-i\epsilon)} + LC (t_{IJ}^{(1)}) \ , 
\end{eqnarray}
where the last term represents the contribution of the left cut and in the right cut we have replaced the perturbative unitarity relation.

The polynomial part is expanded in terms of the pion mass:
\begin{eqnarray}
 C_0 & = & a_o + b_0 + ... \\
 C_1 & =& a_1 + b_1 + ... \\
 C_2 & =  & b_2 + ...
\end{eqnarray}

The inverse amplitude method can be derived writing down a dispersion relation to the inverse amplitude $1/t_{IJ}(s)$, that has the same analytic structure 
as $t_{IJ}(s)$. For convenience one uses the dispersion relation for 
\be G(s)= \frac{t_{IJ}^{(0)2}}{t_{IJ}} \ , \ee
because $t_{IJ}^{(0)2}$ is a real polynomial, it does not change the analytic structure of $1/t_{IJ}(s)$.
Such dispersion relation reads
\be G(s)=G_0 + G_1 s + G_2 s^2 + \frac{s^3}{\pi} \int_{4m^2_{\pi}}^{\infty} \frac{\Im G(s') ds'}{s'^3 (s'-s-i\epsilon)} + LC(G) + PC \ , \ee
where the last contribution represents the pole contribution in $G(s)$, coming from any zeroes of $t(s)$.

In the right cut we use
\be \Im G=t_{IJ}^{(0)2} \Im \frac{1}{t_{IJ}} = t_{IJ}^{(0)2} \frac{1}{|t_{IJ}|^2} \Im t^*_{IJ}=-t_{IJ}^{(0)2} \frac{1}{|t_{IJ}|^2} \Im t_{IJ} \ee
and (\ref{eq:unitar})
\be \Im G = -t_{IJ}^{(0)2} \rho_{ab} \ , \ee

that is an exact substitution from ChPT. On the left cut one cannot calculate $\Im G$ exactly, because (\ref{eq:unitar}) is only valid above
threshold. We use the approximation of order $\mathcal{O}(s^2)$ for the left cut:
\be \frac{\Im t_{IJ}}{|t_{IJ}|^2} = \frac{\Im t_{IJ}^{(0)} + \Im t_{IJ}^{(1)} + \mathcal{O} (s^3)}{|t_{IJ}^{(0)}+t_{IJ}^{(1)}|^2} \simeq \Im t_{IJ}^{(1)} \frac{1}{|t_{IJ}^{(0)}|^2}\ee

\be \Im G = - t_{IJ}^{(0)2} \frac{\Im t_{IJ}}{|t_{IJ}|^2} \simeq - t_{IJ}^{(0)2} \frac{\Im t_{IJ}}{|t^{(0)}_{IJ}|^2} \simeq - \Im t^{(1)}_{IJ} \ee
So that 
\be LC(G) \simeq -LC(t_{IJ}^{(1)}) \ . \ee

Expanding the substraction constants $G_i$ in powers of the pion mass we get:
\be G_0=a_0-b_0 + \cdots \quad G_1=a_1-b_1 + \cdots; \quad G_2 =-b_2 + \cdots  \ee

Finally, the dispersion relation for $G(s)$:
\be G(s) = a_0 - b_0 +a_1 s - b_1 s - b_2 s^2 - \frac{s^3}{\pi} \int_{4m^2_{\pi}}^{\infty} \frac{\rho_{ab} t_{IJ}^{(0)2}(s') ds'}{s'^3 (s'-s-i \epsilon)}-LC (t_{IJ}^{(1)}) \ , \ee
where we have neglected the pole contribution to $G(s)$, coming form zeroes of $t_{IJ}(s)$. These zeros do exist and are called Adler zeroes and the IAM can be modified to include them
if one needs it below threshold~\cite{GomezNicola:2007qj}.

Writing this dispersion relation in terms of the perturbative amplitudes we obtain
\be \frac{t_{IJ}^{(0)2}}{t_{IJ}} \simeq t_{IJ}^{(0)} - t_{IJ}^{(1)} \ . \ee
And therefore the total amplitude is approximated as
\be \label{eq:iamamplit} t_{IJ} \simeq \frac{t_{IJ}^{(0)}}{1-t_{IJ}^{(1)}/t_{IJ}^{(0)}} \ . \ee

One can apply the simple formula~(\ref{eq:iamamplit}) to the ChPT amplitudes at LO $t_{IJ}^{(0)}$ and NLO $t_{IJ}^{(1)}$ to form 
the partial amplitude $t_{IJ}$ that respects exact unitarity~(\ref{eq:unitar}):
\be  \Im t_{IJ} (s)= t_{IJ}^{2(0)} \Im  \frac{1}{t_{IJ}^{(0)}-t_{IJ}^{(1)}}   =\rho_{ab} |t_{IJ}|^2 \ . \ee
Moreover, as befits a rational function of $s$, the possible poles of the amplitude give information about the presence of resonances in that channel.
As an example, consider pion-pion scattering in the three relevant channels at low energy, $IJ=00,11,20$. 

\begin{table}[t]
\begin{center}
\begin{tabular}{|c|c|c|c|} 
\hline
$l_1$ & $l_2$ & $l_3$ & $l_4$ \\
\hline \hline
$-0.27$ & $5.56$ & $3.4$ & $4.3$ \\
\hline
\end{tabular}
\end{center}
\caption{Set of low energy constants used in the $\pi-\pi$ scattering amplitude needed for the calculation of the transport coefficients. The phase-shifts corresponding to these low energy constants\index{low energy constants}
are shown in Figure~\ref{fig:iamsu2}. \label{tab:lecforamp}}
\end{table}

We construct the partial amplitudes for these three channels, apply Eq.~(\ref{eq:iamamplit}) to them and extract the phase-shifts from the relation
\be \label{eq:amplphase}t_{IJ}(s)= \frac{e^{i\delta_{IJ} (s)} \sin \delta_{IJ}(s)}{\rho_{\pi \pi}(s)} \ . \ee

\begin{figure}[t]
\begin{center}
\includegraphics[scale=0.35]{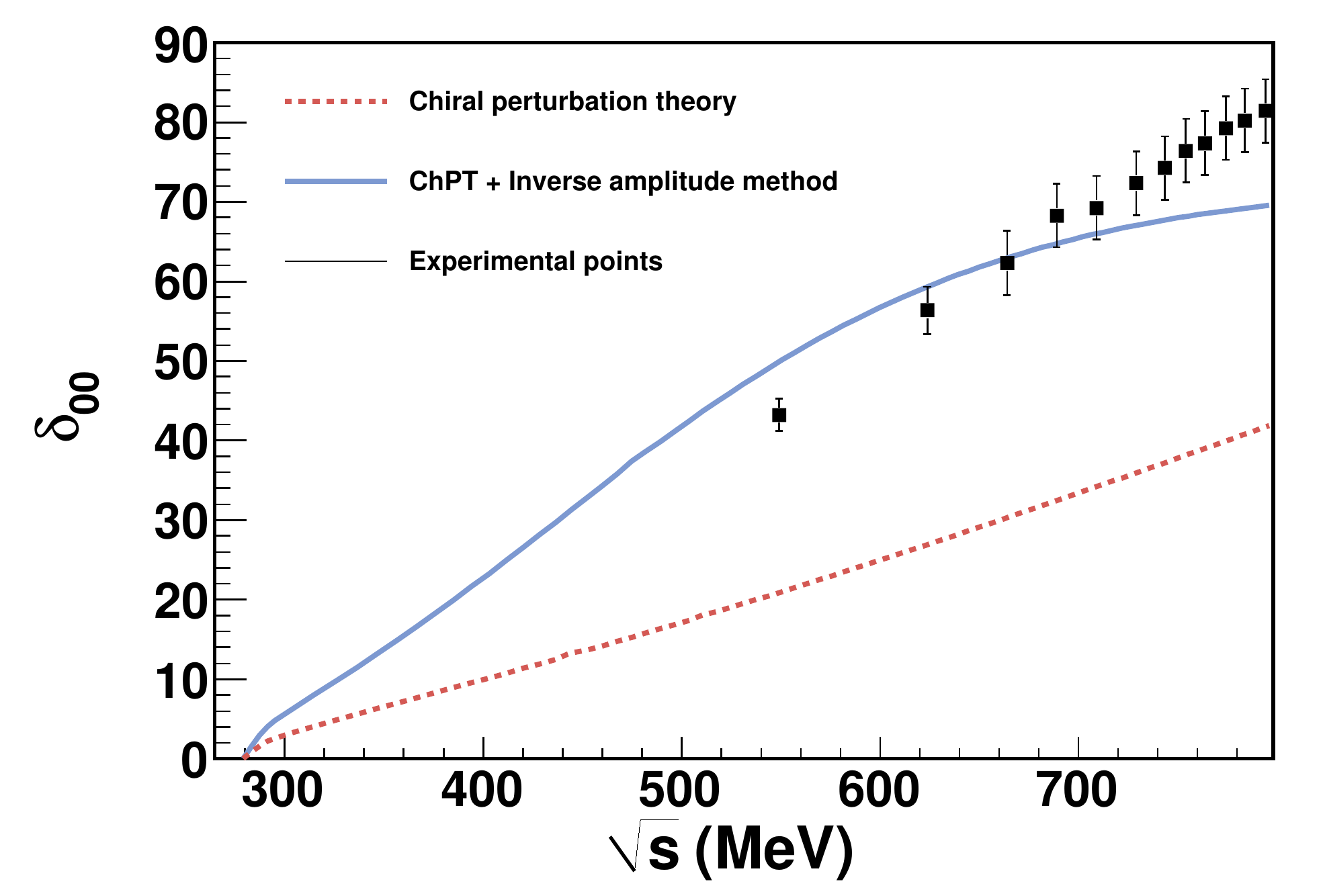}
\includegraphics[scale=0.35]{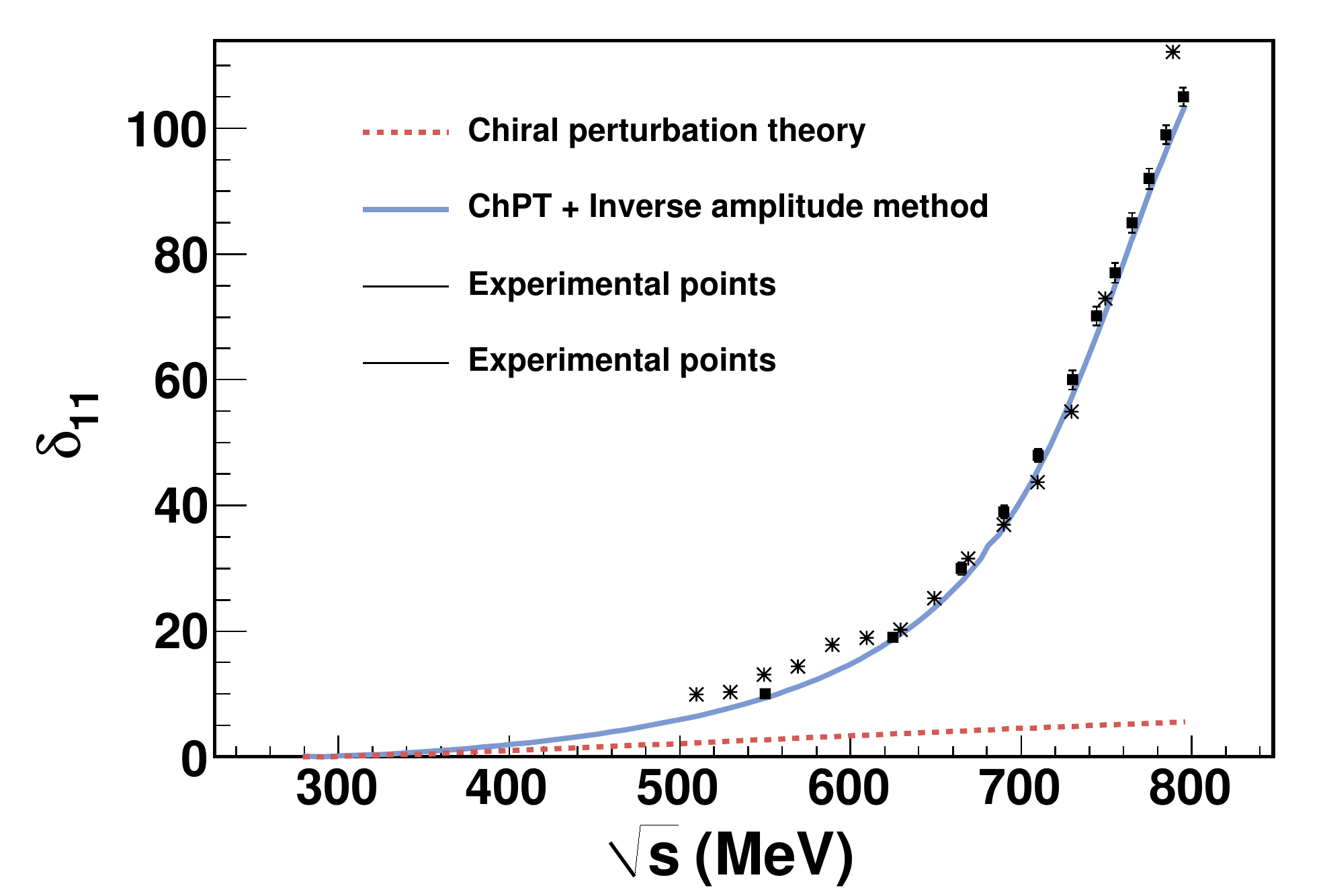}
\includegraphics[scale=0.35]{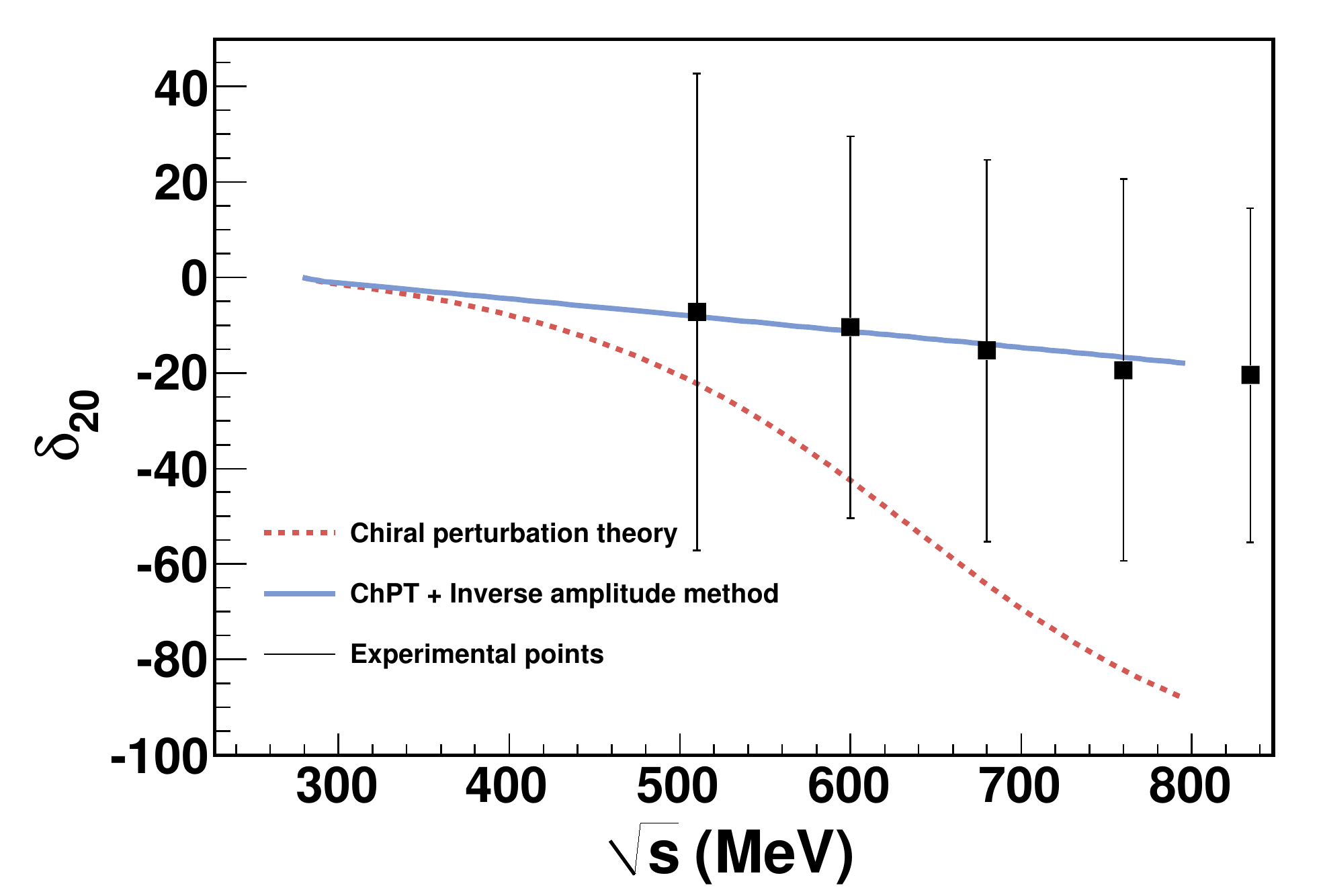}    
\caption{\label{fig:iamsu2} Results for the pion-pion phase-shifts obtained from the perturbative $SU(2)$ ChPT amplitudes (dashed line) and from those obtained using the inverse amplitude method (solid line).
We plot the three relevant isospin-spin channels at low energy. Data points are obtained from \cite{Protopopescu:1973sh}, \cite{Estabrooks:1974vu} and \cite{Losty:1973et}.}
\end{center}
\end{figure}

In Fig.~\ref{fig:iamsu2} we show the results from the standard $SU(2)$ ChPT in dashed line and after using the inverse amplitude method in solid line. We use the set of low energy constants
given in Table \ref{tab:lecforamp}. We compare the two results with the experimental data in \cite{Protopopescu:1973sh}, \cite{Estabrooks:1974vu} and \cite{Losty:1973et}. As can be seen, the description
of the experimental data above few hundred MeV is only acceptable after proper unitarization\index{unitarity} of the scattering amplitudes\index{scattering amplitudes}.

\chapter{Moments of the Distribution Function \label{app:moments}}

In Chapters \ref{ch:3.shear}, \ref{ch:4.bulk} and \ref{ch:5.conductivities}  we have defined three sets of functions $I_i,J_i$ and $K_i$ that are
integrals over an appropriate measure containing the function $n_p(1+n_p)$. One can generalize all these functions by studying
the moments of the equilibrium distribution function \cite{Muronga:2006zx}. There are two sets of moments. The first one is useful for the thermodynamical quantities
defined at equilibrium and they are the moments of the distribution function $n_p$:
\be \mathcal{I}^{\ \alpha_1 \alpha_2 ... \alpha_n} (x) = g \int \frac{d^3 p}{(2 \pi)^3 E_p} \ n_p(x) \ p^{\alpha_1} p^{\alpha_2} \  \cdots \ p^{\alpha_n} \ . \ee

These moments define thermodynamical functions in equilibrium, e.g. the first moment is recognized as the particle four-flow and the second moment to be the energy-momentum tensor.

All these moments can be expanded in a finite sum of symmetrized tensors depending on the velocity $u^{\alpha_1}$ and $\Delta^{\alpha_1 \alpha_2} = g^{\alpha_1 \alpha_2} - u^{\alpha_1} u^{\alpha_2}$.
In general, the expansion reads:

\be \label{eq:moment1} \mathcal{I}^{\ \alpha_1 \alpha_2 ... \alpha_n} = \sum_{k=0}^{[n/2]} \left(
\begin{array}{l}
n \\
2k
\end{array} \right) (2k-1)!! \ \mathcal{I}_{n,k} \ \Delta^{(2k} u^{n-2k)} \ , \ee
where $[n/2] = \textrm{Int} (n/2)$ and
\be \label{eq:deltatensor}\Delta^{(2k} u^{n-2k)} \equiv \frac{2^k! k! (n-2k)!}{n!} \sum_{permutations} \Delta^{\alpha_1 \alpha_2} \cdot \Delta^{\alpha_{2k-1} \alpha_{2k}} u^{\alpha_{2k+1}} \cdot u^{\alpha_n} \ . \ee
The coefficients $\mathcal{I}_{n,k}$ are thermodynamical functions depending on $T$ and $\mu$.

For instance, the first three moments read:

\be \label{eq:moments1}  \mathcal{I}^{\ \alpha_1} = \mathcal{I}_{1,0} u^{\alpha_1} \ , \ee
\be \label{eq:moments2}  \mathcal{I}^{\ \alpha_1 \alpha_2} = \mathcal{I}_{2,0} u^{\alpha_1} u^{\alpha_2} - \mathcal{I}_{2,1} \Delta^{\alpha_1 \alpha_2} \ , \ee
\be \label{eq:moments3} \mathcal{I}^{\alpha_1 \alpha_2 \alpha_3} = \mathcal{I}_{3,0} u^{\alpha_1} u^{\alpha_2} u^{\alpha_2}-3 \mathcal{I}_{3,1} \Delta^{(\alpha_1 \alpha_2} u^{\alpha_3 )} \ . \ee
As Eq.~(\ref{eq:moments1}) is nothing but the four-particle flux in equilibrium (\ref{eq:ndensimicro}) and Eq.~(\ref{eq:moments2}) corresponds to the ideal energy-momentum tensor\index{energy momentum tensor} (\ref{eq:energymommicro}), then these expansions give the
 physical interpretation of the coefficients $\mathcal{I}_{nk}$. From the relations (\ref{eq:particlecurrenteq}) and (\ref{eq:energymomentumeq}) one can deduce:
\be \mathcal{I}_{1,0}=n; \quad \mathcal{I}_{2,0}=\epsilon; \quad \mathcal{I}_{2,1}=P \ . \ee

The integral expression of the coefficients $\mathcal{I}_{nk}$ can be obtained by contracting Eq.~(\ref{eq:moment1}) with a tensor
of the shape~(\ref{eq:deltatensor}). The result of this contraction is \cite{Muronga:2006zx} (recall that there is a small typo in Eq. (A8) of this reference):
\be \mathcal{I}_{n,k} (\beta,\mu) = \frac{g}{(2\pi)^3} \frac{1}{(2k+1)!!} \int \frac{d^3p}{E_p} \left[ (p^{\mu}u_{\mu})^2 - p^{\mu} p_{\mu} \right]^k (p^{\nu} u_{\nu})^{n-2k} \frac{1}{e^{\beta (p^{\mu}u_{\mu}-\mu)}-1} \ .   \ee

These integrals are scalars and therefore, one can calculate them in any frame. Using the local rest reference frame and in terms of the variables that we have defined in the main text ($x=E_p/m$ and $y=m/T$)
these coefficients read
\be \mathcal{I}_{n,k} (y,z) = \frac{4 \pi g m^{n+2}}{(2k+1)!! (2\pi)^3} \int_1^{\infty} dx \ x^{n-2k} (x^2-1)^{k+1/2} \frac{1}{z^{-1}e^{y(x-1)}-1} \ . \ee
They satisfy the useful relation:
\be \label{eq:usefulmoments} \mathcal{I}_{n+2,k}=m^2 \mathcal{I}_{n,k} + (2k+3) \mathcal{I}_{n+2,k+1} \ . \ee

This set of functions is not enough when describing some physical quantity that contains the derivative of the distribution function, for example, the susceptibilies $\chi_{xy}$ defined
in (\ref{eq:susceptibilities}) and needed for the bulk viscosity. For this reason, one also defines the moments of $n_p (1+n_p)$, sometimes called ``auxiliary moments'':
\be \label{eq:defauxmoments} \mathcal{J}^{\ \alpha_1 \alpha_2 ... \alpha_n} (x) = g \int \frac{d^3 p}{(2 \pi)^3 E_p} \ n_p(x) [1+n_p (x)]\ p^{\alpha_1} p^{\alpha_2} \  \cdots \ p^{\alpha_n} \ . \ee
These moments can also be expanded in the same basis as before:
\be \label{eq:moment2} \mathcal{J}^{\ \alpha_1 \alpha_2 ... \alpha_n} = \sum_{k=0}^{[n/2]} \left(
\begin{array}{l}
n \\
2k
\end{array} \right) (2k-1)!! \ \mathcal{J}_{n,k} \ \Delta^{(2k} u^{n-2k)} \ , \ee
where now we have defined another set of coefficients $\mathcal{J}_{n,k}$. The expression of these coefficients is
\be \mathcal{J}_{n,k} (\beta,\mu) = \frac{g}{(2\pi)^3} \frac{1}{(2k+1)!!} \int \frac{d^3p}{E_p} \left[ (p^{\mu}u_{\mu})^2 - p^{\mu} p_{\mu} \right]^k (p^{\nu} u_{\nu})^{n-2k} \frac{e^{\beta (p^{\mu}u_{\mu}-\mu)}}{[e^{\beta (p^{\mu}u_{\mu}-\mu)}-1]^2} \ .   \ee
In the local reference frame and in terms of adimensional variables they read:
\be \label{eq:jcoeff}\mathcal{J}_{n,k} (y,z) = \frac{4 \pi g m^{n+2}}{(2k+1)!! (2\pi)^3} \int_1^{\infty} dx \ x^{n-2k} (x^2-1)^{k+1/2} \frac{z^{-1}e^{y(x-1)}}{[z^{-1}e^{y(x-1)}-1]^2} \ . \ee
They fulfill a relation analogous to (\ref{eq:usefulmoments});
\be \mathcal{J}_{n+2,k}=m^2 \mathcal{J}_{n,k} + (2k+3) \mathcal{J}_{n+2,k+1} \ . \ee
There is a useful relation between the $\mathcal{I}_{nk}$ and the $\mathcal{J}_{n,k}$ that can be obtained by the use of integration by parts:
\be \mathcal{J}_{n,k}= T \left[ \mathcal{I}_{n-1,k-1} + (n-2k) \mathcal{I}_{n-1,k} \right] \ . \ee
As for the function $\mathcal{I}_{n,k}$ it is possible to match the coefficients $\mathcal{J}_{n,k}$ with thermodynamical quantities. Using the definition (\ref{eq:defauxmoments}), the equation of state $w=\epsilon+P=Ts+\mu n$ and the Gibbs-Duhem
relation (\ref{eq:gibbs-duhem}) it is straightforward to show that

\be \label{eq:somequan} Tn=\mathcal{J}_{2,1} \ , \quad Tw =T (\epsilon + P)= \mathcal{J}_{3,1} \ , \quad T^2s=\mathcal{J}_{3,1}-\mu \mathcal{J}_{2,1} \ ,\ee
\be \chi_{\mu \mu} \equiv \left( \frac{\pa n}{\pa \mu} \right)_{T} = \frac{\mathcal{J}_{1,0}}{T} \ , \quad \chi_{T\mu} \equiv 
\left( \frac{\pa n}{\pa T} \right)_{\mu} = \frac{\mathcal{J}_{2,0}-\mu \mathcal{J}_{1,0}}{T^2} \ , \ee
\be \chi_{TT} \equiv  \left( \frac{\pa s}{\pa T} \right)_{\mu} = \frac{\mathcal{J}_{3,0} - 2 \mu \mathcal{J}_{2,0} + \mu^2 \mathcal{J}_{1,0}}{T^3} \ . \ee

From the general expression (\ref{eq:jcoeff}) for $\mathcal{J}_{n,k}$ one can make connection to the previously defined integrals $I_i, J_i, K_i$. These relations are:
\begin{eqnarray}
\label{eq:JrekK} \mathcal{J}_{4+i,2} & = & \frac{4\pi}{15} \frac{g m^{6+i}}{(2\pi)^3} K_i \ , \\
\label{eq:JrekI} \mathcal{J}_{i,0} & = & 4\pi \frac{g m^{2+i}}{(2\pi)^3} I_i \ , \\
 \label{eq:JrekJ} \mathcal{J}_{2+i,1} & = & \frac{4\pi}{3} \frac{g m^{4+i}}{(2\pi)^3} J_i \ .
\end{eqnarray}
 
\chapter{Second-Order Relativistic Fluid Dynamics \label{app:second-order}}

In Sec.~\ref{sec:hydrocodes} we have commented the acausality problems that presents the first order fluid dynamics, that amounts
to the presence of instabilities for short wavelenght modes in the hydrodynamic simulations.
We will briefly show the nature of this pathology and motivate why it is essential to introduce the second-order hydrodynamics\index{hydrodynamics!second order} \cite{Israel:1979wp},
at least for numerical calculation purposes. Finally, we discuss how this problem does not affect our theoretical calculation based on 
the Chapman-Enskog expansion\index{Chapman-Enskog expansion}.

We will use the simple case of the Fick's diffusion law\index{Fick's law} and also the case of the Navier-Stokes equation due to the relevance of the shear viscosity in relativistic heavy ion collisions.
We will follow the discussions given in \cite{Romatschke:2009im},\cite{Kelly:1968}.

In the first order relativistic fluid dynamics, the irreversible currents are proportional to the hydrodynamic
gradients. The transport coefficients are just the proportionality constants between the two. In 
Fick's diffusion law, the charge diffusion coefficient $D_x$ is the proportionality constant between the three-current $j^i$ and the charge
concentration $\rho$. For example, in the one dimensional case where the current is taken to be parallel to the $x$ direction:
\be \label{eq:Fickslaw2} j^x = - D_x \frac{\partial \rho}{\partial x} \ . \ee

Using now the current conservation equation $\pa_{\mu} j^{\mu}=0$ one can obtain a second-order differential equation for $\rho$:
\be \label{eq:parequ} \frac{\pa \rho}{\pa t} - D_x \frac{\pa^2 \rho}{\pa x^2} = 0 \ . \ee
Eq.~(\ref{eq:parequ}) is a parabolic differential equation that permits infinite propagation speed, leading to acausality phenomena.

In the case of the shear viscosity we use the Navier-Stokes equation\index{Navier-Stokes equation} (\ref{eq:navier-stokes2}) applied to a fluid moving along the $y$ direction.
We look for  momentum flux along the $x$ direction. Therefore we will need to focus on the $\tau^{xy}$ component of the stress-energy tensor\index{stress-energy tensor}:
\be w \frac{\pa v_y}{\pa t} + \frac{\pa \tau^{xy}}{\pa x}=0 \ , \ee
where $\tau^{xy}$ is taken from Eq.~(\ref{eq:tau})
\be \tau^{xy} = - \eta \ \frac{\pa v_y}{\pa x} \ . \ee
Inserting this equation into the previous one:
\be \label{eq:parabolic} \frac{\pa v_y}{\pa t} - \frac{\eta}{w} \frac{\pa^2 v_y}{\pa x^2}=0 \  , \ee
that is the same kind of parabolic differential equation. 

Consider the general heat equation:
\be \label{eq:parequ2} \frac{\pa n}{\pa t} - A \frac{\pa^2 n}{\pa x^2}=0 \ , \ee
where $A$ represents a normalized transport coefficient (as $D_x$ or $\eta/w$), and $n$ represents the conserved hydrodynamical field that is transported.

\begin{figure}[t]
\begin{center}
\includegraphics[scale=0.5]{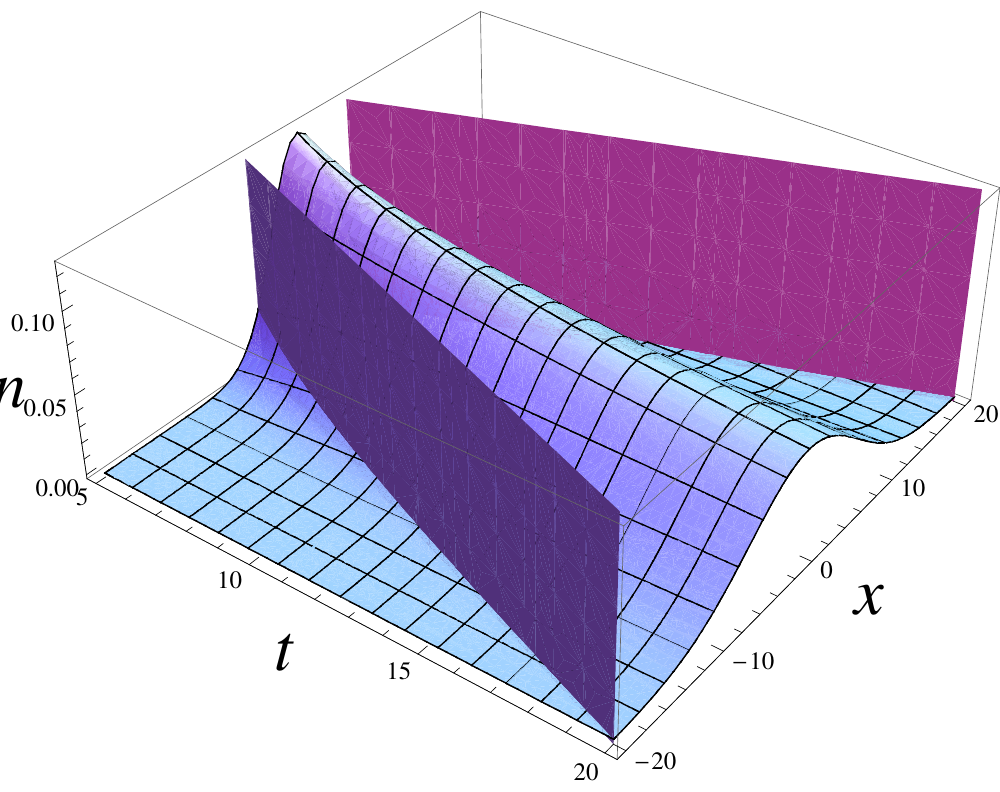}
\includegraphics[scale=0.5]{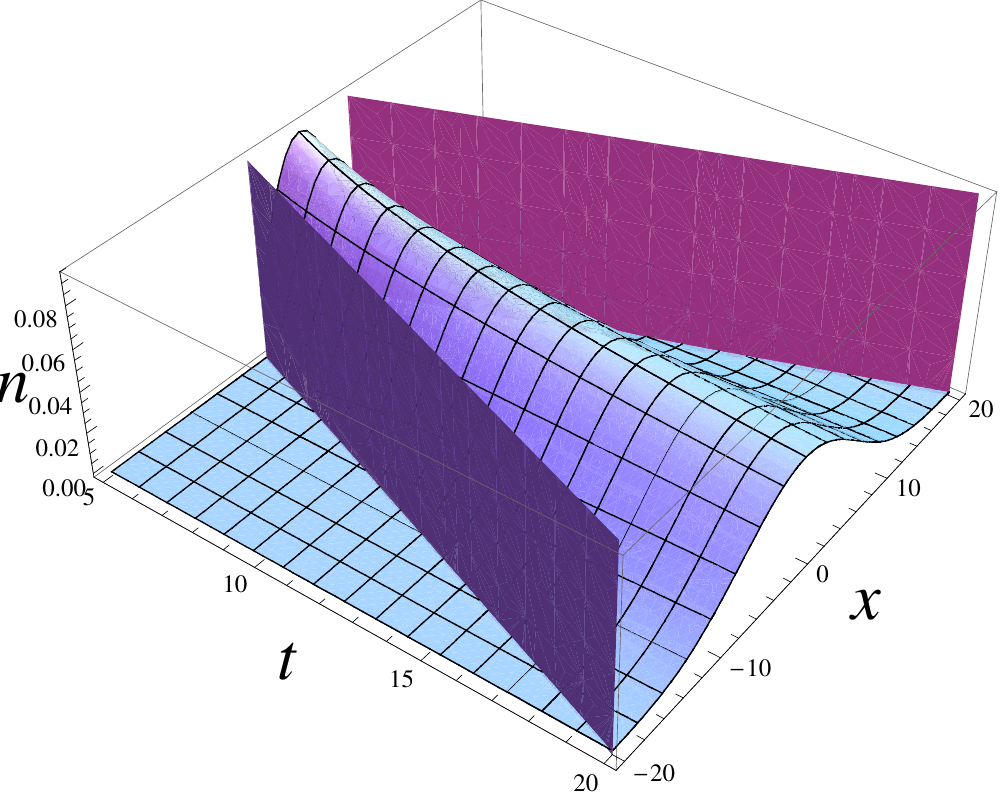}  
\caption{\label{fig:causality} Left panel: Solution of Eq.~(\ref{eq:parabolic}) for $A=1$ . Right panel: solution of Eq.~(\ref{eq:hyperbolic}) for $A=\tau=1$. Note that for early times, the solution
in the left panel has support outside the light cone (represented by the vertical planes) whereas the solution of the right panel is concentrated inside the light cone.}
\end{center}
\end{figure}

Substituting the following {\it ansatz} for $n \propto e^{-\omega t + iqx}$ to obtain the dispersion relation, it gives a group velocity
\be v(q) =\frac{d\omega}{dq}= 2 A q \ . \ee
This velocity can be larger than the speed of light producing an acausal propagation mode. Another way to see this is to calculate
the solution of the equation (\ref{eq:parequ2}) with the initial condition $n(t=0,x)=\delta(x)$. The solution is \cite{Kelly:1968}:
\be n (t,x) = \frac{1}{\sqrt{4\pi At}} \exp \left( - \frac{x^2}{4At}\right) \ . \ee
This equation has support outside of the lightcone $x>t$ as one can see in the left panel of Fig.~\ref{fig:causality}, showing 
that the causality condition is violated in this solution.

A possible way of solving this undesirable effects is to modify the equation by introducing some phenomelogical relaxation time in a second order term:
\be \label{eq:hyperbolic} \tau \frac{\pa^2 n}{\pa t^2} + \frac{\pa n}{\pa t} - A \frac{\pa^2 n}{\pa x^2} = 0 \ , \ee
that converts the differential equation into a hyperbolic-like one (Eq.~(\ref{eq:hyperbolic}) is sometimes called the ``telegrapher equation'' because it 
describes electromagnetic propagation along a lossy line).

Inserting the same {\it ansazt} for the solution we obtain the following propagation velocity:

\be v =\frac{d|\omega|}{dq} = \frac{2 q(\omega) A}{|1-2 \tau \omega|} \ee
when $q \rightarrow \infty$ ($\omega \rightarrow \infty$)
\be v = \sqrt{ \frac{A}{\tau}} \ee
that respects causality as long as
\be \tau \ge A \ .\ee
To see this more clearly, we solve Eq.~(\ref{eq:hyperbolic}) with the initial condition $n(t=0,x)=\delta(x)$ and $\pa n(t,x)/ \pa t |_{t=0}= 0$. The solution is \cite{Kelly:1968}
\be n(t,x)= \left\{
\begin{array}{cc}
\frac{1}{2} \exp(-\frac{t}{2\tau}) \left[ \frac{1}{2 \sqrt{A \tau}} I_0 ( \frac{1}{2\sqrt{A\tau}} (\frac{A}{\tau} t^2 -x^2)) \right. &  \\
 \left. + \frac{t}{2\tau} \frac{1}{\sqrt{A/\tau t^2-x^2}} I_1(\frac{1}{2\sqrt{A\tau}} (\frac{A}{\tau} t^2 -x^2)) \right]  & |x|<\sqrt{\frac{A}{\tau}} t \ , \\
0 &  |x|>\sqrt{\frac{A}{\tau}} t 
\end{array}
\right. \ee

where $I_0,I_1$ are the modified Bessel functions of the first kind. We plot this solution in the right panel of Fig.~\ref{fig:causality} showing that
causality is respected is this solution.

We would like to stress that the acausality problem in first order hydrodynamics is a conceptual problem contained in the equations of motion.
This problem produces some important instabilities in the numerical solution of hydrodynamics for the short wavelenght modes. For these cases, it is sufficient
to include these relaxation times in order to avoid the infinite propagation of high frequency modes. This is usually done in the context of the Israel-Stewart
theory \cite{Israel:1976tn,Israel:1979wp} or in terms of other refinements to that work, as in \cite{Dusling:2007gi,Luzum:2008cw}. 

However, as we have shown, this problem is only present for short wavelenghts, where the hydrodynamics ceases to be valid at some point. Even more, the Chapman-Enskog expansion
is only valid when the mean-free path is much smaller than any other length scale in the system, among them the wavelength (see Section~\ref{sec:chapman-enskog}). This expansion forbids 
large propagation velocities and suppresses the acausality problems. Moreover, it has been shown that the effect of these higher order term can be neglected when one is not far from equilibrium \cite{Heinz:2009xj}.

\chapter{Langevin Equation for Charm Diffusion \label{app:langevin}}

We will derive the relation between the spatial diffusion coefficient \index{diffusion coefficient!in space}$D_x$ (that appears in the Fick's diffusion law\index{Fick's law}, for instance
in Eq.~(\ref{eq:Fickslaw2})) and the momentum diffusion coefficient\index{diffusion coefficient!momentum} $D$, that we have estimated through the Fokker-Planck equation\index{Fokker-Planck equation}.
Moreover we will explain the physical interpretation of the coefficients $F$, $\Gamma_0$ and $\Gamma_1$, and obtain the
expressions for the energy and momentum losses per unit length of the heavy quark.

We start by deriving the Fokker-Planck equation from the Langevin equation.
Then the classical solution to the Langevin equation \index{Langevin equation} will allow us to identify the space-diffusion term and relate it to the Fokker-Planck coefficient of diffusion in momentum space.

The charm quark (that can be understood as a Brownian particle\index{Brownian particle} inside the medium) moves in the pion gas and is diffused because of the collisions with these mesons.
The position and momentum of the charm quark can be regarded as stochastic variables depending on time. The classical, nonrelativistic
stochastic differential equations that govern their motion are:
\ba \frac{dx^i}{dt} & = & \frac{p^i}{m_D} \\
\frac{dp^i}{dt} &= & -F^i(\mathbf{p}) + \xi^i(t) \ , \ea
where the index $i=1,2,3$ labels the space component of $\mathbf{x}$ and $\mathbf{p}$.
This equation is called the Langevin equation. The $F^i (\mathbf{p})$ is a deterministic 
drag force which depends on momentum through the collision processes and $\xi(t)$ is a stochastic term called white noise.
It verifies the following properties
\ba \label{eq:stoc1} \langle \xi^i(t)\rangle & = & 0 \ , \\
 \label{eq:stoc2} \langle \xi^i (t) \xi^j (t') \rangle & = & \Gamma^{ij} (\mathbf{p}) \delta(t-t') \ .\ea

In an isotropic gas one naturally has $\Gamma^{ij} (\mathbf{p}) = \Gamma (\mathbf{p}) \delta^{ij}$.

We now discretize the time variable (in order to simplify some steps) and we will thereafter take again the continuum limit $\delta t\to 0$:
\be 
t_n \equiv n \delta t;\quad  \mathbf{x}_n \equiv \mathbf{x}(t_n); \quad \mathbf{p}_n \equiv \mathbf{p}(t_n); \quad n=0,1,2,... 
\ee
and choose a mid-point discretization for $F^i(\mathbf{p})$~\cite{risken1996fokker} 
\be 
F^i_n (\mathbf{p}) = F^i \left[  \frac{\mathbf{p}_n + \mathbf{p}_{n+1}}{2} \right] \ . 
\ee
The discretized Langevin equation\index{Langevin equation} reads then
\ba 
{\bf x}_{n+1} & = & {\bf x}_n + \frac{{\bf p}_{n}}{m_D} \delta t \ ,\\
{\bf p}_{n+1} &= & {\bf p}_n - {\bf F}_n \delta t +  {\bf L}_n \delta t \ ,\ea
with a time average over the random noise
\be L^i_n = \frac{1}{\delta t} \int_{t_n}^{t_{n+1}} dt \ \xi^i(t) \ . \ee

From Eqs.~(\ref{eq:stoc1}) and (\ref{eq:stoc2}), $L^i_n$ verifies: 
\be \langle L^i_n \rangle =0 \ ,\ee
\be \label{eq:variance} \langle L^i_n L^j_{n'} \rangle= \frac{\Gamma}{\delta t} \ \delta^{ij} \delta_{nn'} \ .
\ee
(From this last relation one deduces that the variable $L^i_n \sim \mathcal{O} (\delta t^{-1/2})$ ).

The average $\langle \rangle$ is taken with respect to the probability associated with the stochastic process. Since the stochastic variables are positions and momenta, this 
probability is nothing but the one-particle classical distribution function, $f(t,\mathbf{x},\mathbf{p})$.
Averages are then computed by means of
\be 
\langle T(t)\rangle_{X,P} \equiv \int d\mathbf{x} d\mathbf{p} T(t,\mathbf{x}_n,\mathbf{p}_n) f(t,\mathbf{x}_n,\mathbf{p}_n) \ ,
\ee
where $T(t,\mathbf{x}_n,\mathbf{p}_n)$ is any function of the stochastic variables and time.

In the Fokker-Planck equation we look for the time evolution of the distribution function itself, so we need to calculate 
the probability that a particle at time $t_{n+1}$ is at $\mathbf{x},\mathbf{p}$
\be \label{eq:probforw} f(t_{n+1},\mathbf{x},\mathbf{p})= \langle \delta^{(3)}(\mathbf{x}_{n+1}-\mathbf{x}) \delta^{(3)}(\mathbf{p}_{n+1}-\mathbf{p})\rangle \ ,
\ee
from the distribution function at a prior time.

We introduce the discretized Langevin equation\index{Langevin equation} inside the deltas in (\ref{eq:probforw}):
\ba 
\delta({\bf x}_{n+1}-{\bf x}) & = & \delta({\bf x}_n  -{\bf x} + \frac{{\bf p}_n}{m_D} \delta t) \ , \\
 \delta({\bf p}_{n+1}-{\bf p}) & = & \delta({\bf p}_n -{\bf p} + \left[ {\bf F}_n + {\bf L}_n \right] \delta t ) \ . 
\ea

Expanding the deltas up to $\mathcal{O} (\delta t)$,
\ba 
\delta(x^i_{n+1}-x^i) & = &\delta(x^i_n -x^i) +\sum_j \frac{\pa}{\pa x^j_n} \delta(x^i_n-x^i) \ \frac{p^j_n}{m_D} \delta t \ , \\ 
\nonumber \delta(p^i_{n+1}-p^i)  & = &  \delta(p^i_n -p^i) +\sum_j \frac{\pa}{\pa p^j_n} \delta(p^i_n-p^i)  \ \left[ F^j (p_n) + L^j_n \right] \delta t  \\
& & +  \frac{1}{2}\sum_{j} \sum_k \frac{\pa^2}{\pa p^j_n \pa p^k_n} \delta(p^i_n-p^i) L^j_n L^k_n  \ (\delta t)^2 \ , 
\ea
and introducing these expansions inside equation (\ref{eq:probforw}), we see that
\ba 
\nonumber f(t_{n+1},\mathbf{x},\mathbf{p})  & = &   \langle \delta^{(3)}(\mathbf{x}_n -\mathbf{x}) \delta^{(3)}(\mathbf{p}_n -\mathbf{p})  \rangle 
 +  \langle \sum_j \frac{\pa}{\pa x^j_n} \delta^{(3)}(\mathbf{x}_n-\mathbf{x}) \ p^j_n \ \delta^{(3)}(\mathbf{p}_n -\mathbf{p}) \rangle \frac{\delta t}{m_D} \\
 & & \nonumber  +  \langle \delta^{(3)}(\mathbf{x}_n-\mathbf{x}) \sum_j \frac{\pa}{\pa p^j_n} \delta^{(3)}(\mathbf{p}_n-\mathbf{p})  \ \left[L^j_n- F^j (p_n)  \right]   \rangle \delta t \\
 & &   +  \frac{1}{2} \langle \delta^{(3)}(\mathbf{x}_n-\mathbf{x}) \sum_{j} \sum_k \frac{\pa^2}{\pa p^j_n \pa p^k_n} \delta^{(3)}(\mathbf{p}_n-\mathbf{p}) L^j_n L^k_n  \rangle  (\delta t)^2 \ .
\nonumber 
\ea

In order to obtain $f(t_n, \mathbf{x}, \mathbf{p})$ in the left-hand side, we introduce the following identity
\be 
\delta^{(3)}(\mathbf{x}_n-\mathbf{x}) \delta^{(3)}(\mathbf{p}_n-\mathbf{p})= \int d\mathbf{z} d\mathbf{q}  \delta^{(3)}(\mathbf{x}_n-\mathbf{z}) 
\delta^{(3)} (\mathbf{z}-\mathbf{x}) \delta^{(3)}(\mathbf{p}_n-\mathbf{q}) \delta^{(3)} (\mathbf{q}-\mathbf{p})
\ee
and replace  the definition in Eq.~(\ref{eq:probforw}) 
\be 
\langle \delta^{(3)} (\mathbf{x}_n -\mathbf{z}) \delta^{(3)}(\mathbf{p}_n-\mathbf{q}) \rangle = f(t_n,\mathbf{z},\mathbf{q})\ .
\ee

One obtains
\ba 
f(t_{n+1},\mathbf{x},\mathbf{p})  &=&   
\int d\mathbf{z} d\mathbf{q} \ \delta^{(3)} (\mathbf{z}-\mathbf{x}) \delta^{(3)} (\mathbf{q}-\mathbf{p}) \ f(t_n,\mathbf{z},\mathbf{q}) 
\\ \nonumber
&&  +\int d\mathbf{z} d\mathbf{q} \  \delta^{(3)}(\mathbf{q}-\mathbf{p}) \sum_i \frac{\pa}{\pa z^i}  \delta^{(3)}(\mathbf{z}-\mathbf{x}) q^i \ f(t_n,\mathbf{z},\mathbf{q}) \frac{\delta t}{m_D} 
\\ \nonumber 
&&  -\int d\mathbf{z} d\mathbf{q} \ \delta^{(3)} (\mathbf{z}-\mathbf{x}) \sum_i \frac{\pa}{\pa q^i}  \delta^{(3)} (\mathbf{q}-\mathbf{p})  F^i (\mathbf{q}) \ f(t_n,\mathbf{z},\mathbf{q})  \delta t 
\\ \nonumber  
&& +\int d\mathbf{z} d\mathbf{q} \ \delta^{(3)} (\mathbf{z}-\mathbf{x}) \sum_{ij} \frac{\pa^2}{\pa q^i \pa q^j}  \delta^{(3)} (\mathbf{q}-\mathbf{p})  \frac{\Gamma^{ij}(\mathbf{q})}{2} f(t_n,\mathbf{z},\mathbf{q})  \delta t 
\ea
where the average operation has been factorized because $p^i_n$ only depend on $L^i_{n'}$ with $n'<n$.

Now integrate by parts and finally, over $\mathbf{z}$ and $\mathbf{q}$:

\ba \nonumber f(\mathbf{x},\mathbf{p}, t_{n+1} ) & = &  f(t_n,\mathbf{x},\mathbf{p}) - \frac{\mathbf{p}}{m_D}  \cdot \frac{\pa}{\pa \mathbf{x}} f(t_n,\mathbf{x},\mathbf{p}) \delta t 
 +  \sum_i \frac{\pa}{\pa p^i} F^i (\mathbf{p}) f(t_n,\mathbf{x},\mathbf{p}) \delta t \\
& &  + \frac{1}{2} \sum_{ij} \frac{\pa^2}{\pa p^i \pa p^j} \Gamma^{ij} (\mathbf{p}) f(t_n,\mathbf{x},\mathbf{p}) \delta t \ . \ea

We return to the continuum limit $\delta t \rightarrow 0$:

\be \frac{\pa f(t,\mathbf{x},\mathbf{p}) }{\pa t} + \frac{\mathbf{p}}{m_D}  \frac{\pa}{\pa \mathbf{x}} f(t,\mathbf{x},\mathbf{p}) = 
\sum_i \frac{\pa}{\pa p^i} F^i (\mathbf{p}) f(t,\mathbf{x},\mathbf{p}) + \frac{1}{2} \sum_{ij} \frac{\pa^2}{\pa p^i \pa p^j} \Gamma^{ij} (\mathbf{p}) f(t,\mathbf{x},\mathbf{p}) \ . \ee

Taking the average in space 
\be \frac{\pa f_c(t,\mathbf{p})}{\pa t} = - \frac{\pa}{\pa p^i} \left[ F^i(\mathbf{p}) f_c(t,\mathbf{p})\right] + \frac{1}{2} \frac{\pa^2}{\pa p^i \pa p^j} \Gamma_{ij} (\mathbf{p}) f_c(t,\mathbf{p}) \ ,
\ee
that coincides with the Fokker-Planck equation in Eq.~(\ref{eq:FKPL}).
We see that the diffusion coefficients $\Gamma_0$, $\Gamma_1$, stem from the random force in the Langevin equations\index{Langevin equation}, and the drag coefficient from the 
deterministic friction force there. 

In the static limit $\mathbf{p} \rightarrow 0$, we can solve the
Langevin (or, in this limit also called Uhlenbeck-Orstein) equation
\be 
\frac{d \mathbf{p}}{dt} = -F\mathbf{p} + \mathbf{\xi} (t) \ , 
\ee
whose solution is
\be 
\label{eq:p_sol} \mathbf{p} (t)= \mathbf{p}_0 e^{-Ft} + e^{-Ft} \int_0^t 
d\tau e^{F\tau} \mathbf{\xi}(\tau) \ .
\ee
Taking the average one can see that due to the drag force, the friction term makes the particle eventually stop in the fluid's rest frame.
\be \langle \mathbf{p} (t) \rangle = \mathbf{p}_0 e^{-Ft} \ .\ee

The second of Hamilton's equations 
\be 
\frac{d \mathbf{x}}{dt} = \frac{\mathbf{p}}{m_D} \  ,
\ee
is then solved by
\be 
\label{eq:x_sol} \mathbf{x}(t)=\mathbf{x}_0 + \int_0^t d\tau \frac{ \mathbf{p} (\tau)}{m_D} \ .
\ee
Taking the average
\be 
\langle \mathbf{x} (t) \rangle =\mathbf{x}_0 + \frac{ \mathbf{p}_0}{F m_D} 
(1-e^{-Ft}).  
\ee

To make the connection with the spatial diffussion coefficient we can show the mean quadratic displacement of the Brownian particle \index{Brownian particle}
($r=\sqrt{x^2+y^2+z^2}$)
\be 
\langle (r(t)-r_0)^2 \rangle=\langle (x(t)-x_0)^2+(y(t)-y_0)^2+(z(t)-z_0)^2 \rangle \ ,
\ee
that,  from Fick's diffusion law, is simply 
\be 
\langle (r(t)-r_0)^2 \rangle = 6 D_x t \ . 
\ee

From the averaged solution to\index{Langevin equation} Langevin's equation~(\ref{eq:x_sol}),
\be 
\langle (x(t)-x_0)^2 \rangle = \frac{1}{m_D^2} \int_0^t \int_0^t d\tau d\tau' \ \langle p^x (\tau) p^x(\tau') \rangle \ . 
\ee

With the help of (\ref{eq:p_sol}) and (\ref{eq:stoc2}) and carefully performing the integral \cite{risken1996fokker} one obtains the leading term of this 
expression when $t \gg F^{-1}$ as
\be 
\langle (x(t)-x_0)^2 \rangle = \frac{2 \Gamma t}{m^2_D F^2} \ ,
\ee
so that
\be 
D_x = \frac{\Gamma}{m^2_D F^2} = \frac{T^2}{\Gamma} \ ,
\ee
where finally we have used Einstein's relation\index{Einstein's relation}. Thus, the calculation of the momentum space diffusion coefficient $\Gamma$ automatically entails an estimate for the space diffusion coefficient $D_x$.

Not all three coefficients $F(p^2)$, $\Gamma_0 (p^2)$ and $\Gamma_1 (p^2)$ appearing in the Fokker-Planck equation are independent, but rather
related by a  fluctuation-dissipation\index{fluctuation-dissipation theorem} relation. Since we consider the $p$-dependence of the three coefficients, the fluctuation-dissipation
relation will be momentum dependent, although we also expose the $p\to 0$ limit. 
A transparent procedure is to  match the asymptotic solution of the Fokker-Planck equation \index{Fokker-Planck equation}to the thermal equilibrium distribution function,
thus guaranteeing energy equipartition.

First of all, the Fokker-Planck equation can be written as an equation of continuity \cite{landau1981physical}:
\be \frac{\pa f_c (t, \mathbf{p})}{\pa t} =- \frac{\pa}{\pa p_i} n_i \ , \ee
where
\be n_i \equiv -F_i (p^2) f_c (t,\mathbf{p}) - \frac{\pa}{\pa p_j} \left[ \Gamma_{ij} (p^2) f_c (t,\mathbf{p}) \right]\ee
is the particle flux density in momentum space.
At statistical equilibrium, this flux is zero, and the equilibrium distribution function is the Bose-Einstein function,
\be f_c \sim \frac{1}{e^{-p^2/2 M T}-1} \ . \ee

Employing the approximation $1 + f_c \approx 1$, valid for small charm-quark number, one can obtain
\be 
F_i (p^2) + \frac{\pa \Gamma_{ij} (p^2)}{\pa p_j} = \frac{1}{MT} \Gamma_{ij}(p^2) p_j \ .
\ee
This momentum-dependent fluctuation-dissipation relation can be recast for the functions $F(p^2), \Gamma_0 (p^2)$ and $\Gamma_1 (p^2)$ as:
\be F (p^2) + \frac{1}{p} \frac{\pa \Gamma_1 (p^2)}{\pa p} + \frac{2}{p^2} \left[ \Gamma_1 (p^2)
- \Gamma_0 (p^2) \right] =  \frac{\Gamma_1 (p^2)}{MT} \ . \ee

For low-momentum charm quarks, $\Gamma_1(p^2), \Gamma_0 (p^2) \rightarrow \Gamma$, $F(p^2) \rightarrow F$. We then recover the
well known Einstein relationship
\be 
\label{eq:Einstein} F = \frac{\Gamma}{MT}  \ . 
\ee
Thus, in the static limit two coefficients take the same value
and the third is obtained from them by Eq.~(\ref{eq:Einstein}), and
we are left with only one independent diffusion coefficient.

\begin{figure}[t]
\centering
\includegraphics[width=8.5cm]{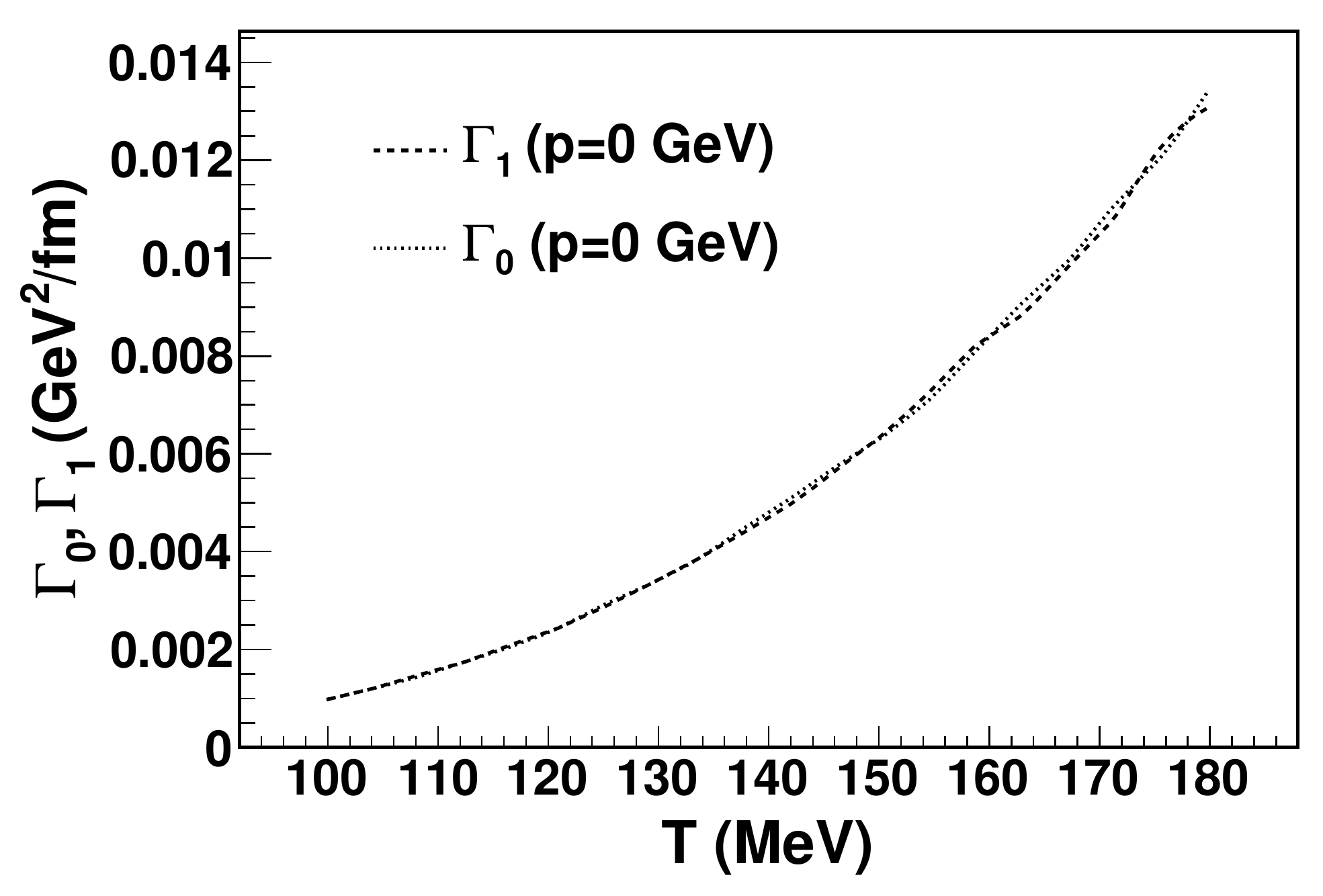}
\caption{\label{fig:momdiff} 
Momentum-space transport coefficients as function of temperature.
Dotted: $\Gamma_0 (p^2 \to 0)$. Dashed: $\Gamma_1 (p^2 \to 0)$.
The very good agreement in our computer programme, as appropriate in this limit, makes the curves barely distinguishable.
}
\end{figure}

\begin{figure}[t]
\centering
\includegraphics[width=8.5cm]{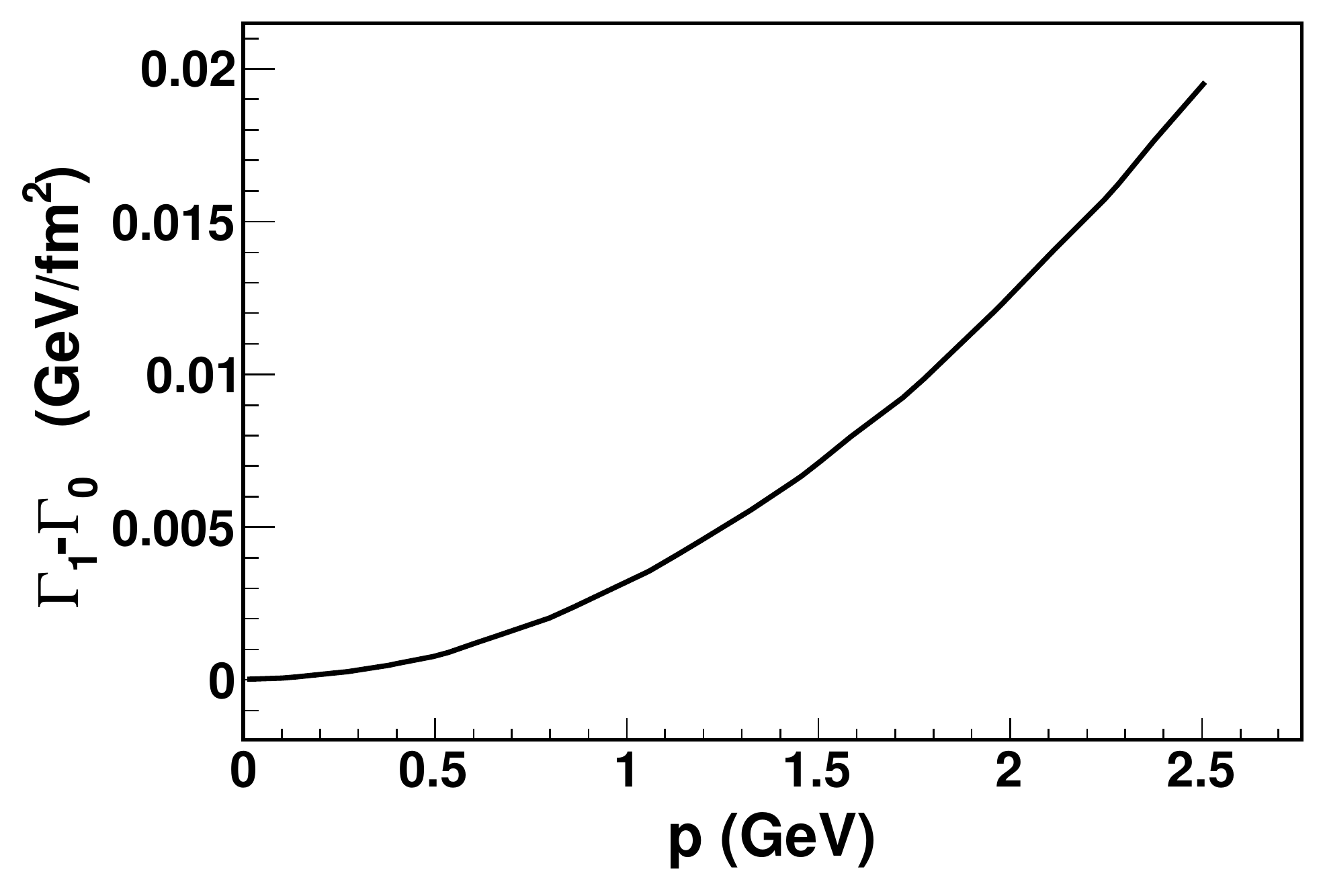}
\caption{\label{fig:momdiff2} 
Momentum-space transport coefficients as function of momentum
at fixed temperature 150 MeV. The two coefficients converge at low $p$.
}
\end{figure}

The equality of the two $\Gamma$ coefficients in the limit of zero momentum is numerically checked in
Figs.~\ref{fig:momdiff} and \ref{fig:momdiff2}.

The Langevin equation\index{Langevin equation} also allows us to directly obtain the classical interpretation of $F$ as a loss of energy per unit length. Ignoring the fluctuating force,
\be
\frac{d\gamma m{\bf v}}{dt} = -{\bf F}
\ee
can be multiplied by ${\bf v}$ to yield
\be
\frac{dm\gamma}{dt} = -{\bf F}\cd {\bf v} 
\ee
and remembering that ${\bf F}=F{\bf p}$ in Eq.~(\ref{eq:defFandG}), the loss of energy per unit length is simply $F|{\bf p}|$, as in the nonrelativistic theory.\\
The loss of momentum per unit length can then be expressed as
\be
\left| \frac{d{\bf p}}{dx} \right| = 
\left| \frac{d{\bf p}}{vdt} \right| =  F E
\ee
in terms of the energy and momentum of the charmed particle.

\chapter{Numerical Evaluation of the Collision Integral \label{app:numerical}}

In this Appendix we summarize the numerical procedure to calculate the collision integrals in Eqs.~(\ref{eq:collision00}), (\ref{eq:c22bulk}) and (\ref{eq:c11thermal}).
An analogous integral in Eq.~(\ref{eq:collstrange}) appears for the strangeness diffusion coefficient. The only difference is that the
masses of the incoming particles are different. 

The one-dimensional integrations needed for the functions (\ref{eq:Kintegrals}), (\ref{eq:Iintegrals}) and (\ref{eq:Jintegrals}) are estimated by a Gaussian quadrature method.

Consider the general multidimensional integral
\be \label{eq:collgeneral} \mathcal{C}_F =  \int \prod_{i=1}^4 \frac{d^3 k_i}{(2\pi)^32E_i} \overline{|T|^2} (2\pi)^4 \delta^{(4)} (k_1+k_2-k_3-p) \ F(k_1,k_2,k_3,p) \ , \ee
where $F(k_1,k_2,k_3,p)$ is an arbitrary function of the momenta (containing Bose-Einstein distribution functions). The three integrals in Eq.~(\ref{Transportintegrals}) are quite similar to that considered in (\ref{eq:collgeneral}) and the method to calculate them
is analogous but somehow simpler as they do not include integration in $p$. 

The masses of the particles will be denoted $m_1=m_3=m_{\pi}$ and $m_2=m_p=m_K$, the case for $\pi-\pi$ dispersion is
easily obtained by doing $m_K \rightarrow m_{\pi}$.

In principle, the integral (\ref{eq:collgeneral}) contains twelve integration variables but they will be reduced to five.
We start by considering the total momentum and total energy variables.
\begin{eqnarray}
 \mb{K} & = & \mb{k}_1 + \mb{k}_2 = \mb{k}_3 + \mb{p} \ , \\
  W & = & E_1 +E_2 = E_3 + E_p \ .
\end{eqnarray}

Without loss of generality take the direction of the total momentum along the $OZ$ axis:
\be \mb{K} = (0,0,K) \ . \ee
Then, take the outgoing momentum $\mb{p}$ to be in the $OZX$ plane, with a polar angle $\theta_p$ with respect to $\mb{K}$.
\be \mb{p} = (p \sin \theta_p,0,p \cos \theta_p) \ . \ee
By momentum conservation the form of $\mb{k}_3$ is fixed
\be \mb{k}_3 = \mb{K} - \mb{p} = (-p \sin \theta_p, 0 , K - p \cos \theta_p) \ . \ee
One of the incoming momenta, say $\mb{k}_1$ is completely arbitrary in space. We will call $\theta_1$ and $\varphi_1$ to its polar and azimuthal angles, respectively. 
\be \mb{k}_1 = (k_1 \sin \theta_1 \cos \varphi_1, k_1 \sin \theta_1 \sin \varphi_1, k_1 \cos \theta_1) \ . \ee
Finally, the last momentum $\mb{k}_2$ is fixed by momentum conservation:
\be \mb{k}_2 = (-k_1 \sin \theta_1 \cos \varphi_1, -k_1 \sin \theta_1 \sin \varphi_1, K-k_1 \cos \theta_1) \ . \ee

Returning to the integral (\ref{eq:collgeneral}), let us change the integration variables from $(\mb{k}_3,\mb{p})$ to $(\mb{K},\mb{p})$. The Jacobian of this change of variables is one.
Note that the Dirac's delta of three-momentum allows for a trivial integration of $\mb{k}_2$. The integral reads

\be \mathcal{C}_F =  \int  d\mb{K} d\mb{k}_1 d\mb{p} \   \frac{1}{(2\pi)^{12} \ 16E_1E_2E_3E_p} \overline{|T|^2} (2\pi)^4 \delta (E_1+E_2-E_3-E_p)   \ F(K,k_1,p) \ . \ee

Then perform the trivial integrations of the angular variables of $\mb{K}$ and the integration over the azimuthal angle of $\mb{p}$:
\begin{eqnarray} \mathcal{C}_F & = &  \int  dK K^2 (4\pi) \  dk_1 k_1^2 d(\cos \theta_1) d\varphi_1 \ dp p^2 d(\cos \theta_p) (2\pi) \\ 
 & & \times \frac{1}{(2\pi)^{8} 16E_1E_2E_3E_p} \overline{|T|^2} \ \delta (E_1+E_2-E_3-E_p)   \ F(K,k_1,p) \ . \nonumber \end{eqnarray}
For simplicity, let us call $x_1 \equiv \cos \theta_1$ and $x_p \equiv \cos \theta_p$. The introduction of the total energy variable can help us to easily perform the two integrations over $x_1$ and $x_p$.
The energy Dirac's delta is expressed as
\be \delta(E_1+E_2-E_3-E_p) = \int dW \ \delta(W-E_1-E_2) \delta(E_3+E_p-W) \ . \ee
Using the properties of the Dirac's delta one can obtain
\begin{eqnarray}
 \delta (W-E_1-E_2(x_1)) &= & \frac{E_2}{k_1 K} \delta(x_1- x_1^0) \ , \\
\delta (W-E_3(x_p)-E_p) & =& \frac{E_3}{pK} \delta(x_p-x_p^0) \ ,
\end{eqnarray}
where
\begin{eqnarray}
 x_1^0 &=& \frac{K^2+ W(2E_1-W)+m^2_K-m^2_{\pi}}{2k_1 K} \ , \\
 x_p^0 &=& \frac{K^2+ W(2E_p-W)-m^2_K+m^2_{\pi}}{2pK} \ . 
\end{eqnarray}

Eq.~(\ref{eq:collgeneral}) is reduced to a five-dimensional integral
\be \label{eq:resultingintegral} \mathcal{C}_F =   \frac{1}{(2\pi)^6} \frac{1}{8}\int  dW dK dk_1 dp\  d\varphi_1  \ \frac{k_1 p}{E_1E_p} \ \overline{|T|^2}  \ F(K,k_1,p) \ . \ee

Finally, we show the relation with the Mandelstam variables (they enter in the expression of the amplitude squared $|T|^2$):

\begin{eqnarray}
s & = & (E_1+E_2)^2 - (\mb{k}_1+\mb{k}_2)^2 = W^2 -K^2 \ , \\ 
t & =& (E_1 - E_3)^2 - (\mb{k}_1- \mb{k}_3)^2 = m^2_{\pi} + m^2_K - 2 E_1 E_3 + 2 k_1 k_3 y \ , \\
u & = & 2m_K^2+2m_{\pi}^2-t -s \ , 
\end{eqnarray}
where $y$ is the cosine of the angle between $\mb{k}_1$ and $\mb{k}_3$:
\be y = \frac{-p \cos \varphi_1 \sqrt{1-(x_1^0)^2} \sqrt{1-(x_p^0)^2}}{\sqrt{p^2 + K^2 - 2Kpx_p^0}} \ . \ee

\section{VEGAS}

The resulting integral (\ref{eq:resultingintegral}) is a five-dimensional over $W,K,k_1,p$ and $\varphi_1$ whose integrand is a very complicated function of these variables. 
The easiest way to perform the integration is to use a Monte Carlo integration routine. In our case we use VEGAS \cite{Lepage:1977sw,Lepage:1980dq}.
In a nutshell, it is an adaptative Monte Carlo method to compute multidimensional integrations. It is adaptative in the sense that the integration mesh is dynamically
modified after each iteration in a way that the random points concentrate where the integrand is maximum in absolute value. 

Suppose that our integral is represented by
\be \mathcal{C}_F = \int_{\Omega} d^5u f(\mb{u}) \ , \ee
where $\mb{u}=(W,K,k_1,p,\varphi_1)$ and $\Omega$ is the volume integral. The integral is aproximated by a sum:
\be \mathcal{C}_F \simeq S = \frac{\Omega}{N} \sum_{i=1}^N \frac{f(\mb{u}_i)}{p (\mb{u}_i)} \ , \ee
where $p(\mb{u})$ is a probability density function normalized to unity:
\be \int_{\Omega} d^5u \ p(\mb{u})= 1 \ . \ee
After $m$ iterations, where the number of random points $N$ is doubled, there are sucessive approximations to the integral, $S_j$, $j=1,...,m$. The weighted average corresponds to the estimate
of the integral:
\be \mathcal{C}_F \simeq \frac{\sum_{j=1}^m \frac{S_j}{\sigma_j^2}}{\sum_{j=1}^m \frac{1}{\sigma^2_j} } \ , \ee
where $\sigma^2_j$ represents the variance of the $S_j$ distribution. For large $m$ it reads:
\be \sigma_j^2 \simeq \frac{1}{N-1} \left[ \frac{\Omega^2}{N} \sum_{i=1}^N \left( \frac{f(\mb{u}_i)}{p(\mb{u}_i)} \right)^2- S^2_j\right] \ . \ee 

The value $\sigma_j$ measures the accuracy of $S_j$ as an approximation for the integral $\mathcal{C}_F$. After each iteration the probability density function $p(\mb{u})$ is modified 
in order to reduce the magnitude of $\sigma_j$. Theoretically, this factor is minimized when the probability density function satisfies:
\be p (\mb{u}) = \frac{|f(\mb{u})|}{\int_{\Omega} d^5 u |f(\mb{u})|} \ . \ee
Therefore the random points must be concentrated where the integrand is largest. More details of the VEGAS routine are given in \cite{Lepage:1977sw,Lepage:1980dq}.

\end{appendix}

\backmatter

\cleardoublepage
\addcontentsline{toc}{chapter}{Bibliography}
\bibliographystyle{alpha}
\bibliography{diss_torresrincon}

\cleardoublepage
\addcontentsline{toc}{chapter}{Index}
\printindex

\cleardoublepage
\addcontentsline{toc}{chapter}{Glossary}
\printglossary

\newpage{\pagestyle{empty}\cleardoublepage}


\end{document}